\documentclass[12pt,a4paper,dvips,fleqn]{report}

\usepackage{a4p}
\usepackage{graphicx}
\usepackage{epsfig}
\usepackage{rotating}
\usepackage{mcite}

\usepackage{multirow}
\usepackage{amsmath}

\usepackage{PhysRep}  %
\usepackage{Lep2Rep}  %

\newcommand{\ZFITTERref} {Bardin:1989di, *Bardin:1990tq, *Bardin:1991fu,
 *Bardin:1991de, *Bardin:1992jc, *Bardin:1999yd, *Arbuzov-1,
 *Arbuzov-2, *Arbuzov:2005ma, *ZFITTER643}
\newcommand{\YFSWWref}{Jadach:1995sp, *Jadach:1996hi, *Jadach:1998tz,
 *Jadach:2000kw, *Jadach:2001uu, *YFSWWack, 4f_bib:kandy}
\newcommand{\RACOONWWref}{Denner:1999gp, *Denner:1999kn, *Denner:2000bj,
 *Denner:2001zp, *RACOONWWack}
\newcommand{\KORALWref}{Skrzypek:1995ur, *Skrzypek:1995wd, *Jadach:1998gi,
 *KORALWack, 4f_bib:kandy} 
\newcommand{\WPHACTref}{Accomando:1996es, *Accomando:2002sz, *WPHACTack}
\newcommand{\GRACEref}{Fujimoto:1996qg, *Fujimoto:1996wj, *Kurihara:1999qz}
\newcommand{\WTOref}{Passarino:1996rc, *Passarino:1999zh, *Passarino:2000mt,
 *WTOack}

\newcommand{\Action}[1]%
{\fbox{\parbox{\linewidth}{\textbf{Action items:}\\#1}}}

\begin{document}

\begin{titlepage}
\begin{center}
\Large {EUROPEAN ORGANIZATION FOR NUCLEAR RESEARCH}
\end{center}
\vspace*{0.4cm}
\begin{flushright}
       CERN-PH-EP/2013-022 \\
       arXiv:1302.3415 [hep-ex] \\
       {\bf February 14th, 2013} \\
\end{flushright}

\vspace*{2cm}

\begin{center}

{\huge {\bf Electroweak Measurements in \\[5mm]
             Electron-Positron Collisions \\[5mm]
           at W-Boson-Pair Energies at LEP\\
}}

\vspace*{4cm}

{\Large {\bf  The ALEPH Collaboration \\
              The DELPHI Collaboration \\
	      The L3 Collaboration \\
	      The OPAL Collaboration \\
              The LEP Electroweak Working Group\footnote{ Web access at
                                 {\tt http://www.cern.ch/LEPEWWG}}\\[1mm]
}}

\vspace{\fill}

{\Large Submitted to PHYSICS REPORTS}

\vskip 1cm

{\Huge February 14th, 2013}

\end{center}

\clearpage

\begin{center}
{\bf Abstract}
\end{center}

Electroweak measurements performed with data taken at the
electron-positron collider LEP at CERN from 1995 to 2000 are reported.
The combined data set considered in this report corresponds to a total
luminosity of about 3~fb$^{-1}$ collected by the four LEP experiments
ALEPH, DELPHI, L3 and OPAL, at centre-of-mass energies ranging from
$130~\GeV$ to $209~\GeV$.

Combining the published results of the four LEP experiments, the
measurements include total and differential cross-sections in
photon-pair, fermion-pair and four-fermion production, the latter
resulting from both double-resonant WW and ZZ production as well as
singly resonant production.  Total and differential cross-sections are
measured precisely, providing a stringent test of the Standard Model
at centre-of-mass energies never explored before in electron-positron
collisions.  Final-state interaction effects in four-fermion
production, such as those arising from colour reconnection and
Bose-Einstein correlations between the two W decay systems arising in
WW production, are searched for and upper limits on the strength of
possible effects are obtained.  The data are used to determine
fundamental properties of the W boson and the electroweak theory.
Among others, the mass and width of the W boson, $\MW$ and $\GW$, the
branching fraction of W decays to hadrons, $B(W\to\mathrm{had})$, and
the trilinear gauge-boson self-couplings $g^Z_1$, $\kappa_\gamma$ and
$\lambda_\gamma$ are determined to be:

\begin{eqnarray*}
\MW                 & = &  80.376   \pm 0.033 ~\GeV \\
\GW                 & = &   2.195   \pm 0.083 ~\GeV \\
B(W\to\mathrm{had}) & = &  67.41    \pm 0.27  ~\%   \\
g^Z_1               & = &   0.984   ^{+0.018}_{-0.020} \\
\kappa_\gamma       & = &   0.982   \pm 0.042  \\
\lambda_\gamma      & = &  -0.022   \pm 0.019  \,.
\end{eqnarray*}

\vfill

\noindent
{\em Keywords:} Electron-positron physics, electroweak interactions,
decays of heavy intermediate gauge bosons, fermion-antifermion
production, precision measurements at W-pair energies, tests of the
Standard Model, radiative corrections, effective coupling constants,
neutral weak current, Z boson, W boson, top quark, Higgs boson.\\

\noindent
{\em PACS:} 12.15.-y, 13.38.-b, 13.66.-a, 14.60.-z, 14.65.-q,
14.70.-e, 14.80.-j.

\vfill

\end{titlepage}

\setcounter{page}{3}

\chapter*{Author Lists}
\label{app:author-list}

The ALEPH, DELPHI, L3, OPAL collaborations have provided the inputs
for the combined results presented in this Report.  The LEP
Electroweak Working Group has performed the combinations.  The Working
Group consists of members of the four LEP collaborations.  The lists
of authors from the collaborations follow.

\section*{The ALEPH Collaboration}

{

\newcommand{\aAachen}     { 1}
\newcommand{\aAnnecy}     { 2}
\newcommand{\aBarcelona}  { 3}
\newcommand{\aBari}       { 4}
\newcommand{\aBeijing}    { 5}
\newcommand{\aCERN}       { 6}
\newcommand{\aAubiere}    { 7}
\newcommand{\aCopenhagen} { 8}
\newcommand{\aAttiki}     { 9}
\newcommand{\aPalaiseau}  {10}
\newcommand{\aFirenze}    {11}
\newcommand{\aTallahasee} {12}
\newcommand{\aFrascati}   {13}
\newcommand{\aGlasgow}    {14}
\newcommand{\aOrem}       {15}
\newcommand{\aHeidelberg} {16}
\newcommand{\aLondon}     {17}
\newcommand{\aInnsbruck}  {18}
\newcommand{\aLancaster}  {19}
\newcommand{\aBelgium}    {20}
\newcommand{\aMainz}      {21}
\newcommand{\aMarseille}  {22}
\newcommand{\aMilano}     {23}
\newcommand{\aMunich}     {24}
\newcommand{\aParis}      {25}
\newcommand{\aPisa}       {26}
\newcommand{\aSurrey}     {27}
\newcommand{\aOxon}       {28}
\newcommand{\aYvette}     {29}
\newcommand{\aSantaCruz}  {30}
\newcommand{\aSheffield}  {31}
\newcommand{\aSiegen}     {32}
\newcommand{\aTrieste}    {33}
\newcommand{\aWashington} {34}
\newcommand{\aWisconsin}  {35}
\newcommand{\aZurich}     {36}

\tolerance=10000
\hbadness=5000
\raggedright
\raggedbottom
\sloppy

\noindent
S.\thinspace Schael,$\!^{\aAachen}$
\nopagebreak
R.\thinspace Barate,$\!^{\aAnnecy}$
R.\thinspace Bruneli\`ere,$\!^{\aAnnecy}$
D.\thinspace Buskulic,$\!^{\aAnnecy}$
I.\thinspace De\thinspace Bonis,$\!^{\aAnnecy}$
D.\thinspace Decamp,$\!^{\aAnnecy}$
P.\thinspace Ghez,$\!^{\aAnnecy}$
C.\thinspace Goy,$\!^{\aAnnecy}$
S.\thinspace J\'ez\'equel,$\!^{\aAnnecy}$
J.-P.\thinspace Lees,$\!^{\aAnnecy}$
% ...

\bigskip

\newcommand{\AlInst}[2]{\item[$^{#1}$] {#2}}

\begin{list}{A}{\itemsep=0pt plus 0pt minus 0pt\parsep=0pt plus 0pt minus 0pt
                \topsep=0pt plus 0pt minus 0pt}
\AlInst{\aAachen}{
Physikalisches Institut der RWTH-Aachen, D-52056 Aachen, Germany}
\AlInst{\aAnnecy}{
Laboratoire de Physique des Particules (LAPP), IN$^{2}$P$^{3}$-CNRS,
F-74019 Annecy-le-Vieux Cedex, France}
\AlInst{\aBarcelona}{
Institut de F\'{i}sica d'Altes Energies, Universitat Aut\`{o}noma
de Barcelona, E-08193 Bellaterra (Barcelona), Spain$^{a7}$}
\AlInst{\aBari}{
Dipartimento di Fisica, INFN Sezione di Bari, I-70126 Bari, Italy}
\AlInst{\aBeijing}{
Institute of High Energy Physics, Academia Sinica, Beijing, The People's
Republic of China$^{a8}$}
\AlInst{\aCERN}{
European Laboratory for Particle Physics (CERN), CH-1211 Geneva 23,
Switzerland}
\AlInst{\aAubiere}{
Laboratoire de Physique Corpusculaire, Universit\'e Blaise Pascal,
IN$^{2}$P$^{3}$-CNRS, Clermont-Ferrand, F-63177 Aubi\`{e}re, France}
\AlInst{\aCopenhagen}{
Niels Bohr Institute, 2100 Copenhagen, DK-Denmark$^{a9}$}
\AlInst{\aAttiki}{
Nuclear Research Center Demokritos (NRCD), GR-15310 Attiki, Greece}
\AlInst{\aPalaiseau}{
Laoratoire Leprince-Ringuet, Ecole Polytechnique, IN$^{2}$P$^{3}$-CNRS,
\mbox{F-91128} Palaiseau Cedex, France}
\AlInst{\aFirenze}{
Dipartimento di Fisica, Universit\`a di Firenze, INFN Sezione di Firenze,
I-50125 Firenze, Italy}
\AlInst{\aTallahasee}{
Supercomputer Computations Research Institute, Florida State University,
Tallahasee, FL-32306-4052, USA}
\AlInst{\aFrascati}{
Laboratori Nazionali dell'INFN (LNF-INFN), I-00044 Frascati, Italy}
\AlInst{\aGlasgow}{
Department of Physics and Astronomy, University of Glasgow,
Glasgow G12 8QQ,United Kingdom$^{a10}$}
\AlInst{\aOrem}{
Utah Valley State College, Orem, UT 84058, U.S.A.}
\AlInst{\aHeidelberg}{
Kirchhoff-Institut f\"ur Physik, Universit\"at Heidelberg, D-69120
Heidelberg, Germany$^{a16}$}
\AlInst{\aLondon}{
Department of Physics, Imperial College, London SW7 2BZ,
United Kingdom$^{a10}$}
\AlInst{\aInnsbruck}{
Institut f\"ur Experimentalphysik, Universit\"at Innsbruck, A-6020
Innsbruck, Austria$^{a18}$}
\AlInst{\aLancaster}{
Department of Physics, University of Lancaster, Lancaster LA1 4YB,
United Kingdom$^{a10}$}
\AlInst{\aBelgium}{
Institut de Physique Nucl\'eaire, D\'epartement de Physique,
Universit\'e Catholique de Louvain, 1348 Louvain-la-Neuve, Belgium}
\AlInst{\aMainz}{
Institut f\"ur Physik, Universit\"at Mainz, D-55099 Mainz, Germany$^{a16}$}
\AlInst{\aMarseille}{
Centre de Physique des Particules de Marseille, Univ M\'editerran\'ee,
IN$^{2}$P$^{3}$-CNRS, F-13288 Marseille, France}
\AlInst{\aMilano}{
Dipartimento di Fisica, Universit\`a di Milano e INFN Sezione di
Milano, I-20133 Milano, Italy.}
\AlInst{\aMunich}{
Max-Planck-Institut f\"ur Physik, Werner-Heisenberg-Institut,
D-80805 M\"unchen, Germany$^{a16}$}
\AlInst{\aParis}{
Laboratoire de l'Acc\'el\'erateur Lin\'eaire, Universit\'e de Paris-Sud,
IN$^{2}$P$^{3}$-CNRS, F-91898 Orsay Cedex, France}
\AlInst{\aPisa}{
Dipartimento di Fisica dell'Universit\`a, INFN Sezione di Pisa,
e Scuola Normale Superiore, I-56010 Pisa, Italy}
\AlInst{\aSurrey}{
Department of Physics, Royal Holloway \& Bedford New College,
University of London, Egham, Surrey TW20 OEX, United Kingdom$^{a10}$}
\AlInst{\aOxon}{
Particle Physics Dept., Rutherford Appleton Laboratory,
Chilton, Didcot, Oxon OX11 OQX, United Kingdom$^{a10}$}
\AlInst{\aYvette}{
CEA, DAPNIA/Service de Physique des Particules,
CE-Saclay, F-91191 Gif-sur-Yvette Cedex, France$^{a17}$}
\AlInst{\aSantaCruz}{
Institute for Particle Physics, University of California at
Santa Cruz, Santa Cruz, CA 95064, USA$^{a22}$}
\AlInst{\aSheffield}{
Department of Physics, University of Sheffield, Sheffield S3 7RH,
United Kingdom$^{a10}$}
\AlInst{\aSiegen}{
Fachbereich Physik, Universit\"at Siegen, D-57068 Siegen, Germany$^{a16}$}
\AlInst{\aTrieste}{
Dipartimento di Fisica, Universit\`a di Trieste e INFN Sezione di Trieste,
I-34127 Trieste, Italy}
\AlInst{\aWashington}{
Experimental Elementary Particle Physics, University of Washington, Seattle,
WA 98195 U.S.A.}
\AlInst{\aWisconsin}{
Department of Physics, University of Wisconsin, Madison, WI 53706,
USA$^{a11}$}
\AlInst{\aZurich}{
Institute for Particle Physics, ETH H\"onggerberg, 8093 Z\"urich,
Switzerland.}
\end{list}

\bigskip

\begin{list}{A}{\itemsep=0pt plus 0pt minus 0pt\parsep=0pt plus 0pt minus 0pt
                \topsep=0pt plus 0pt minus 0pt}
\AlInst{a1}{Also at CERN, 1211 Geneva 23, Switzerland.}
\AlInst{a2}{Now at Fermilab, PO Box 500, MS 352, Batavia, IL 60510, USA}
\AlInst{a3}{Also at Dipartimento di Fisica di Catania and INFN Sezione di
 Catania, 95129 Catania, Italy.}
\AlInst{a4}{Now at University of Florida, Department of Physics, Gainesville, Florida 32611-8440, USA}
\AlInst{a5}{Also IFSI sezione di Torino, INAF, Italy.}
\AlInst{a6}{Also at Groupe d'Astroparticules de Montpellier, Universit\'{e} de Montpellier II, 34095, Montpellier, France.}
\AlInst{a7}{Supported by CICYT, Spain.}
\AlInst{a8}{Supported by the National Science Foundation of China.}
\AlInst{a9}{Supported by the Danish Natural Science Research Council.}
\AlInst{a10}{Supported by the UK Particle Physics and Astronomy Research
Council.}
\AlInst{a11}{Supported by the US Department of Energy, grant
DE-FG0295-ER40896.}
\AlInst{a12}{Now at Departement de Physique Corpusculaire, Universit\'e de
Gen\`eve, 1211 Gen\`eve 4, Switzerland.}
\AlInst{a13}{Also at Department of Physics, Tsinghua University, Beijing, The People's Republic of China.}
\AlInst{a14}{Supported by the Leverhulme Trust.}
\AlInst{a15}{Permanent address: Universitat de Barcelona, 08208 Barcelona,
Spain.}
\AlInst{a16}{Supported by Bundesministerium f\"ur Bildung
und Forschung, Germany.}
\AlInst{a17}{Supported by the Direction des Sciences de la
Mati\`ere, C.E.A.}
\AlInst{a18}{Supported by the Austrian Ministry for Science and Transport.}
\AlInst{a19}{Now at SAP AG, 69185 Walldorf, Germany}
\AlInst{a20}{Now at Groupe d' Astroparticules de Montpellier, Universit\'e de Montpellier II, 34095 Montpellier, France.}
\AlInst{a21}{Now at BNP Paribas, 60325 Frankfurt am Mainz, Germany}
\AlInst{a22}{Supported by the US Department of Energy,
grant DE-FG03-92ER40689.}
\AlInst{a23}{Now at Institut Inter-universitaire des hautes Energies (IIHE), CP 230, Universit\'{e} Libre de Bruxelles, 1050 Bruxelles, Belgique}
\AlInst{a24}{Now at Dipartimento di Fisica e Tecnologie Relative, Universit\`a di Palermo, Palermo, Italy.}
\AlInst{a25}{Deceased.}
\AlInst{a26}{Now at SLAC, Stanford, CA 94309, U.S.A}
\AlInst{a27}{Now at CERN, 1211 Geneva 23, Switzerland}
\AlInst{a28}{Research Fellow of the Belgium FNRS}
\AlInst{a29}{Research Associate of the Belgium FNRS}
\AlInst{a30}{Now at Liverpool University, Liverpool L69 7ZE, United Kingdom}
\AlInst{a31}{Supported by the Federal Office for Scientific, Technical and Cultural Affairs through
the Interuniversity Attraction Pole P5/27}
\AlInst{a32}{Now at Henryk Niewodnicznski Institute of Nuclear Physics, Polish Academy of Sciences, Cracow, Poland}
\AlInst{a33}{Supported by the Commission of the European Communities, contract
ERBFMBICT982894}
\AlInst{a34}{Also Istituto di Fisica Generale, Universit\`a di Torino, 10125 Torino, Italy}

\end{list}

}

\section*{The DELPHI Collaboration}

{

\newcommand{\DpName}[2]{\hbox{#1,$\!^{#2}$}}
\newcommand{\DpNameTwo}[3]{\hbox{#1,$\!^{{#2},{#3}}$}}
\newcommand{\DpNameThree}[4]{\hbox{#1,$\!^{{#2},{#3},{#4}}$}}
\newskip\Bigfill \Bigfill = 0pt plus 1000fill
\newcommand{\DpNameLast}[2]{\hbox{#1.$\!^{#2}$}\hspace{\Bigfill}}

\tolerance=10000
\hbadness=5000
\raggedright

\newcommand{\dAMES}{1}
\newcommand{\dANTWERP}{2}
\newcommand{\dBRUSSELS}{3}
\newcommand{\dATHENS}{4}
\newcommand{\dBERGEN}{5}
\newcommand{\dBOLOGNA}{6}
\newcommand{\dBRASILCBPF}{7}
\newcommand{\dBRASILIFUERJ}{8}
\newcommand{\dCDF}{9}
\newcommand{\dCERN}{10}
\newcommand{\dCRN}{11}
\newcommand{\dDESY}{12}
\newcommand{\dDEMOKRITOS}{13}
\newcommand{\dFZU}{14}
\newcommand{\dGENOVA}{15}
\newcommand{\dGRENOBLE}{16}
\newcommand{\dHELSINKI}{17}
\newcommand{\dJINR}{18}
\newcommand{\dKARLSRUHE}{19}
\newcommand{\dKRAKOWone}{20}
\newcommand{\dKRAKOWtwo}{21}
\newcommand{\dLAL}{22}
\newcommand{\dLANCASTER}{23}
\newcommand{\dLIP}{24}
\newcommand{\dLIVERPOOL}{25}
\newcommand{\dGLASGOW}{26}
\newcommand{\dLPNHE}{27}
\newcommand{\dLUND}{28}
\newcommand{\dLYON}{29}
\newcommand{\dMILANO}{30}
\newcommand{\dMILANOtwo}{31}
\newcommand{\dNC}{32}
\newcommand{\dNIKHEF}{33}
\newcommand{\dNTUATHENS}{34}
\newcommand{\dOSLO}{35}
\newcommand{\dOVIEDO}{36}
\newcommand{\dOXFORD}{37}
\newcommand{\dPADOVA}{38}
\newcommand{\dRAL}{39}
\newcommand{\dROMAtwo}{40}
\newcommand{\dROMAtre}{41}
\newcommand{\dSACLAY}{42}
\newcommand{\dSANTANDER}{43}
\newcommand{\dSERPUKHOV}{44}
\newcommand{\dSLOVENIJAone}{45}
\newcommand{\dSLOVENIJAtwo}{46}
\newcommand{\dSLOVENIJAtre}{47}
\newcommand{\dSTOCKHOLM}{48}
\newcommand{\dTORINO}{49}
\newcommand{\dTORINOTH}{50}
\newcommand{\dTRIESTE}{51}
\newcommand{\dUDINE}{52}
\newcommand{\dUFRJ}{53}
\newcommand{\dUPPSALA}{54}
\newcommand{\dVALENCIA}{55}
\newcommand{\dVIENNA}{56}
\newcommand{\dWARSZAWA}{57}
\newcommand{\dWARWICK}{58}
\newcommand{\dWUPPERTAL}{59}
\newcommand{\dZURICH}{60}

\noindent
\DpName{J.\thinspace Abdallah}{\dLPNHE}
\DpName{P.\thinspace Abreu}{\dLIP}
\DpName{W.\thinspace Adam}{\dVIENNA}
\DpName{P.\thinspace Adzic}{\dDEMOKRITOS}
\DpName{T.\thinspace Albrecht}{\dKARLSRUHE}
\DpName{R.\thinspace Alemany-Fernandez}{\dCERN}
\DpName{T.\thinspace Allmendinger}{\dKARLSRUHE}
\DpName{P.P.\thinspace Allport}{\dLIVERPOOL}
\DpName{U.\thinspace Amaldi}{\dMILANOtwo}
\DpName{N.\thinspace Amapane}{\dTORINO}
% ...

\bigskip

\newcommand{\DpInst}[2]{\item[$^{#2}$] {#1}}

\begin{list}{A}{\itemsep=0pt plus 0pt minus 0pt\parsep=0pt plus 0pt minus 0pt
                \topsep=0pt plus 0pt minus 0pt}
\DpInst{Department of Physics and Astronomy, Iowa State
     University, Ames IA 50011-3160, USA
    }{\dAMES}
\DpInst{Physics Department, Universiteit Antwerpen,
     Universiteitsplein 1, B-2610 Antwerpen, Belgium
    }{\dANTWERP}
\DpInst{IIHE, ULB-VUB,
     Pleinlaan 2, B-1050 Brussels, Belgium
    }{\dBRUSSELS}
\DpInst{Physics Laboratory, University of Athens, Solonos Str.
     104, GR-10680 Athens, Greece
    }{\dATHENS}
\DpInst{Department of Physics, University of Bergen,
     All\'egaten 55, NO-5007 Bergen, Norway
    }{\dBERGEN}
\DpInst{Dipartimento di Fisica, Universit\`a di Bologna and INFN,
     Viale C. Berti Pichat 6/2, IT-40127 Bologna, Italy
    }{\dBOLOGNA}
\DpInst{Centro Brasileiro de Pesquisas F\'{\i}sicas, rua Xavier Sigaud 150,
     BR-22290 Rio de Janeiro, Brazil
    }{\dBRASILCBPF}
\DpInst{Inst. de F\'{\i}sica, Univ. Estadual do Rio de Janeiro,
     rua S\~{a}o Francisco Xavier 524, Rio de Janeiro, Brazil
    }{\dBRASILIFUERJ}
\DpInst{Coll\`ege de France, Lab. de Physique Corpusculaire, IN2P3-CNRS,
     FR-75231 Paris Cedex 05, France
    }{\dCDF}
\DpInst{CERN, CH-1211 Geneva 23, Switzerland
    }{\dCERN}
\DpInst{Institut Pluridisciplinaire Hubert Curien, Universit\'e de Strasbourg,
     IN2P3-CNRS, BP28, FR-67037 Strasbourg \indent~~Cedex~2, France
    }{\dCRN}
\DpInst{Now at DESY-Zeuthen, Platanenallee 6, D-15735 Zeuthen, Germany
    }{\dDESY}
\DpInst{Institute of Nuclear Physics, N.C.S.R. Demokritos,
     P.O. Box 60228, GR-15310 Athens, Greece
    }{\dDEMOKRITOS}
\DpInst{FZU, Inst. of Phys. of the C.A.S. High Energy Physics Division,
     Na Slovance 2, CZ-182 21, Praha 8, Czech Republic
    }{\dFZU}
\DpInst{Dipartimento di Fisica, Universit\`a di Genova and INFN,
     Via Dodecaneso 33, IT-16146 Genova, Italy
    }{\dGENOVA}
\DpInst{Laboratoire de Physique Subatomique et de Cosmologie, Universit\'e
     Joseph Fourier Grenoble 1, CNRS/IN2P3, Institut Polytechnique de Grenoble,
     FR-38026 Grenoble Cedex, France
    }{\dGRENOBLE}
\DpInst{Helsinki Institute of Physics and Department of Physics,
     P.O. Box 64, FIN-00014 University of Helsinki, Finland
    }{\dHELSINKI}
\DpInst{Joint Institute for Nuclear Research, Dubna, Head Post
     Office, P.O. Box 79, RU-101 000 Moscow, Russian Federation
    }{\dJINR}
\DpInst{Institut f\"ur Experimentelle Kernphysik,
     Universit\"at Karlsruhe, Postfach 6980, DE-76128 Karlsruhe,
     Germany
    }{\dKARLSRUHE}
\DpInst{Institute of Nuclear Physics PAN,Ul. Radzikowskiego 152,
     PL-31142 Krakow, Poland
    }{\dKRAKOWone}
\DpInst{Faculty of Physics and Nuclear Techniques, University of Mining
     and Metallurgy, PL-30055 Krakow, Poland
    }{\dKRAKOWtwo}
\DpInst{LAL, Univ Paris-Sud, CNRS/IN2P3, Orsay, France
    }{\dLAL}
\DpInst{School of Physics and Chemistry, University of Lancaster,
     Lancaster LA1 4YB, UK
    }{\dLANCASTER}
\DpInst{LIP, IST, FCUL - Av. Elias Garcia, 14-$1^{o}$,
     PT-1000 Lisboa Codex, Portugal
    }{\dLIP}
\DpInst{Department of Physics, University of Liverpool, P.O.
     Box 147, Liverpool L69 3BX, UK
    }{\dLIVERPOOL}
\DpInst{Dept. of Physics and Astronomy, Kelvin Building,
     University of Glasgow, Glasgow G12 8QQ, UK
    }{\dGLASGOW}
\DpInst{LPNHE, IN2P3-CNRS, Univ.~Paris VI et VII,
     4 place Jussieu, FR-75252 Paris Cedex 05, France
    }{\dLPNHE}
\DpInst{Department of Physics, University of Lund,
     S\"olvegatan 14, SE-223 63 Lund, Sweden
    }{\dLUND}
\DpInst{Universit\'e Claude Bernard de Lyon, IPNL, IN2P3-CNRS,
     FR-69622 Villeurbanne Cedex, France
    }{\dLYON}
\DpInst{Dipartimento di Fisica, Universit\`a di Milano and INFN-MILANO,
     Via Celoria 16, IT-20133 Milan, Italy
    }{\dMILANO}
\DpInst{Dipartimento di Fisica, Univ. di Milano-Bicocca and
     INFN-MILANO, Piazza della Scienza 3, IT-20126 Milan, Italy
    }{\dMILANOtwo}
\DpInst{IPNP of MFF, Charles Univ., Areal MFF,
     V Holesovickach 2, CZ-180 00, Praha 8, Czech Republic
    }{\dNC}
\DpInst{NIKHEF, Postbus 41882, NL-1009 DB
     Amsterdam, The Netherlands
    }{\dNIKHEF}
\DpInst{National Technical University, Physics Department,
     Zografou Campus, GR-15773 Athens, Greece
    }{\dNTUATHENS}
\DpInst{Physics Department, University of Oslo, Blindern,
     NO-0316 Oslo, Norway
    }{\dOSLO}
\DpInst{Dpto. Fisica, Univ. Oviedo, Avda. Calvo Sotelo
     s/n, ES-33007 Oviedo, Spain
    }{\dOVIEDO}
\DpInst{Department of Physics, University of Oxford,
     Keble Road, Oxford OX1 3RH, UK
    }{\dOXFORD}
\DpInst{Dipartimento di Fisica, Universit\`a di Padova and
     INFN, Via Marzolo 8, IT-35131 Padua, Italy
    }{\dPADOVA}
\DpInst{Rutherford Appleton Laboratory, Chilton, Didcot
     OX11 OQX, UK
    }{\dRAL}
\DpInst{Dipartimento di Fisica, Universit\`a di Roma II and
     INFN, Tor Vergata, IT-00173 Rome, Italy
    }{\dROMAtwo}
\DpInst{Dipartimento di Fisica, Universit\`a di Roma III and
     INFN, Via della Vasca Navale 84, IT-00146 Rome, Italy
    }{\dROMAtre}
\DpInst{DAPNIA/Service de Physique des Particules,
     CEA-Saclay, FR-91191 Gif-sur-Yvette Cedex, France
    }{\dSACLAY}
\DpInst{Instituto de Fisica de Cantabria (CSIC-UC), Avda.
     los Castros s/n, ES-39006 Santander, Spain
    }{\dSANTANDER}
\DpInst{Institute for high energy physics, 
     142281 Protvino, Moscow region, Russian Federation
    }{\dSERPUKHOV}
\DpInst{J. Stefan Institute, Jamova 39, SI-1000 Ljubljana, Slovenia
    }{\dSLOVENIJAone}
\DpInst{Laboratory for Astroparticle Physics,
     University of Nova Gorica, Kostanjeviska 16a, SI-5000 Nova Gorica, Slovenia
    }{\dSLOVENIJAtwo}
\DpInst{Department of Physics, University of Ljubljana,
     SI-1000 Ljubljana, Slovenia
    }{\dSLOVENIJAtre}
\DpInst{Fysikum, Stockholm University,
     Box 6730, SE-113 85 Stockholm, Sweden
    }{\dSTOCKHOLM}
\DpInst{Dipartimento di Fisica Sperimentale, Universit\`a di
     Torino and INFN, Via P. Giuria 1, IT-10125 Turin, Italy
    }{\dTORINO}
\DpInst{INFN,Sezione di Torino and Dipartimento di Fisica Teorica,
     Universit\`a di Torino, Via Giuria 1,
     IT-10125 Turin, Italy
    }{\dTORINOTH}
\DpInst{Dipartimento di Fisica, Universit\`a di Trieste and
     INFN, Via A. Valerio 2, IT-34127 Trieste, Italy
    }{\dTRIESTE}
\DpInst{Istituto di Fisica, Universit\`a di Udine and INFN,
     IT-33100 Udine, Italy
    }{\dUDINE}
\DpInst{Univ. Federal do Rio de Janeiro, C.P. 68528
     Cidade Univ., Ilha do Fund\~ao
     BR-21945-970 Rio de Janeiro, Brazil
    }{\dUFRJ}
\DpInst{Department of Radiation Sciences, University of
     Uppsala, P.O. Box 535, SE-751 21 Uppsala, Sweden
    }{\dUPPSALA}
\DpInst{IFIC, Valencia-CSIC, and D.F.A.M.N., U. de Valencia,
     Avda. Dr. Moliner 50, ES-46100 Burjassot (Valencia), Spain
    }{\dVALENCIA}
\DpInst{Institut f\"ur Hochenergiephysik, \"Osterr. Akad.
     d. Wissensch., Nikolsdorfergasse 18, AT-1050 Vienna, Austria
    }{\dVIENNA}
\DpInst{Inst. Nuclear Studies and University of Warsaw, Ul.
     Hoza 69, PL-00681 Warsaw, Poland
    }{\dWARSZAWA}
\DpInst{Now at University of Warwick, Coventry CV4 7AL, UK
    }{\dWARWICK}
\DpInst{Fachbereich Physik, University of Wuppertal, Postfach
     100 127, DE-42097 Wuppertal, Germany 
    }{\dWUPPERTAL}
\DpInst{Now at Physik-Institut der Universit\"at Z\"urich, Z\"urich,
     Switzerland 
    }{\dZURICH}
\DpInst{Deceased}{\dagger}
\end{list}
}

\section*{The L3 Collaboration}

{
\newcount\tutecount  \tutecount=0
\def\tutenum#1{\global\advance\tutecount by 1 \xdef#1{\the\tutecount}}
\def\tute#1{$^{#1}$}

\tutenum\aachen            %
\tutenum\nikhef            %
\tutenum\mich              %
\tutenum\lapp              %
\tutenum\basel             %
\tutenum\lsu               %
\tutenum\beijing           %
\tutenum\bologna           %
\tutenum\tata              %
\tutenum\ne                %
\tutenum\bucharest         %
\tutenum\budapest          %
\tutenum\mit               %
\tutenum\panjab            %
\tutenum\debrecen          %
\tutenum\dublin            %
\tutenum\florence          %
\tutenum\cern              %
\tutenum\wl                %
\tutenum\geneva            %
\tutenum\hamburg           %
\tutenum\hefei             %
\tutenum\lausanne          %
\tutenum\lyon              %
\tutenum\madrid            %
\tutenum\florida           %
\tutenum\milan             %
\tutenum\moscow            %
\tutenum\naples            %
\tutenum\cyprus            %
\tutenum\nymegen           %
\tutenum\caltech           %
\tutenum\perugia           %
\tutenum\peters            %
\tutenum\cmu               %
\tutenum\potenza           %
\tutenum\prince            %
\tutenum\riverside         %
\tutenum\rome              %
\tutenum\salerno           %
\tutenum\ucsd              %
\tutenum\sofia             %
\tutenum\korea             %
\tutenum\taiwan            %
\tutenum\tsinghua          %
\tutenum\purdue            %
\tutenum\psinst            %
\tutenum\zeuthen           %
\tutenum\eth               %

{
\tolerance=10000
\hbadness=5000
\raggedright
\def\r{\rlap,}
\noindent

P.Achard\r\tute\geneva\ 
O.Adriani\r\tute{\florence}\ 
M.Aguilar-Benitez\r\tute\madrid\ 
J.Alcaraz\r\tute{\madrid}\ 
G.Alemanni\r\tute\lausanne\
J.Allaby\r\tute\cern\
A.Aloisio\r\tute\naples\ 
M.G.Alviggi\r\tute\naples\
H.Anderhub\r\tute\eth\ 
V.P.Andreev\r\tute{\lsu,\peters}\
% ...

\bigskip

\begin{list}{A}{\itemsep=0pt plus 0pt minus 0pt\parsep=0pt plus 0pt minus 0pt
                \topsep=0pt plus 0pt minus 0pt}
\item[$^{\aachen}$]
 III. Physikalisches Institut, RWTH, D-52056 Aachen, Germany$^{\S}$
\item[$^{\nikhef}$] National Institute for High Energy Physics, NIKHEF, 
     and University of Amsterdam, NL-1009 DB Amsterdam, The Netherlands
\item[$^{\mich}$] University of Michigan, Ann Arbor, MI 48109, USA
\item[$^{\lapp}$] Laboratoire d'Annecy-le-Vieux de Physique des Particules, 
     LAPP,IN2P3-CNRS, BP 110, F-74941 Annecy-le-Vieux CEDEX, France
\item[$^{\basel}$] Institute of Physics, University of Basel, CH-4056 Basel,
     Switzerland
\item[$^{\lsu}$] Louisiana State University, Baton Rouge, LA 70803, USA
\item[$^{\beijing}$] Institute of High Energy Physics, IHEP, 
  100039 Beijing, China$^{\triangle}$ 
\item[$^{\bologna}$] University of Bologna and INFN-Sezione di Bologna, 
     I-40126 Bologna, Italy
\item[$^{\tata}$] Tata Institute of Fundamental Research, Mumbai (Bombay) 400 005, India
\item[$^{\ne}$] Northeastern University, Boston, MA 02115, USA
\item[$^{\bucharest}$] Institute of Atomic Physics and University of Bucharest,
     R-76900 Bucharest, Romania
\item[$^{\budapest}$] Central Research Institute for Physics of the 
     Hungarian Academy of Sciences, H-1525 Budapest 114, Hungary$^{\ddag}$
\item[$^{\mit}$] Massachusetts Institute of Technology, Cambridge, MA 02139, USA
\item[$^{\panjab}$] Panjab University, Chandigarh 160 014, India
\item[$^{\debrecen}$] KLTE-ATOMKI, H-4010 Debrecen, Hungary$^{\P}$
\item[$^{\dublin}$] UCD School of Physics, University College Dublin, 
 Belfield, Dublin 4, Ireland
\item[$^{\florence}$] INFN Sezione di Firenze and University of Florence, 
     I-50125 Florence, Italy
\item[$^{\cern}$] European Laboratory for Particle Physics, CERN, 
     CH-1211 Geneva 23, Switzerland
\item[$^{\wl}$] World Laboratory, FBLJA  Project, CH-1211 Geneva 23, Switzerland
\item[$^{\geneva}$] University of Geneva, CH-1211 Geneva 4, Switzerland
\item[$^{\hamburg}$] University of Hamburg, D-22761 Hamburg, Germany
\item[$^{\hefei}$] Chinese University of Science and Technology, USTC,
      Hefei, Anhui 230 029, China$^{\triangle}$
\item[$^{\lausanne}$] University of Lausanne, CH-1015 Lausanne, Switzerland
\item[$^{\lyon}$] Institut de Physique Nucl\'eaire de Lyon, 
     IN2P3-CNRS,Universit\'e Claude Bernard, 
     F-69622 Villeurbanne, France
\item[$^{\madrid}$] Centro de Investigaciones Energ{\'e}ticas, 
     Medioambientales y Tecnol\'ogicas, CIEMAT, E-28040 Madrid,
     Spain${\flat}$ 
\item[$^{\florida}$] Florida Institute of Technology, Melbourne, FL 32901, USA
\item[$^{\milan}$] INFN-Sezione di Milano, I-20133 Milan, Italy
\item[$^{\moscow}$] Institute of Theoretical and Experimental Physics, ITEP, 
     Moscow, Russia
\item[$^{\naples}$] INFN-Sezione di Napoli and University of Naples, 
     I-80125 Naples, Italy
\item[$^{\cyprus}$] Department of Physics, University of Cyprus,
     Nicosia, Cyprus
\item[$^{\nymegen}$] Radboud University and NIKHEF, 
     NL-6525 ED Nijmegen, The Netherlands
\item[$^{\caltech}$] California Institute of Technology, Pasadena, CA 91125, USA
\item[$^{\perugia}$] INFN-Sezione di Perugia and Universit\`a Degli 
     Studi di Perugia, I-06100 Perugia, Italy   
\item[$^{\peters}$] Nuclear Physics Institute, St. Petersburg, Russia
\item[$^{\cmu}$] Carnegie Mellon University, Pittsburgh, PA 15213, USA
\item[$^{\potenza}$] INFN-Sezione di Napoli and University of Potenza, 
     I-85100 Potenza, Italy
\item[$^{\prince}$] Princeton University, Princeton, NJ 08544, USA
\item[$^{\riverside}$] University of Californa, Riverside, CA 92521, USA
\item[$^{\rome}$] INFN-Sezione di Roma and University of Rome, ``La Sapienza",
     I-00185 Rome, Italy
\item[$^{\salerno}$] University and INFN, Salerno, I-84100 Salerno, Italy
\item[$^{\ucsd}$] University of California, San Diego, CA 92093, USA
\item[$^{\sofia}$] Bulgarian Academy of Sciences, Central Lab.~of 
     Mechatronics and Instrumentation, BU-1113 Sofia, Bulgaria
\item[$^{\korea}$]  The Center for High Energy Physics, 
     Kyungpook National University, 702-701 Taegu, Republic of Korea
\item[$^{\taiwan}$] National Central University, Chung-Li, Taiwan, China
\item[$^{\tsinghua}$] Department of Physics, National Tsing Hua University,
      Taiwan, China
\item[$^{\purdue}$] Purdue University, West Lafayette, IN 47907, USA
\item[$^{\psinst}$] Paul Scherrer Institut, PSI, CH-5232 Villigen, Switzerland
\item[$^{\zeuthen}$] DESY, D-15738 Zeuthen, Germany
\item[$^{\eth}$] Eidgen\"ossische Technische Hochschule, ETH Z\"urich,
     CH-8093 Z\"urich, Switzerland

\end{list}

\bigskip

\begin{list}{A}{\itemsep=0pt plus 0pt minus 0pt\parsep=0pt plus 0pt minus 0pt
                \topsep=0pt plus 0pt minus 0pt}
\item[$^{\S}$]  Supported by the German Bundesministerium 
        f\"ur Bildung, Wissenschaft, Forschung und Technologie.
\item[$^{\ddag}$] Supported by the Hungarian OTKA fund under contract
numbers T019181, F023259 and T037350.
\item[$^{\P}$] Also supported by the Hungarian OTKA fund under contract
  number T026178.
\item[$^{\flat}$] Supported also by the Comisi\'on Interministerial de Ciencia y 
        Tecnolog{\'\i}a.
\item[$^{\sharp}$] Also supported by CONICET and Universidad Nacional de La Plata,
        CC 67, 1900 La Plata, Argentina.
\item[$^{\triangle}$] Supported by the National Natural Science
  Foundation of China.
\item[$^{\dagger}$] Deceased.
\end{list}
}

}

\section*{The OPAL Collaboration}

{

{
G.\thinspace Abbiendi$^{  2}$,
K.\thinspace Ackerstaff$^{  7}$,
C.\thinspace Ainsley$^{  5}$,
P.F.\thinspace {\AA}kesson$^{  7}$,
G.\thinspace Alexander$^{ 21}$, 
J.\thinspace Allison$^{ 15}$,
N.\thinspace Altekamp$^{  5}$,
K.\thinspace Ametewee$^{ 25}$,
G.\thinspace Anagnostou$^{  1}$,
K.J.\thinspace Anderson$^{  8}$
% ...
}

\bigskip

$^{  1}$School of Physics and Astronomy, University of Birmingham,
Birmingham B15 2TT, UK
\newline
$^{  2}$Dipartimento di Fisica dell' Universit\`a di Bologna and INFN,
I-40126 Bologna, Italy
\newline
% ...
$^{  5}$Cavendish Laboratory, Cambridge CB3 0HE, UK
\newline
% ...
$^{  7}$CERN, European Organisation for Nuclear Research,
CH-1211 Geneva 23, Switzerland
\newline
$^{  8}$Enrico Fermi Institute and Department of Physics,
University of Chicago, Chicago IL 60637, USA
\newline
% ...
$^{ 15}$School of Physics and Astronomy, Schuster Laboratory, The University
of Manchester M13 9PL, UK
\newline
% ...
$^{ 21}$Department of Physics and Astronomy, Tel Aviv University,
Tel Aviv 69978, Israel
\newline
% ...
$^{ 25}$University of Victoria, Department of Physics, P O Box 3055,
Victoria BC V8W 3P6, Canada
\newline
% ...

\bigskip

In addition to the support staff at our own
institutions we are pleased to acknowledge the  \\
Department of Energy, USA, \\
National Science Foundation, USA, \\
Particle Physics and Astronomy Research Council, UK, \\
Natural Sciences and Engineering Research Council, Canada, \\
Israel Science Foundation, administered by the Israel
Academy of Science and Humanities, \\
Benoziyo Center for High Energy Physics,\\
Japanese Ministry of Education, Culture, Sports, Science and
Technology (MEXT) and a grant under the MEXT International
Science Research Program,\\
Japanese Society for the Promotion of Science (JSPS),\\
German Israeli Bi-national Science Foundation (GIF), \\
Bundesministerium f\"ur Bildung und Forschung, Germany, \\
National Research Council of Canada, \\
Hungarian Foundation for Scientific Research, OTKA T-038240,
and T-042864,\\
The NWO/NATO Fund for Scientific Research, the Netherlands.

}

\tableofcontents
\clearpage

\chapter{Introduction}
\label{chap:intro}

The electron-positron collider LEP at CERN increased its collision
centre-of-mass energy, $\sqrt{s}$, from the Z pole (\LEPI) up to
$209~\GeV$ during its second running phase (\LEPII) from 1995 to 2000.
The four LEP experiments ALEPH, DELPHI, L3 and OPAL collected a
combined total integrated luminosity of about 3~fb$^{-1}$ in the
\LEPII\ centre-of-mass energy range above the Z pole, $130~\GeV$ to
$209~\GeV$.  This large data set explores the new energy regime
accessed by \LEPII\ with high precision, allowing new tests of the
electroweak Standard Model of particle physics~\cite{Glashow:1961tr,
*Weinberg:1967tq, *Salam:1968rm} (SM), and searches for new physics
effects at higher mass scales.

Combinations of electroweak measurements performed in
electron-positron collisions at Z-pole centre-of-mass energies, at
\LEPI\ and the SLC, are reported in Reference~\cite{bib-Z-pole}.
Here, the measurements in the electroweak sector of the SM at \LEPII\
centre-of-mass energies are discussed, including, where necessary,
studies of strong-interaction effects.  Photon-pair, fermion-pair and
four-fermion production processes are analysed and the results are
presented in the form of total and differential cross-sections.
Final-state interactions between the decay products in W-boson pair
production are investigated for signals of colour reconnection and
Bose-Einstein correlations.  Pair-production of W bosons yields
measurements of the mass, total decay width and decay branching
fractions of the W boson.  Together with other reactions such as
single-W, single-Z, WW$\gamma$, Z-pair, Z$\gamma$ and Z$\gamma\gamma$
production, the data sample allows stringent tests of the non-Abelian
structure of the electroweak gauge group, by measuring triple and
quartic electroweak gauge boson couplings.

\section{LEP-II Data}

In a circular accelerator such as LEP, the energy loss of the beam
particles due to synchrotron radiation increases with the fourth power
of the Lorentz $\gamma$ factor.  In order to push the LEP
centre-of-mass energy beyond the Z-pole, the warm copper RF cavities
used at \LEPI\ were replaced by superconducting RF cavities to
increase the available RF power. In parallel the \LEPII\
centre-of-mass energy increased in steps up to a maximum of
$209~\GeV$, reached in 2000, the final year of LEP operation.  The
centre-of-mass energies and the corresponding integrated luminosities
collected per experiment are reported in Table~\ref{intro:tab:lumi}.
For some of the analyses described in this report, the data have been
combined in different slices of centre-of-mass energies.  About
0.75~fb$^{-1}$ of integrated luminosity was recorded by each LEP
experiment, for a total of about 3~fb$^{-1}$.

\begin{table}[ht]
\begin{center}
\renewcommand{\arraystretch}{1.25}
\begin{tabular}{|l||c|c|c|}
\hline
Year &       Mean energy & Luminosity \\
     & $\sqrt{s}~[\GeV]$ & [pb$^{-1}$]  \\
\hline
\hline
1995, 1997 & 130.3        &    6 \\
           & 136.3        &    6 \\
           & 140.2        &    1 \\
\hline
1996       & 161.3        &   12 \\
           & 172.1        &   12 \\
\hline
1997       & 182.7        &   60 \\
\hline
1998       & 188.6        &  180 \\
\hline
1999       & 191.6        &   30 \\
           & 195.5        &   90 \\
           & 199.5        &   90 \\
           & 201.8        &   40 \\
\hline
2000       & 204.8        &   80 \\
           & 206.5        &  130 \\
           & 208.0        &    8 \\
\hline
\hline
Total      & $130-209$    &  745 \\
\hline
\end{tabular}
\caption[Centre-of-mass energies and luminosities at {\LEPII}] {
Centre-of-mass energies and integrated luminosities recorded by each
experiment at \LEPII. }
\label{intro:tab:lumi}
\end{center}
\end{table}

\section{Standard-Model Processes}

\begin{figure}[htbp]
\begin{center}
  \mbox{\epsfig{file=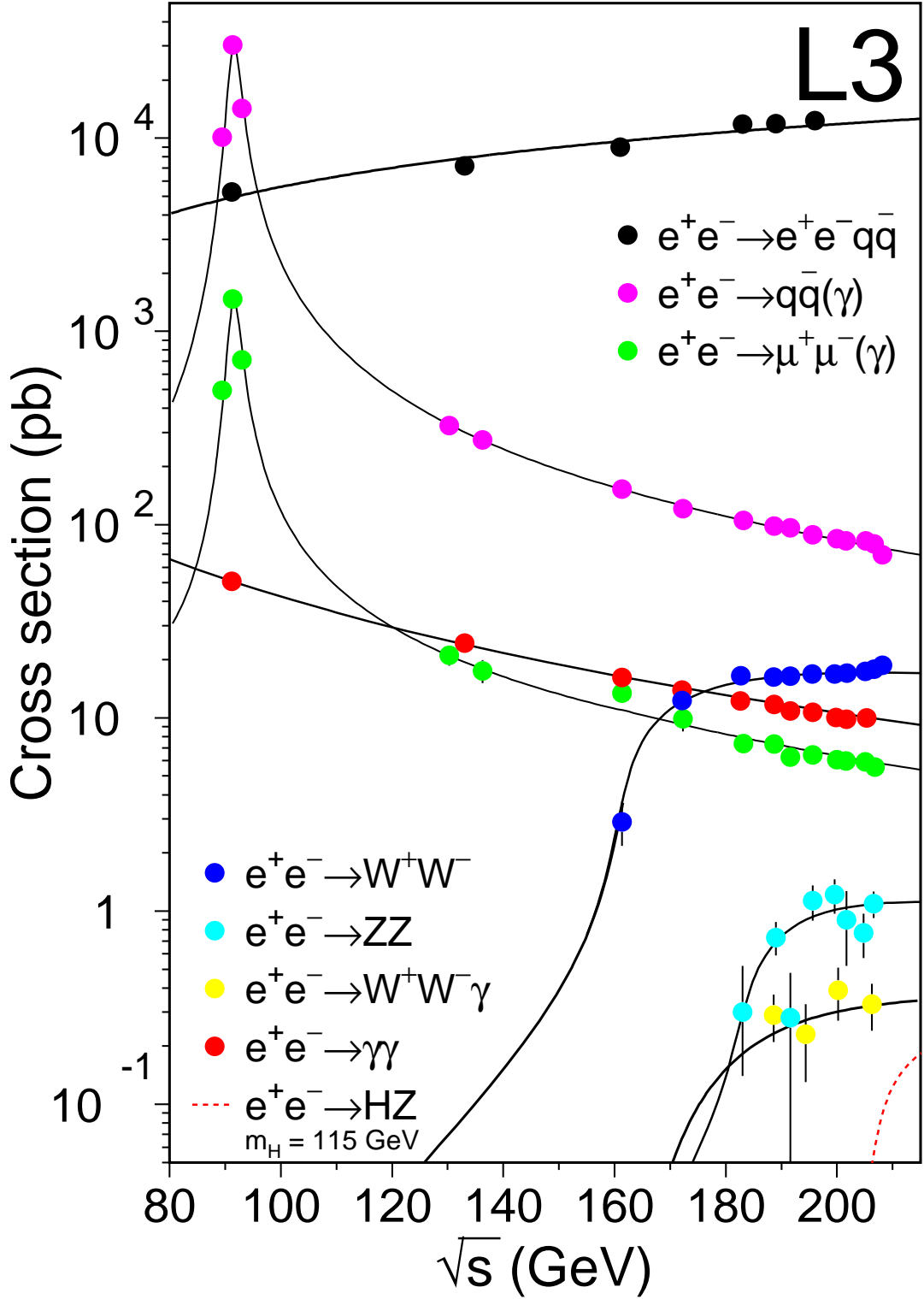,width=0.9\linewidth}}
\caption[Cross-sections of electroweak SM processes] { Cross-sections
of electroweak SM processes. The dots with error bars show the
measurements, while the continuous curves show the theoretical
predictions based on the SM.}
\label{intro:fig:msm-processes}
\end{center}
\end{figure}

The various SM processes occurring at high centre-of-mass energies in
electron-positron collisions and their cross-sections are shown as a
function of the centre-of-mass energy in
Figure~\ref{intro:fig:msm-processes}.

\subsubsection{Photon-Pair Production}

\begin{figure}[t]
\begin{center}
  $ $\hfill
  \mbox{\epsfig{file=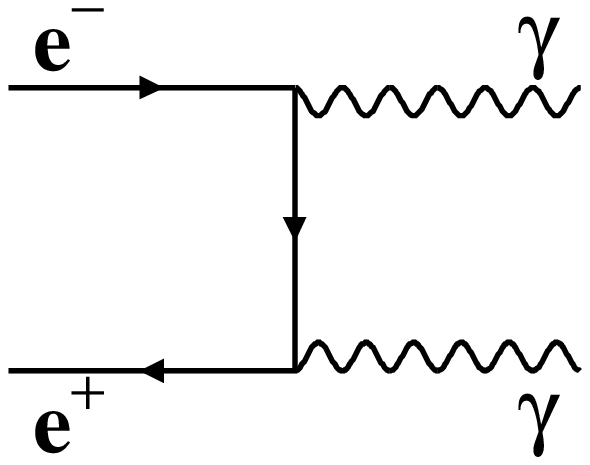,width=0.3\linewidth}}
  \hfill
  \mbox{\epsfig{file=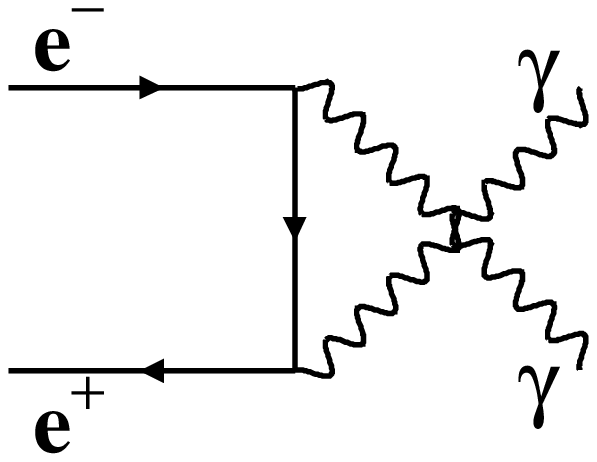,width=0.3\linewidth}}
  \hfill $ $
\caption{Feynman diagrams for the process $\epem\to\gamma\gamma$ at
  the Born level. }
\label{gg:fig:feyn_born}
\end{center}
\end{figure}

The photon-pair production process, $\eeggga$, is dominated by QED
interactions.  The corresponding Feynman diagrams at Born level are
shown in Figure \ref{gg:fig:feyn_born}. Higher-order QED effects play
a significant role but the weak interaction is negligible for the
present data set.  Therefore this reaction is different from the other
processes discussed in this report as it provides a clean test of QED,
independent of other parts of the SM.

\subsubsection{Fermion-Pair Production}

\begin{figure}[t]
\begin{center}
\mbox{\epsfig{file=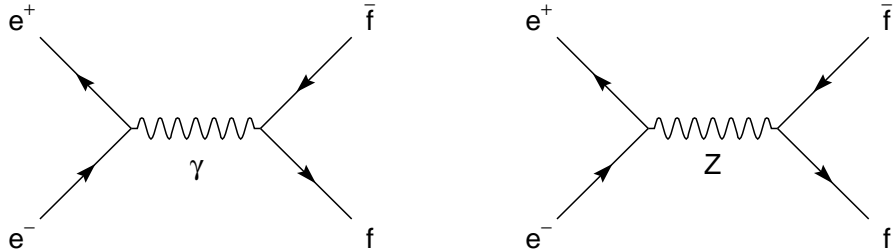,width=0.8\linewidth}}
\caption{Feynman diagrams for the process $\epem\to\ff$
  at the Born level. For $\epem$ final states additional $t$-channel
  diagrams contribute. }
\label{ff:fig:feyn_born}
\end{center}
\end{figure}

Pair-production of fermions proceeds mainly via $s$-channel exchange of
a photon or a Z boson as shown in Figure \ref{ff:fig:feyn_born}. For
energies above the Z resonance, QED radiative corrections are very
large, up to several 100\% of the Born cross-section. This is caused
by hard initial-state radiation of photons, which lowers the
centre-of-mass energy, $\sqrt{s}$, of the hard interaction down to
values $\sqrt{s'}$ close to the Z mass, called radiative return to the
Z.  In order to probe the hard interaction at the nominal energy scale
$\sqrt{s}$, cuts are applied to remove the radiative return to the Z
and only keep the high-$Q^2$ events. Further cuts remove non-resonant
pair corrections arising from four-fermion production not included in
the signal definition.

\subsubsection{WW and ZZ Production}

\begin{figure}[htbp]
\begin{center}
\mbox{\epsfig{file=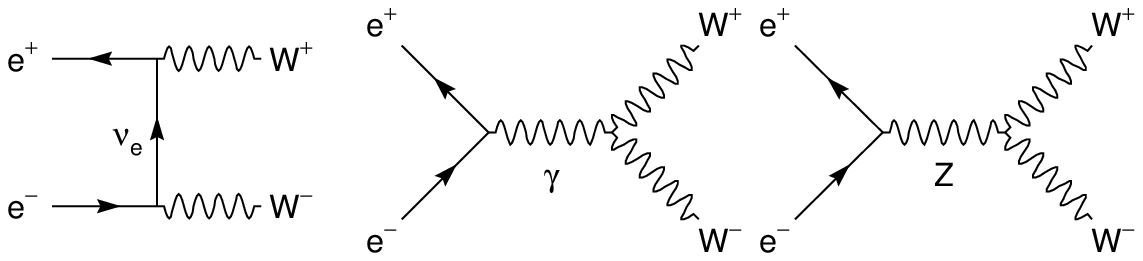,width=0.8\linewidth}}
\caption{Feynman diagrams (CC03) for the process $\epem\to\WW$ at the
  Born level.  }
\label{WW:fig:feyn_born}
\end{center}
\end{figure}

\begin{figure}[htbp]
\begin{center}
\mbox{\epsfig{file=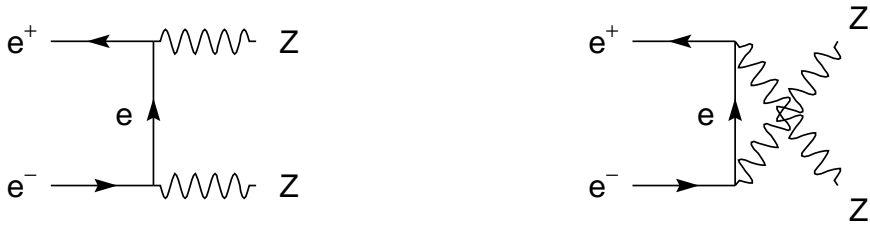,width=0.9\linewidth}}
\caption{Feynman diagrams (NC02) for the process $\epem\to\ZZ$ at the
  Born level. }
\label{ZZ:fig:feyn_born}
\end{center}
\end{figure}

One of the most important processes at \LEPII\ consists of pair
production of on-shell W bosons as shown in
Figure~\ref{WW:fig:feyn_born}.  These events allow a determination of
the mass and total decay width of the W boson. The non-Abelian nature
of the electroweak gauge theory, leading to triple and quartic
gauge-boson vertices such as those appearing in the two $s$-channel WW
production diagrams, is studied and the gauge couplings are measured.
Each W boson decays to a quark-antiquark pair, hadronising into jets,
or to a lepton-neutrino pair, resulting in a four-fermion final state.
The WW events are thus classified into fully hadronic, semileptonic
and purely leptonic events.  At higher centre-of-mass energies,
four-fermion final states are also produced via Z-pair production, as
shown in Figure \ref{ZZ:fig:feyn_born}.

Final-state corrections arising from the interaction between the two W
decay systems, such as colour reconnection and Bose-Einstein
correlations, may lead to a cross-talk effect. Such an effect
potentially spoils the assignment of decay products to decaying weak
bosons in terms of four-momentum, with consequences in the measurement
of the W-boson mass and width in the all-hadronic channel.

\begin{figure}[htbp]
\begin{center}
\mbox{\epsfig{file=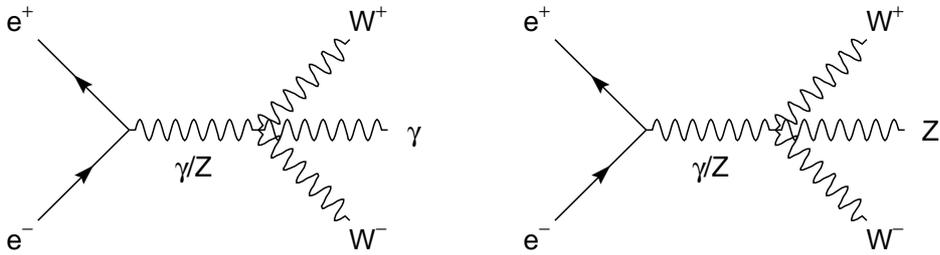,width=0.8\linewidth}}
\caption{Feynman diagrams for the process $\epem\to\WWg$ and WWZ at
  the Born level involving quartic electroweak-gauge-boson vertices.
  }
\label{Q:fig:feyn_born}
\end{center}
\end{figure}

Radiative corrections to W-pair production are particularly
interesting as they allow the study of quartic-gauge-boson vertices as
shown in Figure \ref{Q:fig:feyn_born}.

\begin{figure}[htbp]
\begin{center}
\mbox{\epsfig{file=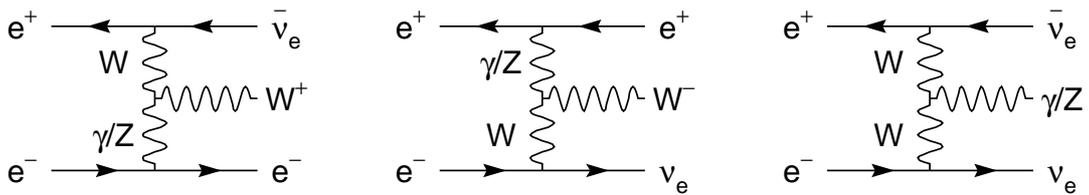,width=0.9\linewidth}}
\caption{Vector-boson fusion diagrams for the single W/Z/$\gamma$
  process at the Born level.  }
\label{V:fig:feyn_born}
\end{center}
\end{figure}

\subsubsection{Four-Fermion Production}

Besides the double-resonant WW and ZZ processes, single-resonant boson
production channels such as those shown in
Figure~\ref{V:fig:feyn_born}, as well as non-resonant diagrams also
contribute to four-fermion production.  Selections are devised to
separate the various four-fermion processes, in particular WW, ZZ,
single-W and single-Z production.  Single-W production is sensitive to
the electromagnetic gauge couplings of the W boson, as the $t$-channel
photon exchange diagram dominates over the $t$-channel Z exchange
diagram at $\LEPII$ energies.  Bremsstrahlung diagrams with radiation
of an on-shell Z boson off an initial- or final-state fermion leg in
Bhabha scattering contribute to single-Z production in the form of Zee
final states.

\chapter{Photon-Pair Production}
\label{chap:gg}

\section{Introduction}
\label{gg:sec:intro}

The differential cross-section for the photon-pair production process
$\eeggga$ is presented here for centre-of-mass energies above
$183~\GeV$.  This process is one of the few channels at LEP energies
with negligible contribution from the weak interaction. Therefore it
provides a clean test of quantum electrodynamics, QED, at high
energies.  The combination is based on the
publications~\cite{Heister:2002ut, Abdallah:2004rc, Achard:2001nt,
Abbiendi:2002je}.

Section \ref{gg:sec:selection} gives a short overview on the event
selections of the four experiments as far as they are relevant for the
determination of the theory uncertainty, which is described in Section
\ref{gg:sec:theory}.  Also the expected cross-sections from QED and
other models are given.  In Section \ref{gg:sec:dxs} the combination
of the differential cross-section is presented.  The total
cross-section given in Section \ref{gg:sec:txs} is derived from the
differential cross-section.  The results are summarised in Section
\ref{gg:sec:results}.

\section{Event Selection}
\label{gg:sec:selection}

The topology of this channel is very clean and the event selection,
which is similar for all experiments, is based on the presence of at
least two energetic clusters in the electromagnetic calorimeters
(ECAL).  A minimum energy of the two highest-energy ECAL clusters is
required.  Restrictions are made either on the acollinearity, $\acol$,
or on the missing longitudinal momentum, $p_z$.  The cuts and the
allowed range in polar angle, $\theta_i$, of the observed clusters are
listed in Table \ref{gg:tab:cuts}.  The clusters are ordered by
decreasing energy.  In order to remove background, especially from
Bhabha events, charged tracks are in general not allowed except when
they can be associated to a photon conversion in one hemisphere.

\begin{table}[htb]
\begin{center}
\renewcommand{\arraystretch}{1.3}
\begin{tabular}{|l@{\hspace{3em}}||c@{\hspace{3em}}c@{\hspace{3em}}c|}
\hline
 Experiment & polar angles & energies & acollinearity \\
\hline
\hline
 ALEPH  & $|\costh_i| < 0.95$ 
        & $E_1, E_2 > 0.5 \cdot E_{\rm beam}$ 
        & $\acol < 20^\circ$  \\
 DELPHI & $25^\circ < \theta_i <155^\circ$ 
        & $E_1, E_2 > 0.3 \cdot E_{\rm beam}$ 
        & $\acol < 50^\circ$  \\
 L3     & $16^\circ < \theta_i <164^\circ$ 
        & $E_1 + E_2 > E_{\rm beam}$
        & $\acol < 165^\circ$  \\
 OPAL   & $|\costh_i| < 0.93$ 
        & $E_1, E_2 > p_z$ 
        & -- \\
\hline
\end{tabular}
\caption[Simplified phase-space definition for $\eeggga$]{Simplified
phase-space definition for the selection of $\eeggga$ events. DELPHI
does not select clusters in the range $[35^\circ,42^\circ]$,
$[88^\circ,92^\circ]$ and $[138^\circ,145^\circ]$. OPAL is sensitive
to additional clusters up to $|\costh_i| < 0.97 \; (i\ge 3$).}
\label{gg:tab:cuts}
\end{center}
\end{table}

Besides limited coverage of the ECAL, selection cuts to reject events
with charged tracks are the main reason for a reduced signal
efficiency.  The effect of the different cuts depends strongly on the
detector geometry. Therefore experimental systematic errors are
considered uncorrelated between the experiments.

\section{Theory}
\label{gg:sec:theory}

\subsection{QED Born Cross-Section}

The differential cross-section for the QED process $\eegg$ in lowest
order is known since a long time~\cite{gg:ref:QED}:

\begin{equation}
\xb = \frac{\alpha^2}{s}\;\left[
\frac{1+\beta^2+\beta^2\sin^2{\theta}}{1-\beta^2\cos^2{\theta}} \; - \;
\frac{2\beta^4\sin^4{\theta}}{(1-\beta^2\cos^2{\theta})^2} \right]
 \; ,
\label{born}
\end{equation}
where $\sqrt{s}$ is the centre-of-mass energy.  Since the emitted
photons are real, with a vanishing invariant mass, the relevant scale
for the fine-structure constant $\alpha$ is zero momentum transfer.
In the following the relativistic limit for the velocity of the
electron $\beta = p/E \to 1$ will be used.  Since the final-state
particles are identical the polar angle $\theta$ is defined such that
$\cost>0$ to avoid double counting. This results in a full phase space
of $\int d\Omega = 2\pi$.

Higher-order QED corrections are relevant but the lowest-order
contribution involving weak couplings is negligible compared to the
current experimental precision of about 1\%.  There is no resonance
effect for this process at energies around the Z mass (\LEPI) since a
spin-one vector or axial-vector particle cannot couple to two real
photons.  However, at the W-pair threshold there is a resonance-like
effect, since the photons can be radiated off an on-shell W loop, with
a dominating contribution from the triangle diagram with $\rm
WW\gamma\gamma$ coupling~\cite{CapdequiPeyranere:1985az,
*Fujimoto:1987xb}.  At such energies, corrections of up to 1.2\% are
expected for $\cost=0$. At the energies considered here the
corrections are smaller, \eg, for a centre-of-mass energy of
$200~\GeV$ they are below 0.2\% at all angles, and will be neglected.

\subsection{Non-QED Models}
\label{gg:sec:noqed}

Various models predict deviations from the QED expectation. The
simplest ansatz is the introduction of cut-off parameters $\Lpm$ as
used for Bhabha and M{\o}ller scattering~\cite{Drell:1958gv,
Low:1965ka}.  With this formalism a short range exponential deviation
is added to the Coulomb potential resulting in a differential
cross-section of the form:

\begin{equation}
\xl   =  \xb \pm \frac{\alpha^2 s}{2\Lambda_\pm^4}(1+\cos^2{\theta}) \,.
\label{lambda}
\end{equation}

New effects can also be described by effective Lagrangian theory
\cite{Eboli:1991ci}.  Here dimension-6 terms lead to anomalous $\rm
ee\gamma$ couplings. The resulting deviations in the differential
cross-section are similar in form to those for cut-off parameters, but
with a slightly different definition of the parameter: $\Lambda_6^4 =
\frac{2}{\alpha}\Lambda_+^4$.  Dimension 7 and 8 Lagrangians introduce
$\rm ee\gamma\gamma$ contact interactions and result in an
angular-independent term added to the Born cross-section:

\begin{equation}
\xq  =  \xb + \frac{s^2}{32\pi}\frac{1}{\Lambda'{}^6} \,.
\end{equation}
The associated parameters are given by $\Lambda_7 = \Lambda'$ and
$\Lambda_8^4 = m_{\rm e} {\Lambda'}^3$ for dimension 7 and dimension 8
couplings, respectively.

Theories of quantum gravity in extra spatial dimensions might solve
the hierarchy problem since gravitons would propagate in a
compactified higher dimensional space, while other Standard Model (SM)
particles are confined to the usual $3 + 1$ space-time
dimensions~\cite{Antoniadis:1998ig}.  While in these models the Planck
mass $M_D$ in $D=n+4$ dimensions is chosen to be at the electroweak
scale, the usual Planck mass $M_{\rm Pl}$ in four dimensions would be
$M_{\rm Pl}^2 = R^n M_D^{n+2}$, where $R$ is the compactification
radius of the additional dimensions.  Since gravitons couple to the
energy-momentum tensor, their interaction with photons is as weak as
that with fermions. However, the huge number of Kaluza-Klein
excitation modes in the extra dimensions may give rise to observable
effects.  These effects depend on the scale $M_s (\sim M_D)$ which may
be as low as ${\cal O}(\rm TeV)$. Model dependences are absorbed in
the parameter $\lambda$ which is expected to be of order 1.  For this
analysis it is assumed that $\lambda = \pm 1$.  The expected
differential cross-section is given by \cite{Agashe:1999qp}:

\begin{equation}
\xg = \xb - \frac{\alpha s}{2\pi} \; \frac{\lambda}{M_s^4}\;(1+\cos^2{\theta})
    + \frac{s^3}{16 \pi^2} \;  \frac{\lambda^2}{M_s^8} \;(1-\cos^4{\theta}) 
    \; , \; \lambda = \pm 1 \,.
\end{equation}

Instead of an ordinary electron an excited electron $\rm e^\ast$
coupling to electron and photon could be exchanged in the t-channel of
the process \cite{Low:1965ka, Mery:1988et}.  In the most general case
$\rm \rm e^\ast e \gamma$ couplings would lead to a large anomalous
magnetic moment of the electron on which strong experimental limits
exist \cite{Brodsky:1980zm, *Renard:1982ij}.  This effect can be
prevented by a chiral magnetic coupling of the form:

\begin{equation}
{\cal L} = \frac{1}{2\Lambda} \bar{e^\ast} \sigma^{\mu\nu}
\left[ g f \frac{\tau}{2}W_{\mu\nu} + g' f' \frac{Y}{2} B_{\mu\nu}
\right] e_L + \mbox{h.c.} \,,
\end{equation}
where $\tau$ are the Pauli matrices and $Y$ is the hypercharge.  The
model parameters are the compositeness scale $\Lambda$ and the
relative couplings $f$ and $f'$ to the gauge fields $W$ and $B$ with
SM couplings $g$ and $g'$.  For the process $\eeggga$, effects vanish
in the case of $f = -f'$.  For $f_\gamma = -\frac{1}{2}(f+f')$ the
following cross-section results~\cite{Vachon:2002dg, *Vachon:2001rh}:

\begin{eqnarray}
\xe = \xb  &+ & 
\frac{\alpha^2}{16}\frac{f_\gamma^4}{\Lambda^4} s \sin^2{\theta}
\left[\frac{p^4}{(p^2-\mestar^2)^2} + \frac{q^4}{(q^2-\mestar^2)^2} \right] \nonumber\\
& - & \frac{\alpha^2}{2 s}\frac{f_\gamma^2}{\Lambda^2}
\left[\frac{p^4}{(p^2-\mestar^2)} + \frac{q^4}{(q^2-\mestar^2)} \right]
\; , 
\end{eqnarray}
with $p^2=-\frac{s}{2}(1-\cost)$ and $q^2=-\frac{s}{2}(1+\cost)$.  In
the following it is assumed that $\Lambda=\mestar$ unless stated
otherwise.

\subsection{Radiative Corrections}

\begin{figure}[t]
\begin{center} 
\mbox{\epsfig{file=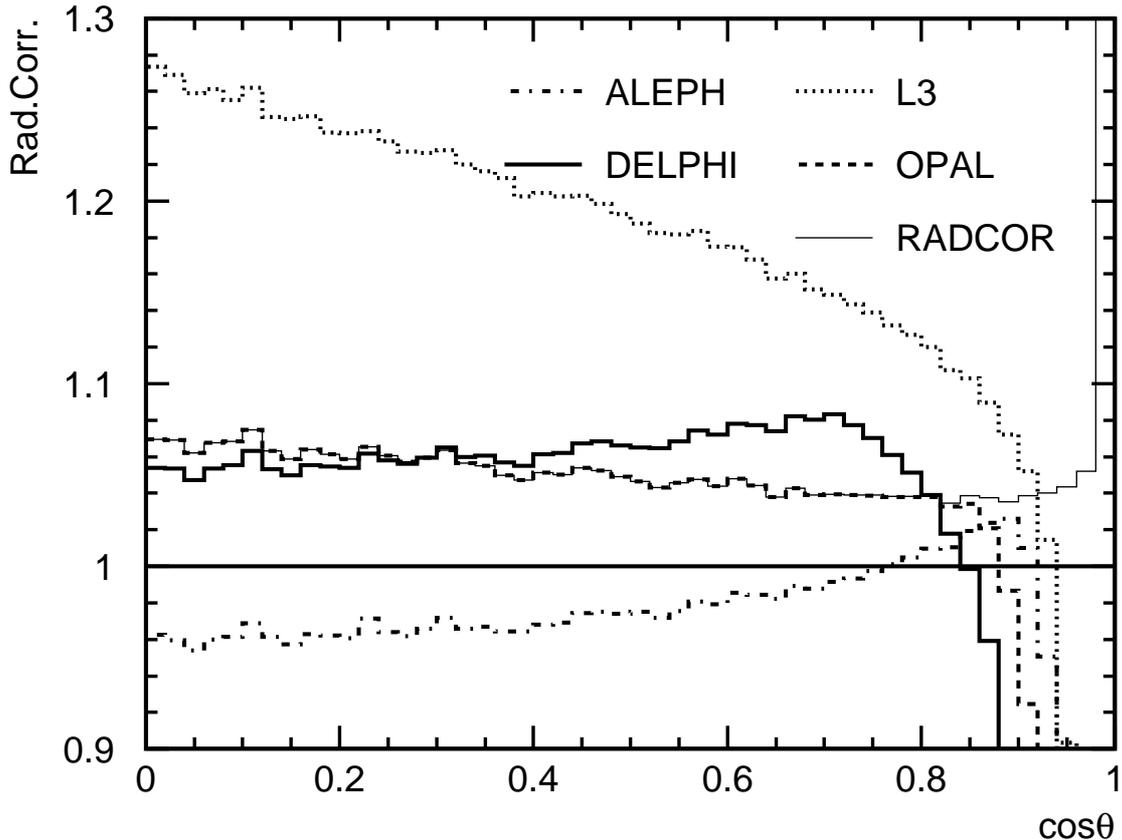,width=0.9\linewidth}}
\end{center}
\caption[Radiative corrections for $\eeggga$] {Radiative corrections
for the four experiments: shown is the ratio of the full third-order
RADCOR Monte-Carlo cross-section calculation with the phase-space cuts
used by each experiment to the Born cross-section. The line labelled
RADCOR is the ratio determined without any phase-space cuts. }
\label{gg:fig:radcor}
\end{figure}

Radiative corrections, \ie, the ratio of the next-to-leading order QED
to Born level, are shown in Figure~\ref{gg:fig:radcor}. They are
determined from Monte-Carlo simulations \cite{Berends:1980px},
implementing a full third-order calculation including electron-mass
effects. In case the third photon is below an energy cut-off, only two
back-to-back photons are generated. Fourth-order effects are not
included.  The event angle $\theta$ is calculated as:

\begin{equation}
\costh ~=~ \left. 
           \left| \sin \left(\frac{\theta_1 - \theta_2}{2}\right) \right| 
           \right/\sin \left(\frac{\theta_1 + \theta_2}{2}\right) \,,
\end{equation} 
to minimise higher order effects, where $\theta_{1,2}$ are the polar
angles of the two highest-energy photons.

The correction labelled RADCOR shown in Figure~\ref{gg:fig:radcor} is
determined from the angles $\theta_{1,2}$ of the two highest-energy
photons generated without restriction on the angle.  The radiative
corrections depend on the selected phase space, which differs between
the four experiments as listed in Table \ref{gg:tab:cuts}.  For OPAL
the radiative corrections are identical to the RADCOR distribution,
apart from the edge effect, since events with a high energy photon
having $|\cost_i| > 0.93$ are rejected due to the cut on the
longitudinal momentum.  Radiative corrections for DELPHI are moderate
and similar to OPAL due to the intermediate restriction on the
acollinearity angle.  L3 on the other hand has a very loose cut on the
acollinearity angle.  Thus events with only one hard photon in the
accepted angle range $|\cost_{2/1}| < 0.96$, the other hard photon
having $0.96 < |\cost_{1/2}| \simeq 1$, are selected. The event angle
is calculated from the angle $\cost_3$ of an observed soft photon
leading to a smaller $\cost$.  Especially in the central region, where
the cross-section is small, this leads to large corrections of up to
30\%.  ALEPH has a very tight cut on the acollinearity angle leading
to a cross-section smaller than the Born cross-section in the central
region.

\subsection{Theory Uncertainty}

For the $\gamma\gamma(\gamma)$ channel, no detailed study of the
theory uncertainty, \ie, the uncertainty of the third-order
Monte-Carlo prediction, exists. For a QED process the higher-order
effect can be estimated to be 10\% ($\simeq\sqrt{\alpha}$) of the
correction due to the highest calculated order.  For each experiment
the theory uncertainty is estimated as 10\% of the radiative
correction, with a minimum of 0.5\%.

A Monte-Carlo study shows that despite different selections the
overlap in the selected phase space is very high, for example, at
$\cost=0.7$ where the third-order DELPHI cross-section is larger than
the OPAL cross-section, all events in the phase space selected by OPAL
are also in the phase space selected by DELPHI. This means that the
common part of the correction is correlated between experiments.

For each $\cost$ bin the theory error is calculated as the luminosity
weighted average over the four experiments taking the correlation into
account.  The resulting error, listed in Tables~\ref{gg:tab:dxs-1}
and~\ref{gg:tab:dxs-2}, varies between 0.5\% and 1.0\%. The first
$\cost$ bin shows a larger error because DELPHI's analysis does not
cover this region and thus the L3 measurements get a larger weight.

To determine limits on non-QED models these correlations are taken
into account in the following way. Obviously the radiative corrections
in neighbouring bins are due to the same effects and hence
correlated. Forward and central region on the other hand are
uncorrelated. A detailed correlation matrix describing this situation
properly is difficult to implement with a log-likelihood fit while for
a $\chi^2$ fit the available statistics are too small. To keep the
log-likelihood fits of the non-QED models simple, just two independent
regions are defined: barrel ($\cost<0.75$) and endcap
($\cost>0.75$). Within each region the theory error is 100\%
correlated, whereas the two regions are treated as uncorrelated. This
simplified treatment is possible, since the theory uncertainty is
smaller than the experimental systematic and statistical
uncertainties.

\section{Combination of the Differential Cross-Section}
\label{gg:sec:dxs}

Apart from ALEPH at $183~\GeV$, all experiments provide the measured
angular distributions in bins of $\cost$, with a bin-width $B=0.05$
for all bins except for the last one which has $B=0.0113$. Only the
$\cost$-range covered differs.  Besides the centre-of-mass energy
$\sqrt{s_k}$ and luminosity ${\cal L}_k$ of each experiment $k$, the
information includes the number of observed events $N^{\rm obs}_k$,
the number of expected events $N^{\rm QED}_k$ or equivalently the
correction $C_k$ with $N^{\rm QED}_k = C_k \left(\frac{d\sigma}{d
\cos{\theta}}\right)_{\rm Born} {\cal L}_k B$, as well as the
experimental systematic error $\delta^{\rm exp}_k$.  The
experiment-dependent terms $C_k$ correct for the different phase-space
cuts reported in Table~\ref{gg:tab:cuts}.  All experiments assume an
experimental systematic error which does not depend on $\cost$ and
hence is correlated between all bins.  The OPAL experiment introduces
an additional uncorrelated experimental error $\delta^{\rm unc}$ for
some bins.  As explained above the experimental systematic error is
uncorrelated between experiments.  The resulting errors on the LEP
combination are reported in Tables~\ref{gg:tab:dxs-1}
and~\ref{gg:tab:dxs-2}.

\begin{table}[p]
{\small
\begin{center}
\renewcommand{\arraystretch}{0.95}
\begin{tabular}{|c||r|c|c|c|c|r||r|c|c|c|c|r|}
\hline
$\cost^\prime$ &  N  &   C    &  theo & exp  &  unc  &  $\frac{d\sigma}{d\cos\theta}$ & 
                  N  &   C    &  theo & exp  &  unc  &  $\frac{d\sigma}{d\cos\theta}$ \\
\hline
\hline
& \multicolumn{6}{|c||}{$\sqrt{s}=182.692~\GeV$ \hfill ${\cal L}=159.4$/pb}
& \multicolumn{6}{|c |}{$\sqrt{s}=188.609~\GeV$ \hfill ${\cal L}=682.6$/pb}\\
\hline %
 0.029 &   23 & 0.7860 & 1.00 & 1.18 & 1.10 &  3.7 &   92 & 0.7853 & 1.00 & 1.00 & 0.84 &  3.4 \\
 0.076 &   39 & 1.0257 & 0.79 & 1.03 & 0.00 &  4.8 &  108 & 0.9321 & 0.79 & 0.80 & 0.00 &  3.4 \\
 0.126 &   32 & 0.9147 & 0.78 & 1.00 & 0.00 &  4.4 &  132 & 0.9718 & 0.78 & 0.83 & 0.00 &  4.0 \\
 0.176 &   32 & 1.0743 & 0.76 & 1.04 & 0.00 &  3.7 &  129 & 0.9473 & 0.76 & 0.81 & 0.00 &  4.0 \\
 0.226 &   33 & 0.9297 & 0.74 & 0.98 & 0.00 &  4.5 &  147 & 0.9210 & 0.74 & 0.80 & 0.00 &  4.7 \\
 0.275 &   41 & 0.9982 & 0.72 & 1.01 & 0.00 &  5.2 &  142 & 0.9539 & 0.72 & 0.82 & 0.00 &  4.4 \\
 0.326 &   44 & 0.9907 & 0.71 & 1.01 & 0.00 &  5.6 &  162 & 0.9308 & 0.71 & 0.81 & 0.00 &  5.1 \\
 0.375 &   37 & 0.9726 & 0.69 & 1.01 & 0.00 &  4.8 &  152 & 0.9207 & 0.69 & 0.82 & 0.00 &  4.8 \\
 0.426 &   39 & 0.9265 & 0.67 & 0.99 & 0.00 &  5.3 &  159 & 0.9301 & 0.67 & 0.81 & 0.00 &  5.0 \\
 0.475 &   37 & 0.9747 & 0.65 & 1.01 & 0.00 &  4.8 &  190 & 0.9351 & 0.65 & 0.80 & 0.00 &  6.0 \\
 0.525 &   55 & 0.9360 & 0.64 & 0.98 & 0.00 &  7.4 &  214 & 0.9523 & 0.64 & 0.79 & 0.00 &  6.6 \\
 0.576 &   55 & 0.9476 & 0.62 & 0.99 & 0.00 &  7.3 &  213 & 0.9380 & 0.62 & 0.80 & 0.00 &  6.7 \\
 0.626 &   73 & 0.9274 & 0.60 & 0.98 & 0.00 &  9.9 &  224 & 0.9240 & 0.60 & 0.79 & 0.00 &  7.1 \\
 0.676 &   70 & 0.9120 & 0.59 & 0.97 & 0.00 &  9.6 &  299 & 0.9198 & 0.59 & 0.79 & 0.00 &  9.5 \\
 0.726 &   44 & 0.4260 & 0.57 & 0.58 & 1.69 & 13.0 &  223 & 0.5398 & 0.57 & 0.88 & 1.01 & 12.1 \\
 0.776 &   53 & 0.4109 & 0.55 & 0.56 & 1.73 & 16.2 &  275 & 0.5295 & 0.55 & 0.89 & 1.02 & 15.2 \\
 0.826 &  104 & 0.5469 & 0.53 & 0.84 & 1.28 & 23.8 &  399 & 0.6400 & 0.53 & 0.89 & 0.83 & 18.3 \\
 0.877 &  197 & 0.7874 & 0.52 & 0.95 & 0.88 & 31.4 &  743 & 0.7959 & 0.52 & 0.82 & 0.66 & 27.4 \\
 0.928 &  133 & 0.3628 & 0.50 & 1.29 & 1.17 & 46.0 &  682 & 0.4409 & 0.50 & 1.10 & 0.73 & 45.3 \\
 0.956 &   35 & 0.2010 & 0.50 & 2.10 & 0.00 & 99.2 &   78 & 0.1426 & 0.50 & 2.10 & 0.00 & 72.8 \\
\hline
\hline
& \multicolumn{6}{|c||}{$\sqrt{s}=191.597~\GeV$ \hfill ${\cal L}=111.8$/pb}
& \multicolumn{6}{|c |}{$\sqrt{s}=195.506~\GeV$ \hfill ${\cal L}=314.0$/pb}\\
\hline %
 0.029 &   13 & 0.6903 & 1.00 & 0.92 & 0.93 &  3.4 &   35 & 0.7437 & 1.00 & 1.00 & 0.80 &  3.0 \\
 0.076 &   22 & 0.9613 & 0.79 & 0.81 & 0.00 &  4.1 &   51 & 0.9882 & 0.79 & 0.84 & 0.00 &  3.3 \\
 0.126 &   14 & 0.9154 & 0.78 & 0.78 & 0.00 &  2.7 &   45 & 0.9061 & 0.78 & 0.79 & 0.00 &  3.2 \\
 0.176 &   18 & 0.9117 & 0.76 & 0.79 & 0.00 &  3.5 &   68 & 0.9401 & 0.76 & 0.84 & 0.00 &  4.6 \\
 0.226 &   12 & 0.9529 & 0.74 & 0.83 & 0.00 &  2.3 &   47 & 1.0174 & 0.74 & 0.83 & 0.00 &  2.9 \\
 0.275 &   30 & 0.9242 & 0.72 & 0.79 & 0.00 &  5.8 &   54 & 0.8987 & 0.72 & 0.80 & 0.00 &  3.8 \\
 0.326 &   21 & 0.9212 & 0.71 & 0.78 & 0.00 &  4.1 &   53 & 0.9260 & 0.71 & 0.82 & 0.00 &  3.6 \\
 0.375 &   26 & 0.9950 & 0.69 & 0.84 & 0.00 &  4.7 &   72 & 0.9005 & 0.69 & 0.80 & 0.00 &  5.1 \\
 0.426 &   28 & 0.9054 & 0.67 & 0.79 & 0.00 &  5.5 &   65 & 0.8896 & 0.67 & 0.81 & 0.00 &  4.7 \\
 0.475 &   29 & 0.9181 & 0.65 & 0.81 & 0.00 &  5.7 &   79 & 0.9573 & 0.65 & 0.81 & 0.00 &  5.3 \\
 0.525 &   27 & 0.8903 & 0.64 & 0.77 & 0.00 &  5.4 &   97 & 0.9172 & 0.64 & 0.80 & 0.00 &  6.7 \\
 0.576 &   29 & 0.9808 & 0.62 & 0.83 & 0.00 &  5.3 &   93 & 0.9437 & 0.62 & 0.82 & 0.00 &  6.3 \\
 0.626 &   46 & 0.9386 & 0.60 & 0.82 & 0.00 &  8.8 &  116 & 0.9216 & 0.60 & 0.81 & 0.00 &  8.0 \\
 0.676 &   41 & 0.9026 & 0.59 & 0.80 & 0.00 &  8.1 &  129 & 0.8611 & 0.59 & 0.78 & 0.00 &  9.5 \\
 0.726 &   34 & 0.5506 & 0.57 & 0.93 & 0.97 & 11.0 &   82 & 0.5200 & 0.57 & 0.92 & 0.96 & 10.0 \\
 0.776 &   43 & 0.5032 & 0.55 & 0.89 & 1.05 & 15.3 &  120 & 0.4941 & 0.55 & 0.92 & 1.00 & 15.5 \\
 0.826 &   75 & 0.6263 & 0.53 & 0.88 & 0.83 & 21.4 &  178 & 0.6082 & 0.53 & 0.91 & 0.80 & 18.6 \\
 0.877 &  108 & 0.7951 & 0.52 & 0.79 & 0.65 & 24.3 &  350 & 0.7900 & 0.52 & 0.79 & 0.61 & 28.2 \\
 0.928 &  117 & 0.4165 & 0.50 & 1.08 & 0.76 & 50.3 &  276 & 0.4203 & 0.50 & 1.11 & 0.70 & 41.8 \\
 0.956 &   16 & 0.1459 & 0.50 & 2.10 & 0.00 & 89.2 &   33 & 0.1492 & 0.50 & 2.10 & 0.00 & 64.0 \\
\hline
\end{tabular}
\end{center}
}
\caption[Combined differential cross-sections for $\eeggga$] {Combined
differential cross-sections for $\eeggga$.  The first two numbers of
each block are the centre-of-mass energy, $\sqrt{s}$, and the total
luminosity, ${\cal L}$.  The following rows list for each bin:
weighted $\cost$, total number of events $N$, correction $C$, theory
error (theo), experimental systematic error (exp) and systematic
uncorrelated error (unc). The errors are relative and given in \%.
The differential cross-section (in pb) is: $d\sigma / d\cost
(\cost,\sqrt{s}) = N / C / B / {\cal L} $.  The value listed for
$\cost^\prime$ corresponds to $d\sigma/d\cost(\cost^\prime) \cdot B \,
= \, \int_{\rm bin} d\sigma/d\cost \; d\cost$.}
\label{gg:tab:dxs-1}
\end{table}

\begin{table}[p]
{\small
\begin{center}
\renewcommand{\arraystretch}{0.95}
\begin{tabular}{|c||r|c|c|c|c|r||r|c|c|c|c|r|}
\hline
$\cost^\prime$ &  N  &   C    &  theo & exp  &  unc  &  $\frac{d\sigma}{d\cos\theta}$ & 
                  N  &   C    &  theo & exp  &  unc  &  $\frac{d\sigma}{d\cos\theta}$ \\
\hline
\hline
& \multicolumn{6}{|c||}{$\sqrt{s}=199.504~\GeV$ \hfill ${\cal L}=315.2$/pb}
& \multicolumn{6}{|c |}{$\sqrt{s}=201.631~\GeV$ \hfill ${\cal L}=157.1$/pb}\\
\hline %
 0.029 &   43 & 0.6607 & 1.00 & 0.92 & 0.93 &  4.1 &   23 & 0.7240 & 1.00 & 0.99 & 0.80 &  4.0 \\
 0.076 &   44 & 0.8989 & 0.79 & 0.76 & 0.00 &  3.1 &   25 & 0.8865 & 0.79 & 0.78 & 0.00 &  3.6 \\
 0.126 &   38 & 0.9171 & 0.78 & 0.78 & 0.00 &  2.6 &   25 & 0.8697 & 0.78 & 0.78 & 0.00 &  3.7 \\
 0.176 &   38 & 0.9480 & 0.76 & 0.78 & 0.00 &  2.5 &   18 & 0.9562 & 0.76 & 0.84 & 0.00 &  2.4 \\
 0.226 &   50 & 0.9385 & 0.74 & 0.76 & 0.00 &  3.4 &   23 & 0.9482 & 0.74 & 0.79 & 0.00 &  3.1 \\
 0.275 &   57 & 0.9574 & 0.72 & 0.80 & 0.00 &  3.8 &   19 & 0.8910 & 0.72 & 0.76 & 0.00 &  2.7 \\
 0.326 &   64 & 0.9220 & 0.71 & 0.78 & 0.00 &  4.4 &   31 & 0.8263 & 0.71 & 0.75 & 0.00 &  4.8 \\
 0.375 &   64 & 0.9122 & 0.69 & 0.80 & 0.00 &  4.5 &   38 & 0.9389 & 0.69 & 0.81 & 0.00 &  5.2 \\
 0.426 &   64 & 0.9186 & 0.67 & 0.80 & 0.00 &  4.4 &   36 & 0.9471 & 0.67 & 0.86 & 0.00 &  4.8 \\
 0.475 &   67 & 0.9311 & 0.65 & 0.77 & 0.00 &  4.6 &   28 & 0.9213 & 0.65 & 0.79 & 0.00 &  3.9 \\
 0.525 &   77 & 0.9137 & 0.64 & 0.78 & 0.00 &  5.3 &   43 & 0.8979 & 0.64 & 0.80 & 0.00 &  6.1 \\
 0.576 &   94 & 0.9057 & 0.62 & 0.77 & 0.00 &  6.6 &   48 & 0.9472 & 0.62 & 0.82 & 0.00 &  6.5 \\
 0.626 &  104 & 0.9226 & 0.60 & 0.80 & 0.00 &  7.2 &   52 & 0.9153 & 0.60 & 0.81 & 0.00 &  7.2 \\
 0.676 &  111 & 0.8897 & 0.59 & 0.77 & 0.00 &  7.9 &   62 & 0.8703 & 0.59 & 0.78 & 0.00 &  9.1 \\
 0.726 &   70 & 0.5447 & 0.57 & 0.96 & 0.94 &  8.2 &   52 & 0.5281 & 0.57 & 0.98 & 0.91 & 12.5 \\
 0.776 &  108 & 0.5174 & 0.55 & 0.94 & 0.98 & 13.2 &   53 & 0.5151 & 0.55 & 0.97 & 0.93 & 13.1 \\
 0.826 &  160 & 0.5807 & 0.53 & 0.90 & 0.86 & 17.5 &   92 & 0.5886 & 0.53 & 0.93 & 0.80 & 19.9 \\
 0.877 &  307 & 0.8001 & 0.52 & 0.77 & 0.62 & 24.3 &  152 & 0.7988 & 0.52 & 0.79 & 0.58 & 24.2 \\
 0.928 &  279 & 0.4092 & 0.50 & 1.10 & 0.74 & 43.3 &  115 & 0.4240 & 0.50 & 1.12 & 0.67 & 34.5 \\
 0.956 &   28 & 0.1231 & 0.50 & 2.10 & 0.00 & 65.6 &   11 & 0.1197 & 0.50 & 2.10 & 0.00 & 53.2 \\
\hline
\hline
& \multicolumn{6}{|c||}{$\sqrt{s}=205.279~\GeV$ \hfill ${\cal L}=393.3$/pb}
& \multicolumn{6}{|c |}{$\sqrt{s}=206.671~\GeV$ \hfill ${\cal L}=462.9$/pb}\\
\hline %
 0.029 &   44 & 0.5596 & 1.00 & 0.96 & 0.89 &  4.0 &   59 & 0.8530 & 1.00 & 0.99 & 0.85 &  3.0 \\
 0.076 &   64 & 0.9151 & 0.79 & 0.74 & 0.00 &  3.6 &   68 & 1.0029 & 0.79 & 0.89 & 0.00 &  2.9 \\
 0.126 &   53 & 0.9524 & 0.78 & 0.72 & 0.00 &  2.8 &   70 & 1.0074 & 0.78 & 0.91 & 0.00 &  3.0 \\
 0.176 &   51 & 0.9325 & 0.76 & 0.75 & 0.00 &  2.8 &   66 & 0.9777 & 0.76 & 0.87 & 0.00 &  2.9 \\
 0.226 &   65 & 0.9267 & 0.74 & 0.72 & 0.00 &  3.6 &   74 & 1.0103 & 0.74 & 0.88 & 0.00 &  3.2 \\
 0.275 &   50 & 0.9477 & 0.72 & 0.73 & 0.00 &  2.7 &   67 & 0.9818 & 0.72 & 0.87 & 0.00 &  2.9 \\
 0.326 &   71 & 0.8851 & 0.71 & 0.72 & 0.00 &  4.1 &   94 & 0.9437 & 0.71 & 0.87 & 0.00 &  4.3 \\
 0.375 &   63 & 0.9136 & 0.69 & 0.75 & 0.00 &  3.5 &   72 & 0.9200 & 0.69 & 0.92 & 0.00 &  3.4 \\
 0.426 &   72 & 0.9104 & 0.67 & 0.72 & 0.00 &  4.0 &   88 & 0.9542 & 0.67 & 0.90 & 0.00 &  4.0 \\
 0.475 &   62 & 0.9108 & 0.65 & 0.72 & 0.00 &  3.5 &   98 & 0.9776 & 0.65 & 0.88 & 0.00 &  4.3 \\
 0.525 &   91 & 0.8862 & 0.64 & 0.71 & 0.00 &  5.2 &  122 & 0.9286 & 0.64 & 0.87 & 0.00 &  5.7 \\
 0.576 &   97 & 0.9212 & 0.62 & 0.72 & 0.00 &  5.4 &  126 & 0.9500 & 0.62 & 0.88 & 0.00 &  5.7 \\
 0.626 &  102 & 0.8721 & 0.60 & 0.72 & 0.00 &  5.9 &  144 & 0.9281 & 0.60 & 0.87 & 0.00 &  6.7 \\
 0.676 &  150 & 0.8650 & 0.59 & 0.71 & 0.00 &  8.8 &  206 & 0.9089 & 0.59 & 0.86 & 0.00 &  9.8 \\
 0.726 &   89 & 0.4266 & 0.57 & 0.92 & 0.97 & 10.6 &  147 & 0.6288 & 0.57 & 0.92 & 0.97 & 10.1 \\
 0.776 &  105 & 0.3995 & 0.55 & 0.89 & 1.03 & 13.4 &  166 & 0.5891 & 0.55 & 0.90 & 1.02 & 12.2 \\
 0.826 &  154 & 0.4833 & 0.53 & 0.89 & 0.84 & 16.2 &  227 & 0.7137 & 0.53 & 0.89 & 0.83 & 13.7 \\
 0.877 &  345 & 0.7747 & 0.52 & 0.71 & 0.52 & 22.6 &  431 & 0.8173 & 0.52 & 0.86 & 0.72 & 22.8 \\
 0.928 &  252 & 0.3169 & 0.50 & 1.07 & 0.77 & 40.4 &  418 & 0.4780 & 0.50 & 1.09 & 0.75 & 37.8 \\
 0.956 &   24 & 0.0960 & 0.50 & 2.10 & 0.00 & 57.8 &   61 & 0.1490 & 0.50 & 2.10 & 0.00 & 80.4 \\
\hline
\end{tabular}
\end{center}
}
\caption[Combined differential cross-sections for $\eeggga$] {Combined
differential cross-sections for $\eeggga$.  The first two numbers of
each block are the centre-of-mass energy, $\sqrt{s}$, and the total
luminosity, ${\cal L}$.  The following rows list for each bin:
weighted $\cost$, total number of events $N$, correction $C$, theory
error (theo), experimental systematic error (exp) and systematic
uncorrelated error (unc). The errors are relative and given in \%.
The differential cross-section (in pb) is: $d\sigma / d\cost
(\cost,\sqrt{s}) = N / C / B / {\cal L} $.  The value listed for
$\cost^\prime$ corresponds to $d\sigma/d\cost(\cost^\prime) \cdot B \,
= \, \int_{\rm bin} d\sigma/d\cost \; d\cost$.}
\label{gg:tab:dxs-2}
\end{table}

\begin{figure}[hbtp]
\begin{center}
\mbox{\epsfig{file=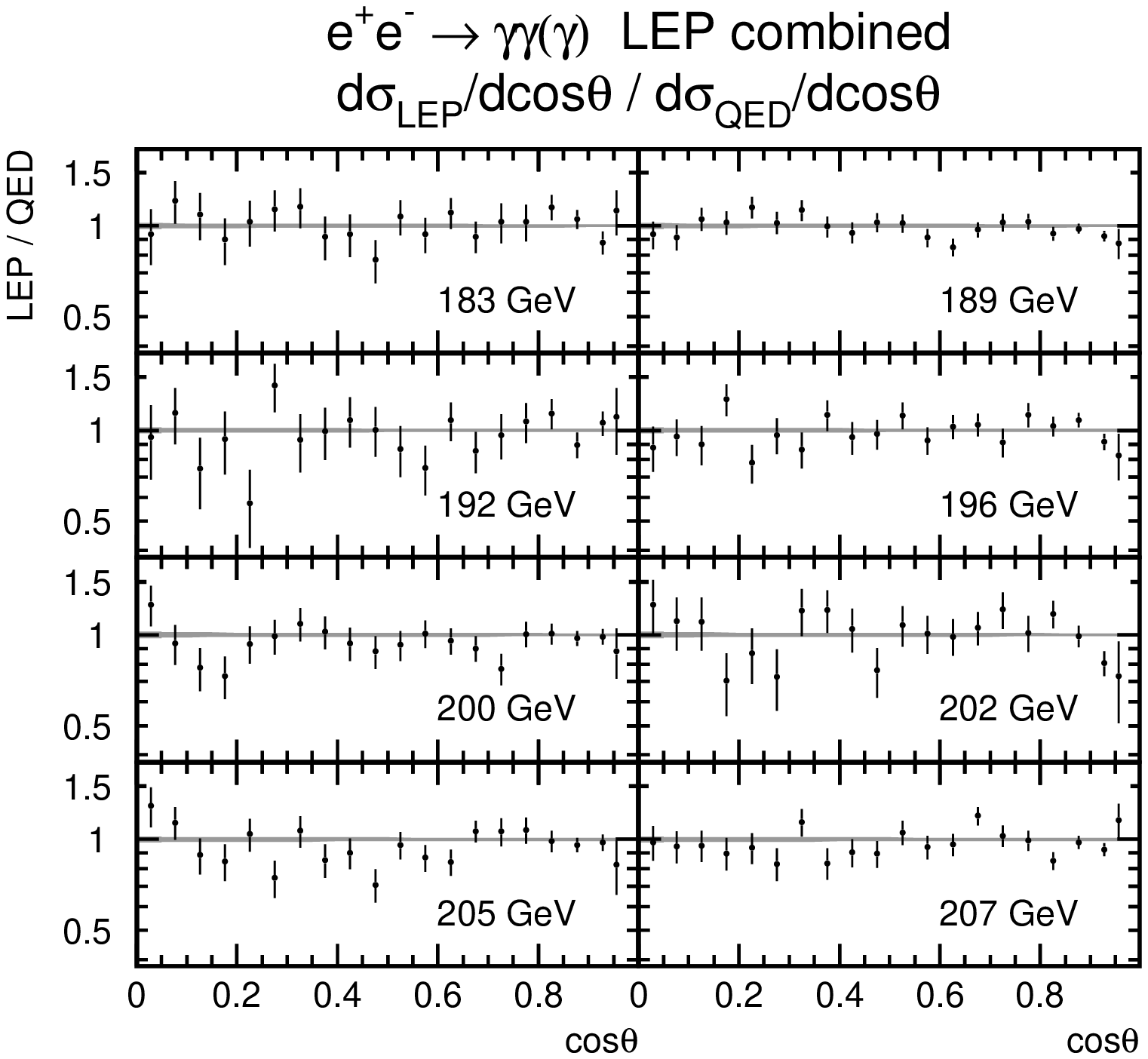,width=0.9\linewidth}}
\caption[Combined differential cross-sections for $\eeggga$] {Combined
differential cross-sections relative to the QED expectation. The error
bars shown include the statistical and systematic experimental errors.
The theory uncertainty is small, decreasing from 1.0\% to 0.5\% for
increasing $|\cos\theta|$.  }
\label{gg:fig:difxs}
\end{center}
\end{figure}

\begin{figure}[hbtp]
\begin{center}
\mbox{\epsfig{file=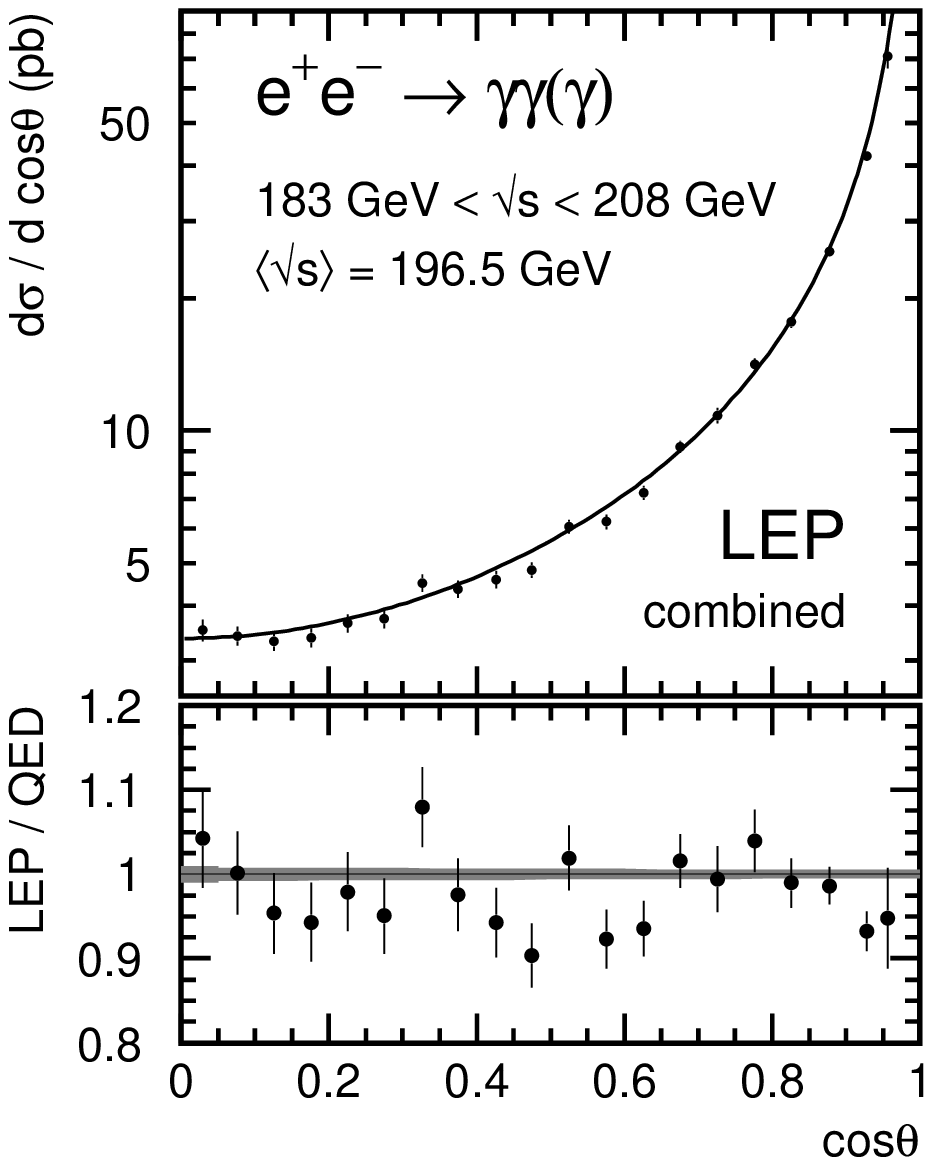,width=0.9\linewidth}}
\caption[Differential cross-section combined for all energies] {The
differential cross-section combined for all energies compared to the
expectation from QED. The lower plot shows the ratio of measured and
expected cross-section, with the band indicating the theory error.  }
\label{gg:fig:avg}
\end{center}
\end{figure}

The effective centre-of-mass energy, $\sqrt{s}$, is determined as the
luminosity weighted average, taking into account that the
cross-section is proportional to $1/s$:

\begin{equation}
 s = {\cal L} \left/ \sum_k \frac{{\cal L}_k}{s_k} \right. \,,
\end{equation} 
where ${\cal L} = \sum_k {\cal L}_k$.  The average correction $C$ at a
given angle and energy is calculated as:

\begin{equation}
 C = \left.
\sum_k \frac{d\sigma}{d \cos{\theta}}(\cost_k,s_k) {\cal L}_k C_k \, \right/ \,
\frac{d\sigma}{d \cos{\theta}}(\cost,s) {\cal L} \, .
\end{equation} 
Similarly the systematic errors are calculated, adding the
contributions in quadrature.  The results are given in
Tables~\ref{gg:tab:dxs-1} and~\ref{gg:tab:dxs-2}.  The combined
differential cross-section in each bin, corrected to Born level, is
then calculated as:

\begin{equation}
\frac{d\sigma}{d \cost} = \frac{N}{C B {\cal L}}
\end{equation}
where $N=\sum_kN^{obs}_k$. The ratio of the combined cross-section and
the expected Born cross-section is shown in Figure \ref{gg:fig:difxs}.
For illustration the differential cross-section combined for all
energies is shown in Figure \ref{gg:fig:avg}.  On average, the
cross-section is slightly below the QED expectation.

\section{Combined Total Cross-Section}
\label{gg:sec:txs}

\begin{figure}[htbp]
\begin{center}
\mbox{\epsfig{file=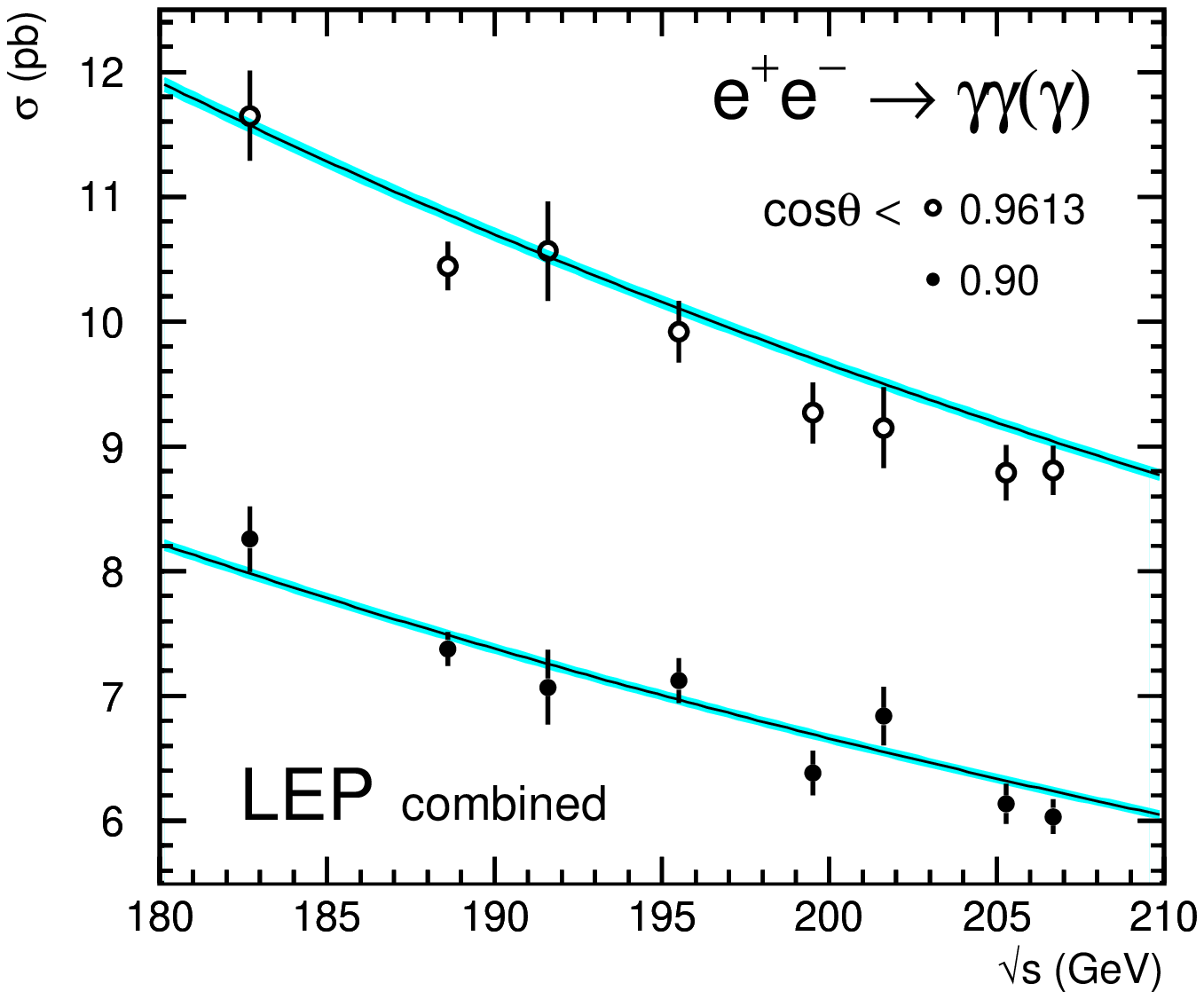,width=0.9\linewidth}}
\end{center}
\caption[Total cross-section as a function of energy] {The total
cross-section as a function of energy for two regions in $\cost$. The
error includes statistical and systematic experimental error.  The
theory error is shown as the band on the QED prediction.  }
\label{gg:fig:xsn}
\end{figure}

The total cross-section is derived by integrating the combined
differential cross-section.  Since the coverage in the scattering
angle varies between experiments, the total cross-section is given for
two ranges, $\cost < 0.9613$ and $\cost<0.90$.  The latter range is
covered by all four experiments.  The results are shown in
Figure~\ref{gg:fig:xsn} and are summarised in Table~\ref{gg:tab:xsn}.
For the theory error the contributions in barrel and endcap are added
in quadrature.  The total cross-section (especially for $\cost <
0.9613$) is dominated by the very forward region, where the
cross-section is strongly increasing.

\begin{table}[htbp]
\begin{center}
\begin{tabular}{|c||r@{$\pm$}l@{$\pm$}lr@{$\pm$}l|r@{$\pm$}l@{$\pm$}lr@{$\pm$}l|}
\hline
$\sqrt{s}$ & \multicolumn{5}{c|}{$\cost<0.90$} & 
\multicolumn{5}{c|}{$\cost<0.9613$}\\
(GeV) & \multicolumn{3}{c}{LEP} & \multicolumn{2}{c|}{QED} & 
\multicolumn{3}{c}{LEP} & \multicolumn{2}{c|}{QED} \\
\hline
\hline
182.7 & 8.26&0.26&0.08 & 7.98&0.04 & 11.65&0.34&0.13 & 11.57&0.05\\
188.6 & 7.38&0.12&0.06 & 7.49&0.03 & 10.44&0.15&0.11 & 10.86&0.04\\
191.6 & 7.07&0.28&0.06 & 7.26&0.03 & 10.56&0.39&0.11 & 10.52&0.04\\
195.5 & 7.12&0.17&0.06 & 6.97&0.03 &  9.92&0.22&0.10 & 10.10&0.04\\
199.5 & 6.38&0.16&0.06 & 6.69&0.03 &  9.27&0.21&0.10 &  9.70&0.04\\
201.6 & 6.84&0.24&0.06 & 6.55&0.03 &  9.15&0.30&0.10 &  9.50&0.04\\
205.3 & 6.13&0.15&0.05 & 6.32&0.03 &  8.79&0.20&0.09 &  9.16&0.04\\
206.7 & 6.03&0.13&0.06 & 6.24&0.03 &  8.81&0.17&0.10 &  9.04&0.04\\
\hline
\end{tabular}
\end{center}
\caption[Total cross-section for $\eeggga$] {The total cross-section
(in pb) for $\eeggga$.  For the measured cross-sections (LEP) the
statistical and systematic errors are given. The theory error of
0.45\% (0.41\%) for $\cost < 0.90$ ($0.9613$) is quoted for the QED
expectation. }
\label{gg:tab:xsn}
\end{table}

\section{Interpretation}
\label{gg:sec:results}

Limits on the parameters describing the non-QED models discussed in
Section~\ref{gg:sec:noqed} are determined from log-likelihood fits to
the combined differential cross-section.  Where possible the fit
parameters are chosen such that the likelihood function is
approximately Gaussian.  The results of the fits are given in
Table~\ref{gg:tab:results}.  The values of the fit parameters are
about 1.5 standard deviations below the expectation, reflecting the
low cross-section in the central region.
  
\begin{figure}[p]
\begin{center}
\mbox{\epsfig{file=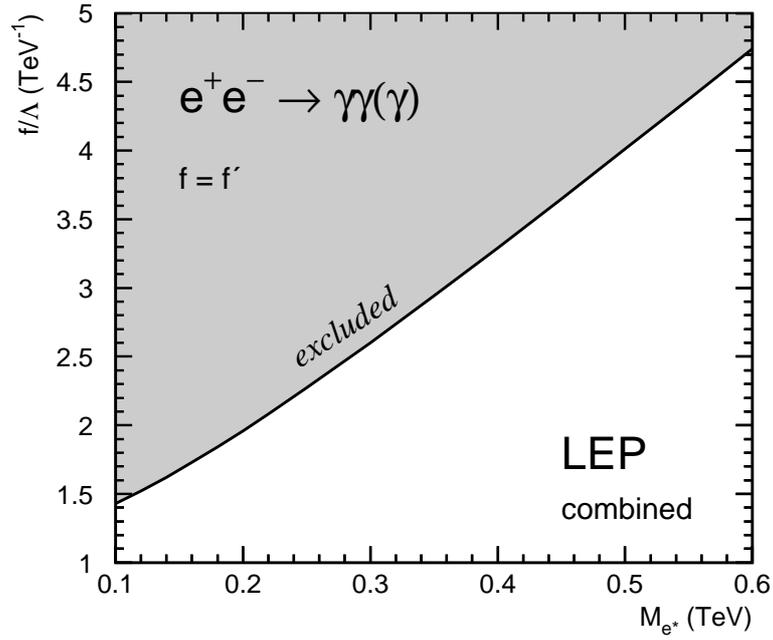,width=0.6\linewidth}}
\end{center}
\caption[95\% CL limits on $f_\gamma/\Lambda$ as a function of
$\mestar$] {95\% CL limits on $f_\gamma/\Lambda$ as a function of
$\mestar$.  In the case of $f=f'$ it follows that $f_\gamma = -f$.  It
is assumed that $\Lambda=\mestar$.}
\label{gg:fig:estar}
\end{figure}

\begin{figure}[p]
\begin{center}
\mbox{\epsfig{file=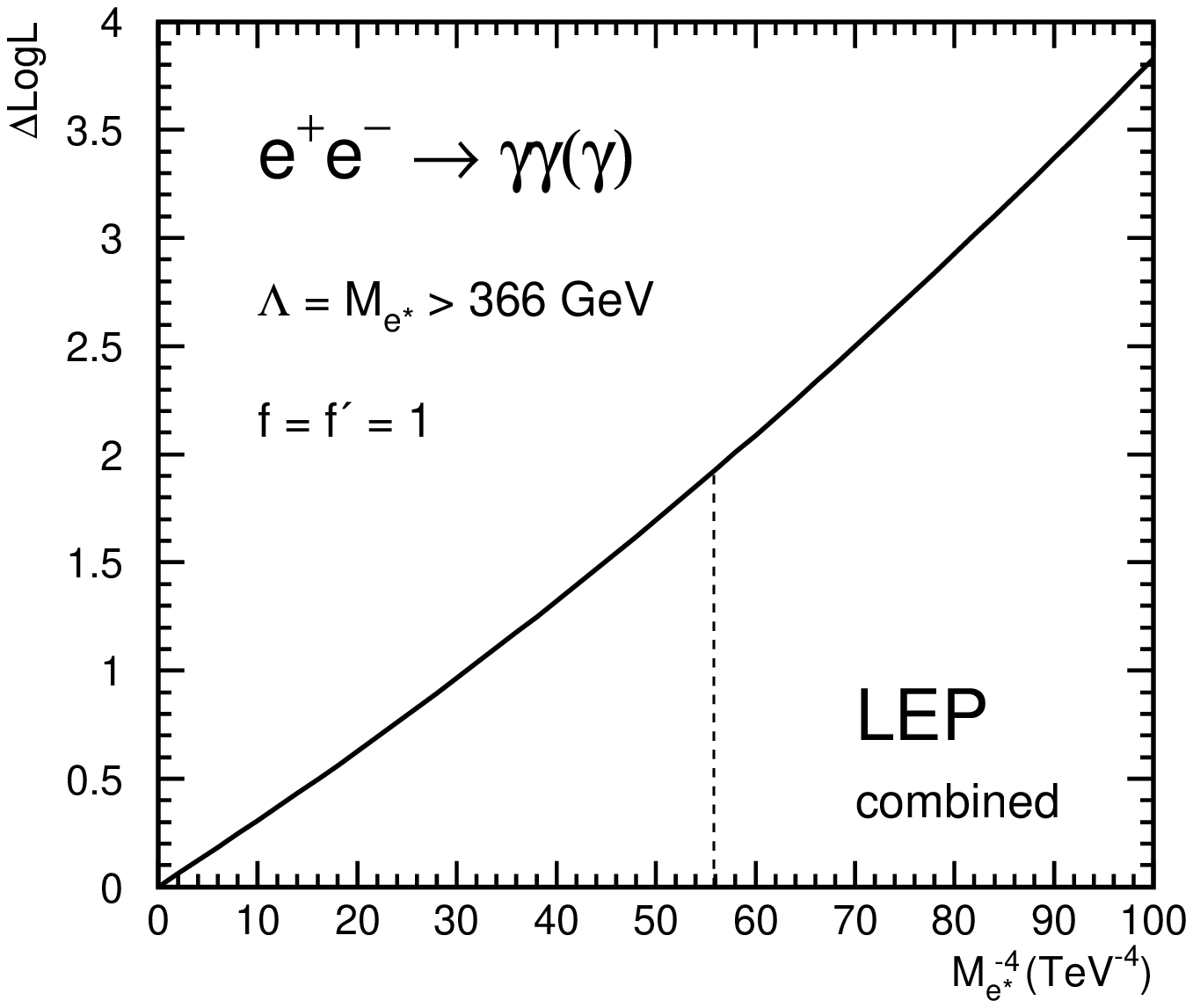,width=0.6\linewidth}}
\end{center}
\caption[Log likelihood difference as a function of $\mestar^{-4}$]
{Log likelihood difference $\Delta\mbox{LogL} = -\ln{\cal L}+\ln{\cal
L}_{\rm max}$ as a function of $\mestar^{-4}$. The coupling is fixed
at $f = f' = 1$.  The value corresponding to $\Delta\mbox{LogL} =
1.92$ is $\mestar^{-4} = 55.8$ \TeV$^{-4} \to \mestar = 366$ \GeV.}
\label{gg:fig:ll}
\end{figure}

Since no significant deviations with respect to the QED expectations
are found -- all the parameters are compatible with zero -- 95\%
confidence level limits are obtained by renormalising the probability
distribution of the fit parameter to the physically allowed region,
$\epsilon \ge 0$ for each $\Lambda^+$ limit and $\epsilon \le 0$ for
$\Lambda^-$ limits.  For limits on the coupling of an excited electron
$f_\gamma/\Lambda$ a scan over the mass $\mestar$ is performed and
presented in Figure~\ref{gg:fig:estar}.  The cross-section is
nonlinear in the fit parameter only for $\mestar$.  The obtained
negative log likelihood distribution is shown in
Figure~\ref{gg:fig:ll} and the limit is determined at 1.92 units above
the minimum.

\section{Conclusion}

The differential cross-section for the photon-pair production process
$\eeggga$ was measured and found in agreement with the expectation
from QED. Limits on new physics were obtained for various models. They
supersede by large factors previous limits on cut-off parameters
obtained from data collected at electron-positron colliders of lower
centre-of-mass energies~\cite{gg-VENUS, *gg-ALEPH, *gg-DELPHI,
*gg-L3-1, *gg-L3-2, *gg-L3-3, *gg-OPAL-1, *gg-OPAL-2, *gg-EBOLI}.

\begin{table}[t]
\begin{center}
\renewcommand{\arraystretch}{1.25}
\begin{tabular}{|c||c|r@{ }l|}
\hline
Model and     &            &                    &                        \\ %
Fit parameter & Fit result & \multicolumn{2}{c|}{95\%\ CL limit (\GeV )} \\
\hline
\hline
Cut-off parameters&        &                    &                        \\
 &  & $\Lambda_+ >$ &  431 \\
 \raisebox{2.2ex}[-2.2ex]{$\Lpm^{-4}$} &
 \raisebox{2.2ex}[-2.2ex]{$
 \left(-37\apm{24}{23}\right)\cdot 10^{-12}$ \GeV$^{-4}$}
 & $\Lambda_- > $&  339  \\\hline
effective Lagrangian & & &\\ %
 dimension 7 & & &\\
$\Lambda_7^{-6}$ & $ \left(-2.8\apm{1.8}{1.7}\right)\cdot
                                  10^{-18}$ \GeV$^{-6}$
& $\Lambda_7 > $&  880  \\ \hline
effective Lagrangian & derived from $\Lambda_+$ & $\Lambda_6 > $&  1752   \\
dimension 6 and 8    & derived from $\Lambda_7$ & $\Lambda_8 > $&  24.3   \\
\hline
quantum gravity & & &\\
 &  & $\lambda = +1$: $M_s >$ &  868  \\
 \raisebox{2.2ex}[-2.2ex]{$\lambda/M_s^4$} &
 \raisebox{2.2ex}[-2.2ex]{ $ \left(-0.85\apm{0.54}{0.55}\right)\cdot
                                         10^{-12}$ \GeV$^{-4} $ }
 & $\lambda = -1$: $M_s >$&  1108 \\ \hline
excited electrons& & &\\
 $\mestar (f_\gamma=1) $ & see Figure \ref{gg:fig:ll} & $\mestar >$& 366 \\
 $f_\gamma^2 (\mestar=200$ \GeV) & 
  $-0.17\apm{0.12}{0.12 }$ & 
 \multicolumn{2}{c|}{$f_\gamma/\Lambda <  7.0$ \TeV$^{-1}$} \\ \hline
\end{tabular}
\caption[Fit-results for $\eeggga$] {Results of the fits to the
differential cross-section for $\eeggga$ and the 95\% confidence level
limits on the model parameters.}
\label{gg:tab:results} 
\end{center} 
\end{table}

\chapter{Fermion-Pair Production}
\label{chap:ff}

\newcommand {\dsdc}   {\frac{{\mathrm{d}}\sigma}{{\mathrm{d}}\cos\theta}}
\newcommand {\zprime} {\mathrm{Z^{\prime}}}
\newcommand {\thtzzp} {\mbox{$\Theta_{\mathrm{Z} \mathrm{Z}^{'}}$}}
\newcommand {\MZp}    {\mbox{$\mathrm{M}_{\mathrm{Z}^{'}}$}}
\newcommand {\MZplim} {\mbox{$\mathrm{M}_{\mathrm{Z}^{'}}^{limit}$}}

\section{Introduction}
\label{sec:ff:intro}

The $\LEPII$ data were taken at centre-of-mass energies, $\sqrt{s}$,
increasing from 130 $\GeV$ to 209 $\GeV$. These energies are well
above the Z-pole and the cross-sections for $\eeff$ are significantly
smaller than those at the Z-pole.  The four LEP experiments have made
measurements of the $\eeff$ process over this range of
energies~\cite{Schael:2006wu, Abdallah:2005ph, Achard:2005nb,
Abbiendi:2003dh}, and a combination of these data is discussed in this
section.
 
Initial-state photon radiation is very important in analysing
$\eeff$. If an initial-state photon (or photons) is emitted then the
effective $\ee$ centre-of-mass energy is reduced from $\sqrt{s}$ to a
lower value $\sqrt{\spr}$. The rate of events at a given effective
energy is given by the probability to emit photons times the
cross-section of $\eeff$ at the reduced centre-of-mass energy
$\sqrt{\spr}$.  For the case when $\sqrt{\spr}\simeq\MZ$,
corresponding to a photon energy of $E_\gamma =
(s-\MZ^2)/(2\sqrt{s})$, the rate becomes very large. This part, which
is called radiative return to the Z, is thus important in both the
event selection and the analysis of $\eeff$. For the studies reported
in this section only events with a small amount of initial state
radiation, \ie, large $\sqrt{\spr/s}$, are retained.

The cross-section for $\eeee$ is considerably larger than those of
$\eemumu$ and $\eetautau$ because of the additional Feynman diagrams
involving $t$-channel photon and Z exchange. The low angle $\eeee$
Bhabha scattering process is used to determine the luminosity.

In the years 1995 through 1999 LEP delivered luminosity at a number of
distinct centre-of-mass energy points. In 2000 most of the luminosity
was delivered close to two distinct energies, but there was also a
significant fraction of the luminosity delivered in more or less a
continuum of energies. To facilitate the combination of the
fermion-pair measurements, the four LEP experiments divided the data
collected in 2000 into two energy bins: from 202.5 to 205.5 $\GeV$,
and above 205.5 $\GeV$.  The nominal and actual centre-of-mass
energies to which the LEP data are averaged for each year are given in
Table~\ref{tab:ff:ecms}.

\begin{table}[ht]
 \begin{center}
 \begin{tabular}{|c||c|c|c|}
  \hline
   Year & Nominal Energy & Actual Energy & Luminosity \\
        &     $\GeV$     &    $\GeV$     &  pb$^{-1}$ \\
  \hline
  \hline
   1995 &      130       &    130.1      & $  3 $ \\
        &      136       &    136.1      & $  3 $ \\
  \hline
   1996 &      161       &    161.3      & $ 10 $ \\
        &      172       &    172.1      & $ 10 $ \\
  \hline
   1997 &      130       &    130.1      & $  2 $ \\
        &      136       &    136.1      & $  2 $ \\
        &      183       &    182.7      & $ 50 $ \\
  \hline
   1998 &      189       &    188.6      & $170 $ \\
  \hline
   1999 &      192       &    191.6      & $ 30 $ \\
        &      196       &    195.5      & $ 80 $ \\
        &      200       &    199.5      & $ 80 $ \\
        &      202       &    201.7      & $ 40 $ \\
  \hline
   2000 &      205       &    204.9      & $ 80 $ \\
        &      207       &    206.5      & $140 $ \\
  \hline
 \end{tabular}
 \end{center}
 \caption{The nominal and actual average centre-of-mass energies for
          data collected during $\LEPII$ operation in each year. The
          approximate average integrated luminosity analysed per
          experiment at each energy is also shown.  }
 \label{tab:ff:ecms}
\end{table}

A number of measurements of the process $\eeff$ exist and are
combined.  The averages of cross-section and forward-backward
asymmetry measurements are discussed in Section~\ref{sec:ff:ave-xsc-afb}.  In Section~\ref{sec:ff:dsdc} the averages
of the differential cross-section measurements, $\dsdc$, for the
channels $\eemumu$ and $\eetautau$ are presented; similar averages for
differential cross-sections for $\eeee$ are given in
Section~\ref{sec:ff:dsdc-ee}.
In Section~\ref{sec:ff:interp} the combined results are interpreted in
terms of contact interactions, the exchange of $\zprime$ bosons, the
exchange of leptoquarks or squarks and the exchange of gravitons in
large-extra-dimensions scenarios.  The results are summarised in
Section~\ref{sec:ff:conc}.

The uncorrelated systematic errors on the input measurements have been
separated from the statistical errors, allowing the decomposition of
the errors on the averages into statistical and systematic components.
Multiplicative corrections have been used to correct measurements to
the full solid angle or full $s'$ region of the common signal
definition.  Additional errors have been included to account for
uncertainties in these corrections.

Where comparisons with Standard Model (SM) predictions are performed,
the predictions are calculated using ZFITTER~\cite{\ZFITTERref}
version 6.36 with the following input parameters:

\begin{eqnarray}
  \MZ           &= &91.1875~\GeV \\
  \Mt           &= &170.9~\GeV \\
  \MH           &= &150~\GeV \\
  \dalhad       &= &0.02758 \\
  \alfas(\MZ)   &= &0.118 \,.
\end{eqnarray}

\section{Averages for Cross-Sections and Asymmetries}
\label{sec:ff:ave-xsc-afb}

In this section the results of the combination of cross-sections and
asymmetries are given. The individual experiments' analyses of
cross-sections and forward-backward asymmetries are presented in a
number of publications~\cite{bib:ff:aleph_130-183,
*bib:ff:aleph_189-209, bib:ff:delphi_130-172, *bib:ff:delphi_183-189,
*bib:ff:delphi_130-207, bib:ff:l3_161-172, *bib:ff:l3_130-189,
*bib:ff:l3_192-208, bib:ff:opal_161, *bib:ff:opal_130-172,
*bib:ff:opal_183, *bib:ff:opal_189, *bib:ff:opal_189-209}.
Cross-section results are combined for the $\eeqq$, $\eemumu$ and
$\eetautau$ channels, forward-backward asymmetry measurements are
combined for the $\mumu$ and $\tautau$ final states.  Events are
classified according to the effective centre-of-mass energy,
$\sqrt{\spr}$.  The averages are made for the samples of events with
high effective centre-of-mass energies.

Individual experiments study different \ff\ signal definitions;
corrections are applied to bring the measurements to the common signal
definition:\footnote{ZFITTER flags BOXD=2, CONV=2, FINR=0, INTF=0,
ALEM=2.}

\begin{figure}[ht]
 \begin{center}
 \mbox{\epsfig{file=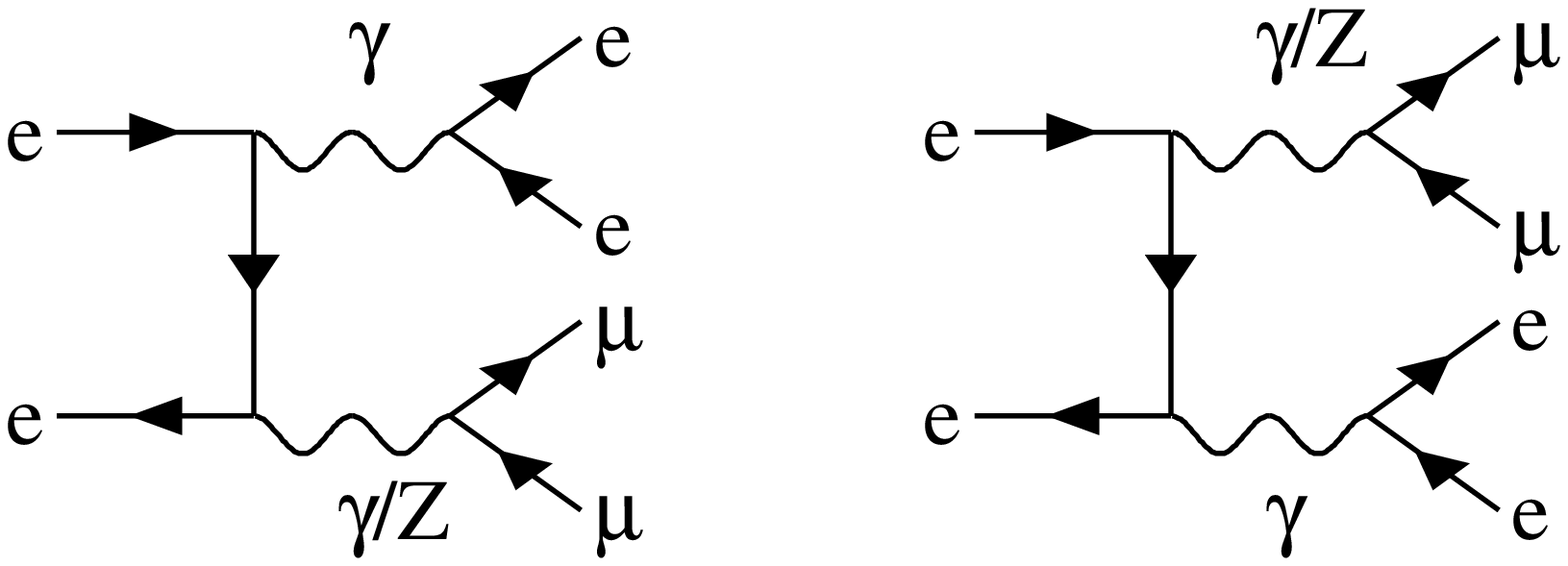,width=0.8\textwidth}}
 \end{center}
 \caption{Diagrams leading to the production of initial state
          non-singlet electron-positron pairs in $\eemumu$, which are
          considered as signal in the common signal definition.}
\label{fig:ff:isnspairs}
\end{figure}

\begin{itemize}
\item $\sqrt{\spr}$ is taken to be the mass of the $s$-channel
      propagator, with the $\ff$ signal being defined by the cut
      $\sqrt{\spr/s} > 0.85$.

\item ISR-FSR photon interference is subtracted to render the
      propagator mass unambiguous.

\item Results are given for the full $4\pi$ angular acceptance.

\item Initial state non-singlet diagrams \cite{bib:ff:lepffwrkshp},
      see for example Figure~\ref{fig:ff:isnspairs}, which lead to
      events containing additional fermion pairs are considered as
      part of the two-fermion signal. In such events, the additional
      fermion pairs are typically lost down the beampipe of the
      experiments, such that the visible event topologies are usually
      similar to difermion events with photons radiated from the
      initial state.
\end{itemize}
The corrections to the common signal definition were applied in two
stages. First, for any measurement which used a restricted angular
range or $s'$ cut different from the default, a multiplicative
correction was applied to the measurement, the associated errors, and
the associated SM prediction to correct the acceptance to $4\pi$ and
to the common $s'$ cut.  These corrections were calculated with
ZFITTER for each centre-of-mass energy value. Although these
corrections are sizeable, up to 14\%, they are expected to be well
modelled.  In the second stage an additive correction was used to
correct for any other differences in signal definition (\eg, use of a
different $s'$ definition, inclusion of interference between initial-
and final-state radiation, treatment of four-fermion contribution) and
centre-of-mass energy.  The additive correction is simply the
difference between the SM prediction calculated using the common
signal definition, at the mean centre-of-mass energy of the
measurements, and that provided by the experiment (corrected for
acceptance where necessary).

Uncertainties derived from a comparison of ZFITTER with
KK2f~\cite{bib:ff:KK, *bib:ff:CEEX} are included; these are shown in
Table~\ref{tab:ff:zfkkerr}.  Additional errors are also included to
account for those cases where the SM prediction provided by the
experiment had used a version of ZFITTER other than the default one,
or different parameters; these are shown in Table~\ref{tab:ff:zferr}.
The inclusion of these errors has a very small effect on the averages.
The hadronic cross-sections change by less than $0.02\%$, the leptonic
cross-sections by less than $0.1\%$ and typically $0.05\%$ and the
lepton asymmetries by 0.001.

\begin{table}[t]
\begin{center}
\begin{tabular}{|l||l|l|l|l|l|}
\hline
                     &$\sigma(\qq)$     &$\sigma(\mumu)$  &$\sigma(\tautau)$  
                     &$\Afb(\mumu)$       &$\Afb(\tautau)$ \\ 
\hline
\hline
\cost\ cut           &0.0005    &0.0005   &0.0005    &--        &--    \\
$s'$ cut             &0.0015    &0.0005   &0.0005    &--        &--    \\
$s'$ definition      &0.002     &0.001    &0.001     &0.0002    &0.0002 \\
ISR-FSR interference &0.002     &0.002    &0.002     &0.005     &0.005  \\
\hline
\end{tabular}
\caption{Errors derived from a comparison between ZFITTER and KK2f for
         variations on the standard signal definition.  Values for
         cross-sections are given as a fraction of the corresponding
         cross-section; those for asymmetries are absolute.}
\label{tab:ff:zfkkerr}
\end{center}
\end{table}

\begin{table}[t]
\begin{center}
\begin{tabular}{|l||c|c|c|c|c|c|}
\hline
Expt. &Energies  &$\sigma(\qq)$  &$\sigma(\mumu)$  &$\sigma(\tautau)$  
                 &$\Afb(\mumu)$     &$\Afb(\tautau)$ \\ 
\hline
\hline
ALEPH   &130--183   &0.002   &0.005   &0.005   &0.001   &0.001   \\
        &189--207   &0.002   &0.003   &0.003   &0.0006  &0.0006  \\ \hline
DELPHI  &130--207   &0.00015 &0.00007 &0.00007 &0.00002 &0.00002 \\ \hline
L3      &130--189   &0.002   &0.005   &0.005   &0.005   &0.005   \\
        &192--207   &0.002   &0.003   &0.003   &0.002   &0.002   \\ \hline
OPAL    &130--207   &0.00005 &0.00005 &0.00005 &--      &--      \\ \hline
\end{tabular}
\caption{Errors applied to account for uncertainties on the ZFITTER
         predictions quoted by each experiment, depending on ZFITTER
         version and parameter settings used by each experiment.
         Values for cross-sections are given as a fraction of the
         corresponding cross-section; those for asymmetries are
         absolute.}
\label{tab:ff:zferr}
\end{center}
\end{table}

Theoretical uncertainties associated with the SM predictions for each
of the measurements are not included during the averaging procedure,
but must be included when assessing the compatibility of the data with
theoretical predictions.  The theoretical uncertainties on the SM
predictions amount to $0.26\%$ on $\sigma(\qq)$, $0.4\%$ on
$\sigma(\mumu)$ and $\sigma(\tautau)$, $2\%$ on $\sigma(\ee)$, and
0.004 on the leptonic forward-backward
asymmetries~\cite{bib:ff:lepffwrkshp}.

The average is performed using the best linear unbiased estimator
technique (BLUE)~\cite{BLUE:1988, *BLUE:2003}, which is equivalent to
a $\chi^2$ minimisation.  All data from the nominal centre-of-mass
energy points are averaged at the same time.

Particular care is taken to ensure that the correlations between the
hadronic cross-sections are reasonably estimated.  The errors are
broken down into six categories, with the ensuing correlations
accounted for in the combinations:
\begin{itemize}
\item The statistical uncertainty.
\item The systematic uncertainty for the final state X which is fully
correlated between energy points for that experiment.
\item The systematic uncertainty for experiment Y which is fully
correlated between different final states for this energy point.
\item The systematic uncertainty for the final state X which is fully
correlated between energy points and between different experiments.
\item The systematic uncertainty which is fully correlated between
energy points and between different experiments for all final states.
\item The uncorrelated systematic uncertainty.
\end{itemize}
The measurements used in the combination are presented in
Appendix~\ref{app:2F:inputs}, using this decomposition of the
uncertainties.  Uncertainties in the hadronic cross-sections arising
from fragmentation models and modelling of ISR are treated as fully
correlated between experiments. Despite some differences between the
models used and the methods of evaluating the errors in the different
experiments, there are significant common elements in the estimation
of these sources of uncertainty.

Table~\ref{tab:ff:xsafbres} gives the averaged cross-sections and
forward-backward asymmetries for all energies.  The $\chidf$ for the
average of the $\LEPII$ $\ff$ data is $163/180$, corresponding to a
$\chi^2$ probability of 81\%.  Most correlations are rather small,
with the largest components at any given pair of energies being those
between the hadronic cross-sections. The other off-diagonal terms in
the correlation matrix are smaller than $10\%$. The correlation matrix
between the averaged hadronic cross-sections at different
centre-of-mass energies is given in Table~\ref{tab:ff:hadcorrel}.

\begin{figure}[p]
 \begin{center}
 \mbox{\epsfig{file=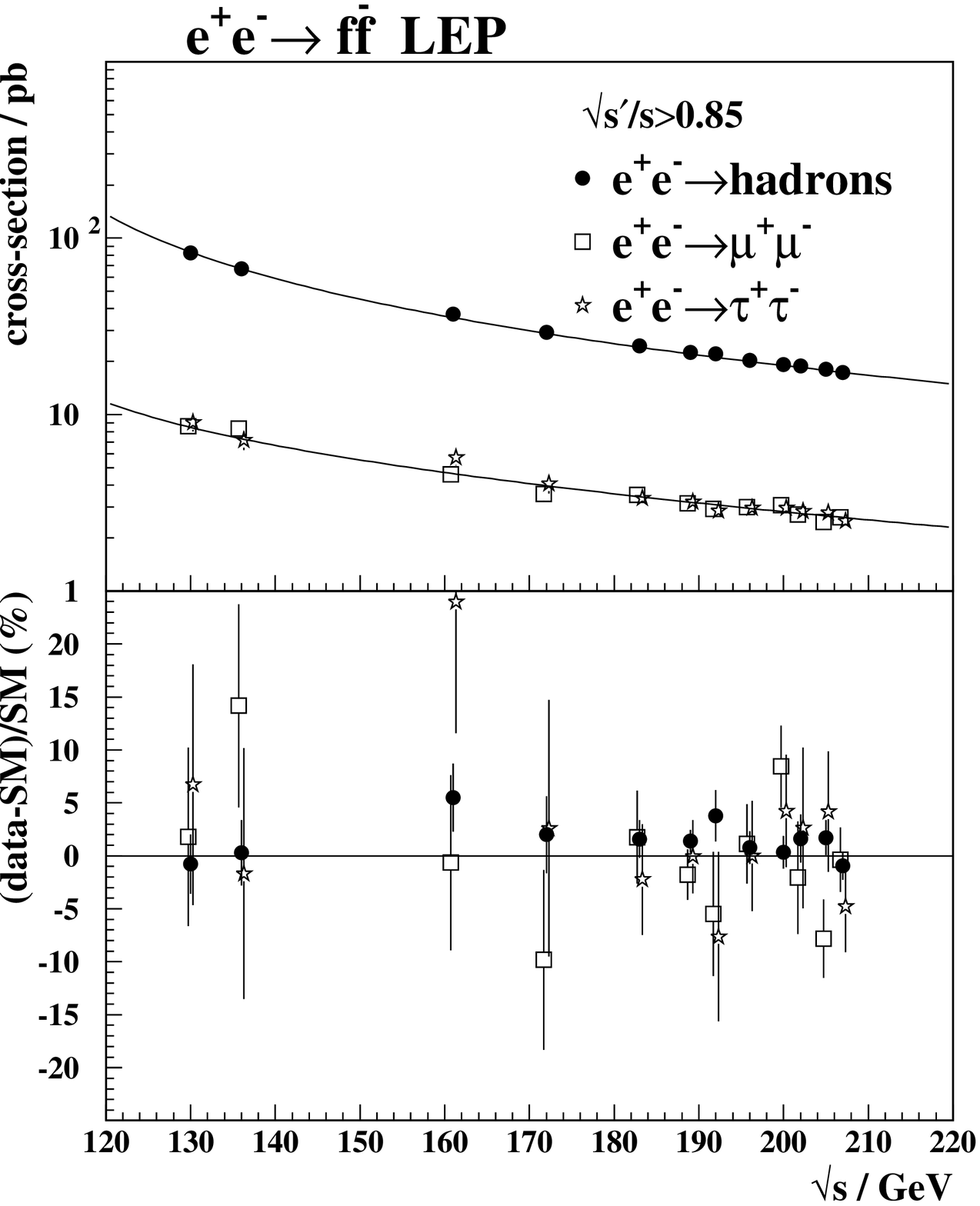,width=0.9\textwidth}}
 \end{center}
 \caption{Combined LEP results on the cross-sections for 
          $\qq$, $\mumu$ and $\tautau$ final states, as a function of 
          centre-of-mass energy. The expectations of the SM, 
          computed with ZFITTER, are shown as curves.
          The lower plot shows the difference between the data and the SM.}
\label{ff:fig-xs_lep}
\end{figure}

\begin{figure}[p]
 \begin{center}
 \mbox{\epsfig{file=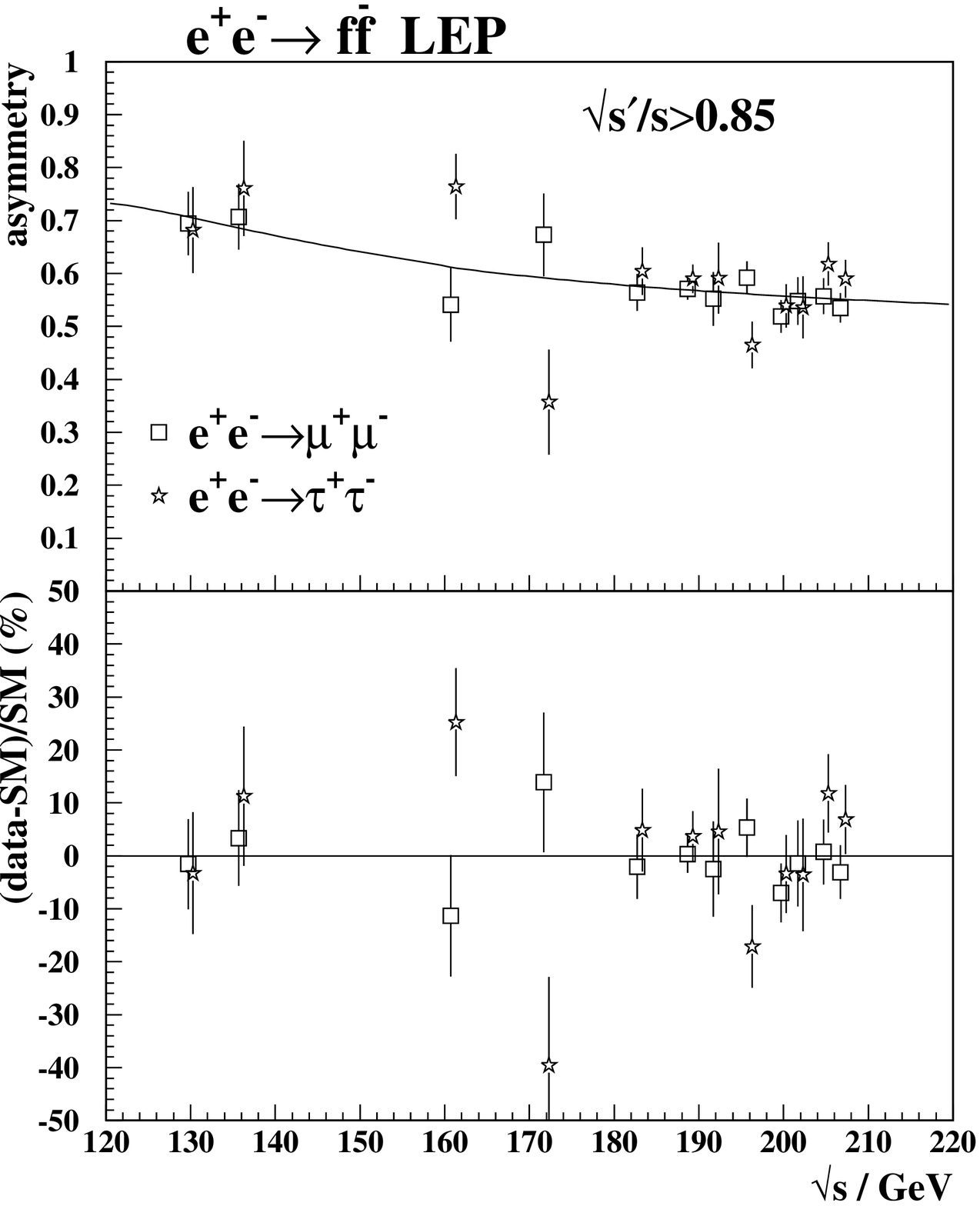,width=0.9\textwidth}}
 \end{center}
 \caption{Combined LEP results on the forward-backward 
          asymmetry for $\mumu$  and $\tautau$ final states as a function of 
          centre-of-mass energy. The expectations of the SM 
          computed  with ZFITTER, are shown as 
          curves. The lower plot shows differences between the data 
          and the SM.}
 \label{ff:fig-afb_lep}
\end{figure}

Figures~\ref{ff:fig-xs_lep} and~\ref{ff:fig-afb_lep} show the LEP
averaged cross-sections and asymmetries, respectively, as a function
of the centre-of-mass energy, together with the SM predictions.  There
is good agreement between the SM expectations and the measurements of
the individual experiments and the combined averages.  The ratios of
the measured cross-sections and asymmetries to the SM expectations,
averaged over all energies and taking into account the correlations
between the data points and the theoretical errors on the SM
predictions, are given in Table~\ref{tab:ff:smrat}.  It is concluded
that there is no evidence in the results of the combinations of the
cross-sections and lepton forward-backward asymmetries for physics
beyond the SM in the process $\eeff$, for f = q, $\mu$ or $\tau$.

\begin{table}[p]
\begin{center}
 \begin{tabular}{|l||c|r@{$\pm$}r@{$\pm$}r|r||c|r@{$\pm$}r@{$\pm$}r|r|}
 \hline
 & $\sqrt{s}$ &\multicolumn{3}{c|}{Average} &  & $\sqrt{s}$  &\multicolumn{3}{c|}{Average} & \\
 Quantity & ($\GeV$) & \multicolumn{3}{c|}{value}  & \multicolumn{1}{c||}{SM} & ($\GeV$) &\multicolumn{3}{c|}{value}  & \multicolumn{1}{c|}{SM} \\
\hline
\hline
 $\sigma(q\overline{q})$              &  130 &   82.445 &  2.197 &  0.766 & 83.090  &   192    & 22.064 &  0.507 &  0.107 & 21.259 \\
 $\sigma(\mu^{+}\mu^{-})$             &  130 &    8.606 &  0.699 &  0.131 &  8.455  &   192    &  2.926 &  0.181 &  0.018 &  3.096 \\
 $\sigma(\tau^{+}\tau^{-})$           &  130 &    9.020 &  0.944 &  0.175 &  8.452  &   192    &  2.860 &  0.246 &  0.032 &  3.096 \\
 $\mathrm{A_{fb}}(\mu^{+}\mu^{-})$    &  130 &    0.694 &  0.059 &  0.012 &  0.705  &   192    &  0.551 &  0.051 &  0.007 &  0.566 \\
 $\mathrm{A_{fb}}(\tau^{+}\tau^{-})$  &  130 &    0.682 &  0.079 &  0.016 &  0.705  &   192    &  0.590 &  0.067 &  0.008 &  0.565 \\
\hline
 $\sigma(q\overline{q})$              &  136 &   66.984 &  1.954 &  0.630 & 66.787  &   196    & 20.307 &  0.294 &  0.096 & 20.148 \\
 $\sigma(\mu^{+}\mu^{-})$             &  136 &    8.325 &  0.692 &  0.109 &  7.292  &   196    &  2.994 &  0.110 &  0.018 &  2.961 \\
 $\sigma(\tau^{+}\tau^{-})$           &  136 &    7.167 &  0.851 &  0.143 &  7.290  &   196    &  2.961 &  0.152 &  0.029 &  2.961 \\
 $\mathrm{A_{fb}}(\mu^{+}\mu^{-})$    &  136 &    0.707 &  0.061 &  0.011 &  0.684  &   196    &  0.592 &  0.030 &  0.005 &  0.562 \\
 $\mathrm{A_{fb}}(\tau^{+}\tau^{-})$  &  136 &    0.761 &  0.089 &  0.013 &  0.684  &   196    &  0.464 &  0.044 &  0.008 &  0.561 \\
\hline
 $\sigma(q\overline{q})$              &  161 &   37.166 &  1.063 &  0.398 & 35.234  &   200    & 19.170 &  0.283 &  0.095 & 19.105 \\
 $\sigma(\mu^{+}\mu^{-})$             &  161 &    4.580 &  0.376 &  0.062 &  4.610  &   200    &  3.072 &  0.108 &  0.018 &  2.833 \\
 $\sigma(\tau^{+}\tau^{-})$           &  161 &    5.715 &  0.553 &  0.139 &  4.610  &   200    &  2.952 &  0.148 &  0.029 &  2.832 \\
 $\mathrm{A_{fb}}(\mu^{+}\mu^{-})$    &  161 &    0.542 &  0.069 &  0.012 &  0.610  &   200    &  0.519 &  0.031 &  0.005 &  0.558 \\
 $\mathrm{A_{fb}}(\tau^{+}\tau^{-})$  &  161 &    0.764 &  0.061 &  0.013 &  0.610  &   200    &  0.539 &  0.041 &  0.007 &  0.558 \\
\hline
 $\sigma(q\overline{q})$              &  172 &   29.350 &  0.989 &  0.336 & 28.775  &   202    & 18.873 &  0.408 &  0.098 & 18.569 \\
 $\sigma(\mu^{+}\mu^{-})$             &  172 &    3.562 &  0.331 &  0.058 &  3.950  &   202    &  2.709 &  0.146 &  0.017 &  2.766 \\
 $\sigma(\tau^{+}\tau^{-})$           &  172 &    4.053 &  0.469 &  0.092 &  3.950  &   202    &  2.838 &  0.208 &  0.022 &  2.765 \\
 $\mathrm{A_{fb}}(\mu^{+}\mu^{-})$    &  172 &    0.673 &  0.077 &  0.012 &  0.591  &   202    &  0.547 &  0.045 &  0.005 &  0.556 \\
 $\mathrm{A_{fb}}(\tau^{+}\tau^{-})$  &  172 &    0.357 &  0.098 &  0.013 &  0.591  &   202    &  0.535 &  0.058 &  0.009 &  0.556 \\
\hline
 $\sigma(q\overline{q})$              &  183 &   24.599 &  0.393 &  0.182 & 24.215  &   205    & 18.137 &  0.282 &  0.087 & 17.832 \\
 $\sigma(\mu^{+}\mu^{-})$             &  183 &    3.505 &  0.145 &  0.042 &  3.444  &   205    &  2.464 &  0.098 &  0.015 &  2.673 \\
 $\sigma(\tau^{+}\tau^{-})$           &  183 &    3.367 &  0.174 &  0.049 &  3.444  &   205    &  2.783 &  0.149 &  0.028 &  2.672 \\
 $\mathrm{A_{fb}}(\mu^{+}\mu^{-})$    &  183 &    0.564 &  0.034 &  0.008 &  0.576  &   205    &  0.556 &  0.034 &  0.004 &  0.553 \\
 $\mathrm{A_{fb}}(\tau^{+}\tau^{-})$  &  183 &    0.604 &  0.044 &  0.011 &  0.576  &   205    &  0.618 &  0.040 &  0.008 &  0.553 \\
\hline
 $\sigma(q\overline{q})$              &  189 &   22.492 &  0.206 &  0.119 & 22.184  &   207    & 17.316 &  0.212 &  0.083 & 17.482 \\
 $\sigma(\mu^{+}\mu^{-})$             &  189 &    3.150 &  0.075 &  0.016 &  3.207  &   207    &  2.618 &  0.078 &  0.014 &  2.628 \\
 $\sigma(\tau^{+}\tau^{-})$           &  189 &    3.204 &  0.107 &  0.032 &  3.206  &   207    &  2.502 &  0.109 &  0.029 &  2.628 \\
 $\mathrm{A_{fb}}(\mu^{+}\mu^{-})$    &  189 &    0.571 &  0.020 &  0.005 &  0.569  &   207    &  0.535 &  0.028 &  0.004 &  0.552 \\
 $\mathrm{A_{fb}}(\tau^{+}\tau^{-})$  &  189 &    0.590 &  0.026 &  0.007 &  0.569  &   207    &  0.590 &  0.034 &  0.010 &  0.552 \\
\hline
\end{tabular}
\end{center}
\caption{Combined LEP results for the $\eeff$ cross-sections (in pb)
 and forward-backward asymmetries; in each case the first error is
 statistical and the second systematic.  The SM predictions are from
 ZFITTER.}
\label{tab:ff:xsafbres}
\end{table}
\begin{table}[htbp]
{\small
 \begin{center}
 \begin{tabular}{|c||c|c|c|c|c|c|c|c|c|c|c|c|}
 \hline
 $\roots$ & \multicolumn{12}{|c|}{$\roots~(\GeV)$} \\
 \cline{2-13}
 $(\GeV)$ 
     & 130    & 136    & 161    & 172    & 183    & 189    & 192    & 196    & 200    & 202    & 205    & 207    \\
 \hline\hline
 130 &  1.000 &  0.060 &  0.065 &  0.058 &  0.104 &  0.112 &  0.043 &  0.065 &  0.065 &  0.045 &  0.061 &  0.076 \\
 136 &  0.060 &  1.000 &  0.061 &  0.055 &  0.098 &  0.104 &  0.040 &  0.061 &  0.061 &  0.042 &  0.057 &  0.071 \\
 161 &  0.065 &  0.061 &  1.000 &  0.060 &  0.108 &  0.117 &  0.044 &  0.067 &  0.068 &  0.047 &  0.063 &  0.078 \\
 172 &  0.058 &  0.055 &  0.060 &  1.000 &  0.096 &  0.103 &  0.039 &  0.060 &  0.060 &  0.041 &  0.056 &  0.069 \\
 183 &  0.104 &  0.098 &  0.108 &  0.096 &  1.000 &  0.205 &  0.078 &  0.120 &  0.121 &  0.084 &  0.114 &  0.140 \\
 189 &  0.112 &  0.104 &  0.117 &  0.103 &  0.205 &  1.000 &  0.097 &  0.149 &  0.151 &  0.105 &  0.141 &  0.174 \\
 192 &  0.043 &  0.040 &  0.044 &  0.039 &  0.078 &  0.097 &  1.000 &  0.060 &  0.061 &  0.042 &  0.057 &  0.071 \\
 196 &  0.065 &  0.061 &  0.067 &  0.060 &  0.120 &  0.149 &  0.060 &  1.000 &  0.094 &  0.066 &  0.089 &  0.110 \\
 200 &  0.065 &  0.061 &  0.068 &  0.060 &  0.121 &  0.151 &  0.061 &  0.094 &  1.000 &  0.067 &  0.090 &  0.112 \\
 202 &  0.045 &  0.042 &  0.047 &  0.041 &  0.084 &  0.105 &  0.042 &  0.066 &  0.067 &  1.000 &  0.063 &  0.079 \\
 205 &  0.061 &  0.057 &  0.063 &  0.056 &  0.114 &  0.141 &  0.057 &  0.089 &  0.090 &  0.063 &  1.000 &  0.106 \\
 207 &  0.076 &  0.071 &  0.078 &  0.069 &  0.140 &  0.174 &  0.071 &  0.110 &  0.112 &  0.079 &  0.106 &  1.000 \\
 \hline
 \end{tabular}
 \end{center}
}
\caption{The correlation coefficients between averaged hadronic cross-sections
         at different energies.}
\label{tab:ff:hadcorrel}
\end{table}
\begin{table}[htbp]
\begin{center}
\begin{tabular}{|l||r@{$\pm$}l|c|}
\hline
Channel   &\multicolumn{2}{c|}{Ratio}    &Deviation \\
\hline
\hline
$\sigma(\qq)$     &1.0092 &0.0076  &+1.21   \\
$\sigma(\mumu)$   &0.9936 &0.0141  &$-$0.45 \\
$\sigma(\tautau)$ &1.0005 &0.0203  &+0.02   \\
$\Afb(\mumu)$     &0.9925 &0.0212  &$-$0.35 \\
$\Afb(\tautau)$   &1.0246 &0.0274  &+0.90   \\
\hline
\end{tabular}
\caption{Comparison of measurements to SM predictions for each
         channel. The second column gives the mean ratio of data to
         prediction; the third column gives the numbers of standard
         deviations of the ratio from unity.}
\label{tab:ff:smrat}
\end{center}
\end{table}

\section{Differential Cross-Sections for Muon- and Tau-Pair Final States}
\label{sec:ff:dsdc}

The LEP experiments have measured the differential cross-section,
$\dsdc$, for the $\eemumu$ and $\eetautau$ channels for samples of
events with high effective centre-of-mass energy, $\sqrt{s'/s}>0.85$.
A combination of these results is made using the BLUE technique.  For
some bins the number of observed events is very small, so the
statistical error associated with each measurement is taken as the
expected statistical error on the differential cross-section, computed
from the expected number of events in each bin for each experiment.
Using a Monte-Carlo simulation it has been shown that this method
provides a good approximation to the exact likelihood method based on
Poisson statistics.

The combination includes data from $183~\GeV$ to $207~\GeV$ from
DELPHI and OPAL, data at $189~\GeV$ from L3 and data from $189~\GeV$
to $207~\GeV$ from ALEPH.  Each experiment's data are binned in 10
bins of $\cos\theta$ at each energy, using their own signal
definition. The polar scattering angle, $\theta$, is the angle of the
outgoing negative lepton with respect to the incoming electron
direction in the detector coordinate system. The outer acceptances of
the most forward and most backward bins for which the experiments
present their data are different. This was accounted for as part of
the correction to a common signal definition. The ranges in
$\cos\theta$ for the measurements of the individual experiments and
the average are given in Table~\ref{tab:ff:acpt}.  The signal
definition used corresponded to the definition given in Section
\ref{sec:ff:ave-xsc-afb}.

Correlated systematic errors between different experiments, channels
and energies, arising from uncertainties on the overall normalisation,
are considered in the averaging procedure.  All data from all energies
are combined in a single fit to obtain averages at each centre-of-mass
energy.

The results of the averages are reported in
Tables~\ref{tab:ff:dsdcmmres} and~\ref{tab:ff:dsdcttres} and shown in
Figures~\ref{fig:ff:dsdc-res-mm} and~\ref{fig:ff:dsdc-res-tt}, with
more details summarised in Appendix~\ref{app:2F:mt}.  The correlations
between bins in the average are less that $2\%$ of the total error on
the averages in each bin.  The combination results in a $\chi^2$ of
$352.2$ for $320$ degrees of freedom, corresponding to a probability
of 10.4\%.

\begin{table}[t]
 \begin{center}
 \begin{tabular}{|l||c|c|}
  \hline
  Experiment                   & $\cos\theta_{min}$ & $\cos\theta_{max}$ \\
  \hline
  \hline 
   ALEPH                       &    $-0.95$         &     $0.95$         \\
   DELPHI ($\eemumu$)          &    $-0.97$         &     $0.97$         \\
   DELPHI ($\eetautau$)        &    $-0.96$         &     $0.96$         \\
   L3                          &    $-0.90$         &     $0.90$         \\
   OPAL                        &    $-1.00$         &     $1.00$         \\
  \hline
   Average                     &    $-1.00$         &     $1.00$         \\
  \hline
 \end{tabular}
 \end{center}
 \caption{The acceptances in $\cos\theta$ for which the experimental
          measurements at all energies are presented for combination,
          and the acceptance for the LEP average.  For DELPHI the
          acceptance is shown for the different channels. For ALEPH,
          L3 and OPAL the acceptance is the same for muon and
          tau-lepton channels. }
 \label{tab:ff:acpt}
\end{table}

\begin{table}[p]
\begin{center}
\begin{tabular}{|c||c|r@{$\pm$}r|c||c|r@{$\pm$}r|c|}
 \hline
                 & $\sqrt{s}$ & \multicolumn{2}{c|}{Average} &                         & $\sqrt{s}$ &\multicolumn{2}{c|}{Average} & \\
$\cos\theta$ bin &  ($\GeV$)  & \multicolumn{2}{c|}{value}   & \multicolumn{1}{c||}{SM} & ($\GeV$)   &\multicolumn{2}{c|}{value}   & \multicolumn{1}{c|}{SM} \\
\hline
\hline
  $[-1.00, -0.80]$        &  183 &      0.197   &  0.183   &   0.547  &  200 &   0.558   &  0.113   &   0.501  \\
  $[-0.80, -0.60]$        &  183 &      0.589   &  0.163   &   0.534  &  200 &   0.376   &  0.098   &   0.478  \\
  $[-0.60, -0.40]$        &  183 &      0.807   &  0.174   &   0.627  &  200 &   0.799   &  0.105   &   0.541  \\
  $[-0.40, -0.20]$        &  183 &      1.033   &  0.197   &   0.823  &  200 &   0.817   &  0.118   &   0.689  \\
  $[-0.20, 0.00]$         &  183 &      1.178   &  0.236   &   1.121  &  200 &   1.105   &  0.139   &   0.922  \\
  $[0.00, 0.20]$          &  183 &      1.778   &  0.276   &   1.521  &  200 &   1.462   &  0.162   &   1.239  \\
  $[0.20, 0.40]$          &  183 &      2.143   &  0.315   &   2.020  &  200 &   1.849   &  0.185   &   1.640  \\
  $[0.40, 0.60]$          &  183 &      2.690   &  0.367   &   2.619  &  200 &   2.122   &  0.211   &   2.126  \\
  $[0.60, 0.80]$          &  183 &      2.916   &  0.420   &   3.314  &  200 &   2.947   &  0.239   &   2.694  \\
  $[0.80, 1.00]$          &  183 &      4.368   &  0.529   &   4.096  &  200 &   3.474   &  0.306   &   3.336  \\
\hline
  $[-1.00, -0.80]$        &  189 &      0.614   &  0.080   &   0.532  &  202 &   1.137   &  0.162   &   0.495  \\
  $[-0.80, -0.60]$        &  189 &      0.420   &  0.065   &   0.514  &  202 &   0.295   &  0.139   &   0.471  \\
  $[-0.60, -0.40]$        &  189 &      0.530   &  0.069   &   0.595  &  202 &   0.506   &  0.149   &   0.531  \\
  $[-0.40, -0.20]$        &  189 &      0.651   &  0.077   &   0.772  &  202 &   0.455   &  0.169   &   0.674  \\
  $[-0.20, 0.00]$         &  189 &      1.064   &  0.089   &   1.044  &  202 &   0.860   &  0.197   &   0.900  \\
  $[0.00, 0.20]$          &  189 &      1.313   &  0.111   &   1.411  &  202 &   1.010   &  0.230   &   1.208  \\
  $[0.20, 0.40]$          &  189 &      2.038   &  0.123   &   1.872  &  202 &   1.749   &  0.264   &   1.599  \\
  $[0.40, 0.60]$          &  189 &      2.158   &  0.139   &   2.426  &  202 &   1.844   &  0.299   &   2.072  \\
  $[0.60, 0.80]$          &  189 &      2.954   &  0.158   &   3.072  &  202 &   2.268   &  0.339   &   2.627  \\
  $[0.80, 1.00]$          &  189 &      3.795   &  0.216   &   3.799  &  202 &   3.396   &  0.435   &   3.254  \\
\hline
  $[-1.00, -0.80]$        &  192 &      0.481   &  0.198   &   0.524  &  205 &   0.621   &  0.113   &   0.485  \\
  $[-0.80, -0.60]$        &  192 &      0.384   &  0.173   &   0.504  &  205 &   0.385   &  0.098   &   0.461  \\
  $[-0.60, -0.40]$        &  192 &      0.788   &  0.186   &   0.579  &  205 &   0.382   &  0.104   &   0.517  \\
  $[-0.40, -0.20]$        &  192 &      0.581   &  0.212   &   0.748  &  205 &   0.443   &  0.118   &   0.654  \\
  $[-0.20, 0.00]$         &  192 &      1.324   &  0.248   &   1.008  &  205 &   0.891   &  0.137   &   0.870  \\
  $[0.00, 0.20]$          &  192 &      1.187   &  0.292   &   1.360  &  205 &   1.205   &  0.160   &   1.166  \\
  $[0.20, 0.40]$          &  192 &      1.932   &  0.334   &   1.803  &  205 &   1.614   &  0.183   &   1.542  \\
  $[0.40, 0.60]$          &  192 &      2.080   &  0.379   &   2.337  &  205 &   1.663   &  0.209   &   1.998  \\
  $[0.60, 0.80]$          &  192 &      3.003   &  0.429   &   2.960  &  205 &   2.097   &  0.237   &   2.534  \\
  $[0.80, 1.00]$          &  192 &      3.083   &  0.552   &   3.662  &  205 &   3.318   &  0.306   &   3.140  \\
\hline
  $[-1.00, -0.80]$        &  196 &      0.535   &  0.119   &   0.512  &  207 &   0.518   &  0.087   &   0.481  \\
  $[-0.80, -0.60]$        &  196 &      0.485   &  0.103   &   0.491  &  207 &   0.496   &  0.075   &   0.456  \\
  $[-0.60, -0.40]$        &  196 &      0.668   &  0.111   &   0.560  &  207 &   0.473   &  0.079   &   0.510  \\
  $[-0.40, -0.20]$        &  196 &      0.484   &  0.126   &   0.718  &  207 &   0.781   &  0.089   &   0.643  \\
  $[-0.20, 0.00]$         &  196 &      0.802   &  0.147   &   0.964  &  207 &   0.795   &  0.104   &   0.855  \\
  $[0.00, 0.20]$          &  196 &      1.507   &  0.172   &   1.298  &  207 &   0.995   &  0.121   &   1.145  \\
  $[0.20, 0.40]$          &  196 &      1.657   &  0.197   &   1.720  &  207 &   1.630   &  0.139   &   1.515  \\
  $[0.40, 0.60]$          &  196 &      2.303   &  0.223   &   2.229  &  207 &   2.247   &  0.159   &   1.963  \\
  $[0.60, 0.80]$          &  196 &      2.949   &  0.253   &   2.824  &  207 &   2.491   &  0.179   &   2.489  \\
  $[0.80, 1.00]$          &  196 &      3.272   &  0.325   &   3.495  &  207 &   2.995   &  0.231   &   3.086  \\
\hline
\end{tabular}
\end{center}
\caption{Combined LEP results for the $\eemumu$ differential
 cross-sections, in pb divided by $\Delta(\cos\theta)$.  The combined
 statistical and systematic error is shown.  The SM predictions are
 from ZFITTER.}
\label{tab:ff:dsdcmmres}
\end{table}

\begin{table}[p]
\begin{center}
\begin{tabular}{|c||c|r@{$\pm$}r|c||c|r@{$\pm$}r|c|}
\hline
                 & $\sqrt{s}$ & \multicolumn{2}{c|}{Average} &                          & $\sqrt{s}$ & \multicolumn{2}{c|}{Average} & \\
$\cos\theta$ bin &  ($\GeV$)  & \multicolumn{2}{c|}{value}   & \multicolumn{1}{c||}{SM} &  ($\GeV$)  & \multicolumn{2}{c|}{value}   & \multicolumn{1}{c|}{SM} \\
\hline
\hline
   $[-1.00, -0.80]$        &  183 &     0.302   &  0.351   &  0.548   &  200 &      0.489   &  0.201   &  0.501   \\
   $[-0.80, -0.60]$        &  183 &     0.206   &  0.240   &  0.535   &  200 &      0.619   &  0.141   &  0.478   \\
   $[-0.60, -0.40]$        &  183 &     0.198   &  0.230   &  0.627   &  200 &      0.528   &  0.137   &  0.541   \\
   $[-0.40, -0.20]$        &  183 &     0.542   &  0.254   &  0.823   &  200 &      0.628   &  0.155   &  0.689   \\
   $[-0.20, 0.00]$         &  183 &     1.364   &  0.302   &  1.121   &  200 &      1.067   &  0.186   &  0.922   \\
   $[0.00, 0.20]$          &  183 &     1.519   &  0.350   &  1.521   &  200 &      1.130   &  0.214   &  1.239   \\
   $[0.20, 0.40]$          &  183 &     1.583   &  0.389   &  2.020   &  200 &      1.871   &  0.240   &  1.640   \\
   $[0.40, 0.60]$          &  183 &     2.296   &  0.450   &  2.619   &  200 &      2.043   &  0.274   &  2.125   \\
   $[0.60, 0.80]$          &  183 &     3.954   &  0.574   &  3.313   &  200 &      2.777   &  0.339   &  2.694   \\
   $[0.80, 1.00]$          &  183 &     4.156   &  0.919   &  4.095   &  200 &      3.437   &  0.523   &  3.336   \\
\hline
   $[-1.00, -0.80]$        &  189 &     0.389   &  0.123   &  0.532   &  202 &      0.968   &  0.287   &  0.495   \\
   $[-0.80, -0.60]$        &  189 &     0.379   &  0.093   &  0.515   &  202 &      0.322   &  0.189   &  0.471   \\
   $[-0.60, -0.40]$        &  189 &     0.485   &  0.089   &  0.595   &  202 &      0.420   &  0.194   &  0.531   \\
   $[-0.40, -0.20]$        &  189 &     0.809   &  0.100   &  0.772   &  202 &      0.731   &  0.220   &  0.674   \\
   $[-0.20, 0.00]$         &  189 &     0.848   &  0.118   &  1.044   &  202 &      0.922   &  0.263   &  0.900   \\
   $[0.00, 0.20]$          &  189 &     1.323   &  0.139   &  1.411   &  202 &      0.789   &  0.300   &  1.208   \\
   $[0.20, 0.40]$          &  189 &     1.989   &  0.154   &  1.872   &  202 &      1.953   &  0.341   &  1.599   \\
   $[0.40, 0.60]$          &  189 &     2.445   &  0.179   &  2.426   &  202 &      1.838   &  0.386   &  2.072   \\
   $[0.60, 0.80]$          &  189 &     2.467   &  0.225   &  3.071   &  202 &      3.129   &  0.479   &  2.626   \\
   $[0.80, 1.00]$          &  189 &     4.111   &  0.357   &  3.798   &  202 &      3.186   &  0.747   &  3.254   \\
\hline
   $[-1.00, -0.80]$        &  192 &     0.014   &  0.325   &  0.524   &  205 &      0.363   &  0.203   &  0.486   \\
   $[-0.80, -0.60]$        &  192 &     0.355   &  0.247   &  0.505   &  205 &      0.562   &  0.137   &  0.461   \\
   $[-0.60, -0.40]$        &  192 &     0.479   &  0.245   &  0.580   &  205 &      0.603   &  0.135   &  0.517   \\
   $[-0.40, -0.20]$        &  192 &     0.762   &  0.278   &  0.748   &  205 &      0.443   &  0.154   &  0.654   \\
   $[-0.20, 0.00]$         &  192 &     0.816   &  0.331   &  1.008   &  205 &      0.397   &  0.179   &  0.870   \\
   $[0.00, 0.20]$          &  192 &     1.609   &  0.385   &  1.360   &  205 &      1.242   &  0.209   &  1.166   \\
   $[0.20, 0.40]$          &  192 &     1.810   &  0.433   &  1.803   &  205 &      1.522   &  0.237   &  1.542   \\
   $[0.40, 0.60]$          &  192 &     2.059   &  0.491   &  2.337   &  205 &      1.846   &  0.268   &  1.998   \\
   $[0.60, 0.80]$          &  192 &     2.643   &  0.599   &  2.959   &  205 &      2.045   &  0.330   &  2.533   \\
   $[0.80, 1.00]$          &  192 &     2.575   &  0.935   &  3.661   &  205 &      4.671   &  0.520   &  3.140   \\
\hline
   $[-1.00, -0.80]$        &  196 &     0.810   &  0.211   &  0.513   &  207 &      0.272   &  0.145   &  0.481   \\
   $[-0.80, -0.60]$        &  196 &     0.738   &  0.147   &  0.491   &  207 &      0.412   &  0.106   &  0.456   \\
   $[-0.60, -0.40]$        &  196 &     0.524   &  0.141   &  0.560   &  207 &      0.534   &  0.104   &  0.510   \\
   $[-0.40, -0.20]$        &  196 &     0.688   &  0.162   &  0.718   &  207 &      0.563   &  0.118   &  0.644   \\
   $[-0.20, 0.00]$         &  196 &     0.976   &  0.195   &  0.964   &  207 &      0.683   &  0.140   &  0.855   \\
   $[0.00, 0.20]$          &  196 &     0.977   &  0.225   &  1.298   &  207 &      1.443   &  0.161   &  1.145   \\
   $[0.20, 0.40]$          &  196 &     1.648   &  0.252   &  1.719   &  207 &      1.351   &  0.181   &  1.514   \\
   $[0.40, 0.60]$          &  196 &     1.965   &  0.289   &  2.228   &  207 &      1.761   &  0.207   &  1.962   \\
   $[0.60, 0.80]$          &  196 &     2.269   &  0.357   &  2.823   &  207 &      1.655   &  0.255   &  2.489   \\
   $[0.80, 1.00]$          &  196 &     3.346   &  0.557   &  3.494   &  207 &      3.597   &  0.399   &  3.085   \\
\hline
\end{tabular}
\end{center}
\caption{Combined LEP results for the $\eetautau$ differential
 cross-sections, in pb divided by $\Delta(\cos\theta)$.  The combined
 statistical and systematic error is shown.  The SM predictions are
 from ZFITTER.}
\label{tab:ff:dsdcttres}
\end{table}

\begin{figure}[p]
 \begin{center}
  \epsfig{file=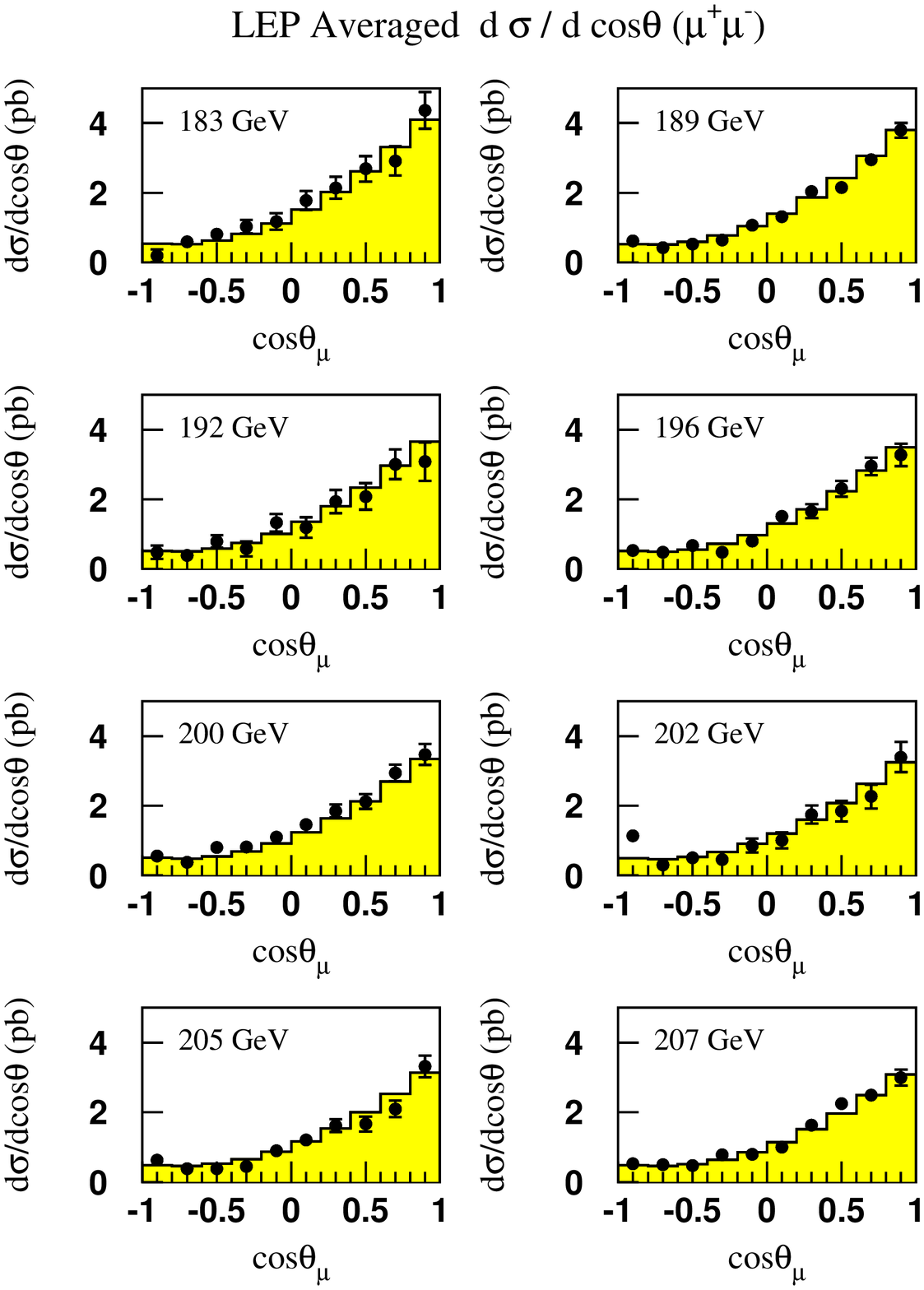,width=0.9\textwidth}
 \end{center}
 \caption{LEP averaged differential cross-sections for $\eemumu$ at
          energies of 183--207 $\GeV$. The SM predictions, shown as
          solid histograms, are computed with ZFITTER.}
 \label{fig:ff:dsdc-res-mm}
 \vskip 2cm 
\end{figure}
\begin{figure}[p]
 \begin{center}
  \epsfig{file=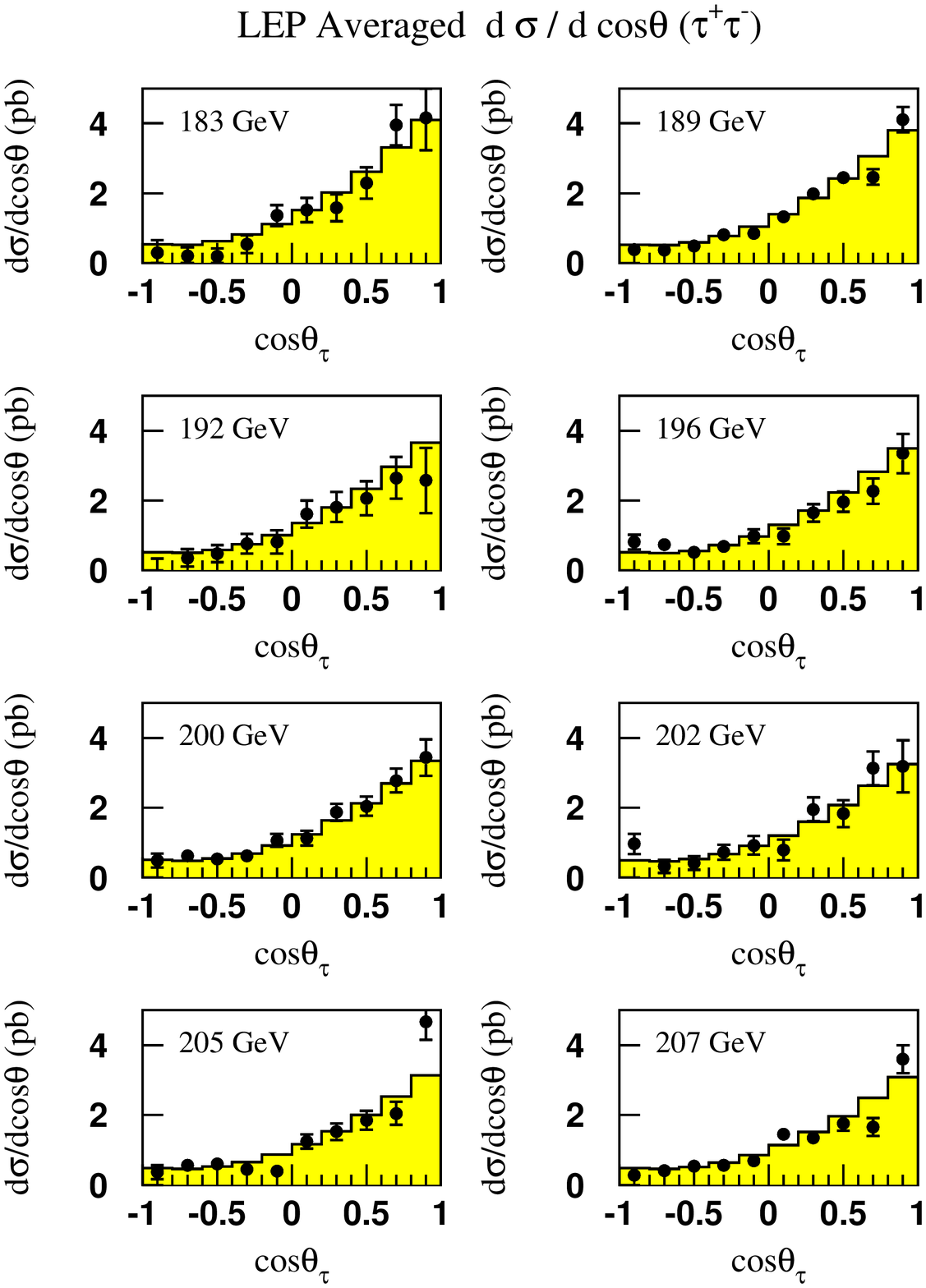,width=0.9\textwidth}
 \end{center}
 \caption{LEP averaged differential cross-sections for $\eetautau$ at
          energies of 183--207 $\GeV$. The SM predictions, shown as
          solid histograms, are computed with ZFITTER.}
 \label{fig:ff:dsdc-res-tt}
 \vskip 2cm 
\end{figure}

\section{Differential Cross-Sections for Electron-Positron Final States}
\label{sec:ff:dsdc-ee}

The LEP experiments have measured the differential cross-section,
$\dsdc$ for the process $\eeee$ with different acollinearity
cuts~\cite{Schael:2006wu, Abdallah:2005ph, Achard:2005nb,
Abbiendi:2003dh}. The results are combined using a $\chi^2$ fit to the
measured differential cross-sections, using the experimental errors as
given by the experiments. In contrast to the muon and tau-lepton
channels, the higher statistics makes the use of expected errors, as
discussed in Section~\ref{sec:ff:dsdc}, unnecessary here.

The combination includes data from 189 to 207 $\GeV$, provided by
ALEPH, DELPHI and OPAL.  Each experiment's data are binned according
to an agreed common definition, which takes into account the large
forward peak of Bhabha scattering:
\begin{itemize}
 \item 10 bins for $\cos\theta$ between  0.0  and 0.90 and
 \item  5 bins for $\cos\theta$ between -0.90 and 0.0
\end{itemize}
at each energy, where the polar scattering angle, $\theta$, is the
angle of the outgoing electron with respect to the incoming electron
direction in the lab coordinate system. Apart from the common binning
in $\cos\theta$, each experiment uses its own signal definition.  The
ranges in $\cos\theta$ covered by the individual experiments and the
range used for the combination are given in
Table~\ref{tab:ff:dsdc-acpt-ee}. The signal definition used for the
LEP average corresponds to an acollinearity cut of $\rm 10^{\circ}$.

Correlated systematic errors between different experiments, energies
and bins at the same energy, arising from uncertainties on the overall
normalisation, and from migration of events between forward and
backward bins with the same absolute value of $\cos\theta$ due to
uncertainties in the corrections for charge confusion, were considered
in the averaging procedure.

An average for all energies between 189 and 207 $\GeV$ was performed.
The results of the averages are reported in
Tables~\ref{tab:ff:dsdceeres1} and~\ref{tab:ff:dsdceeres2} and shown
in Figures~\ref{fig:ff:dsdc-res-ee} and~\ref{fig:ff:dsdc-ratio-ee},
with more details summarised in Appendix~\ref{app:2F:ee}.  The
$\chidf$ for the average is $199.4/189$, corresponding to a
probability of $28.8\%$.

The correlations between bins in the average are well below $5\%$ of
the total error on the averages in each bin for most of the cases, and
around $10\%$ for bins close to the edges of the acceptance.  The
agreement between the averaged data and the predictions from the
Monte-Carlo generator BHWIDE~\cite{bib:ff:BHWIDE} is good, with a
$\chi^2$ of $85$ for $90$ degrees of freedom, corresponding to a
probability of $63\%$. In conclusion, the combined results for the
$\eeee$ channel are compatible with the SM.

\begin{table}[t]
 \begin{center}
 \begin{tabular}{|l||c|c|}
  \hline
  Experiment                       & $\cos\theta_{min}$ & $\cos\theta_{max}$ \\
  \hline
  \hline 
   ALEPH  ($\sqrt{s'/s}>0.85$)     &    $-0.90$         &     $0.90$         \\
   DELPHI (acol. $<\ 20^{\circ}$)  &    $-0.72$         &     $0.72$         \\
   OPAL   (acol. $<\ 10^{\circ}$)  &    $-0.90$         &     $0.90$         \\
  \hline
   Average (acol. $<\ 10^{\circ}$) &    $-0.90$         &     $0.90$         \\
  \hline
 \end{tabular}
 \end{center}
 \caption[Acceptances for combined differential cross-sections for \eeee.]
         {The acceptances for which experimental data are presented 
          for the $\eeee$ channel
          and the acceptance for the LEP average.}
 \label{tab:ff:dsdc-acpt-ee}
\end{table}

\begin{table}[p]
\begin{center}
\begin{tabular}{|c||c|r@{$\pm$}r|r||c|r@{$\pm$}r|r|}
\hline
                 & $\sqrt{s}$ & \multicolumn{2}{c|}{Average} &                          & $\sqrt{s}$ & \multicolumn{2}{c|}{Average} & \\
$\cos\theta$ bin &  ($\GeV$)  & \multicolumn{2}{c|}{value}   & \multicolumn{1}{c||}{SM} &  ($\GeV$)  & \multicolumn{2}{c|}{value}   & \multicolumn{1}{c|}{SM} \\
\hline
\hline
   $[-0.90, -0.72]$        &  189 &      1.401   &   0.161  &    1.590  &  196 &     1.470   &   0.261  &   1.483   \\
   $[-0.72, -0.54]$        &  189 &      2.030   &   0.160  &    1.816  &  196 &     1.527   &   0.221  &   1.695   \\
   $[-0.54, -0.36]$        &  189 &      2.162   &   0.170  &    2.162  &  196 &     2.058   &   0.250  &   2.000   \\
   $[-0.36, -0.18]$        &  189 &      2.298   &   0.186  &    2.681  &  196 &     2.788   &   0.284  &   2.498   \\
   $[-0.18, 0.00]$         &  189 &      4.321   &   0.230  &    3.906  &  196 &     3.646   &   0.318  &   3.610   \\
   $[0.00, 0.09]$          &  189 &      4.898   &   0.348  &    5.372  &  196 &     5.887   &   0.521  &   4.999   \\
   $[0.09, 0.18]$          &  189 &      6.090   &   0.404  &    6.892  &  196 &     6.233   &   0.591  &   6.406   \\
   $[0.18, 0.27]$          &  189 &      8.838   &   0.476  &    9.610  &  196 &     9.016   &   0.694  &   8.832   \\
   $[0.27, 0.36]$          &  189 &     12.781   &   0.576  &   13.345  &  196 &    13.444   &   0.856  &  12.326   \\
   $[0.36, 0.45]$          &  189 &     19.586   &   0.707  &   19.445  &  196 &    18.568   &   0.977  &  18.039   \\
   $[0.45, 0.54]$          &  189 &     30.598   &   0.895  &   30.476  &  196 &    27.056   &   1.223  &  28.300   \\
   $[0.54, 0.63]$          &  189 &     50.488   &   1.135  &   51.012  &  196 &    49.391   &   1.619  &  47.362   \\
   $[0.63, 0.72]$          &  189 &     95.178   &   1.520  &   95.563  &  196 &    88.163   &   2.154  &  88.473   \\
   $[0.72, 0.81]$          &  189 &    211.427   &   2.900  &  212.390  &  196 &   197.369   &   4.121  & 198.250   \\
   $[0.81, 0.90]$          &  189 &    679.146   &   5.773  &  689.989  &  196 &   637.846   &   8.003  & 642.688   \\
				   \hline
   $[-0.90, -0.72]$        &  192 &      1.300   &   0.364  &   1.539   &  200 &     1.483   &   0.245  &   1.420   \\
   $[-0.72, -0.54]$        &  192 &      2.099   &   0.419  &   1.754   &  200 &     1.638   &   0.214  &   1.623   \\
   $[-0.54, -0.36]$        &  192 &      1.871   &   0.385  &   2.091   &  200 &     2.068   &   0.227  &   1.885   \\
   $[-0.36, -0.18]$        &  192 &      1.808   &   0.422  &   2.604   &  200 &     2.362   &   0.250  &   2.409   \\
   $[-0.18, 0.00]$         &  192 &      3.800   &   0.519  &   3.778   &  200 &     4.251   &   0.313  &   3.435   \\
   $[0.00, 0.09]$          &  192 &      5.015   &   0.891  &   5.205   &  200 &     5.244   &   0.506  &   4.770   \\
   $[0.09, 0.18]$          &  192 &      5.695   &   0.976  &   6.692   &  200 &     5.888   &   0.571  &   6.157   \\
   $[0.18, 0.27]$          &  192 &      9.239   &   1.175  &   9.242   &  200 &     8.244   &   0.667  &   8.471   \\
   $[0.27, 0.36]$          &  192 &     12.941   &   1.414  &  12.800   &  200 &     9.506   &   0.736  &  11.773   \\
   $[0.36, 0.45]$          &  192 &     20.761   &   1.807  &  18.776   &  200 &    16.376   &   0.920  &  17.262   \\
   $[0.45, 0.54]$          &  192 &     26.466   &   2.074  &  29.471   &  200 &    27.000   &   1.214  &  27.117   \\
   $[0.54, 0.63]$          &  192 &     49.382   &   2.671  &  49.338   &  200 &    44.614   &   1.537  &  45.607   \\
   $[0.63, 0.72]$          &  192 &     89.676   &   3.615  &  92.079   &  200 &    86.454   &   2.060  &  85.143   \\
   $[0.72, 0.81]$          &  192 &    204.579   &   6.760  & 206.087   &  200 &   190.962   &   3.941  & 190.786   \\
   $[0.81, 0.90]$          &  192 &    655.724   &  12.588  & 669.173   &  200 &   604.986   &   7.608  & 617.718   \\
\hline
\end{tabular}
\end{center}
\caption{Combined LEP results for the $\eeee$ differential
 cross-sections, in pb divided by $\Delta(\cos\theta)$, for $\sqrt{s}$
 between $189~\GeV$ and $200~\GeV$.  The combined statistical and
 systematic error is shown.  The SM predictions are from BHWIDE}
\label{tab:ff:dsdceeres1}
\end{table}

\begin{table}[p]
\begin{center}
\begin{tabular}{|c||c|r@{$\pm$}r|r||c|r@{$\pm$}r|r|}
\hline
                 & $\sqrt{s}$ & \multicolumn{2}{c|}{Average} &                          & $\sqrt{s}$ & \multicolumn{2}{c|}{Average} & \\
$\cos\theta$ bin &  ($\GeV$)  & \multicolumn{2}{c|}{value}   & \multicolumn{1}{c||}{SM} &  ($\GeV$)  & \multicolumn{2}{c|}{value}   & \multicolumn{1}{c|}{SM} \\
\hline
\hline
    $[-0.90, -0.72]$        &  202 &    1.568    &   0.368  &    1.401  &  207 &    1.440    &   0.196  &    1.339  \\
    $[-0.72, -0.54]$        &  202 &    1.344    &   0.276  &    1.579  &  207 &    1.426    &   0.163  &    1.517  \\
    $[-0.54, -0.36]$        &  202 &    2.107    &   0.345  &    1.836  &  207 &    1.889    &   0.177  &    1.745  \\
    $[-0.36, -0.18]$        &  202 &    3.240    &   0.406  &    2.361  &  207 &    2.156    &   0.198  &    2.240  \\
    $[-0.18, 0.00]$         &  202 &    2.911    &   0.394  &    3.356  &  207 &    3.215    &   0.233  &    3.194  \\
    $[0.00, 0.09]$          &  202 &    4.603    &   0.628  &    4.669  &  207 &    4.434    &   0.357  &    4.380  \\
    $[0.09, 0.18]$          &  202 &    6.463    &   0.861  &    6.017  &  207 &    6.393    &   0.463  &    5.729  \\
    $[0.18, 0.27]$          &  202 &    7.457    &   0.957  &    8.320  &  207 &    6.951    &   0.481  &    7.972  \\
    $[0.27, 0.36]$          &  202 &   11.032    &   1.113  &   11.554  &  207 &   11.221    &   0.615  &   11.019  \\
    $[0.36, 0.45]$          &  202 &   16.428    &   1.338  &   16.891  &  207 &   15.933    &   0.739  &   16.053  \\
    $[0.45, 0.54]$          &  202 &   27.153    &   1.643  &   26.583  &  207 &   25.676    &   0.923  &   25.254  \\
    $[0.54, 0.63]$          &  202 &   46.490    &   2.214  &   44.786  &  207 &   42.075    &   1.188  &   42.456  \\
    $[0.63, 0.72]$          &  202 &   87.253    &   2.887  &   83.473  &  207 &   77.611    &   1.569  &   79.639  \\
    $[0.72, 0.81]$          &  202 &  189.026    &   5.516  &  186.904  &  207 &  173.825    &   3.002  &  178.042  \\
    $[0.81, 0.90]$          &  202 &  599.860    &  10.339  &  605.070  &  207 &  573.637    &   6.024  &  576.688  \\
\hline
    $[-0.90, -0.72]$        &  205 &    1.102    &   0.205  &    1.355  & \multicolumn{4}{c}{ } \\
    $[-0.72, -0.54]$        &  205 &    1.470    &   0.195  &    1.539  & \multicolumn{4}{c}{ } \\
    $[-0.54, -0.36]$        &  205 &    2.050    &   0.231  &    1.786  & \multicolumn{4}{c}{ } \\
    $[-0.36, -0.18]$        &  205 &    2.564    &   0.255  &    2.280  & \multicolumn{4}{c}{ } \\
    $[-0.18, 0.00]$         &  205 &    3.410    &   0.300  &    3.253  & \multicolumn{4}{c}{ } \\
    $[0.00, 0.09]$          &  205 &    5.308    &   0.472  &    4.479  & \multicolumn{4}{c}{ } \\
    $[0.09, 0.18]$          &  205 &    5.836    &   0.571  &    5.820  & \multicolumn{4}{c}{ } \\
    $[0.18, 0.27]$          &  205 &    7.996    &   0.635  &    8.077  & \multicolumn{4}{c}{ } \\
    $[0.27, 0.36]$          &  205 &   10.607    &   0.764  &   11.200  & \multicolumn{4}{c}{ } \\
    $[0.36, 0.45]$          &  205 &   14.729    &   0.874  &   16.322  & \multicolumn{4}{c}{ } \\
    $[0.45, 0.54]$          &  205 &   26.189    &   1.157  &   25.722  & \multicolumn{4}{c}{ } \\
    $[0.54, 0.63]$          &  205 &   43.124    &   1.497  &   43.217  & \multicolumn{4}{c}{ } \\
    $[0.63, 0.72]$          &  205 &   79.255    &   1.976  &   80.939  & \multicolumn{4}{c}{ } \\
    $[0.72, 0.81]$          &  205 &  179.842    &   3.838  &  180.878  & \multicolumn{4}{c}{ } \\
    $[0.81, 0.90]$          &  205 &  587.999    &   7.527  &  586.205  & \multicolumn{4}{c}{ } \\
\cline{1-5}
\end{tabular}
\end{center}
\caption{ Combined LEP results for the $\eeee$ differential
 cross-sections (continued), in pb divided by $\Delta(\cos\theta)$,
 for $\sqrt{s}$ larger than $200~\GeV$.  The combined statistical and
 systematic error is shown.  The SM predictions are from BHWIDE.}
\label{tab:ff:dsdceeres2}
\end{table}

\begin{figure}[p]
 \begin{center}
  \epsfig{file=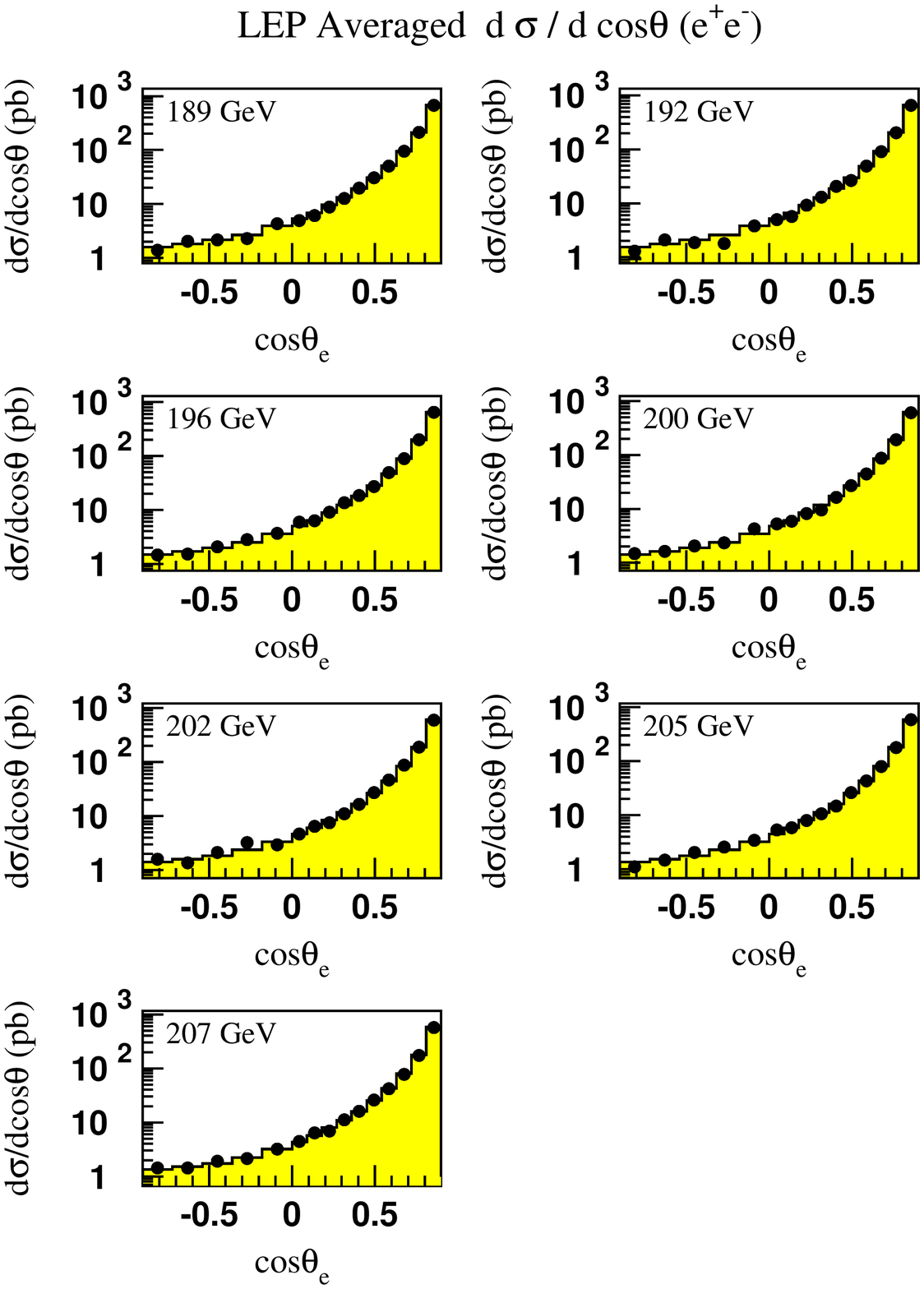,width=0.9\textwidth}
 \end{center}
 \caption[Combined differential cross-sections for \eeee.]  {LEP
         averaged differential cross-sections for $\eeee$ at energies
         of 189--207 $\GeV$. The SM predictions, shown as solid
         histograms, are computed with BHWIDE.}
 \label{fig:ff:dsdc-res-ee}
 \vskip 2cm 
\end{figure}
\begin{figure}[p]
 \begin{center}
  \epsfig{file=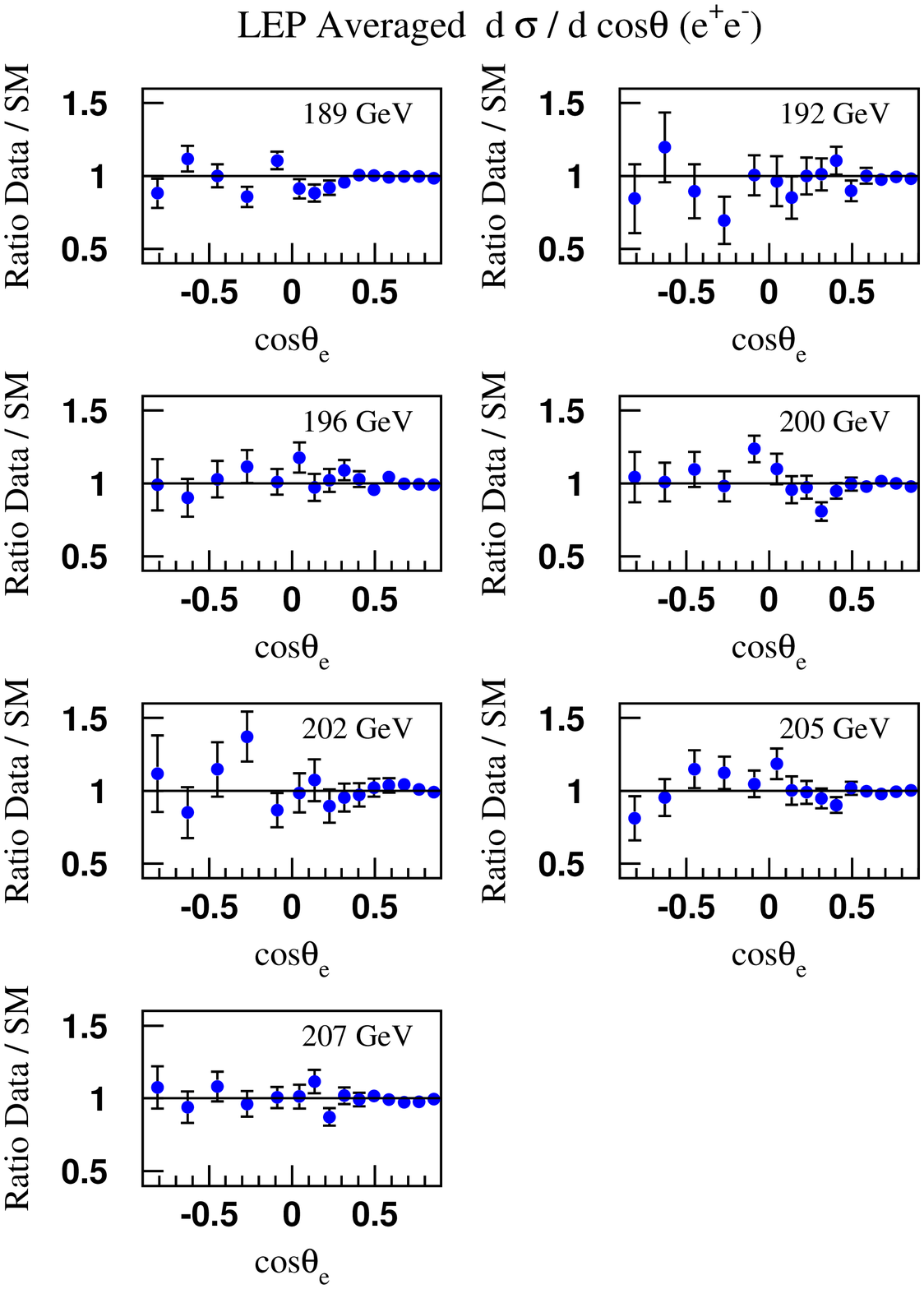,width=0.9\textwidth}
 \end{center}
 \caption[Combined differential cross-sections for \eeee.]  {Ratio of
         the LEP averaged differential cross-sections for $\eeee$ at
         energies of 189--207 $\GeV$ to the SM predictions, as
         computed with BHWIDE.}
 \label{fig:ff:dsdc-ratio-ee}
 \vskip 2cm 
\end{figure}

\section{Interpretation}
\label{sec:ff:interp}

The combined cross-section and asymmetry results are interpreted in a
variety of models.  They are used to place limits on the mass of a
possible additional heavy neutral boson, $\zprime$, under different
assumptions.  Limits on contact interactions between leptons and
between leptons and quarks are obtained.
The former results are of particular interest since they are
inaccessible to ${\mathrm{p\bar{p}}}$, pp or ep colliders.  Limits are
also provided on the masses of leptoquarks.  The \eeee\ channel is
used to constrain the scale of gravity in models with extra
dimensions.

\subsection{Models with Z$'$ Bosons}
\label{sec:ff:interp_zp}

The combined hadronic and leptonic cross-sections and the leptonic
forward-backward asymmetries are used to fit the data to models
including an additional, heavy, neutral boson, $\zprime$.

New gauge bosons in the intermediate TeV scale are motivated by
several theoretical approaches~\cite{Langacker:2009im,
*Langacker:2008yv, *Leike:1998wr, *Hewett:1988xc}.  For instance, the
breaking of Grand Unifying Theories (GUTs) based on SO(10) or E$_6$
symmetries may leave one or several U(1) remnants unbroken down to TeV
energies, before the symmetry reduces to the SM symmetry.  In the case
of the ${\rm E_6}$ model, one has the possible breaking pattern:

\begin{equation}
\mathrm{E}_6
 \rightarrow \mathrm{SO}(10) \times \mathrm{U}(1)_\psi
 \rightarrow \mathrm{SU}(5) \times \mathrm{U}(1)_\chi \times \mathrm{U}(1)_\psi
 \rightarrow  \mathrm{SM} \times \mathrm{U}(1)^\prime\,, 
\end{equation}
and the new $\zprime$ boson corresponding to the final ${\rm
U(1)^\prime}$ remnant is a linear combination of the gauge bosons of
the two U(1) groups, U(1)$_{\chi}$ and U(1)$_{\psi}$, generated in the
two-step symmetry breaking, $\zprime = \zprime_\chi \cos\beta\!
+\!\zprime_\psi \sin\beta$. The value $\beta= {\rm
arctan}(-\sqrt{5/3})$ would correspond to a $\zprime_\eta$ originating
from the direct breaking of ${\rm E_6}$ to a rank-5 group in
superstring inspired models.  Other options are left-right (L-R)
models, based on the group ${\rm SU(2)_R \times SU(2)_L \times
U(1)_{B-L}}$ in which the new $\zprime_{LR}$ will couple to a linear
combination of the right-handed and B-L currents with a parameter:

\begin{equation}
\alpha_{\rm LR}^2 = \frac{\swsq}{\cwsq}\frac{g_{\rm R}^2}{g_{\rm L}^2}-1 \,.
\end{equation}

Below the resonance, new gauge bosons appear as deviations from the
SM predictions due to $\gamma-\zprime$ and Z$-\zprime$
interference terms.  Fits are made to the mass of a $\zprime$, $\MZp$,
for $\zprime$ models varying the parameters $\beta$ and $\alpha_{\rm
LR}$ including four special models referred to as $\chi$, $\psi$,
$\eta$ and L-R~\cite{bib:ff:zpthry1984, *bib:ff:zpthry1986,
*bib:ff:zpthry1974, *bib:ff:zpthry1974-e, *bib:ff:zpthry1975} and the
Sequential Standard Model (SSM)~\cite{bib:ff:zpsqsm,
*bib:ff:zpsqsm-e}, which proposes the existence of a $\zprime$ with
exactly the same coupling to fermions as the standard Z.

The $\LEPII$ data alone do not significantly constrain the mixing
angle between the Z and $\zprime$ fields, $\thtzzp$.  However, results
from a single experiment in which $\LEPI$ data are used in the fit
show that the mixing is consistent with zero (see for example
Reference~\cite{bib:ff:zplep1}, giving limits of 30~mrad or less
depending on the model).  Hence, for these fits $\thtzzp$ is fixed to
zero. The calculation of $\zprime$ contributions is implemented in an
extension of the ZFITTER program~\cite{Leike:1991if, *Leike:1992uf}.

The predictions from the $\zprime$ models are fitted to the combined
{\LEPII} cross-section and forward-backward asymmetry measurements.
In this approach the absence of $\zprime$ bosons is equivalent to
infinite $\zprime$ mass or zero coupling.

\begin{figure}[h]
\begin{center}
  {\epsfig{file=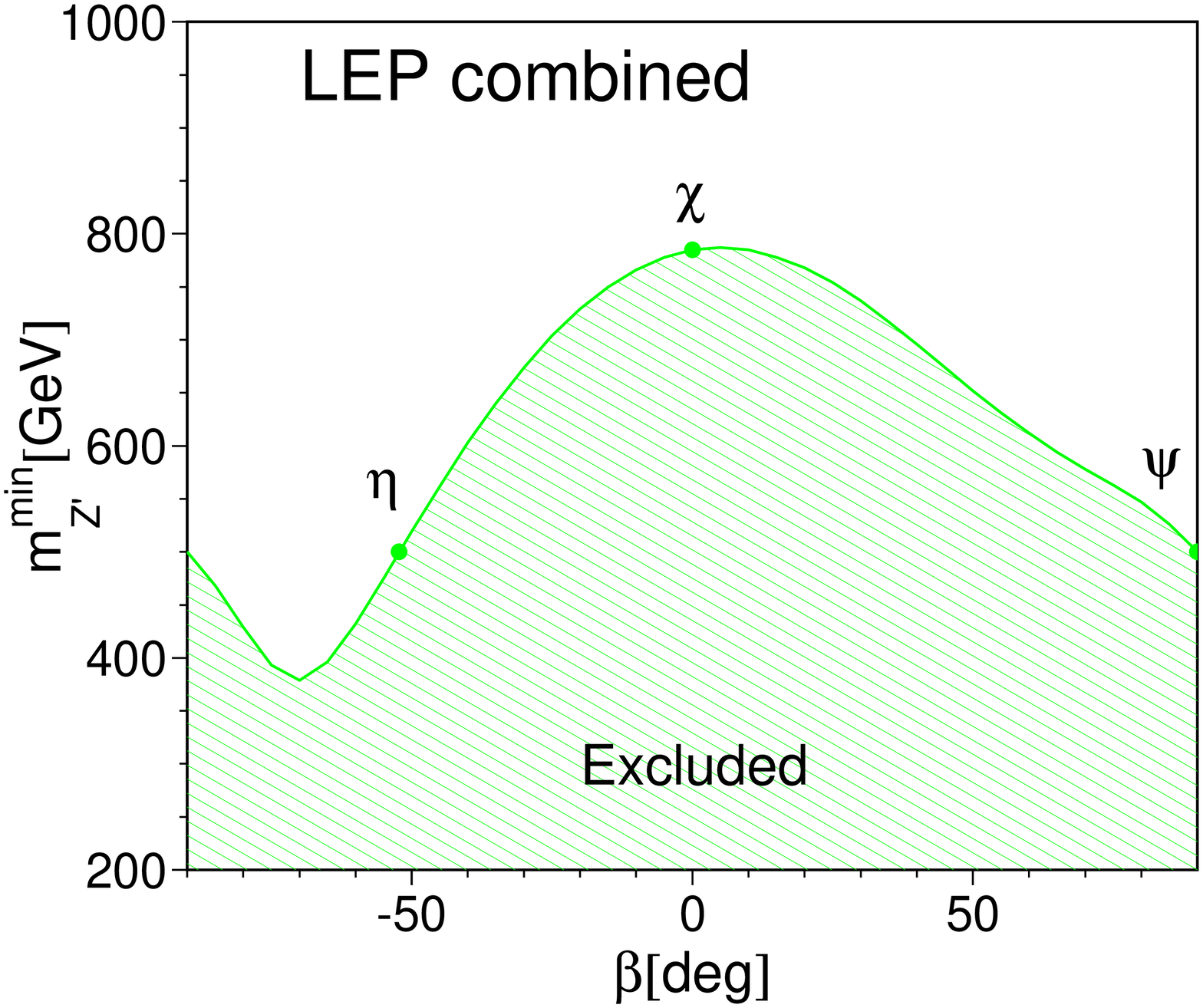,width=0.49\textwidth}}
  \hfill
  {\epsfig{file=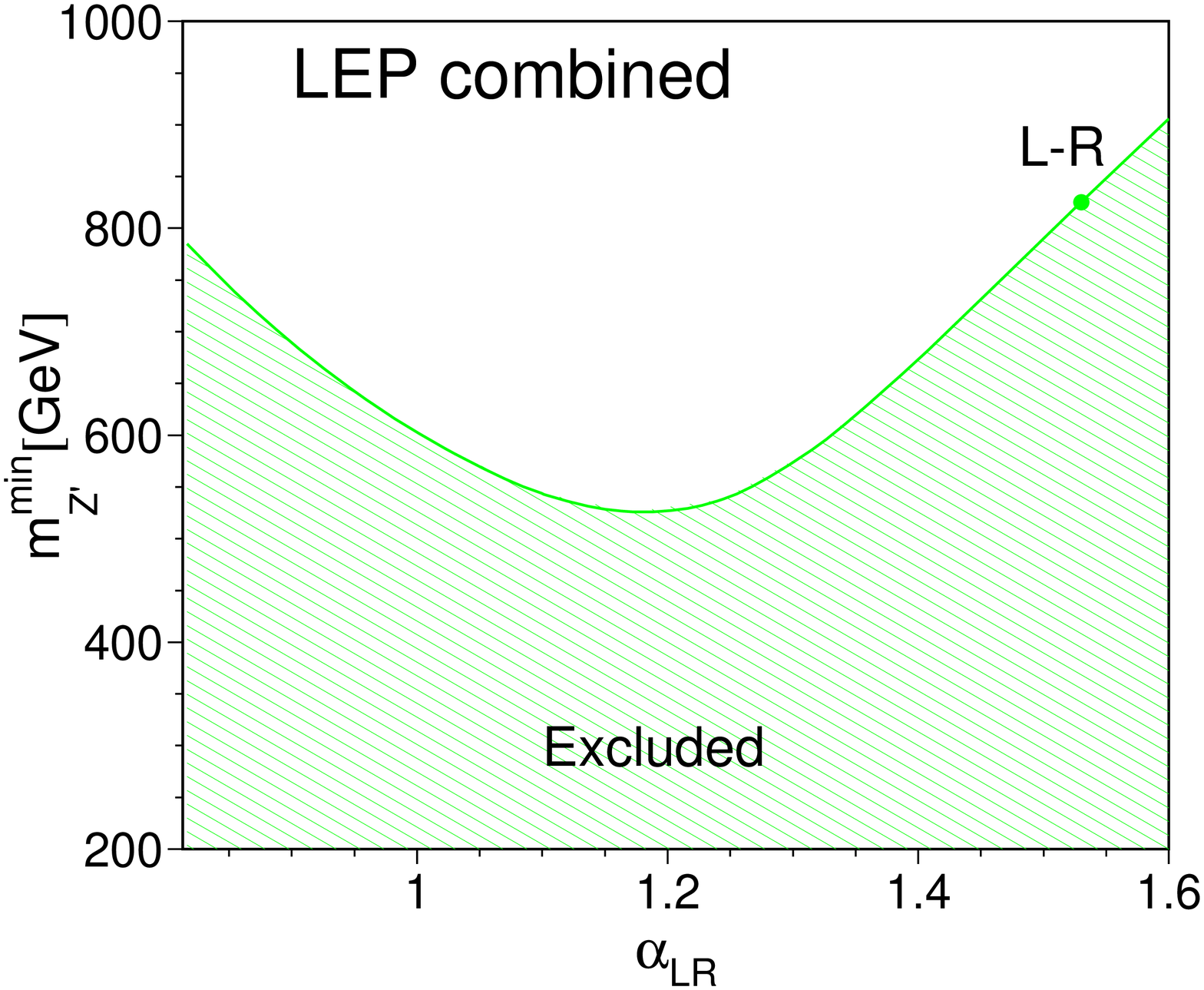,width=0.49\textwidth}}
\end{center}
\caption[Lower limits on the $\zprime$ mass.]  {Lower limits on the
        $\zprime$ mass at the 95\% C.L.  for $\zprime$ models based on
        the symmetry breaking of E$_6$ GUT models (left plot) and on
        left-right models (right plot).}
\label{fig:ff:zprime_mass_limit} 
\end{figure}

No significant evidence is found for the existence of a $\zprime$
boson in any of the models. In its absence, $95\%$ confidence level
lower limits on $\MZp$ are obtained with a Bayesian method with the
assumption of a flat prior in the physically allowed region.
The lower limits on the ${\zprime}$ mass are summarised in
Table~\ref{tab:ff:zprime_mass_lim} and shown in
Figure~\ref{fig:ff:zprime_mass_limit}.

\begin{table}[ht]
\begin{center}
\renewcommand{\arraystretch}{1.5}
\begin{tabular}{|r||c|c|c|c|c|}
\hline
 Model                    & $\chi$  & $\psi$ & $\eta$ & L-R  & SSM   \\
\hline \hline
 $\MZplim$ ($\GeV$) & 785     &  500    & 500    &  825 & 1760  \\
\hline
\end{tabular}
\end{center}
\caption[Limits on $\zprime$ masses.]
        {The $95\%$ confidence level lower limits on the $\zprime$ mass in the
         $\chi$, $\psi$, $\eta$, L-R and SSM models.}
\label{tab:ff:zprime_mass_lim}
\end{table}

\subsection{Contact Interactions}
\label{sec:ff:interp_cntclep}

The averaged differential cross-sections for electron-pairs, the
averaged cross-sections and forward-backward asymmetries for
muon-pairs and tau-lepton pairs, and the hadron cross-sections are
used to search for contact interactions between leptons and between
leptons and quarks.

Following Reference~\cite{bib:ff:ELPthr}, contact interactions are
parametrised by an effective Lagrangian, $\cal{L}_{\mathrm{eff}}$,
which is added to the SM Lagrangian and has the form:

\begin{equation}
 \mbox{$\cal{L}$}_{\mathrm{eff}} ~= ~
                        \frac{g^{2}}{(1+\delta)\Lambda^{2}_{\pm}} 
                          \sum_{i,j=L,R} \eta_{ij} 
                           \overline{e}_{i} \gamma_{\mu} e_{i}
                            \overline{f}_{j} \gamma^{\mu} f_{j}\,,
\end{equation}
where $g^{2}/{4\pi}$ is taken to be 1 by convention, $\delta=1 (0)$
for $f=e ~(f \neq e)$, $\eta_{ij}=\pm 1$ or $0$, $\Lambda_{\pm}$ is
the scale of the contact interactions, $e_{i}$ and $f_{j}$ are left or
right-handed spinors.  By assuming different helicity coupling between
the initial state and final state currents, a set of different models
can be defined from this Lagrangian~\cite{bib:ff:Kroha}, with either
constructive ($+$) or destructive ($-$) interference between the SM
process and the contact interactions. The models and corresponding
choices of $\eta_{ij}$ are given in Table~\ref{tab:ff:cntcdef}.  The
models LL, RR, VV, AA, LR, RL, V0, A0, A1 are considered here since
these models lead to large deviations in the $\eemumu$ and $\eetautau$
channels.
Potential deviations between SM predictions and measurements of the
hadronic cross-section can be interpreted as new interactions
occurring between electrons and a single quark flavour only, or as
interaction between electrons and all quark flavours at the same
time. In the former case the scale of the contact interaction is
denoted by $\Lambda_{uu}$ for a flavour of up type (u,c) and by
$\Lambda_{dd}$ for a flavour of down type (d,s,b), while for the
latter the scale of the single contact interaction is denoted by
$\Lambda_{qq}$.

For the purpose of fitting contact interaction models to the data, a
new parameter $\varepsilon_{\pm}=1/\Lambda^{2}_{\pm}$ is defined;
$\varepsilon=0$ in the limit that there are no contact interactions.
This parameter is allowed to take both positive and negative values in
the fits.  Theoretical uncertainties on the SM predictions are taken
from Reference~\cite{bib:ff:lepffwrkshp}, see above.

\begin{figure}[htbp]
 \begin{center}
  {\epsfig{file=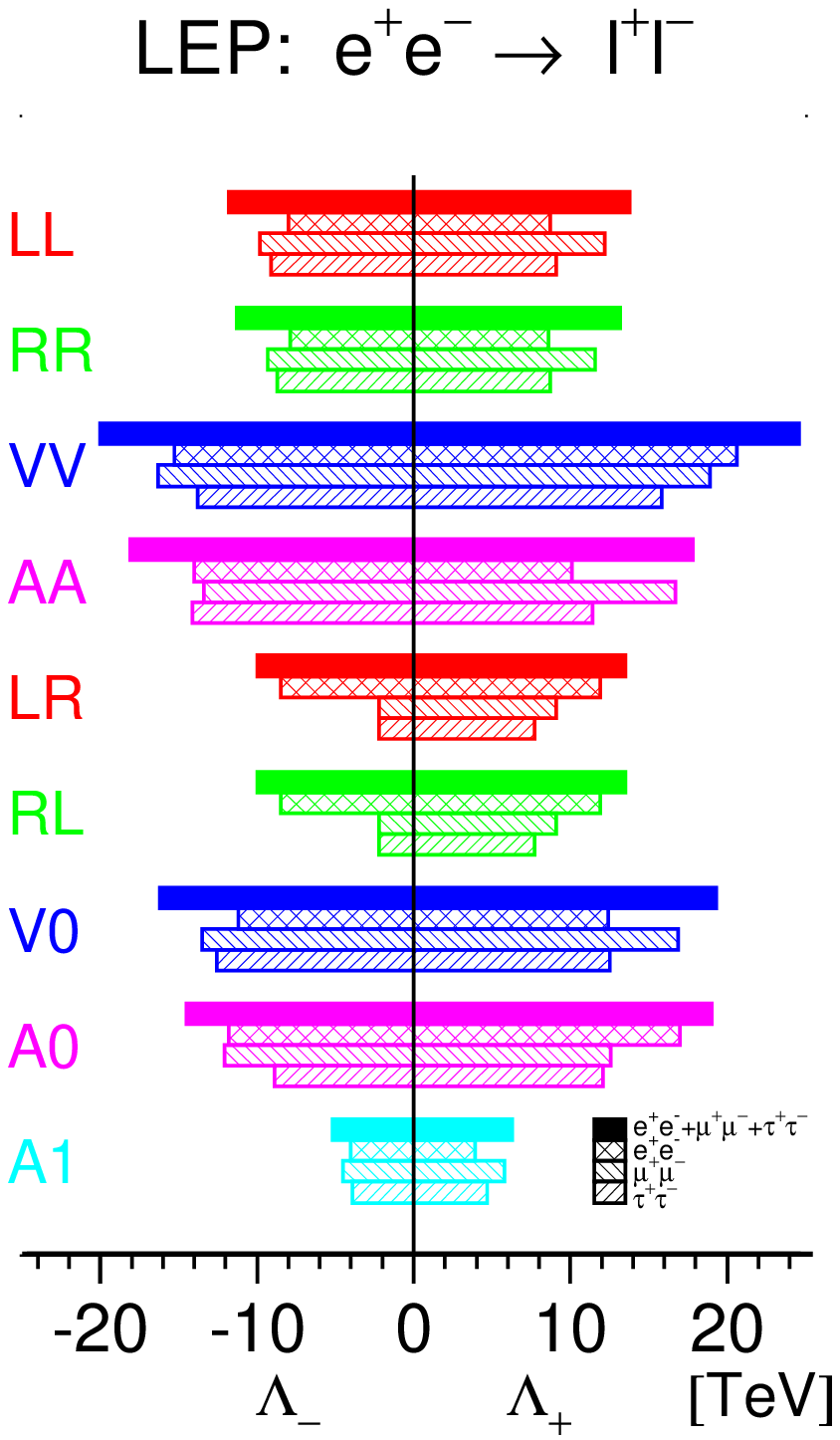,width=0.49\textwidth}}
  \hfill
  {\epsfig{file=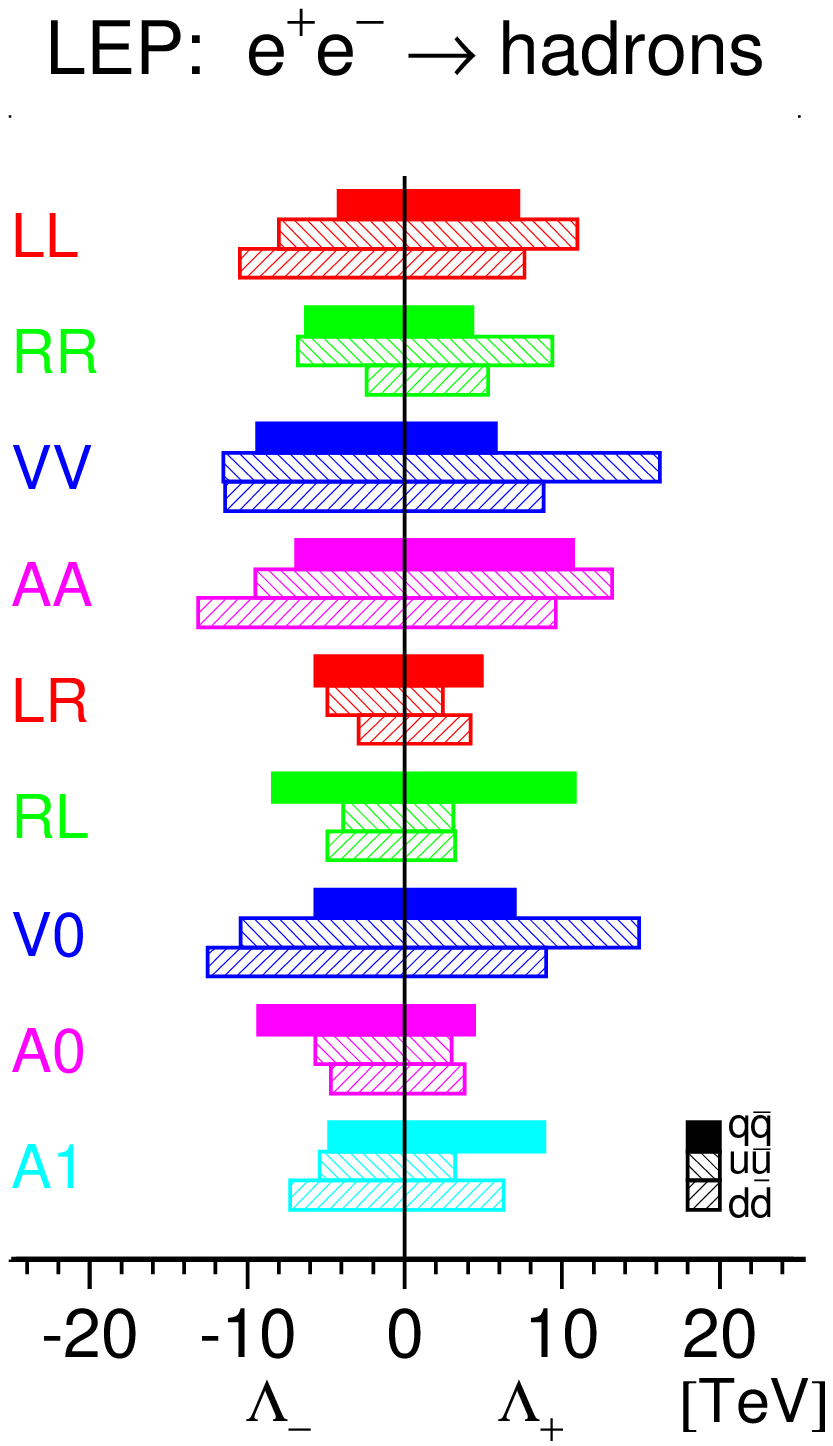,width=0.49\textwidth}}
\end{center}
 \caption[Limits on the scale of contact interactions.]  {The $95\%$
          confidence limits on $\Lambda_{\pm}$, for constructive ($+$)
          and destructive interference ($-$) with the SM, for the
          contact interaction models discussed in the text. Results
          are shown for $\eeee$, $\eemumu$, and $\eetautau$ as well as
          for $\ee \rightarrow \uu$, $\ee \rightarrow \dd$ and $\ee
          \rightarrow \qq$. For $\ee \rightarrow \leptlept$,
          universality in the contact interactions between leptons is
          assumed.}

 \label{fig:ff:cntc-all}
\end{figure}

The values of $\varepsilon$ extracted for each model are all
compatible with the SM expectation $\varepsilon=0$ within at most two
standard deviations.  The fitted values of $\varepsilon$ are converted
into $95\%$ confidence level lower limits on $\Lambda_{\pm}$.  The
limits are obtained with a Bayesian method with the assumption of a
flat prior in the physically allowed region,
$\varepsilon \ge 0$ for each $\Lambda_{+}$ limit and $\varepsilon \le
0$ for $\Lambda_{-}$ limits. The results are shown in
Table~\ref{tab:ff:cntc-all} and illustrated in
Figure~\ref{fig:ff:cntc-all}.  The parameters $\Lambda$ given in the
last column of Table~\ref{tab:ff:cntc-all} are derived from the
$\Lambda_{\ee}$ values combined with the results on $\Lambda$ from a
combined fit to the $\mumu$ and $\tautau$ cross-sections and
asymmetries.

\begin{table}[t]
 \begin{center}
  \begin{tabular}{|c||c|c|c|c|}
   \hline
   Model      & $\eta_{LL}$ & $\eta_{RR}$ & $\eta_{LR}$ & $\eta_{RL}$ \\
   \hline\hline
   LL$^{\pm}$ &   $\pm 1$   &      0      &      0      &      0      \\
   RR$^{\pm}$ &      0      &   $\pm 1$   &      0      &      0      \\
   VV$^{\pm}$ &   $\pm 1$   &   $\pm 1$   &   $\pm 1$   &   $\pm 1$   \\
   AA$^{\pm}$ &   $\pm 1$   &   $\pm 1$   &   $\mp 1$   &   $\mp 1$   \\
   LR$^{\pm}$ &      0      &      0      &   $\pm 1$   &      0      \\
   RL$^{\pm}$ &      0      &      0      &      0      &   $\pm 1$   \\
   V0$^{\pm}$ &   $\pm 1$   &   $\pm 1$   &      0      &      0      \\
   A0$^{\pm}$ &      0      &      0      &  $\pm 1$    &   $\pm 1$   \\
   A1$^{\pm}$ &    $\pm 1$  &   $\mp 1$   &      0      &      0      \\
   \hline
  \end{tabular}
 \end{center}
 \caption[Choices of $\eta_{ij}$ for different contact interaction models.]
         {Choices of $\eta_{ij}$ for different contact interaction models.}
 \label{tab:ff:cntcdef}.
\end{table}

The full correlation matrix of the differential cross-sections for
electron pairs, obtained in the averaging procedure, is used in the
fits.  Some aspects of the combination of the LEP data on Bhabha
scattering are discussed in References~\cite{bib:ff:Bourilkov:1999,
bib:ff:Bourilkov:2000, bib:ff:Bourilkov:2001}).  For the VV model with
positive interference and assuming electromagnetic coupling strength
instead of $g^{2}/{4\pi} = 1$~\cite{bib:ff:Bourilkov:2000}, the scale
$\Lambda$ can be converted to an upper limit on the electron size:

\begin{equation}
r_e < 1.1 \cdot 10^{-19}\; \mathrm{m}.
\end{equation}
Models with stronger couplings will make this upper limit even stronger.

\begin{table}[htbp]
 \begin{center}
 \renewcommand{\arraystretch}{1.1}
  \begin{tabular}{|c||rr|rr|rr|rr|}
   \hline
   \multicolumn{9}{|c|}{\boldmath $\mathrm{e^{+}e^{-}}\rightarrow \leptlept$\unboldmath} \\
   \hline
   Model  & \multicolumn{2}{|c|}
            { $\Lambda^{-}_{\mathrm{ee}}~(\TeV)\,\,\Lambda^{+}_{\mathrm{ee}}$} 
          & \multicolumn{2}{|c|}
            { $\Lambda^{-}_{\mu\mu}~(\TeV)\,\,\Lambda^{+}_{\mu\mu}$} 
          & \multicolumn{2}{|c|}
            {$\Lambda^{-}_{\tau\tau}~(\TeV)\,\,\Lambda^{+}_{\tau\tau}$} 
          & \multicolumn{2}{|c|}
            {$\Lambda^{-}_{\leptlept}(\TeV)\,\,\Lambda^{+}_{\leptlept}$}\\
   \hline
   \hline
   LL &  8.0 &  8.7  &   9.8 &   12.2 &   9.1 &   9.1  & 11.8 & 13.8 \\
   RR &  7.9 &  8.6  &   9.3 &   11.6 &   8.7 &   8.7  & 11.3 & 13.2 \\
   VV & 15.3 & 20.6  &  16.3 &   18.9 &  13.8 &  15.8  & 20.0 & 24.6 \\
   AA & 14.0 & 10.1  &  13.4 &   16.7 &  14.1 &  11.4  & 18.1 & 17.8 \\
   LR &  8.5 & 11.9  &   2.2 &    9.1 &   2.2 &   7.7  & 10.0 & 13.5 \\
   RL &  8.5 & 11.9  &   2.2 &    9.1 &   2.2 &   7.7  & 10.0 & 13.5 \\
   V0 & 11.2 & 12.4  &  13.5 &   16.9 &  12.6 &  12.5  & 16.2 & 19.3 \\
   A0 & 11.8 & 17.0  &  12.1 &   12.6 &   8.9 &  12.1  & 14.5 & 19.0 \\
   A1 &  4.0 &  3.9  &   4.5 &    5.8 &   3.9 &   4.7  &  5.2 &  6.3 \\
   \hline
  \end{tabular}

  \begin{tabular}{|c||cc|cc|cc|}
   \hline
   \multicolumn{7}{|c|}{\boldmath $\mathrm{e^{+}e^{-}} \rightarrow \mathrm{q}\bar{\mathrm{q}}$\unboldmath} \\
   \hline
   Model  &   $\Lambda^{-}_{\mathrm{u\bar{u}}} (\TeV)$  & $\Lambda^{+}_{\mathrm{u\bar{u}}} (\TeV)$ 
          &   $\Lambda^{-}_{\mathrm{d\bar{d}}} (\TeV)$  & $\Lambda^{+}_{\mathrm{d\bar{d}}} (\TeV)$   
          &   $\Lambda^{-}_{\mathrm{q\bar{q}}} (\TeV)$  & $\Lambda^{+}_{\mathrm{q\bar{q}}} (\TeV)$ 
\\
   \hline
   \hline
   LL &   8.0 &   11.0 &  10.5 &   7.6 &   4.2 &   7.2 \\
   RR &   6.8 &    9.4 &   2.4 &   5.3 &   6.3 &   4.3 \\
   VV &  11.5 &   16.2 &  11.4 &   8.8 &   9.4 &   5.8 \\
   AA &   9.5 &   13.2 &  13.1 &   9.6 &   6.9 &  10.7 \\
   LR &   4.9 &    2.4 &   2.9 &   4.2 &   5.7 &   4.9 \\
   RL &   3.9 &    3.1 &   4.9 &   3.2 &   8.4 &  10.8 \\
   V0 &  10.4 &   14.9 &  12.5 &   9.0 &   5.7 &   7.0 \\
   A0 &   5.7 &    3.0 &   4.7 &   3.8 &   9.3 &   4.4 \\
   A1 &   5.4 &    3.2 &   7.3 &   6.3 &   4.8 &   8.9 \\
   \hline
  \end{tabular}
 \end{center}
 \caption[Limits on the scale of contact interactions between leptons,
         and between electrons and quarks.]  {The $95\%$ confidence
         limits on the scale, $\Lambda^{\pm}$, for constructive ($+$)
         and destructive interference ($-$) with the SM, for the
         contact interaction models discussed in the text. Results are
         given for $\eemumu$, $\eetautau$ and $\eeee$ as well as for
         $\ee \rightarrow \uu$, $\ee \rightarrow \dd$ and $\ee
         \rightarrow \qq$. For $\ee \rightarrow \leptlept$,
         universality in the contact interactions between leptons is
         assumed.}
 \label{tab:ff:cntc-all}
\end{table}

\subsection{Large Extra Dimensions}
\label{sec:ff:interp_led}

An approach to the solution of the hierarchy problem has been proposed
in~\cite{bib:ff:ADD, *bib:ff:ADD2, *bib:ff:ADD3}, which brings close
the electroweak scale $\rm m_{EW} \sim 1\; TeV$ and the Planck scale
$\rm M_{Pl} = \frac{1}{\sqrt{G_N}} \sim 10^{15}\; TeV$.  In this
framework the effective 4 dimensional $\rm M_{Pl}$ is connected to a
new $\rm M_{Pl(4+n)}$ scale in a (4+n) dimensional theory:

\begin{equation}
\rm M_{Pl}^2 \sim M_{Pl(4+n)}^{2+n} R^n \,,
\end{equation}
where there are $n$ extra compact spatial dimensions of radius $R$.

In the production of fermion- or boson-pairs in $\ee$ collisions this
class of models can manifest itself through virtual effects due to the
exchange of gravitons (Kaluza-Klein excitations). As discussed
in~\cite{bib:ff:Hewett, bib:ff:Rizzo, bib:ff:Giudice, bib:ff:Lykken,
*bib:ff:Shrock}, the exchange of spin-2 gravitons modifies in a unique
way the differential cross-sections for fermion pairs, providing clear
signatures. These models introduce an effective scale (ultraviolet
cut-off).  We will adopt the notation from~\cite{bib:ff:Hewett} and
call the gravitational mass scale $M_s$.  The cut-off scale is
supposed to be of the order of the fundamental gravity scale in $4+n$
dimensions.

The parameter $\varepsilon$ is defined as:

\begin{equation}
\varepsilon = \frac{\lambda}{M_s^4}\,,
\end{equation}
where the coefficient $\rm \lambda$ is of order 1 and cannot be
calculated explicitly without knowledge of the full quantum gravity
theory. In the following analysis we will assume that $\rm \lambda =
\pm 1$ in order to study both the cases of positive and negative
interference.  To compute the deviations from the SM due to virtual
graviton exchange we use the calculations~\cite{bib:ff:Giudice,
bib:ff:Rizzo}.

\begin{figure}[p]
 \begin{center}
  \epsfig{file=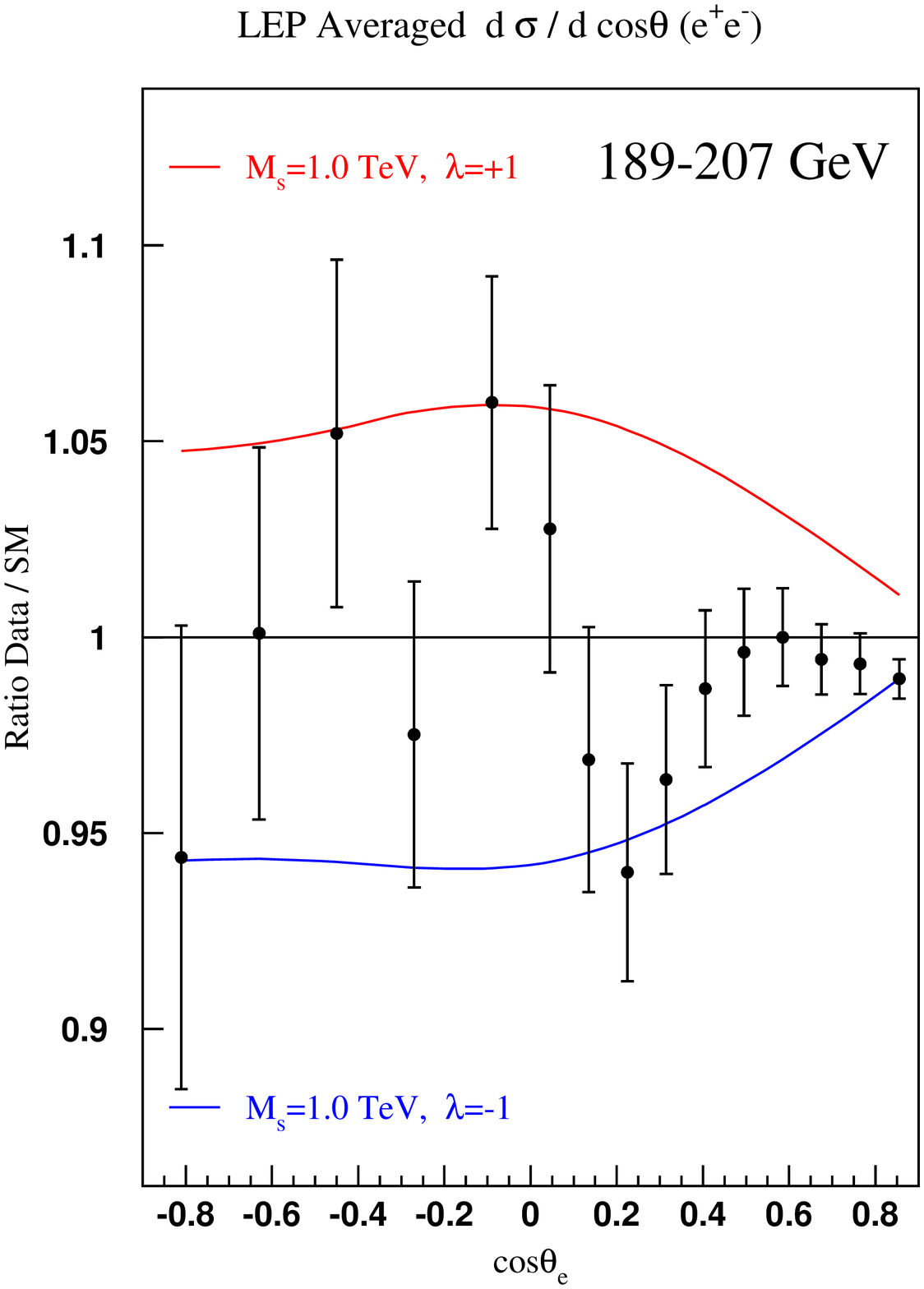,width=0.9\textwidth}
 \end{center}
 \caption{Ratio of the LEP averaged differential cross-section for
          $\eeee$ compared to the SM prediction. The effects expected
          from virtual graviton exchange are also shown.}
 \label{fig:ff:dsdc-ee-lsg}
 \vskip 2cm 
\end{figure}

A fit to the $\eeee$ differential cross-section is performed; this
channel has by far the highest sensitivity.  The fitted values of
$\varepsilon$ agree well with the SM expectation, and are used to
derive limits on the gravitational mass scale $M_s$ at 95~\% CL:

\begin{eqnarray}
\rm M_s & > & 1.09~\TeV\ \mathrm{for}\ \lambda = +1\,, \\
\rm M_s & > & 1.25~\TeV\ \mathrm{for}\ \lambda = -1\,.
\end{eqnarray}
An example of the analysis is shown in
Figure~\ref{fig:ff:dsdc-ee-lsg}.

The interference of virtual graviton exchange amplitudes with both
$t$-channel and $s$-channel Bhabha scattering amplitudes makes this
the most sensitive search channel at LEP. The results obtained here
would not be strictly valid if the luminosity measurements of the LEP
experiments, based on the very same process, is also be affected by
graviton exchange.  However, as shown in~\cite{bib:ff:Bourilkov:1999},
the effect on the cross-section in the luminosity angular range is so
small that it can safely be neglected in this analysis.

\subsection{Leptoquarks } %
\label{sec:ff:interp_lq}

Leptoquarks (LQ) mediate quark-lepton transitions. They carry fermion
numbers, $F=L+3B$. Following the notations in
References~\cite{Buchmuller:1986zs, *Buchmuller:1986zs-e} and
\cite{Kalinowski:1997fk}, scalar leptoquarks, $S_I$, and vector
leptoquarks, $V_I$, are indicated based on spin and isospin
$I$. Isomultiplets with different hypercharges are denoted by an
additional tilde.  It is assumed that leptoquark couplings to
quark-lepton pairs are flavour-diagonal and preserve baryon- and
lepton-number.  The couplings refer to $g_L,~g_R$, according to the
chirality of the lepton.  In the process $\eeqq$ leptoquarks can be
exchanged in $u$- or $t$-channel, with $F=0$ or $|F|=2$.

For convenience, one type of leptoquarks is assumed to be much lighter
than the others. Further, experimental constraints on the product $g_L
g_R$ allow separate studies of $g_L \neq 0$ or $g_R \neq 0$.

Assuming a coupling of electromagnetic strength,
$g=\sqrt{4\pi\alpha}$, where $\alpha$ is the fine structure constant,
limits on the masses of leptoquarks coupling to electrons and the
first generation of quarks are derived with a Bayesian method with the
assumption of a flat prior in the physically allowed region
from comparisons of the theoretical predictions for the total hadronic
cross-section to the $\LEPII$ averaged measurements.

The $95\%$ confidence level lower limits on masses $m_{LQ}$ are
summarised in Table~\ref{tab:ff:lq-mass}.

\begin{table}[th]
  \renewcommand{\arraystretch}{1.2}
  \begin{center}
   \begin{tabular}{|l|c||l|c|}
     \hline
        LQ type  & $m_{\mathrm{LQ}}^{min}$($\GeV$)
      & LQ type  & $m_{\mathrm{LQ}}^{min}$($\GeV$) \\
      \hline
      \hline
      $S_0(\mathrm{L})\rightarrow \mathrm{e u}$                      & 646 &%
      $V_{1/2}(\mathrm{L})\rightarrow \mathrm{e d}$                   & 348 \\
      $S_0(\mathrm{R})\rightarrow \mathrm{e u}$                      & 516 &%
      $V_{1/2}(\mathrm{R})\rightarrow \mathrm{e u,~ e d}$             & 238 \\
      $\tilde{S}_0(\mathrm{R})\rightarrow \mathrm{e d}$              & 256 &%
      $\tilde{V}_{1/2} (\mathrm{L}) \rightarrow \mathrm{e u}$         & 186 \\
      $S_1(\mathrm{L})\rightarrow \mathrm{e u,~ ed}$                 & 429 &%
      $V_0(\mathrm{L})\rightarrow \mathrm{e \bar{d}}$                & 897 \\
      $S_{1/2}(\mathrm{L})\rightarrow \mathrm{e \bar{u}}$             & 228 &%
      $V_0(\mathrm{R})\rightarrow \mathrm{e \bar{d}}$                & 482 \\
      $S_{1/2}(\mathrm{R})\rightarrow \mathrm{e \bar{u},~e \bar{d}}$  & 285 &%
      $\tilde{V}_0(\mathrm{R})\rightarrow \mathrm{e \bar{u}}$        & 577 \\
      $\tilde{S}_{1/2}(\mathrm{L})\rightarrow \mathrm{e \bar{d}}$     & --   &
      $V_1(\mathrm{L})\rightarrow \mathrm{e \bar{u},~e \bar{d}} $    & 765 \\
     \hline     
  \end{tabular}
  \end{center}
  \caption[Lower limits on LQ masses.]  {The $95\%$ confidence level
          lower limits on the LQ mass assuming $g_{L,R}=\sqrt{4\pi
          \alpha}$. For $ \tilde{S}_{1/2}$(L) no limit can be set
          because the contribution from this leptoquark type to the
          hadronic cross-section is not observable with the precision
          of the measurements. }
   \label{tab:ff:lq-mass}
\end{table}

\section{Summary}
\label{sec:ff:conc}

A combination of the $\LEPII$ $\eeff$ cross-sections (for hadron, muon
and tau-lepton final states) and forward-backward asymmetries (for
muon and tau-lepton final states) from LEP running at energies from
130 to 209~$\GeV$ is made.  The results from the four LEP experiments
are in good agreement with each other.  The averages for all energies
are shown in Table~\ref{tab:ff:xsafbres}.  The use of the combined
fermion-pair results in an S-Matrix analysis is discussed in
Appendix~\ref{chap:s-matrix}.
Differential cross-sections, $\dsdc$, for $\eemumu$, $\eetautau$ and
$\eeee$ are also combined. Results are shown in
Figures~\ref{fig:ff:dsdc-res-mm}, \ref{fig:ff:dsdc-res-tt}
and~\ref{fig:ff:dsdc-res-ee}.
All results are in good agreement with the predictions of the SM.

The averaged cross-section, forward-backward asymmetry and
differential cross-section results
are interpreted in a variety of models. The $\LEPII$ averaged
cross-sections and lepton asymmetries are used to obtain lower limits
on the mass of a possible $\zprime$ boson in different models. Limits
range from $500$ to $1760$ $\GeV$ depending on the model.  Limits on
the scale of contact interactions between leptons
and between electrons and quarks are determined.  A full set of limits
are reported in Table~\ref{tab:ff:cntc-all}.
Limits on the scale of gravity in models with extra dimensions ranging
from 1.09 to 1.25 $\TeV$ are obtained.  Limits on the masses of
leptoquarks are derived from the hadronic cross-sections. The limits
range from 186 to 897 $\GeV$ depending on the type of leptoquark.

\chapter{Final-State Interconnection Effects}
\label{chap:fsi}

At \LEPII, Final-State Interconnection (FSI) effects may exist when
two colourless W or Z bosons decay hadronically, close in space-time
to one another.  Two phenomena are considered: Colour Reconnection
(CR) and Bose-Einstein Correlations (BEC).  The former is expected to
appear as a consequence of the strong interaction described by
non-perturbative QCD, while the latter is due to the quantum
mechanical properties of those particles in the hadronic final state
which follow Bose statistics. Both were observed in other physical
systems~\cite{bib:fsi:hanbury-brown-twiss, *bib:fsi:goldhaber-pion,
*bib:fsi:interferometry, *bib:fsi:baym, bib:fsi:argus-bpsi,
*bib:fsi:cleo-bpsi}.  An additional motivation for the study of FSI
effects is that they introduce potentially large systematic
uncertainties in the measurement of the mass of the W boson using
fully hadronic W-pair decays.  The studies described here allow a
better understanding of CR and BEC at \LEPII\ and, by constraining
models and their parameters, impose limits on their quantitative
effect in the W-boson mass measurement.

\section{Colour Reconnection}
\label{sec:CR}

\subsection{Introduction}
 
In \WWtoqqqq\ events, the products of the two colour singlet W decays
have in general a significant space-time overlap, because the
separation of their decay vertices, $\tau_W \sim 1/\Gamma_W\approx
0.1$~fm, is small compared to characteristic hadronic distance scales
of $\sim 1$~fm.  Colour reconnection, also known as colour
rearrangement (CR), was first introduced in~\cite{bib:cr:GPZ} and
refers to a reorganisation of the colour flow between the decay
products of the two W bosons.  A precedent for such effects is set by
colour suppressed B meson decays, \eg\ $B \rightarrow J/\psi K$, where
there is ``cross-talk'' between the two original colour singlets,
$\bar{\mathrm c}$+s and c+spectator~\cite{bib:cr:GPZ,
bib:cr:SK_MODELS}.
 
QCD interference effects between the colour singlets in \WW\ decays
during the perturbative phase are expected to be small, affecting the
W mass by $\sim (\frac{\alfas}{\pi N_{\mathrm{colours}}})^2 \Gamma_W$
$\sim \cal{O}(\mathrm{1~\MeV})$~\cite{bib:cr:SK_MODELS}. In contrast,
non-perturbative effects involving soft gluons with energies less than
\GW\ may be significant, with effects on \MW\ of $\sim
\cal{O}(\mathrm{10~\MeV})$.  To estimate the impact of this
phenomenon, a variety of phenomenological models have been
developed~\cite{bib:cr:SK_MODELS, bib:cr:ARIADNECR_MODEL, HERWIG6,
bib:cr:GH_MODEL, bib:cr:EG_MODEL, bib:cr:RATHSMAN}. These models
differ mainly in the detailed mechanism of CR and hadronisation, and
in the fraction of reconnected events.

Some of the models can also be tested at the Z peak in three-jet
events.  The analyses~\cite{bib:cr:ALEPH_CRZ, bib:cr:L3_CRZ,
bib:cr:OPAL_CRZ} showed that the \Ariadne\ model type
1~\cite{bib:cr:ARIADNECR_MODEL}, and similar the Rathsman/GAL
model~\cite{bib:cr:RATHSMAN} with default parameter settings, is not
consistent with the data.  Colour rearrangement in W-pair events
could, however, also be caused by additional reconnection
mechanisms. The combination presented here concentrates on the SK1
model~\cite{bib:cr:SK_MODELS} in which the probability for
reconnection to occur in an event is given by
$p_{\mathrm{reco}}=1-\exp(-{\cal I}k_I)$. The quantity ${\cal I}$ is
the space-time overlap integral between the colour flux tubes that are
stretched between quarks and gluons originating from the perturbative
phase of the two hadronic W decays, and $k_I$ is an adjustable
parameter of the SK1 model, thus allowing to vary the fraction of
reconnected events in the Monte-Carlo simulation.
Figure~\ref{fsi:cr:fig:preco-sk1} shows the reconnection probability,
$p_{\mathrm{reco}}$ as a function of the model parameter $k_I$, for an
SK1 Monte-Carlo event sample generated at a centre-of-mass energy of
$189~\GeV$, and used by all LEP experiments as part of the combination
procedure. By varying $k_I$, the SK1 model results can be compared to
other models which have a fixed reconnection probability, such as the
\Ariadne\ model type 2~\cite{bib:cr:ARIADNECR_MODEL} and
\Herwig~\cite{HERWIG6}. In the context of W mass measurements, it is
observed~\cite{bib:cr:ALEPH_MW, bib:cr:DELPHI_MW, bib:cr:L3_MW,
bib:cr:OPAL_MW} that all models behave similarly when adjusted to the
same reconnection fraction. The \Herwig\ CR model assumes a
reconnection probability of 1/9 counting the possible colour
rearrangements, while the \Ariadne -2 reconnection probability is
about 22\% at a centre-of-mass energy of 189~\GeV.

Many observables have been studied in the search for an experimental
signature of colour reconnection.  The inclusive properties of events
such as the mean charged particle multiplicity, distributions of
thrust, rapidity, transverse momentum and $\ln(1/x_p)$, where $x_p$ is
the scaled particle momentum, are found to have limited
sensitivity~\cite{bib:cr:OPAL_CR, bib:cr:DELPHI_HADRONSWW}.  The
effects of CR are predicted to be numerically larger in these
observables when only higher mass hadrons such as kaons and protons
are considered~\cite{bib:cr:SK_HEAVYHAD}.  However, experimental
investigations~\cite{bib:cr:DELPHI_HADRONSWW}
find no significant gain in sensitivity due to the low production rate
of such particles in W decays.

Eventually, two methods were developed which yield a sensitive handle
on CR effects in hadronic W decays: the so-called ``particle-flow''
method~\cite{bib:cr:pflow2, *bib:cr:pflow2bis, bib:cr:OXFORD_WS}, and
the influence of CR on the W-boson mass reconstructed as a function of
the particle momentum threshold and when applying different jet
algorithms.  These two are described in the following and their
combined results are presented.

\subsection{Particle-Flow Measurements}

In the analogy with the ``string effect'' analysis in 3-jet
$\eeqq\mathrm{g}$ events~\cite{bib:cr:JADE_STRING2,
*bib:cr:JADE_STRING3, *bib:cr:JADE_STRING4, *bib:cr:JADE_STRING5,
*bib:cr:TPC2GAM_STRING1, *bib:cr:TPC2GAM_STRING2,
*bib:cr:TASSO_STRING1}, the particle-flow method has been investigated
by the \Delphi, \Ltre\ and \Opal\
collaborations~\cite{bib:cr:DELPHI_CR, bib:cr:L3_CR, bib:cr:OPAL_CR}.
In these analyses, pairs of jets in \WWtoqqqq\ events are associated
with the decay of a W, after which four jet-jet regions are chosen:
two corresponding to jets sharing the same W parent (intra-W), and two
in which the parents differ (inter-W).  As there is a two-fold
ambiguity in the assignment of inter-W regions, the configuration
having the smaller sum of inter-W angles is chosen.
 
Particles are projected onto the planes defined by these jet pairs and
the particle density constructed as a function of $\phi$, the
projected angle relative to one jet in each plane.  To account for the
variation in the opening angles, $\phi_0$, of the jet-jet pairs
defining each plane, the particle densities in $\phi$ are constructed
as functions of normalised angles, $\phi_r=\phi/\phi_0$, by a simple
rescaling of the projected angles for each particle, event by event.
Particles having projected angles $\phi$ smaller than $\phi_0$ in at
least one of the four planes are considered further.  This gives
particle densities, $\frac{1}{\Nevt}\dndphir$, in four regions with
$\phi_r$ in the range from 0 to 1, and where $n$ and \Nevt\ are the
number of particles and events, respectively.
 
As the particle density reflects the colour flow in an event, CR
models predict a change in the relative particle densities between
inter-W and intra-W regions.  On average, colour reconnection is
expected to affect the particle densities of both inter-W regions in
the same way and so they are added together, as are the two intra-W
regions.  The observable used to quantify such changes, \Rn, is
defined:

\begin{equation}
 \Rn =
  \frac{\frac{1}{\Nevt}\int^{0.8}_{0.2} \dndphir (\mathrm{intra-W}) \dphir}
       {\frac{1}{\Nevt}\int^{0.8}_{0.2} \dndphir (\mathrm{inter-W}) \dphir} \,.
 \label{fsi:cr:eq:Rn}
\end{equation}
As the effects of CR are expected to be enhanced for low momentum
particles far from the jet axes, the range of integration excludes jet
cores ($\phi_r\approx 0$ and $\phi_r\approx 1$).  The precise upper
and lower limits are optimised by model studies of predicted
sensitivity.
 
The \Delphi, \Ltre\ and \Opal\ experiments have developed their own
variation on this analysis, differing primarily in the selection of
\WWtoqqqq\ events.  In \Delphi~\cite{bib:cr:DELPHI_CR} and
\Ltre~\cite{bib:cr:L3_CR}, events are selected in a very particular
configuration (``topological selection'') by imposing restrictions on
the jet-jet angles and on the jet-resolution parameter for the three-
to four-jet transition (Durham~\cite{bib:mw:durham-1,
*bib:mw:durham-2, *bib:mw:durham-3, *bib:mw:durham-3e} or
Luclus~\cite{bib:mw:luclus-1, *bib:mw:luclus-2, *bib:mw:luclus-3,
*bib:mw:luclus-4, *bib:mw:luclus-5} schemes).  This leads to more
planar events than those in an inclusive \WWtoqqqq\ sample and the
association between jet pairs and W bosons is given by the relative
angular separation of the jets.  The overall efficiency for selecting
signal events ranges between 12\% and 22\% with purities of
70-85\%. The efficiency to assign the correct jets to the parent W's
amounts to 70-91\%. Data samples with small signal efficiency
typically have the highest purity and best efficiency for correct jet
assignment. The \Opal~\cite{bib:cr:OPAL_CR} event selection is based
on their W mass analysis.  Assignment of pairs of jets to W's follows
the procedure used in measuring \MW, using a multivariate
algorithm~\cite{bib:cr:OPAL_MW} with an overall efficiency for
selecting \WWtoqqqq\ events of 40\%, a signal purity of 86\%, and an
efficiency for correctly assigning jets to parent W's of 90\%, albeit
with a less planar topology and hence a more complicated colour flow.
 
The data are corrected bin-by-bin for background contamination in the
inter-W and intra-W regions separately.  The possibility of CR effects
existing in background processes is neglected because the background
is dominated by $\eeqq$ events and the \ZZtoqqqq\ background, in which
CR effects may also be present, is at the level of 2\% only.

The measured values of \Rn\ are compared after they have been
normalised using a common sample of Monte-Carlo events, processed
using the detector simulation and particle-flow analysis of each
experiment.  The ratio, $r$, is constructed:

\begin{equation}
          r = \frac{\Rndata}{\Rnnocr} \,,
\end{equation}
where \Rndata\ and \Rnnocr\ are the values of \Rn\ measured by each
experiment in data and in a common sample of events simulated without
CR.  In the absence of CR, all experiments should find $r$ consistent
with unity.  The default no-CR sample used for this normalisation
consists of \eeWW\ events produced using the
\KoralW~\cite{bib:fsi:KORALW} event generator and hadronised using the
\Jetset~\cite{JETSET} model.

The common Monte-Carlo samples used in the combination are only
available at a single centre-of-mass energy, $E_{\mathrm{cm}}$, of
$188.6~\GeV$.  The \Rn\ are however measured at each centre-of-mass
energy separately, in both real data and Monte-Carlo simulations.  The
predicted variation of \Rn\ with centre-of-mass energy is determined
by each experiment using its own samples of simulated \eeWW\ events,
with hadronisation performed using the no-CR \Jetset\ model. The
evolutions of $\Rn$ are parametrised by second order polynomial
functions in $E_{\mathrm{cm}}$ and are detailed in
References~\cite{bib:cr:DELPHI_CR, bib:cr:L3_CR, bib:cr:OPAL_CR}.  The
\Rn\ measured in data are subsequently extrapolated by each experiment
to the reference energy of $188.6~\GeV$.

Input from a particle-flow measurement is provided by \Ltre\ and
\Opal\ in terms of measured \Rn\ and corresponding $\Delta\Rn$ for
different systematic variations of the analysis or different Monte
Carlo modelling~\cite{bib:cr:L3_CR, bib:cr:OPAL_CR}. They are shown in
Table~\ref{fsi:cr:tab:pflow}. \Delphi\ provides their results in terms
of likelihood functions, which are discussed below. Systematic
uncertainties due to Bose-Einstein correlations are limited to the
level which is compatible with the LEP measurement of BEC (see
Chapter~\ref{sec:BEC}). Scale uncertainties on the main background
processes \eeqq\ and \ZZtoqqqq, and hadronisation uncertainties, which
are derived from the spread of \Rn\ for the \Jetset, \Ariadne\ and
\Herwig\ hadronisation models are also taken into account. For these
uncertainties the smallest of each systematic uncertainty of \Ltre\
and \Opal\ is taken as correlated, the remaining part as
uncorrelated. Detector effects and the extrapolation to a single
centre-of-mass energy, as well as the uncertainty of the 4-jet
background shape of \eeqq\ events with multi-gluon emission, are
assumed to be uncorrelated.

\begin{table}[t]
\begin{center}
\begin{tabular}{| c || c | c |}
\hline
  & \multicolumn{2}{c |}{Experiment} \\
\hline
\hline
\Rn & \Ltre & \Opal \\
\hline
 Data & $0.8436\pm 0.0217$ & $1.2426\pm 0.0248$ \\
 \Jetset & $0.8622\pm 0.0037$ & $1.2910\pm 0.0028$ \\
 SK1 (100\%) & $0.7482\pm 0.0033$ & $1.0780\pm 0.0028$ \\
 \Herwig  & $0.8822\pm 0.0038$ & $1.3110\pm 0.0029$ \\
 \Ariadne & $0.8754\pm 0.0037$ & $1.2860\pm 0.0028$ \\
\hline
\hline
Systematics &  \Ltre & \Opal \\
\hline
 Intra-W BEC & 0.0017 & 0.0017 \\
 \eeqq\ shape & 0.0086 & 0.0104 \\
 $\sigma(\eeqq)$ & 0.0071 & 0.0024 \\
 \ZZtoqqqq\ shape  &
 \multirow{2}{*}{$\left.\begin{array}{c} \\
                        \end{array}\right\} 0.0020$} & 0.0018 \\
 $\sigma(\ZZtoqqqq)$ & & 0.0009 \\
 Detector effects & 0.0016 & 0.0142 \\
 $E_\mathrm{cm}$ dependence & 0.0020 & 0.0005 \\      
\hline
\end{tabular}
\caption[Particle-Flow Measurements of L3 and OPAL]{
\label{fsi:cr:tab:pflow}
Particle-flow measurements compared to Monte-Carlo predictions for the
SK1 CR model and different hadronisation models, together with
systematic uncertainties, provided by \Ltre\ and \Opal\ for the CR
combination.  }
\end{center}
\end{table}

The scaled measurements of \Ltre\ and \Opal,
$r_1^{\mathrm{data}}=r_L^{\mathrm{data}}$ and
$r_2^{\mathrm{data}}=r_O^{\mathrm{data}}$, are combined by minimising
a $\chi^2$ function which depends on the model parameter $k_I$ through
the model dependence of $r_i(k_I)$:

\begin{eqnarray}
  \chi^2_{r}(k_I,c_1,c_2)&=&\sum\limits_{i,j=1,2}\left\{\left(r_i^{\mathrm{data}}-r_i(k_I)+c_i
  \delta_{i,r}\right)
  \left(C_r^{-1}\right)_{ij}\right.\nonumber\\ &&\hspace*{3cm}\left.\cdot\left(r_j^{\mathrm{data}}-r_j(k_I)+c_j
  \delta_{j,r}\right)\right\}\nonumber\\ &&+\sum\limits_{m,n=1,2} c_m
  \left(C_c^{-1}\right)_{mn}c_n \,.
\end{eqnarray}
The covariance matrix, $C_r$, is constructed from only the
uncorrelated uncertainties and is actually diagonal. Correlated
uncertainties are introduced by varying each measurement
$r_i^{\mathrm{data}}$ with an additive term $c_i \delta_{i,r}$, where
$\delta_{i,r}$ is the correlated part of the uncertainty on
$r_i^{\mathrm{data}}$, and $c_i$ are auxiliary variables. The second
term in the $\chi^2_r$ function introduces correlations between the
auxiliary variables, so that the systematic uncertainties
$\delta_{i,r}$ also become effectively correlated.  This procedure is
equivalent to the so-called profile likelihood method with correlated
nuisance parameters, see, \eg,~\cite{bib:cr:rolke} and references
therein.
The correlation matrix $\left(C_c^{-1}\right)_{mn}$ is constructed
such that the uncertainty and central value of $k_I$ is exactly
identical to the result obtained with a classical and full covariance
matrix $C_r$. The best agreement is found for a correlation
coefficient of 0.50 between the auxiliary parameters. This more
complicated prescription is used to combine this result with other CR
inputs, which are provided in terms of likelihood functions.

To be able to vary $k_I$ continuously in the minimisation, the SK1
model predictions of $r_i(k_I)$ are described by a parametrised,
phenomenological function:

\begin{eqnarray}
r_i(k_I)&=&1+a_{i,1} \frac{k_I}{k_I+b_i}+
a_{i,2} \left(\frac{k_I}{k_I+b_i}\right)^2+
a_{i,3} \left(\frac{k_I}{k_I+b_i}\right)^4\nonumber\\
&&+\frac{a_{i,4}}{(1+k_I)}-a_{i,4}\;.
\label{eq:fsi:cr:pheno}
\end{eqnarray}
By construction, $r_i(k_I)$ is equal to 1 in the limit $k_I\to 0$,
\ie, when no CR effects are present.  The parameters, $a_{i,j}$ and
$b_i$ ($i=1,2$,$j=1,\ldots,4$), of the function are adjusted to fit
the $r_i(k_I)$ dependence determined in the SK1 Monte-Carlo simulation
by \Ltre\ and \Opal, which are shown in
Table~\ref{fsi:cr:tab:pflowsk1} of Appendix~\ref{fsi:cr:appendix}.
The terms $\frac{k_I}{k_I+b}$ are motivated by the approximate
description of the functional shape of the reconnection probability,
$p_{\mathrm{reco}}(k_I)$.

\begin{figure}[p]
\begin{center}
\epsfig{file=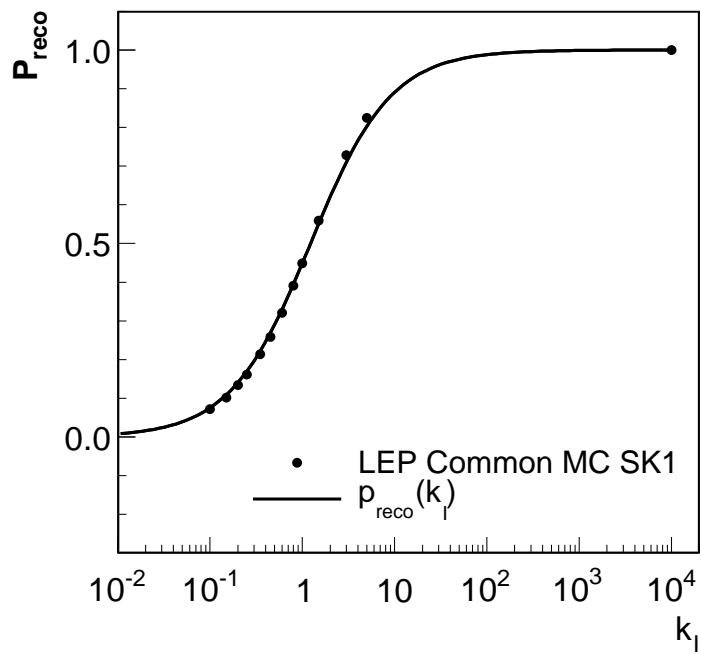,width=0.6\textwidth}
\epsfig{file=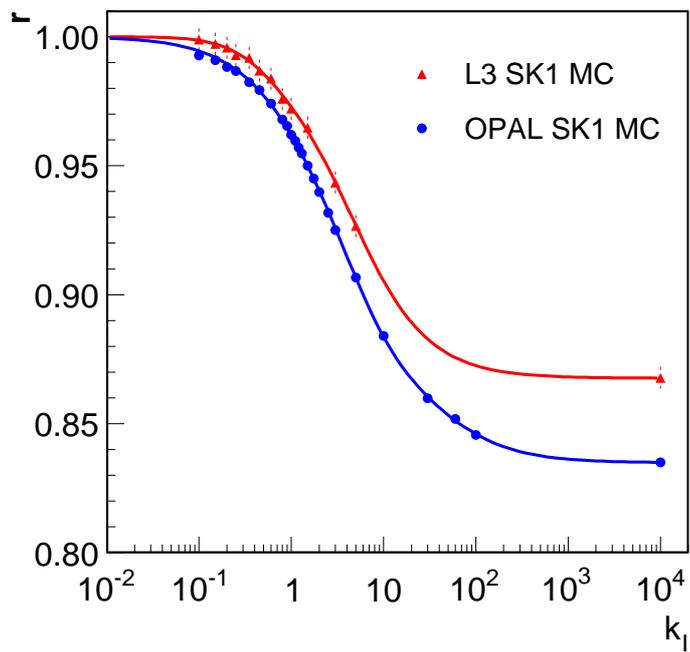,width=0.6\textwidth}
\end{center}
\caption[SK1 Colour Reconnection Model.]{ Top: Reconnection
probability as a function of the SK1 model parameter, $k_I$, together
with an approximate curve $p_{\mathrm{reco}}(k_I)$ to guide the
eye. Bottom: Monte-Carlo calculation and parametrisation of the
particle-flow ratio, $r(k_I)$, for \Ltre\ and \Opal, shown as
triangles and circles, respectively.}
\label{fsi:cr:fig:preco-sk1}
\end{figure}

With this parameter set, the function and the Monte-Carlo simulations
agree within less than one statistical standard deviation, as shown in
Figure~\ref{fsi:cr:fig:preco-sk1}. The best fitting parameter values
are listed in Table~\ref{fsi:cr:tab:rparams}.

\begin{figure}[p]
  \centerline{
    \epsfig{file=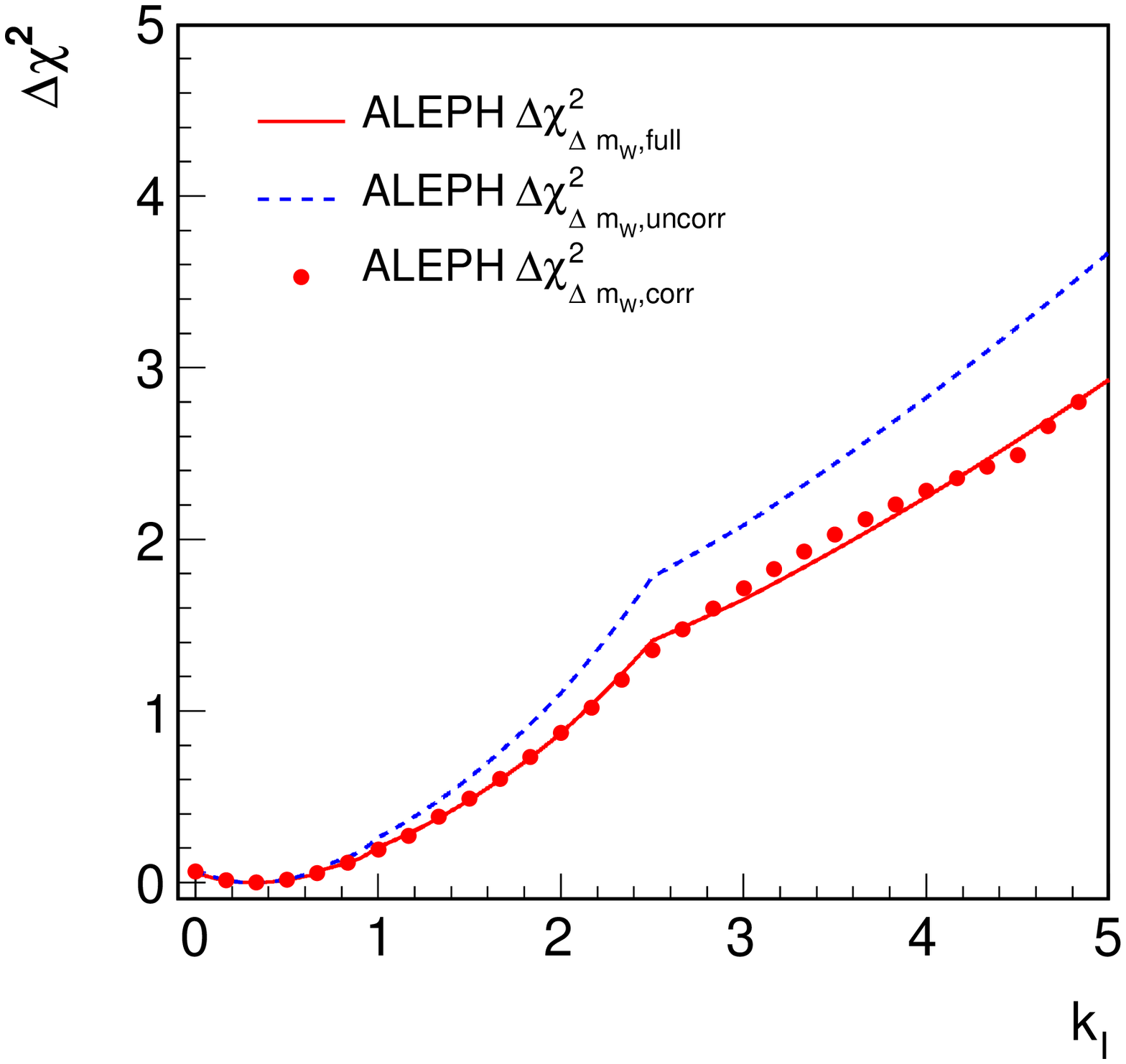,width=0.5\textwidth}
    \epsfig{file=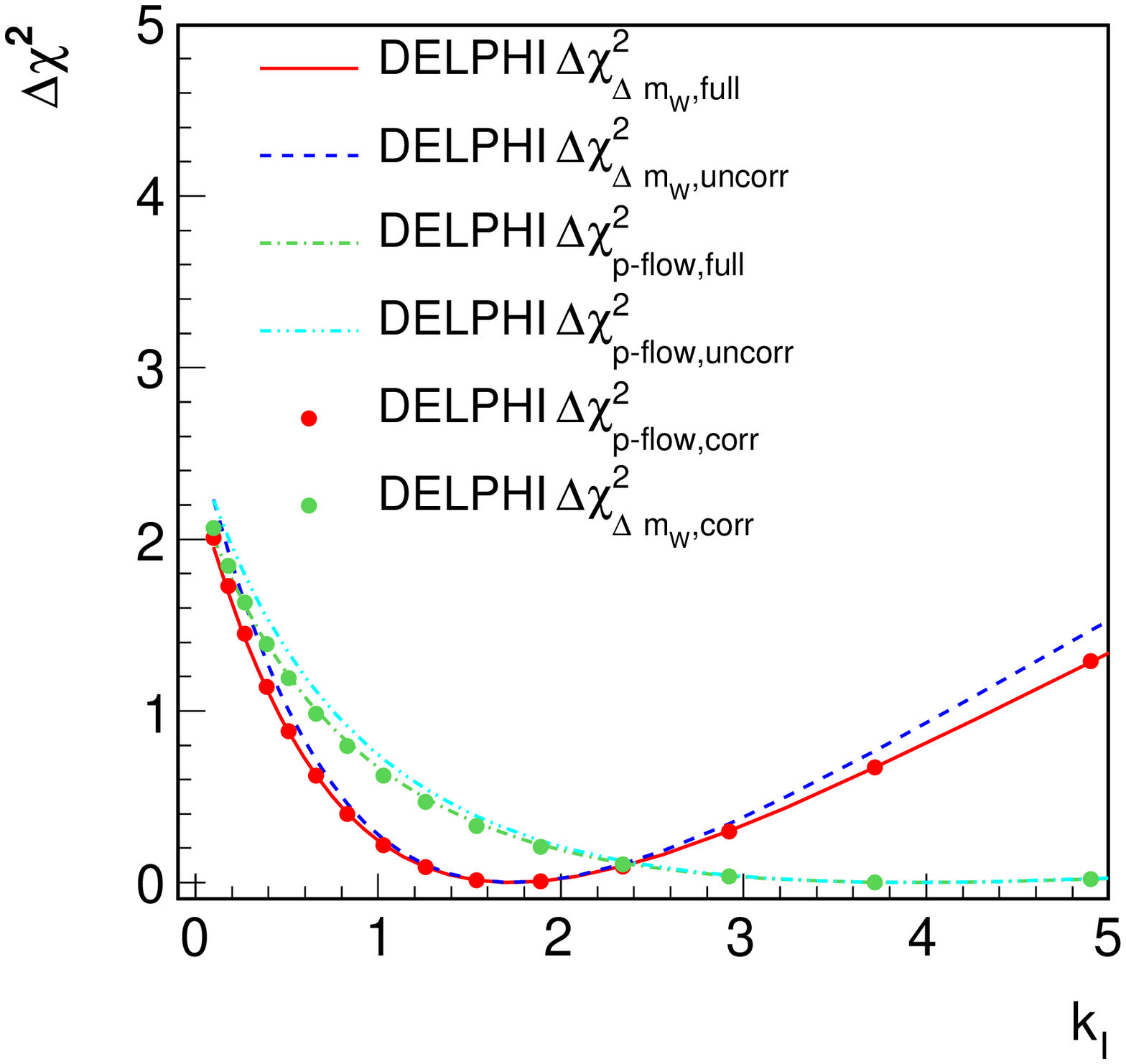,width=0.5\textwidth}
  }
  \centerline{
    \epsfig{file=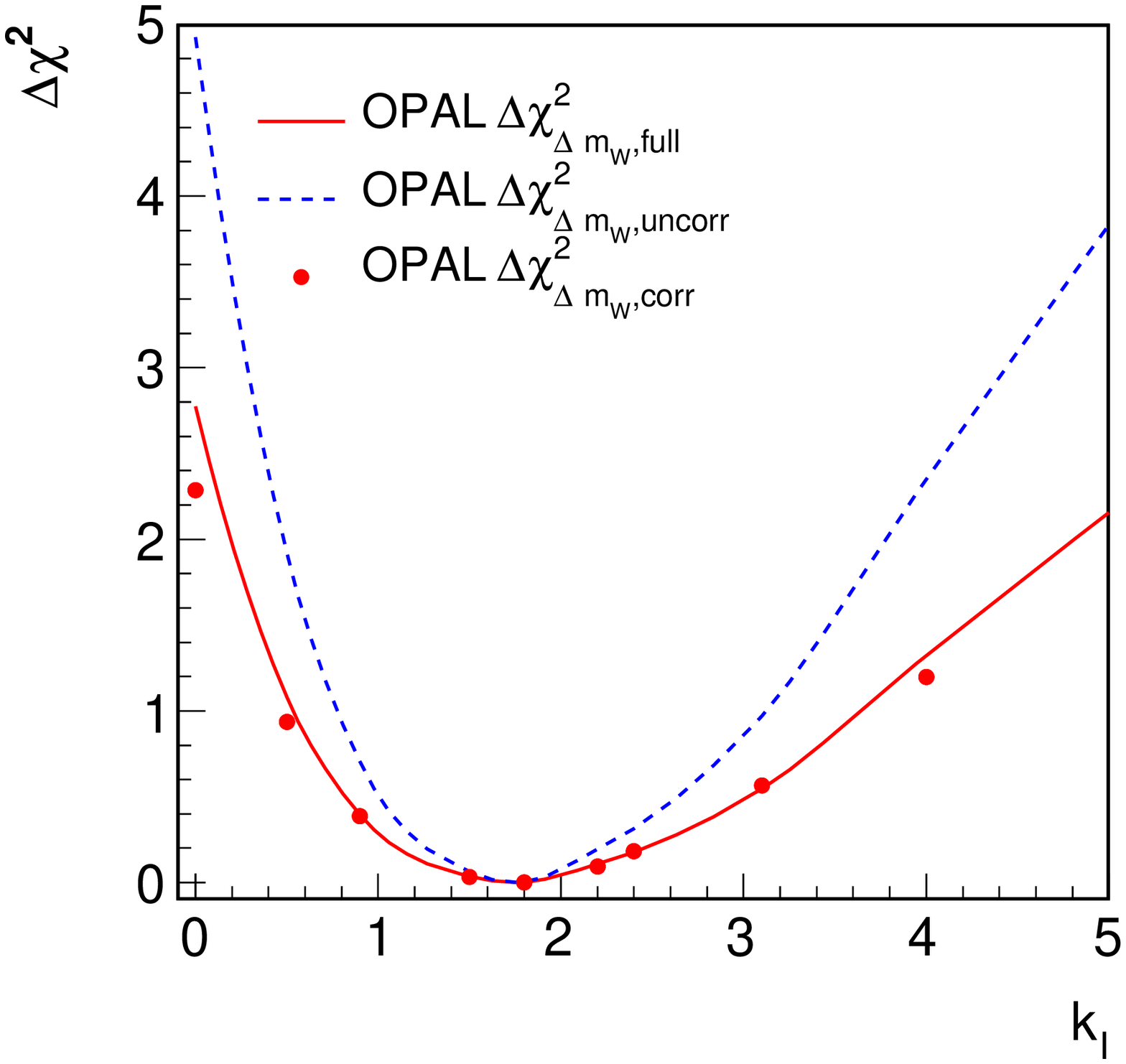,width=0.5\textwidth}
  }
\caption[Parametrisation of Correlated Uncertainties in CR.]  {
Comparison of $\Delta\chi^2$ distributions for CR measurements from
particle-flow and mass estimator differences, $\Delta\MW$, by the
\Aleph, \Delphi\ and \Opal\ experiments. Distributions are shown when
all uncertainties (continuous lines) and only uncorrelated
uncertainties (dashed lines) are taken into account. The
full-uncertainty curves are compared to the $\Delta\chi^2$
distribution when the variation of the parametrised uncertainty
$\delta(k_I)$ is used to introduce the correlated part of the
systematic uncertainties (circles).}
\label{fsi:cr:fig:chi2ki}
\end{figure}

The \Delphi\ experiment also performed a particle-flow
analysis~\cite{bib:cr:DELPHI_CR}.  The result is represented in terms
of two likelihood functions,
$L_{\mathrm{p-flow},D,\mathrm{full}}(k_I)$ and
$L_{\mathrm{p-flow},D,\mathrm{uncorr}}(k_I)$, where the former
contains all systematic uncertainties and the latter only uncorrelated
systematic uncertainties. These likelihoods are transformed into
$\Delta\chi^2(k_I)=-2\log L(k_I)$ values, which are smoothed by cubic
splines and then used in the combination. To treat correlations with
other inputs properly, a systematic variation,
$\delta_{\mathrm{p-flow},D}(k_I)$, of $k_I$ is introduced such that
the full $\Delta\chi^2_{\mathrm{p-flow},D,\mathrm{full}}(k_I)$ can be
reproduced in the following way from the uncorrelated
$\Delta\chi^2_{\mathrm{p-flow},D,\mathrm{uncorr}}(k_I)$ using an
auxiliary variable $c_3$:

\begin{equation}
  \Delta\chi^2_{\mathrm{p-flow},D,\mathrm{corr}}(k_I) = \min\limits_{c_3}\left\{\Delta\chi^2_{\mathrm{p-flow},D,\mathrm{uncorr}}(k_I+c_3 \delta_{\mathrm{p-flow},D}(k_I)) + c_3^2\right\} \,.
\end{equation}
The combined minimisation of
$\Delta\chi^2_{\mathrm{p-flow},D,\mathrm{corr}}(k_I,c_3)$ with respect
to $k_I$ and $c_3$ is equivalent to a minimisation of
$\Delta\chi^2_{\mathrm{p-flow},D,\mathrm{full}}(k_I)$ with respect to
$k_I$ only. The best agreement between the full description and this
procedure is obtained for $\delta_{\mathrm{p-flow},D}(k_I)=0.246 +
(0.754)^2 k_I$, which is shown in Figure~\ref{fsi:cr:fig:chi2ki}. The
advantage of this method is again the possibility to correlate $c_3$
with systematic uncertainties from other CR inputs.

\subsection{Determination of CR Effects Using W Mass Estimators}

A second very sensitive observable for CR is the variation of the
reconstructed W-boson mass in fully hadronic events when applying
different particle momentum thresholds and jet algorithms at event
reconstruction. As pointed out before, CR influences mostly the
particle-flow between jets and hence the low momentum component of the
hadronic jets.  Thus, estimators of $\MW$ in which the jet-defining
parameters are chosen to enhance or reduce the weight given to such
low momentum particles allow an observable to be constructed which is
sensitive to the presence or absence of CR.  To measure the effect of
CR, the mass difference, $\Delta\MW$, of two estimators is determined
in data and compared to the mass difference predicted by a certain CR
model. Since only mass {\it differences} are used to measure CR, the
correlation with the actual W mass measurement is small, in the order
of 10\%~\cite{bib:cr:ALEPH_MW, bib:cr:DELPHI_CR, bib:cr:OPAL_MW}.
 
The \Aleph\ experiment studied the dependence of $\MW$ as a function
of a momentum threshold, $p_{\mathrm{cut}}$, of the jet particles and
of the cone radius $R$ of the jets, which were constructed using the
Durham-PE algorithm~\cite{bib:mw:durham-1, *bib:mw:durham-2,
*bib:mw:durham-3, *bib:mw:durham-3e}. The $p_{\mathrm{cut}}$
thresholds were varied between 1~\GeV\ and 3~\GeV\ and the radius
between 0.4~rad and 1.0~rad. \Delphi\ compared the $\MW$ value from
the standard W mass analysis with alternative estimators applying a
cone cut at $R=0.5~\mathrm{rad}$ and a particle momentum cut at
$2~\GeV$, applying again the Durham jet clustering
algorithm~\cite{bib:mw:durham-1, *bib:mw:durham-2, *bib:mw:durham-3,
*bib:mw:durham-3e} in combination with an iterative cone algorithm in
order to estimate the direction of the modified jets. \Opal\ found
that their optimal CR sensitivity is for the comparison between an
analysis with a particle momentum cut at $2.5~\GeV$ and an alternative
one in which the jet particles are weighted according to a factor
$p^\kappa$, with $\kappa=-0.5$. The Durham jet clustering
algorithm~\cite{bib:mw:durham-1, *bib:mw:durham-2, *bib:mw:durham-3,
*bib:mw:durham-3e} is used to calculate the modified jet directions.

The \Aleph, \Delphi\ and \Opal\ inputs are provided in terms of
$\Delta\chi^2$ curves with complete systematic uncertainties and with
only the uncorrelated part,
$\Delta\chi^2_{\Delta\MW,i,\mathrm{full}}(k_I)$ and
$\Delta\chi^2_{\Delta\MW,i,\mathrm{uncorr}}(k_I)$, respectively.
Cubic splines are used to interpolate between the provided data
points. Correlations are again not taken directly from the input
function, $\Delta\chi^2_{\Delta\MW,i,\mathrm{full}}(k_I)$, but are
introduced by varying $k_I$ with additional uncertainties
$\pm\delta(k_I)$ using auxiliary variables $c_i$:

\begin{eqnarray}
  \Delta\chi^2_{\Delta\MW,A,\mathrm{corr}}(k_I) &=& \min\limits_{c_4}\left\{\Delta\chi^2_{\Delta\MW,A,\mathrm{uncorr}}(k_I+c_4 \delta_{\Delta\MW,A}(k_I)) + c_4^2\right\} \,, \\
  \Delta\chi^2_{\Delta\MW,D,\mathrm{corr}}(k_I) &=& \min\limits_{c_5}\left\{\Delta\chi^2_{\Delta\MW,D,\mathrm{uncorr}}(k_I+c_5 \delta_{\Delta\MW,D}(k_I)) + c_5^2\right\} \,, \\
  \Delta\chi^2_{\Delta\MW,O,\mathrm{corr}}(k_I) &=& \min\limits_{c_6}\left\{\Delta\chi^2_{\Delta\MW,O,\mathrm{uncorr}}(k_I+c_6 \delta_{\Delta\MW,O}(k_I)) + c_6^2\right\} \,.
\end{eqnarray}
The parametrisations of $\delta_{\Delta\MW,i}(k_I)$ follow step-wise
linear functions and are listed in Appendix~\ref{fsi:cr:appendix}. The
original input of \Aleph, \Delphi\ and \Opal\ is shown in
Figure~\ref{fsi:cr:fig:chi2ki} and compared to the
$\Delta\chi^2_{\Delta\MW,i,\mathrm{corr}}(k_I)$ functions using the
prescription described above. Good agreement is observed.

The main correlated systematic uncertainties which are taken into
account are from comparisons of hadronisation models, background scale
and shape uncertainties, as well as Bose-Einstein
correlations. Detector effects and corrections of the 4-jet background
are taken as uncorrelated. The original \Aleph\
analysis~\cite{bib:cr:ALEPH_MW} does not consider uncertainties due to
the BEC effect. Therefore, the corresponding
$\delta_{\Delta\MW,A}(k_I))$ values are scaled up by 11\%, which is
derived from an additional dedicated systematic study.

\subsection{Combination of LEP CR Measurements}

The LEP measurements of CR using the particle-flow method and the mass
estimator differences are combined using the following total
$\Delta\chi^2$ function:

\begin{eqnarray}
  \Delta\chi^2(k_I,c_1,\ldots,c_6)
  & = &
  \sum\limits_{i,j=1,2}\left\{
  (r_i^{\mathrm{data}}-r_i(k_I)+c_i \delta_{i,r})
  \left(C_r^{-1}\right)_{ij}
  (r_j^{\mathrm{data}}-r_j(k_I)+c_j \delta_{j,r})
  \right\} \nonumber\\
  &&+\Delta\chi^2_{\mathrm{p-flow},D,\mathrm{uncorr}}(k_I+c_3 \delta_{\mathrm{p-flow},D}(k_I))\nonumber\\
  &&+\Delta\chi^2_{\Delta\MW,A,\mathrm{uncorr}}(k_I+c_4 \delta_{\Delta\MW,A}(k_I))\nonumber\\
  &&+\Delta\chi^2_{\Delta\MW,D,\mathrm{uncorr}}(k_I+c_5 \delta_{\Delta\MW,D}(k_I))\nonumber\\
  &&+\Delta\chi^2_{\Delta\MW,O,\mathrm{uncorr}}(k_I+c_6 \delta_{\Delta\MW,O}(k_I))\nonumber\\
  &&+\sum\limits_{m,n=1}^{6} c_m \left(C_c^{-1}\right)_{mn}c_n \,,
\end{eqnarray}
which is constructed from the ingredients presented above. It is
minimised with respect to $k_I$ and the auxiliary parameters
$c_1,\ldots,c_6$, which are correlated through the covariance matrix
$C_c$. In the LEP combination, the correlation coefficients are set to
0.5, motivated by the full covariance matrix of the \Rn\ measurements,
where the correlated systematic uncertainties are reduced to only the
common part in each pair of measurements.

As a cross-check, the CR measurements of each collaboration are
combined, and the best $k_I$ values as well as their uncertainties are
extracted using the procedure described above. All results of the
individual experiments~\cite{bib:cr:ALEPH_MW, bib:cr:DELPHI_CR,
bib:cr:L3_CR, bib:cr:OPAL_MW} could be adequately reproduced, with
small deviations being attributed to known systematic effects covered
by the assigned uncertainties. More details can be found in the
Appendix~\ref{fsi:cr:appendix}.

\begin{figure}[tbh]
\begin{center}
\epsfig{file=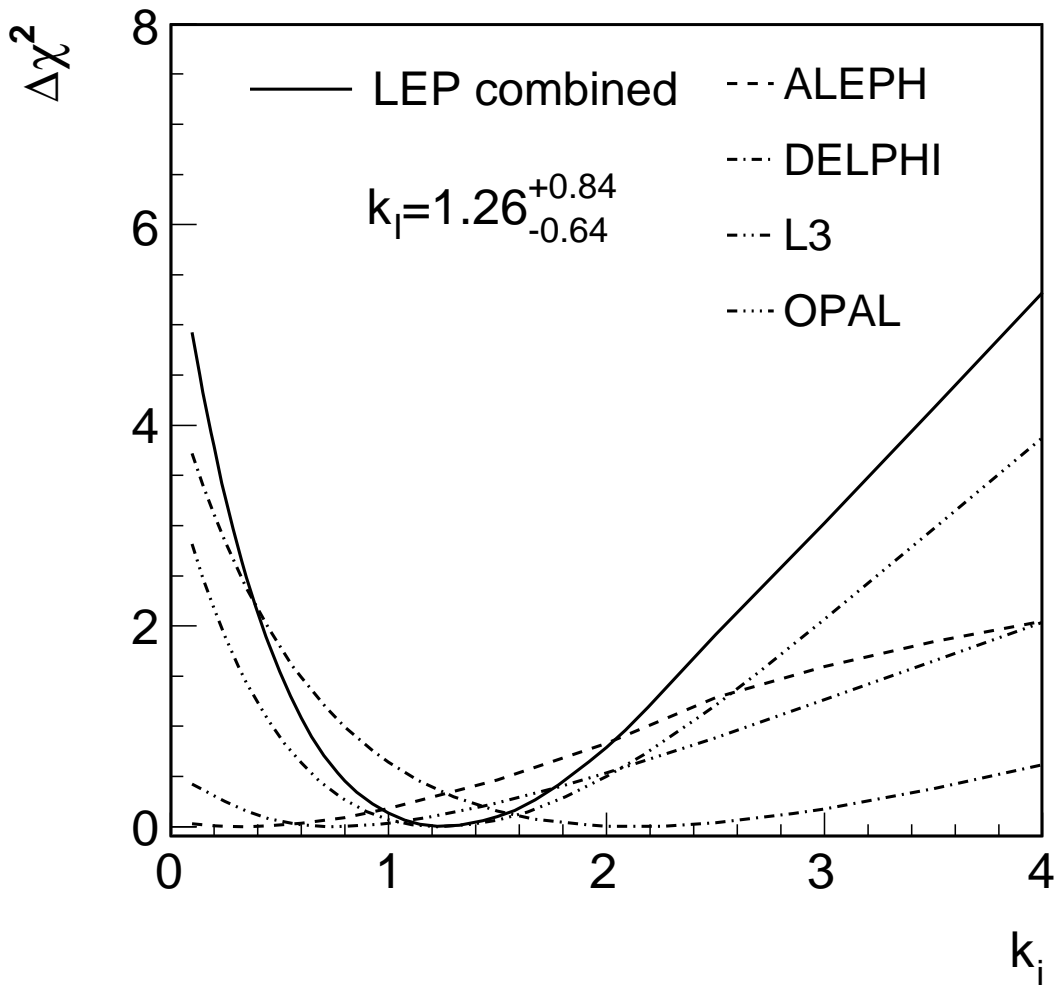,width=0.9\textwidth}
\end{center}
\caption[LEP Colour Reconnection Measurement.] { Individual and LEP
combined $\Delta\chi^2$ curves for the measurement of the CR parameter
$k_I$ in the SK1 model. }
\label{fsi:cr:fig:lep-final}
\end{figure}

The $\Delta\chi^2$ curves obtained for each experiment are shown in
Figure~\ref{fsi:cr:fig:lep-final}, together with the LEP
result. Combining all LEP data yields:

\begin{eqnarray}
k_I&=&1.26^{+0.84}_{-0.64}\,.
\end{eqnarray}
This result corresponds to a preferred reconnection probability of
51\% at a centre-of-mass energy of $189~\GeV$ in the SK1 model.
Absence of CR cannot be excluded, but is disfavoured by LEP at more
than two standard deviations.

\subsection{Summary}
 
A combination of the LEP particle-flow and W-mass estimator results is
presented, using the entire \LEPII\ data sample.  The data exclude
with 6.9 standard deviations an extreme version of the \SKI\ model in
which colour reconnection has been forced to occur in essentially all
events.  The combination procedure has been generalised to the \SKI\
model as a function of its reconnection probability.  The combined
data are described best by the model in which 51\% of events at
$189~\GeV$ are reconnected, corresponding to $\kI =1.26$.  The LEP
data disfavour the no-CR hypothesis at 99.5\% confidence level,
deviating from it by 2.8 standard deviations.  The 68\% confidence
level range for \kI\ is determined to be $0.62\le\kI\le2.10$.

\section{Bose-Einstein Correlations}
\label{sec:BEC}

\subsection{Introduction}

The LEP experiments have studied the strength of particle correlations
between two hadronic systems obtained from W-pair decays occurring
close in space-time at \LEPII.  The work presented in this section is
focused on so-called Bose-Einstein correlations (BEC), \ie, the
enhanced probability of production of pairs (multiplets) of identical
mesons close together in phase space. The effect is readily observed
in particle physics, in particular in hadronic decays of the Z boson,
and is qualitatively understood as a result of quantum-mechanical
interference originating from the symmetry of the amplitude of the
particle production process under exchange of identical mesons.

The presence of correlations between hadrons coming from the decay of
a W-pair, in particular those between hadrons originating from
different W bosons, can affect the direct reconstruction of the mass
of the initial W bosons.  The measurement of the strength of these
correlations can be used to constrain the corresponding systematic
uncertainty in the W mass measurement.

\subsection{Methods}

The principal method~\cite{be:chekanov}, called ``mixing method'',
used in the measurement is based on the direct comparison of
2-particle spectra from genuine hadronic W-pair events, $\mathrm{WW
\rightarrow q\bar{q}q\bar{q}}$, and from mixed WW events.  The latter
are constructed by mixing the hadronic sides of two semileptonic
W-pair events, $\mathrm{WW \rightarrow q\bar{q}\ell\nu}$, first used
in~\cite{be:DELPHI97}.  Such a reference sample has the advantage of
reproducing the correlations between particles belonging to the same W
boson, while the particles from different W bosons are uncorrelated by
construction.

This method gives a model-independent estimate of the interplay
between the two hadronic systems, for which BEC and also colour
reconnection are considered as dominant sources. The possibility of
establishing the strength of inter-W correlations in a
model-independent way is rather unique; most correlations do carry an
inherent model dependence on the reference sample. In the present
measurement, the model dependence is limited to the background
subtraction.

\subsection{Distributions}
\label{sec:FSI:BE:Distributions}

The two-particle correlations are evaluated using two-particle
densities defined in terms of the 4-momentum transfer
$Q=\sqrt{-(p_{1}-p_{2})^{2}}$, where $p_1,p_2$ are the 4-momenta of
the two particles:

\begin{equation}
\rho_{2}(Q)=\frac{1}{N_{ev}}\frac{dn_{pairs}}{dQ} \,.
\end{equation}
Here $n_{pairs}$ stands for the number of like-sign (unlike-sign)
2-particle permutations.\footnote{For historical reasons, the number
of particle permutations rather than combinations is used in formulas,
leading to a factor 2 in front of $\rho_{2}^{mix}$ in
Equation~\ref{eq:be:rho-mix}. The experimental statistical errors are,
however, based on the number of particle pairs, \ie, 2-particle
combinations.}  In the case of two stochastically independent
hadronically decaying W bosons the two-particle inclusive density is
given by:

\begin{equation}
 \rho_{2}^{WW}~=~\rho_{2}^{W^{+}}+\rho_{2}^{W^{-}}+2 \rho_{2}^{mix},
\label{eq:be:rho-mix}
\end{equation}
where $\rho_{2}^{mix}$ can be expressed via the single-particle
inclusive density $\rho_1(p)$ as:

\begin{equation}
\rho_2^{mix}(Q)~=~\int d^{4}p_{1}d^{4}p_{2}\rho^{W^{+}}(p_1)\rho^{W^{-}}(p_2)\delta(Q^{2}+(p_{1}-p_{2})^{2})\delta(p_1^2-m_{\pi}^2)\delta(p_2^2-m_{\pi}^2).
\end{equation}
Assuming further that:

\begin{equation}
\rho_{2}^{W^{+}}(Q)~=~\rho_{2}^{W^{-}}(Q)~=~\rho_{2}^{W}(Q),
\end{equation}
one obtains for the case of two stochastically independent
hadronically decaying W bosons:

\begin{equation}
\rho_{2}^{WW}(Q)~=~2\rho_{2}^{W}(Q)+2\rho_2^{mix}(Q).
\end{equation}
In the mixing method, $\rho_2^{mix}$ is obtained by combining two
hadronic W systems from two different semileptonic W-pair events.  The
direct search for inter-W BEC is done using the difference of
2-particle densities:

\begin{equation}
\Delta \rho(Q) ~=~ \rho_{2}^{WW}(Q)-2\rho_{2}^{W}(Q)-2\rho_2^{mix}(Q),
\label{eq:be:dr}
\end{equation}
or, alternatively, their ratio:

\begin{equation}
    D(Q)~=~\frac{\rho_{2}^{WW}(Q)}{2\rho_{2}^{W}(Q)+2\rho^{mix}(Q)}
        ~=~ 1 + \frac{ \Delta \rho(Q) }{2\rho_{2}^{W}(Q)+2\rho^{mix}(Q)} .
\label{eq:be:D}
\end{equation}
Given the definition of the genuine inter-W correlations function
$\delta_I(Q)$~\cite{be:DeWolf}, it can be shown that

\begin{equation}
     \delta_I(Q)~=~\frac{\Delta \rho(Q)}{2\rho_2^{mix}(Q)}.
\end{equation}
To disentangle the BEC effects from other possible correlation sources
(such as energy-momentum conservation or colour reconnection), which
are supposed to be the same for like-sign and unlike-sign charge
pairs, the double difference:

 \begin{equation}
 \delta \rho(Q) =  \Delta \rho^{like-sign}(Q)- \Delta \rho^{unlike-sign}(Q),
\end{equation}
or the double ratio,

 \begin{equation}
    d(Q) =  D^{like-sign}(Q)/D^{unlike-sign}(Q),
\end{equation}
is analysed.

The event mixing procedure may introduce artificial distortions, or
may not fully account for some detector effects or for correlations
other than BEC.  Most of these possible effects are treated in the
Monte-Carlo simulation without inter-W BEC.  Therefore they are
reduced by using the double ratio or the double difference:

\begin{equation}
    D'(Q)~=~\frac{D(Q)_{data}}{D(Q)_{MC,no inter}}\hspace{0.2cm} ,\hspace{1cm}
\Delta \rho'(Q) ~=~\Delta \rho(Q)_{data} - \Delta \rho(Q)_{MC,no inter}
\hspace{0.2cm} ,
\end{equation}
where $D(Q)_{MC,no inter}$ and $\Delta \rho(Q)_{MC,no inter}$ are
derived from a MC without inter-W BEC.

In addition to the mixing method, ALEPH~\cite{be:ALEPH00} also uses
the double ratio of like-sign pairs ($N_{\pi}^{++,--}(Q)$) and
unlike-sign pairs $N_{\pi}^{+-}(Q)$ corrected with Monte-Carlo
simulations without BEC effects:

\begin{equation}
R^*(Q) ~=~ \left.\left(\frac{ N_{\pi}^{++,--}(Q) }
{ N_{\pi}^{+-}(Q) } \right)^{data}\right/
\left(\frac{ N_{\pi}^{++,--}(Q) }
{ N_{\pi}^{+-}(Q) }\right)^{MC}_{noBE}.
\end{equation}
In analyses based on $\Delta\rho(Q)$, $\delta \rho(Q)$ or
$\delta_I(Q)$, a deviation from zero indicates the presence of inter-W
correlations, whereas for studies of $D(Q)$, $D'(Q)$ or $d(Q)$, the
corresponding signature is a deviation from unity.  For $R^*(Q)$, a
difference between data and the Monte-Carlo prediction excluding
inter-W BEC is studied.

\subsection{Results}

\begin{figure}[htbp]
\begin{center}
\epsfig{file=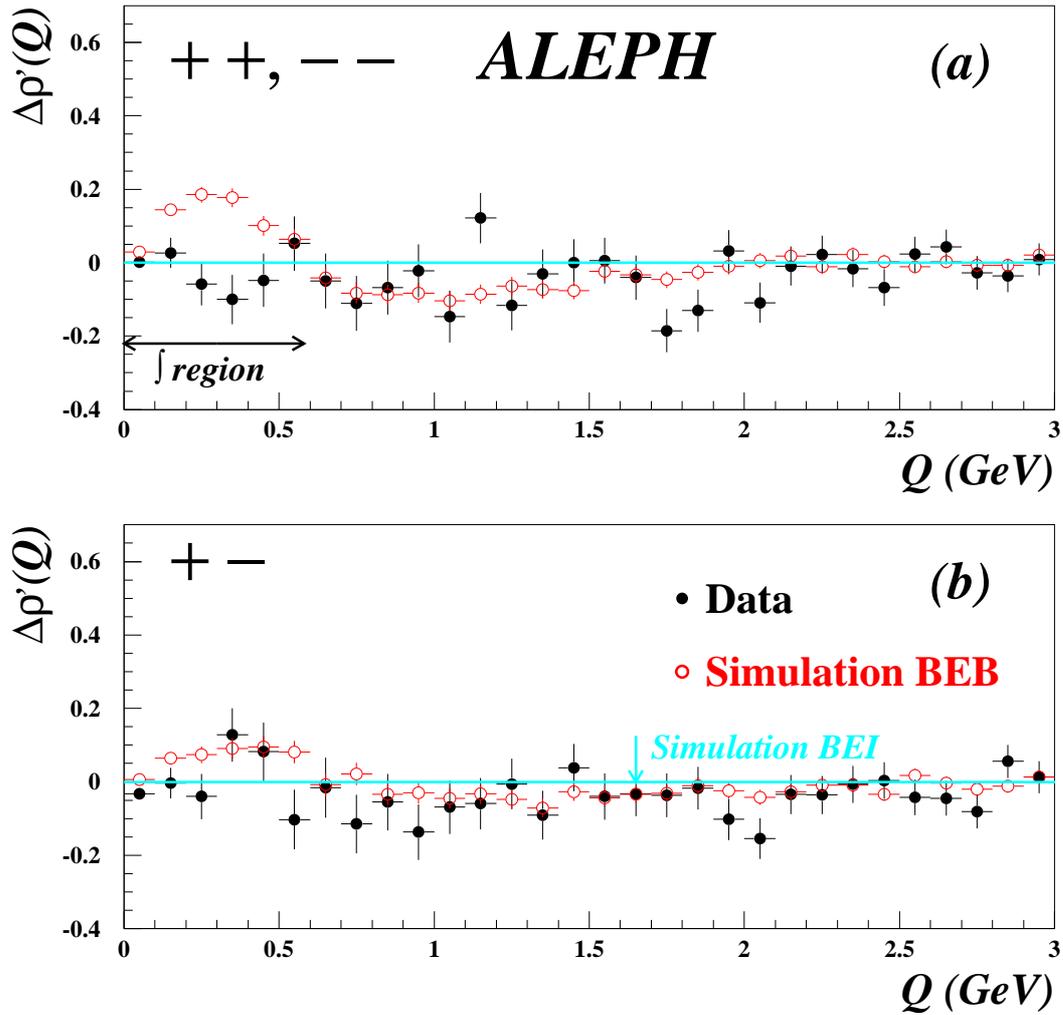,width=0.9\linewidth}
\end{center}
\caption[BE distribution measured by ALEPH]{ Distribution of the
quantity $\Delta\rho'$ for like- and unlike-sign pairs as a function
of $Q$ as measured by the ALEPH collaboration~\cite{be:ALEPH05}.  BEI
stands for the case in which Bose-Einstein correlations do not occur
between decay products of different W bosons, and BEB if they do.}
\label{be:fig:aleph}
\end{figure}

\begin{figure}[htbp]
\begin{center}
\epsfig{file=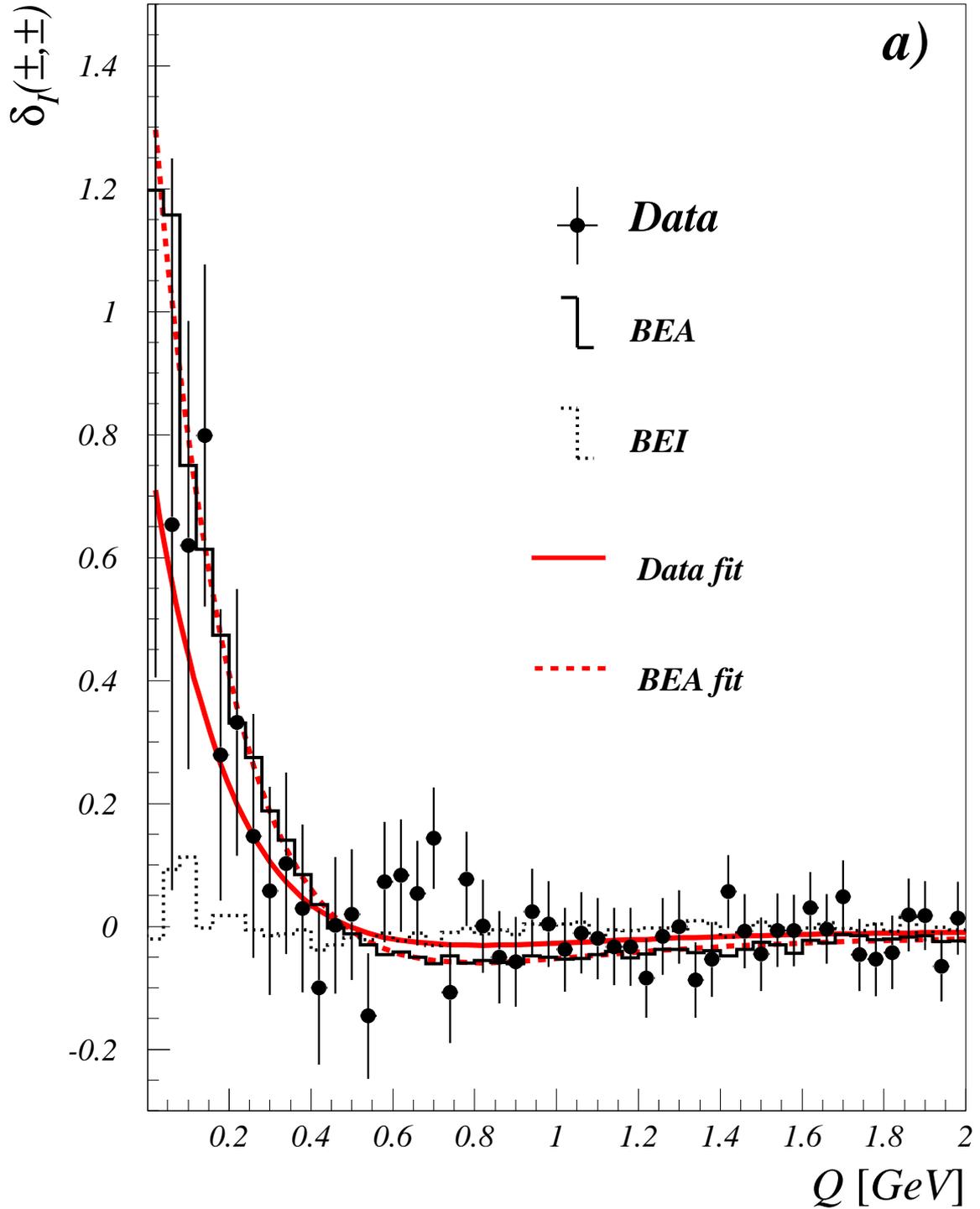,width=0.9\linewidth}
\end{center}
\caption[BE distribution measured by DELPHI]{Distributions of the
quantity $\delta_I$ for like-sign pairs as a function of $Q$ as
measured by the DELPHI collaboration~\cite{be:DELPHI05}.  The solid
line shows the fit results. BEI stands for the case in which
Bose-Einstein correlations do not occur between decay products of
different W bosons, and BEA if they do. }
\label{be:fig:delphi}
\end{figure}

\begin{figure}[htbp]
\begin{center}
\epsfig{file=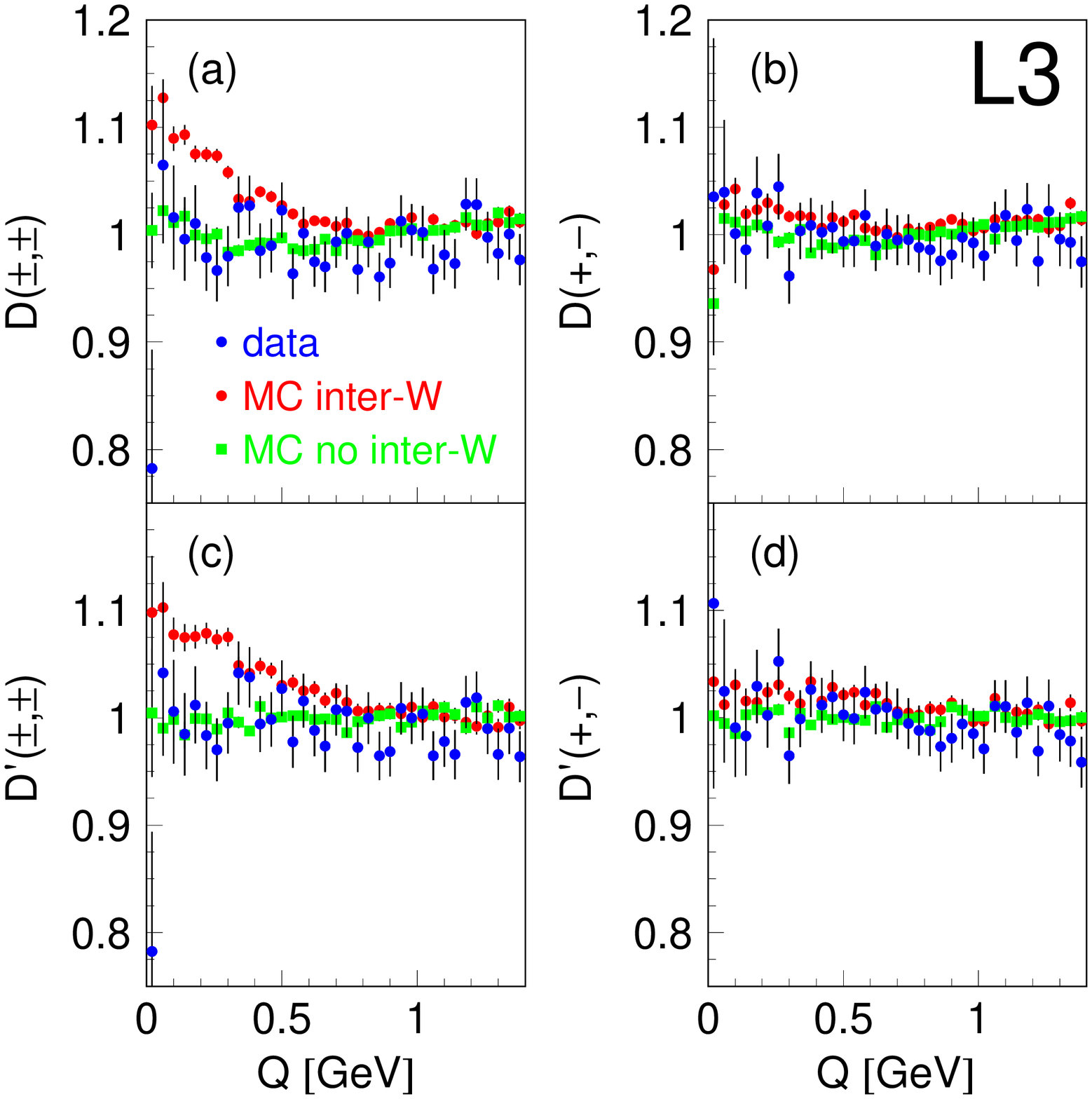,width=\linewidth}
\end{center}
\caption[BE distribution measured by L3]{Distributions of the quantity
$D$ and $D'$ for like- and unlike-sign pairs as a function of $Q$ as
measured by the L3 collaboration~\cite{be:L302}.}
\label{be:fig:l3}
\end{figure}

\begin{figure}[htbp]
\begin{center}
\epsfig{file=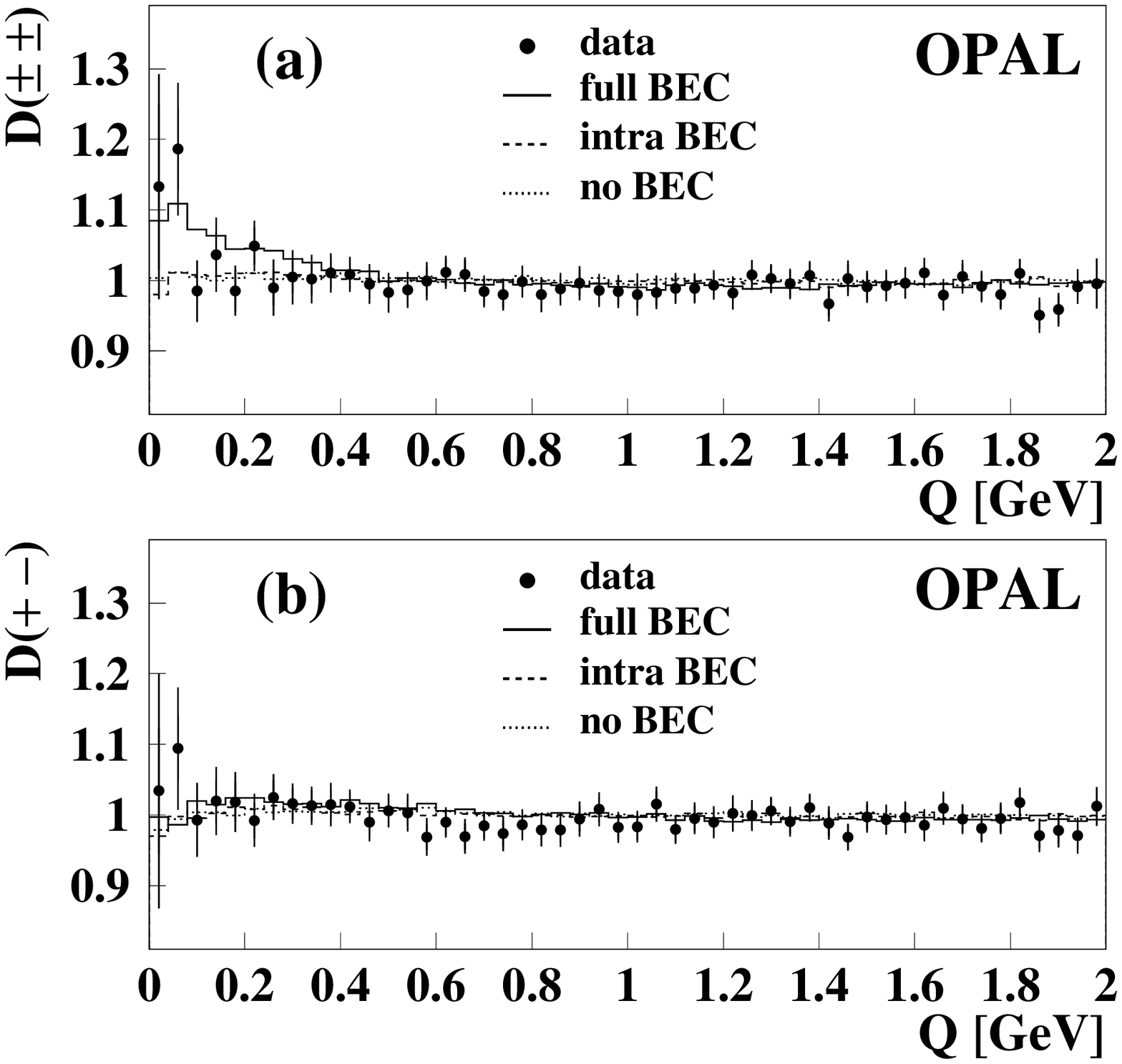,width=0.9\linewidth}
\end{center}
\caption[BE distribution measured by OPAL]{Distribution of the
quantity $D$ for like- and unlike-sign pairs as a function of $Q$ as
measured by the OPAL collaboration~\cite{be:OPAL05}. }
\label{be:fig:opal}
\end{figure}

The four LEP experiments have published results applying the mixing
method to the full \LEPII\ data sample.  As examples, the
distributions of $\Delta\rho'$ measured by ALEPH~\cite{be:ALEPH05},
$\delta_I$ measured by DELPHI~\cite{be:DELPHI05}, $D$ and $D'$
measured by L3~\cite{be:L302} and $D$ measured by
OPAL~\cite{be:OPAL05} are shown in Figures~\ref{be:fig:aleph},
\ref{be:fig:delphi}, \ref{be:fig:l3} and~\ref{be:fig:opal},
respectively.  In addition ALEPH have published results using
$R^*$~\cite{be:ALEPH00}. The centre-of-mass energies, luminosities and
the number of events used for the different measurements are listed in
Table~\ref{be:table:lumi}.

\begin{table}[htb]   
\begin{center}
\vspace*{0.4cm}
\begin{tabular}{|l||c|c|r|r|}
\hline
& $\sqrt{s}$  & Luminosity          & \multicolumn{2}{c|}{Number of events} \\
&    [GeV]              & [pb$^{-1}$]        & $\mathrm{WW \rightarrow q\bar{q}q\bar{q}\,}$      & $\mathrm{WW \rightarrow q\bar{q}\ell \nu\,}$ \\ 
\hline
\hline
ALEPH&  183-209    & 683 &  6155  & 4849  \\ 
DELPHI& 189-209    & 550 &  3252  & 2567  \\  
L3    & 189-209    & 629 &  5100  & 3800  \\
OPAL  & 183-209    & 680 &  4470  & 4533  \\
ALEPH R$^*$ & 172-189 & 242 & 2021 & -    \\\hline
\end{tabular}
\caption[Data samples used for BEC measurements]{The centre-of-mass
energies, luminosities and the number of events used for the different
measurements. }
\label{be:table:lumi}
\end{center}   
\end{table}

A simple combination procedure is available through a $\chi^2$ average
of the numerical results of each experiment~\cite{be:ALEPH00,
be:ALEPH05, be:DELPHI05, be:L302, be:OPAL05} with respect to a
specific BEC model under study, here based on comparisons with various
tuned versions of the LUBOEI model~\cite{JETSET, mw:bib:LUBOEI}.  The
tuning is performed by adjusting the parameters of the model to
reproduce correlations in samples of Z and semileptonic W decays, and
applying identical parameters to the modelling of inter-W correlations
(so-called ``fullBE'' scenario).  In this way the tuning of each
experiment takes into account detector systematic uncertainties in the
track measurements.

An important advantage of the combination procedure used here is that
it allows the combination of results obtained using different
analyses.  The combination procedure assumes a linear dependence of
the observed size of BEC on various estimators used to analyse the
different distributions.  It is also verified that there is a linear
dependence between the measured W mass shift and the values of these
estimators~\cite{bib:cr:L3_MW}.  The estimators are: the integral of
the $\Delta\rho(Q)$ distribution (ALEPH, L3, OPAL); the parameter
$\Lambda$ when fitting the function $N(1+\delta
Q)(1+\Lambda\exp(-k^2Q^2))$ to the $D'(Q)$ distribution, with $N$
fixed to unity (L3), or $\delta$ fixed to zero and $k$ fixed to the
value obtained from a fit to the full BEC sample (ALEPH); the
parameter $\Lambda$ when fitting the function $N(1+\delta
Q)(1+\Lambda\exp(-Q/R))$ to the $D(Q)$, $D(Q)'$ and $d$ distributions,
with $R$ fixed to the value obtained from a fit to the full BEC sample
(OPAL); the parameter $\Lambda$ when fitting the function
$\Lambda\exp(-RQ)(1+\epsilon RQ)+\delta(1+\frac{\rho_{2}^{W}}{
\rho_{2}^{mix}})$ to the $\delta_I$ distribution, with $R$ and
$\epsilon$ fixed to the value obtained from a fit to the full BEC
sample (DELPHI); and finally the integral of the term describing the
BEC part, $\int \lambda\exp(-\sigma^2Q^2)$, when fitting the function
$\kappa(1+\epsilon Q)(1+\lambda\exp(-\sigma^2Q^2))$ to the $R^*(Q)$
distribution (ALEPH).

The size of the correlations for like-sign pairs of particles measured
in terms of these estimators is compared with the values expected in
the model with and without inter-W correlations in
Table~\ref{be:table:bei}.  Table~\ref{be:table:rel} summarises the
normalised fractions of the model seen.

\begin{table}[p]
\begin{center}
 \begin{tabular}{|l||l|l|l|l|l|c|}
 \hline
Analysis & Data--noBE &  stat. & syst. & corr. syst. &fullBE--noBE & Ref. \\ \hline
\hline
{\bf ALEPH (fit to $D'$)}  &{\bf  $-$0.004} &{\bf 0.011} &{\bf 0.014} & {\bf 0.003} & {\bf 0.081} &\cite{be:ALEPH05} \\
ALEPH (integral of $\Delta\rho$)  &  $-$0.127 & 0.143 & 0.199 & 0.044 &  0.699 &\cite{be:ALEPH05} \\
ALEPH (fit to $R^*$)  &  $-$0.004 & 0.0062 & 0.0036 & negligible &  0.0177 &\cite{be:ALEPH00} \\
{\bf DELPHI (fit to  $\delta_I$)} &{\bf $+$0.72} &{\bf 0.29} &{\bf 0.17} &{\bf 0.070} &{\bf 1.40} &\cite{be:DELPHI05} \\
{\bf L3 (fit to $D'$)}  &{\bf $+$0.008} &{\bf 0.018} &{\bf 0.012} &{\bf 0.0042} &{\bf  0.103} & \cite{be:L302} \\
L3 (integral of $\Delta\rho$) &   $+$0.03 & 0.33 &  0.15 &  0.055 & 1.38 & \cite{be:L302} \\
OPAL (integral of $\Delta\rho$) &   $-$0.01 & 0.27 &  0.23 &  0.06 & 0.77 & \cite{be:OPAL05} \\
{\bf OPAL (fit to $D$)} &{\bf $+$0.040} &{\bf 0.038} &{\bf  0.038} &{\bf  0.017} &{\bf 0.120} & \cite{be:OPAL05} \\
OPAL (fit to $D'$) &   $+$0.042 & 0.042 &  0.047 &  0.019 & 0.123 & \cite{be:OPAL05} \\
OPAL (fit to $d$) &   $-$0.017 & 0.055 &  0.050 &  0.003 & 0.133 & \cite{be:OPAL05} \\
\hline
\end{tabular}
\caption[Results of BEC analyses]{An overview of the results from
 different measurements described in
 Section~\ref{sec:FSI:BE:Distributions}: the difference between the
 measured correlations and the model without inter-W correlations
 (data--noBE), the corresponding statistical (stat.)~and total
 systematic (syst.)~errors, the correlated systematic error
 contribution (corr.~syst.), and the difference between ``fullBE'' and
 ``noBE'' scenario. The measurements used in the combination are
 highlighted. }
\label{be:table:bei}
\end{center}
\end{table}

\begin{table}[p]
\begin{center}
 \begin{tabular}{|l||c|c|c|}
 \hline
Analysis & Fraction of model &  stat. & syst.  \\ 
\hline
\hline
{\bf ALEPH (fit to $D'$)} & {\bf $-$0.05} & {\bf 0.14} &{\bf 0.17}  \\
ALEPH (integral of $\Delta\rho$) &  $-$0.18 & 0.20 & 0.28  \\
ALEPH (fit to $R^*$) &  $-$0.23 & 0.35 & 0.20  \\
{\bf DELPHI (fit to $\delta_I$)} & {\bf  $+$0.51} &{\bf 0.21} &{\bf 0.12}  \\
{\bf L3  (fit to $D'$)} & {\bf $+$0.08} &{\bf 0.17} &{\bf 0.12}  \\
L3 (integral of $\Delta\rho$) & $+$0.02 & 0.24 & 0.11 \\
OPAL (integral of $\Delta\rho$) &   $-$0.01 & 0.35 & 0.30  \\
{\bf OPAL  (fit to $D$)} & {\bf  $+$0.33} &{\bf 0.32} &{\bf 0.32}  \\
OPAL  (fit to $D'$) &   $+$0.34 & 0.34 & 0.38  \\
OPAL  (fit to $d$) &   $-$0.13 & 0.41 & 0.38  \\
\hline
 \end{tabular}
\caption[Fractions of BEC model seen]{ The measured size of BEC
  expressed as the relative fraction of the model with inter-W
  correlations (see Equation~\ref{eq:be:model-fra} and
  Table~\ref{be:table:bei}). The measurements used in the combination
  are highlighted.}
\label{be:table:rel}
\end{center}
\end{table}

\begin{figure}[htbp]
\begin{center}
\epsfig{file=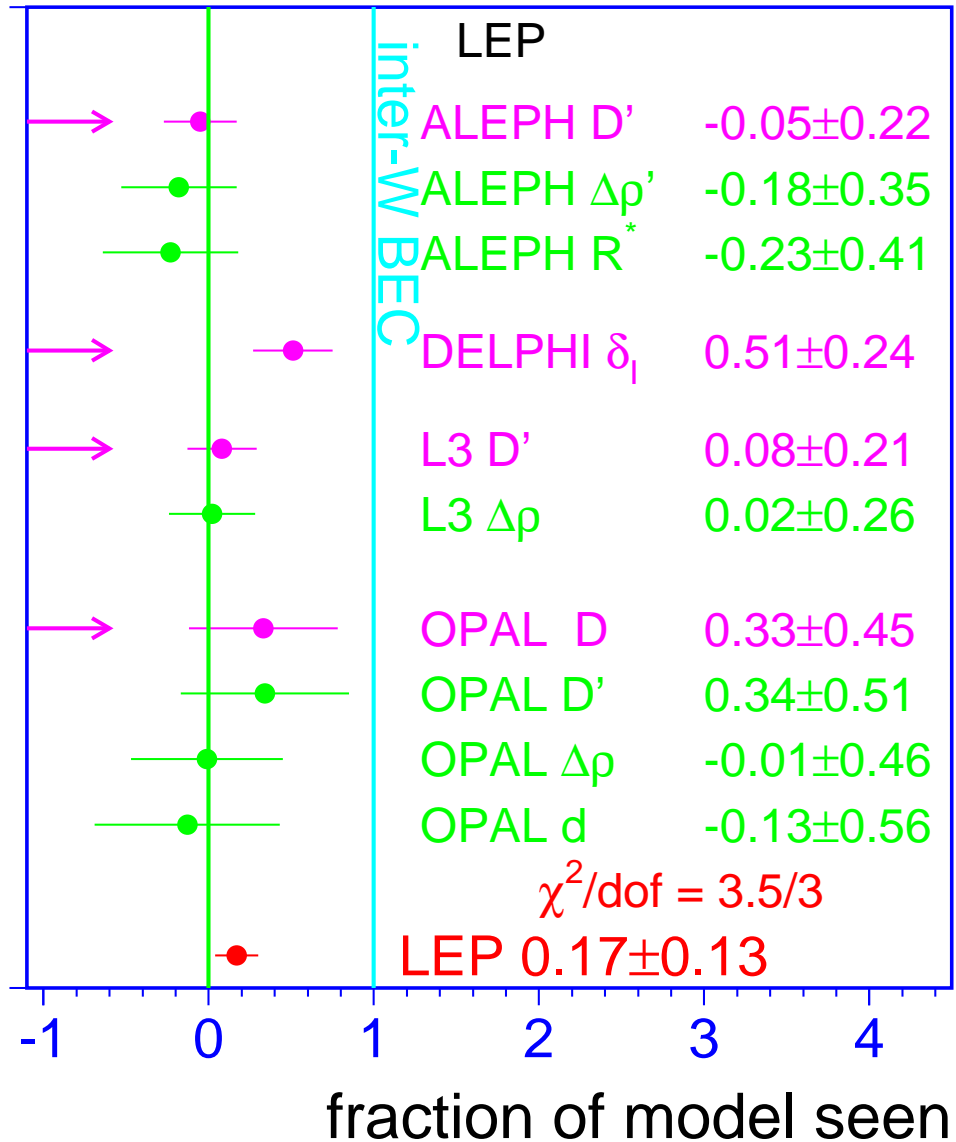,width=0.9\linewidth}
\end{center}
\caption[Fractions of BEC model seen]{Measured BEC expressed as the
relative fraction of the model with inter-W correlations. The arrows
indicate the measurements used in the combination. The LEP combination
is shown at the bottom. }
\label{be:chi-comb}
\end{figure}

For the combination of the above measurements one has to take into
account correlations between them. Correlations between results of the
same experiment are strong and are not available. Varying these
correlations and combining the three ALEPH measurements, for example,
one obtains normalised fractions of the model seen very close to those
of the most precise measurement.  Therefore, for simplicity, the
combination of the most precise measurements of each experiment is
made here: $D'$ from ALEPH, $\delta_I$ from DELPHI, $D'$ from L3 and
$D$ from OPAL.  In this combination only the uncertainties in the
understanding of the background contribution in the data are treated
as correlated between experiments (denoted as ``corr. syst.''  in
Table~\ref{be:table:bei}).  The combination via a fit using MINUIT
gives:

\begin{equation}
\frac{\mathrm{data - model(noBE)}}{\mathrm{model(fullBE)-model(noBE)}}
 ~ = ~ 0.17 \pm 0.095(stat.)\pm 0.085(sys.) ~ = ~ 0.17 \pm 0.13~\,,~~
\label{eq:be:model-fra}
\end{equation}
where ``noBE'' includes correlations between decay products of each W,
but not the ones between decay products of different Ws and ``fullBE''
includes all the correlations.  A $\chidf=3.5/3$ of the fit is
observed.  The measurements and their average are shown in
Figure~\ref{be:chi-comb}. The measurements used in the combination are
marked with an arrow. The results of LEP experiments are in good
agreement.

\subsection{Conclusions}

The LUBOEI model of BEC between pions from different W bosons is
disfavoured. The 68\% confidence level (CL) upper limit on these
correlations is $0.17+0.13=0.30$.  This result can be translated into
a 68\% CL upper limit on the shift of the W mass measurements due to
the BEC between particles from different Ws, $\Delta\MW$, assuming a
linear dependence of $\Delta\MW$ on the size of the correlation. For
the specific BE model investigated, LUBOEI, a shift of $-35~\MeV$ in
the W mass is obtained at full BEC strength.  The W mass analysis
techniques applied are, however, optimised to reduce colour
reconnection effects on $\MW$ which also has the effect of reducing
the mass shift due to BEC. A combination of the reduced shifts
reported by the LEP experiments~\cite{bib:mw:a-mw, bib:mw:d-mw,
bib:mw:l-mw, bib:mw:o-mw} gives a shift of $-23~\MeV$ in the W mass at
full BEC strength.  Thus the 68\% CL upper limit on the magnitude of
the mass shift within the LUBOEI model is: $|\Delta\MW| = 0.30 \times
23~\MeV = 7~\MeV$.

\newcommand{\sww}        {\sigma_\mathrm{WW}}
\newcommand{\szz}        {\sigma_\mathrm{ZZ}}
\newcommand{\swent}      {\sigma_\mathrm{We\nu\,tot}}
\newcommand{\swenh}      {\sigma_\mathrm{We\nu\,had}}
\newcommand{\szee}       {\sigma_\mathrm{Zee}}
\newcommand{\sWWg}       {\sigma_\mathrm{WW\gamma}}
\newcommand{\sZgg}       {\sigma_\mathrm{Z\gamma\gamma}}

\newcommand{\rww}        {{\cal{R}}_\mathrm{WW}} 
\newcommand{\rzz}        {{\cal{R}}_\mathrm{ZZ}} 
\newcommand{\rwev}       {{\cal{R}}_\mathrm{We\nu}} 
\newcommand{\rzee}       {{\cal{R}}_\mathrm{Zee}} 

\newcommand{\wwdsdt}     {d\sigma/d\theta_{W-}}
\newcommand{\wwbr}       {\mbox{$\mathcal{B}$}}
\newcommand{\oa}         {{\cal{O}}(\alpha)}
\newcommand{\Wtolnu}     {\mbox{$\mathrm{W}\rightarrow
                                 \ell\overline{\nu}_{\ell}$}}
\newcommand{\Wtoenu}     {\mbox{$\mathrm{W}\rightarrow
                                 \mathrm{e\overline{\nu}_{e}}$}}
\newcommand{\Wtomnu}     {\mbox{$\mathrm{W}\rightarrow
                                 \mu\overline{\nu}_{\mu}$}}
\newcommand{\Wtotnu}     {\mbox{$\mathrm{W}\rightarrow
                                 \tau\overline{\nu}_{\tau}$}}
\newcommand{\BWtoenu}    {\mbox{$\mathcal{B}(\mathrm{W}\rightarrow
                                 \mathrm{e\overline{\nu}_{e}})$}}
\newcommand{\BWtomnu}    {\mbox{$\mathcal{B}(\mathrm{W}\rightarrow
                                 \mu\overline{\nu}_{\mu})$}}
\newcommand{\BWtotnu}    {\mbox{$\mathcal{B}(\mathrm{W}\rightarrow
                                 \tau\overline{\nu}_{\tau})$}}
\newcommand{\BWtolnu}    {\mbox{$\mathcal{B}(\mathrm{W}\rightarrow
                                 \ell\overline{\nu}_{\ell})$}}
\newcommand{\phs}        {\phantom{^*}}
\newcommand{\phzs}       {\phantom{^{0*}}}
\newcommand{\phz}        {\phantom{0}}
\newcommand{\phzz}       {\phantom{00}}
\newcommand{\phm}        {\phantom{-}}
\newcommand{\CoM}        {centre-of-mass}
\newcommand{\pbinv}      {\mbox{$\mathrm{pb}^{-1}$}}
\newcommand{\Conf}       {\mbox{CONF}}
\newcommand{\Grace}      {\mbox{\tt grc4f}}
\newcommand{\Gentle}     {\mbox{GENTLE}}
\newcommand{\YFSZZ}      {\mbox{YFSZZ}}
\newcommand{\ZZTO}       {\mbox{ZZTO}}
\newcommand{\WTO}        {\mbox{WTO}}
\newcommand{\WPHACT}     {\mbox{WPHACT}}
\newcommand{\KandY}      {\mbox{KandY}}
\newcommand{\SWAP}       {\mbox{SWAP}}
\newcommand{\EEWWG}      {\mbox{EEWWG}}

\newcommand{\costw}       {\mbox{$\cos\theta_{\mathrm{W}}$}}
\newcommand{\cosstw} {\mbox{$\cos^2\theta_{\mathrm{W}^-}$}}
\newcommand{\cosetw} {\mbox{$\cos^8\theta_{\mathrm{W}^-}$}}
\newcommand{\sigWW}{\mbox{$\sigma_{_{\mathrm{WW}}}$}}

\newcommand{\lnu} {\ensuremath{\ell      \nu_{\mathrm{e}}}}

\newcommand{\qqen}{\mbox{\qq e$\nu$}}
\newcommand{\qqmn}{\mbox{\qq$\mu\nu$}}
\newcommand{\qqtn}{\mbox{\qq$\tau\nu$}}
\newcommand{\qqlv}{\mbox{\qq\lv}}
\newcommand{\qqln}{\mbox{\qq\lv}}

\chapter{Boson-Pair and Four-Fermion Processes}
\label{chap:4f}

\section{Introduction and Signal Definitions}
\label{introduction}

Cross-section measurements at {\LEPII} are essential because they
allow many direct and indirect tests of the Standard Model (SM).
W-pair production and decay, certainly the most interesting
manifestation of four-fermion production, is directly related to
fundamental parameters of the model, such as the W-boson mass at the
production threshold energy and to the non-Abelian gauge structure of
the theory. The W-decay branching fractions and the value of
$|\mathrm{V}_{\mathrm{cs}}|$ can be directly extracted from a
cross-section measurement. A broader investigation of four-fermion
production in several regions of phase space also brings additional
information on the boson-fermion coupling structure, on the effect of
radiative corrections and on the possible presence of new physics.

This chapter summarises the combination of final results of the four
LEP experiments on four-fermion production cross-sections.  The
signals, with the exception of WW and ZZ, are defined on the basis of
their final states together with cuts to enhance certain regions of
phase space. For WW and ZZ, a diagrammatic definition is used for
the sake of simplicity, even though this corresponds to a non gauge
invariant definition.  In what follows we will use terms such as
``singly resonant'' or ``doubly resonant'', indicating regions of the
phase space rather than a process itself.

The most interesting regions of the four-fermion phase space that are
measured at LEP and for which a combination is performed, are
summarised as:
\begin{itemize}
\item {\bf WW:} defined as the CC03 component of the four-fermion
processes, involving $s$-channel $\gamma$ and $Z$ exchange and
$t$-channel $\nu$ exchange (see Figure~\ref{WW:fig:feyn_born}).
\item {\bf ZZ:} in analogy with the definition of W-pair production,
Z-pair production is defined as the subset of NC02 Feynman diagrams
having two resonant Z bosons (see Figure~\ref{ZZ:fig:feyn_born}).
\item {\bf Z$\gamma^*$:} defined for final states with two
fermion-antifermion pairs, at least one being leptonic (electrons or
muons). To properly consider only singly resonant regions, it is
required that one and only one of the invariant masses of the couples
satisfies:
\mbox{$|\mathrm{m}_{\mathrm{ff'}}-\MZ|<2\GZ$},
where m$_{\mathrm{ff'}}$ is the invariant mass of the two same-flavour
fermions.  In case of four identical leptons all oppositely charged
couples have to be considered.  Moreover the following final state
dependent phase-space cuts have been introduced:
\begin{itemize}
\item{eeqq, $\mu\mu$qq: $|\cos{\theta_{\ell}}|<$0.95, m$_{\ell\ell}>5~\GeV$, m$_{\mathrm{qq}}>10~\GeV$, $\ell=$e,$\mu$}
\item{$\nu\nu$qq: m$_{\mathrm{qq}}>10~\GeV$}
\item{$\nu\nu\ell\ell$: m$_{\ell\ell}>10~\GeV$, m$_{\ell\nu}>90~\GeV$ or m$_{\ell\nu}<70~\GeV$, $\ell=$e,$\mu$}
\item{$\ell_1\ell_1\ell_2\ell_2$: $|\cos{\theta_{\ell_1\ell_2}}|<$0.95, m$_{\ell_1\ell_1}>5~\GeV$, m$_{\ell_2\ell_2}>5~\GeV$, $\ell=$e,$\mu$}
\end{itemize}
\item {\bf We$\nu$:} considered as the complete $t$-channel subset of
Feynman diagrams contributing to e$\nu_\mathrm{e}$f$\bar{\mathrm{f}}'$
final states, with additional cuts to exclude the regions of phase
space dominated by multiperipheral diagrams, where the cross-section
calculation is affected by large uncertainties. The phase space cuts
are (charge conjugation is implicit):
\mbox{$m_{q\bar{q}}>45~\GeV$} for the $e\nu q\bar{q}$ final
states, \mbox{$E_\ell>20~\GeV$} for the $e\nu l\nu$ final states with
$\ell=\mu$ or $\tau$, and finally
\mbox{$|\cos\theta_\mathrm{e^-}|>0.95$},
\mbox{$|\cos\theta_\mathrm{e^+}|<0.95$} and
\mbox{$E_\mathrm{e^+}>20~\GeV$} for the $e\nu e\nu$ final states (see
Figure~\ref{V:fig:feyn_born}).
\item {\bf Zee:} defined considering only the ee$q\bar{q}$ and
ee$\mu\mu$ final states. The following phase space cuts are applied to
enhance the signal: \mbox{$m_{q\bar{q}}(m_{\mu\mu})>60~\GeV$},
and in addition: \mbox{$\theta_\mathrm{e^-}<12^\circ$},
\mbox{$12^\circ<\theta_\mathrm{e^+}<120^\circ$} and
\mbox{$E_\mathrm{e^+}>3~\GeV$} when the positron is visible, or
\mbox{$\theta_\mathrm{e^+}>168^\circ$},
\mbox{$60^\circ<\theta_\mathrm{e^-}<168^\circ$} and
\mbox{$E_\mathrm{e^-}>3~\GeV$} when the electron is visible.
This definition assumes the electron direction to be $+z$ and the
positron direction to be $-z$.
\item {\bf WW$\gamma$:} it is defined as the part of the
$\eeWW\gamma$ process compatible with charged currents,
\ie, including the final states $udud\gamma$, $cscs\gamma$,
$udcs\gamma$, $ud\ell\nu\gamma$, $cs\ell\nu\gamma$,
$\ell\nu\ell\nu\gamma$, assuming a diagonal CKM matrix. The following
phase-space cuts are applied to enhance the signal:
\mbox{$E_{\gamma}>5~\GeV$}, \mbox{$|\cos\theta_{\gamma}|<0.95$},
\mbox{$\cos\theta_{\gamma, f}<0.9$}, the invariant mass of the W-decay
fermion pairs between $\MW-2\GW$ and $\MW+2\GW$ (see
Figure~\ref{Q:fig:feyn_born}).
\end{itemize}
The cross-sections are determined from a fit to the number of observed
events in data, knowing the signal efficiencies corresponding to the
above signal definitions, and the expected accepted backgrounds, from
Monte-Carlo simulations.

The LEP cross-section combination is performed in a fit to the
N=N$_{exp}\times$N$_{\sqrt{s}}$ experimental inputs, where N$_{exp}$
identifies the number of LEP experiments providing input for the
measurement and N$_{\sqrt{s}}$ is the number of energy points
provided. The method used is the Best Linear Unbiased Estimate method
(BLUE) described in~\cite{BLUE:1988, *BLUE:2003}.  In the fits,
inter-experiment and inter-energy correlations of the systematic
errors are taken into account, dividing the sources according to their
correlation and assuming for each either 0\% or 100\% correlation
strength for simplicity. After building the appropriate N$\times$N
correlation matrix for the measurements, the $\chi^2$ minimisation fit
is performed by matrix algebra and is cross-checked with the use of
Minuit~\cite{MINUIT}.

The numbers shown here represent the combination of cross-section
values and derived quantities such as branching fractions or
differential distributions. For each measurement, the collaborations
provided input in agreement with the conventions used to define the
signal and to split the systematic uncertainties: small differences
may therefore appear between the values quoted here and those
published by the experiments.  The combinations are performed for the
whole \LEPII\ period, that includes data from $e^+e^-$ collisions from
$\sqrt{s}=183~\GeV$ up to $\sqrt{s}=207~\GeV$. The energy binning
chosen and the corresponding average integrated luminosity per
experiment at each energy point are shown in Table~\ref{4f_tab:lumi};
they result from a combination of the luminosity in the hadronic and
leptonic channels, therefore small changes from the values published
by individual experiments may be present.

\begin{table}[tbp]
\begin{center}
\renewcommand{\arraystretch}{1.25}
\begin{small}
\begin{tabular}{|c||r|r|r|r|r|r|r|r||r|} 
\hline
 & \multicolumn{2}{|c|}{ALEPH} & \multicolumn{2}{|c|}{DELPHI}
 & \multicolumn{2}{|c|}{L3}    & \multicolumn{2}{|c|}{OPAL} & LEP \\
\hline
Year & $\sqrt{s}$ & $\cal{L}$
     & $\sqrt{s}$ & $\cal{L}$
     & $\sqrt{s}$ & $\cal{L}$
     & $\sqrt{s}$ & $\cal{L}$
     & $\sqrt{s}$ \\
     & $[\GeV]$ & $[\pb]$
     & $[\GeV]$ & $[\pb]$
     & $[\GeV]$ & $[\pb]$
     & $[\GeV]$ & $[\pb]$
     & $[\GeV]$ \\
\hline
\hline
 1997 & 182.66 &  56.81 & 182.65 &  52.08 & 182.68 &  55.46 & 182.68 &  57.38 & 182.67 \\
 1998 & 188.63 & 174.21 & 188.63 & 154.07 & 188.64 & 176.77 & 188.63 & 183.04 & 188.63 \\
 1999 & 191.58 &  28.93 & 191.58 &  24.84 & 191.60 &  29.83 & 191.61 &  29.33 & 191.59 \\
 1999 & 195.52 &  79.86 & 195.51 &  74.04 & 195.54 &  84.15 & 195.54 &  76.41 & 195.53 \\
 1999 & 199.52 &  86.28 & 199.51 &  82.31 & 199.54 &  83.31 & 199.54 &  76.58 & 199.52 \\
 1999 & 201.63 &  41.89 & 201.64 &  40.01 & 201.75 &  37.18 & 201.65 &  37.68 & 201.67 \\
 2000 & 204.86 &  81.41 & 204.86 &  75.66 & 204.82 &  79.01 & 204.88 &  81.91 & 204.85 \\
 2000 & 206.53 & 133.21 & 206.55 & 129.95 & 206.57 & 139.12 & 206.56 & 138.54 & 206.55 \\
\hline
\end{tabular}
\caption[Luminosity and CM Energy]{Summary of luminosity and
luminosity-weighted centre-of-mass energies, per year, of the four LEP
experiments, and final LEP averaged energy. 
}
\label{4f_tab:lumi}
\end{small}
\end{center}
\end{table}

The structure of this chapter is as follows: in Section~\ref{WWintro}
the W-pair production is described and the combined results on
cross-sections, W branching fractions, $|\mathrm{V}_{\mathrm{cs}}|$ and
W polar-angle distributions are presented. Sections~\ref{ZZxsec}
and~\ref{Zgxsec} concern neutral current boson production and present
combined results on ZZ and Z$\gamma^*$ cross-sections, 
respectively.  The combination of single boson production
cross-sections is reported in Section~\ref{Singlebosonintro}.  All the
presented results are compared to recent theoretical predictions, many
of which were developed in the framework of the \LEPII\ Monte-Carlo
workshop~\cite{4f_bib:lep2mcws}.

\section{W-Pair Production}
\label{WWintro}

The signal definition has been given in Section~\ref{introduction}.
W-pair production is investigated via all possible final states
arising in the decay of the two W bosons. According to the different
decays, three topologically different final states can arise: the
fully hadronic, where both W bosons decay into quarks, characterised
by high multiplicity of the final state; the mixed hadronic-leptonic,
also called semileptonic, with the presence of an isolated and
energetic lepton and hadronic decay products of the second W; and the
fully leptonic, with the production of two acoplanar leptons.  The SM
branching fractions for these final states are, respectively, 0.456,
0.349, and 0.105~\cite{\ZFITTERref}.  There are ten experimentally
distinguishable final states: $\qq\qq$, $\qq\mu^+\nu$,
$\qq\mathrm{e}^+\nu$, $\qq\tau^+\nu$, $\mu^+\nu\tau^-\nu$,
$\mathrm{e}^+\nu\tau^-\nu$, $\tau^+\nu\tau^-\nu$, $\mu^+\nu
\mathrm{e}^-\nu$, $\mu^+\nu\mu^-\nu$, $\mathrm{e}^+\nu \mathrm{e}^-\nu$. 
Charge conjugation is assumed everywhere.

Event selections are generally based on Neural Network approaches to
separate the signals from the major backgrounds, which arise mainly
from $q\bar{q}(\gamma)$ events in the fully hadronic final state,
while four-fermion backgrounds are also important in the other
channels.  Typical selection efficiencies range from 80\% to 90\% in
the fully hadronic channel, from 70\% to 90\% in the various
semileptonic channels and about 70\% in the fully leptonic ones.  The
purest channels (95\%) are the semileptonic ones with electrons or
muons in the final state.  Details on the event selections and
experimental performances can be found in~\cite{4f_bib:aleww,
4f_bib:delww, 4f_bib:ltrww, 4f_bib:opaww}.

\subsection{Total Cross-Section Measurement}
\label{WWxsec}

From the number of WW selected events in data, knowing the expected
background and the efficiency on the signal, the production
cross-section is extracted through likelihood fits.  Cross-sections
are then combined, accounting for the correlation of the systematic
errors as shown in Table~\ref{4f_tab:systclassification}. The inputs
used for the combinations are given in Table~\ref{4f_tab:WWmeasADLO},
with the details on the composition of the systematic error in terms
of correlated components shown in Table~\ref{4f_tab:WWmeasLEP}.  For
this analysis, the SM W-decay branching fractions are assumed; see
Section~\ref{WWderived} for the measurement of these branching
fractions.

\begin{table}[thbp]
\begin{center}
\renewcommand{\arraystretch}{1.25}
\begin{tabular}{|c||c|c|} 
\hline
Source & LEP correlation & Energy correlation \\
\hline
\hline
Theory uncertainties on signal           & Yes & Yes \\
Theory uncertainties on backgrounds      & Yes & Yes \\ 
Theory uncertainty on luminosity         & Yes & Yes \\ 
Experimental uncertainties on luminosity & No & Yes \\ 
Detector effects                         & No & Yes \\ 
Monte-Carlo statistics                   & No & No \\ 
\hline
\end{tabular}
\caption[Systematic uncertainties and their correlations]{ Grouping of
systematic uncertainties into those correlated among experiments and those
correlated among different energies.
The theory
uncertainties on the signal include fragmentation effects, radiative
corrections and final state interaction effects.  }
\label{4f_tab:systclassification}
\end{center}
\end{table}

The measured statistical errors are used for the combination; after
building the full 32$\times$32 covariance matrix for the measurements,
the $\chi^2$ minimisation fit is performed by matrix algebra, as
described in Ref.~\cite{BLUE:1988, *BLUE:2003}, and is cross-checked
using Minuit~\cite{MINUIT}.

The results from each experiment for the W-pair production
cross-section are shown in Table~\ref{4f_tab:wwxsec}, together with
the LEP combination at each energy.  All measurements assume SM values
for the W decay branching fractions.  The combined LEP cross-sections
at the eight energies are all positively correlated, see
Table~\ref{4f_tab:WWcorr}, with correlations ranging from 6\% to 19\%.

\begin{table}[htb]
\begin{center}
\renewcommand{\arraystretch}{1.25}
\begin{tabular}{|c||c|c|c|c||c|} 
\hline
\roots & \multicolumn{5}{|c|}{WW Cross-Section [pb]} \\
\cline{2-6} 
$[\GeV]$   & \Aleph\                & \Delphi\               &
             \Ltre\                 & \Opal\                 &
             LEP                                             \\
\hline
\hline
161.3      & $\phz4.23\pm0.75$    & 
             $\phz3.61^{\phz+\phz0.99\phz}_{\phz-\phz0.87}$& 
             $\phz2.89^{\phz+\phz0.82\phz}_{\phz-\phz0.71}$& 
             $\phz3.62^{\phz+\phz0.94\phz}_{\phz-\phz0.84}$& 
             $\phz3.69\pm0.45\phs$                             \\
172.1      & $11.7\phz\pm1.3\phz$ & $11.4\phz\pm1.4\phz$ &
             $12.3\phz\pm1.4\phz$ & $12.3\phz\pm1.3\phz$ &
             $12.0\phz\pm0.7\phz\phs$                          \\
182.7      & $15.86\pm0.63\phs$       & $16.07\pm0.70\phs$       &
             $16.53\pm0.72\phs$       & $15.45\pm0.62\phs$       &
             $15.92\pm0.34\phs$                                \\
188.6      & $15.78\pm0.36\phs$   & $16.09\pm0.42\phs$     &
             $16.17\pm0.41\phs$   & $16.24\pm0.37\phs$     &
             $16.05\pm0.21\phs$                                \\
191.6      & $17.10\pm0.90\phs$   & $16.64\pm1.00\phs$     &
             $16.11\pm0.92\phs$   & $15.93\pm0.86\phs$     &
             $16.42\pm0.47\phs$                                \\
195.5      & $16.60\pm0.54\phs$   & $17.04\pm0.60\phs$     &
             $16.22\pm0.57\phs$   & $18.27\pm0.58\phs$     &
             $16.99\pm0.29\phs$                                \\
199.5      & $16.93\pm0.52\phs$   & $17.39\pm0.57\phs$     &
             $16.49\pm0.58\phs$   & $16.29\pm0.55\phs$     &
             $16.77\pm0.29\phs$                                \\
201.6      & $16.63\pm0.71\phs$   & $17.37\pm0.82\phs$     &
             $16.01\pm0.84\phs$   & $18.01\pm0.82\phs$     &
             $16.98\pm0.40\phs$                                \\
204.9      & $16.84\pm0.54\phs$   & $17.56\pm0.59\phs$     &
             $17.00\pm0.60\phs$   & $16.05\pm0.53\phs$     &
             $16.81\pm0.29\phs$                                \\
206.6      & $17.42\pm0.43\phs$   & $16.35\pm0.47\phs$     &
             $17.33\pm0.47\phs$   & $17.64\pm0.44\phs$     &
             $17.20\pm0.24\phs$                                \\
\hline
\end{tabular}
\caption[W-pair production cross-sections]{
 W-pair production cross-section from the four LEP experiments and
combined values at all recorded centre-of-mass energies.  The
measurements above $175~\GeV$ have been combined in a single fit,
taking into account inter-experiment as well as inter-energy
correlations of systematic errors, with a $\chidf$ of 26.6/24. The fit
at $161.3~\GeV$ ($172.1~\GeV$) has a $\chidf$ of 1.3/3 (0.22/3).  }
\label{4f_tab:wwxsec}
\end{center}
\end{table}

\begin{figure}[p]
\begin{center}
\epsfig{figure=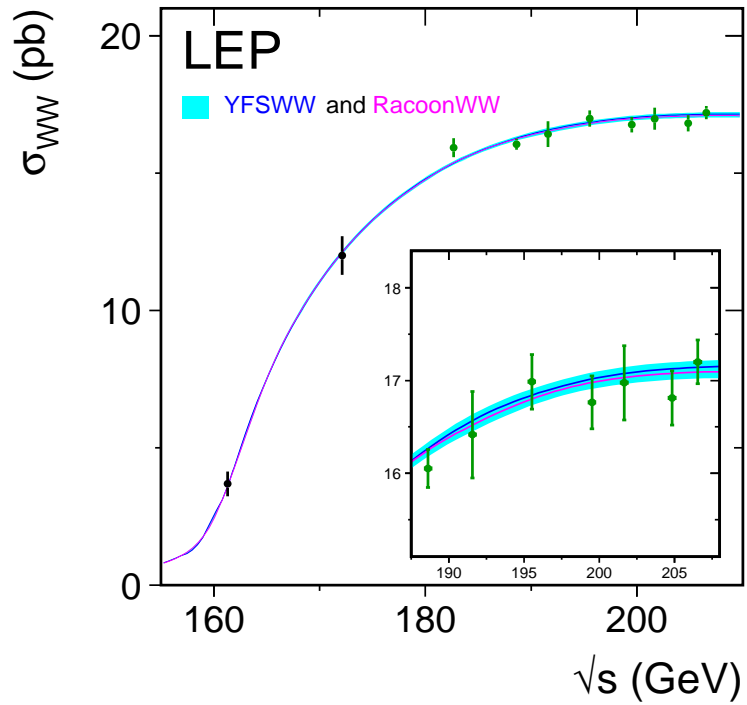,width=0.6\textwidth}
\vskip -1cm
\epsfig{figure=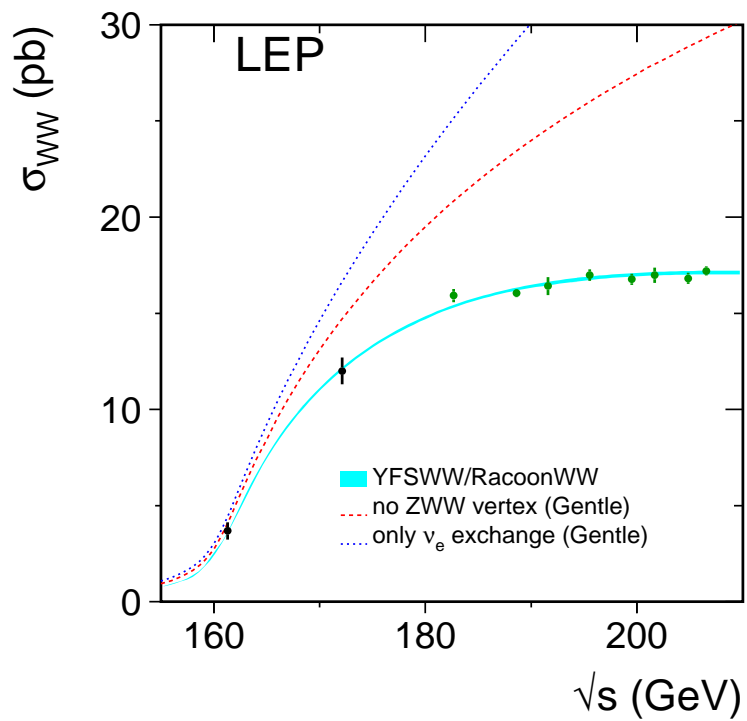,width=0.6\textwidth}
\vskip -1cm
\end{center}
\caption[W-pair cross-sections]{ Measurements of the W-pair production
cross-section, compared to the predictions of
\RacoonWW~\protect\cite{\RACOONWWref} and
\YFSWW~\protect\cite{\YFSWWref}.  The shaded area represents the
uncertainty on the theoretical predictions, estimated as $\pm$2\% for
$\roots\!<\!170$ GeV and ranging from 0.7 to 0.4\% above 170~GeV. The
W mass is fixed at $80.35~\GeV$; its uncertainty is expected to give a
significant contribution only at threshold energies.  }
\label{4f_fig:sww_vs_sqrts}
\end{figure}

Figure~\ref{4f_fig:sww_vs_sqrts} shows the combined LEP W-pair
cross-section measured as a function of the \CoM\ energy.  The
experimental results are compared with the theoretical calculations
from
\YFSWW~\cite{\YFSWWref} and \RacoonWW~\cite{\RACOONWWref} between 155
and 215 $\GeV$ using $\Mw=80.35$~GeV.  The two programs have been
extensively compared and agree at a level better than 0.5\% at the
\LEPII\ energies~\cite{4f_bib:lep2mcws}.  The calculations above
$170~\GeV$, based for the two programs on the so-called
leading-pole~(LPA) or double-pole~(DPA)
approximations~\cite{4f_bib:lep2mcws}, have theoretical uncertainties
decreasing from 0.7\% at $170~\GeV$ to about 0.4\% at \CoM\ energies
larger than $200~\GeV$\footnote{ The theoretical uncertainty
$\Delta\sigma/\sigma$ on the W-pair production cross-section
calculated in the LPA/DPA above $170~\GeV$ can be parametrised as
$\Delta\sigma/\sigma=(0.4\oplus0.072\cdot t_1\cdot t_2)\%$, where
$t_1=(200-2\cdot\Mw)/(\roots-2\cdot\Mw)$ and $t_2=(1-(\frac{2\cdot
M_{\mathrm{W}}}{200})^2) / (1-(\frac{2\cdot
M_{\mathrm{W}}}{\sqrt{s}})^2)$.  }, while in the threshold region,
where the programs use an improved Born approximation, a larger
theoretical uncertainty of 2\% is assigned.  This theoretical
uncertainty is represented by the blue band in the figure.  The
cross-sections are sensitive to the W-boson mass, such that an error
of 50 MeV on the W mass would translate into additional errors of
0.1\% (3.0\%) on the cross-section predictions at $200~\GeV$
($161~\GeV$), respectively.  All results, up to the highest \CoM\
energies, are in agreement with the two theoretical predictions
considered and listed in Table~\ref{4f_tab:WWtheo}.  In the lower part
of the figure, the data are also compared with hypothetical
predictions for which W-pair production happens in absence of one or
two of the CC03 diagrams. The need for the diagram with a ZWW vertex
is a spectacular confirmation of the non-Abelian nature of the
electroweak SM.

The agreement between the measured W-pair cross-section,
$\sww^\mathrm{meas}$, and its expectation according to a given
theoretical model, $\sww^\mathrm{theo}$, can be expressed
quantitatively in terms of their ratio:

\begin{equation}
\rww = \frac{\sww^\mathrm{meas}}{\sww^\mathrm{theo}} ,
\end{equation}
averaged over the measurements performed by the four experiments at
different energies in the $\LEPII$ region.  The above procedure has
been used to compare the measurements at the eight energies between
$183~\GeV$ and $207~\GeV$ with the predictions of
\Gentle~\cite{bib:GentleV2}, \KoralW~\cite{\KORALWref},
\YFSWW~\cite{\YFSWWref} and \RacoonWW~\cite{\RACOONWWref}.  The
measurements at $161~\GeV$ and $172~\GeV$ have not been used in the
combination because they were performed using data samples of low
statistics and because of the high sensitivity of the cross-section to
the value of the W mass at these energies.

The combination of the ratio $\rww$ is performed using as input from
the four experiments the 32 cross-sections measured at each of the
eight energies.  These are then converted into 32 ratios by dividing
them by the theoretical predictions listed in
Table~\ref{4f_tab:WWtheo}.  The full 32$\times$32 covariance matrix
for the ratios is built taking into account the same sources of
systematic errors used for the combination of the W-pair
cross-sections at these energies.

The small statistical errors on the theoretical predictions at the
various energies, taken as fully correlated for the four experiments
and uncorrelated between different energies, are also translated into
errors on the individual measurements of $\rww$.  The theoretical
errors on the predictions, due to the physical and technical precision
of the generators used, are not propagated to the individual ratios
but are used when comparing the combined values of $\rww$ to unity.
For each of the four models considered, two fits are performed: in the
first, eight values of $\rww$ at the different energies are extracted,
averaged over the four experiments; in the second, only one value of
$\rww$ is determined, representing the global agreement of measured
and predicted cross-sections over the whole energy range.

\begin{table}[t]
\begin{center}
\renewcommand{\arraystretch}{1.25}
\begin{tabular}{|c||c|c|} 
\hline
\roots~$[\GeV]$
  & $\rww^{\footnotesize\YFSWW}$ & $\rww^{\footnotesize\RacoonWW}$ \\
\hline
\hline
182.7      & $1.037\pm0.022$ & $1.036\pm0.023$ \\
188.6      & $0.987\pm0.013$ & $0.988\pm0.013$ \\
191.6      & $0.991\pm0.028$ & $0.994\pm0.029$ \\
195.5      & $1.009\pm0.018$ & $1.011\pm0.019$ \\
199.5      & $0.985\pm0.017$ & $0.987\pm0.018$ \\
201.6      & $0.994\pm0.023$ & $0.997\pm0.024$ \\
204.9      & $0.982\pm0.017$ & $0.984\pm0.018$ \\
206.6      & $1.003\pm0.014$ & $1.007\pm0.015$ \\
\hline
$\chidf$   & 26.6/24         & 26.6/24        \\
\hline
\hline
Average    & $0.995\pm0.008$ & $0.997\pm0.008$ \\
\hline
$\chidf$   & 32.2/31         & 32.0/31        \\
\hline
\end{tabular}
\caption[W-pair cross-section ratios experiment/theory]{ Ratios of LEP
combined W-pair cross-section measurements to the expectations
according to \YFSWW~\protect\cite{\YFSWWref} and
\RacoonWW~\protect\cite{\RACOONWWref}.  For each of the two models,
two fits are performed, one to the LEP combined values of $\rww$ at
the eight energies between $183~\GeV$ and $207~\GeV$, and another to
the LEP combined average of $\rww$ over all energies.  The results of
the fits are given in the table together with the resulting $\chidf$.
The fits take into account inter-experiment as well as inter-energy
correlations of systematic errors.  }
\label{4f_tab:wwratio}
\end{center}
\end{table}

\begin{figure}[ht]
\begin{center}
\epsfig{figure=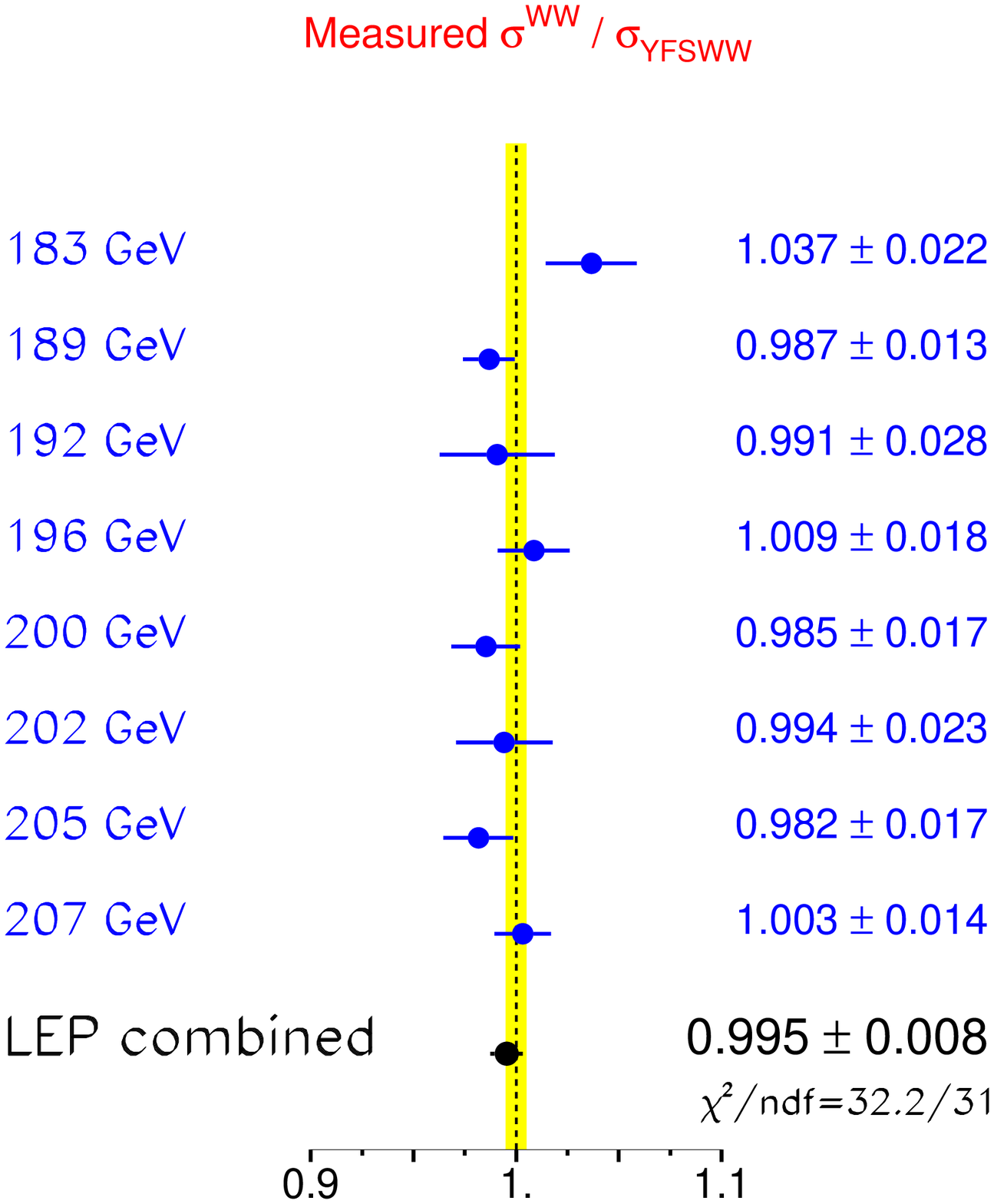,width=0.48\textwidth}
\hfill
\epsfig{figure=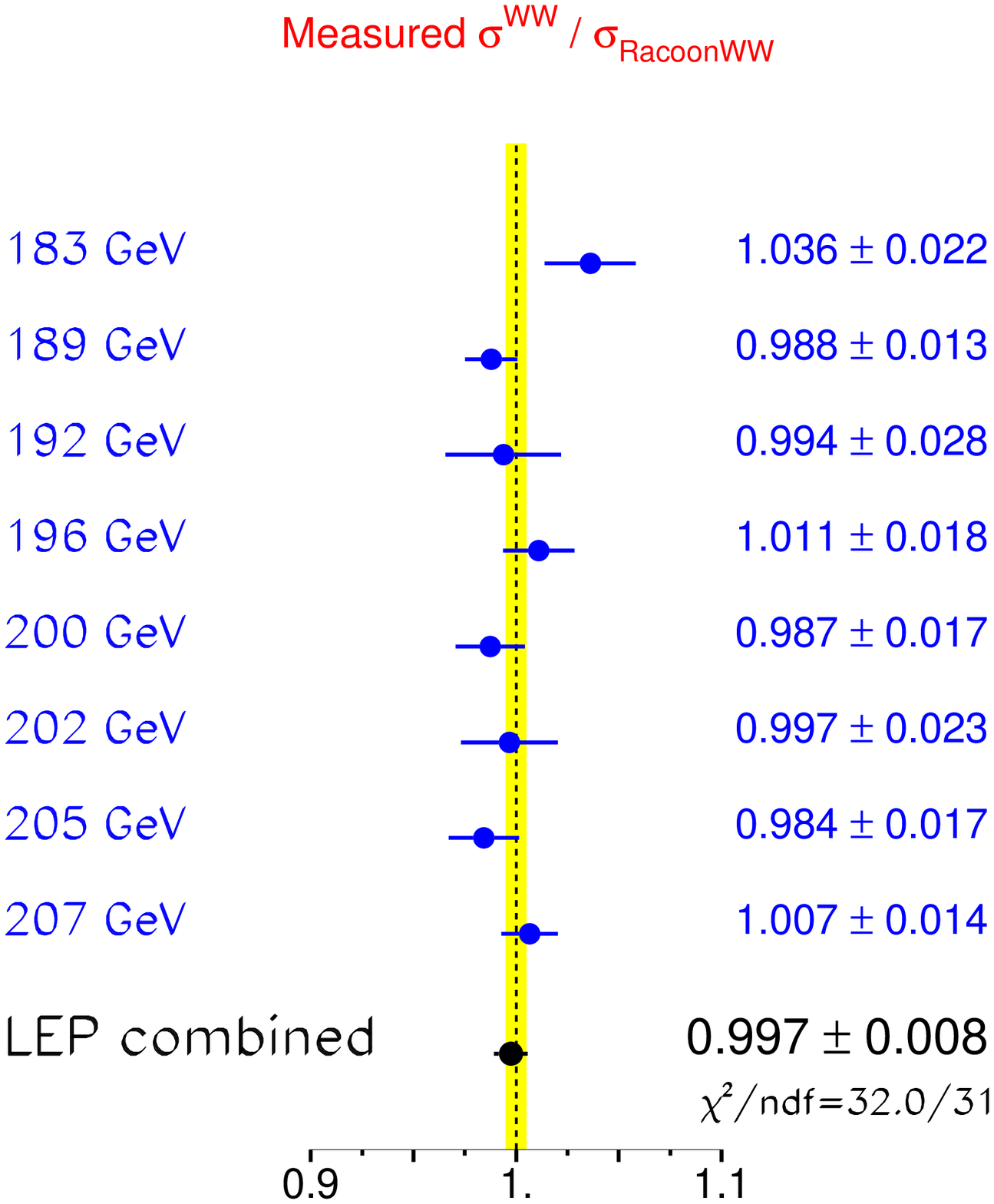,width=0.48\textwidth}
\caption[W-pair cross-section ratios experiment/theory]{ Ratios of LEP
combined W-pair cross-section measurements to the expectations
calculated with \YFSWW~\protect\cite{\YFSWWref} and
\RacoonWW~\protect\cite{\RACOONWWref} The yellow bands represent
constant relative errors of 0.5\% on the two cross-section
predictions.  }
\label{4f_fig:rww}
\end{center}
\end{figure} 

The results of the fits to $\rww$ for \YFSWW\ and \RacoonWW\ are given
in Table~\ref{4f_tab:wwratio}, with more details given in
Table~\ref{4f_tab:rWWmeas}.  As already qualitatively noted from
Figure~\ref{4f_fig:sww_vs_sqrts}, the LEP measurements of the W-pair
cross-section above threshold are in very good agreement with the
predictions and can test the theory at the level of better than 1\%.
In contrast, the predictions from \Gentle\ and \KoralW\ are about 3\%
too high with respect to the measurements due to the lack of LPA/DPA
corrections; the equivalent values of $\rww$ in those cases are,
respectively, $0.970\pm0.008$ and $0.976\pm0.008$.  The results of the
fits for \YFSWW\ and \RacoonWW\ are also shown in
Figure~\ref{4f_fig:rww}, where relative errors of 0.5\% on the
cross-section predictions have been assumed.  For simplicity the
energy dependence of the theory error on the W-pair cross-section has
been neglected in the figure.  The main differences between the
predictions of \YFSWW/\RacoonWW\ and \Gentle/\KoralW\ arise from
non-leading $\oa$ electroweak radiative corrections to the W-pair
production process and non-factorisable corrections, which are
included (in the LPA/DPA leading-pole/double-pole
approximation~\cite{4f_bib:lep2mcws}) in both
\YFSWW\ and \RacoonWW, but not in \Gentle\ and \KoralW.  The data
clearly prefer the computations which more precisely include $\oa$
radiative corrections.

\subsection{Derived Quantities}
\label{WWderived}

From the cross-sections of the individual WW decay channels, each
experiment determined the values of the W branching fractions, with
and without the assumption of lepton universality~\footnote{In what
follows any effects from lepton masses on W partial widths are
neglected given their small size.}.  In the fit with lepton
universality, the branching fraction to hadrons is determined from
that to leptons by constraining the sum to unity.  In building the
full 12$\times$12 covariance matrix, the same correlations of the
systematic errors as used for the cross-section measurements are
assumed.  The detailed inputs to LEP and the correlation matrices are
reported in Table~\ref{4f_tab:Wbrmeas}.

\begin{figure}[p]
\begin{center}
\epsfig{figure=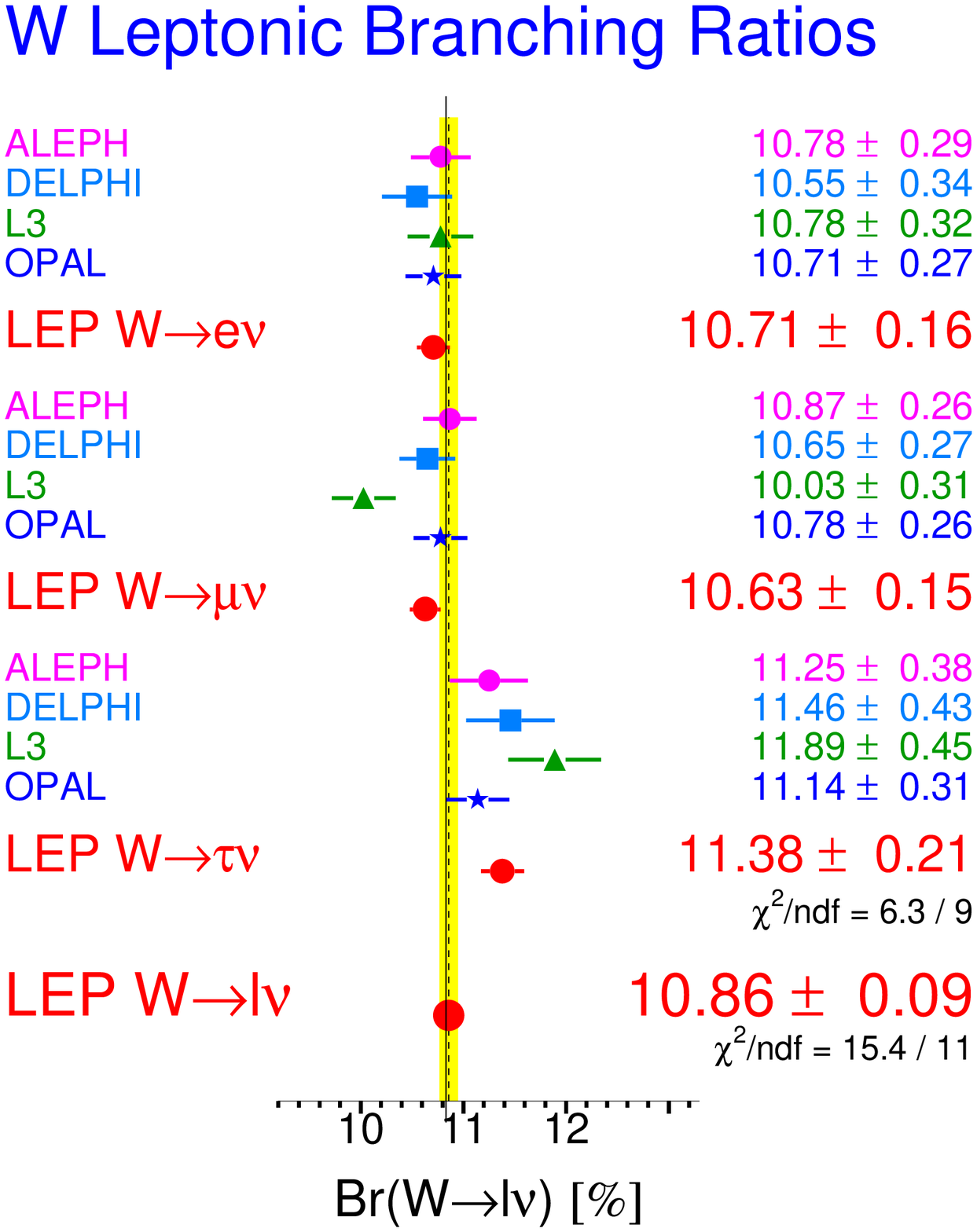,width=0.48\textwidth}
\hfill
\epsfig{figure=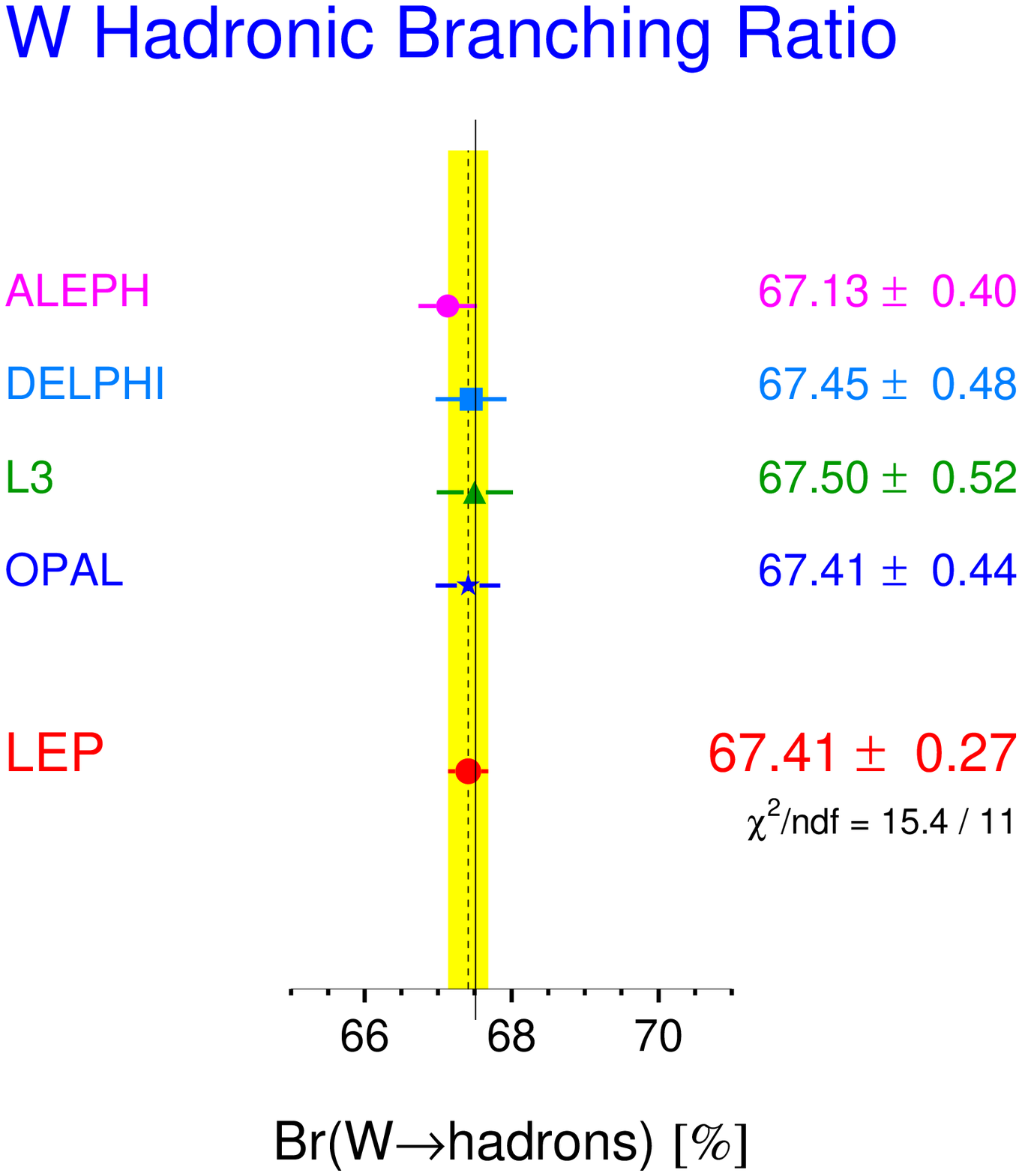,width=0.48\textwidth}
\caption[W branching fractions]{Leptonic and hadronic W branching
fractions, as measured by the experiments, and the LEP combined values
according to the procedures described in the text.}
\label{4f_fig:brw}
\end{center}
\end{figure}

The results from each experiment are reported in Table~\ref{tab:wwbra}
together with the LEP combination and shown in
Figure~\ref{4f_fig:brw}.  The results of the fit which does not assume
lepton universality show a negative correlation of 20.1\% (12.2\%)
between the \Wtotnu\ and \Wtoenu\ (\Wtomnu) branching fractions, while
between the electron and muon decay channels there is a positive
correlation of 13.5\%.

\begin{table}[p]
\begin{center}
\renewcommand{\arraystretch}{1.25}
\begin{tabular}{|c||c|c|c|c|} 
\hline
         & \multicolumn{3}{|c|}{Lepton} & Lepton \\
         & \multicolumn{3}{|c|}{non--universality} & universality \\
\hline
Experiment 
         & \wwbr(\Wtoenu) & \wwbr(\Wtomnu) & \wwbr(\Wtotnu)  
         & \wwbr({\mbox{$\mathrm{W}\rightarrow\mathrm{hadrons}$}}) \\
         & [\%] & [\%] & [\%] & [\%]  \\
\hline
\hline
\Aleph\  & $10.78\pm0.29$ & $10.87\pm0.26$ & $11.25\pm0.38$ & $67.13\pm0.40$ \\
\Delphi\ & $10.55\pm0.34$ & $10.65\pm0.27$ & $11.46\pm0.43$ & $67.45\pm0.48$ \\
\Ltre\   & $10.78\pm0.32$ & $10.03\pm0.31$ & $11.89\pm0.45$ & $67.50\pm0.52$ \\
\Opal\   & $10.71\pm0.27$ & $10.78\pm0.26$ & $11.14\pm0.31$ & $67.41\pm0.44$ \\
\hline
LEP      & $10.71\pm0.16$ & $10.63\pm0.15$ & $11.38\pm0.21$ & $67.41\pm0.27$ \\
\hline
$\chidf$ & \multicolumn{3}{|c|}{6.3/9} & 15.4/11 \\
\hline
\end{tabular}
\caption[W branching fractions] {Summary of W branching fractions
derived from W-pair production cross-sections measurements up to 207
GeV \CoM\ energy.}
\label{tab:wwbra} 
\end{center}
\end{table}

From the results on the leptonic branching fractions an excess of the
branching fraction \Wtotnu\ with respect to the other leptons is
evident.  The excess can be quantified by the pair-wise ratios of the
branching fractions, which represent a test of lepton universality in
the decay of on-shell W bosons:

\begin{eqnarray}
\wwbr\mathrm{(\Wtomnu)} \, / \, \wwbr\mathrm{(\Wtoenu)} \,
& = & 0.993 \pm 0.019 \, ,\\
\wwbr\mathrm{(\Wtotnu)} \; / \, \wwbr\mathrm{(\Wtoenu)} \,
& = & 1.063 \pm 0.027 \, ,\\
\wwbr\mathrm{(\Wtotnu)} \, / \, \wwbr\mathrm{(\Wtomnu)} 
& = & 1.070 \pm 0.026 \, .
\end{eqnarray}
The branching fraction of W into taus with respect to that into
electrons and muons differs by more than two standard deviations,
where the correlations have been taken into account. The branching
fractions of W into electrons and into muons agree well.  Assuming
only partial lepton universality the ratio between the tau fractions
and the average of electrons and muons can also be computed:

\begin{eqnarray}
2\wwbr\mathrm{(\Wtotnu)} \, / \, 
(\wwbr\mathrm{(\Wtoenu)}+\wwbr\mathrm{(\Wtomnu)}) \,
& = & 1.066 \pm 0.025 \,
\end{eqnarray}
resulting in an agreement at the level of 2.6 standard deviations
only, with all correlations included.

If overall lepton universality is assumed (in the massless
assumption), the hadronic branching fraction is determined to be
$67.41\pm0.18\mathrm{(stat.)}\pm0.20\mathrm{(syst.)}\%$, while the
leptonic branching fraction is
$10.86\pm0.06\mathrm{(stat.)}\pm0.07\mathrm{(syst.)}\%$.  These
results are consistent with the SM expectations of 67.51\% and
10.83\%~\cite{\ZFITTERref}, respectively.  The systematic error
receives equal contributions from the correlated and uncorrelated
sources.

Within the SM, the branching fractions of the W boson depend on the
six matrix elements $|\mathrm{V}_{\mathrm{qq'}}|$ of the
Cabibbo--Kobayashi--Maskawa (CKM) quark mixing matrix not involving
the top quark.  In terms of these matrix elements, the leptonic
branching fraction of the W boson $\mathcal{B}(\Wtolnu)$ is given by

\begin{equation}
  \frac{1}{\mathcal{B}(\Wtolnu)}\quad = \quad 3 
  \Bigg\{ 1 + 
          \bigg[ 1 + \frac{\alpha_{\mathrm{s}}(\mathrm{M}^2_{\mathrm{W}})}{\pi} 
          \bigg] 
          \sum_{\tiny\begin{array}{c}i=(u,c),\\j=(d,s,b)\\\end{array}}
          |\mathrm{V}_{ij}|^2 
  \Bigg\},
\end{equation} 
where $\alpha_{\mathrm{s}}(\mathrm{M}^2_{\mathrm{W}})$ is the strong
coupling constant and fermion mass effects are negligible.  Taking
$\alpha_{\mathrm{s}}(\mathrm{M}^2_{\mathrm{W}}) =
0.119\pm0.002$~\cite{PDG2010}, and using the experimental knowledge of
the sum
$|\mathrm{V}_{\mathrm{ud}}|^2+|\mathrm{V}_{\mathrm{us}}|^2+|\mathrm{V}_{\mathrm{ub}}|^2+
|\mathrm{V}_{\mathrm{cd}}|^2+|\mathrm{V}_{\mathrm{cb}}|^2=1.0544\pm0.0051$~\cite{PDG2010},
the above result can be interpreted as a measurement of
$|\mathrm{V}_{\mathrm{cs}}|$ which is the least well determined of
these matrix elements:

\begin{equation*}
  |\mathrm{V}_{\mathrm{cs}}|\quad=\quad0.969\,\pm\,0.013.
\end{equation*}
The error includes a contribution of $0.0006$ from the uncertainty on
$\alpha_{\mathrm{s}}$ and a $0.003$ contribution from the
uncertainties on the other CKM matrix elements, the largest of which
is that on $|\mathrm{V}_{\mathrm{cd}}|$.  These uncertainties are
negligible in the error of this determination of
$|\mathrm{V}_{\mathrm{cs}}|$, which is dominated by the experimental
error of $0.013$ arising from the measurement of the W branching
fractions.

\subsection{W Angular Distribution}
\label{WWdiffxsec}

In addition to measuring the total $\WW$ cross-section, the LEP
experiments produce results for the differential cross-section,
$\mathrm{d}\sigma_{\mathrm{WW}}/\mathrm{d}\costw$, where
$\theta_{\mathrm{W}}$ is the polar angle of the produced
$\mathrm{W}^-$ with respect to the $\mathrm{e}^-$ beam direction.  The
LEP combination of these measurements will allow future theoretical
models which predict deviations in this distribution to be tested
against the LEP data in a direct and, as far as possible,
model-independent manner. To reconstruct the $\costw$ distribution it
is necessary to identify the charges of the decaying W bosons. This
can only be performed without significant ambiguity when one of
W-boson decays via $\mathrm{W}\rightarrow e \nu$ or
$\mathrm{W}\rightarrow\mu\nu$, in which case the lepton provides the
charge tag. Consequently, the combination of the differential
cross-section measurements is performed for the $\qqen$ and $\qqmn$
channels combined.  Selected $\qqtn$ events are not considered due to
the larger backgrounds and difficulties in determining the tau lepton
charge in the case where not all charged decay products are detected.

The measured $\qqen$ and $\qqmn$ differential cross-sections are
corrected to correspond to the CC03 set of diagrams with the
additional constraint that the charged lepton is more than $20^\circ$
away from the $\epem$ beam direction, $|\theta_{\ell^\pm}|>20^\circ$.
This angular requirement corresponds approximately to the experimental
acceptance of the four LEP experiments and also greatly reduces the
difference between the full $4f$ cross-section and the CC03
cross-section by reducing the contribution of $t$-channel diagrams in
the $\qqen$ final state\footnote{With this requirement the difference
between the total four-fermion (CC20~\cite{4f_bib:lep2mcws}) and
double-resonant (CC03) $\qqen$ cross-section is approximately
3.5\,\%, as opposed to 24.0\,\% without the lepton angle
requirement. For the $\qqmn$ channel the difference between the total
four-fermion (CC10~\cite{4f_bib:lep2mcws}) and double-resonant
(CC03) cross-section is less than 1\,\% in both cases.}. The
angle $\costw$ is reconstructed from the four-momenta of the fermions
from the ${\mathrm{W}^-}$ decay using the {\sc ECALO5} photon
recombination scheme\cite{4f_bib:lep2mcws}, a prescription for
combining photons to a close-by charged fermion.

The LEP combination is performed in ten bins of $\costw$. Because the
differential cross-section distribution evolves with $\roots$,
reflecting the changing relative $s-$ and $t-$ channel contributions,
the LEP data are divided into four $\roots$ ranges: $180~\GeV
-184~\GeV$, $184~\GeV - 194~\GeV$, $194~\GeV - 204~\GeV$, and
$204~\GeV - 210~\GeV$.  It has been verified for each $\roots$ range
that the differences in the differential cross-sections at the mean
value of $\roots$ compared to the luminosity-weighted sum of the
differential cross-sections reflecting the actual distribution of the
data across $\roots$ are negligible compared to the statistical
errors.

The experimental resolution in LEP on the reconstructed values of
$\costw$ is typically 0.15-0.20. When simulating W-pair production, a
significant migration between generated and reconstructed bins of
$\costw$ is observed. The effects of bin-to-bin migration are not
explicitly unfolded, instead each experiment obtains the cross-section
in the $i${th bin of the differential distribution, $\sigma_i$, from

\begin{eqnarray}
      \sigma_i & = & {{N_i-b_i}\over{\epsilon_i\cal{L}}}, 
\end{eqnarray}
where: 
\begin{itemize}
   \item[ $N_i$ ] is the observed number of $\qqen$/$\qqmn$ events
                  reconstructed in the $i$th bin of the $\costw$
                  distribution.
   \item[ $b_i$ ] is the expected number of background events in bin
                  $i$. The contribution from four-fermion background
                  is treated as in each of the experiments' $\WW$
                  cross-section analyses.
   \item[ $\epsilon_i$ ] is the Monte-Carlo efficiency in bin $i$,
                  defined as $\epsilon_i=S_i/G_i$ where $S_i$ is the
                  number of selected CC03 MC $\qqln$ events
                  reconstructed in bin $i$ and $G_i$ is the number of
                  MC CC03 $\qqen$/$\qqmn$ events with generated
                  \costw\ (calculated using the ECALO5 recombination
                  scheme) lying in the $i$th bin
                  ($|\theta_{\ell^\pm}|>20^\circ$).  Selected $\qqtn$
                  events are included in the numerator of the
                  efficiency.
\end{itemize}
This bin-by-bin efficiency correction method has the advantages of
simplicity and that the resulting $\sigma_i$ are uncorrelated.  The
main disadvantage of this procedure is that bin-by-bin migrations
between generated and reconstructed $\costw$ are corrected purely on
the basis of the SM expectation and may potentially be biased.  The
validity of the simple correction procedure was tested by considering
a range of deviations from the SM. Specifically, the SM $\costw$
distribution was reweighted, in turn, by factors of $1+0.1(\costw-1)$,
$1-0.2\cosstw$, $1+0.2\cosstw $ and $1-0.4\cosetw$, and data samples
generated corresponding to the combined LEP luminosity. These
reweighting functions represent deviations which are large compared to
the statistics of the combined LEP measurements. The bin-by-bin
correction method was found to result in good $\chi^2$ distributions
when the extracted $\costw$ distributions were compared with the
underlying generated distribution ({\em e.g.}  the worst case gave a
mean $\chi^2$ of 11.3 for the 10 degrees of freedom corresponding to
the ten $\costw$ bins), and no significant bias was found in these
tests.

For the LEP combination the systematic uncertainties on measured
differential cross-sections are broken down into two terms:
uncertainties which are fully correlated between bins and experiments
and errors which are correlated between bins but uncorrelated between
experiments. This procedure reflects the fact that the dominant
systematic errors affect the overall normalisation of the measured
distributions rather than the shape.

The detailed inputs provided by the four LEP experiments are reported
in Tables~\ref{4f_tab:dsdcost_aleph}, \ref{4f_tab:dsdcost_delphi},
\ref{4f_tab:dsdcost_l3} and \ref{4f_tab:dsdcost_opal}.
Table~\ref{4f_tab:dsdcost} presents the combined LEP results.  In the
table the bin-by-bin error breakdown is also reported.  The result is
also shown in Figure~\ref{4f_fig:dsdcost}, where the combined data are
superimposed on the four-fermion theory predictions calculated with
\KandY~\cite{4f_bib:kandy} and \RacoonWW~\cite{\RACOONWWref}, which
are indistinguishable on the plot scale.  The agreement of data and
calculations is generally very good, with an apparent under-fluctuation
of data with respect to the central values of the theory predictions
in the last bin of the $194~\GeV - 204~\GeV$ energy range.

\begin{table}[hbtp]
\begin{center}
\renewcommand{\arraystretch}{1.25}
\begin{small}
\begin{tabular}{|c|c|c|c|c|c|c|c|c|c|c|}
\hline
cos$\theta_{\mathrm{W}-}$ bin $i$ & 1 & 2 & 3 & 4 & 5 & 6 & 7 & 8 & 9 & 10 \\
\hline
\hline
\multicolumn{11}{|c|}{$\sqrt{s}$ range: $180-184~\GeV$ \hfill ${\cal L} = 163.90$ pb$^{-1}$ \hfill Weighted $\sqrt{s} = 182.66~\GeV$} \\
\hline
$\sigma_i$  (pb)            & 0.502 & 0.705 & 0.868 & 1.281 & 1.529 & 2.150 & 2.583 & 2.602 & 4.245 & 5.372 \\
$\delta\sigma_i$  (pb)      & 0.114 & 0.129 & 0.143 & 0.203 & 0.195 & 0.244 & 0.270 & 0.254 & 0.367 & 0.419 \\
$\delta\sigma_i$(stat) (pb) & 0.112 & 0.128 & 0.142 & 0.202 & 0.194 & 0.241 & 0.267 & 0.249 & 0.362 & 0.413 \\
$\delta\sigma_i$(syst) (pb) & 0.016 & 0.017 & 0.018 & 0.023 & 0.024 & 0.036 & 0.040 & 0.049 & 0.059 & 0.073 \\
\hline
\hline
\multicolumn{11}{|c|}{$\sqrt{s}$ range: $184-194~\GeV$ \hfill ${\cal L} = 587.95$ pb$^{-1}$ \hfill Weighted $\sqrt{s} = 189.09~\GeV$} \\
\hline
$\sigma_i$  (pb)            & 0.718 & 0.856 & 1.009 & 1.101 & 1.277 & 1.801 & 2.215 & 2.823 & 4.001 & 5.762 \\
$\delta\sigma_i$  (pb)      & 0.074 & 0.079 & 0.086 & 0.088 & 0.094 & 0.123 & 0.140 & 0.151 & 0.179 & 0.223 \\
$\delta\sigma_i$(stat) (pb) & 0.073 & 0.078 & 0.084 & 0.085 & 0.091 & 0.119 & 0.135 & 0.144 & 0.169 & 0.208 \\
$\delta\sigma_i$(syst) (pb) & 0.015 & 0.015 & 0.018 & 0.023 & 0.023 & 0.031 & 0.035 & 0.046 & 0.060 & 0.081 \\
\hline
\hline
\multicolumn{11}{|c|}{$\sqrt{s}$ range: $194-204~\GeV$ \hfill ${\cal L} = 605.05$ pb$^{-1}$ \hfill Weighted $\sqrt{s} = 198.38~\GeV$} \\
\hline
$\sigma_i$  (pb)            & 0.679 & 0.635 & 0.991 & 1.087 & 1.275 & 1.710 & 2.072 & 2.866 & 4.100 & 6.535 \\
$\delta\sigma_i$  (pb)      & 0.079 & 0.065 & 0.084 & 0.088 & 0.096 & 0.116 & 0.126 & 0.158 & 0.185 & 0.236 \\
$\delta\sigma_i$(stat) (pb) & 0.078 & 0.064 & 0.083 & 0.085 & 0.094 & 0.112 & 0.122 & 0.152 & 0.175 & 0.220 \\
$\delta\sigma_i$(syst) (pb) & 0.012 & 0.013 & 0.016 & 0.021 & 0.021 & 0.029 & 0.033 & 0.043 & 0.059 & 0.085 \\
\hline
\hline
\multicolumn{11}{|c|}{$\sqrt{s}$ range: $204-210~\GeV$ \hfill ${\cal L} = 630.51$ pb$^{-1}$ \hfill Weighted $\sqrt{s} = 205.92~\GeV$} \\
\hline
$\sigma_i$  (pb)            & 0.495 & 0.602 & 0.653 & 1.057 & 1.240 & 1.707 & 2.294 & 2.797 & 4.481 & 7.584 \\
$\delta\sigma_i$  (pb)      & 0.058 & 0.066 & 0.069 & 0.094 & 0.093 & 0.115 & 0.140 & 0.143 & 0.187 & 0.262 \\
$\delta\sigma_i$(stat) (pb) & 0.057 & 0.065 & 0.068 & 0.091 & 0.090 & 0.111 & 0.137 & 0.136 & 0.175 & 0.244 \\
$\delta\sigma_i$(syst) (pb) & 0.012 & 0.013 & 0.015 & 0.021 & 0.022 & 0.030 & 0.033 & 0.045 & 0.064 & 0.096 \\
\hline
\end{tabular}
\end{small}
\caption[W differential cross-sections]{ Combined W$^{-}$ differential
angular cross-section in the 10 angular bins for the four chosen
energy intervals.  For each energy range, the sum of the measured
integrated luminosities and the luminosity-weighted centre-of-mass
energy is reported.  The results per angular bin in each of the energy
interval are then presented: $\sigma_{i}$ indicates the average of
d[$\sigma_{\mathrm{WW}}$(BR$_{e\nu}$+BR$_{\mu\nu}$)]/dcos$\theta_{\mathrm{W}^-}$
in the $i$-th bin of cos$\theta_{\mathrm{W}^-}$, with a bin width of
0.2.  For each bin, the values of the total, statistical and
systematic errors are reported.  All values are given in pb. }
\label{4f_tab:dsdcost} 
\end{center}
\end{table}

\begin{figure}[p]
\begin{center}
\epsfig{figure=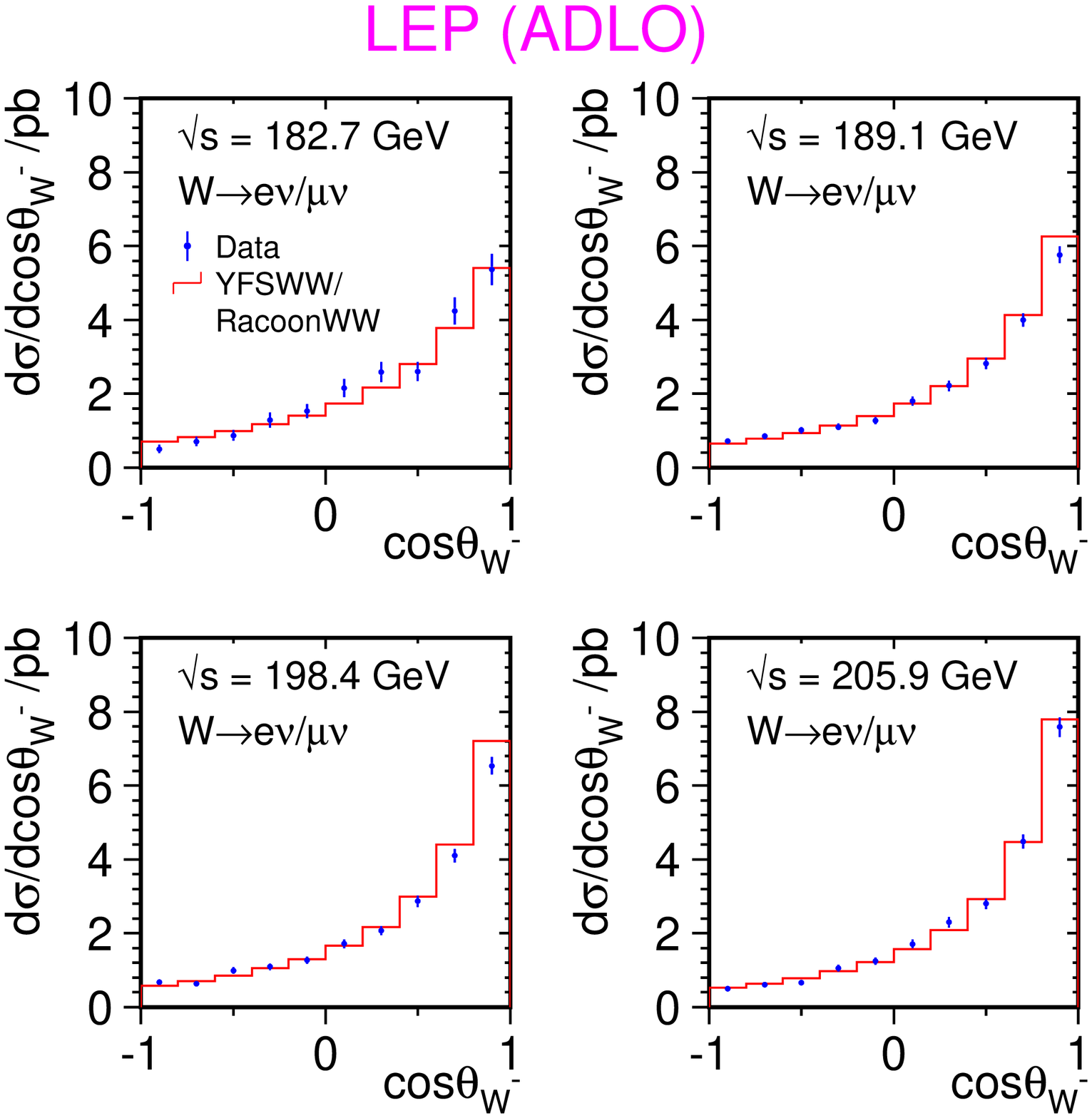,width=0.9\textwidth}
\caption[W differential cross-sections]{ LEP combined
d[$\sigma_{\mathrm{WW}}$(BR$_{e\nu}$+BR$_{\mu\nu}$)]/dcos$\theta_{\mathrm{W}^-}$
distributions for the four chosen energy intervals. The combined
values (points) are superimposed on the four-fermion predictions from
\KandY\ and \RacoonWW.  }
\label{4f_fig:dsdcost}
\end{center}
\end{figure}

\section{Z-Pair Production}
\label{ZZxsec}

The signal definition has been given in Section~\ref{introduction}.
Z-pair production shows several similarities to W-pair production.
The different final states depend on the decay of the heavy bosons: it
is possible to have four quarks, two quarks and two leptons or four
leptons in the final state.  The signatures are very clean and the
main background is represented by WW production.

The approaches used by the experiments for the selection are based on
Neural Network techniques. The final states studied involve both the
hadronic and leptonic decays of the Z boson, where invisible decays
are included when accompanied by a charged decay.  The selection
efficiencies depend significantly on the final state, ranging from
25\% to 60\%, with purities from 30\% to 70\%~\cite{4f_bib:alezz,
4f_bib:delzz, 4f_bib:ltrzz, 4f_bib:opazz}.  The main backgrounds
include four-fermion production, di-leptonic and QCD final states.

The LEP combination is performed applying the same technique as used
for the WW cross-section measurement. The symmetrised expected
statistical error of each analysis is used, to avoid biases due to the
limited number of selected events.  The detailed inputs from the
experiments are reported in Table~\ref{4f_tab:ZZmeasADLO}.  The
results of the individual experiments are summarised in
Table~\ref{4f_tab:zzxsec}, together with the LEP averages.  The
composition of the systematic error in terms of correlated components
is shown in Table~\ref{4f_tab:ZZmeasLEP}.  The cross-sections used for
the combination are determined by the experiments using the
frequentist approach, without assuming any prior for the value of the
cross-section itself.

The measurements are shown in Figure~\ref{4f_fig:szz_vs_sqrts} as a
function of the LEP \CoM\ energy, where they are compared to the
\YFSZZ~\cite{ref:YFSZZ} and \ZZTO~\cite{4f_bib:zzto} predictions as
listed in Table~\ref{4f_tab:ZZtheo}.  Both these calculations have an
estimated uncertainty of $2\%$~\cite{4f_bib:lep2mcws}.  The data do
not show any significant deviation from the theoretical expectations.

\begin{table}[p]
\begin{center}
\renewcommand{\arraystretch}{1.25}
\begin{tabular}{|c||c|c|c|c||c|} 
\hline
\roots & \multicolumn{5}{|c|}{ZZ cross-section [pb]} \\
\cline{2-6} 
(GeV) & \Aleph\ & \Delphi\ & \Ltre\ & \Opal\ & LEP \\ 
\hline
\hline
182.7 & 
$0.11^{\phz+\phz0.16\phz}_{\phz-\phz0.12\phz}$ & 
$0.35^{\phz+\phz0.20\phz}_{\phz-\phz0.15\phz}$ & 
$0.31\pm0.17$ & 
$0.12^{\phz+\phz0.20\phz}_{\phz-\phz0.18\phz}$ &
$0.22\pm0.08\phs$ \\
188.6 & 
$0.67^{\phz+\phz0.14\phz}_{\phz-\phz0.13\phz}$ & 
$0.52^{\phz+\phz0.12\phz}_{\phz-\phz0.11\phz}$ & 
$0.73\pm0.15$ & 
$0.80^{\phz+\phz0.15\phz}_{\phz-\phz0.14\phz}$ &
$0.66\pm0.07\phs$ \\
191.6 & 
$0.62^{\phz+\phz0.40\phz}_{\phz-\phz0.33\phz}$ &
$0.63^{\phz+\phz0.36\phz}_{\phz-\phz0.30\phz}$ &
$0.29\pm0.22$ & 
$1.29^{\phz+\phz0.48\phz}_{\phz-\phz0.41\phz}$ &
$0.67\pm0.18\phs$ \\
195.5 & 
$0.73^{\phz+\phz0.25\phz}_{\phz-\phz0.22\phz}$ & 
$1.05^{\phz+\phz0.25\phz}_{\phz-\phz0.22\phz}$ & 
$1.18\pm0.26$ & 
$1.13^{\phz+\phz0.27\phz}_{\phz-\phz0.25\phz}$ &
$1.00\pm0.12\phs$ \\
199.5 & 
$0.91^{\phz+\phz0.25\phz}_{\phz-\phz0.22\phz}$ & 
$0.75^{\phz+\phz0.20\phz}_{\phz-\phz0.18\phz}$ & 
$1.25\pm0.27$ & 
$1.05^{\phz+\phz0.26\phz}_{\phz-\phz0.23\phz}$ &
$0.95\pm0.12\phs$ \\
201.6 & 
$0.71^{\phz+\phz0.32\phz}_{\phz-\phz0.27\phz}$ & 
$0.85^{\phz+\phz0.33\phz}_{\phz-\phz0.28\phz}$ & 
$0.95\pm0.39$ & 
$0.79^{\phz+\phz0.36\phz}_{\phz-\phz0.30\phz}$ &
$0.81\pm0.18\phs$ \\
204.9 & 
$1.20^{\phz+\phz0.28\phz}_{\phz-\phz0.26\phz}$ & 
$1.03^{\phz+\phz0.23\phz}_{\phz-\phz0.20\phz}$ & 
$0.77^{\phz+\phz0.21\phz}_{\phz-\phz0.19\phz}$ & 
$1.07^{\phz+\phz0.28\phz}_{\phz-\phz0.25\phz}$ &
$0.98\pm0.13\phs$ \\
206.6 & 
$1.05^{\phz+\phz0.22\phz}_{\phz-\phz0.21\phz}$ & 
$0.96^{\phz+\phz0.16\phz}_{\phz-\phz0.15\phz}$ & 
$1.09^{\phz+\phz0.18\phz}_{\phz-\phz0.17\phz}$ & 
$0.97^{\phz+\phz0.20\phz}_{\phz-\phz0.19\phz}$ &
$1.00\pm0.09\phs$ \\
\hline
\end{tabular}
\caption[Z-pair cross-sections]{
 Z-pair production cross-sections from the four LEP experiments and
combined values for the eight centre-of-mass energies between 183~GeV
and 207~GeV. The $\chidf$ of the combined fit is 14.5/24.}
\label{4f_tab:zzxsec}
\end{center}
\end{table}

\begin{figure}[p]
\begin{center}
\epsfig{figure=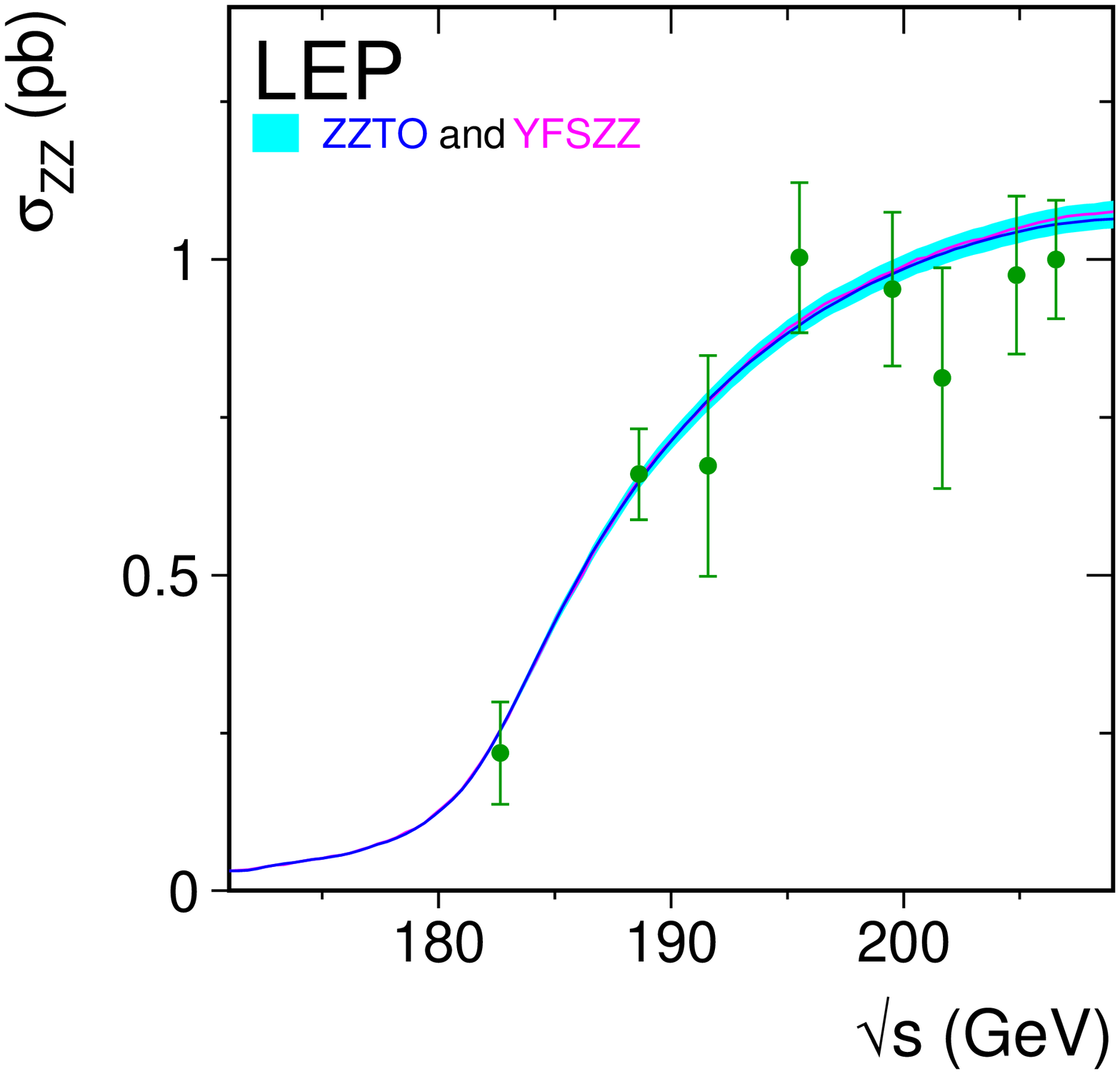,width=0.7\textwidth}
\caption[Z-pair cross-sections]{ Measurements of the Z-pair production
cross-section, compared to the predictions of
\YFSZZ~\protect\cite{ref:YFSZZ} and \ZZTO~\protect\cite{4f_bib:zzto}.
The shaded area represents the $\pm2$\% uncertainty on the
predictions.  }
\label{4f_fig:szz_vs_sqrts}
\end{center}
\end{figure} 

In analogy with the W-pair cross-section, a value for $\rzz$ can also
be determined: its definition and the procedure of the combination
follows those described for $\rww$.  The data are compared with the
\YFSZZ\ and \ZZTO\ predictions; Table~\ref{4f_tab:zzratio}, with more
details given in Table~\ref{4f_tab:rZZmeas}, reports the combined
values of $\rzz$ at each energy and combined, and
Figure~\ref{4f_fig:rzz} shows them in comparison to unity, where the
$\pm$2\% error on the theoretical ZZ cross-section is shown as a
yellow band. The experimental accuracy on the combined value of $\rzz$
is about 5\%.

\begin{table}[p]
\begin{center}
\renewcommand{\arraystretch}{1.25}
\begin{tabular}{|c||c|c|} 
\hline
\roots (GeV)      & $\rzz^{\footnotesize\ZZTO}$ 
           & $\rzz^{\footnotesize\YFSZZ}$ \\
\hline
\hline
182.7             & $0.857\pm0.320$ & $0.857\pm0.320$  \\
188.6             & $1.017\pm0.113$ & $1.007\pm0.111$  \\
191.6             & $0.865\pm0.226$ & $0.859\pm0.224$  \\
195.5             & $1.118\pm0.134$ & $1.118\pm0.134$  \\
199.5             & $0.974\pm0.126$ & $0.970\pm0.126$  \\
201.6             & $0.805\pm0.174$ & $0.800\pm0.174$  \\
204.9             & $0.934\pm0.122$ & $0.928\pm0.121$  \\
206.6             & $0.948\pm0.092$ & $0.938\pm0.091$  \\
\hline
$\chidf$          & 14.5/24         & 14.5/24         \\
\hline
\hline
Average           & $0.966\pm0.052$ & $0.960\pm0.052$  \\
\hline
$\chidf$          & 17.4/31         & 17.4/31        \\
\hline
\end{tabular}
\caption[Z-pair cross-section ratio experiment/theory]{ Ratios of LEP
combined Z-pair cross-section measurements to the expectations
according to \ZZTO~\protect\cite{4f_bib:zzto} and
\YFSZZ~\protect\cite{ref:YFSZZ}.  The results of the combined fits are
given together with the resulting $\chidf$.  Both fits take into
account inter-experiment as well as inter-energy correlations of
systematic errors.  }
\label{4f_tab:zzratio}
\end{center}
\end{table}

\begin{figure}[p]
\begin{center}
\epsfig{figure=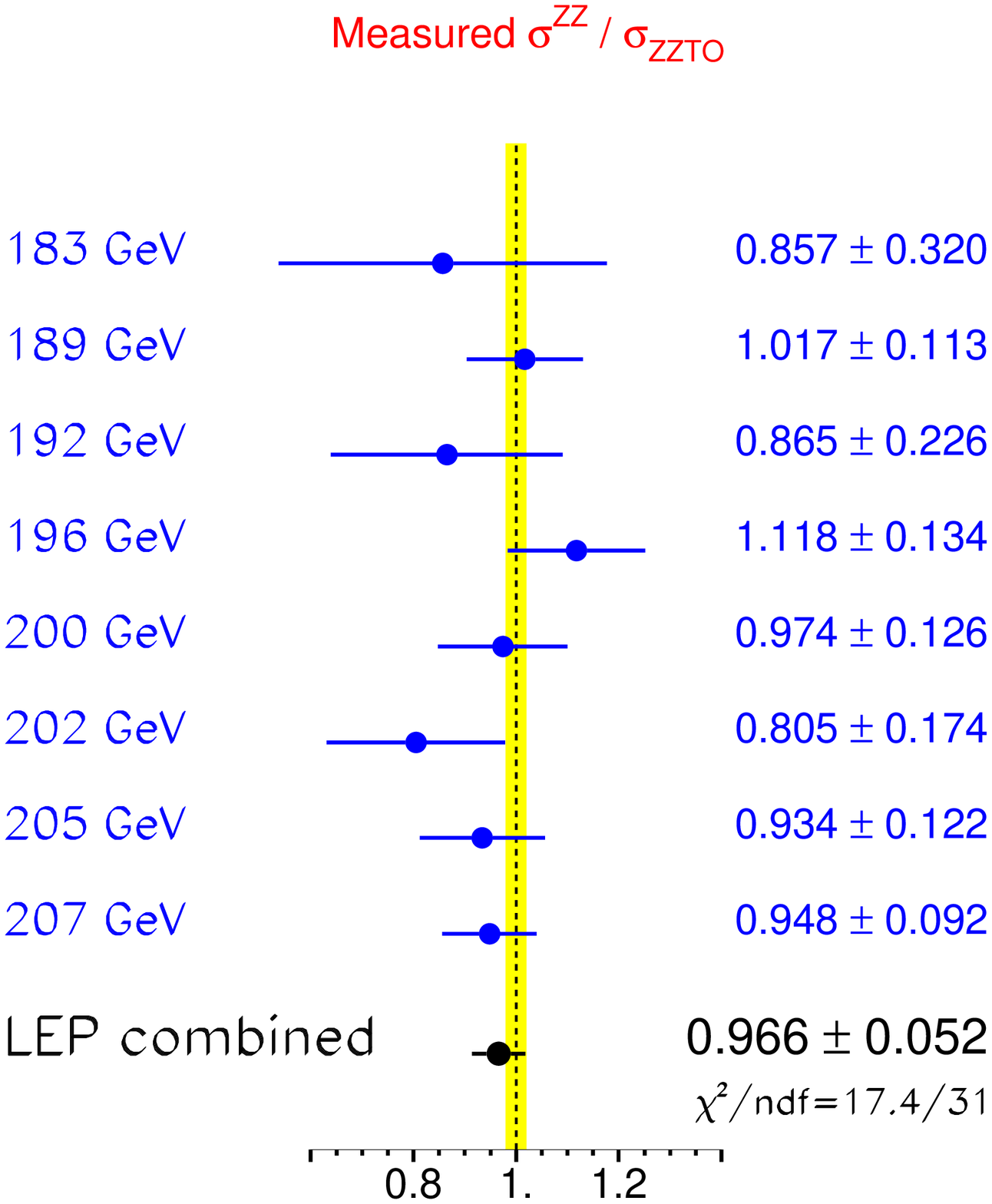,width=0.48\textwidth}
\hfill
\epsfig{figure=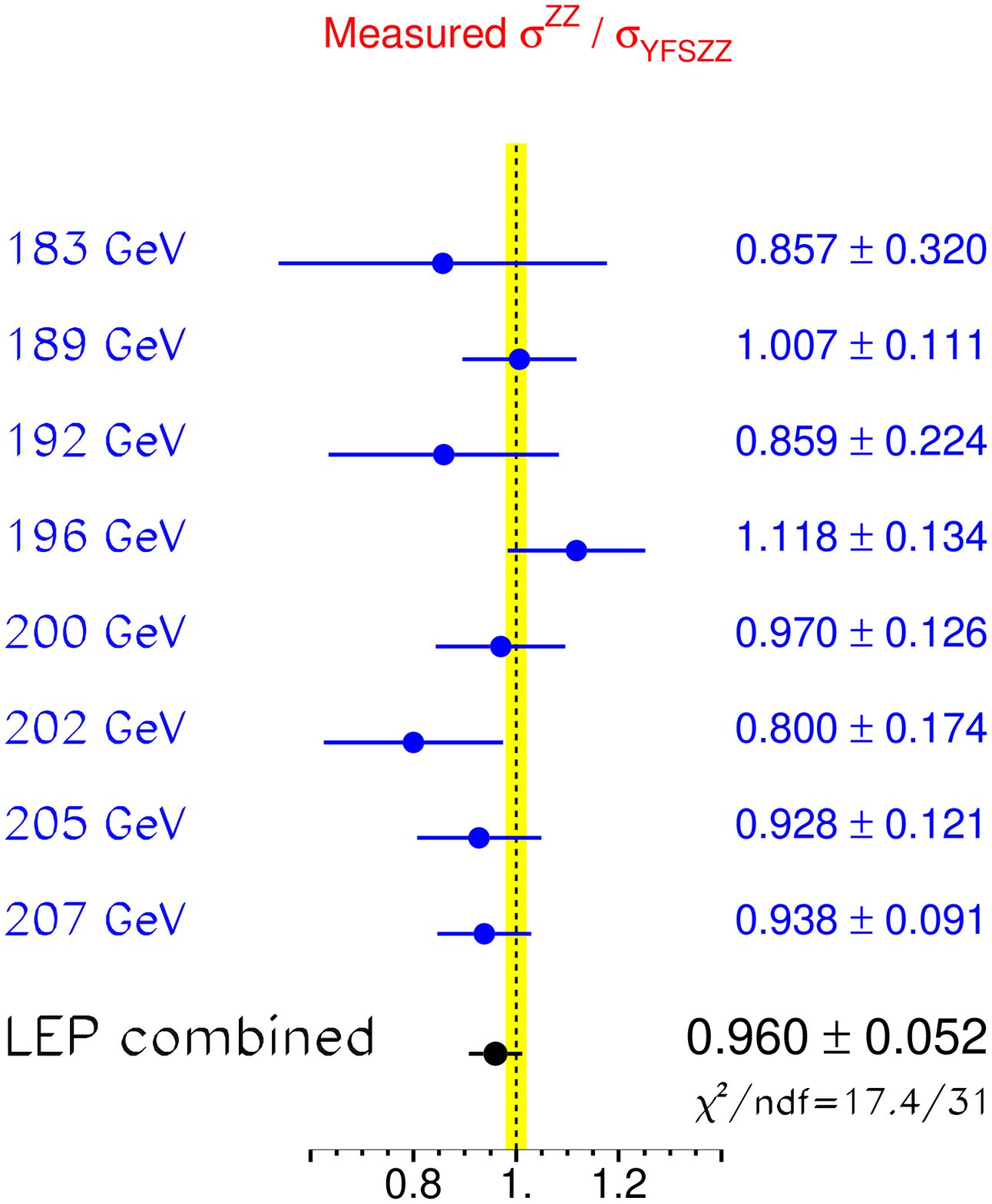,width=0.48\textwidth}
\caption[Z-pair cross-section ratio experiment/theory]{ Ratios of LEP
combined Z-pair cross-section measurements to the expectations
according to \ZZTO~\protect\cite{4f_bib:zzto} and
\YFSZZ~\protect\cite{ref:YFSZZ} The yellow bands represent constant
relative errors of 2\% on the two cross-section predictions.}
\label{4f_fig:rzz}
\end{center}
\end{figure}

\section{Z-$\gamma^*$ Production}
\label{Zgxsec}

The signal definition has been given in Section~\ref{introduction}.
The study of these final states is also relevant for the measurement
of neutral gauge couplings.  The LEP collaborations did not provide a
complete analysis of all possible Z$\gamma^*$ final states.  While
\Aleph\ and \Opal\ did not present any results on Z$\gamma^*$,
\Delphi\ provided results for the $\nu\nu$qq, $\ell\ell$qq,
$\ell\ell\ell\ell$ and qqqq final states~\cite{4f_bib:delzgstar}, and
\Ltre\ provided results for the $\nu\nu$qq, $\ell\ell$qq,
$\ell\ell\nu\nu$, and $\ell\ell\ell\ell$
channels~\cite{4f_bib:ltrzgstar}, where $\ell=\mathrm{e},\mu$. Final
states containing $\tau$ leptons were not studied.  The combination
reported here has been performed using data from the final states
provided by both \Delphi\ and \Ltre, namely $\nu\nu$qq, $\mu\mu$qq and
eeqq.
 
To increase the statistics the cross-sections were determined using
the full data sample at an average $\LEPII$ centre-of-mass energy.
Table~\ref{4f_tab:zgstarxsec} presents the measured cross-sections,
where the expected statistical errors were used for the combination.
As noted in Section~\ref{introduction}, the $Z\gamma^*$ signal has
been defined by mass and angular cuts specific to each of the
contributing channels, and the comparison of the combined LEP
cross-section with the theoretical prediction, calculated with
\Grace~\cite{\GRACEref} and shown in the last section of
Table~\ref{4f_tab:zgstarxsec}, has been made by imposing the same cuts
on each of the experimental and simulated samples included in the
combination.  The results agree well with the expectations.

\begin{table}[t]
\begin{center}
\renewcommand{\arraystretch}{1.25}
\begin{tabular}{|c||c|c|c|c|c|c|c|} 
\hline
        & $\sqrt{s}$       & L                & $\sigma$      & 
$\delta\sigma_{\mathrm{stat}}$      & 
$\delta\sigma_{\mathrm{syst}}^{\mathrm{unc}}$      &  
$\delta\sigma_{\mathrm{syst}}^{\mathrm{cor}}$      & 
$\delta\sigma_{\mathrm{MC}}$      \\
Channel &            [GeV] &      [pb$^{-1}$] &          [pb] & 
                               [pb] & 
                                              [pb] &  
                                              [pb] & 
                             [pb] \\
\hline
\hline
\multicolumn{8}{|c|}{\Delphi} \\
\hline
$\nu\nu$qq & 197.1 & 666.7 & 0.042 & $^{+0.022}_{-0.014}$ & 0.008 & 0.002 & 0.042 \\
$\mu\mu$qq & 197.1 & 666.7 & 0.031 & $^{+0.013}_{-0.011}$ & 0.004 & 0.001 & 0.016 \\
eeqq       & 197.1 & 666.7 & 0.063 & $^{+0.018}_{-0.016}$ & 0.009 & 0.001 & 0.016 \\
\hline
\multicolumn{8}{|c|}{\Ltre} \\
\hline
$\nu\nu$qq & 196.7 & 679.4 & 0.072 & $^{+0.047}_{-0.041}$ & 0.004 & 0.016 & 0.046 \\
$\mu\mu$qq & 196.7 & 681.9 & 0.040 & $^{+0.018}_{-0.016}$ & 0.002 & 0.003 & 0.017 \\
eeqq       & 196.7 & 681.9 & 0.100 & $^{+0.024}_{-0.022}$ & 0.004 & 0.007 & 0.020 \\
\hline
\multicolumn{8}{|c|}{LEP combined} \\
\hline
        & $\sqrt{s}$       & L                & $\sigma$     & $\delta\sigma_{\mathrm{stat}}$      & 
$\delta\sigma_{\mathrm{syst}}$      & $\delta\sigma_{\mathrm{tot}}$      & $\sigma_{\mathrm{theory}}$      \\
Channel &            [GeV] &      [pb$^{-1}$] &         [pb] &                                [pb] & 
                               [pb] &                               [pb] &                            [pb] \\
\hline
$\nu\nu$qq & 196.9 & 679.4 & 0.055 & 0.031 & 0.008 & 0.032 & 0.083 \\
$\mu\mu$qq & 196.9 & 681.9 & 0.035 & 0.012 & 0.003 & 0.012 & 0.042 \\
eeqq       & 196.9 & 681.9 & 0.079 & 0.012 & 0.005 & 0.013 & 0.059 \\
\hline
\end{tabular}
\end{center}
\caption[Z$\gamma^*$ cross-sections]{Z$\gamma^*$ measurements by the
experiments and combined LEP measurements.  The columns show,
respectively, the channel, the luminosity-weighted centre-of-mass
energy, the luminosity, the measured cross-section, the measured
statistical error, the systematic contribution uncorrelated between
experiments, the systematic contribution correlated between
experiments and the expected statistical error from the simulation.
For the LEP combination the full systematic error and the total error
are given and the last column presents the theory expectation with
\Grace~\protect\cite{\GRACEref}.}
\label{4f_tab:zgstarxsec}
\end{table}

\section{Single-Boson Production}
\label{Singlebosonintro}

The study of singly resonant final states finds its motivations in the
comparison with SM calculations in a delicate region of the 4-$f$
phase space, where the treatment of ISR or fermion loop corrections
can induce large corrections, up to several percent, to the total
cross-section.  These processes are also very sensitive to the value
of $\alpha_{QED}$.  Moreover, single W production also brings
information on possible anomalous WW$\gamma$ couplings.

Single boson production at LEP is mostly realised via $t$-channel
processes, where either the incident electron or positron maintains
its direction, escaping undetected along the beam and thus generating
missing momentum along the $z$ axis.  Single W and single Z production
then proceed dominantly via the vector boson fusion process
illustrated in Figure~\ref{V:fig:feyn_born} or via Bremsstrahlung
processes. In the case of single W production in the \Wtoenu\ final
state, the W is detected either by its hadronic decay producing two
jets, or by its leptonic decay producing a single charged lepton;
single Z production in the {\Ztoee} final state is identified from an
electron recoiling against two fermions (quarks or leptons) coming
from the Z decay.

The selection of these events is particularly difficult because of the
relatively low cross-section of the signal and because of the presence
of large backgrounds in these phase space regions.  Particularly large
backgrounds arise from radiative $q\bar{q}$ production or
$\gamma\gamma$ scattering.  The analyses, mostly based on sequential
cuts on kinematic variables, have an efficiency which depends on the
considered final state and ranges typically from 35\% to
60\%~\cite{4f_bib:alesw, 4f_bib:delsw, 4f_bib:ltrsw-1, 4f_bib:ltrsw-2,
4f_bib:ltrsw-3}.  These references describe results on single-boson
production using selection criteria which are specific to the
individual experiments.  The results shown below are derived from a
common selection procedure using the criteria listed in
Section~\ref{introduction}.

\subsection{We$\nu$ Cross-Section Measurement}
\label{Wevxsec}

The signal definition has been given in Section~\ref{introduction}.
The LEP combination of the single-W production cross-section is
performed using the expected statistical errors, given the limited
statistical precision of the single measurements.  The correlation of
the systematic errors in energy and among experiments is properly
taken into account.  The hadronic and the total single-W
cross-sections are combined independently, as the former is less
contaminated by $\gamma\gamma$ interaction contributions.  The details
on the input measurements are summarised in
Tables~\ref{4f_tab:WevHADmeas} and~\ref{4f_tab:WevTOTmeas}.

The hadronic single-W results and combinations are reported in
Table~\ref{4f_tab:swxsechad} and Figure~\ref{4f_fig:swen_had}.  The
total single-W results, for all decay modes and combinations are
listed in Table~\ref{4f_tab:swxsectot} and
Figure~\ref{4f_fig:swen_all}.  In the two figures, the measurements
are compared with the expected values from \WPHACT~\cite{\WPHACTref}
and \Grace~\cite{\GRACEref}, listed in Table~\ref{4f_tab:Wentheo}.  In
Figure~\ref{4f_fig:swen_had}, the predictions of \WTO~\cite{\WTOref},
which includes fermion-loop corrections for the hadronic final states,
have also been included.  As discussed more in detail
in~\cite{4f_bib:lep2mcws}, the theoretical predictions are scaled
upward to correct for the implementation of QED radiative corrections
at the wrong energy scale {\it s}.  The full correction of 4\%,
derived from comparison with the theoretical predictions from
\SWAP~\cite{4f_bib:swap}, is conservatively taken as a systematic
error.  This uncertainty dominates the $\pm$5\% theoretical error
currently assigned to these predictions, represented by the shaded
area in Figures~\ref{4f_fig:swen_had} and~\ref{4f_fig:swen_all}.  All
results, up to the highest \CoM\ energies, are in agreement with the
theoretical predictions.

\begin{table}[p]
\begin{center}
\renewcommand{\arraystretch}{1.25}
\begin{tabular}{|c||c|c|c||c|} 
\hline
\roots & \multicolumn{4}{|c|}{Single-W hadronic cross-section (pb)} \\
\cline{2-5} 
(GeV) & \Aleph\ & \Delphi\ & \Ltre\ & LEP \\
\hline
\hline
182.7 & $0.44^{\phz+\phz0.29}_{\phz-\phz0.24}\phs$ & $0.11^{\phz+\phz0.31}_{\phz-\phz0.14}\phs$ & 
$0.58^{\phz+\phz0.23\phz}_{\phz-\phz0.20}$ & $0.42\pm0.15\phs$ \\
188.6 & $0.33^{\phz+\phz0.16}_{\phz-\phz0.15}\phs$ & $0.57^{\phz+\phz0.21}_{\phz-\phz0.20}\phs$ &
$0.52^{\phz+\phz0.14\phz}_{\phz-\phz0.13}$ & $0.47\pm0.09\phs$  \\
191.6 & $0.52^{\phz+\phz0.52}_{\phz-\phz0.40}\phs$ & $0.30^{\phz+\phz0.48}_{\phz-\phz0.31}\phs$ &
$0.84^{\phz+\phz0.44\phz}_{\phz-\phz0.37}\phs$ & $0.56\pm0.25\phs$ \\
195.5 & $0.61^{\phz+\phz0.28}_{\phz-\phz0.25}\phs$ & $0.50^{\phz+\phz0.30}_{\phz-\phz0.27}\phs$ &
$0.66^{\phz+\phz0.25\phz}_{\phz-\phz0.23}\phs$ & $0.60\pm0.14\phs$ \\
199.5 & $1.06^{\phz+\phz0.30}_{\phz-\phz0.27}\phs$ & $0.57^{\phz+\phz0.28}_{\phz-\phz0.26}\phs$ &
$0.37^{\phz+\phz0.22\phz}_{\phz-\phz0.20}\phs$ & $0.65\pm0.14\phs$ \\
201.6 & $0.72^{\phz+\phz0.39}_{\phz-\phz0.33}\phs$ & $0.67^{\phz+\phz0.40}_{\phz-\phz0.36}\phs$ &
$1.10^{\phz+\phz0.40\phz}_{\phz-\phz0.35}\phs$ & $0.82\pm0.20\phs$ \\
204.9 & $0.34^{\phz+\phz0.24}_{\phz-\phz0.21}\phs$ & $0.99^{\phz+\phz0.33}_{\phz-\phz0.31}\phs$ & 
$0.42^{\phz+\phz0.25\phz}_{\phz-\phz0.21}\phs$ & $0.54\pm0.15\phs$ \\
206.6 & $0.64^{\phz+\phz0.21}_{\phz-\phz0.19}\phs$ & $0.81^{\phz+\phz0.23}_{\phz-\phz0.22}\phs$ & 
$0.66^{\phz+\phz0.20\phz}_{\phz-\phz0.18}\phs$ & $0.69\pm0.12\phs$ \\
\hline
\end{tabular}
\end{center}
\caption[Single-W hadronic cross-sections]{ Single-W hadronic
production cross-section from the LEP experiments and combined
values for the eight energies between 183 and 207~GeV, in the hadronic
decay channel of the W boson. The $\chidf$ of the combined fit is 13.2/16.}
\label{4f_tab:swxsechad}
\end{table}

\begin{figure}[p]
\begin{center}
\epsfig{figure=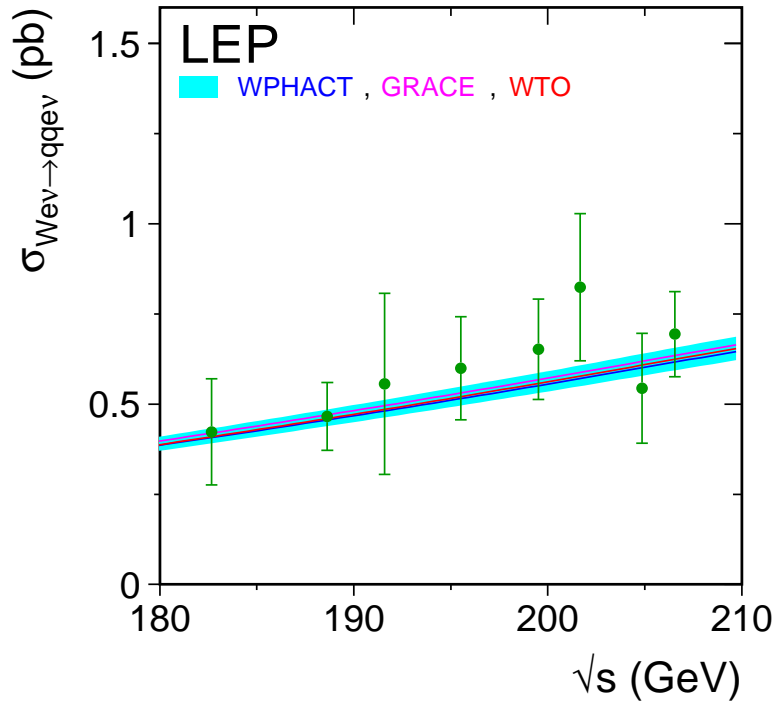,width=0.6\textwidth}
\end{center}
\caption[Single-W hadronic cross-sections]{ Measurements of the
single-W production cross-section in the hadronic decay channel of the
W boson, compared to the predictions of \WTO~\protect\cite{\WTOref},
\WPHACT~\protect\cite{\WPHACTref} and \Grace~\protect\cite{\GRACEref}.
The shaded area represents the $\pm5$\% uncertainty on the
predictions.  }
\label{4f_fig:swen_had}
\end{figure}

\begin{table}[p]
\begin{center}
\renewcommand{\arraystretch}{1.25}
\begin{tabular}{|c||c|c|c||c|} 
\hline
\roots & \multicolumn{4}{|c|}{Single-W total cross-section (pb)} \\
\cline{2-5} 
(GeV) & \Aleph\ & \Delphi\ & \Ltre\ & LEP \\
\hline
\hline
182.7 & $0.60^{\phz+\phz0.32}_{\phz-\phz0.26}\phs$ & $0.69^{\phz+\phz0.42}_{\phz-\phz0.25}\phs$ &
$0.80^{\phz+\phz0.28\phz}_{\phz-\phz0.25}$ & $0.70\pm0.17\phs$ \\
188.6 & $0.55^{\phz+\phz0.18}_{\phz-\phz0.16}\phs$ & $0.75^{\phz+\phz0.23}_{\phz-\phz0.22}\phs$ &
$0.69^{\phz+\phz0.16\phz}_{\phz-\phz0.15}$ & $0.66\pm0.10\phs$ \\

191.6 & $0.89^{\phz+\phz0.58}_{\phz-\phz0.44}\phs$ & $0.40^{\phz+\phz0.55}_{\phz-\phz0.33}\phs$ &
$1.11^{\phz+\phz0.48\phz}_{\phz-\phz0.41}\phs$ & $0.81\pm0.28\phs$ \\

195.5 & $0.87^{\phz+\phz0.31}_{\phz-\phz0.27}\phs$ & $0.68^{\phz+\phz0.34}_{\phz-\phz0.38}\phs$ &
$0.97^{\phz+\phz0.27\phz}_{\phz-\phz0.25}\phs$ & $0.85\pm0.16\phs$ \\

199.5 & $1.31^{\phz+\phz0.32}_{\phz-\phz0.29}\phs$ & $0.95^{\phz+\phz0.34}_{\phz-\phz0.30}\phs$ &
$0.88^{\phz+\phz0.26\phz}_{\phz-\phz0.24}\phs$ & $1.05\pm0.16\phs$ \\

201.6 & $0.80^{\phz+\phz0.42}_{\phz-\phz0.35}\phs$ & $1.24^{\phz+\phz0.52}_{\phz-\phz0.43}\phs$ &
$1.50^{\phz+\phz0.45\phz}_{\phz-\phz0.40}\phs$ & $1.17\pm0.23\phs$ \\

204.9 & $0.65^{\phz+\phz0.27}_{\phz-\phz0.23}\phs$ & $1.06^{\phz+\phz0.37}_{\phz-\phz0.32}\phs$ & 
$0.78^{\phz+\phz0.29\phz}_{\phz-\phz0.25}\phs$ & $0.80\pm0.17\phs$ \\

206.6 & $0.81^{\phz+\phz0.22}_{\phz-\phz0.20}\phs$ & $1.14^{\phz+\phz0.28}_{\phz-\phz0.25}\phs$ & 
$1.08^{\phz+\phz0.21\phz}_{\phz-\phz0.20}\phs$ & $1.00\pm0.14\phs$  \\
\hline
\end{tabular}
\end{center}
\caption[Single-W cross-sections]{ Single-W total production
cross-section from the LEP experiments and combined values for
the eight energies between 183 and 207~GeV. The $\chidf$ of the
combined fit is 8.1/16.}
\label{4f_tab:swxsectot}
\end{table}

\begin{figure}[p]
\begin{center}
\epsfig{figure=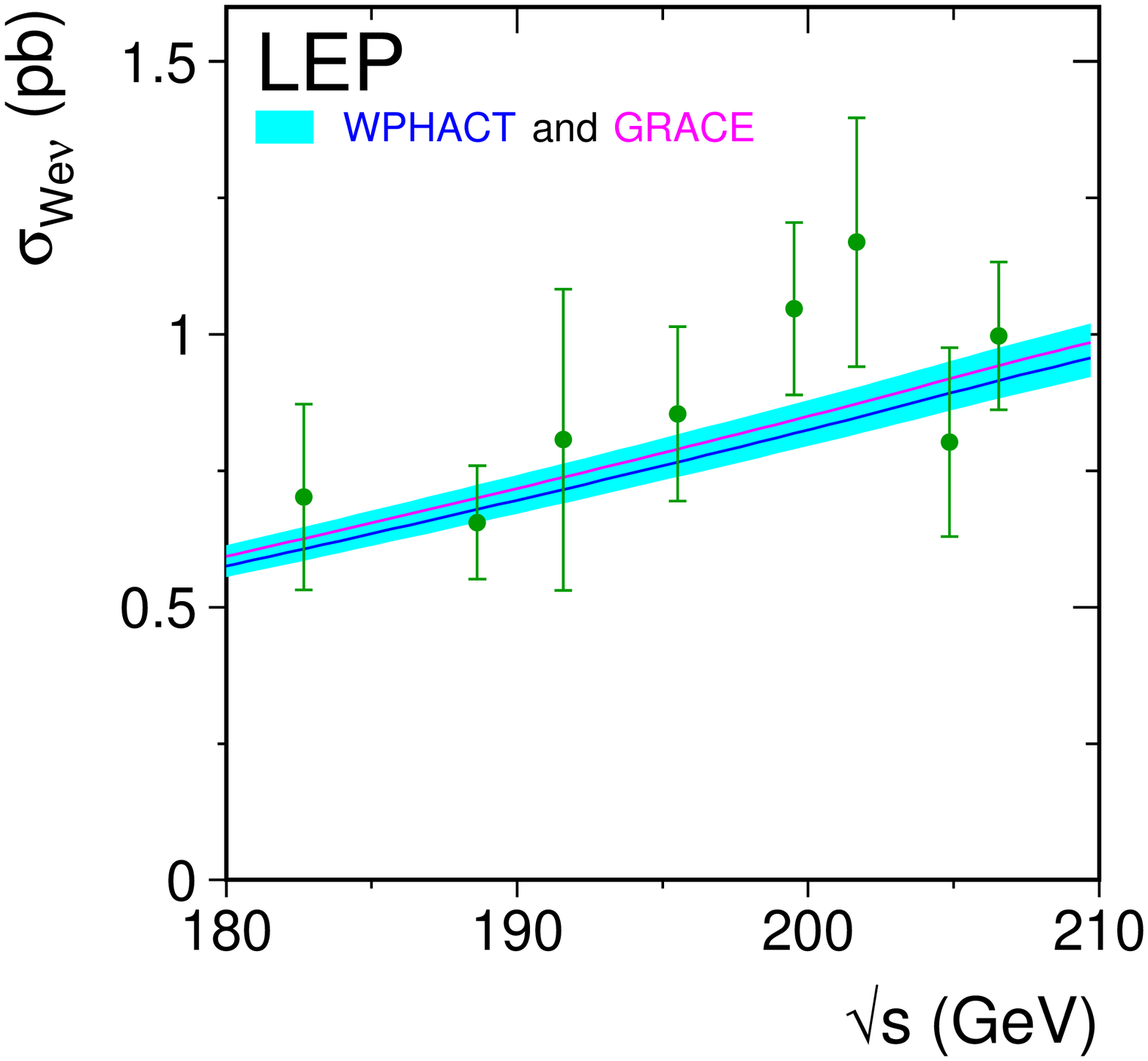,width=0.6\textwidth}
\end{center}
\caption[Single-W cross-sections]{ Measurements of the single-W total
production cross-section, compared to the predictions of
\WPHACT~\protect\cite{\WPHACTref} and
\Grace~\protect\cite{\GRACEref}.  The shaded area represents the $\pm5$\% uncertainty on the
predictions.  }
\label{4f_fig:swen_all}
\end{figure}

The agreement can also be appreciated in Table~\ref{4f_tab:wevratio},
where the values of the ratio between measured and expected
cross-section values according to the computations of \Grace\ and
\WPHACT\, are reported, with additional details listed in
Table~\ref{4f_tab:rwenmeas}.  The combination is performed accounting
for the energy and experiment correlations of the systematic sources.
The results are also presented in Figure~\ref{4f_fig:rwev}.

\begin{table}[p]
\begin{center}
\renewcommand{\arraystretch}{1.25}
\begin{tabular}{|c||c|c|} 
\hline 
\roots (GeV) & $\rwev^{\footnotesize\Grace}$ & $\rwev^{\footnotesize\WPHACT}$ \\
\hline
\hline
182.7             & $1.122\pm0.272$ & $1.157\pm0.281$  \\
188.6             & $0.936\pm0.149$ & $0.965\pm0.154$  \\
191.6             & $1.094\pm0.373$ & $1.128\pm0.385$  \\
195.5             & $1.081\pm0.203$ & $1.115\pm0.210$  \\
199.5             & $1.242\pm0.187$ & $1.280\pm0.193$  \\
201.6             & $1.340\pm0.261$ & $1.380\pm0.269$  \\
204.9             & $0.873\pm0.189$ & $0.899\pm0.195$  \\
206.6             & $1.058\pm0.143$ & $1.089\pm0.148$  \\
\hline
$\chidf$          & 8.1/16         & 8.1/16         \\
\hline
Average           & $1.058\pm0.078$ & $1.090\pm0.080$  \\
\hline
$\chidf$          & 12.1/23         & 12.1/23        \\
\hline
\end{tabular}
\caption[Single-W cross-section ratios experiment/theory]{ Ratios of
LEP combined total single-W cross-section measurements to the
expectations according to \Grace~\protect\cite{\GRACEref} and
\WPHACT~\protect\cite{\WPHACTref}.  The resulting averages over
energies are also given.  The averages take into account
inter-experiment as well as inter-energy correlations of systematic
errors.  }
\label{4f_tab:wevratio}
\end{center}
\vspace*{-0.3cm}
\end{table}

\begin{figure}[p]
\begin{center}
\epsfig{figure=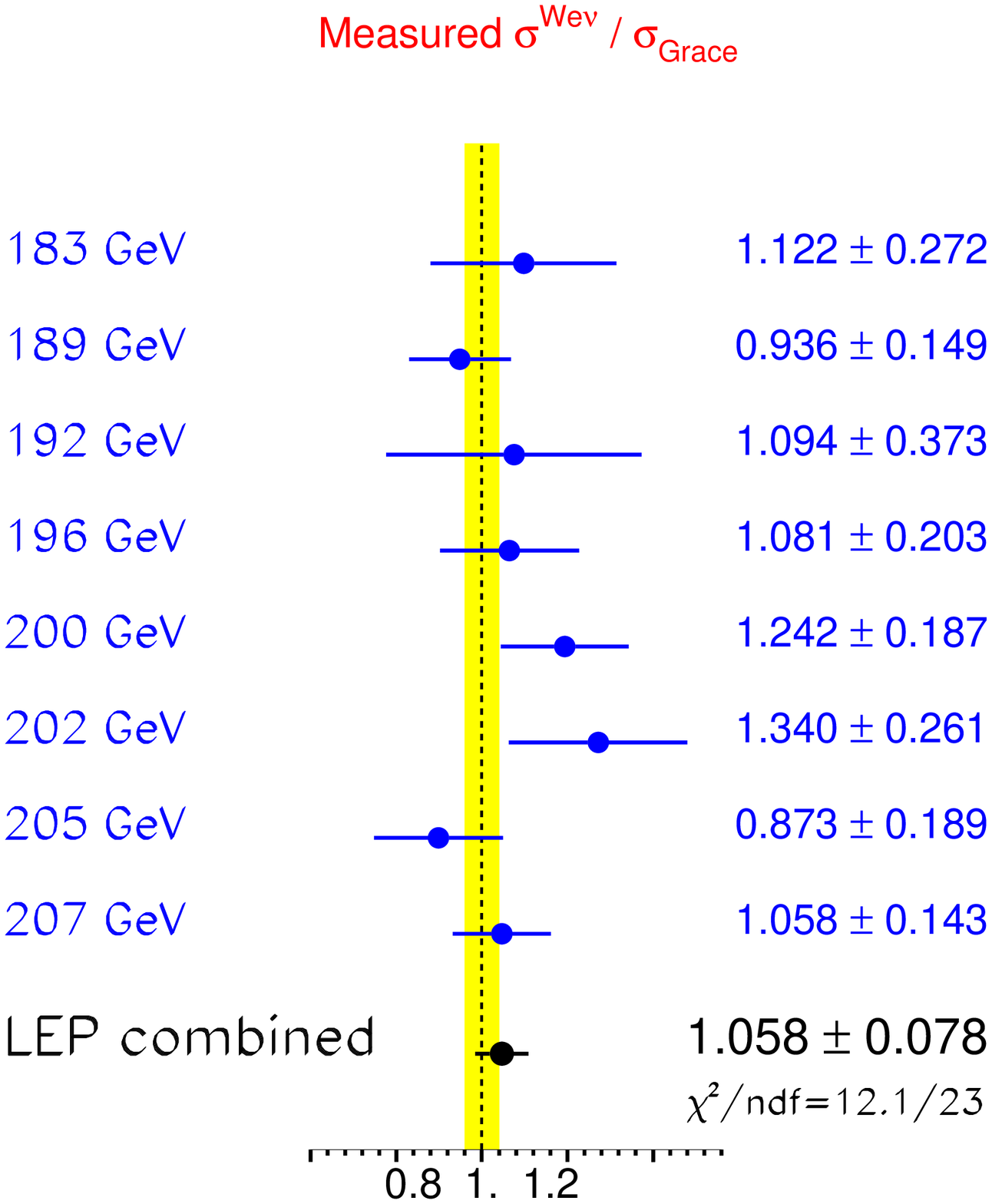,width=0.48\textwidth}
\hfill
\epsfig{figure=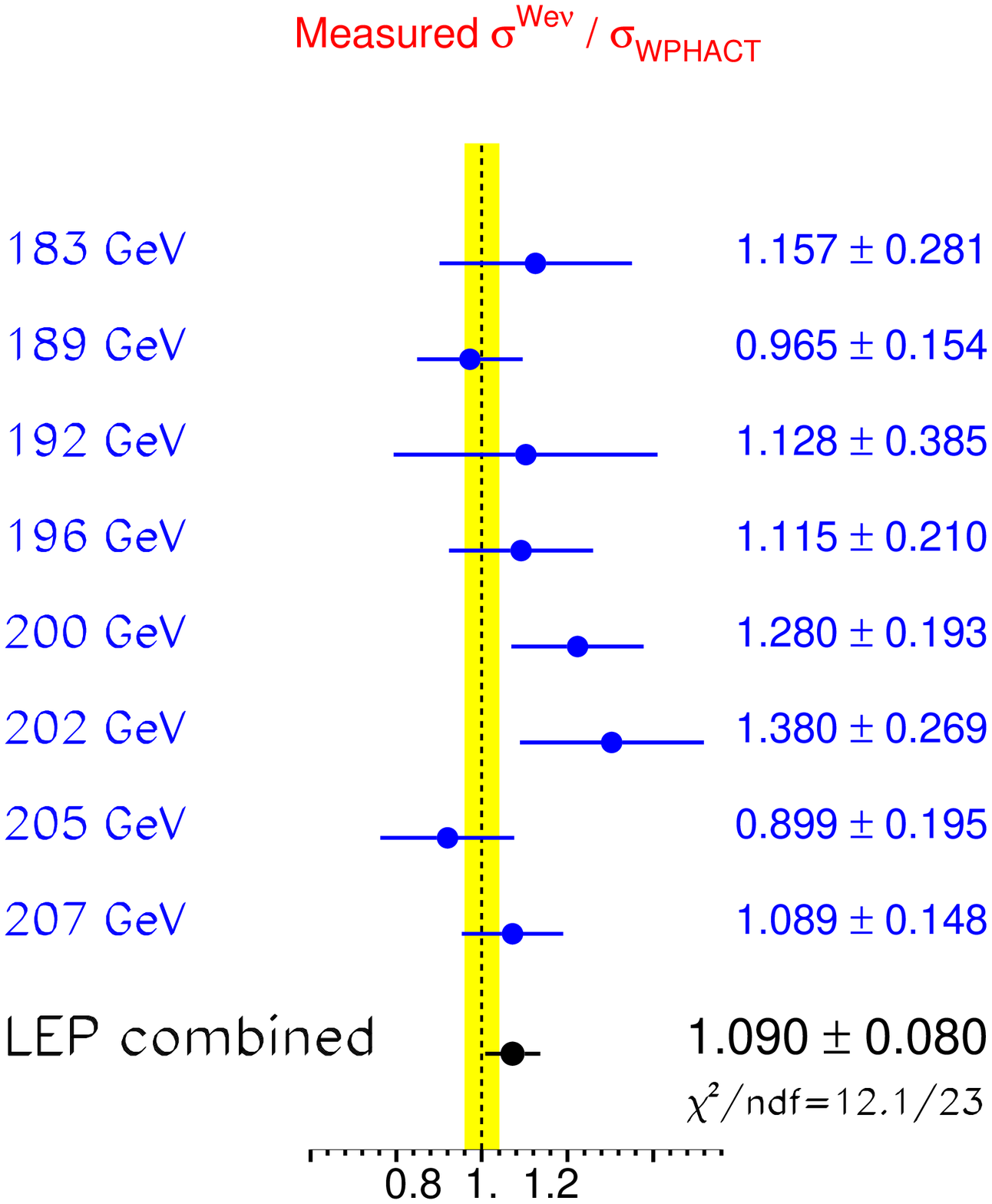,width=0.48\textwidth}
\end{center}
\caption[Single-W cross-section ratios experiment/theory]{ Ratios of
LEP combined total single-W cross-section measurements to the
expectations according to \Grace~\protect\cite{\GRACEref} and
\WPHACT~\protect\cite{\WPHACTref}.  The yellow bands represent
constant relative errors of 5\% on the two cross-section predictions.
}
\label{4f_fig:rwev}
\end{figure}

\subsection{Zee Cross-Section Measurement}
\label{Zeexsec}

The signal definition has been given in Section~\ref{introduction}.
The combination of results is performed with the same technique used
for the other channels.  The results include the hadronic and the
leptonic channels and all the \CoM\ energies from 183 to 209~GeV from
the \Aleph~\cite{4f_bib:alesw}, \Delphi~\cite{4f_bib:delsw} and
\Ltre~\cite{4f_bib:ltrzee} Collaborations. The \Opal\ 
results~\cite{Opal-Zee-183, *Opal-Zee-189} are not included in the
combination as they were not provided according to the common signal
definition.

Tables~\ref{4f_tab:szxsecmm} and~\ref{4f_tab:szxsecqq}, with details
summarised in Table~\ref{4f_tab:Zeemeas}, present the inputs from the
experiments and the corresponding LEP combinations in the muon and
hadronic channel, respectively.  The $ee\mu\mu$ cross-section is
already combined in energy by the individual experiments to increase
the statistics of the data.  The combination takes into account the
correlation of the energy and experimental systematic errors.  The
results in the hadronic channel are compared with the predictions of
{\WPHACT} and {\Grace}, listed in Table~\ref{4f_tab:Zeetheo}, in
Figure~\ref{4f_fig:szee} as a function of the \CoM\ energy.

The same data are expressed as ratios of the measured to the predicted
cross-section, listed in Table~\ref{4f_tab:zeeratio}, with details on
the decomposition of the systematic error reported in
Table~\ref{4f_tab:rzeemeas}, and shown in Figure~\ref{4f_fig:rzee}.
The accuracy of the combined ratio is about 7\% with three experiments
contributing to the average.

\begin{table}[p]
\begin{center}
\renewcommand{\arraystretch}{1.25}
\begin{tabular}{|c||c|c|c||c|} 
\hline
  & \multicolumn{4}{|c|}{Single-Z cross-section into muons(pb)} \\
\cline{2-5} 
    & \Aleph\ & \Delphi\ & \Ltre\ & LEP  \\
\hline
\hline
Av. \roots (GeV) & 196.67 & 197.10 & 196.60 & 196.79   \\
$\sigma_{\mathrm{Zee}\rightarrow \mu\mu \mathrm{ee}}$ 
& $0.055\pm0.016\phs$ &  $0.070^{\phz+\phz0.023\phz}_{\phz-\phz0.019}$ & 
  $0.043\pm0.013\phs$ & $0.057\pm0.009\phs$  \\
\hline
\end{tabular}
\end{center}
\caption[Single Z muon cross-sections]{ Energy averaged
single-Z production cross-section into muons from the LEP
experiments and the LEP combined value.}
\label{4f_tab:szxsecmm}
\end{table}

\begin{table}[p]
\begin{center}
\renewcommand{\arraystretch}{1.25}
\begin{tabular}{|c||c|c|c||c|} 
\hline
\roots & \multicolumn{4}{|c|}{Single-Z hadronic cross-section (pb)} \\
\cline{2-5} 
(GeV) & \Aleph\ & \Delphi\ & \Ltre\ & LEP \\
\hline
\hline
182.7 & $0.27^{\phz+\phz0.21\phz}_{\phz-\phz0.16}$ & $0.56^{\phz+\phz0.28\phz}_{\phz-\phz0.23}$ &
        $0.51^{\phz+\phz0.19\phz}_{\phz-\phz0.16}$ & $0.45\pm0.11\phs$ \\
188.6 & $0.42^{\phz+\phz0.14}_{\phz-\phz0.12}\phs$ & 
        $0.64^{\phz+\phz0.16}_{\phz-\phz0.14}\phs$ &
        $0.55^{\phz+\phz0.11\phz}_{\phz-\phz0.10}$ & $0.53\pm0.07\phs$ \\

191.6 & $0.61^{\phz+\phz0.39}_{\phz-\phz0.29}\phs$ & 
        $0.63^{\phz+\phz0.40}_{\phz-\phz0.30}\phs$ &
        $0.60^{\phz+\phz0.26}_{\phz-\phz0.21}\phs$ & $0.61\pm0.15\phs$ \\

195.5 & $0.72^{\phz+\phz0.24}_{\phz-\phz0.20}\phs$ & 
        $0.66^{\phz+\phz0.22}_{\phz-\phz0.19}\phs$ &
        $0.40^{\phz+\phz0.13}_{\phz-\phz0.11}\phs$ & $0.55\pm0.10\phs$ \\

199.5 & $0.60^{\phz+\phz0.21}_{\phz-\phz0.18}\phs$ & 
        $0.57^{\phz+\phz0.20}_{\phz-\phz0.17}\phs$ &
        $0.33^{\phz+\phz0.13}_{\phz-\phz0.11}\phs$ & $0.47\pm0.10\phs$ \\

201.6 & $0.89^{\phz+\phz0.35}_{\phz-\phz0.28}\phs$ & 
        $0.19^{\phz+\phz0.21}_{\phz-\phz0.16}\phs$ &
        $0.81^{\phz+\phz0.27}_{\phz-\phz0.23}\phs$ & $0.67\pm0.13\phs$ \\

204.9 & $0.42^{\phz+\phz0.17}_{\phz-\phz0.15}\phs$ & 
        $0.37^{\phz+\phz0.18}_{\phz-\phz0.15}\phs$ & 
        $0.56^{\phz+\phz0.16}_{\phz-\phz0.14}\phs$ & $0.47\pm0.10\phs$ \\

206.6 & $0.70^{\phz+\phz0.17}_{\phz-\phz0.15}\phs$ & 
        $0.69^{\phz+\phz0.16}_{\phz-\phz0.14}\phs$ & 
        $0.59^{\phz+\phz0.12}_{\phz-\phz0.11}\phs$ & $0.65\pm0.08\phs$ \\
\hline
\end{tabular}
\end{center}
\caption[Single-Z hadronic cross-sections]{ Single-Z hadronic
production cross-section from the LEP experiments and combined
values for the eight energies between 183 and 207~GeV. The $\chidf$ of
the combined fit is 13.0/16.}
\label{4f_tab:szxsecqq}
\vspace*{-0.1cm}
\end{table}

\begin{figure}[p]
\begin{center}
\epsfig{figure=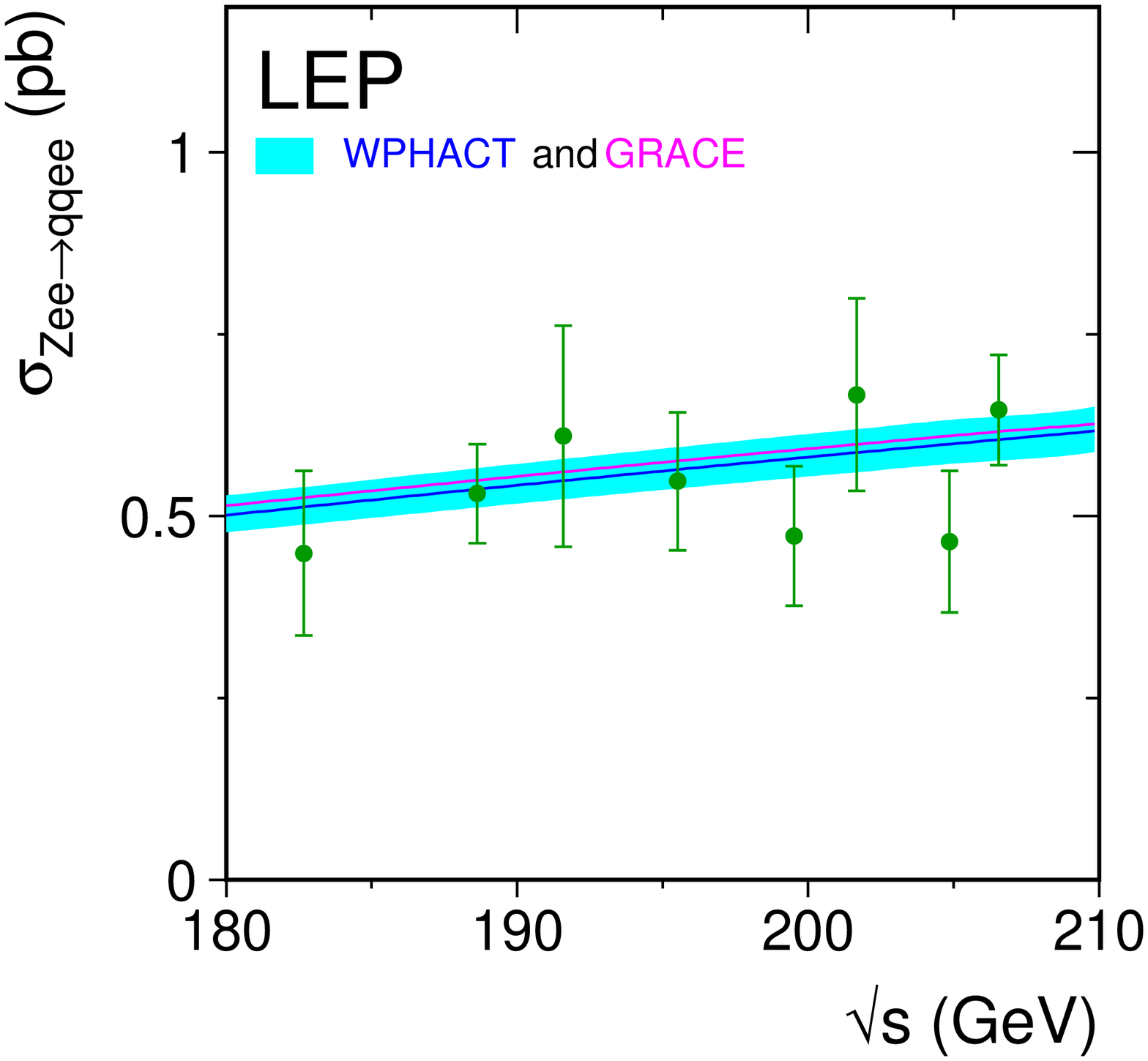,width=0.7\textwidth}
\end{center}
\caption[Single Z hadronic cross-sections]{ Measurements of the
single-Z hadronic production cross-section, compared to the
predictions of \WPHACT~\protect\cite{\WPHACTref} and
\Grace~\protect\cite{\GRACEref}.  The shaded area represents the
$\pm5$\% uncertainty on the predictions.  }
\label{4f_fig:szee}
\end{figure}

\begin{table}[p]
\begin{center}
\renewcommand{\arraystretch}{1.25}
\begin{tabular}{|c||c|c|} 
\hline
\roots (GeV) & $\rzee^{\footnotesize\Grace}$ & $\rzee^{\footnotesize\WPHACT}$ \\
\hline
\hline
182.7             & $0.871\pm0.219$ & $0.876\pm0.220$  \\
188.6             & $0.982\pm0.126$ & $0.990\pm0.127$  \\
191.6             & $1.104\pm0.275$ & $1.112\pm0.277$  \\
195.5             & $0.964\pm0.167$ & $0.972\pm0.168$  \\
199.5             & $0.809\pm0.165$ & $0.816\pm0.167$  \\
201.6             & $1.126\pm0.222$ & $1.135\pm0.224$  \\
204.9             & $0.769\pm0.160$ & $0.776\pm0.162$  \\
206.6             & $1.062\pm0.124$ & $1.067\pm0.125$  \\
\hline
$\chidf$    &  13.0/16          & 13.0/16         \\
\hline
\hline
Average           & $0.955\pm0.065$ & $0.962\pm0.065$  \\
\hline
$\chidf$    & 17.1/23         & 17.0/23        \\
\hline
\end{tabular}
\caption[Single Z hadronic cross-section ratios experiment/theory]{
Ratios of LEP combined single-Z hadronic cross-section measurements to
the expectations according to \Grace~\protect\cite{\GRACEref} and
\WPHACT~\protect\cite{\WPHACTref}.  The resulting averages over
energies are also given.  The averages take into account
inter-experiment as well as inter-energy correlations of systematic
errors.  }
\label{4f_tab:zeeratio}
\end{center}
\vspace*{-6mm}
\end{table}
\renewcommand{\arraystretch}{1.}

\begin{figure}[p]
\begin{center}
\epsfig{figure=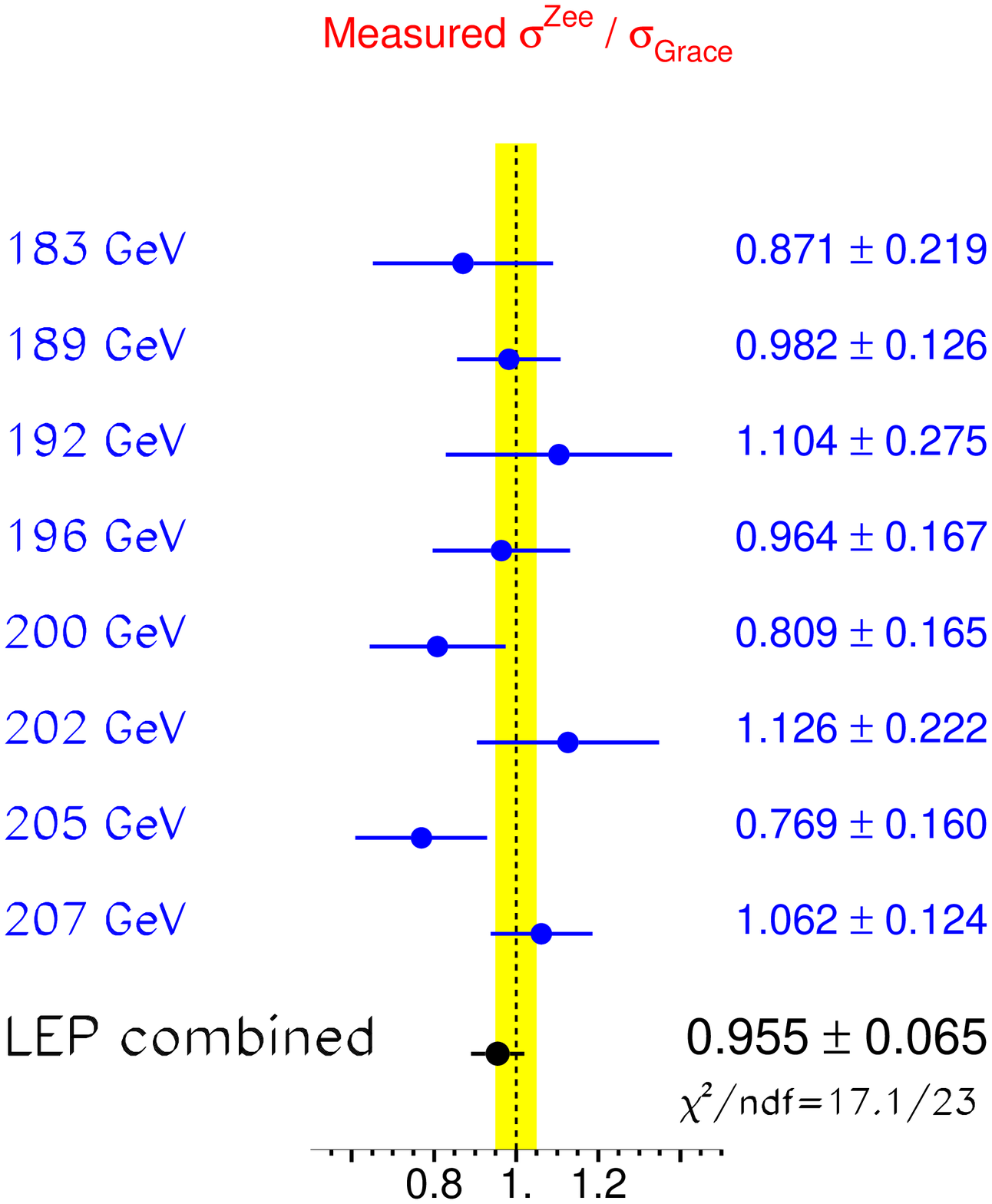,width=0.48\textwidth}
\hfill
\epsfig{figure=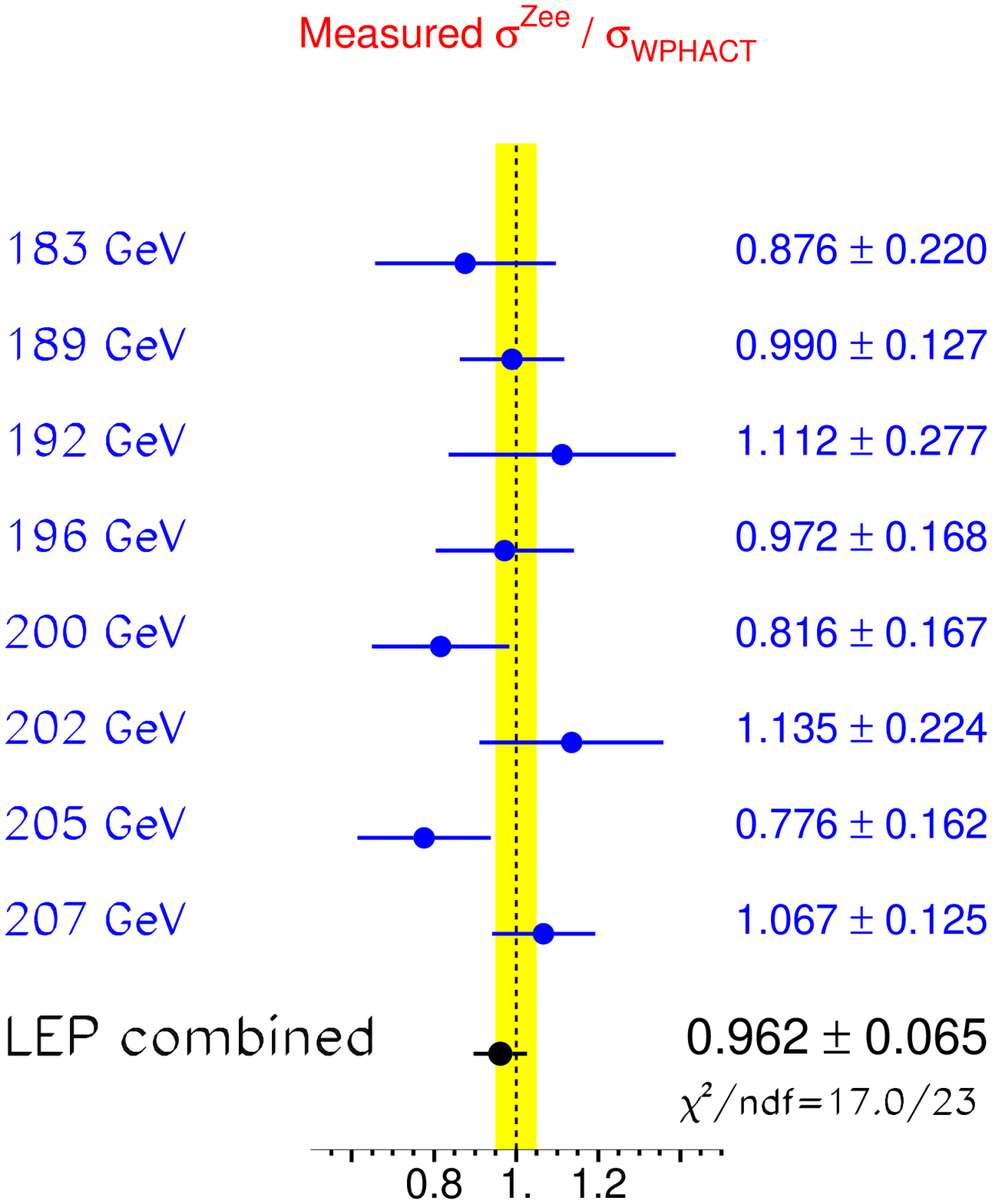,width=0.48\textwidth}
\end{center}
\caption[Single Z hadronic cross-section ratios experiment/theory]{
Ratios of LEP combined single-Z hadronic cross-section measurements to
the expectations according to \Grace~\protect\cite{\GRACEref} and
\WPHACT~\protect\cite{\WPHACTref}.  The yellow bands represent
constant relative errors of 5\% on the two cross-section predictions.
}
\label{4f_fig:rzee}
\end{figure}

\section{WW$\gamma$ Production}
\label{4f_sec:wwgxsec}

The signal definition has been given in Section~\ref{introduction}.
The study of photon production in association with a W-boson pair is
important for testing the sector of quartic gauge couplings.  In order
to increase the statistical accuracy, the LEP combination is performed
in energy intervals rather than at each energy point; they are defined
according to the $\LEPII$ running periods where more statistics were
accumulated.  The luminosity-weighted \CoM\ energy per interval is
determined in each experiment and then combined to obtain the
corresponding value for each energy interval.
Table~\ref{4f_tab:wwgxsec} reports those energies and the
cross-sections measured by the experiments that are used in this
combination~\cite{4f_bib:delwwg, 4f_bib:ltrwwg, 4f_bib:opawwg},
together with the combined LEP values.

Figure~\ref{4f_fig:swwg} shows the combined data points compared with
the cross-section calculated with \EEWWG~\cite{4f_bib:eewwg} and
\RacoonWW. The \RacoonWW\ prediction is shown in the figure without
any theory error band.

\begin{table}[p]
\begin{center}
\renewcommand{\arraystretch}{1.25}
\begin{tabular}{|c||c|c|c||c|} 
\hline
\roots & \multicolumn{4}{|c|}{WW$\gamma$ cross-section (pb)}  \\
\cline{2-5} 
(GeV) & \Delphi\ & \Ltre\ & \Opal\ & LEP  \\
\hline
\hline
188.6 & $0.05\pm0.08\phs$ & $0.20\pm0.09\phs$ & $0.16\pm0.04\phs$ & $0.15\pm0.03\phs$ \\
194.4 & $0.17\pm0.12\phs$ & $0.17\pm0.10\phs$ & $0.17\pm0.06\phs$ & $0.17\pm0.05\phs$ \\
200.2 & $0.34\pm0.12\phs$ & $0.43\pm0.13\phs$ & $0.21\pm0.06\phs$ & $0.27\pm0.05\phs$ \\
206.1 & $0.18\pm0.08\phs$ & $0.13\pm0.08\phs$ & $0.30\pm0.05\phs$ & $0.24\pm0.04\phs$ \\
\hline
\end{tabular}
\end{center}
\caption[WW$\gamma$ cross-sections]{ WW$\gamma$ production
cross-section from the LEP experiments and combined values for
the four energy bins.}
\label{4f_tab:wwgxsec}
\end{table}

\begin{figure}[p]
\begin{center}
\epsfig{figure=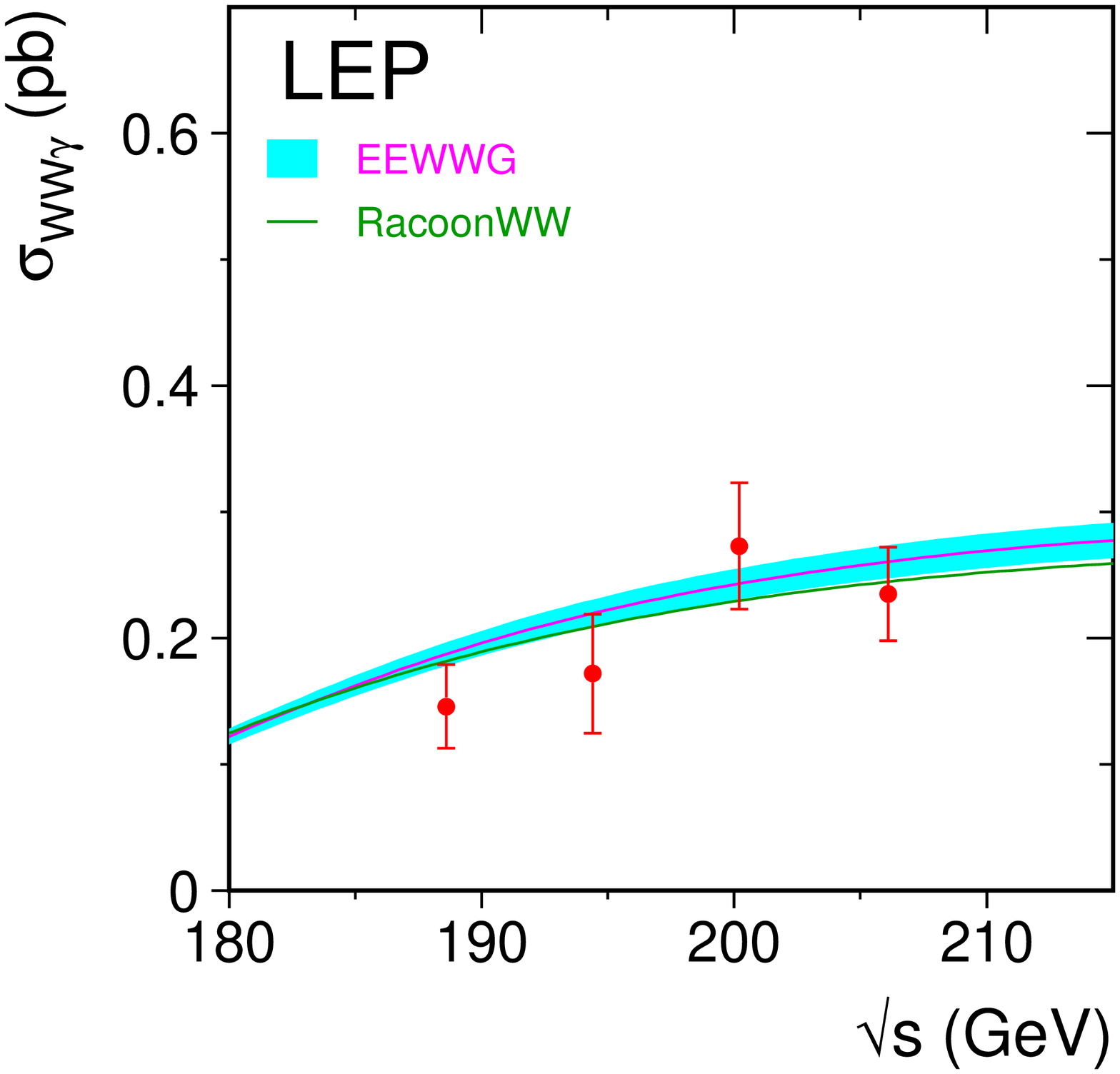,width=0.7\textwidth}
\caption[WW$\gamma$ cross-sections]{ Measurements of the WW$\gamma$
production cross-section, compared to the predictions of
\EEWWG~\protect\cite{4f_bib:eewwg} and
\RacoonWW~\protect\cite{\RACOONWWref}.  The shaded area in the
\EEWWG\ curve represents the $\pm5$\% uncertainty on the predictions.
}
\label{4f_fig:swwg}
\end{center}
\end{figure}

\section{Summary}
\label{summary}

This chapter has summarised the final LEP results in terms of
four-fermion cross-sections and derived quantities.  The WW
cross-section has been measured precisely at $\LEPII$ energies. The
measurements clearly favour those theoretical predictions which
properly include O($\alpha$) electroweak corrections, thus showing
that the SM can be tested at the loop level at $\LEPII$.

In general the results are in good agreement with the SM predictions,
both in the charged current and in the neutral current sector.  A
small anomaly in the W decay branching fractions, favouring W decays
into $\tau\nu_{\tau}$ compared to the other lepton families, is
observed in the data. This excess is above two standard deviations in
the measured branching fractions into both $e\nu_{e}$ and
$\mu\nu_{\mu}$.

\newcommand{\fFourgMinusOneSigmaLEP}{-0.09}
\newcommand{\fFourgMinusOneSigmaOPAL}{-0.20}
\newcommand{\fFourgMinusOneSigmaLThree}{-0.17}
\newcommand{\fFourgMinusOneSigmaDELPHI}{-0.12}
\newcommand{\fFourgMinusOneSigmaALEPH}{-0.23}
\newcommand{\fFourgPlusOneSigmaLEP}{0.12}
\newcommand{\fFourgPlusOneSigmaOPAL}{0.23}
\newcommand{\fFourgPlusOneSigmaLThree}{0.17}
\newcommand{\fFourgPlusOneSigmaDELPHI}{0.16}
\newcommand{\fFourgPlusOneSigmaALEPH}{0.23}
\newcommand{\fFourgMinusTwoSigmaLEP}{-0.17}
\newcommand{\fFourgMinusTwoSigmaOPAL}{-0.32}
\newcommand{\fFourgMinusTwoSigmaLThree}{-0.28}
\newcommand{\fFourgMinusTwoSigmaDELPHI}{-0.23}
\newcommand{\fFourgMinusTwoSigmaALEPH}{-0.32}
\newcommand{\fFourgPlusTwoSigmaLEP}{0.19}
\newcommand{\fFourgPlusTwoSigmaOPAL}{0.33}
\newcommand{\fFourgPlusTwoSigmaLThree}{0.28}
\newcommand{\fFourgPlusTwoSigmaDELPHI}{0.25}
\newcommand{\fFourgPlusTwoSigmaALEPH}{0.33}
\newcommand{\fFourzMinusOneSigmaLEP}{-0.14}
\newcommand{\fFourzMinusOneSigmaOPAL}{-0.17}
\newcommand{\fFourzMinusOneSigmaLThree}{-0.30}
\newcommand{\fFourzMinusOneSigmaDELPHI}{-0.23}
\newcommand{\fFourzMinusOneSigmaALEPH}{-0.38}
\newcommand{\fFourzPlusOneSigmaLEP}{0.22}
\newcommand{\fFourzPlusOneSigmaOPAL}{0.42}
\newcommand{\fFourzPlusOneSigmaLThree}{0.27}
\newcommand{\fFourzPlusOneSigmaDELPHI}{0.26}
\newcommand{\fFourzPlusOneSigmaALEPH}{0.38}
\newcommand{\fFourzMinusTwoSigmaLEP}{-0.28}
\newcommand{\fFourzMinusTwoSigmaOPAL}{-0.45}
\newcommand{\fFourzMinusTwoSigmaLThree}{-0.48}
\newcommand{\fFourzMinusTwoSigmaDELPHI}{-0.40}
\newcommand{\fFourzMinusTwoSigmaALEPH}{-0.53}
\newcommand{\fFourzPlusTwoSigmaLEP}{0.32}
\newcommand{\fFourzPlusTwoSigmaOPAL}{0.58}
\newcommand{\fFourzPlusTwoSigmaLThree}{0.46}
\newcommand{\fFourzPlusTwoSigmaDELPHI}{0.42}
\newcommand{\fFourzPlusTwoSigmaALEPH}{0.54}
\newcommand{\fFivezMinusOneSigmaLEP}{-0.19}
\newcommand{\fFivezMinusOneSigmaOPAL}{-0.67}
\newcommand{\fFivezMinusOneSigmaLThree}{0.00}
\newcommand{\fFivezMinusOneSigmaDELPHI}{-0.14}
\newcommand{\fFivezMinusOneSigmaALEPH}{-0.85}
\newcommand{\fFivezPlusOneSigmaLEP}{0.17}
\newcommand{\fFivezPlusOneSigmaOPAL}{0.00}
\newcommand{\fFivezPlusOneSigmaLThree}{0.06}
\newcommand{\fFivezPlusOneSigmaDELPHI}{0.36}
\newcommand{\fFivezPlusOneSigmaALEPH}{0.31}
\newcommand{\fFivezMinusTwoSigmaLEP}{-0.34}
\newcommand{\fFivezMinusTwoSigmaOPAL}{-0.94}
\newcommand{\fFivezMinusTwoSigmaLThree}{-0.35}
\newcommand{\fFivezMinusTwoSigmaDELPHI}{-0.38}
\newcommand{\fFivezMinusTwoSigmaALEPH}{-1.18}
\newcommand{\fFivezPlusTwoSigmaLEP}{0.35}
\newcommand{\fFivezPlusTwoSigmaOPAL}{0.25}
\newcommand{\fFivezPlusTwoSigmaLThree}{1.03}
\newcommand{\fFivezPlusTwoSigmaDELPHI}{0.62}
\newcommand{\fFivezPlusTwoSigmaALEPH}{1.19}
\newcommand{\fFivegMinusOneSigmaLEP}{-0.23}
\newcommand{\fFivegMinusOneSigmaOPAL}{-0.52}
\newcommand{\fFivegMinusOneSigmaLThree}{-0.20}
\newcommand{\fFivegMinusOneSigmaDELPHI}{-0.33}
\newcommand{\fFivegMinusOneSigmaALEPH}{-0.54}
\newcommand{\fFivegPlusOneSigmaLEP}{0.16}
\newcommand{\fFivegPlusOneSigmaOPAL}{0.23}
\newcommand{\fFivegPlusOneSigmaLThree}{0.30}
\newcommand{\fFivegPlusOneSigmaDELPHI}{0.26}
\newcommand{\fFivegPlusOneSigmaALEPH}{0.52}
\newcommand{\fFivegMinusTwoSigmaLEP}{-0.35}
\newcommand{\fFivegMinusTwoSigmaOPAL}{-0.71}
\newcommand{\fFivegMinusTwoSigmaLThree}{-0.39}
\newcommand{\fFivegMinusTwoSigmaDELPHI}{-0.52}
\newcommand{\fFivegMinusTwoSigmaALEPH}{-0.73}
\newcommand{\fFivegPlusTwoSigmaLEP}{0.32}
\newcommand{\fFivegPlusTwoSigmaOPAL}{0.59}
\newcommand{\fFivegPlusTwoSigmaLThree}{0.47}
\newcommand{\fFivegPlusTwoSigmaDELPHI}{0.48}
\newcommand{\fFivegPlusTwoSigmaALEPH}{0.74}
\newcommand{\fFourgMinusOneSigmaLEPTwoD}{-0.12}
\newcommand{\fFourzMinusOneSigmaLEPTwoD}{-0.15}
\newcommand{\fFourgMinusOneSigmaOPALTwoD}{-0.21}
\newcommand{\fFourzMinusOneSigmaOPALTwoD}{-0.22}
\newcommand{\fFourgMinusOneSigmaLThreeTwoD}{-0.17}
\newcommand{\fFourzMinusOneSigmaLThreeTwoD}{-0.31}
\newcommand{\fFourgMinusOneSigmaALEPHTwoD}{-0.16}
\newcommand{\fFourzMinusOneSigmaALEPHTwoD}{-0.22}
\newcommand{\fFourgPlusOneSigmaLEPTwoD}{0.10}
\newcommand{\fFourzPlusOneSigmaLEPTwoD}{0.19}
\newcommand{\fFourgPlusOneSigmaOPALTwoD}{0.22}
\newcommand{\fFourzPlusOneSigmaOPALTwoD}{0.42}
\newcommand{\fFourgPlusOneSigmaLThreeTwoD}{0.17}
\newcommand{\fFourzPlusOneSigmaLThreeTwoD}{0.27}
\newcommand{\fFourgPlusOneSigmaALEPHTwoD}{0.12}
\newcommand{\fFourzPlusOneSigmaALEPHTwoD}{0.23}
\newcommand{\fFourgMinusTwoSigmaLEPTwoD}{-0.20}
\newcommand{\fFourzMinusTwoSigmaLEPTwoD}{-0.29}
\newcommand{\fFourgMinusTwoSigmaOPALTwoD}{-0.32}
\newcommand{\fFourzMinusTwoSigmaOPALTwoD}{-0.47}
\newcommand{\fFourgMinusTwoSigmaLThreeTwoD}{-0.28}
\newcommand{\fFourzMinusTwoSigmaLThreeTwoD}{-0.48}
\newcommand{\fFourgMinusTwoSigmaALEPHTwoD}{-0.29}
\newcommand{\fFourzMinusTwoSigmaALEPHTwoD}{-0.43}
\newcommand{\fFourgPlusTwoSigmaLEPTwoD}{0.18}
\newcommand{\fFourzPlusTwoSigmaLEPTwoD}{0.32}
\newcommand{\fFourgPlusTwoSigmaOPALTwoD}{0.33}
\newcommand{\fFourzPlusTwoSigmaOPALTwoD}{0.58}
\newcommand{\fFourgPlusTwoSigmaLThreeTwoD}{0.28}
\newcommand{\fFourzPlusTwoSigmaLThreeTwoD}{0.46}
\newcommand{\fFourgPlusTwoSigmaALEPHTwoD}{0.25}
\newcommand{\fFourzPlusTwoSigmaALEPHTwoD}{0.44}
\newcommand{\fFivegMinusOneSigmaLEPTwoD}{-0.22}
\newcommand{\fFivezMinusOneSigmaLEPTwoD}{-0.35}
\newcommand{\fFivegMinusOneSigmaOPALTwoD}{-0.44}
\newcommand{\fFivezMinusOneSigmaOPALTwoD}{-0.68}
\newcommand{\fFivegMinusOneSigmaLThreeTwoD}{-0.27}
\newcommand{\fFivezMinusOneSigmaLThreeTwoD}{0.00}
\newcommand{\fFivegMinusOneSigmaALEPHTwoD}{-0.30}
\newcommand{\fFivezMinusOneSigmaALEPHTwoD}{-0.49}
\newcommand{\fFivegPlusOneSigmaLEPTwoD}{0.20}
\newcommand{\fFivezPlusOneSigmaLEPTwoD}{0.12}
\newcommand{\fFivegPlusOneSigmaOPALTwoD}{0.36}
\newcommand{\fFivezPlusOneSigmaOPALTwoD}{0.00}
\newcommand{\fFivegPlusOneSigmaLThreeTwoD}{0.40}
\newcommand{\fFivezPlusOneSigmaLThreeTwoD}{0.07}
\newcommand{\fFivegPlusOneSigmaALEPHTwoD}{0.28}
\newcommand{\fFivezPlusOneSigmaALEPHTwoD}{0.37}
\newcommand{\fFivegMinusTwoSigmaLEPTwoD}{-0.40}
\newcommand{\fFivezMinusTwoSigmaLEPTwoD}{-0.56}
\newcommand{\fFivegMinusTwoSigmaOPALTwoD}{-0.67}
\newcommand{\fFivezMinusTwoSigmaOPALTwoD}{-0.95}
\newcommand{\fFivegMinusTwoSigmaLThreeTwoD}{-0.53}
\newcommand{\fFivezMinusTwoSigmaLThreeTwoD}{-0.47}
\newcommand{\fFivegMinusTwoSigmaALEPHTwoD}{-0.59}
\newcommand{\fFivezMinusTwoSigmaALEPHTwoD}{-0.90}
\newcommand{\fFivegPlusTwoSigmaLEPTwoD}{0.38}
\newcommand{\fFivezPlusTwoSigmaLEPTwoD}{0.36}
\newcommand{\fFivegPlusTwoSigmaOPALTwoD}{0.62}
\newcommand{\fFivezPlusTwoSigmaOPALTwoD}{0.33}
\newcommand{\fFivegPlusTwoSigmaLThreeTwoD}{0.62}
\newcommand{\fFivezPlusTwoSigmaLThreeTwoD}{1.39}
\newcommand{\fFivegPlusTwoSigmaALEPHTwoD}{0.57}
\newcommand{\fFivezPlusTwoSigmaALEPHTwoD}{0.78}
\newcommand{\fFourgfFourg}{1.00}
\newcommand{\fFourgfFourz}{-0.33}
\newcommand{\fFivegfFiveg}{1.00}
\newcommand{\fFivegfFivez}{-0.20}

\newcommand{\hOnegMinusOneSigmaLEP}{-0.03}
\newcommand{\hOnegMinusOneSigmaOPAL}{-0.06}
\newcommand{\hOnegMinusOneSigmaLThree}{-0.03}
\newcommand{\hOnegMinusOneSigmaDELPHI}{-0.10}
\newcommand{\hOnegPlusOneSigmaLEP}{0.00}
\newcommand{\hOnegPlusOneSigmaOPAL}{0.00}
\newcommand{\hOnegPlusOneSigmaLThree}{0.00}
\newcommand{\hOnegPlusOneSigmaDELPHI}{0.00}
\newcommand{\hOnegMinusTwoSigmaLEP}{-0.05}
\newcommand{\hOnegMinusTwoSigmaOPAL}{-0.11}
\newcommand{\hOnegMinusTwoSigmaLThree}{-0.06}
\newcommand{\hOnegMinusTwoSigmaDELPHI}{-0.14}
\newcommand{\hOnegPlusTwoSigmaLEP}{0.05}
\newcommand{\hOnegPlusTwoSigmaOPAL}{0.11}
\newcommand{\hOnegPlusTwoSigmaLThree}{0.06}
\newcommand{\hOnegPlusTwoSigmaDELPHI}{0.14}
\newcommand{\hTwogMinusOneSigmaLEP}{-0.03}
\newcommand{\hTwogMinusOneSigmaOPAL}{-0.04}
\newcommand{\hTwogMinusOneSigmaLThree}{-0.03}
\newcommand{\hTwogPlusOneSigmaLEP}{0.00}
\newcommand{\hTwogPlusOneSigmaOPAL}{0.00}
\newcommand{\hTwogPlusOneSigmaLThree}{0.00}
\newcommand{\hTwogMinusTwoSigmaLEP}{-0.04}
\newcommand{\hTwogMinusTwoSigmaOPAL}{-0.08}
\newcommand{\hTwogMinusTwoSigmaLThree}{-0.05}
\newcommand{\hTwogPlusTwoSigmaLEP}{0.02}
\newcommand{\hTwogPlusTwoSigmaOPAL}{0.08}
\newcommand{\hTwogPlusTwoSigmaLThree}{0.02}
\newcommand{\hThreegMinusOneSigmaLEP}{-0.04}
\newcommand{\hThreegMinusOneSigmaOPAL}{-0.12}
\newcommand{\hThreegMinusOneSigmaLThree}{-0.04}
\newcommand{\hThreegMinusOneSigmaDELPHI}{-0.02}
\newcommand{\hThreegPlusOneSigmaLEP}{0.00}
\newcommand{\hThreegPlusOneSigmaOPAL}{0.00}
\newcommand{\hThreegPlusOneSigmaLThree}{0.00}
\newcommand{\hThreegPlusOneSigmaDELPHI}{0.00}
\newcommand{\hThreegMinusTwoSigmaLEP}{-0.05}
\newcommand{\hThreegMinusTwoSigmaOPAL}{-0.16}
\newcommand{\hThreegMinusTwoSigmaLThree}{-0.06}
\newcommand{\hThreegMinusTwoSigmaDELPHI}{-0.05}
\newcommand{\hThreegPlusTwoSigmaLEP}{-0.00}
\newcommand{\hThreegPlusTwoSigmaOPAL}{-0.01}
\newcommand{\hThreegPlusTwoSigmaLThree}{0.00}
\newcommand{\hThreegPlusTwoSigmaDELPHI}{0.04}
\newcommand{\hFourgMinusOneSigmaLEP}{0.02}
\newcommand{\hFourgMinusOneSigmaOPAL}{0.04}
\newcommand{\hFourgMinusOneSigmaLThree}{0.01}
\newcommand{\hFourgPlusOneSigmaLEP}{0.04}
\newcommand{\hFourgPlusOneSigmaOPAL}{0.10}
\newcommand{\hFourgPlusOneSigmaLThree}{0.03}
\newcommand{\hFourgMinusTwoSigmaLEP}{0.01}
\newcommand{\hFourgMinusTwoSigmaOPAL}{0.01}
\newcommand{\hFourgMinusTwoSigmaLThree}{-0.00}
\newcommand{\hFourgPlusTwoSigmaLEP}{0.05}
\newcommand{\hFourgPlusTwoSigmaOPAL}{0.13}
\newcommand{\hFourgPlusTwoSigmaLThree}{0.04}
\newcommand{\hOnezMinusOneSigmaLEP}{-0.06}
\newcommand{\hOnezMinusOneSigmaOPAL}{-0.10}
\newcommand{\hOnezMinusOneSigmaLThree}{-0.08}
\newcommand{\hOnezMinusOneSigmaDELPHI}{-0.16}
\newcommand{\hOnezPlusOneSigmaLEP}{0.00}
\newcommand{\hOnezPlusOneSigmaOPAL}{0.00}
\newcommand{\hOnezPlusOneSigmaLThree}{0.00}
\newcommand{\hOnezPlusOneSigmaDELPHI}{0.00}
\newcommand{\hOnezMinusTwoSigmaLEP}{-0.12}
\newcommand{\hOnezMinusTwoSigmaOPAL}{-0.19}
\newcommand{\hOnezMinusTwoSigmaLThree}{-0.15}
\newcommand{\hOnezMinusTwoSigmaDELPHI}{-0.23}
\newcommand{\hOnezPlusTwoSigmaLEP}{0.11}
\newcommand{\hOnezPlusTwoSigmaOPAL}{0.19}
\newcommand{\hOnezPlusTwoSigmaLThree}{0.14}
\newcommand{\hOnezPlusTwoSigmaDELPHI}{0.23}
\newcommand{\hTwozMinusOneSigmaLEP}{-0.04}
\newcommand{\hTwozMinusOneSigmaOPAL}{-0.07}
\newcommand{\hTwozMinusOneSigmaLThree}{-0.05}
\newcommand{\hTwozPlusOneSigmaLEP}{0.00}
\newcommand{\hTwozPlusOneSigmaOPAL}{0.00}
\newcommand{\hTwozPlusOneSigmaLThree}{0.00}
\newcommand{\hTwozMinusTwoSigmaLEP}{-0.07}
\newcommand{\hTwozMinusTwoSigmaOPAL}{-0.13}
\newcommand{\hTwozMinusTwoSigmaLThree}{-0.09}
\newcommand{\hTwozPlusTwoSigmaLEP}{0.07}
\newcommand{\hTwozPlusTwoSigmaOPAL}{0.13}
\newcommand{\hTwozPlusTwoSigmaLThree}{0.08}
\newcommand{\hThreezMinusOneSigmaLEP}{-0.13}
\newcommand{\hThreezMinusOneSigmaOPAL}{-0.18}
\newcommand{\hThreezMinusOneSigmaLThree}{-0.14}
\newcommand{\hThreezMinusOneSigmaDELPHI}{-0.24}
\newcommand{\hThreezPlusOneSigmaLEP}{0.00}
\newcommand{\hThreezPlusOneSigmaOPAL}{0.00}
\newcommand{\hThreezPlusOneSigmaLThree}{0.00}
\newcommand{\hThreezPlusOneSigmaDELPHI}{0.00}
\newcommand{\hThreezMinusTwoSigmaLEP}{-0.19}
\newcommand{\hThreezMinusTwoSigmaOPAL}{-0.27}
\newcommand{\hThreezMinusTwoSigmaLThree}{-0.22}
\newcommand{\hThreezMinusTwoSigmaDELPHI}{-0.30}
\newcommand{\hThreezPlusTwoSigmaLEP}{0.06}
\newcommand{\hThreezPlusTwoSigmaOPAL}{0.12}
\newcommand{\hThreezPlusTwoSigmaLThree}{0.11}
\newcommand{\hThreezPlusTwoSigmaDELPHI}{0.16}
\newcommand{\hFourzMinusOneSigmaLEP}{-0.00}
\newcommand{\hFourzMinusOneSigmaOPAL}{-0.02}
\newcommand{\hFourzMinusOneSigmaLThree}{-0.02}
\newcommand{\hFourzPlusOneSigmaLEP}{0.00}
\newcommand{\hFourzPlusOneSigmaOPAL}{0.00}
\newcommand{\hFourzPlusOneSigmaLThree}{0.00}
\newcommand{\hFourzMinusTwoSigmaLEP}{-0.04}
\newcommand{\hFourzMinusTwoSigmaOPAL}{-0.08}
\newcommand{\hFourzMinusTwoSigmaLThree}{-0.07}
\newcommand{\hFourzPlusTwoSigmaLEP}{0.13}
\newcommand{\hFourzPlusTwoSigmaOPAL}{0.17}
\newcommand{\hFourzPlusTwoSigmaLThree}{0.15}

\chapter{Electroweak Gauge Boson Self Couplings}
\label{chap:gc}

\section{Introduction}
\label{sec:gc_introduction}
 
The measurement of gauge boson couplings and the search for possible
anomalous contributions due to the effects of new physics beyond the
Standard Model (SM) are among the principal physics aims at
\LEPII~\cite{gc_bib:LEP2YR}.  Combined results on triple gauge boson
couplings are presented here.

The W-pair production process, $\mathrm{e^+e^-\rightarrow\WW}$,
involves the charged triple gauge boson vertices between the $\WW$ and
the Z or photon shown in Figure~\ref{WW:fig:feyn_born}.  During
\LEPII\ operation, about 10,000 W-pair events were collected by each
experiment.  Single W ($\enw$) and single photon ($\nng$) production
at LEP are also sensitive to the $\WWg$ vertex, see
Figure~\ref{V:fig:feyn_born}. Results from these channels are also
included in the combination for some experiments; the individual
references should be consulted for details.  The Monte-Carlo
calculations, RacoonWW~\cite{\RACOONWWref} and YFSWW~\cite{\YFSWWref},
incorporate an improved treatment of $O(\alpha_{em})$ corrections to
WW production. The corrections affect the measurements of the charged
TGCs in W-pair production.  Results including these $O(\alpha_{em})$
corrections have been submitted from all four LEP collaborations
ALEPH~\cite{gc_bib:ALEPH-cTGC}, DELPHI~\cite{gc_bib:DELPHI-cTGC},
L3~\cite{Achard:2004ji} and OPAL~\cite{gc_bib:OPAL-cTGC3}.

At centre-of-mass energies exceeding twice the Z boson mass, pair
production of Z bosons is kinematically allowed. Here, one searches
for the possible existence of triple vertices involving only neutral
electroweak gauge bosons. Such vertices could also contribute to
Z$\gamma$ production.  In contrast to triple gauge boson vertices with
two charged gauge bosons, purely neutral gauge boson vertices do not
occur at tree level in the SM of electroweak interactions.

The expected total and differential cross-sections of these processes
depend on the values of the triple gauge couplings, allowing their
measurements by comparing Monte-Carlo simulations to the data.  In
contrast to the analysis of electroweak gauge boson self-couplings
performed at hadron colliders, no form-factor term scaling the bare
couplings is applied in the analysis of the LEP data.

\subsection{Charged Triple Gauge Boson Couplings}
\label{sec:gc_cTGCs}

The parametrisation of the charged triple gauge boson vertices is
described in References~\cite{gc_bib:LEP2YR, gc_bib:GAEMERS,
gc_bib:Hagiwara1987vm, gc_bib:HAGIWARA-1, *gc_bib:HAGIWARA-2,
*gc_bib:HAGIWARA-3, *gc_bib:BILENKY-1, *gc_bib:BILENKY-2,
*gc_bib:KUSS, *gc_bib:PAPADOPOULOSCP}.  The most general Lorentz
invariant Lagrangian which describes the triple gauge boson
interaction has fourteen independent complex couplings, seven
describing the WW$\gamma$ vertex and seven describing the WWZ vertex.
Assuming electromagnetic gauge invariance as well as C and P
conservation, the number of independent TGCs reduces to five.  A
common set is \{$\gz, \kg, \kz, \lg$, $\lz$\}, with $\SM$ values of
$\gz = \kg = \kz = 1$ and $\lg = \lz = 0$.  The parameters proposed
in~\cite{gc_bib:LEP2YR} and used by the LEP experiments are $\gz$,
$\kg$ and $\lg$ with the gauge constraints:

\begin{eqnarray}
\kz & = & \gz - (\kg - 1) \twsq \,, \\
\lz & = & \lg \,,
\end{eqnarray}                               
where $\theta_W$ is the weak mixing angle. In an effective Lagrangian
approach, all three of the remaining independent couplings, $\gz$,
$\kg$ and $\lg$, receive contributions from operators of dimension six
or greater.  The couplings are considered as real, with the imaginary
parts fixed to zero.  Note that the measured coupling values
themselves and not their deviation from the SM values are
quoted.  LEP combinations are made in single-parameter fits, in each
case setting the other TGCs to their SM value.

The coupling $\gz$ describes the overall strength of the WWZ coupling.
The photonic couplings $\kg$ and $\lg$ are related to the magnetic and
electric properties of the W boson. One can write the lowest order
terms for a multipole expansion describing the W-$\gamma$ interaction
as a function of $\kg$ and $\lg$. For the magnetic dipole moment,
$\mu_{W}$, and the electric quadrupole moment, $q_{W}$, one obtains:

\begin{eqnarray}
  \label{eq:moments}
   \mu_{W} & = & \frac{e}{2m_{W}} \left( 1 + \kappa_{\gamma} +
   \lambda_{\gamma} \right) \,, \\
   q_{W} & = & - \frac{e}{m_{W}^{2}} \left( \kappa_{\gamma} -
   \lambda_{\gamma} \right) \,.
\end{eqnarray}
The inclusion of $O(\alpha_{em})$ corrections in the Monte-Carlo
calculations has a considerable effect on the charged TGC
measurement. Both the total cross-section and the differential
distributions are affected. The cross-section is reduced by $1-2$\%
depending on the energy.  For the differential distributions, the
effects are naturally more complex. The polar W$^-$ production angle
carries most of the information on the TGC parameters; its shape is
modified to be more forwardly peaked. In a fit to data, the
$O(\alpha_{em})$ effect manifests itself as a negative shift of the
obtained TGC values with a magnitude of typically $-0.015$ for \lg\
and \gz\, and $-0.04$ for \kg.

\subsection{Neutral Triple Gauge Boson Couplings}
\label{sec:gc_nTGCs}

There are two classes of Lorentz invariant structures associated with
neutral TGC vertices which preserve $U(1)_{em}$ and Bose symmetry, as
described in~\cite{gc_bib:Hagiwara1987vm, gc_bib:Gounaris2000tb}.

The first class refers to anomalous Z$\gamma\gamma^*$ and $\rm Z\gamma
\rm Z^*$ couplings which are accessible at LEP in the process
$\mathrm{e^{+} e^{-}} \rightarrow {\rm Z} \gamma$. The parametrisation
contains eight couplings: $h_i^{V}$ with $i=1,...,4$ and $V=\gamma$,Z.
The superscript $V=\gamma$ refers to Z$\gamma\gamma^*$ couplings and
superscript $V=$Z refers to $\rm Z\gamma \rm Z^*$ couplings.  The
photon and the Z boson in the final state are considered as on-shell
particles, while the third boson at the vertex, the $s$-channel
internal propagator, is off shell.  The couplings $h_{1}^{V}$ and
$h_{2}^{V}$ are CP-odd while $h_{3}^{V}$ and $h_{4}^{V}$ are CP-even.

The second class refers to anomalous $\rm{ZZ}\gamma^*$ and
$\rm{ZZZ}^*$ couplings which are accessible at \LEPII\ in the process
$\mathrm{e^{+} e^{-}} \rightarrow$ ZZ.  This anomalous vertex is
parametrised in terms of four couplings: $f_{i}^{V}$ with $i=4,5$ and
$V=\gamma$,Z.  The superscript $V=\gamma$ refers to ZZ$\gamma^*$
couplings and the superscript $V=\mathrm{Z}$ refers to $\rm{ZZZ}^*$
couplings.  Both Z bosons in the final state are assumed to be
on-shell, while the third boson at the triple vertex, the $s$-channel
internal propagator, is off-shell.  The couplings $f_{4}^{V}$ are
CP-odd whereas $f_{5}^{V}$ are CP-even.

In an effective Lagrangian approach, the couplings $h_1^V$, $h_3^V$,
$f_4^V$, $f_5^V$ receive contributions from operators of dimension six
or greater, while the lowest-dimension operators contributing to
$h_2^V$ and $h_4^V$ have dimension eight.  Note that the $h_i^{V}$ and
the $f_{i}^{V}$ couplings are independent of each other.  They are
assumed to be real and they vanish at tree level in the SM.  Results
on neutral gauge boson couplings are reported for single- and
two-parameter fits.

\section{Combination Procedure}
\label{sec:gc_combination}

The combination is based on the individual likelihood functions from
the four LEP experiments.  Each experiment provides the negative log
likelihood, $\LL$, as a function of the coupling parameters to be
combined.  The single-parameter analyses are performed fixing all
other parameters to their SM values.  For the charged
TGCs, the gauge constraints listed in Section~\ref{sec:gc_cTGCs} are
always enforced. Either the $\LL$ curves were available in numerical
form or they have been treated as parabolic according to the
respective publication.  Details of the individual measurements
entering the combination are summarised below.

The $\LL$ functions from each experiment include statistical as well
as those systematic uncertainties which are considered as uncorrelated
between experiments.  In all combinations, the individual $\LL$
functions are combined.  It is necessary to use the $\LL$ functions
directly in the combination, since in some cases they are not
parabolic, and hence it is not possible to properly combine the
results by simply taking weighted averages of the measurements.

The main contributions to the systematic uncertainties that are
uncorrelated between experiments arise from detector effects,
background in the selected signal samples, limited Monte-Carlo
statistics and the fitting method.  Their importance varies for each
experiment and the individual references should be consulted for
details.

In the neutral TGC sector, the main correlated systematic
uncertainties arise from the theoretical cross-section prediction in
ZZ and Z$\gamma$-production, about $2\%$ for ZZ and about $1\%$
($2\%$) in the $\qq\gamma$ ($\nng$) channel. The effect of a
correlated treatment has been estimated in earlier measurements to be
negligible.  Hence this and all other correlated sources of systematic
errors, such as those arising from the LEP beam energy, are for
simplicity treated as uncorrelated. The combination is performed by
adding the $\LL$ curves of the individual experiments.

In the charged TGC sector, systematic uncertainties considered
correlated between the experiments are summarised in
Table~\ref{tab:gc_cTGC-syst}: the theoretical cross-section
prediction, $\sigma$, which is $0.5\%$ for W-pair production and $5\%$
for single W production, hadronisation effects (HAD), the final state
interactions, namely Bose-Einstein correlations (BEC) and colour
reconnection (CR), and the uncertainty in the radiative corrections
themselves (LPA).  The latter was the dominant systematic error in
previous combinations, where we used a conservative estimate, namely
the full effect from applying the $O(\alpha_{em})$ corrections.
Analyses on the subject are available from several LEP experiments,
based on comparisons of fully simulated events using two different
leading-pole approximation schemes (LPA-A and
LPA-B,~\cite{gc_bib:LPA_A-B} and references therein).  In addition,
the availability of comparisons of the generators incorporating
$O(\alpha_{em})$ corrections, RacoonWW and YFSWW~\cite{\RACOONWWref,
\YFSWWref}, makes it possible to perform a more realistic estimation
of this effect and its uncertainty.  In general, the TGC shift
measured in the comparison of the two generators is found to be larger
than the effect from the different LPA schemes.  This improved
estimation, while still being conservative, reduces the systematic
uncertainty from $O(\alpha_{em})$ corrections by about a third for
$\gz$ and $\lg$ and roughly halves it for $\kg$, compared to the full
$O(\alpha_{em})$ correction. The application of this reduced
systematic error renders the charged TGC measurements statistics
dominated.  In case of the charged TGCs, the systematic uncertainties
considered correlated between the experiments amount to 32\% of the
combined statistical and uncorrelated uncertainties for $\lg$ and
$\gz$, while for $\kg$ they amount to 57\%, indicating again that the
measurements of $\lg$, $\gz$ and $\kg$ are limited by data statistics.

\begin{table}[t]
\begin{center}
\renewcommand{\arraystretch}{1.3}
\begin{tabular}{|l||r|r|r|} 
\hline
Source  & \gz  & \kg   & \lg  \\
\hline
\hline
$\sigma_{WW}$ prediction        & 0.003  & 0.018 & 0.002 \\ 
$\sigma_{W}$ prediction         &  -     & 0.003 & 0.001 \\ 
Hadronisation                   & 0.003  & 0.005 & 0.004 \\ 
Bose-Einstein Correlation       & 0.002  & 0.003 & 0.002 \\ 
Colour Reconnection             & 0.003  & 0.005 & 0.002 \\
$O(\alpha_{em})$ correction     & 0.002  & 0.014 & 0.002 \\ 
\hline
\end{tabular}
\caption[Systematic uncertainties]{ The systematic uncertainties
  considered correlated between the LEP experiments in the charged TGC
  combination and their effect on the combined fit results for the
  charged TGC parameters.}
\label{tab:gc_cTGC-syst}
\end{center}
\end{table}

The combination procedure~\cite{gc_bib:Alcaraz} used for the charged
TGCs allows the combination of statistical and correlated systematic
uncertainties, independently of the analysis method chosen by the
individual experiments.  The combination uses the likelihood curves
and correlated systematic errors submitted by each of the four
experiments.  The procedure is based on the introduction of an
additional free parameter to take into account the systematic
uncertainties, which are treated as shifts on the fitted TGC value,
and are assumed to have a Gaussian distribution.  A simultaneous
minimisation of both parameters, TGC and systematic error, is
performed.

In detail, the combination proceeds in the following way: the set of
measurements from the LEP experiments ALEPH, DELPHI, OPAL and L3 is
given with statistical and uncorrelated systematic uncertainties in
terms of likelihood curves: $-\log{\mathcal L}^A_{stat}(x)$,
$-\log{\mathcal L}^D_{stat}(x)$ $-\log{\mathcal L}^L_{stat}(x)$ and
$-\log{\mathcal L}^O_{stat}(x)$, respectively, where $x$ is the
coupling parameter in question.  Also given are the shifts for each of
the five totally correlated sources of uncertainty mentioned above;
each source $S$ leads to systematic errors $\sigma^S_A$, $\sigma^S_D$,
$\sigma^S_L$ and $\sigma^S_O$.

Additional parameters $\Delta^S$ are included in order to take into 
account a Gaussian distribution for each of the systematic uncertainties.
The procedure then consists in minimising the function:

\begin{eqnarray}
-\log {\mathcal L}_{total} = 
\sum_{E=A,D,L,O} \log {\mathcal L}^E_{stat} 
(x-\sum_{S}(\sigma^S_E \Delta^S))
 + \sum_{S} {\displaystyle \frac{(\Delta^S)^{2}}{2}} \\ \nonumber
\end{eqnarray}
where $x$ and $\Delta_S$ are the free parameters, and the sums run
over the four experiments $E$ and the correlated systematic errors $S$
discussed above and listed in Table~\ref{tab:gc_cTGC-syst}.  The
resulting uncertainty on $x$ takes into account all sources of
uncertainty, yielding a measurement of the coupling with the error
representing statistical and systematic sources.  The projection of
the minima of the log-likelihood as a function of $x$ gives the
combined log-likelihood curve including statistical and systematic
uncertainties.  The advantage over the scaling method used previously
is that it treats systematic uncertainties that are correlated between
the experiments correctly, while not forcing the averaging of these
systematic uncertainties into one global LEP systematic uncertainties
scaling factor. In other words, the (statistical) precision of each
experiment gets reduced by its own correlated systematic errors,
instead of an averaged LEP systematic error.  The method has been
cross-checked against the scaling method, and was found to give
comparable results.  The inclusion of the systematic uncertainties
leads to small differences, as expected by the improved treatment of
correlated systematic errors.  A similar behaviour is seen in
Monte-Carlo comparisons of these two combination methods
~\cite{gc_bib:renaud}. Furthermore, it was shown that the
minimisation-based combination method used for the charged TGCs agrees
with the method based on optimal observables, where systematic effects
are included directly in the mean values of the optimal observables
(see~\cite{gc_bib:renaud}), for any realistic ratio of statistical to
systematic uncertainties.  Further details on the combination method
can be found in~\cite{gc_bib:Alcaraz}.

In the following, single-parameter fits are presented for the TGC
parameters $\gz$, $\kg$, $\lg$, $h_i^V$ and $f_i^V$, while results
from two-parameter fits are also given for
$(f_4^{\gamma},f_4^{\mathrm{Z}})$ and
$(f_5^{\gamma},f_5^{\mathrm{Z}})$.  For results quoted in numerical
form, the one standard deviation uncertainties (68\% confidence level)
are obtained by taking the coupling values for which $\Delta\LL=+0.5$
above the minimum.  The 95\% confidence level (C.L.)  limits are given
by the coupling values for which $\Delta\LL=+1.92$ above the minimum.
For multi-parameter analyses, the two dimensional 68\%~C.L.~contour
curves for any pair of couplings are obtained by requiring
$\Delta\LL=+1.15$, while for the 95\% C.L.~contour curves
$\Delta\LL=+3.0$ is required.  Since the results on the different
parameters and parameter sets are obtained from the same data sets,
they cannot themselves be combined when looking at models establishing
additional relations between these couplings.

\section{Measurements}
\label{sec:gc_data}

The combined results presented here are obtained from charged and
neutral electroweak gauge boson coupling measurements as discussed
above.  The individual references should be consulted for details
about the data samples used.

The charged TGC analyses of ALEPH, DELPHI, L3 and OPAL use data
collected at \LEPII\ with centre-of-mass energies up to
$209~\GeV$. These analyses use different channels, typically the
semileptonic and fully hadronic W-pair decays~\cite{gc_bib:ALEPH-cTGC,
gc_bib:DELPHI-cTGC, gc_bib:L3-cTGC-1, gc_bib:L3-cTGC-2,
gc_bib:L3-cTGCsingleW, Achard:2004ji, gc_bib:OPAL-cTGC3}.  The full
data set is analysed by all four experiments.  Anomalous TGCs affect
both the total production cross-section and the shape of the
differential cross-section as a function of the polar W$^-$ production
angle.  The relative contributions of each helicity state of the W
bosons are also changed, which in turn affects the distributions of
their decay products.  The analyses presented by each experiment make
use of different combinations of each of these quantities.  In
general, however, all analyses use at least the expected variations of
the total production cross-section and the W$^-$ production
angle. Results from $\enw$ and $\nng$ production are included by some
experiments.  Single-W production is particularly sensitive to \kg,
thus providing information complementary to that from W-pair
production.

The $h$-coupling analyses of ALEPH, DELPHI and L3 use data collected
at \LepII\ with centre-of-mass energies of up to $209~\GeV$. The OPAL
measurements use the data at $189~\GeV$ only. The results of the
$f$-couplings are obtained from the whole data set above the
ZZ-production threshold by all experiments.  The experiments already
pre-combine different processes and final states for each of the
couplings.  All analyses use measurements of the total cross-sections
of Z$\gamma$ and ZZ production and the differential distributions in
the determination of the $h_i^V$ couplings~\cite{gc_bib:ALEPH-nTGC,
gc_bib:DELPHI-nTGC, gc_bib:L3-hTGC, gc_bib:OPAL-hTGC} and the $f_i^V$
couplings~\cite{gc_bib:ALEPH-nTGC, gc_bib:DELPHI-nTGC, gc_bib:L3-fTGC,
gc_bib:OPAL-fTGC}, while DELPHI also includes $Z\gamma^*$ data in the
determination of both sets of couplings.

\section{Results}

We present results from the four LEP experiments on the various
electroweak gauge boson couplings, and their combination.  The results
quoted for each individual experiment are calculated using the methods
described in Section~\ref{sec:gc_combination}.  Therefore they may
differ slightly from those reported in the individual references, as
the experiments in general use other methods to combine the data from
different channels and to include systematic uncertainties.  In
particular for the charged couplings, experiments using a combination
method based on optimal observables (ALEPH, OPAL) obtain results with
small differences compared to the values given by our combination
technique. These small differences have been studied in Monte-Carlo
tests and are well understood~\cite{gc_bib:renaud}.  For the
$h$-coupling results from OPAL and DELPHI, a slightly modified
estimate of the systematic uncertainty due to the theoretical
cross-section prediction is responsible for slightly different limits
compared to the published results.

\subsection{Charged Triple Gauge Boson Couplings}

The individual analyses and results of the experiments for the charged
couplings are described in~\cite{gc_bib:ALEPH-cTGC,
gc_bib:DELPHI-cTGC, gc_bib:L3-cTGC-1, gc_bib:L3-cTGC-2,
gc_bib:L3-cTGCsingleW, gc_bib:OPAL-cTGC3}.  The results of
single-parameter fits from each experiment are shown in
Table~\ref{tab:gc_cTGC-1-ADLO}, where the errors include both
statistical and systematic effects. The individual $\LL$ curves and
their sum are shown in Figure~\ref{fig:gc_cTGC-1}.  The results of the
combination are given in Table~\ref{tab:gc_cTGC-1-LEP}. A list of the
systematic errors treated as fully correlated between the LEP
experiments, and their shift on the combined fit result were given in
Table~\ref{tab:gc_cTGC-syst}. The combined results agree well with the
SM expectation.

\begin{table}[hp]
\begin{center}
\renewcommand{\arraystretch}{1.3}
\begin{tabular}{|l||r|r|r|r||r|} 
\hline
Parameter  & ALEPH   & DELPHI  &  L3   & OPAL  & SM \\
\hline
\hline
\gz       & $0.996\apm{0.030}{0.028}$ & $0.975\apm{0.035}{0.032}$ 
           & $0.965\apm{0.038}{0.037}$ & $0.985\apm{0.035}{0.034}$ & 1 \\ 
\hline
\kg       & $0.983\apm{0.060}{0.060}$ & $1.022\apm{0.082}{0.084}$  
           & $1.020\apm{0.075}{0.069}$ & $0.899\apm{0.090}{0.084}$ & 1 \\  
\hline
\lg        & $-0.014\apm{0.029}{0.029}$ & $0.001\apm{0.036}{0.035}$  
           & $-0.023\apm{0.042}{0.039}$ & $-0.061\apm{0.037}{0.036}$ & 0 \\
\hline
\end{tabular}
\caption[Charged TGCs]{ The measured central values and one standard
  deviation errors obtained by the four LEP experiments for the
  charged TGC parameters.  In each case the parameter listed is varied
  while the remaining two are fixed to their SM values
  (also shown). Both statistical and systematic errors are
  included. The values given here differ slightly from the ones quoted
  in the individual contributions from the four LEP experiments, as a
  different combination method is used. See text in section
  \ref{sec:gc_combination} for details. }
\label{tab:gc_cTGC-1-ADLO}
\end{center}
\end{table}

\begin{table}[hp]
\begin{center}
\renewcommand{\arraystretch}{1.3}
\begin{tabular}{|l||r|c||c|} 
\hline
Parameter  & 68\% C.L.   & 95\% C.L.  & SM    \\
\hline
\hline
$\gz$     & $+0.984\apm{0.018}{0.020} $  & [$0.946,~1.021$]  & 1 \\ 
\hline
$\kg$     & $+0.982\apm{0.042}{0.042} $  & [$0.901,~1.066$]  & 1 \\ 
\hline
$\lg$     & $-0.022\apm{0.019}{0.019} $  & [$-0.059,~0.017$] & 0 \\ 
\hline
\end{tabular}
\caption[Charged TGCs]{ The combined results for the 68\% C.L. errors
  and 95\% C.L. intervals obtained for the charged TGC parameters from
  the four LEP experiments.  In each case the parameter listed is
  varied while the other two are fixed to their SM values
  (also shown).  Both statistical and systematic errors are included.
  }
 \label{tab:gc_cTGC-1-LEP}
\end{center}
\end{table}

\begin{figure}[p]
\begin{center}
\includegraphics[width=0.9\linewidth]{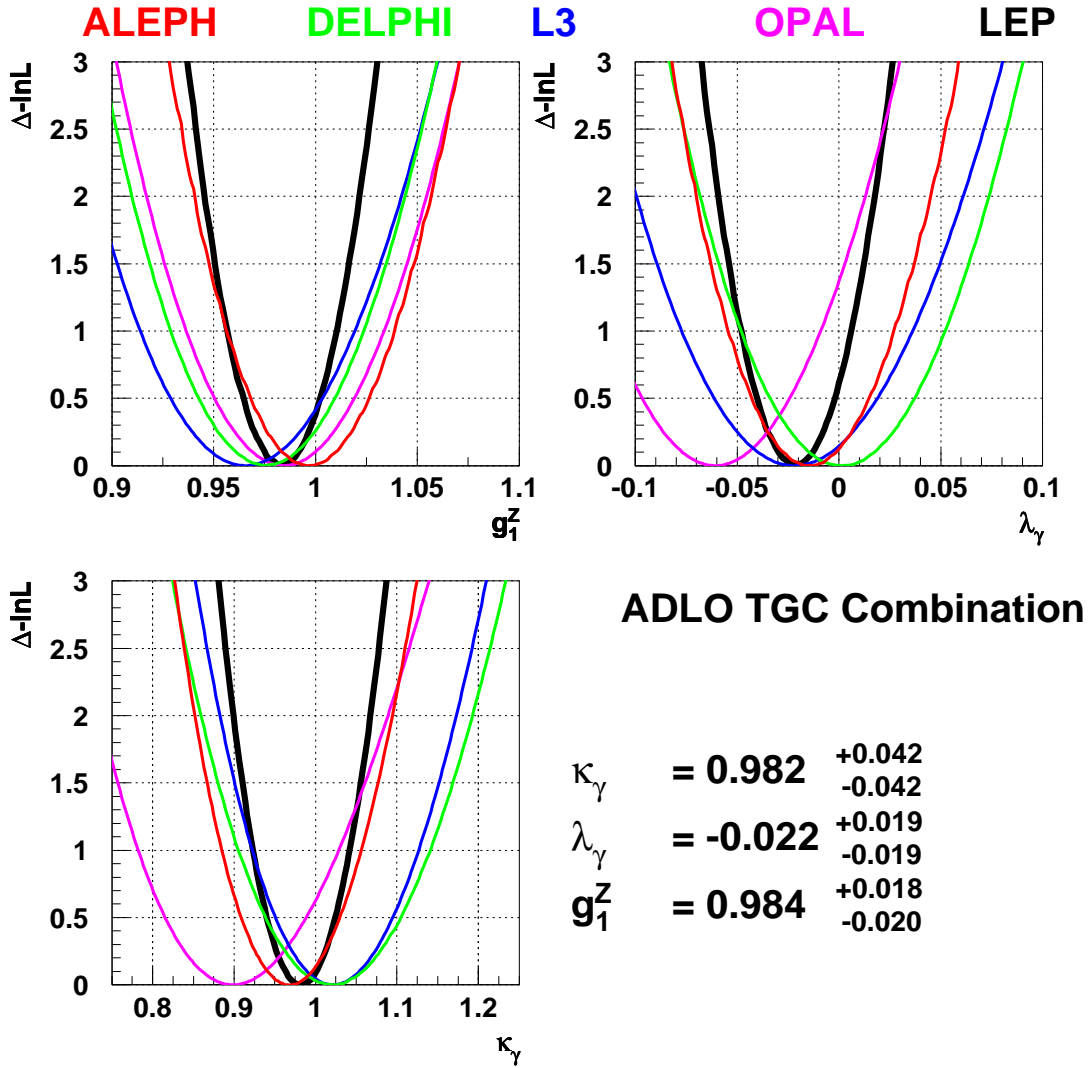}
\caption[Charged TGCs]{
  The $\LL$ curves of the four experiments (thin lines) and the LEP
  combined curve (black line) for the three charged TGCs $\gz$,
  $\kg$ and $\lg$.  In each case, the minimal $\LL$ value is subtracted.  }
\label{fig:gc_cTGC-1}
\end{center}
\end{figure}

\clearpage

\subsection{Neutral Triple Gauge Boson Couplings}

The individual analyses and results of the experiments for the
$h$-couplings are described in~\cite{gc_bib:DELPHI-nTGC,
gc_bib:L3-hTGC, gc_bib:OPAL-hTGC}.  The results from DELPHI, L3 and
OPAL, and the LEP combination, are shown in
Table~\ref{tab:gc_hTGC-1-ADLO}, where the errors include both
statistical and systematic uncertainties.  The individual $\LL$ curves
and their sum are shown in Figures~\ref{fig:gc_hgTGC-1}
and~\ref{fig:gc_hzTGC-1}.  The results agree with the SM expectation.

\begin{table}[bh]
\begin{center}
\renewcommand{\arraystretch}{1.3}
\begin{tabular}{|l||r|r|r||r|} 
\hline
Parameter  & DELPHI  &  L3  & OPAL & LEP \\
\hline
\hline
$h_1^\gamma$ & 
    [$\hOnegMinusTwoSigmaDELPHI,~\hOnegPlusTwoSigmaDELPHI$] & 
    [$\hOnegMinusTwoSigmaLThree,~\hOnegPlusTwoSigmaLThree$] & 
    [$\hOnegMinusTwoSigmaOPAL,  ~\hOnegPlusTwoSigmaOPAL$]   &
    [$\hOnegMinusTwoSigmaLEP,~\hOnegPlusTwoSigmaLEP$]       \\
\hline                             
$h_2^\gamma$ & 
& %
    [$\hTwogMinusTwoSigmaLThree,~\hTwogPlusTwoSigmaLThree$] & 
    [$\hTwogMinusTwoSigmaOPAL,  ~\hTwogPlusTwoSigmaOPAL$]   &
    [$\hTwogMinusTwoSigmaLEP,~\hTwogPlusTwoSigmaLEP$]       \\ 
\hline                             
$h_3^\gamma$ & 
    [$\hThreegMinusTwoSigmaDELPHI,~\hThreegPlusTwoSigmaDELPHI$] & 
    [$\hThreegMinusTwoSigmaLThree,~\hThreegPlusTwoSigmaLThree$] & 
    [$\hThreegMinusTwoSigmaOPAL,  ~\hThreegPlusTwoSigmaOPAL$]   &
    [$\hThreegMinusTwoSigmaLEP,~\hThreegPlusTwoSigmaLEP$]    \\ 
\hline                             
$h_4^\gamma$ & 
& %
    [$\hFourgMinusTwoSigmaLThree,~\hFourgPlusTwoSigmaLThree$] & 
    [$\hFourgMinusTwoSigmaOPAL,  ~\hFourgPlusTwoSigmaOPAL$]   &
    [$\hFourgMinusTwoSigmaLEP,~\hFourgPlusTwoSigmaLEP$]       \\ 
\hline                             
$h_1^Z$ & 
    [$\hOnezMinusTwoSigmaDELPHI,~\hOnezPlusTwoSigmaDELPHI$] & 
    [$\hOnezMinusTwoSigmaLThree,~\hOnezPlusTwoSigmaLThree$] & 
    [$\hOnezMinusTwoSigmaOPAL,  ~\hOnezPlusTwoSigmaOPAL$]   &
    [$\hOnezMinusTwoSigmaLEP,~\hOnezPlusTwoSigmaLEP$]       \\ 
\hline                             
$h_2^Z$ & 
& %
    [$\hTwozMinusTwoSigmaLThree,~\hTwozPlusTwoSigmaLThree$] & 
    [$\hTwozMinusTwoSigmaOPAL,  ~\hTwozPlusTwoSigmaOPAL$]   &
    [$\hTwozMinusTwoSigmaLEP,~\hTwozPlusTwoSigmaLEP$]       \\
\hline                             
$h_3^Z$ & 
    [$\hThreezMinusTwoSigmaDELPHI,~\hThreezPlusTwoSigmaDELPHI$] & 
    [$\hThreezMinusTwoSigmaLThree,~\hThreezPlusTwoSigmaLThree$] & 
    [$\hThreezMinusTwoSigmaOPAL,  ~\hThreezPlusTwoSigmaOPAL$]   &
    [$\hThreezMinusTwoSigmaLEP,~\hThreezPlusTwoSigmaLEP$]       \\ 
\hline                             
$h_4^Z$ & 
& %
    [$\hFourzMinusTwoSigmaLThree,~\hFourzPlusTwoSigmaLThree$] & 
    [$\hFourzMinusTwoSigmaOPAL,  ~\hFourzPlusTwoSigmaOPAL$]   &
    [$\hFourzMinusTwoSigmaLEP,~\hFourzPlusTwoSigmaLEP$]       \\ 
\hline
\end{tabular}
\caption[Neutral TGCs]{The 95\% C.L. intervals ($\Delta\LL=1.92$) in
  the neutral TGC parameters $h_i^V$ measured by the DELPHI, L3 and
  OPAL, and the LEP combined values.  In each case the parameter
  listed is varied while the remaining ones are fixed to their
  SM values ($h_i^V=0$).  Both statistical and systematic
  uncertainties are included. DELPHI did not interpret its
  measurements in terms of neutral gauge couplings of dimension 8
  operators, hence does not enter in the combination for $h_{2/4}^V$.
}
\label{tab:gc_hTGC-1-ADLO}
\end{center}
\end{table}

\begin{figure}[htbp]
\begin{center}
\includegraphics[width=\linewidth]{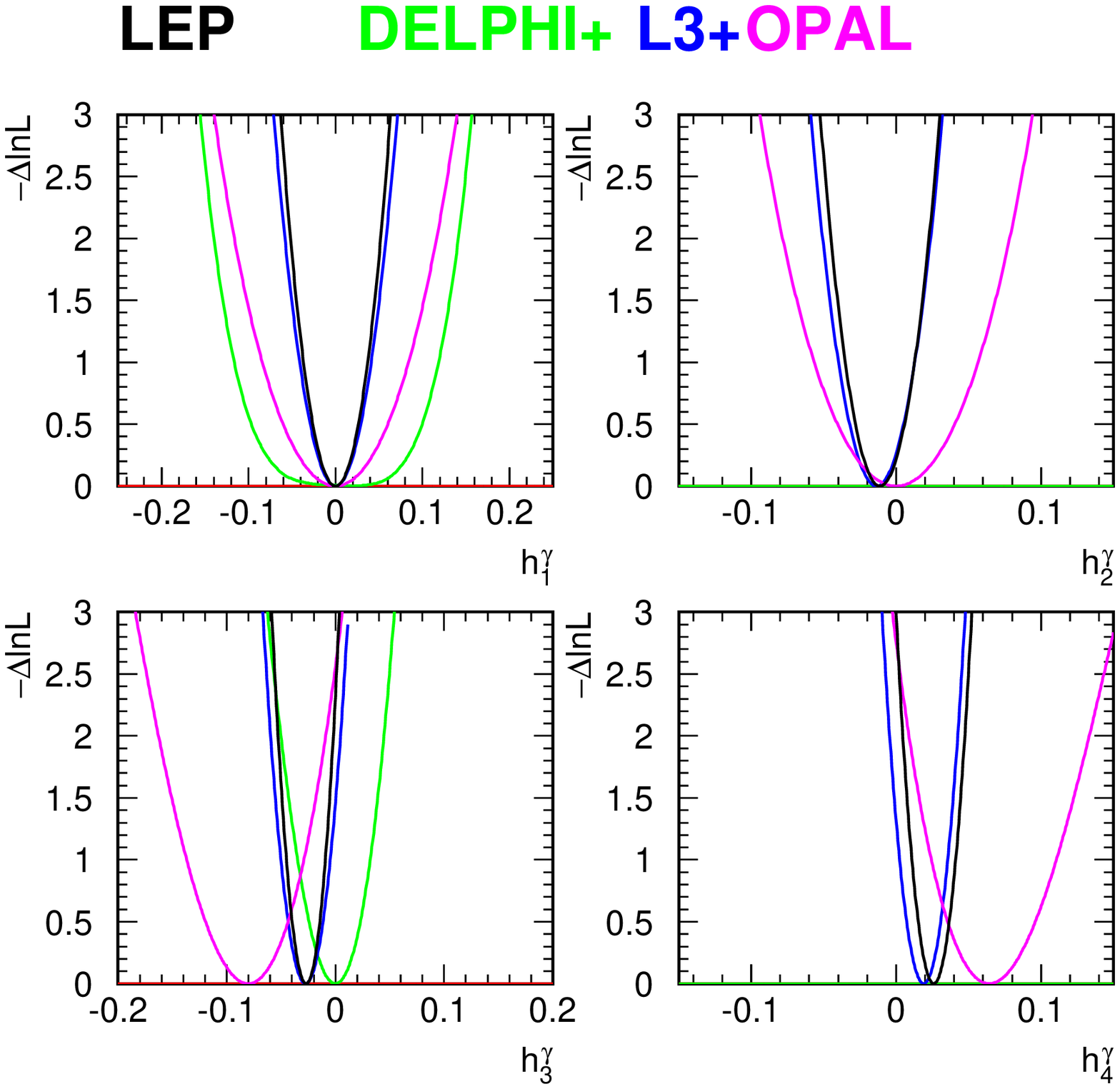}
\caption[Neutral TGCs]{ The $\LL$ curves of DELPHI, L3, and OPAL
  experiments and the LEP combined curve for the four neutral TGCs
  $h_i^\gamma,~i=1,2,3,4$. In each case, the minimal value is
  subtracted. Note, DELPHI did not interpret its measurements in terms
  of neutral gauge couplings of dimension 8 operators, hence does not
  enter in the combination for $h_{2/4}^V$.
}
\label{fig:gc_hgTGC-1}
\end{center}
\end{figure}

\begin{figure}[htbp]
\begin{center}
\includegraphics[width=\linewidth]{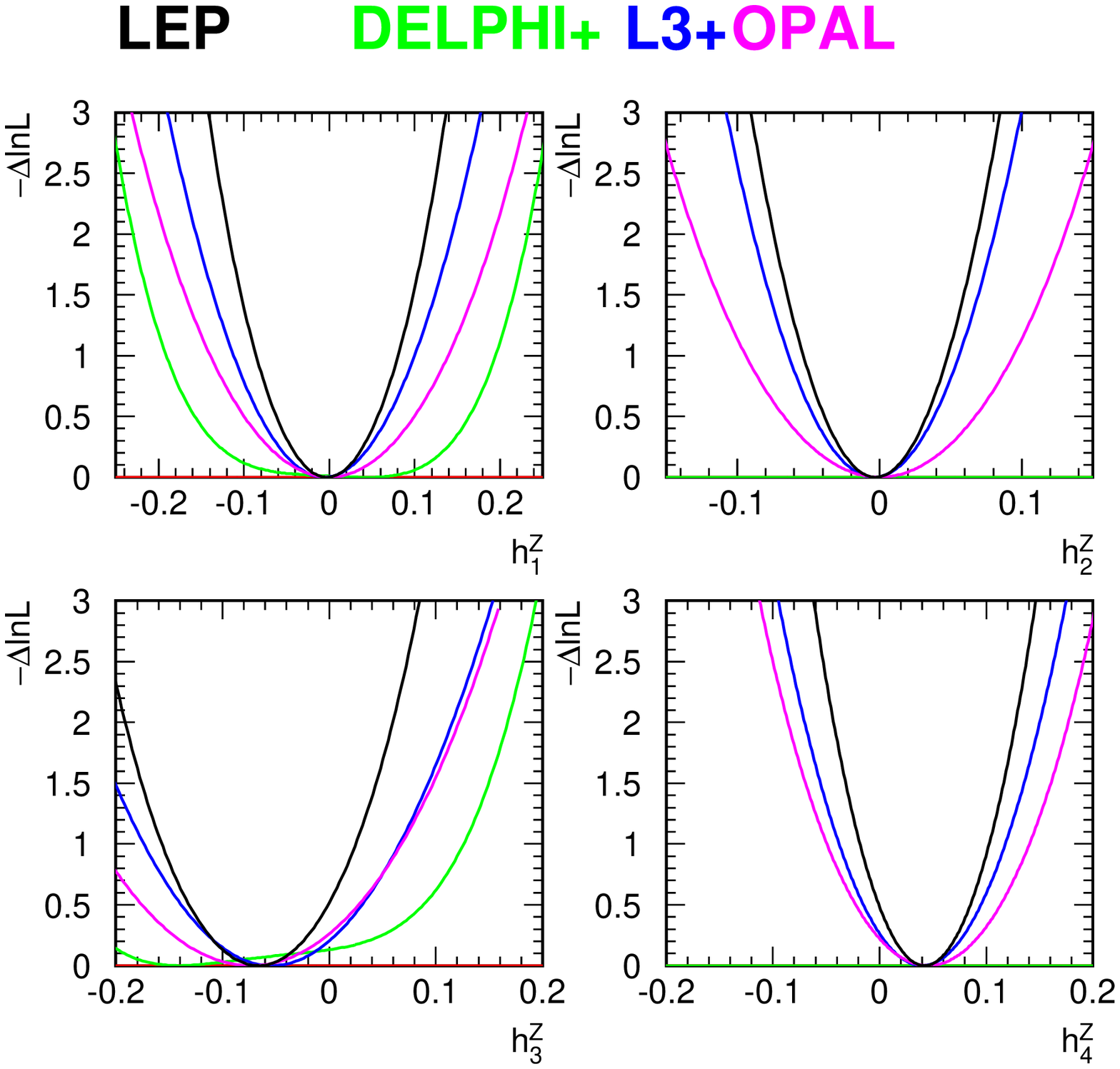}
\caption[Neutral TGCs]{ The $\LL$ curves of the DELPHI, L3, OPAL
  experiments and the LEP combined curve for the four neutral TGCs
  $h_i^Z,~i=1,2,3,4$.  In each case, the minimal value is subtracted.
  Note, DELPHI did not interpret its measurements in terms of neutral
  gauge couplings of dimension 8 operators, hence does not enter in
  the combination for $h_{2/4}^V$.
}
\label{fig:gc_hzTGC-1}
\end{center}
\end{figure}

The individual analyses and results of the experiments for the
$f$-couplings are described
in~\cite{gc_bib:ALEPH-nTGC,gc_bib:DELPHI-nTGC,gc_bib:L3-fTGC,gc_bib:OPAL-fTGC}.
The single-parameter results for each experiment and the LEP
combination are shown in Table~\ref{tab:gc_fTGC-1-ADLO}, where the
errors include both statistical and systematic uncertainties.  The
individual $\LL$ curves and their sum are shown in
Figure~\ref{fig:gc_fTGC-1}.  Three experiments, ALEPH, L3 and OPAL,
contributed data to two-parameter fits to the TGC pairs
$(f_4^{\gamma}, f_4^Z)$ and $(f_5^{\gamma}, f_5^Z)$.  The
two-parameter results including the LEP combination are shown in
Table~\ref{tab:gc_fTGC-2-ADLO}, where the errors include both
statistical and systematic uncertainties.  The 68\% C.L. and 95\%
C.L. contour curves resulting from the combinations of the
two-dimensional likelihood curves are shown in
Figures~\ref{fig:gc_fTGC-4} and~\ref{fig:gc_fTGC-5}.  The couplings
agree with the SM expectation.

\begin{table}[htbp]
\begin{center}
\renewcommand{\arraystretch}{1.3}
\begin{tabular}{|l||r|r|r|r||r|} 
\hline
Parameter  & ALEPH & DELPHI  &  L3   & OPAL & LEP \\
\hline
\hline
$f_4^\gamma$ & 
    [$\fFourgMinusTwoSigmaALEPH,~\fFourgPlusTwoSigmaALEPH$] & 
    [$\fFourgMinusTwoSigmaDELPHI,~\fFourgPlusTwoSigmaDELPHI$] & 
    [$\fFourgMinusTwoSigmaLThree,~\fFourgPlusTwoSigmaLThree$] & 
    [$\fFourgMinusTwoSigmaOPAL,~\fFourgPlusTwoSigmaOPAL$] &
    [$\fFourgMinusTwoSigmaLEP,~\fFourgPlusTwoSigmaLEP$]  \\ 
\hline
$f_4^Z$ & 
    [$\fFourzMinusTwoSigmaALEPH,~\fFourzPlusTwoSigmaALEPH$] & 
    [$\fFourzMinusTwoSigmaDELPHI,~\fFourzPlusTwoSigmaDELPHI$] & 
    [$\fFourzMinusTwoSigmaLThree,~\fFourzPlusTwoSigmaLThree$] & 
    [$\fFourzMinusTwoSigmaOPAL,~\fFourzPlusTwoSigmaOPAL$] &
    [$\fFourzMinusTwoSigmaLEP,~\fFourzPlusTwoSigmaLEP$]  \\ 
\hline
$f_5^\gamma$ & 
    [$\fFivegMinusTwoSigmaALEPH,~\fFivegPlusTwoSigmaALEPH$] & 
    [$\fFivegMinusTwoSigmaDELPHI,~\fFivegPlusTwoSigmaDELPHI$] & 
    [$\fFivegMinusTwoSigmaLThree,~\fFivegPlusTwoSigmaLThree$] & 
    [$\fFivegMinusTwoSigmaOPAL,~\fFivegPlusTwoSigmaOPAL$] &
    [$\fFivegMinusTwoSigmaLEP,~\fFivegPlusTwoSigmaLEP$]  \\ 
\hline
$f_5^Z$ & 
    [$\fFivezMinusTwoSigmaALEPH,~\fFivezPlusTwoSigmaALEPH$] & 
    [$\fFivezMinusTwoSigmaDELPHI,~\fFivezPlusTwoSigmaDELPHI$] & 
    [$\fFivezMinusTwoSigmaLThree,~\fFivezPlusTwoSigmaLThree$] & 
    [$\fFivezMinusTwoSigmaOPAL,~\fFivezPlusTwoSigmaOPAL$] &
    [$\fFivezMinusTwoSigmaLEP,~\fFivezPlusTwoSigmaLEP$]  \\ 
\hline
\end{tabular}
\caption[Neutral TGCs]{The 95\% C.L. intervals ($\Delta\LL=1.92$) in
  the neutral TGC parameters $f_i^V$ measured by ALEPH, DELPHI, L3 and
  OPAL, and the LEP combined values.  In each case the parameter
  listed is varied while the remaining ones are fixed to their
  SM values ($f_i^V=0$).  Both statistical and systematic
  uncertainties are included.  }
\label{tab:gc_fTGC-1-ADLO}
\end{center}
\end{table}

\begin{table}[htbp]
\begin{center}
\renewcommand{\arraystretch}{1.3}
\begin{tabular}{|l||r|r|r||r|rr|} 
\hline
Parameter  & ALEPH &  L3   & OPAL & LEP & \multicolumn{2}{|c|}{Correlations} \\
\hline
\hline
$f_4^\gamma$ & 
    [$\fFourgMinusTwoSigmaALEPHTwoD,~\fFourgPlusTwoSigmaALEPHTwoD$] & 
    [$\fFourgMinusTwoSigmaLThreeTwoD,~\fFourgPlusTwoSigmaLThreeTwoD$] & 
    [$\fFourgMinusTwoSigmaOPALTwoD,~\fFourgPlusTwoSigmaOPALTwoD$] &
    [$\fFourgMinusTwoSigmaLEPTwoD,~\fFourgPlusTwoSigmaLEPTwoD$] 
    & $\fFourgfFourg$& $\fFourgfFourz$  \\ 
$f_4^Z$ & 
    [$\fFourzMinusTwoSigmaALEPHTwoD,~\fFourzPlusTwoSigmaALEPHTwoD$] & 
    [$\fFourzMinusTwoSigmaLThreeTwoD,~\fFourzPlusTwoSigmaLThreeTwoD$] & 
    [$\fFourzMinusTwoSigmaOPALTwoD,~\fFourzPlusTwoSigmaOPALTwoD$] &
    [$\fFourzMinusTwoSigmaLEPTwoD,~\fFourzPlusTwoSigmaLEPTwoD$]  
    & $\fFourgfFourz$& $\fFourgfFourg$  \\ 
\hline                                              
$f_5^\gamma$ & 
    [$\fFivegMinusTwoSigmaALEPHTwoD,~\fFivegPlusTwoSigmaALEPHTwoD$] & 
    [$\fFivegMinusTwoSigmaLThreeTwoD,~\fFivegPlusTwoSigmaLThreeTwoD$] & 
    [$\fFivegMinusTwoSigmaOPALTwoD,~\fFivegPlusTwoSigmaOPALTwoD$] &
    [$\fFivegMinusTwoSigmaLEPTwoD,~\fFivegPlusTwoSigmaLEPTwoD$]   
    & $\fFivegfFiveg$& $\fFivegfFivez$  \\ 
$f_5^Z$ & 
    [$\fFivezMinusTwoSigmaALEPHTwoD,~\fFivezPlusTwoSigmaALEPHTwoD$] & 
    [$\fFivezMinusTwoSigmaLThreeTwoD,~\fFivezPlusTwoSigmaLThreeTwoD$] & 
    [$\fFivezMinusTwoSigmaOPALTwoD,~\fFivezPlusTwoSigmaOPALTwoD$] &
    [$\fFivezMinusTwoSigmaLEPTwoD,~\fFivezPlusTwoSigmaLEPTwoD$]  
    & $\fFivegfFivez$& $\fFivegfFiveg$  \\ 
\hline
\end{tabular}
\caption[Neutral TGCs]{The 95\% C.L. intervals ($\Delta\LL=1.92$) in
  the neutral TGC parameters $f_i^V$ in two-parameter fits measured by
  ALEPH, L3 and OPAL, and the LEP combined values.  In each case the
  two parameters listed are varied while the remaining ones are fixed
  to their SM values ($f_i^V=0$).  Both statistical and
  systematic uncertainties are included. Since the shape of the
  log-likelihood is not parabolic, there is some ambiguity in the
  definition of the correlation coefficients and the values quoted
  here are approximate.  }
\label{tab:gc_fTGC-2-ADLO}
\end{center}
\end{table}

\begin{figure}[htbp]
\begin{center}
\includegraphics[width=\linewidth]{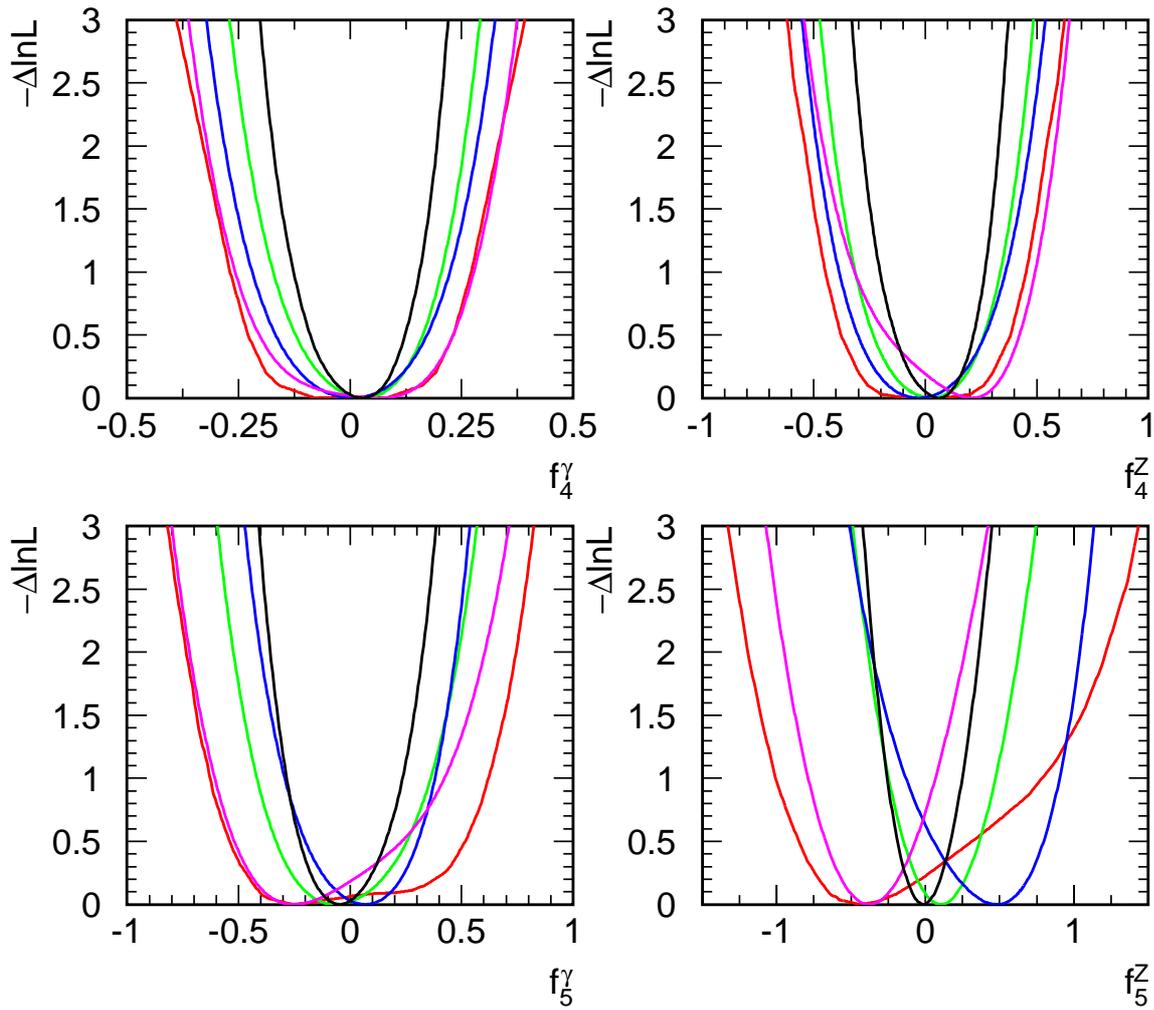}
\caption[Neutral TGCs]{
  The $\LL$ curves of the four experiments, and the LEP combined curve
  for the four neutral TGCs $f_i^V,~V=\gamma,Z,~i=4,5$.  In each case,
  the minimal value is subtracted.  }
\label{fig:gc_fTGC-1}
\end{center}
\end{figure}

\begin{figure}[htbp]
\begin{center}
\includegraphics[width=0.53\linewidth]{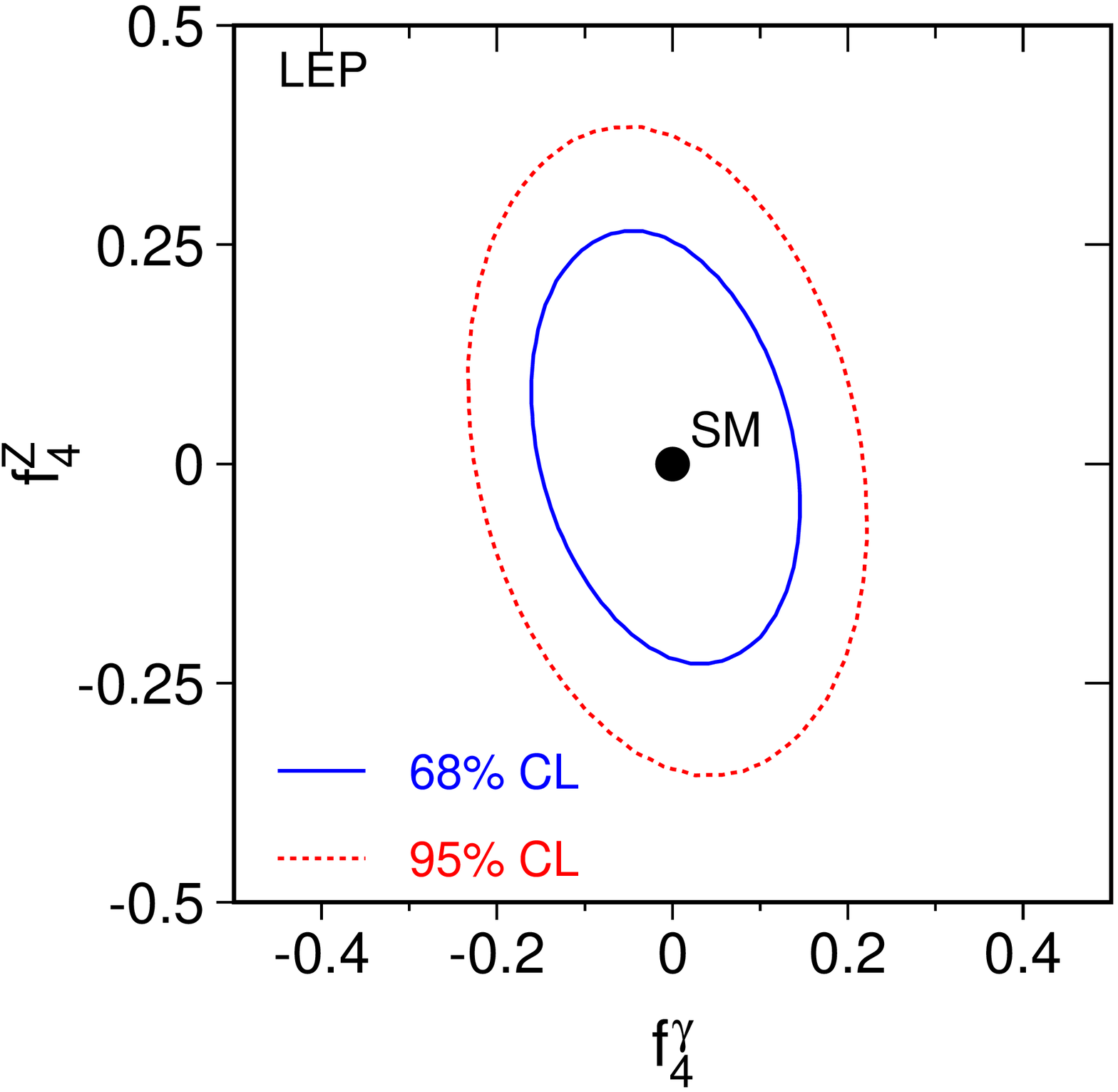}
\caption[Neutral TGCs]{ Contour curves of 68\% C.L. and 95\% C.L. in
  the plane of the neutral TGC parameters $(f_4^\gamma,f_4^Z)$ showing
  the LEP combined result to which ALEPH, L3 and OPAL contributed.
  }
\label{fig:gc_fTGC-4}
\end{center}
\end{figure}

\begin{figure}[htbp]
\begin{center}
\includegraphics[width=0.53\linewidth]{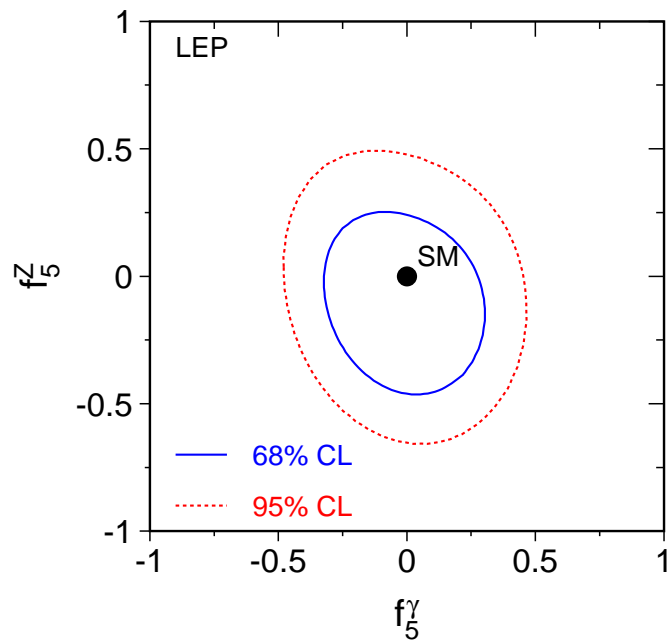}
\caption[Neutral TGCs]{ Contour curves of 68\% C.L. and 95\% C.L. in
  the plane of the neutral TGC parameters $(f_5^\gamma,f_5^Z)$ showing
  the LEP combined result to which ALEPH, L3 and OPAL contributed.
  }
\label{fig:gc_fTGC-5}
\end{center}
\end{figure}

\clearpage

\section{Summary and Conclusions}

Combinations of charged and neutral triple gauge boson couplings
were made, based on results from the four LEP experiments ALEPH,
DELPHI, L3 and OPAL.  No deviation from the SM prediction
is seen for any of the electroweak gauge boson couplings studied.
While the existence of charged TGCs was experimentally verified
already early on by the measurement of the total WW cross-section, see
also Chapter~\ref{chap:4f}, their values have now been measured with
an accuracy of 0.02 to 0.04, and found to be in agreement with the SM
expectation.  As an example, these data allow the Kaluza-Klein
theory~\cite{gc_bib:klein}, in which $\kg = -2$, to be
excluded~\cite{gc_bib:maiani}.  No evidence of the existence of
neutral TGCs are found, limiting their magnitude to less than 0.05 to
0.35 depending on coupling.

\chapter{Mass and Width of the W~Boson}
\label{chap:mw}

\section{Introduction}

The mass of the W~boson is a fundamental parameter in particle
physics.  Together with the Z-boson mass, it sets the energy scale of
electroweak symmetry breaking.  Both masses are closely related to the
weak mixing angle. At LEP, the W-boson mass is determined by measuring
the cross-section of W-boson pairs at the production threshold, from
the leptonic decay spectrum of the W boson, and by directly
reconstructing W~boson decays. The latter method is the more precise
one. It also allows a determination of the total decay width of the
W~boson.  Direct measurements of W-boson mass and width are also
performed at the Tevatron $\mathrm{p\bar{p}}$ collider~\cite{CDF2GW,
D02GW, CDF2MWPRL2012, D0-MW:PRL2012}.

\section{Determination of the W Mass at the W-Pair Production Threshold}
\label{sec:mw:threshold}

The SM cross-section of the reaction $\eeWW$ shows a typical threshold
behaviour close to a centre-of-mass energy that corresponds to twice
the W mass. In the threshold region the cross-section rises in
proportion to the velocity of the W bosons produced, which is
approximately given by $\beta=\sqrt{1-4\MW^2/s}$, neglecting radiative
corrections and finite width effects.  Thus, a measurement of the
production cross-section at a given centre-of-mass energy is directly
related to the W~boson mass.  The intrinsic precision of this method
is similar to the direct-reconstruction method, described below.
However, since LEP predominantly operated at higher centre-of-mass
energies in order to search for new physics as well as to make precise
electroweak measurements, the data collected at threshold energies
corresponds to only 3\% of the full data set (see
Table~\ref{intro:tab:lumi}).

Using Monte-Carlo simulations, the centre-of-mass energy where the
cross-section is most sensitive to $\MW$ was determined to be
$\sqrt{s}=161~\GeV$, but data at 172-183~$\GeV$ were also analysed to
extract $\MW$ from the measured cross-section.  Each LEP experiment
compared the measured cross-sections at each centre-of-mass energy to
the $\MW$ dependent SM prediction calculated using the GENTLE
program~\cite{bib:GentleV2}.  The results of the four LEP experiments
combined for the different centre-of-mass
energies~\cite{bib:mw:a-threshold2, bib:mw:d-mw,
  bib:mw:l-threshold, bib:mw:o-threshold} are shown in
Table~\ref{mw:tab:wmass-threshold}.  Owing to the dependence of the
theory cross-section on the mass for a given centre-of-mass energy,
both the extracted mass and its uncertainty decrease with increasing
measured cross-sections.

\begin{table}[htbp]
\begin{center}
\renewcommand{\arraystretch}{1.25}
\begin{tabular}{|l||c|}
\hline
\multicolumn{2}{|c|}{Threshold Analysis} \\
\hline
 Experiment &   \MW [\GeV]     \\ 
\hline
\hline
   \Aleph         & $80.20\pm 0.34$         \\ %
   \Delphi        & $80.45^{+0.45}_{-0.41}$    \\
   \Ltre          & $80.78^{+0.48}_{-0.42}$    \\
   \Opal          & $80.40^{+0.46}_{-0.43}$    \\ \hline
\end{tabular}
\caption[Results on the W mass using the cross-section]{ W mass
  measurements from the $\WW$ threshold cross-section at
  $\roots=161-183~\GeV$~\cite{bib:mw:a-threshold2, bib:mw:d-mw,
    bib:mw:l-threshold, bib:mw:o-threshold}. The uncertainties include
  statistical and systematic contributions.}
\label{mw:tab:wmass-threshold}
\end{center}
\end{table} 

Systematic uncertainties from hadronisation and fragmentation effects
in hadronically decaying W bosons, radiative corrections, final-state
interactions are all negligible compared to the statistical
uncertainty of the measurement. Combining all LEP W-pair threshold
data yields:

\begin{equation}
  \MW(\mbox{threshold})=80.42\pm 0.20 \pm 0.03 (E_\mathrm{LEP})~\GeV\;,
\end{equation}
where the uncertainty due to the LEP centre-of-mass
energy~\cite{bib:mw:lep-energy-1, *bib:mw:lep-energy-2} is given
separately.  The treatment of systematic uncertainties is further
detailed below.

\section{Measurement of Mass and Width by Direct Reconstruction}

\subsection{Mass Reconstruction} 

The mass and total decay width of the W~boson is determined with high
precision by reconstructing directly the decay products of the two
W~bosons, mainly in the fully hadronic, $\WWtoqqqq$, and
semi-leptonic, $\WWtoqqlv$, decay channels.

The $\WWtolvlv$ decay also contains information on $\MW$ when
analysing the leptonic energy spectrum or reconstructing an
approximated mass of the decaying W~bosons, as performed by the \Opal\
collaboration~\cite{bib:mw:opal-lvlv}.  However, the intrinsic
statistical precision dominates the total uncertainty and \Opal\
determines a value of:

\begin{equation}
  \MW(\lv\lv)=80.41\pm 0.41 (\mbox{stat.}) \pm 0.13 (\mbox{syst.})~\GeV\;,
\end{equation}
analysing data at centre-of-mass energies between $183~\GeV$ and
$209~\GeV$. It is interesting to compare this result with those from
the other decay channels, since systematic uncertainties from hadronic
W decays are absent. Within the given precision it agrees well with
the W mass measurements in $\WWtoqqqq$ and $\WWtoqqlv$ events
discussed below.  For the purpose of the LEP combination, \Opal\
combines the measurements in the fully leptonic channel at each run
period with the semi-leptonic results.

The $\WWtoqqqq$ decays are reconstructed from hadronic jets observed
in the final state, formed from measured particle tracks and energy
depositions in the calorimeters. Different jet clustering algorithms
are applied, \eg, the Durham~\cite{bib:mw:durham-1, *bib:mw:durham-2,
*bib:mw:durham-3, *bib:mw:durham-3e}, Diclus~\cite{bib:mw:diclus}, and
Cambridge~\cite{bib:mw:cambridge-1, *bib:mw:cambridge-2} algorithms.
Depending on the choice of clustering parameters, additional gluon
radiation may be resolved, so that not only pairs of jets, but also
five-jet topologies are reconstructed. Similarly, initial-state photon
radiation (ISR) and final-state photon radiation (FSR) may be detected
by a calorimetric cluster consistent with an electromagnetic shower
shape and without a matched track in a given angular cone around the
photon candidate.  Such reconstruction methods improve the detailed
knowledge of the event kinematics and therefore the resolution in the
reconstructed masses of the decaying W bosons. The correct
reconstruction of the fully hadronic final state is further
complicated by combinatorial ambiguities to pair the reconstructed
jets to the W decays. In case of four jets there are three possible
combinations. For five-jet topologies this number increases to 15. The
ambiguity is treated differently by the four LEP experiments. \Aleph\
selects only one combination in their analyses, using a pairing
probability that is based on the CC03 matrix element evaluated for the
reconstructed jets~\cite{bib:mw:a-mw}. The other experiments use a
W-mass estimator which combines all pairings that have a high
probability to be correct~\cite{bib:mw:d-mw, bib:mw:l-mw,
bib:mw:o-mw}. The pairings are weighted accordingly in the combined
mass likelihood. In this way, a maximum of information is retained for
the subsequent mass extraction method. For \Delphi, the weights are
based on the polar angle of the reconstructed W boson, the sum of jet
charges of each jet combination and the transverse momentum of the
gluon jet in five-jet events~\cite{bib:mw:d-mw}.  \Ltre\ exploits the
probability of a kinematic fit~\cite{bib:mw:l-mw}, while \Opal\ uses a
neural network trained with the above-mentioned variables and the
reconstructed mass differences of the W bosons~\cite{bib:mw:o-mw}. The
fully hadronic data samples are furthermore separated into 4-jet and
5-jet sub-samples (\Ltre), or all possible jet configurations, also
with different clustering schemes, and properly weighted in the final
$\MW$ and $\GW$ analysis.

Semi-leptonic W-pair decays, $\WW\to\qq\ev$, $\WW\to\qq\mv$ and
$\WW\to\qq\tv$, are reconstructed as a pair of hadronic jets, possibly
with a third jet from gluon radiation, and an isolated electron, muon
or tau lepton. Photons from initial state radiation are detected in
about 5\% of the events and excluded from the jet clustering. The mass
of the hadronically decaying W is determined directly from the jet
system. In the leptonic $\mathrm{W}\to\ev$ and $\mathrm{W}\to\mv$
decays, the missing momentum vector is calculated applying total
momentum conservation and is assigned to the momentum of the
neutrino. The masses of both W~decays can thus be reconstructed. In
case of the $\qq\tv$ final state, only the hadronically decaying W
contains useable W-mass information due to the presence of a second
neutrino from the tau decay.

\subsection{Kinematic Fitting}

The di-jet mass resolution is mainly determined by the precision of
jet energy measurements.  The jet energy is carried by charged
particles ($\sim 62$\% on average), photons ($\sim 27$\%) and neutral
hadrons ($\sim 10$\%), which are measured using the tracking and
calorimetric devices of the detectors.  Even with the help of
sophisticated energy-flow algorithms which combine tracks and
calorimetric clusters in order to reduce effects of double counting of
particles, the best jet energy resolutions achieved are typically
$\Delta E/E \approx 60-80\%/\sqrt{E/\mathrm{GeV}}$.  The corresponding
di-jet mass resolution for W-boson decays is in the order of
$8-9~\GeV$.

The mass resolution is substantially improved by imposing the
constraint that the total energy in the event should equal the known
LEP centre-of-mass energy~\cite{bib:mw:lep-energy-1,
*bib:mw:lep-energy-2}, or that the energy of each W boson should be
equal to the LEP beam energy. In practice, this is most commonly
implemented by means of a kinematic fit. In such a fit, the measured
parameters of the jets and leptons are adjusted, taking account of
their measurement uncertainties in such a way as to satisfy the
constraints of energy and momentum conservation. In case of hadronic
jets, the jet three-momenta are varied while keeping the jet velocity
constant, as systematic effects cancel in the ratio of jet momentum
and jet energy. For leptons, the energy for electrons and momentum for
muons, together with the polar and azimuthal angles, are considered in
the fit. The lepton masses are set to their nominal values. For
$\qq\tv$ final states, an energy rescaling of the hadronic system to
the beam energy is practically equivalent to a kinematic fit, due to
the lack of further kinematic constraints.

In the $\qq\qq$ case, the improved kinematic reconstruction is
referred to as a 4C fit, because there are four energy and momentum
constraints.  In the $\qq\mathrm{e}\nu$ and $\qq\mv$ channels it is
referred to as a 1C or one constraint fit, because the three momentum
components of the neutrino have to be determined, eliminating three of
the constraints. It is often useful to impose the additional
constraint that the masses of the two W bosons are equal, leading to a
5C or 2C fit, in which case the kinematic fit provides a single
estimate of the average W mass in each event.  Although the equal-mass
assumption is not fulfilled in an individual event, it is valid on
average. Since the intrinsic total width of the W is much smaller than
the mass resolution, the equal-mass assumption further improves the
mass resolution.  The corresponding probabilities of fits in terms of
a $\chi^2$ variable are used to reject background and to resolve
combinatorial ambiguities in the $\qq\qq$ channel.

The resolution on the W-boson mass varies slightly from experiment to
experiment.  Typical values\footnote{The resolutions quoted here are
estimated from the distributions of the difference between the fitted
W mass and the average of the two true W masses in each event. These
resolution functions are not Gaussian, and the values quoted represent
RMS values computed in a range $\pm 10~\GeV$ around zero.  In order to
estimate the intrinsic mass resolution, events with significant ISR
are excluded, and Monte-Carlo information is used to identify the
correct jet-pairings in the $\qq\qq$ channel.}, after use of kinematic
fitting, are $2.5~\GeV$ for the $\WW\to\qq\ev$ and $\WW\to\qq\mv$
channels, $3.1~\GeV$ for the $\WW\to\qq\tv$ channel and $1.5~\GeV$ for
the $\WWtoqqqq$ channel, at $\sqrt{s}=189~\GeV$.  These resolutions
increase to $2.9~\GeV$, $3.4~\GeV$ and $1.7~\GeV$, respectively, at
$\sqrt{s}=207~\GeV$.

The use of a kinematic fit or an equivalent kinematic constraint
implies that the scale of the W mass measurement is directly linked to
the knowledge of the LEP beam energy.  Checks on the determination of
the LEP energy are discussed in Appendix~\ref{mw:zret:appendix}.  It
should also be emphasised that the kinematic fit technique neglects
the effects of initial-state radiation (ISR) if it is not measured
directly in the detector.  The average energy radiated in ISR in
$\ee\to\WW$ events is $2.2~\GeV$ at $\sqrt{s}=189~\GeV$, rising to
$\sim 3.5~\GeV$ at $\sqrt{s}=207~\GeV$, which is substantially smaller
than the intrinsic resolution of the jet energies and hence of the W
mass, and therefore cannot be resolved by kinematic fitting.  Any
remaining bias due to unmeasured ISR photons is taken into account in
the W mass and width extraction methods based on MC simulations of
radiative effects.

\subsection{Techniques for Determining the W-Boson Mass and Width}

In the direct reconstruction method, the mass of the W boson is
obtained by comparing data to simulated $\ee\to\WW$ event samples
generated with known values of $\MW$ and $\GW$, in order to obtain
those which describe the data best. These Monte-Carlo samples are of
large statistics, typically $10^6$ events. Since the generation of
event samples for all possible parameter values is very computing time
intensive, different methods are used to perform the $\MW$ and $\GW$
extraction in a more efficient, but still precise way.

The Monte-Carlo simulation programs used to generate the signal
process, \KandY~\cite{4f_bib:kandy}, \RacoonWW~\cite{\RACOONWWref},
and \Wphact~\cite{\WPHACTref}, include all relevant diagrams leading
to the same 4-fermion final state and full $O(\alpha)$ electroweak
radiative corrections.  Real ISR photons are calculated in
$O(\alpha^3)$, and FSR photons to higher order leading-log
approximation. The underlying mass and width of the W boson are
defined using a relativistic Breit-Wigner propagator with
$s$-dependent width which is also the convention adopted to quote the
measured values.  Tau decays are simulated using the
\Tauola~\cite{TAUOLA-1, *TAUOLA-2} package. The fragmentation and
hadronisation of quark and gluon jets is described by the
\Jetset~\cite{JETSET}, \Herwig~\cite{HERWIG6}, and
\Ariadne~\cite{ARIADNE} programs, which are compared to estimate the
corresponding systematic uncertainties. The default fragmentation
parameters exclude any FSI effects from Bose-Einstein correlations
(BEC) or colour reconnection (CR). For the latter, a dedicated
procedure is developed to suppress mass biases in the $\WWtoqqqq$
channel, which is detailed below.

The background, mainly from $\ee\to\qq(\gamma)$ with additional gluon
radiation and pair production of Z bosons, amounts to $2-15$\% in the
$\qq\lv$ channels, depending on the selected $\WW$ final state, and
about 30\% in the $\qq\qq$ channel. The background is simulated using
Monte-Carlo programs which include radiative corrections with higher
order ISR and FSR. Dedicated control samples of 2-fermion and
4-fermion events are studied by the LEP experiments to ensure the
agreement of the Monte-Carlo simulations with data, concerning jet and
lepton resolutions, event shape variables, and detector response.  Any
remaining differences are taken into account as systematic
uncertainties.

The methods that are applied to extract the W mass and width results
are based on unbinned maximum likelihood fits to the measured
data. Different procedures are employed to construct the likelihood
functions and to describe their dependence on the underlying $\MW$ and
$\GW$ values.  For the final results, \Aleph\ and \Ltre\ apply a
reweighting method, while \Opal\ and \Delphi\ use a convolution
technique.  The \Opal\ collaboration also performs fits of an
analytical description of the Breit-Wigner resonance curves and
background shapes to data, in order to access systematic uncertainties
of the mass and width extraction method.  Since the W-boson width,
$\GW$, depends on the mass $\MW$, the SM dependence of $\GW$ on $\MW$
is assumed when performing the fit to the data to determine $\MW$. In
fits for $\GW$, both $\MW$ and $\GW$ are varied independently. The
$\MW$ values obtained in the two-parameter fits are consistent within
the given uncertainty with the one-parameter fit for $\MW$ only.  The
methods used are described in the following.

\subsubsection{Monte-Carlo reweighting}

In the reweighting method, a multi-dimensional probability density is
calculated using different mass estimators. These estimators are the
masses from the 5C and 4C kinematic fit in the $\qq\qq$ channel, and
those of the 2C and 1C fit for $\qq\ev$ and $\qq\mv$ events. To
further improve the sensitivity, \Aleph\ also includes the uncertainty
on the 5C and 2C masses. The $\qq\tv$ sample contributes only with the
rescaled hadronic mass. The probability densities are determined from
distributions of the corresponding multi-differential cross-sections,
including $\MW$ and $\GW$ dependent signal predictions and background
contributions.  This is done either using binned distributions or a
local sampling of the phase-space density determined from Monte-Carlo
simulations.  Since the signal Monte-Carlo sample is generated with
pre-defined underlying W mass and width values, the $\MW$ and $\GW$
dependence is introduced by reweighting of Monte-Carlo events. Each
signal event is given a weight according to the ratio of the absolute
values of the matrix element squared for the $\ee\to\WW\to
f\bar{f}f\bar{f}(\gamma)$ process, calculated for the $\MW$ and $\GW$
values that are to be determined and for the nominal $\MW$ and $\GW$
used in the simulation.  The total likelihood functions of the
different data samples are maximised with respect to $\MW$ and
$\GW$. This method is applied for the final \Aleph\ and
\Ltre\ results, and by the \Opal\ collaboration to evaluate systematic
uncertainties of the extraction method.

\subsubsection{Convolution method}

In this method, a probability density function is computed for each
event, giving the probability that this event, with a set of
reconstructed mass estimators $m_{i,\mathrm{rec}}$ ($i=1,\ldots,n$),
originated from a sample with true W mass and width, $\MW$ and $\GW$,
of the following schematic form:

\begin{equation}
  P_s(\MW,\GW,m_{i,\mathrm{rec}})=S(\MW,\GW,m_i,s')\otimes ISR(s',s)
  \otimes R(m_i,m_{i,\mathrm{rec}})
\end{equation}
In this expression, $S(\MW,\GW,m_i,s')$ is the true distribution of
the mass estimators, folded with the radiator function $ISR(s',s)$ and
the detector resolution function, $R(m,m_{\mathrm{rec}})$, which is
determined from Monte-Carlo simulations and describes the probability
that an event of true mass estimator $m_i$ would be reconstructed with
mass estimators $m_{i,\mathrm{rec}}$.  The likelihood for the data is
then constructed as the product of $f_s
P_s(\MW,\GW,m_{i,\mathrm{rec}}) + f_b P_b(m_{i,\mathrm{rec}})$ over
all events, where $f_s$ and $f_b$ are the probabilities that the event
originates from signal and background processes, respectively, and
$P_b(m_{i,\mathrm{rec}})$ is a parametrisation of the background
distribution.  The parameters of interest, $\MW$ and $\GW$, are
estimated by maximising the total likelihood. In this approach, the
resolution function may take account of the uncertainties in the
reconstructed mass, which are likely to vary from event to event, and
thus better measured events are given greater weight. This procedure
is used for the final \Opal\ and \Delphi\ results.

\subsection{Combination Procedure}

The maximum likelihood fits are performed for each of the data sets at
the different centre-of-mass energies and for each W-pair decay
channel separately. Table~\ref{mw:tab:wmass-experiments} shows the
final results on $\MW$ obtained by the four LEP experiments with the
direct reconstruction method in the $\WWtoqqlv$ and $\WWtoqqqq$ final
states. For the LEP combination, each experiment individually combines
the results of the three $\qq\lv$ channels. The \Opal\ collaboration
also includes the $\lv\lv$ measurements in these results. Input to the
combination procedure are thus the $\MW$ and $\GW$ central values and
uncertainties from the four LEP experiments in the $\qq\qq$ (4q) and
$\qq\lv+\lv\lv$ (non-4q) final states for five centre-of-mass energy
bins corresponding to the five years of data taking. These inputs
combine the data collected in 1996 at $172~\GeV$, in 1997 at
$183~\GeV$, in 1998 at $189~\GeV$, in 1999 at $192-202~\GeV$, and in
2000 at $205-209~\GeV$.

The combination of the measurements is performed and the evaluation of
the components of the total measurement uncertainty is assessed using
the Best Linear Unbiased Estimate (BLUE) technique~\cite{BLUE:1988,
*BLUE:2003}.  In this way, statistical and systematic uncertainties of
each measurement are properly taken into account, including
correlations between them. The LEP combination procedure as described
here is also applied to combine the measurements of each LEP
experiment for comparison with the combined measurement published by
each experiment in Table~\ref{mw:tab:wmass-experiments}. The observed
differences are mainly due to a different assessment of FSI
uncertainties, which affects the fully hadronic channel, as discussed
below.  The changes of the semi-leptonic results are due to systematic
uncertainties correlated between the $\qq\qq$ and $\qq\lv$ channels.

\begin{table}[h]
\begin{center}
\renewcommand{\arraystretch}{1.25}
\begin{tabular}{|l||c|c|c|}
\hline
\multicolumn{4}{|c|}{Direct Reconstruction } \\
\hline
           & \WWtoqqlv         & \WWtoqqqq         & Combined         \\   
Experiment & \MW [\GeV]        & \MW [\GeV]        & \MW [\GeV]       \\ 
\hline
\hline
\multicolumn{4}{|c|}{Published}\\
\hline
\Aleph
           & $80.429 \pm 0.060$ & $80.475 \pm 0.080$ & $80.444 \pm 0.051$ \\ 
\Delphi
           & $80.339 \pm 0.075$ & $80.311 \pm 0.137$ & $80.336 \pm 0.067$ \\ 
\Ltre
           & $80.212 \pm 0.071$ & $80.325 \pm 0.080$ & $80.270 \pm 0.055$ \\ 
\Opal
           & $80.449 \pm 0.063$ & $80.353 \pm 0.083$ & $80.416 \pm 0.053$  \\ 
\hline
\hline
\multicolumn{4}{|c|}{LEP combination}\\
\hline
\Aleph
           & $80.429 \pm 0.059$ & $80.477 \pm 0.082$ & $80.444 \pm 0.051$ \\ 
\Delphi
           & $80.339 \pm 0.076$ & $80.310 \pm 0.101$ & $80.330 \pm 0.064$ \\ 
\Ltre
           & $80.217 \pm 0.071$ & $80.324 \pm 0.090$ & $80.254 \pm 0.058$ \\ 
\Opal
           & $80.449 \pm 0.062$ & $80.353 \pm 0.081$ & $80.415 \pm 0.052$  \\ 
\hline
\end{tabular}
\caption[Results on ta W mass using direct reconstruction]{ W mass
 measurements from direct reconstruction ($\roots=172-209$~\GeV).
 Results are given for the semi-leptonic, fully-hadronic channels and
 the combined value. The top part of the table shows the results as
 published by the experiments~\cite{bib:mw:a-mw, bib:mw:d-mw,
 bib:mw:l-mw, bib:mw:o-mw}, using their individual evaluations of FSI
 effects; these results are final.  The bottom part of the table shows
 the results of the experiments when propagating the common LEP
 estimates of FSI effects to the mass, which also affects the
 $\WWtoqqlv$ results through correlations due to other systematic
 uncertainties.  The $\WWtoqqlv$ results from the \Opal\ collaboration
 include mass information from the $\WWtolvlv$ channel.  }
\label{mw:tab:wmass-experiments}
\end{center}
\end{table}

\subsection{Overview of Systematic Uncertainties}

There are several sources of systematic uncertainties affecting the
measurements of $\MW$ and $\GW$.  Table~\ref{mw:tab:errors} summarises
the systematic and statistical uncertainties on the W mass and width
measurements evaluated for the combined LEP data using the direct
reconstruction method. For the W mass determination, the uncertainties
are also given separately for the $\qq\lv$ and $\qq\qq$ final states,
and for their combination.  The main contributions are discussed in
the following.

\begin{table}[h]
\begin{center}
\renewcommand{\arraystretch}{1.25}
\begin{tabular}{|l||r|r|r|r|}\hline
 Source  &  \multicolumn{4}{|c|}{Systematic Uncertainty in $\MeV$}\\
\hline
& \multicolumn{3}{|c|}{on \MW} & \multicolumn{1}{|c|}{on \GW}  \\
\hline
         &  \qq\lv & \qq\qq  & Combined &   \\ 
\hline   
\hline   
 ISR/FSR                                &  8 &  5 &  7 &  6 \\
 Hadronisation                          & 13 & 19 & 14 & 40 \\
 Detector effects                       & 10 &  8 &  9 & 23 \\
 LEP energy                             &  9 &  9 &  9 &  5 \\
 Colour reconnection                    & $-$& 35 &  8 & 27 \\
 Bose-Einstein Correlations             & $-$&  7 &  2 &  3 \\
 Other                                  &  3 & 10 &  3 & 12 \\ 
\hline
 Total systematic                       & 21 & 44 & 22 & 55 \\ 
\hline
 Statistical                            & 30 & 40 & 25 & 63 \\ 
 Statistical in absence of systematics  & 30 & 31 & 22 & 48 \\ 
\hline
\hline
 Total                                  & 36 & 59 & 34 & 83 \\ 
\hline
\end{tabular}
\caption[Uncertainties in the W-mass and W-width combination]{Error
  decomposition for the combined LEP W mass and width results using
  the direct reconstruction method. Information from cross-section
  measurements at the W-pair production threshold are not included in
  the W-mass uncertainties.  Detector effects include uncertainties in
  the jet and lepton energy scales and resolution. The `Other'
  category refers to errors, all of which are uncorrelated between
  experiments, arising from: simulation statistics, background
  estimation, four-fermion treatment, fitting method and event
  selection. The error decomposition in the $\qq\lv$ and $\qq\qq$
  channels refers to the independent fits to the results from the two
  channels separately. Large correlated uncertainties, mainly from
  FSI, lead to a reduced weight of measurements contributing to the
  average result and thus an increased statistical uncertainty both in
  the $\qq\qq$ channel and for the LEP combination.}
 \label{mw:tab:errors}
\end{center}
\end{table}

\subsubsection{LEP centre-of-mass energy}

Since the LEP centre-of-mass energy is used as a constraint in order
to improve the W mass resolution, uncertainties in the centre-of-mass
energy translate directly into uncertainties on $\MW$. These can
approximately be obtained by scaling the LEP centre-of-mass energy
uncertainties with the ratio $\MW/(\sqrt{s}/2)$.  The W width is less
affected. At W-pair threshold energies, the calibration of the LEP
centre-of-mass energy yields precisions of $25-27~\MeV$, and at
energies between $182.7~\GeV$ up to $201.6~\GeV$ the uncertainty is
$20-24~\MeV$. Since in the last LEP runs in the year 2000 horizontal
corrector magnets were used to spread the magnetic field over a larger
bending section in order to eventually increase the LEP beam energy to
its absolute maximum, the related additional systematic effects
reduced the centre-of-mass energy precision to $37-42~\MeV$.
  
A cross-check of the LEP energy determination is performed by
analysing $\ee\to\mathrm{Z}+\gamma\to\ff+\gamma$ events with hard ISR
photons, mostly emitted at small polar angles with respect to the beam
directions. In these events with a so-called radiative return to the
Z, the mass of the 2-fermion system is calculated from the fermion
production angles only, assuming energy-momentum conservation. The
mass spectrum exhibits a peak around the Z mass value. Comparing the Z
mass, $\MZ^\ff$, determined from this spectrum with the precise value
of $\MZ$ measured at Z pole energies~\cite{bib-Z-pole} is equivalent
to a test of the LEP centre-of-mass energy (see
Appendix~\ref{mw:zret:appendix} for further details):

\begin{equation}
  \Delta\sqrt{s}=\sqrt{s}-\sqrt{s}_{LEP}=\sqrt{s}\frac{\MZ^\ff-\MZ}{\MZ}\;,
\end{equation}
with the nominal value of $\sqrt{s}_{LEP}$~\cite{bib:mw:lep-energy-1,
  *bib:mw:lep-energy-2} provided by the LEP energy working group.
When combining all available LEP data~\cite{bib:mw:a-mw,
  bib:mw:d-zreturn, bib:mw:l-zreturn, bib:mw:o-zreturn} with Z decays
to hadrons, and to electron, muon, and tau pairs, the difference is
found to be

\begin{equation}
  \Delta\sqrt{s} = -54 \pm 54~\MeV\,,
\end{equation}
in good agreement with no shift with respect to the more precise
standard LEP energy calibration.

The properly calibrated LEP centre-of-mass energy is used in the W
mass and width analysis on event-by-event basis. Uncertainties on
$\MW$ and $\GW$ are determined by detailed Monte-Carlo studies, and
also the effect of the LEP energy spread is taken into account. When
combining the LEP W mass and width results the correlations between
the LEP energy uncertainties at the different energies are properly
included. They are derived from the LEP energy
model~\cite{bib:mw:lep-energy-1, *bib:mw:lep-energy-2} and listed in
Table~\ref{mw:tab:lep-energy}.  The overall LEP energy uncertainty is
$9~\MeV$ on \MW\ and $5~\MeV$ on \GW.

\begin{table}[t]
\begin{center}
\renewcommand{\arraystretch}{1.25}
\begin{tabular}{| c || c  c  c  c  c  c |}
\hline
 & \multicolumn{6}{c |}{LEP energy correlations} \\
\hline
 $\sqrt{s}$ $[\GeV]$ & 161 & 172 & 183 & 189 & 192-202 & 205-209 \\
\hline
\hline
        161     & 1.00 &  1.00 &  0.57 &  0.56 &  0.58 &  0.36 \\
        172     & 1.00 &  1.00 &  0.58 &  0.57 &  0.58 &  0.37 \\
        183     & 0.57 &  0.58 &  1.00 &  0.94 &  0.95 &  0.53 \\
        189     & 0.56 &  0.57 &  0.94 &  1.00 &  0.94 &  0.53 \\
        192-202 & 0.58 &  0.58 &  0.95 &  0.94 &  1.00 &  0.55 \\
        205-209 & 0.36 &  0.37 &  0.53 &  0.53 &  0.55 &  1.00 \\
\hline
\end{tabular}
\caption[LEP energy correlation matrix]{ Correlation between the LEP
 centre-of-mass energy measurements in the six run
 periods~\cite{bib:mw:lep-energy-1, *bib:mw:lep-energy-2}.  }
\label{mw:tab:lep-energy}
\end{center}
\end{table}

\subsubsection{Detector effects}

The effects of detector performance as well as of identification and
reconstruction efficiencies for final state leptons, jets and photons
are studied in dedicated control data samples. Energy and momentum
calibration, as well as detector alignment and angular measurements,
very important for the mass reconstructed, were
studied~\cite{bib:mw:a-mw, bib:mw:d-mw, bib:mw:l-mw, bib:mw:o-mw}.
Since Monte-Carlo samples are compared to data to extract \MW\ and
\GW, all effects are modelled in detail in the simulation and
remaining differences to data result in corresponding systematic
uncertainties. The LEP experiments provide separate uncertainties for
lepton and jet measurements. These are considered uncorrelated between
measurements from different experiments, but correlated for $\MW$ and
$\GW$ measurements from the same experiment at different LEP energy
points. The total systematic uncertainty from detector effects is
$10~\MeV$ and $8~\MeV$ on \MW\ in the $\qq\lv$ and $\qq\qq$
channels. The W width systematic uncertainties due to finite precision
in modelling jet and lepton measurements is $23~\MeV$, combining all
final states.

\subsubsection{Fragmentation and hadronisation}

Since the \MW\ and \GW\ extraction methods rely on the comparison of
Monte-Carlo simulations to data the modelling of the fragmentation and
hadronisation process subsequent to the $\mathrm{W}\to\qq$ decay is
essential. The calibration of the reconstructed jets is very sensitive
to the fractions of the different final state hadrons inside the
jets. Furthermore, the jet reconstruction usually cannot resolve each
individual hadron, so that the same particle masses are assumed
(usually the pion mass) when tracks and clusters are combined to form
quark and gluon jets. To assess systematic uncertainties due to
fragmentation and hadronisation, different Monte-Carlo models are
compared, whose parameters are adjusted to describe high-statistic
data samples of $\mathrm{Z}\to\qq$ decays at the Z pole. These Z
decays are depleted in b-quarks, to resemble the hadronic decays of W
bosons. The systematic uncertainty is derived from the relative shifts
in W mass and width values determined in Monte-Carlo samples using the
\Jetset/\Pythia~\cite{JETSET}, \Herwig~\cite{HERWIG}, and
\Ariadne~\cite{ARIADNE} fragmentation models. In addition, the
fraction of certain hadrons, like kaons and protons, is directly
measured in $\mathrm{W}\to\qq$ decays and compared to the
fragmentation models. The measurement uncertainties on these fractions
are also taken into account in the fragmentation and hadronisation
systematic uncertainties for the \MW\ and \GW\ determination.

Since all four LEP experiments study the same fragmentation models,
the systematic uncertainty is taken as fully correlated for all
measurements of the W mass and width. Eventually, the systematic
effect on \MW\ is estimated to be $13~\MeV$ and $19~\MeV$ in the
$\qq\lv$ and $\qq\qq$ final states. In the W width determination, the
corresponding systematic uncertainties contribute with $40~\MeV$ to
the combined W width measurement.

\subsubsection{Colour reconnection}

A particular systematic uncertainty arises in the $\WWtoqqqq$ channel,
where the two W bosons decay close in phase space so that FSI effects
may play a significant role. Indeed, colour reconnection (CR) effects
leads to shifts of the extracted W mass up to about
$100~\MeV$~\cite{bib:mw:a-mw, bib:mw:d-mw, bib:mw:l-mw, bib:mw:o-mw}
if nominal jet reconstruction is applied and data are compared to
Monte-Carlo models with and without colour reconnection. These large
shifts are observed even if the measured constraints on the
reconnection parameters, which are discussed in
section~\ref{chap:fsi}, are applied. The LEP collaborations therefore
developed new techniques in the $\qq\qq$ channel. It is observed that
colour reconnection effects on \MW\ as implemented in the
\Ariadne~\cite{bib:cr:ARIADNECR_MODEL}, SK~\cite{bib:cr:SK_MODELS},
and \Herwig~\cite{HERWIG6} models are reduced when the jet
reconstruction is modified.  This is achieved by either rejecting
particles inside jets with energies or momenta lower than a given
threshold or by reweighting their energies and momenta to suppress
soft particles, which are mainly in the inter-jet and
reconnection-sensitive region.  The four LEP experiments applied
thresholds and weights which are optimised individually for the colour
reconnection constraints of the \SKI\ model~\cite{bib:cr:SK_MODELS}
which are measured by each experiment separately. In the optimisation
process the overall uncertainty on $\MW$ is minimised, again
individually, trading a reduced statistical precision due to a
modified jet reconstruction for an improved FSI systematic
uncertainty. For the LEP combined analysis, the threshold values and
weights of each experiment are however not always optimal when the LEP
combined upper limit on the \SKI\ parameter, $k_I<2.10$, is used as
reference for the CR uncertainty in the LEP \MW\ combination.
Although this reduces the relative weight of some \MW\ measurements in
the LEP combination, a further optimisation is not performed.

For the final LEP combination, the central value of the W mass is
determined using Monte-Carlo samples without colour reconnection. The
systematic uncertainties are evaluated from the mass differences
observed when data is compared to the $\SKI$ model with
$k_I=2.10$. The systematic uncertainties are evaluated at each
centre-of-mass energy independently since the colour reconnection
effects are energy dependent. The systematic uncertainties are taken
as symmetric in the combination procedure and correlated between all
measurements in the $\qq\qq$ channel at the different centre-of-mass
energies and by the four LEP experiments. They contribute $35~\MeV$ to
the total uncertainty in the fully hadronic final state.

When the W width is extracted, the optimisation of the jet
reconstruction is not applied by the LEP collaborations, and the
standard jet measurement is used. The reason is the relatively large
statistical uncertainty of the W width measurement, which does not
require a modification of the standard $\qq\qq$ analysis. The
corresponding CR uncertainty is evaluated using the LEP upper limit on
the \SKI\ parameter, $k_I<2.10$, like in the W mass determination, and
corresponds to $27~\MeV$ on the combined width result.

\subsubsection{Bose-Einstein correlations}

A further source of uncertainty connected with FSI in the $\WWtoqqqq$
channel is the possibility of Bose-Einstein correlations (BEC) between
identical mesons in the decay of different W bosons.  The measurement
of these correlations is discussed in detail in
section~\ref{chap:fsi}.  For the final LEP results, Bose-Einstein
correlations between particles from inside each hadronically decaying
W are implemented in the Monte-Carlo simulation according to the
$\mathrm{BE}_{32}$ model~\cite{JETSET}, which describes $\WWtoqqlv$
data well. However, the combined analysis of LEP data yields an upper
limit on the strength of Bose-Einstein correlations between mesons
from different W bosons of 30\% of the full correlation in the
$\mathrm{BE}_{32}$ model. The systematic effect on the W mass and
width in the $\WWtoqqqq$ channel is effectively reduced by the
modified jet reconstruction algorithms, which were originally
introduced for controlling systematic uncertainties from
CR. Therefore, the uncertainties due to Bose-Einstein correlations on
the W mass in $\WWtoqqqq$ events is $7~\MeV$, while it is just
$3~\MeV$ on the combined width result.

\subsubsection{Initial state radiation and $\mathcal{O}(\alpha)$ effects}

Photon radiation influences the reconstructed W mass spectra. The
Monte-Carlo programs used to extract \MW\ and \GW, \KandY, \RacoonWW\
and \Wphact, include ISR effects in the YFS exponentiation scheme to
$O(\alpha^3)$, full $O(\alpha)$ electroweak corrections, including
interference between ISR, FSR and photon radiation of the W boson, as
well as screened Coulomb corrections. These describe Coulomb
interactions between the W bosons, which are potentially large but
screened due to the limited lifetime of the W bosons. Higher-order
leading-log FSR corrections are included using PHOTOS for leptons and
\Pythia\ for quarks. ISR effects on \MW\ are estimated by comparing
the $O(\alpha^3)$ with the $O(\alpha^2)$ calculation, yielding small
shifts of about $1~\MeV$~\cite{bib:mw:order-alpha-ew}. The effect of
Coulomb screening are estimated by taking half of the difference
between Monte-Carlo samples with screened Coulomb effect and without
any Coulomb effect, which amounts to about $7~\MeV$. To evaluate the
uncertainty on the non-leading $O(\alpha)$ electroweak corrections, a
direct comparison of the \RacoonWW\ and the \KandY\ generators is
performed. The observed differences are in the order of $10~\MeV$ for
$\qq\lv$ and $5~\MeV$ for $\qq\qq$. Some systematic studies overlap,
however, and the experiments apply different strategies to assess
them. The total LEP uncertainty on the W mass due to radiative
corrections is $8~\MeV$ in the semi-leptonic channel and $5~\MeV$ in
the fully hadronic channel. Full correlation between all data sets is
assumed. In case of the W width, the corresponding uncertainties
amount to $6~\MeV$ when combining all final states.

\subsubsection{Other sources of systematic uncertainties}

The contribution of background to the selected W-pair samples arises
mainly from 4-fermion and hadronic 2-fermion events.  All LEP
experiments study the event shapes of the different background
contributions using control samples to best describe the data. The
systematic effect of the background on \MW\ and \GW\ are derived by
varying the overall scale on the production cross-sections of the
background processes, mainly $\ee\to\qq(\gamma,g)$ and
$\ee\to\mathrm{ZZ}$, within the measured uncertainty. Effects on the
mass spectrum which do not scale with the overall production rate are
studied by varying, for example, the slope of the background spectra.

In addition, uncertainties due to limited Monte-Carlo statistics, from
the mass and width extraction techniques, and due to the event
selection are considered. Early analyses at \LEPII\ used Monte-Carlo
simulations based on CC03 matrix elements to simulate $\WW$
production.  In this case, systematic biases of the W mass and width
may arise because four-fermion diagrams are neglected which might
interfere with W-pair production.
  
All these categories of systematic uncertainties are taken as
uncorrelated in the LEP combination and contribute on the mass with
$3~\MeV$ in the semi-leptonic channel and $10~\MeV$ in the fully
hadronic channel, and $12~\MeV$ on \GW.

\section{LEP Combined W-Boson Mass }

The combined LEP W mass from direct reconstruction data alone is:

\begin{eqnarray}
   \MW(\mathrm{direct}) = 
   80.375\pm0.025(\mathrm{stat.})\pm0.022(\mathrm{syst.})~\GeV \,,
\end{eqnarray}
with a total uncertainty of $34~\MeV$. The combination has a $\chidf$
of 47.7/37, corresponding to a probability of 11.1\%.  The weight of
the fully-hadronic channel in the combination amounts to just 22\% due
to significant FSI systematic uncertainties.

The largest contribution to the systematic error originates from
hadronisation uncertainties, which are fully correlated between all
measurements. In the absence of any systematic effects the current LEP
statistical precision on $\MW$ would be $22~\MeV$. The statistical
error contribution in the LEP combination is larger than this,
$25~\MeV$, due to the reduced weight of the fully-hadronic channel,
mainly due to FSI systematic uncertainties.

When the threshold measurements (Section~\ref{sec:mw:threshold})
are combined with the precise results obtained from direct
reconstruction one achieves a W mass measurement of:

\begin{eqnarray}
   \MW = 80.376\pm0.025(\mathrm{stat.})\pm0.022(\mathrm{syst.})~\GeV\,,
\end{eqnarray}
with a slightly improved total uncertainty of $33~\MeV$. The
combination has a $\chidf$ of 48.9/41, corresponding to a probability
of 18.5\%.  The LEP energy uncertainty is the only correlated
systematic error source between the threshold and direct
reconstruction measurements.  The threshold measurements have a weight
of only 2\% in the combined fit.  This LEP combined result is compared
with the final results of the four LEP experiments in
Figure~\ref{mw:fig:mw-final}.

\begin{figure}[ht]
\begin{center}
\mbox{\epsfig{file=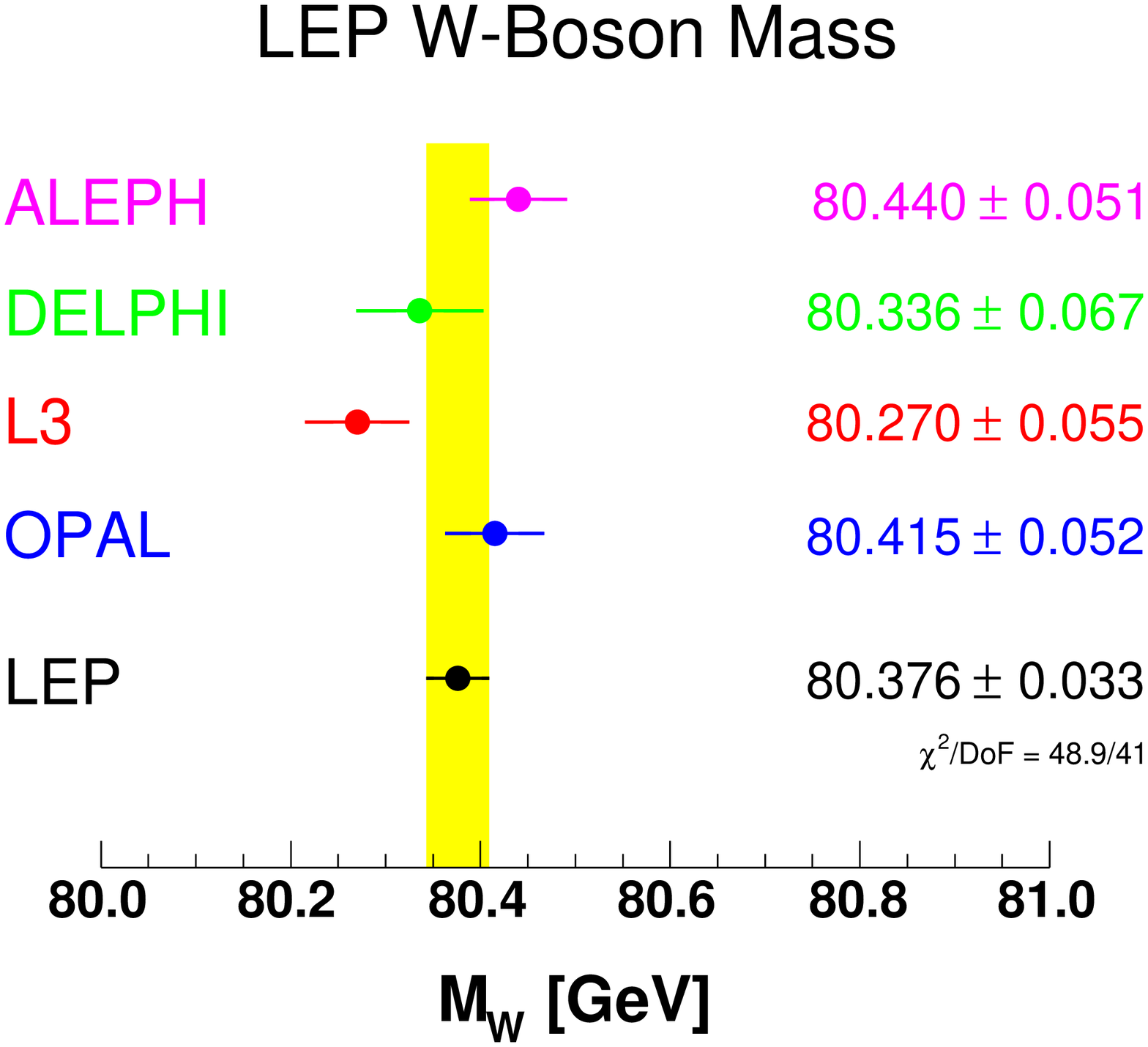,width=0.9\textwidth}}
\caption[Measurements of the W-boson mass]{ The measurements of the
 W-boson mass obtained by the four LEP collaborations (as published)
 together with the LEP combined result. The combined value includes
 correlations between experiments, between different energy points,
 and between the $\qq\lv$ and $\qq\qq$ channels. A revised estimation
 of systematic uncertainties due to colour reconnection and
 Bose-Einstein correlations is applied to the input of the individual
 measurements to the LEP combined results in order to take the direct
 determination of FSI parameters into account.}
\label{mw:fig:mw-final} 
\end{center}
\end{figure}

\section{Consistency Checks}

The masses from the two channels with all uncertainties and
correlations included are:

\begin{eqnarray}
 \MW(\WWtoqqlv) = 80.372\pm0.030(\mathrm{stat.})\pm0.021(\mathrm{syst.})~\GeV, \\
 \MW(\WWtoqqqq) = 80.387\pm0.040(\mathrm{stat.})\pm0.044(\mathrm{syst.})~\GeV.  
\end{eqnarray}
The two results are correlated with a correlation coefficient of 0.20.
These results and the correlation between them can be used to combine
the two measurements or to form the mass difference. The LEP combined
results from the two channels are compared with those quoted by the
individual experiments in Figure~\ref{mw:fig:mw-hadlep-final}. When
combining the $\MW$ measurements in the $\qq\lv$ and $\qq\qq$ channels
separately and neglecting any correlations between these final states,
results consistent within $2~\MeV$ with the correlated averages above
are obtained.

The difference between the combined W-boson mass measurements obtained
from the fully-hadronic and semi-leptonic channels,
$\Delta\MW(\qq\qq-\qq\lv)$ is also determined. Since $\Delta\MW$ is
primarily of interest as a check of the possible effects of final
state interactions, the uncertainties from Bose-Einstein correlation
and colour reconnection are set to zero in its determination.  A fit
imposing otherwise the same correlations as those for the results
given in the previous sections yields:

\begin{eqnarray}
 \Delta\MW(\qq\qq-\qq\lv) = -12\pm45~\MeV.
\end{eqnarray}
Note that this mass difference has a different value and opposite sign
compared to the difference between the $\qq\qq$ and $\qq\lv$ mass
values presented above, because the BEC and CR uncertainties are not
included in its determination. A significant non-zero value for
$\Delta\MW$ could indicate that such Bose-Einstein correlation or
colour reconnection effects are biasing the value of \MW\ determined
from \WWtoqqqq\ events.  The consistency of the mass difference with
zero shows that such FSI effects are well suppressed by the modified
jet reconstruction in the fully hadronic channel.

\begin{figure}[p]
\begin{center}
\mbox{\epsfig{file=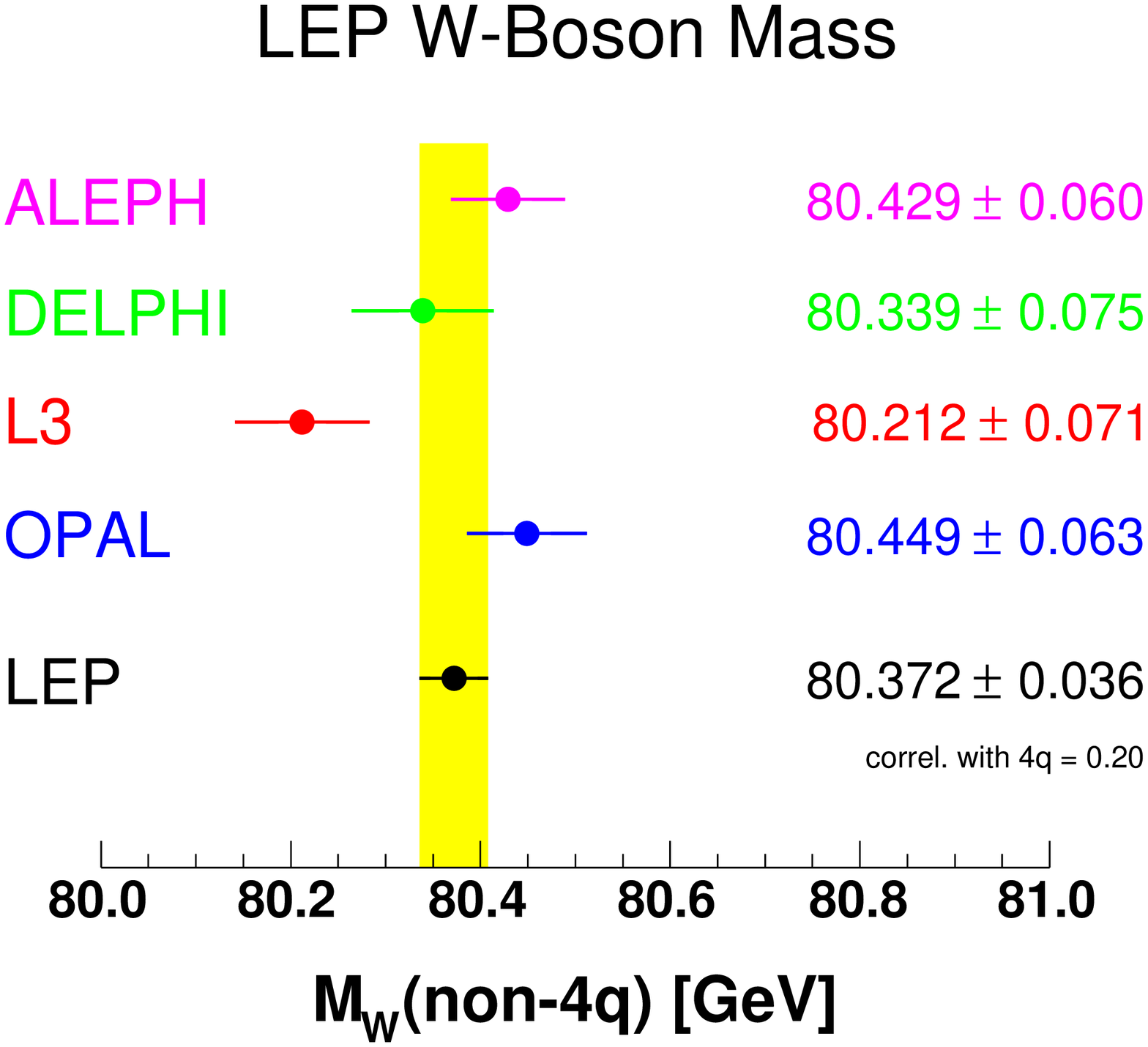,width=0.6\textwidth}}
\mbox{\epsfig{file=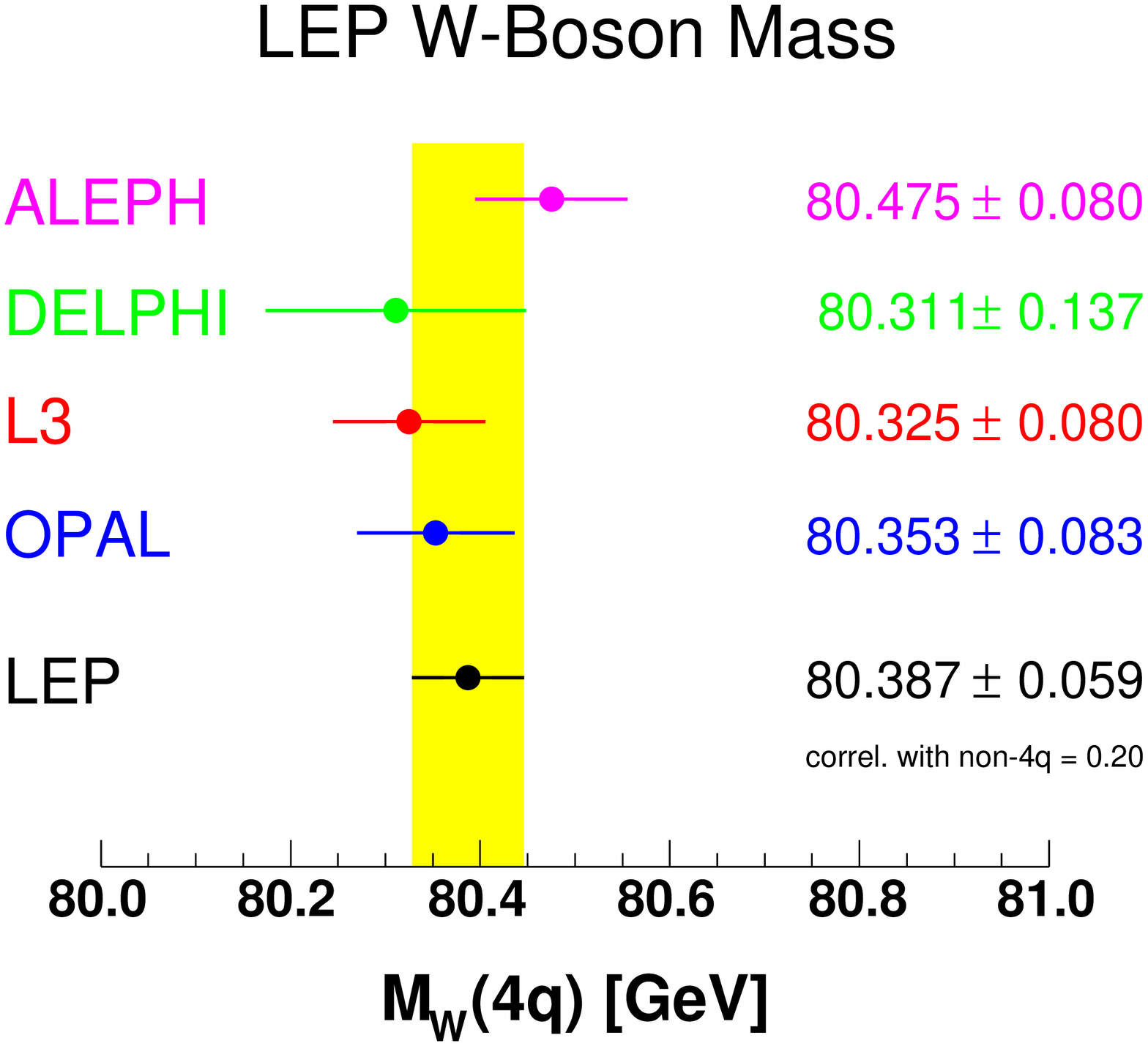,width=0.6\textwidth}}
\vskip -0.5cm
\caption[Measurements of the W-boson mass in different channels]{ The
 W mass measurements in the $\WWtoqqlv$ channels (top), and the
 $\WWtoqqqq$ channel (bottom) obtained by the four LEP collaborations
 (as published) compared to the combined value.  Correlations between
 experiments and between measurements at different energy points are
 properly taken into account.  The combined non-4q and 4q results are
 correlated since they are obtained from a fit to both channels taking
 into account inter-channel correlations. For the LEP combination, the
 assessment of systematic uncertainties due to colour reconnection and
 Bose-Einstein correlations for the individual measurements of the
 four experiments is revised with respect to the direct LEP
 measurements of FSI.}
\label{mw:fig:mw-hadlep-final} 
\end{center}
\end{figure}

\section{LEP Combined W-Boson Width}

The method of direct reconstruction is also well suited to the direct
measurement of the total decay width of the W boson. The published
results of the four LEP experiments are shown in Table
\ref{mw:tab:wwidth_experiments} and in Figure~\ref{mw:fig:gw-final}.

\begin{table}[htbp]
\begin{center}
\renewcommand{\arraystretch}{1.25}
\begin{tabular}{|c||c|c|}
\hline
Experiment &  Published & LEP combination \\ 
           & \GW\ [\GeV] & \GW\ [\GeV] \\ 
\hline
\hline
ALEPH      & $2.14\pm0.11$&$2.14\pm0.11$ \\ 
DELPHI     & $2.40\pm0.17$&$2.39\pm0.17$ \\
L3         & $2.18\pm0.14$&$2.24\pm0.15$ \\
OPAL       & $2.00\pm0.14$&$2.00\pm0.14$ \\ 
\hline
\end{tabular}
\caption[Measurements of the W-boson width]{ W width measurements
 ($\roots=172-209$~\GeV) from the individual experiments.  The column
 labelled ``published'' shows the results as published by the
 experiments, using their individual evaluations of FSI effects. The
 column labelled ``LEP combination'' shows the results of the
 experiments when propagating the LEP measurements of FSI effects to
 the W width. }
\label{mw:tab:wwidth_experiments}
\end{center}
\end{table}

For the LEP combination, each experiment provided a W width
measurement for both $\WWtoqqlv$ and $\WWtoqqqq$ channels for each of
the data taking periods that were analysed, and using the same error
categories as for the mass.  The BEC and CR uncertainties supplied by
the experiments were based on studies of phenomenological models of
these effects, using the same estimates of such FSI effects as for the
mass and propagating them to the width.  Note that the W width results
of the experiments do not use the techniques introduced to reduce
sensitivity to FSI effects used for the mass analysis.  A simultaneous
fit to the results of the four LEP collaborations is performed in the
same way as for the $\MW$ measurement. Correlated systematic
uncertainties are taken into account and the combination yields:

\begin{eqnarray}
      \GW = 2.195\pm0.063(\mathrm{stat.})\pm0.055 (\mathrm{syst.})~\GeV,
\end{eqnarray}
for a total error of $83~\MeV$. The combination has a $\chidf$ of
37.4/33, corresponding to a probability of 27.3\%.

\section{Summary}

The final results of the four LEP experiments on the mass and width of
the W boson are combined taking into account correlated systematic
uncertainties, with the result:

\begin{eqnarray}
      \MW & = & 80.376\pm0.033~\GeV\,,  \\
      \GW & = &  2.195\pm0.083~\GeV\,.
\end{eqnarray}
The correlations between mass and width are found to be less than 5\%
and thus negligible. These values correspond to the theoretical
definition of a W-boson propagator with $s$-dependent width.  The
results of the mass and width determined by the LEP collaborations are
in good agreement with the measurements at hadron
colliders~\cite{CDF2GW, D02GW, CDF2MWPRL2012, D0-MW:PRL2012}.  Updated
constraints on SM parameters using the mass and width results are
presented in Appendix~\ref{chap:sm}.

\begin{figure}[h]
\begin{center}
\mbox{\epsfig{file=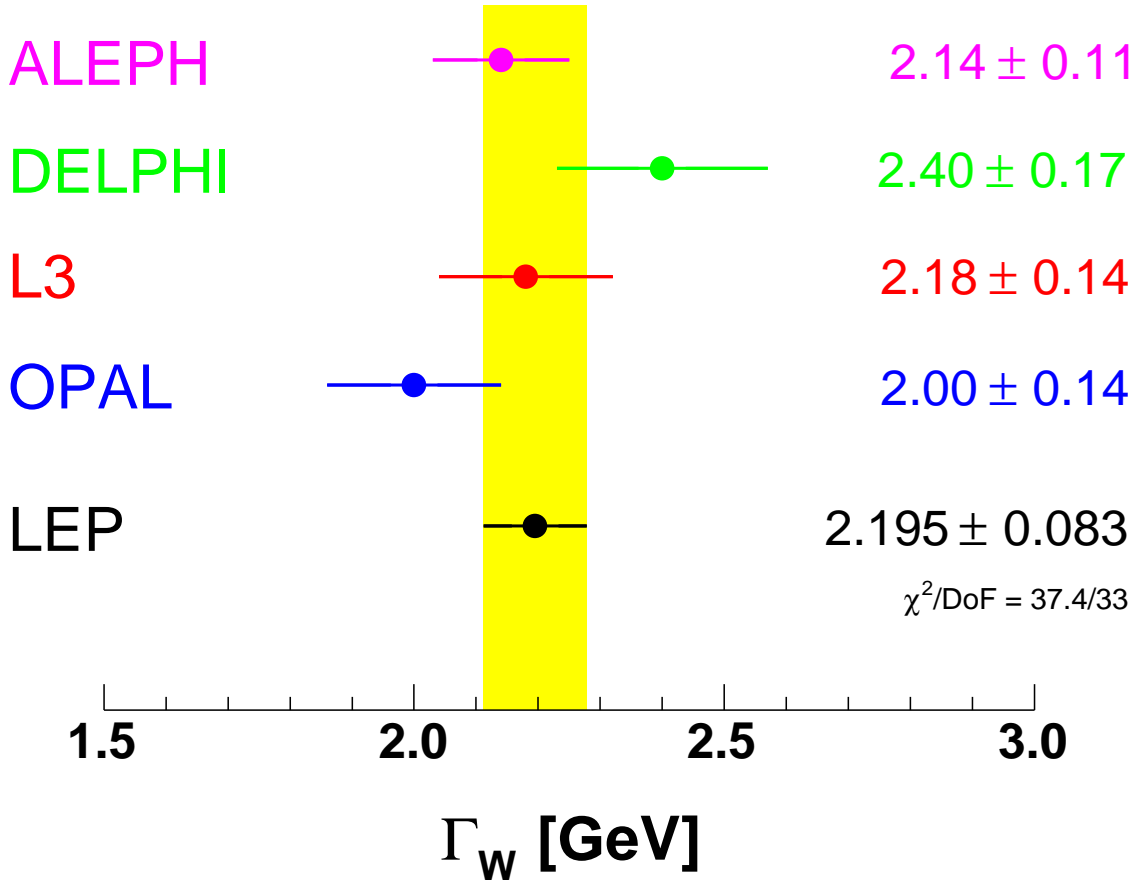,width=0.9\textwidth}}
\caption[Measurements of the W-boson width]{ The measurements of the
 W-boson width obtained by the four LEP collaborations (as published)
 together with the LEP combined result. The combined value includes
 correlations between experiments, between different energy points,
 and between the $\qq\lv$ and $\qq\qq$ channels. A revised estimation
 of systematic uncertainties due to colour reconnection and
 Bose-Einstein correlations is applied to the input of the individual
 measurements to the LEP combined results in order to take the direct
 determination of FSI parameters into account.}
\label{mw:fig:gw-final}
\end{center}
\end{figure}

\chapter{Summary and Conclusions}
\label{chap:sum}

The four LEP experiments ALEPH, DELPHI, L3 and OPAL performed
measurements in electron-positron collisions at centre-of-mass
energies above the mass of the Z boson, ranging from $130~\GeV$,
crossing the W-pair production threshold at $160~\GeV$, up to
$209~\GeV$. Based on about 0.75~fb$^{-1}$ of luminosity collected by
each experiment, yielding a total of 3~fb$^{-1}$, many precision
measurements are summarised in this report.

The combinations of precise electroweak results yield stringent
constraints on the Standard Model (SM) and its free parameters, for
example:

\begin{eqnarray*}
\MW                 & = &  80.376   \pm 0.033 ~\GeV \\
\GW                 & = &   2.195   \pm 0.083 ~\GeV \\
B(W\to\mathrm{had}) & = &  67.41    \pm 0.27  ~\%   \\
g^Z_1               & = &   0.984   ^{+0.018}_{-0.020} \\
\kappa_\gamma       & = &   0.982   \pm 0.042  \\
\lambda_\gamma      & = &  -0.022   \pm 0.019  \,.
\end{eqnarray*}
The results, together with measurements performed in electron-positron
collisions at the Z-pole and in hadron collider experiments, test the
SM with unprecedented precision at the highest interaction energies.
The measurements agree well with the SM predictions.

Overall, the $\SM$ is verified to be a good theory up to the
$200~\GeV$ scale, see also the studies presented in
Appendix~\ref{chap:sm}.  The data impose very tight constraints on any
new physics beyond the $\SM$, and are well compatible with a
$125-126~\GeV$ SM Higgs boson~\cite{ATLAS-HIGGS-2012-Summer,
*CMS-HIGGS-2012-Summer}.  Any extended theory must be consistent with
the $\SM$ or one or more Higgs doublet models such as super-symmetry.

\vfill

\chapter*{Acknowledgements}
\addcontentsline{toc}{chapter}{Acknowledgements}

We would like to thank the CERN accelerator divisions for the
efficient operation of the LEP accelerator, the precise information on
the beam energy scale and their close cooperation with the four
experiments.  We would also like to thank members of the SLD, CDF,
D\O, NuTeV and E-158 collaborations for making results available to us
and for useful discussions concerning their combination.  Finally, the
results and their interpretation within the SM would not have been
possible without the close collaboration of many theorists.

\appendix

\newcommand{\smxhad}{S_{\mathrm{had}}}
\newcommand{\Rsmxe}{R^{\mathrm{smx}}_{\mathrm{e}}}
\newcommand{\Rsmxm}{R^{\mathrm{smx}}_{\mu}}
\newcommand{\Rsmxt}{R^{\mathrm{smx}}_{\tau}}
\newcommand {\Afbsmxe} {A_{\mathrm{FB}}^{\mathrm{smx,e}}}
\newcommand {\Afbsmxm} {A_{\mathrm{FB}}^{\mathrm{smx},\mu}}
\newcommand {\Afbsmxt} {A_{\mathrm{FB}}^{\mathrm{smx},\tau}}
\newcommand {\Afbsmxl} {A_{\mathrm{FB}}^{\mathrm{smx},\ell}}

\chapter{S-Matrix}
\label{chap:s-matrix}

\section{Introduction}

The S-Matrix ansatz provides a coherent way of describing the
measurements of the cross-section and forward-backward asymmetries in
$s$-channel $\eeff$ processes at centre-of-mass energies around the
$\Zzero$ resonance and the measurements at centre-of-mass energies
from $130~\GeV$ to $209~\GeV$ from the {\LEPII} program.  This chapter
describes the combination of results from the full {\LEPI} data sets
of the four LEP experiments, to obtain a LEP combined result on the
parameters of the S-Matrix ansatz describing the Z lineshape.

The standard description of the measurements at the $\Zzero$
resonance~\cite{bib-Z-pole} makes use of nine parameters ($\MZ$,
$\GZ$, $\shad$, $\Rl$, $\Afbzl$, for $\ell=\mathrm{e},\mu,\tau$) which
are reduced to five in case lepton universality is assumed.  The
S-Matrix formalism utilises an extra three parameters (assuming lepton
universality) or seven parameters (not assuming lepton universality).
The additional parameters describe the contributions to the
cross-sections and forward-backward asymmetries of the interference
between the exchange of a $\Zzero$ and a photon.  The Z-pole data
alone cannot tightly constrain these interference terms, in particular
the interference term for cross-sections, since their contributions
are small around the $\Zzero$ resonance and change sign at the pole.
Owing to strong correlations between the size of the hadronic
cross-section interference term and the mass of the $\Zzero$, this
leads to a larger error on the extracted mass of the $\Zzero$ compared
to the standard five and nine parameter analyses where the hadronic
interference term is fixed to the value predicted in the Standard
Model (SM).  However, using the {\LEPII} data leads to a significant
improvement in the constraints on the interference terms and a
corresponding reduction in the uncertainty on the mass of the
$\Zzero$, expected to result in a measurement of {$\MZ$} which is
almost as precise but without having to constrain the $\gamma/$Z
interference to the SM prediction.

The LEP combination is a two-step procedure: first a combination of
the ${\LEPI}$ based results, and then including the ${\LEPII}$ data.
For the {\LEPI} data, an average of the individual experiments'
results on the S-Matrix parameters is made. Such a combination at
parameter level, similar to the method used to combine the Z lineshape
results in terms of the five and nine parameters~\cite{bib-Z-pole}, is
presented here.  To include the {\LEPII} data, a fit of the S-Matrix
parameters to the combined {\LEPII} measurements of cross-sections and
asymmetries as presented in Chapter~\ref{chap:ff} is envisaged,
including in the $\chi^2$ the {\LEPI} based combination of S-Matrix
parameters with uncertainties and correlations as additional
constraints.\footnote{Based on preliminary LEP measurements,
Reference~\cite{bib-EWEP-06} contains a partial {\LEPI}+{\LEPII}
combination along these lines, which shows the vast improvement made
possible by including the {\LEPII} measurements.}

In Section~\ref{smat:sec:ansatz} the parameters of the S-Matrix ansatz
are explained in detail. In Section~\ref{smat:sec:comb} the average of
the {\LEPI} data is described, preparing for the inclusion of the
{\LEPII} measurements in the future.  The results are discussed in
Section~\ref{smat:sec:discuss} while the detailed combination tables
are listed in Section~\ref{smat:comb:appendix}.

\section{The S-Matrix Ansatz}
\label{smat:sec:ansatz}

The S-Matrix
ansatz~\cite{smat:ref:bcms,*smat:ref:rgs,*smat:ref:lrr,*smat:ref:tr}
is a rigorous approach to describe the cross-sections and
forward-backward asymmetries in $s$-channel $\ee$ annihilations under
the basic assumption that the processes can be parametrised as the
exchange of a massless and a massive vector boson, in which the
couplings of the bosons including their interference are treated as
free and independent parameters.  In this model, the cross-sections
are parametrised as follows:

\begin{equation}
\sigma^0_\mathrm{tot, f}(s)=\frac{4}{3}\pi\alpha^2
\left[
      \frac{\gf^{tot}}{s}
     +\frac{\jf^{tot} (s-\MZbar^2) + \rf^{tot} \, s}
           {(s-\MZbar^2)^2 + \MZbar^2 \GZbar^2}
\right]
\,\,\,\mathrm{with}\,\,\mathrm{f=had,e,\mu,\tau}\,,
\label{smat:eqn:eq1}
\end{equation}
while the forward-backward asymmetries are given by:

\begin{equation}
A^0_\mathrm{fb, f}(s)=\pi\alpha^2
\left[
 \frac{\gf^{fb}}{s} +
 \frac{\jf^{fb} (s-\MZbar^2) + \rf^{fb} \, s}
           {(s-\MZbar^2)^2 + \MZbar^2 \GZbar^2}
\right]
                       / {\sigma^0_{\mathrm{tot, f}}(s)}\,,
\label{smat:eqn:eq2}
\end{equation}
where $\sqrt{s}$ is the centre-of-mass energy.  The parameters $\rf$
and $\jf$ scale the $\Zzero$ exchange and the $\gamma/$Z interference
contributions to the total cross-section and forward-backward
asymmetries.  The contribution $\gf$ of the pure $\gamma$ exchange is
fixed to the value predicted by QED.  Neither the hadronic charge
asymmetry nor the flavour-tagged quark forward-backward asymmetries
are considered here, which leaves 16 S-Matrix parameters to describe
the LEP data: the mass and total width of the $\Zzero$ resonance, and
14 $\rf$ and $\jf$ parameters.  Applying the constraint of
neutral-current lepton universality reduces the number of parameters
from 16 to 8.

In the SM the $\Zzero$ exchange term, the $\gamma/$Z
interference term and the photon exchange term are given in terms of
the fermion charges and their effective vector and axial-vector couplings to
the $\Zzero$ by:

\begin{eqnarray}
\rtotf & = & \kappa^2
             \left[\gae^2+\gve^2\rule{0mm}{4mm}\right]
             \left[\gaf^2+\gvf^2\rule{0mm}{4mm}\right]
            -2\kappa\,\gve\,\gvf C_{Im}         \\
\jtotf & = & 2\kappa\,\gve\,\gvf \left(C_{Re}+C_{Im}\right) \\
\gtotf & = & Q^2_{\mathrm{e}}Q^2_{\mathrm{f}}
             \left|F_A(\MZ)\rule{0mm}{4mm}\right|^2 \\
\rfbf  & = & 4\kappa^2\gae\,\gve\,\gaf\,\gvf
            -2\kappa\,\gae\,\gaf C_{Im}         \\
\jfbf  & = & 2\kappa\,\gae\,\gaf \left(C_{Re}+C_{Im}\right) \\
\gfbf  & = & 0 \,,
\end{eqnarray}
with the following definitions:

\begin{eqnarray}
\displaystyle
\kappa & = & \dfrac{G_F\MZ^2}{2\sqrt{2\,}\pi\alpha} \approx 1.50\\
C_{Im} & = & \dfrac{\GZ}{\MZ}  \left.Q_{\mathrm{e}}Q_{\mathrm{f}}\right.
                                \mathrm{Im} \left\{F_A(\MZ)\right\} \\
C_{Re} & = &                 \left.Q_{\mathrm{e}}Q_{\mathrm{f}}\right.
                                \mathrm{Re} \left\{F_A(\MZ)\right\} \\
F_A(\MZ)&= & \dfrac{\alpha(\MZ)}{\alpha} \,,
\end{eqnarray}
where $\alpha(\MZ)$ is the complex fine-structure constant, and
$\alpha\equiv\alpha(0)$.  The expressions of the S-Matrix parameters
in terms of the effective vector and axial-vector couplings given
above neglect the imaginary parts of the effective couplings.  The
photonic virtual and bremsstrahlung corrections are included through
the convolution of Equations~\ref{smat:eqn:eq1} and~\ref{smat:eqn:eq2}
with the same radiator functions as used in the five and nine
parameter Z-lineshape fits~\cite{bib-Z-pole}.

In the S-Matrix framework, the parameters mass ($\MZbar$)and total
width ($\GZbar$) of the Z boson are defined in terms of a relativistic
Breit-Wigner with $s$-independent width. These definitions are related
to the usual definitions of the mass $\MZ$ and width $\GZ$ of a
Breit-Wigner resonance with $s$-dependent width, used
in~\cite{bib-Z-pole}, as follows:

\begin{eqnarray}
\MZ & \equiv  & \MZbar\sqrt{1+\GZbar^2/\MZbar^2}  \approx  \MZbar  +  34.20~\MeV\,,\\
\GZ & \equiv  & \GZbar\sqrt{1+\GZbar^2/\MZbar^2}  \approx  \GZbar  +   0.94~\MeV\,.
\label{smat:eqn:smat-ls}
\end{eqnarray}

The predictions of the S-Matrix ansatz for cross-sections and
asymmetries are calculated using SMATASY~\cite{smat:ref:smatasy},
which in turn uses ZFITTER~\cite{\ZFITTERref} to calculate the QED
convolution of the electroweak kernel.  In case of the $\ee$ final
state, $t$-channel and $s/t$ interference contributions are added to
the $s$-channel ansatz~\cite{bib-Z-pole}.

\section{LEP-I Combination}
\label{smat:sec:comb}

The LEP experiments have determined the 16 S-Matrix parameters using
their full {\LEPI} data set~\cite{smat:ref:alephlep1+2,
smat:ref:delphilep1+2, smat:ref:l3lep1, smat:ref:opallep1}.  These
results are averaged using the BLUE technique~\cite{BLUE:1988,
*BLUE:2003}.  Sources of systematic uncertainty correlated between the
experiments have been investigated using techniques described in
Reference~\cite{bib-Z-pole} and are accounted for in the averaging
procedure.

The main problem in the combination is the proper treatment of the
common systematic uncertainties.  The LEP experiments provide their
results in terms of the standard S-Matrix parametrisation.  This
parameter set is not well suited for the determination of common
systematic uncertainties since common errors such as the theory error
for luminosity affect many parameters.  Using a transformed parameter
set, which is defined as similar as possible to the standard LEP nine
parameter set, facilitates the study of common systematic errors as
well as cross checks with the LEP nine-parameter
combination~\cite{bib-Z-pole}.  The experiments' results are
transformed to this parameter set, combined, and the final results
transformed back to the standard S-Matrix parameter set.  The
transformed S-Matrix parameters are defined as follows:

\begin{eqnarray}
 \smxhad    & \equiv  & \frac{\rtoth} { \GZ^2} \\
 \Rsmxe     & \equiv  & \frac{\rtoth}{\rtote} \\
 \Rsmxm     & \equiv  & \frac{\rtoth}{\rtotm} \\
 \Rsmxt     & \equiv  & \frac{\rtoth}{\rtott} \\
 \Afbsmxe   & \equiv  & \frac{3}{4} \frac{\rfbe}{\rtote} \\
 \Afbsmxm   & \equiv  & \frac{3}{4} \frac{\rfbm}{\rtotm} \\
 \Afbsmxt   & \equiv  & \frac{3}{4} \frac{\rfbt}{\rtott} 
\end{eqnarray}

Table~\ref{smat:tab:lepinput:trans} gives the input of the four LEP
experiments for the 16 transformed S-Matrix parameters.  The
corresponding correlation matrices are given in
Tables~\ref{smat:apx:tab:corra},~\ref{smat:apx:tab:corrd},~\ref{smat:apx:tab:corrl}
and~\ref{smat:apx:tab:corro}.

Table~\ref{smat:tab:lepinput:corr-energy} shows the common systematic
uncertainty of the transformed S-Matrix parameters due to the
uncertainties in the LEP centre-of-mass energy.  The parameters $\MZ$
and $\jtoth$ are the most sensitive of all 16 S-Matrix parameters to
the inclusion of the {\LEPII} data, and are also the most interesting
ones in comparison to the five and nine parameter fits. For these
parameters the most significant source of systematic error correlated
between experiments arises from the uncertainty on the $\ee$
centre-of-mass energy. These errors amount to $\pm 3.2~\MeV$ on $\MZ$
and $\pm 0.16$ on $\jtoth$, with a correlation coefficient of $-0.86$.
Table~\ref{smat:tab:lepinput:corr-tch} specifies the common
uncertainties due to theoretical uncertainties in the calculation of
the $t$-channel contributions for Bhabha scattering.  In this case the
determination of the common error was complicated by the fact that the
experiments choose different procedures for the $t$-channel
correction, which yield different common errors.  We used the common
$t$-channel errors as determined by ALEPH~\cite{bib-Z-pole} as basis
for the combination since these result in the smallest common errors.
As a cross-check the combination was repeated with common $t$-channel
errors based on OPAL's analysis which yields the largest common
errors.  The effect on the combined result is small, the shift of
central values is below 20\,\% of its uncertainty.  In this
parametrisation, the luminosity theory uncertainty affects only the
parameter $\smxhad$. The uncertainties are $0.061\,\%$ for ALEPH,
DELPHI and L3, and $0.054\,\%$ for OPAL.

The result of the {\LEPI} combination in terms of the transformed
S-Matrix parameters is listed in Table~\ref{smat:tab:lepcomb:trans},
Table~\ref{smat:tab:lepcomb:trans-corr} shows the corresponding
correlation matrix.  Transforming this result back to the standard
S-Matrix parameter set, the combination is reported in
Tables~\ref{smat:tab:lepcomb:std} and~\ref{smat:tab:lepcomb:std-corr}.
The $\chidf$ for the average of all 16 parameters is 59.8/48,
corresponding to a probability of $12\%$.

\section{Discussion}
\label{smat:sec:discuss}

In the {\LEPI} combination the measured values of the Z boson mass
$\MZ = 91.1929 \pm 0.0059~\GeV$ agrees well with the results of the
standard nine parameter fit, $91.1876 \pm 0.0021~\GeV$, albeit with a
significantly larger error, resulting from the correlation with the
large uncertainty on $\jtoth$. This uncertainty is the dominant source
of uncertainty on $\MZ$ in the S-Matrix fits.  The measured value of
$\jtoth = -0.10 \pm 0.33 $ also agrees with the prediction of the
SM, $0.2201^{+0.0032}_{-0.0137}$.

\clearpage

\begin{table}[p]
\begin{center}
\renewcommand{\arraystretch}{1.3}
\begin{tabular}{|l||r|r|r|r|}
\hline
Parameter  & ALEPH & DELPHI & L3 & OPAL  \\
\hline
\hline
$\MZ$ (GeV)  & $91.2143  \pm   0.0120 $ & $  91.1939  \pm    0.0112    $  & $  91.1893  \pm    0.0112 $     & $  91.1903 \pm     0.0114 $  \\ 
$\GZ$ (GeV)  & $ 2.4900  \pm   0.0052 $ & $   2.4861  \pm    0.0048    $  & $   2.5028  \pm    0.0046 $     & $   2.4935 \pm     0.0047 $  \\
$\smxhad$    & $ 0.47736 \pm   0.00068$ & $   0.47713 \pm    0.00080   $  & $   0.47660 \pm    0.00063$     & $   0.47629\pm     0.00064$  \\ 
$\jtoth$     & $ -1.2618 \pm   0.6500 $ & $  -0.2067  \pm    0.6364    $  & $   0.2109  \pm    0.6370 $     & $   0.0017 \pm     0.6419 $  \\
$\Rsmxe$     & $ 20.8010 \pm   0.0830 $ & $  20.9270  \pm    0.1200    $  & $  20.8528  \pm    0.0977 $     & $  20.9718 \pm     0.0945 $  \\
$\Rsmxm$     & $ 20.8360 \pm   0.0580 $ & $  20.6600  \pm    0.0773    $  & $  20.8790  \pm    0.0982 $     & $  20.8484 \pm     0.0589 $  \\
$\Rsmxt$     & $ 20.6860 \pm   0.0640 $ & $  20.8250  \pm    0.1277    $  & $  20.7546  \pm    0.1339 $     & $  20.8255 \pm     0.0918 $  \\
$\jtote$     & $ -0.0531 \pm   0.0500 $ & $  -0.0939  \pm    0.0750    $  & $  -0.0293  \pm    0.0542 $     & $  -0.0856 \pm     0.0528 $  \\
$\jtotm$     & $ -0.0646 \pm   0.0430 $ & $   0.0561  \pm    0.0421    $  & $   0.0355  \pm    0.0459 $     & $  -0.0131 \pm     0.0415 $  \\
$\jtott$     & $ -0.0449 \pm   0.0440 $ & $   0.0040  \pm    0.0464    $  & $   0.0729  \pm    0.0476 $     & $  -0.0073 \pm     0.0442 $  \\
$\Afbsmxe$   & $  0.0164 \pm   0.0034 $ & $   0.0163  \pm    0.0048    $  & $   0.0091  \pm    0.0059 $     & $   0.0071 \pm     0.0046 $  \\
$\Afbsmxm$   & $  0.0178 \pm   0.0027 $ & $   0.0145  \pm    0.0026    $  & $   0.0179  \pm    0.0034 $     & $   0.0140 \pm     0.0024 $  \\
$\Afbsmxt$   & $  0.0180 \pm   0.0030 $ & $   0.0215  \pm    0.0038    $  & $   0.0238  \pm    0.0049 $     & $   0.0126 \pm     0.0031 $  \\
$\jfbe$      & $  0.8599 \pm   0.0570 $ & $   0.8021  \pm    0.0748    $  & $   0.6983  \pm    0.0797 $     & $   0.7640 \pm     0.0715 $  \\ 
$\jfbm$      & $  0.8196 \pm   0.0400 $ & $   0.7110  \pm    0.0366    $  & $   0.8192  \pm    0.0474 $     & $   0.7319 \pm     0.0363 $  \\ 
$\jfbt$      & $  0.8481 \pm   0.0430 $ & $   0.7070  \pm    0.0472    $  & $   0.7536  \pm    0.0550 $     & $   0.7394 \pm     0.0420 $  \\ 
\hline
\end{tabular}
\end{center}
\caption{Transformed {\LEPI} S-Matrix input parameters
 of the four LEP experiments}
\label{smat:tab:lepinput:trans}
\end{table}

\begin{table}[p]
{\small
\begin{center}
\begin{tabular}{|r||rrrrrrrrrr|}
\hline
\multicolumn{11}{|c|}{Parameters} \\
\hline
& 1 & 2 & 3 & 4 & 5 & 8 & 9 & 10 & 11 & 14 \\
&    $\MZ$       
&    $\GZ$       
&    $\smxhad$   
&    $\jtoth$    
&    $\Rsmxe$    
&    $\jtote$    
&    $\jtotm$    
&    $\jtott$    
&    $\Afbsmxe$  
&    $\jfbe$     \\
\hline
  1   &3.2e-03& -1.4e-03& 1.2e-04& -2.1e-02& 5.1e-03&  -4.4e-03& -4.4e-03& -4.5e-03& -8.3e-04& 1.3e-03  \\  
  2   &-1.4e-03& 1.4e-03& -3.2e-04& 9.2e-03& -3.0e-03& 1.8e-03& 2.0e-03& 2.0e-03& 4.4e-04    & -6.9e-04 \\  
  3   &1.2e-04& -3.2e-04& 1.3e-04& -1.2e-03& 9.7e-04&  8.4e-05& -2.2e-04& -2.5e-04& -1.2e-04 & 2.0e-04  \\  
  4   &-2.1e-02& 9.2e-03& -1.2e-03& 1.6e-01& -3.6e-02& 3.3e-02& 3.3e-02& 3.4e-02& 5.7e-03    & -9.3e-03 \\  
  5   &5.1e-03& -3.0e-03& 9.7e-04& -3.6e-02& 1.6e-02&  -7.3e-03& -7.5e-03& -7.6e-03& -2.6e-03& 3.5e-03  \\
  8   &-4.4e-03& 1.8e-03& 8.4e-05& 3.3e-02& -7.3e-03&  7.0e-03& 7.1e-03& 7.2e-03& 1.2e-03    & -1.8e-03 \\  
  9   &-4.4e-03& 2.0e-03& -2.2e-04& 3.3e-02& -7.5e-03& 7.1e-03& 7.0e-03& 7.2e-03& 1.2e-03    & -2.0e-03 \\  
 10   &-4.5e-03& 2.0e-03& -2.5e-04& 3.4e-02& -7.6e-03& 7.2e-03& 7.2e-03& 7.3e-03& 1.2e-03    & -2.0e-03 \\ 
 11   &-8.3e-04& 4.4e-04& -1.2e-04& 5.7e-03& -2.6e-03& 1.2e-03& 1.2e-03& 1.2e-03& 4.3e-04    & -5.4e-04 \\  
 14   &1.3e-03& -6.9e-04& 2.0e-04& -9.3e-03& 3.5e-03& -1.8e-03& -2.0e-03& -2.0e-03& -5.4e-04&  1.4e-03  \\
\hline
\end{tabular}
\end{center}
\caption{Signed square-root of {\LEPI} covariance matrix for common energy errors}
\label{smat:tab:lepinput:corr-energy}
}
\end{table}

\begin{table}[p]
\begin{center}
\begin{tabular}{|rl||rrrr|}
\hline
\multicolumn{2}{|l||}{Parameters} & 5 & 8 & 11 & 14  \\
\hline
\hline
  5 &      $\Rsmxe$   &   2.4e-02   & -3.20e-03  & -5.00e-03  &  -3.20e-03    \\
  8 &      $\jtote$   &  -3.20e-03  &  0.89e-02  &  0.00000   &   0.99e-02    \\
 11 &      $\Afbsmxe$ &  -5.00e-03  &  0.00000   &  1.00e-03  &  -0.32e-03    \\
 14 &      $\jfbe$    &  -3.20e-03  &  0.99e-02  & -0.32e-03  &   1.10e-02    \\
\hline                                                                                                                             
\end{tabular}
\caption{Signed square-root of {\LEPI} covariance matrix for common $t$-channel errors }
\label{smat:tab:lepinput:corr-tch}
\end{center}    
\end{table}                                                                                            

\begin{table}[p]
\begin{center}
\begin{tabular}{|l||r|r|r|r|}
\hline
Parameter & {\LEPI} \\
\hline
\hline
$\MZ$ (GeV)                & $  91.1929  \pm 0.0059  $  \\ 
$\GZ$ (GeV)                & $   2.4940  \pm 0.0026  $   \\
$\smxhad$                  & $   0.47676 \pm 0.00043  $  \\ 
$\jtoth$                   & $  -0.10    \pm 0.33  $   \\
$\Rsmxe$                   & $  20.865   \pm 0.052  $   \\
$\Rsmxm$                   & $  20.811   \pm 0.034  $   \\
$\Rsmxt$                   & $  20.746   \pm 0.045  $   \\
$\jtote$                   & $ -0.054    \pm 0.029  $    \\
$\jtotm$                   & $  0.013    \pm 0.022  $   \\
$\jtott$                   & $  0.014    \pm 0.023  $   \\
$\Afbsmxe$                 & $  0.0132   \pm 0.0023  $   \\
$\Afbsmxm$                 & $  0.0153   \pm 0.0014  $   \\
$\Afbsmxt$                 & $  0.0170   \pm 0.0017  $  \\
$\jfbe$                    & $  0.792    \pm 0.037  $  \\ 
$\jfbm$                    & $  0.763    \pm 0.020  $  \\ 
$\jfbt$                    & $  0.766    \pm 0.023  $  \\ 
\hline
$\chidf$                   & $ 59.96 / 48 $ \\
\hline                                   
\end{tabular}
\caption{ {\LEPI} combination result for transformed S-Matrix parameters }
\label{smat:tab:lepcomb:trans}
\end{center}    
\end{table}

\begin{table}[p]
{\small
\begin{sideways}                                                                                                                    
\begin{minipage}[b]{\textheight}                                                                                                    
\begin{center}
\begin{tabular}{|r||rrrrrrrrrrrrrrrr|}                                                                                
\hline                                                                                                                             
\multicolumn{17}{|c|}{Parameters} \\
\hline
& 1 & 2 & 3 & 4 & 5 & 6 & 7 & 8 & 9 & 10 & 11 & 12 & 13 & 14 & 15 & 16\\                                                                                                    
&      $\MZ$      
&      $\GZ$      
&      $\smxhad$  
&      $\jtoth$   
&      $\Rsmxe$   
&      $\Rsmxm$   
&      $\Rsmxt$   
&      $\jtote$   
&      $\jtotm$   
&      $\jtott$   
&      $\Afbsmxe$ 
&      $\Afbsmxm$ 
&      $\Afbsmxt$ 
&      $\jfbe$    
&      $\jfbm$    
&      $\jfbt$    \\
\hline                                                                                                                             
\hline                                                                                                                             
  1  &   1.000 & -0.435 &  0.083 & -0.936 &  0.330 & -0.007 & -0.006 & -0.597 & -0.665 & -0.630 & -0.128 &  0.221 &  0.182 & -0.009 & -0.006 &  0.005 \\
  2  &  -0.435 &  1.000 & -0.307 &  0.442 & -0.164 &  0.006 &  0.004 &  0.254 &  0.319 &  0.301 &  0.062 & -0.096 & -0.079 &  0.011 &  0.041 &  0.030  \\
  3  &   0.083 & -0.307 &  1.000 & -0.081 &  0.134 &  0.130 &  0.093 & -0.056 & -0.065 & -0.063 & -0.013 &  0.026 &  0.022 & -0.003 & -0.004 & -0.002 \\ 
  4  &  -0.936 &  0.442 & -0.081 &  1.000 & -0.317 &  0.014 &  0.011 &  0.604 &  0.679 &  0.645 &  0.121 & -0.221 & -0.182 &  0.010 &  0.007 & -0.004  \\
  5  &   0.330 & -0.164 &  0.134 & -0.317 &  1.000 &  0.053 &  0.035 & -0.276 & -0.228 & -0.215 & -0.407 &  0.082 &  0.067 & -0.020 & -0.002 &  0.002  \\
  6  &  -0.007 &  0.006 &  0.130 &  0.014 &  0.053 &  1.000 &  0.059 &  0.005 & -0.128 &  0.005 &  0.002 & -0.008 & -0.002 & -0.000 & -0.045 & -0.000  \\
  7  &  -0.006 &  0.004 &  0.093 &  0.011 &  0.035 &  0.059 &  1.000 &  0.005 &  0.005 & -0.109 &  0.002 & -0.002 &  0.000 &  0.000 & -0.000 & -0.057  \\
  8  &  -0.597 &  0.254 & -0.056 &  0.604 & -0.276 &  0.005 &  0.005 &  1.000 &  0.433 &  0.408 &  0.084 & -0.148 & -0.123 &  0.221 &  0.003 & -0.004  \\
  9  &  -0.665 &  0.319 & -0.065 &  0.679 & -0.228 & -0.128 &  0.005 &  0.433 &  1.000 &  0.460 &  0.086 & -0.137 & -0.131 &  0.007 & -0.034 & -0.003  \\
 10  &  -0.630 &  0.301 & -0.063 &  0.645 & -0.215 &  0.005 & -0.109 &  0.408 &  0.460 &  1.000 &  0.081 & -0.150 & -0.107 &  0.007 &  0.005 & -0.046  \\
 11  &  -0.128 &  0.062 & -0.013 &  0.121 & -0.407 &  0.002 &  0.002 &  0.084 &  0.086 &  0.081 &  1.000 & -0.024 & -0.019 &  0.092 &  0.001 & -0.001  \\
 12  &   0.221 & -0.096 &  0.026 & -0.221 &  0.082 & -0.008 & -0.002 & -0.148 & -0.137 & -0.150 & -0.024 &  1.000 &  0.061 & -0.005 &  0.198 &  0.002  \\
 13  &   0.182 & -0.079 &  0.022 & -0.182 &  0.067 & -0.002 &  0.000 & -0.123 & -0.131 & -0.107 & -0.019 &  0.061 &  1.000 & -0.004 & -0.001 &  0.181  \\
 14  &  -0.009 &  0.011 & -0.003 &  0.010 & -0.020 & -0.000 &  0.000 &  0.221 &  0.007 &  0.007 &  0.092 & -0.005 & -0.004 &  1.000 &  0.001 &  0.000  \\
 15  &  -0.006 &  0.041 & -0.004 &  0.007 & -0.002 & -0.045 & -0.000 &  0.003 & -0.034 &  0.005 &  0.001 &  0.198 & -0.001 &  0.001 &  1.000 &  0.002  \\
 16  &   0.005 &  0.030 & -0.002 & -0.004 &  0.002 & -0.000 & -0.057 & -0.004 & -0.003 & -0.046 & -0.001 &  0.002 &  0.181 &  0.000 &  0.002 &  1.000  \\
\hline                                                                                                                             
\end{tabular}\end{center}   
\caption{Correlation matrix for transformed {\LEPI} S-Matrix parameters }          
\label{smat:tab:lepcomb:trans-corr}
\end{minipage}
\end{sideways}                                                                                                      
}
\end{table}

\begin{table}[p]
\begin{center}
\begin{tabular}{|l||r|r|r|r|}
\hline
Parameter  & {\LEPI} \\
\hline
\hline
$\MZ$ (GeV)                & $  91.1929  \pm 0.0059  $  \\ 
$\GZ$ (GeV)                & $   2.4940  \pm 0.0026  $   \\
$\rtoth$                   & $   2.9654  \pm 0.0060  $  \\ 
$\jtoth$                   & $  -0.10    \pm 0.33  $   \\
$\rtote$                   & $   0.14214 \pm 0.00049  $   \\
$\rtotm$                   & $   0.14249 \pm 0.00036  $   \\
$\rtott$                   & $   0.14294 \pm 0.00042  $   \\
$\jtote$                   & $  -0.054   \pm 0.029  $    \\
$\jtotm$                   & $   0.013   \pm 0.022  $   \\
$\jtott$                   & $   0.014   \pm 0.023  $   \\
$\rfbe$                    & $   0.00251 \pm 0.00045  $   \\
$\rfbm$                    & $   0.00291 \pm 0.00026  $   \\
$\rfbt$                    & $   0.00324 \pm 0.00033  $  \\
$\jfbe$                    & $   0.792   \pm 0.036  $  \\ 
$\jfbm$                    & $   0.763   \pm 0.020  $  \\ 
$\jfbt$                    & $   0.766   \pm 0.023  $  \\ 
\hline
$\chidf$                   & $ 59.84 / 48 $ \\
\hline                                   
\end{tabular}
\caption{ {\LEPI} combination result for standard S-Matrix parameters }
\label{smat:tab:lepcomb:std}
\end{center}    
\end{table}                                                                                            

\begin{table}[p]
{\small
\begin{sideways}                                                                                                                    
\begin{minipage}[b]{\textheight}                                                                                                    
\begin{center}
\begin{tabular}{|r||rrrrrrrrrrrrrrrr|}                                                                                
\hline                                                                                                                             
\multicolumn{17}{|c|}{Parameters} \\
\hline
& 1 & 2 & 3 & 4 & 5 & 6 & 7 & 8 & 9 & 10 & 11 & 12 & 13 & 14 & 15 & 16 \\
& $\MZ$ 
& $\GZ$ 
& $\rtoth$ 
& $\jtoth$ 
& $\rtote$ 
& $\rtotm$ 
& $\rtott$ 
& $\jtote$ 
& $\jtotm$ 
& $\jtott$ 
& $\rfbe$  
& $\rfbm$  
& $\rfbt$  
& $\jfbe$  
& $\jfbm$  
& $\jfbt$  \\
\hline                                                                                                                             
\hline                                                                                                                             
  1    &   1.000 & -0.434 & -0.416 & -0.936 & -0.493 & -0.330 & -0.285 & -0.597 & -0.664 & -0.630 & -0.138 &  0.212 &  0.174 & -0.008 & -0.006 &  0.005  \\
  2    &  -0.434 &  1.000 &  0.905 &  0.441 &  0.660 &  0.725 &  0.628 &  0.254 &  0.319 &  0.300 &  0.076 & -0.075 & -0.060 &  0.012 &  0.041 &  0.030   \\
  3    &  -0.416 &  0.905 &  1.000 &  0.424 &  0.678 &  0.764 &  0.663 &  0.240 &  0.303 &  0.285 &  0.073 & -0.066 & -0.053 &  0.011 &  0.041 &  0.031  \\ 
  4    &  -0.936 &  0.441 &  0.424 &  1.000 &  0.488 &  0.332 &  0.287 &  0.605 &  0.678 &  0.645 &  0.131 & -0.212 & -0.174 &  0.009 &  0.007 & -0.004   \\
  5    &  -0.493 &  0.660 &  0.678 &  0.488 &  1.000 &  0.546 &  0.472 &  0.347 &  0.349 &  0.329 &  0.356 & -0.098 & -0.079 &  0.022 &  0.026 &  0.017   \\
  6    &  -0.330 &  0.725 &  0.764 &  0.332 &  0.546 &  1.000 &  0.534 &  0.190 &  0.327 &  0.226 &  0.058 & -0.037 & -0.041 &  0.009 &  0.062 &  0.025   \\
  7    &  -0.285 &  0.628 &  0.663 &  0.287 &  0.472 &  0.534 &  1.000 &  0.163 &  0.207 &  0.280 &  0.049 & -0.045 & -0.021 &  0.007 &  0.028 &  0.064   \\
  8    &  -0.597 &  0.254 &  0.240 &  0.605 &  0.347 &  0.190 &  0.163 &  1.000 &  0.433 &  0.408 &  0.091 & -0.143 & -0.118 &  0.219 &  0.003 & -0.004   \\
  9    &  -0.664 &  0.319 &  0.303 &  0.678 &  0.349 &  0.327 &  0.207 &  0.433 &  1.000 &  0.460 &  0.093 & -0.128 & -0.125 &  0.007 & -0.034 & -0.003   \\
 10    &  -0.630 &  0.300 &  0.285 &  0.645 &  0.329 &  0.226 &  0.280 &  0.408 &  0.460 &  1.000 &  0.087 & -0.143 & -0.099 &  0.007 &  0.005 & -0.046   \\
 11    &  -0.138 &  0.076 &  0.073 &  0.131 &  0.356 &  0.058 &  0.049 &  0.091 &  0.093 &  0.087 &  1.000 & -0.025 & -0.020 &  0.093 &  0.001 & -0.000   \\
 12    &   0.212 & -0.075 & -0.066 & -0.212 & -0.098 & -0.037 & -0.045 & -0.143 & -0.128 & -0.143 & -0.025 &  1.000 &  0.059 & -0.005 &  0.200 &  0.003   \\
 13    &   0.174 & -0.060 & -0.053 & -0.174 & -0.079 & -0.041 & -0.021 & -0.118 & -0.125 & -0.099 & -0.020 &  0.059 &  1.000 & -0.004 &  0.000 &  0.183   \\
 14    &  -0.008 &  0.012 &  0.011 &  0.009 &  0.022 &  0.009 &  0.007 &  0.219 &  0.007 &  0.007 &  0.093 & -0.005 & -0.004 &  1.000 &  0.001 &  0.000   \\
 15    &  -0.006 &  0.041 &  0.041 &  0.007 &  0.026 &  0.062 &  0.028 &  0.003 & -0.034 &  0.005 &  0.001 &  0.200 &  0.000 &  0.001 &  1.000 &  0.002   \\
 16    &   0.005 &  0.030 &  0.031 & -0.004 &  0.017 &  0.025 &  0.064 & -0.004 & -0.003 & -0.046 & -0.000 &  0.003 &  0.183 &  0.000 &  0.002 &  1.000   \\
\hline                                                                                                                             
\end{tabular}\end{center}   
\caption{Correlation matrix for standard {\LEPI} S-Matrix parameters }          
\label{smat:tab:lepcomb:std-corr}
\end{minipage}
\end{sideways}                                                                                                      
}
\end{table}

\clearpage

\section{S-Matrix Combination Tables}
\label{smat:comb:appendix}

See Tables~\ref{smat:apx:tab:corra}--\ref{smat:apx:tab:corro}.

\begin{table}[htbp]
{\small
\begin{sideways}                                                                                                                    
\begin{minipage}[b]{\textheight}                                                                            
\begin{center}
\begin{tabular}{|r||rrrrrrrrrrrrrrrr|}                                                                                
\hline                                                                                                                             
\multicolumn{17}{|c|}{Parameters} \\
\hline
& 1 & 2 & 3 & 4 & 5 & 6 & 7 & 8 & 9 & 10 & 11 & 12 & 13 & 14 & 15 & 16\\                                                                                                    
&      $\MZ$      
&      $\GZ$      
&      $\smxhad$  
&      $\jtoth$   
&      $\Rsmxe$   
&      $\Rsmxm$   
&      $\Rsmxt$   
&      $\jtote$   
&      $\jtotm$   
&      $\jtott$   
&      $\Afbsmxe$ 
&      $\Afbsmxm$ 
&      $\Afbsmxt$ 
&      $\jfbe$    
&      $\jfbm$    
&      $\jfbt$    \\
\hline                                                                                                                             
\hline                                                                                                                             
  1  &   1.000&  -0.537&   0.243&  -0.963&   0.449&  -0.004&  -0.015&  -0.592&  -0.685&  -0.676&  -0.209&   0.313&   0.296&   0.005&  -0.023&   0.003 \\
  2  &  -0.537&   1.000&  -0.436&   0.547&  -0.234&   0.008&   0.008&   0.324&   0.391&   0.385&   0.106&  -0.169&  -0.160&   0.014&   0.056&   0.040  \\
  3  &   0.243&  -0.436&   1.000&  -0.225&   0.219&   0.160&   0.143&  -0.144&  -0.171&  -0.168&  -0.041&   0.087&   0.082&   0.000&  -0.012&  -0.005 \\ 
  4  &  -0.963&   0.547&  -0.225&   1.000&  -0.426&   0.011&   0.021&   0.593&   0.685&   0.676&   0.197&  -0.307&  -0.290&  -0.003&   0.024&  -0.002  \\
  5  &   0.449&  -0.234&   0.219&  -0.426&   1.000&   0.070&   0.051&  -0.400&  -0.307&  -0.301&  -0.413&   0.139&   0.131&  -0.047&  -0.011&   0.001  \\
  6  &  -0.004&   0.008&   0.160&   0.011&   0.070&   1.000&   0.089&   0.002&  -0.171&   0.003&   0.001&  -0.008&  -0.001&  -0.001&  -0.036&   0.000  \\
  7  &  -0.015&   0.008&   0.143&   0.021&   0.051&   0.089&   1.000&   0.011&   0.011&  -0.142&   0.003&  -0.005&  -0.007&   0.000&   0.000&  -0.038  \\
  8  &  -0.592&   0.324&  -0.144&   0.593&  -0.400&   0.002&   0.011&   1.000&   0.422&   0.411&   0.133&  -0.189&  -0.179&   0.159&   0.014&  -0.002  \\
  9  &  -0.685&   0.391&  -0.171&   0.685&  -0.307&  -0.171&   0.011&   0.422&   1.000&   0.481&   0.141&  -0.198&  -0.206&  -0.002&  -0.015&  -0.002  \\
 10  &  -0.676&   0.385&  -0.168&   0.676&  -0.301&   0.003&  -0.142&   0.411&   0.481&   1.000&   0.139&  -0.215&  -0.193&  -0.002&   0.017&  -0.050  \\
 11  &  -0.209&   0.106&  -0.041&   0.197&  -0.413&   0.001&   0.003&   0.133&   0.141&   0.139&   1.000&  -0.055&  -0.053&   0.159&   0.005&   0.000  \\
 12  &   0.313&  -0.169&   0.087&  -0.307&   0.139&  -0.008&  -0.005&  -0.189&  -0.198&  -0.215&  -0.055&   1.000&   0.105&   0.000&   0.231&   0.002  \\
 13  &   0.296&  -0.160&   0.082&  -0.290&   0.131&  -0.001&  -0.007&  -0.179&  -0.206&  -0.193&  -0.053&   0.105&   1.000&   0.000&  -0.006&   0.202  \\
 14  &   0.005&   0.014&   0.000&  -0.003&  -0.047&  -0.001&   0.000&   0.159&  -0.002&  -0.002&   0.159&   0.000&   0.000&   1.000&   0.001&   0.001  \\
 15  &  -0.023&   0.056&  -0.012&   0.024&  -0.011&  -0.036&   0.000&   0.014&  -0.015&   0.017&   0.005&   0.231&  -0.006&   0.001&   1.000&   0.003  \\
 16  &   0.003&   0.040&  -0.005&  -0.002&   0.001&   0.000&  -0.038&  -0.002&  -0.002&  -0.050&   0.000&   0.002&   0.202&   0.001&   0.003&   1.000  \\
\hline                                                                                                                             
\end{tabular}\end{center}                                                                                   
\caption{Correlation matrix of transformed {\LEPI} S-Matrix input parameters for ALEPH.}
\label{smat:apx:tab:corra}
\end{minipage}                                                                                                                      
\end{sideways}                                                                                                                      
}
\end{table}

\begin{table}[htbp]
{\small
\begin{sideways}                                                                                                                    
\begin{minipage}[b]{\textheight}
\begin{center}
\begin{tabular}{|r||rrrrrrrrrrrrrrrr|}                                                                                
\hline                                                                                                                             
\multicolumn{17}{|c|}{Parameters} \\
\hline
& 1 & 2 & 3 & 4 & 5 & 6 & 7 & 8 & 9 & 10 & 11 & 12 & 13 & 14 & 15 & 16\\                                                                                                    
&      $\MZ$      
&      $\GZ$      
&      $\smxhad$  
&      $\jtoth$   
&      $\Rsmxe$   
&      $\Rsmxm$   
&      $\Rsmxt$   
&      $\jtote$   
&      $\jtotm$   
&      $\jtott$   
&      $\Afbsmxe$ 
&      $\Afbsmxm$ 
&      $\Afbsmxt$ 
&      $\jfbe$    
&      $\jfbm$    
&      $\jfbt$    \\
\hline                                                                                                                             
\hline                                                                                                                             
  1   &   1.000 & -0.504 &  0.123 & -0.966 &  0.034 & -0.030 &  0.002 & -0.804 & -0.702 & -0.640 &  0.133 &  0.253 &  0.173 & -0.029 & -0.002 & -0.003 \\
  2   &  -0.504 &  1.000 & -0.285 &  0.528 & -0.018 &  0.008 & -0.004 &  0.403 &  0.385 &  0.350 & -0.069 & -0.125 & -0.086 &  0.040 &  0.043 &  0.034  \\
  3   &   0.123 & -0.285 &  1.000 & -0.112 &  0.124 &  0.185 &  0.113 & -0.098 & -0.092 & -0.085 &  0.018 &  0.033 &  0.022 & -0.003 &  0.003 &  0.002 \\ 
  4   &  -0.966 &  0.528 & -0.112 &  1.000 & -0.027 &  0.037 &  0.002 &  0.786 &  0.695 &  0.634 & -0.131 & -0.247 & -0.169 &  0.030 &  0.004 &  0.005  \\
  5   &   0.034 & -0.018 &  0.124 & -0.027 &  1.000 &  0.053 &  0.033 & -0.061 & -0.023 & -0.021 & -0.100 &  0.009 &  0.006 & -0.066 & -0.000 & -0.000  \\
  6   &  -0.030 &  0.008 &  0.185 &  0.037 &  0.053 &  1.000 &  0.051 &  0.025 & -0.086 &  0.019 & -0.005 & -0.013 & -0.006 &  0.001 & -0.056 & -0.000  \\
  7   &   0.002 & -0.004 &  0.113 &  0.002 &  0.033 &  0.051 &  1.000 & -0.002 & -0.001 & -0.089 &  0.000 &  0.000 & -0.002 & -0.000 & -0.000 & -0.079  \\
  8   &  -0.804 &  0.403 & -0.098 &  0.786 & -0.061 &  0.025 & -0.002 &  1.000 &  0.571 &  0.521 & -0.081 & -0.205 & -0.140 &  0.102 &  0.001 &  0.003  \\
  9   &  -0.702 &  0.385 & -0.092 &  0.695 & -0.023 & -0.086 & -0.001 &  0.571 &  1.000 &  0.461 & -0.095 & -0.158 & -0.123 &  0.022 & -0.038 &  0.004  \\
 10   &  -0.640 &  0.350 & -0.085 &  0.634 & -0.021 &  0.019 & -0.089 &  0.521 &  0.461 &  1.000 & -0.086 & -0.164 & -0.090 &  0.020 &  0.003 & -0.035  \\
 11   &   0.133 & -0.069 &  0.018 & -0.131 & -0.100 & -0.005 &  0.000 & -0.081 & -0.095 & -0.086 &  1.000 &  0.044 &  0.029 &  0.087 &  0.001 & -0.000  \\
 12   &   0.253 & -0.125 &  0.033 & -0.247 &  0.009 & -0.013 &  0.000 & -0.205 & -0.158 & -0.164 &  0.044 &  1.000 &  0.053 & -0.008 &  0.196 & -0.000  \\
 13   &   0.173 & -0.086 &  0.022 & -0.169 &  0.006 & -0.006 & -0.002 & -0.140 & -0.123 & -0.090 &  0.029 &  0.053 &  1.000 & -0.005 &  0.001 &  0.176  \\
 14   &  -0.029 &  0.040 & -0.003 &  0.030 & -0.066 &  0.001 & -0.000 &  0.102 &  0.022 &  0.020 &  0.087 & -0.008 & -0.005 &  1.000 &  0.002 &  0.001  \\
 15   &  -0.002 &  0.043 &  0.003 &  0.004 & -0.000 & -0.056 & -0.000 &  0.001 & -0.038 &  0.003 &  0.001 &  0.196 &  0.001 &  0.002 &  1.000 &  0.002  \\
 16   &  -0.003 &  0.034 &  0.002 &  0.005 & -0.000 & -0.000 & -0.079 &  0.003 &  0.004 & -0.035 & -0.000 & -0.000 &  0.176 &  0.001 &  0.002 &  1.000  \\
\hline                                                                                                                             
\end{tabular}\end{center}                                                                                   
\caption{Correlation matrix of transformed {\LEPI} S-Matrix input parameters for DELPHI.}
\label{smat:apx:tab:corrd}
\end{minipage}                                                                                                                      
\end{sideways}                                                                                                                      
}
\end{table}
                   
\begin{table}[htbp]
{\small
\begin{sideways}                                                                                                                    
\begin{minipage}[b]{\textheight}
\begin{center}
\begin{tabular}{|r||rrrrrrrrrrrrrrrr|}                                                                                
\hline                                                                                                                             
\multicolumn{17}{|c|}{Parameters} \\
\hline
& 1 & 2 & 3 & 4 & 5 & 6 & 7 & 8 & 9 & 10 & 11 & 12 & 13 & 14 & 15 & 16\\                                                                                                    
&      $\MZ$      
&      $\GZ$      
&      $\smxhad$  
&      $\jtoth$   
&      $\Rsmxe$   
&      $\Rsmxm$   
&      $\Rsmxt$   
&      $\jtote$   
&      $\jtotm$   
&      $\jtott$   
&      $\Afbsmxe$ 
&      $\Afbsmxm$ 
&      $\Afbsmxt$ 
&      $\jfbe$    
&      $\jfbm$    
&      $\jfbt$    \\
\hline                                                                                                                             
\hline                                                                                                                             
  1   &  1.000 & -0.378 &  0.024 & -0.959 &  0.418 & -0.010 & -0.013 & -0.528 & -0.627 & -0.600 & -0.200 &  0.226 &  0.150 &  0.011 & -0.006 &  0.008 \\
  2   & -0.378 &  1.000 & -0.331 &  0.410 & -0.165 &  0.002 &  0.006 &  0.196 &  0.271 &  0.262 &  0.076 & -0.087 & -0.057 & -0.009 &  0.042 &  0.028  \\
  3   &  0.024 & -0.331 &  1.000 & -0.020 &  0.076 &  0.079 &  0.055 & -0.029 & -0.025 & -0.025 &  0.010 &  0.011 &  0.007 & -0.008 & -0.002 & -0.001 \\ 
  4   & -0.959 &  0.410 & -0.020 &  1.000 & -0.403 &  0.015 &  0.017 &  0.528 &  0.627 &  0.600 &  0.195 & -0.220 & -0.146 & -0.009 &  0.007 & -0.006  \\
  5   &  0.418 & -0.165 &  0.076 & -0.403 &  1.000 &  0.024 &  0.016 & -0.274 & -0.267 & -0.256 & -0.202 &  0.107 &  0.070 &  0.027 & -0.003 &  0.003  \\
  6   & -0.010 &  0.002 &  0.079 &  0.015 &  0.024 &  1.000 &  0.021 &  0.006 & -0.104 &  0.006 &  0.002 & -0.007 & -0.002 &  0.000 & -0.068 & -0.001  \\
  7   & -0.013 &  0.006 &  0.055 &  0.017 &  0.016 &  0.021 &  1.000 &  0.007 &  0.008 & -0.078 &  0.002 & -0.003 & -0.004 &  0.000 & -0.000 & -0.080  \\
  8   & -0.528 &  0.196 & -0.029 &  0.528 & -0.274 &  0.006 &  0.007 &  1.000 &  0.346 &  0.331 &  0.097 & -0.121 & -0.080 &  0.166 &  0.002 & -0.005  \\
  9   & -0.627 &  0.271 & -0.025 &  0.627 & -0.267 & -0.104 &  0.008 &  0.346 &  1.000 &  0.393 &  0.127 & -0.119 & -0.096 & -0.006 & -0.041 & -0.004  \\
 10   & -0.600 &  0.262 & -0.025 &  0.600 & -0.256 &  0.006 & -0.078 &  0.331 &  0.393 &  1.000 &  0.122 & -0.138 & -0.075 & -0.006 &  0.005 & -0.039  \\
 11   & -0.200 &  0.076 &  0.010 &  0.195 & -0.202 &  0.002 &  0.002 &  0.097 &  0.127 &  0.122 &  1.000 & -0.051 & -0.034 &  0.026 &  0.001 & -0.002  \\
 12   &  0.226 & -0.087 &  0.011 & -0.220 &  0.107 & -0.007 & -0.003 & -0.121 & -0.119 & -0.138 & -0.051 &  1.000 &  0.038 &  0.003 &  0.170 &  0.002  \\
 13   &  0.150 & -0.057 &  0.007 & -0.146 &  0.070 & -0.002 & -0.004 & -0.080 & -0.096 & -0.075 & -0.034 &  0.038 &  1.000 &  0.002 & -0.001 &  0.150  \\
 14   &  0.011 & -0.009 & -0.008 & -0.009 &  0.027 &  0.000 &  0.000 &  0.166 & -0.006 & -0.006 &  0.026 &  0.003 &  0.002 &  1.000 & -0.001 & -0.000  \\
 15   & -0.006 &  0.042 & -0.002 &  0.007 & -0.003 & -0.068 & -0.000 &  0.002 & -0.041 &  0.005 &  0.001 &  0.170 & -0.001 & -0.001 &  1.000 &  0.002  \\
 16   &  0.008 &  0.028 & -0.001 & -0.006 &  0.003 & -0.001 & -0.080 & -0.005 & -0.004 & -0.039 & -0.002 &  0.002 &  0.150 & -0.000 &  0.002 &  1.000  \\
\hline                                                                                                                             
\end{tabular}\end{center}                                                                                   
\caption{Correlation matrix of transformed {\LEPI} S-Matrix input parameters for L3.}
\label{smat:apx:tab:corrl}
\end{minipage}                                                                                                                      
\end{sideways}                                                                                                                      
}
\end{table}

\begin{table}[htbp]
{\small
\begin{sideways}                                                                                                                    
\begin{minipage}[b]{\textheight}
\begin{center}
\begin{tabular}{|r||rrrrrrrrrrrrrrrr|}                                                                                
\hline                                                                                                                             
\multicolumn{17}{|c|}{Parameters} \\
\hline
& 1 & 2 & 3 & 4 & 5 & 6 & 7 & 8 & 9 & 10 & 11 & 12 & 13 & 14 & 15 & 16\\                                                                                                    
&      $\MZ$      
&      $\GZ$      
&      $\smxhad$  
&      $\jtoth$   
&      $\Rsmxe$   
&      $\Rsmxm$   
&      $\Rsmxt$   
&      $\jtote$   
&      $\jtotm$   
&      $\jtott$   
&      $\Afbsmxe$ 
&      $\Afbsmxm$ 
&      $\Afbsmxt$ 
&      $\jfbe$    
&      $\jfbm$    
&      $\jfbt$    \\
\hline                                                                                                                             
\hline                                                                                                                             
  1   &  1.000 & -0.446 &  0.120 & -0.963 &  0.442 &  0.012 &  0.013 & -0.525 & -0.703 & -0.651 & -0.244 &  0.299 &  0.262 &  0.025 &  0.001 &  0.013 \\
  2   & -0.446 &  1.000 & -0.360 &  0.462 & -0.194 &  0.008 &  0.001 &  0.224 &  0.338 &  0.315 &  0.110 & -0.131 & -0.114 & -0.012 &  0.043 &  0.032  \\
  3   &  0.120 & -0.360 &  1.000 & -0.110 &  0.188 &  0.221 &  0.141 & -0.067 & -0.090 & -0.084 & -0.018 &  0.039 &  0.037 & -0.002 & -0.007 & -0.004 \\ 
  4   & -0.963 &  0.462 & -0.110 &  1.000 & -0.428 & -0.005 & -0.009 &  0.525 &  0.701 &  0.650 &  0.239 & -0.293 & -0.256 & -0.024 &  0.000 & -0.012  \\
  5   &  0.442 & -0.194 &  0.188 & -0.428 &  1.000 &  0.085 &  0.043 & -0.278 & -0.315 & -0.292 & -0.298 &  0.151 &  0.131 &  0.023 &  0.001 &  0.008  \\
  6   &  0.012 &  0.008 &  0.221 & -0.005 &  0.085 &  1.000 &  0.056 & -0.006 & -0.133 & -0.007 & -0.004 & -0.002 &  0.004 &  0.001 & -0.037 &  0.002  \\
  7   &  0.013 &  0.001 &  0.141 & -0.009 &  0.043 &  0.056 &  1.000 & -0.008 & -0.009 & -0.100 & -0.003 &  0.004 &  0.017 & -0.000 &  0.001 & -0.060  \\
  8   & -0.525 &  0.224 & -0.067 &  0.525 & -0.278 & -0.006 & -0.008 &  1.000 &  0.383 &  0.354 &  0.125 & -0.160 & -0.140 &  0.218 & -0.002 & -0.008  \\
  9   & -0.703 &  0.338 & -0.090 &  0.701 & -0.315 & -0.133 & -0.009 &  0.383 &  1.000 &  0.473 &  0.174 & -0.193 & -0.187 & -0.017 & -0.041 & -0.009  \\
 10   & -0.651 &  0.315 & -0.084 &  0.650 & -0.292 & -0.007 & -0.100 &  0.354 &  0.473 &  1.000 &  0.161 & -0.198 & -0.157 & -0.016 &  0.001 & -0.056  \\
 11   & -0.244 &  0.110 & -0.018 &  0.239 & -0.298 & -0.004 & -0.003 &  0.125 &  0.174 &  0.161 &  1.000 & -0.083 & -0.072 &  0.056 & -0.000 & -0.004  \\
 12   &  0.299 & -0.131 &  0.039 & -0.293 &  0.151 & -0.002 &  0.004 & -0.160 & -0.193 & -0.198 & -0.083 &  1.000 &  0.090 &  0.008 &  0.179 &  0.005  \\
 13   &  0.262 & -0.114 &  0.037 & -0.256 &  0.131 &  0.004 &  0.017 & -0.140 & -0.187 & -0.157 & -0.072 &  0.090 &  1.000 &  0.007 &  0.001 &  0.175  \\
 14   &  0.025 & -0.012 & -0.002 & -0.024 &  0.023 &  0.001 & -0.000 &  0.218 & -0.017 & -0.016 &  0.056 &  0.008 &  0.007 &  1.000 & -0.000 &  0.000  \\
 15   &  0.001 &  0.043 & -0.007 &  0.000 &  0.001 & -0.037 &  0.001 & -0.002 & -0.041 &  0.001 & -0.000 &  0.179 &  0.001 & -0.000 &  1.000 &  0.002  \\
 16   &  0.013 &  0.032 & -0.004 & -0.012 &  0.008 &  0.002 & -0.060 & -0.008 & -0.009 & -0.056 & -0.004 &  0.005 &  0.175 &  0.000 &  0.002 &  1.000  \\
\hline                                                                                                                             
\end{tabular}\end{center}                                                                                   
\caption{Correlation matrix of transformed {\LEPI} S-Matrix input parameters for OPAL.}
\label{smat:apx:tab:corro}
\end{minipage}                                                                                                                      
\end{sideways}                                                                                                                      
}
\end{table}

\chapter{Two-Fermion Combination Details}
\section{Input Measurements}
\label{app:2F:inputs}

In this section, the experimental measurements of total cross-sections
and forward-backward asymmetries as used in the combination are
reported.  For each result, the ZFITTER prediction, followed by the
measured value and the six error components as described in
Section~\ref{sec:ff:ave-xsc-afb}, are listed.  The results are
extrapolated to $4\pi$ acceptance ($|\cos\theta| \le 1$) except for
ALEPH ($|\cos\theta|<0.95$).
 
\subsection*{ALEPH}
 
\subsubsection*{ALEPH results at 130 GeV}
 
\begin{verbatim}
* E_CM = 130.200 GeV
*
XSEC_QQ       71.15    71.6    3.8     0.64      0.82      0.29     0.19  0.22
XSEC_MUMU      6.987    7.9    1.22    0.041     0.008     0.04     0.02  0.077
XSEC_TAUTAU    7.234   10.9    1.79    0.152     0.22      0.29     0.03  0.137
AFB_MUMU       0.698    0.83   0.09    0.004     0.026     0.004    0.0   0.01
AFB_TAUTAU     0.697    0.56   0.12    0.011     0.035     0.004    0.0   0.01
\end{verbatim}
 
\subsubsection*{ALEPH results at 136 GeV}
 
\begin{verbatim}
* E_CM = 136.200 GeV
*
XSEC_QQ       57.64     58.8    3.5     0.53      0.67      0.23     0.15  0.18
XSEC_MUMU      6.053     6.9    1.1     0.04      0.007     0.034    0.02  0.076
XSEC_TAUTAU    6.267     5.6    1.3     0.073     0.11      0.15     0.01  0.1
AFB_MUMU       0.678     0.63   0.105   0.004     0.024     0.004    0.0   0.01
AFB_TAUTAU     0.677     0.65   0.14    0.009     0.028     0.004    0.0   0.012
\end{verbatim}
 
\subsubsection*{ALEPH results at 161 GeV}
 
\begin{verbatim}
* E_CM = 161.314 GeV
*                                                       
XSEC_QQ       30.88     29.9    1.8     0.21      0.29      0.16     0.08  0.09
XSEC_MUMU      3.857     4.5   	0.69    0.03      0.008     0.027    0.01  0.06
XSEC_TAUTAU    3.992     5.75   0.94    0.08      0.13      0.17     0.01  0.17
AFB_MUMU       0.609     0.63   0.11    0.004     0.026     0.004    0.0   0.009
AFB_TAUTAU     0.608     0.48   0.14    0.009     0.029     0.004    0.0   0.008
\end{verbatim}
 
\subsubsection*{ALEPH results at 172 GeV}
 
\begin{verbatim}
* E_CM = 172.086 GeV
*
XSEC_QQ       25.22    26.4    1.7     0.18      0.30     0.18     0.06   0.08
XSEC_MUMU      3.30     2.64   0.53    0.042     0.008    0.021    0.006  0.04
XSEC_TAUTAU    3.415    3.26   0.74    0.04      0.07     0.04     0.008  0.07
AFB_MUMU       0.593    0.72   0.14    0.005     0.034    0.005    0.0    0.01
AFB_TAUTAU     0.592    0.44   0.16    0.009     0.029    0.004    0.0    0.01
\end{verbatim}
 
\subsubsection*{ALEPH results at 183 GeV}
 
\begin{verbatim}
* E_CM = 183.00  GeV
*
XSEC_QQ       21.24    21.71   0.70    0.13      0.12    0.126   0.06   0.07
XSEC_MUMU      2.871    2.98   0.24    0.045     0.004   0.019   0.012  0.05
XSEC_TAUTAU    2.974    2.90   0.29    0.048     0.067   0.011   0.012  0.06
AFB_MUMU       0.579    0.54   0.06    0.02      0.01    0.004   0.0    0.008
AFB_TAUTAU     0.579    0.52   0.08    0.03      0.02    0.004   0.0    0.009
\end{verbatim}
 
\subsubsection*{ALEPH results at 189 GeV}
 
\begin{verbatim}
* E_CM = 189 GeV
*
XSEC_QQ       20.580  20.800   0.380   0.156  0.108  0.021  0.052  0.021
XSEC_MUMU      2.831   2.879   0.134   0.007  0.014  0.000  0.007  0.004
XSEC_TAUTAU    2.910   2.787   0.198   0.020  0.014  0.000  0.007  0.020
AFB_MUMU       0.570   0.576   0.036   0.001  0.000  0.000  0.000  0.009
AFB_TAUTAU     0.570   0.598   0.046   0.007  0.000  0.000  0.000  0.010
\end{verbatim}
 
\subsubsection*{ALEPH results at 192 GeV}
 
\begin{verbatim}
* E_CM = 192 GeV
*
XSEC_QQ       19.720   20.070  0.920   0.151  0.111  0.020  0.050  0.040
XSEC_MUMU      2.729   2.862   0.333   0.008  0.013  0.000  0.004  0.004
XSEC_TAUTAU    2.811   2.600   0.467   0.062  0.011  0.000  0.003  0.020
AFB_MUMU       0.567   0.580   0.088   0.001  0.000  0.000  0.000  0.009
AFB_TAUTAU     0.567   0.490   0.124   0.006  0.000  0.000  0.000  0.009 
\end{verbatim}
 
\subsubsection*{ALEPH results at 196 GeV}
 
\begin{verbatim}
* E_CM = 196 GeV
*
XSEC_QQ       18.670  18.930   0.540   0.144  0.115  0.015  0.047  0.038
XSEC_MUMU      2.611   2.704   0.193   0.014  0.012  0.000  0.003  0.004
XSEC_TAUTAU    2.69    2.551   0.289   0.012  0.012  0.000  0.003  0.020
AFB_MUMU       0.563   0.553   0.057   0.001  0.000  0.000  0.000  0.006 
AFB_TAUTAU     0.563   0.543   0.075   0.007  0.000  0.000  0.000  0.010
\end{verbatim}
 
\subsubsection*{ALEPH results at 200 GeV}
 
\begin{verbatim}
* E_CM = 200 GeV
*
XSEC_QQ       17.690  17.940   0.510   0.138  0.113  0.014  0.045  0.036
XSEC_MUMU      2.502   2.991   0.195   0.015  0.012  0.000  0.004  0.005  
XSEC_TAUTAU    2.571   2.881   0.293   0.012  0.012  0.000  0.003  0.021 
AFB_MUMU       0.560   0.442   0.056   0.003  0.000  0.000  0.000  0.006
AFB_TAUTAU     0.560   0.445   0.073   0.005  0.000  0.000  0.000  0.009
\end{verbatim}
 
\subsubsection*{ALEPH results at 202 GeV}
 
\begin{verbatim}
*E_CM = 202 GeV  
*
XSEC_QQ       17.210  17.560   0.710   0.137  0.133  0.012  0.044  0.035
XSEC_MUMU      2.442   2.639   0.262   0.015  0.011  0.000  0.003  0.005  
XSEC_TAUTAU    2.512   2.832   0.411   0.012  0.011  0.000  0.003  0.021 
AFB_MUMU       0.558   0.573   0.078   0.001  0.000  0.000  0.000  0.010
AFB_TAUTAU     0.557   0.654   0.090   0.008  0.000  0.000  0.000  0.012
\end{verbatim}
 
\subsubsection*{ALEPH results at 205 GeV}
 
\begin{verbatim}
*E_CM = 205 GeV    
*
XSEC_QQ       16.510  16.940   0.520   0.129  0.100  0.012  0.042  0.034
XSEC_MUMU      2.358   1.918   0.162   0.014  0.011  0.000  0.003  0.005 
XSEC_TAUTAU    2.434   2.430   0.290   0.016  0.010  0.000  0.003  0.020 
AFB_MUMU       0.555   0.572   0.066   0.003  0.000  0.000  0.000  0.008
AFB_TAUTAU     0.555   0.593   0.075   0.007  0.000  0.000  0.000  0.011
\end{verbatim}
 
\subsubsection*{ALEPH results at 207 GeV}
 
\begin{verbatim}
*E_CM = 207 GeV 
*
XSEC_QQ       16.160  16.340   0.380   0.124  0.087  0.011  0.041  0.033
XSEC_MUMU      2.318   2.458   0.143   0.014  0.010  0.000  0.003  0.005 
XSEC_TAUTAU    2.383   2.101   0.212   0.015  0.010  0.000  0.003  0.021 
AFB_MUMU       0.554   0.572   0.066   0.001  0.000  0.000  0.000  0.009
AFB_TAUTAU     0.554   0.568   0.062   0.007  0.000  0.000  0.000  0.011
\end{verbatim}
 
\subsection*{DELPHI}
 
\subsubsection*{DELPHI results at 130 GeV}
 
\begin{verbatim}
* Centre-of-mass energy used: 130.200 GeV
*
XSEC_QQ       82.506   82.400   5.200   0.411     0.296     0.000    0.098  2.509
XSEC_MUMU      8.107    9.700   1.900   0.015     0.000     0.000    0.000  0.359
XSEC_TAUTAU    8.312   10.200   3.100   0.009     0.037     0.000    0.012  0.714
AFB_MUMU       0.719    0.670   0.150   0.000     0.000     0.000    0.000  0.003
AFB_TAUTAU     0.719    0.730   0.170   0.000     0.000     0.000    0.000  0.020
\end{verbatim}
 
\subsubsection*{DELPHI results at 136 GeV}
 
\begin{verbatim}
* Centre-of-mass energy used: 136.20 GeV
*
XSEC_QQ       66.362   65.300   4.700   0.326     0.241     0.000    0.078  2.010
XSEC_MUMU      6.997    6.600   1.600   0.010     0.000     0.000    0.000  0.244
XSEC_TAUTAU    7.173    8.800   3.000   0.008     0.033     0.000    0.011  0.616
AFB_MUMU       0.699    0.740   0.160   0.000     0.000     0.000    0.000  0.003
AFB_TAUTAU     0.699    0.490   0.230   0.000     0.000     0.000    0.000  0.020
\end{verbatim}
 
\subsubsection*{DELPHI results at 161 GeV}
 
\begin{verbatim}
* Centre-of-mass energy used: 161.30 GeV
*
XSEC_QQ       35.119   41.000   2.100   0.215     0.162     0.000    0.051  1.223
XSEC_MUMU      4.426    3.600   0.700   0.019     0.000     0.000    0.000  0.126
XSEC_TAUTAU    4.538    5.100   1.200   0.025     0.016     0.000    0.006  0.357
AFB_MUMU       0.629    0.430   0.160   0.000     0.000     0.000    0.000  0.003
AFB_TAUTAU     0.628    0.920   0.080   0.000     0.000     0.000    0.000  0.020
\end{verbatim}
 
\subsubsection*{DELPHI results at 172 GeV}
 
\begin{verbatim}
* Centre-of-mass energy used: 172.10 GeV
*
XSEC_QQ       28.745   30.400   1.900   0.176     0.159     0.000    0.042  0.932
XSEC_MUMU      3.790    3.600   0.700   0.019     0.000     0.000    0.000  0.122
XSEC_TAUTAU    3.886    4.500   1.100   0.023     0.020     0.000    0.005  0.315
AFB_MUMU       0.610    0.940   0.140   0.000     0.000     0.000    0.000  0.003
AFB_TAUTAU     0.610    0.130   0.200   0.000     0.000     0.000    0.000  0.020
\end{verbatim}
 
\subsubsection*{DELPHI results at 183 GeV}
 
\begin{verbatim}
* Centre-of-mass energy used: 182.65 GeV
*
XSEC_QQ       24.154   25.500   0.796   0.272     0.057     0.026    0.137  0.056
XSEC_MUMU      3.304    3.605   0.284   0.027     0.000     0.000    0.000  0.011
XSEC_TAUTAU    3.387    3.292   0.376   0.071     0.006     0.011    0.013  0.000
AFB_MUMU       0.596    0.588   0.064   0.001     0.000     0.000    0.000  0.001
AFB_TAUTAU     0.596    0.671   0.080   0.011     0.000     0.002    0.002  0.000
\end{verbatim}
 
\subsubsection*{DELPHI results at 189 GeV}
 
\begin{verbatim}
* Centre-of-mass energy used: 188.63 GeV
*
XSEC_QQ       22.099   22.630   0.452   0.257     0.034     0.023    0.136  0.040
XSEC_MUMU      3.072    3.071   0.150   0.023     0.000     0.000    0.000  0.008
XSEC_TAUTAU    3.150    3.105   0.215   0.065     0.003     0.011    0.013  0.000
AFB_MUMU       0.589    0.600   0.039   0.001     0.000     0.000    0.000  0.001
AFB_TAUTAU     0.589    0.697   0.048   0.011     0.000     0.002    0.002  0.000
\end{verbatim}
 
\subsubsection*{DELPHI results at 192 GeV}
 
\begin{verbatim}
* Centre-of-mass energy used: 191.58 GeV
*
XSEC_QQ       21.191   22.140   1.119   0.255     0.098     0.022    0.136  0.072
XSEC_MUMU      2.967    2.822   0.357   0.021     0.000     0.000    0.000  0.006
XSEC_TAUTAU    3.042    2.497   0.479   0.053     0.007     0.008    0.011  0.000
AFB_MUMU       0.586    0.636   0.098   0.001     0.000     0.000    0.000  0.001
AFB_TAUTAU     0.586    0.578   0.150   0.011     0.000     0.002    0.002  0.000
\end{verbatim}
 
\subsubsection*{DELPHI results at 196 GeV}
 
\begin{verbatim}
* Centre-of-mass energy used: 195.51 GeV
*
XSEC_QQ       20.075   21.180   0.634   0.249     0.058     0.021    0.136  0.053
XSEC_MUMU      2.837    2.763   0.207   0.020     0.000     0.000    0.000  0.006
XSEC_TAUTAU    2.908    2.895   0.301   0.062     0.006     0.010    0.012  0.000
AFB_MUMU       0.582    0.586   0.061   0.001     0.000     0.000    0.000  0.000
AFB_TAUTAU     0.582    0.465   0.083   0.011     0.000     0.002    0.002  0.000
\end{verbatim}
 
\subsubsection*{DELPHI results at 200 GeV}
 
\begin{verbatim}
* Centre-of-mass energy used: 199.51 GeV
*
XSEC_QQ       19.035   19.450   0.591   0.240     0.054     0.020    0.135  0.051
XSEC_MUMU      2.713    3.080   0.207   0.023     0.000     0.000    0.000  0.007
XSEC_TAUTAU    2.781    2.614   0.270   0.056     0.005     0.009    0.011  0.000
AFB_MUMU       0.578    0.548   0.056   0.001     0.000     0.000    0.000  0.000
AFB_TAUTAU     0.578    0.540   0.080   0.011     0.000     0.002    0.002  0.000
\end{verbatim}
 
\subsubsection*{DELPHI results at 202 GeV}
 
\begin{verbatim}
* Centre-of-mass energy used: 201.64 GeV
*
XSEC_QQ       18.517   18.880   0.843   0.237     0.077     0.019    0.135  0.066
XSEC_MUMU      2.650    2.464   0.268   0.018     0.000     0.000    0.000  0.005
XSEC_TAUTAU    2.717    2.550   0.380   0.054     0.007     0.009    0.011  0.000
AFB_MUMU       0.577    0.544   0.090   0.001     0.000     0.000    0.000  0.001
AFB_TAUTAU     0.576    0.464   0.122   0.011     0.000     0.002    0.002  0.000
\end{verbatim}
 
\subsubsection*{DELPHI results at 205 GeV}
 
\begin{verbatim}
* Centre-of-mass energy used: 204.87 GeV
*
XSEC_QQ       17.775   17.670   0.580   0.230     0.053     0.018    0.135  0.042
XSEC_MUMU      2.560    2.345   0.188   0.017     0.000     0.000    0.000  0.005
XSEC_TAUTAU    2.625    2.803   0.282   0.059     0.006     0.010    0.012  0.000
AFB_MUMU       0.574    0.642   0.061   0.001     0.000     0.000    0.000  0.001
AFB_TAUTAU     0.574    0.709   0.068   0.011     0.000     0.002    0.002  0.000
\end{verbatim}
 
\subsubsection*{DELPHI results at 207 GeV}
 
\begin{verbatim}
* Centre-of-mass energy used: 206.55 GeV
*
XSEC_QQ       17.408   17.040   0.444   0.228     0.040     0.017    0.135  0.033
XSEC_MUMU      2.515    2.475   0.145   0.018     0.000     0.000    0.000  0.004
XSEC_TAUTAU    2.578    2.534   0.210   0.055     0.004     0.009    0.011  0.000
AFB_MUMU       0.573    0.558   0.048   0.001     0.000     0.000    0.000  0.001
AFB_TAUTAU     0.572    0.666   0.059   0.011     0.000     0.002    0.002  0.000
\end{verbatim}
 
\subsection*{L3}
 
\subsubsection*{L3 results at 130 GeV}
 
\begin{verbatim}
* Exact centre-of-mass energy: 130.0  GeV
*
XSEC_QQ       83.5    84.2     4.4    0.96   0.05   0.27   0.03   0.0
XSEC_MUMU      8.5     8.2     1.4    0.200  0.006  0.012  0.004  0.0
XSEC_TAUTAU    8.5     9.8     1.9    0.300  0.006  0.010  0.004  0.0
AFB_MUMU       0.707   0.67    0.11   0.020  0.0    0.004  0.0    0.0
AFB_TAUTAU     0.707   0.78    0.16   0.020  0.0    0.004  0.0    0.0
\end{verbatim}
 
\subsubsection*{L3 results at 136 GeV}
 
\begin{verbatim}
* Exact centre-of-mass energy: 136.1  GeV
*
XSEC_QQ       66.9    66.6     3.9    0.77   0.05   0.22   0.03   0.0
XSEC_MUMU      7.3     6.9     1.4    0.300  0.006  0.012  0.004  0.0
XSEC_TAUTAU    7.3     7.5     1.8    0.300  0.006  0.010  0.004  0.0
AFB_MUMU       0.686   0.75    0.11   0.050  0.0    0.004  0.0    0.0
AFB_TAUTAU     0.686   0.96    0.17   0.030  0.0    0.004  0.0    0.0
\end{verbatim}
 
\subsubsection*{L3 results at 161 GeV}
 
\begin{verbatim}
* Exact centre-of-mass energy: 161.3  GeV
*
XSEC_QQ       35.4    37.3     2.2    0.69   0.05   0.12   0.03   0.0
XSEC_MUMU      4.70    4.59    0.84   0.180  0.006  0.012  0.004  0.0
XSEC_TAUTAU    4.7     4.6     1.1    0.300  0.006  0.010  0.004  0.0
AFB_MUMU       0.619   0.59    0.15   0.050  0.0    0.004  0.0    0.0
AFB_TAUTAU     0.619   0.97    0.25   0.020  0.0    0.004  0.0    0.0
\end{verbatim}
 
\subsubsection*{L3 results at 172 GeV}
 
\begin{verbatim}
* Exact centre-of-mass energy: 172.1  GeV
*
XSEC_QQ       28.8    28.2     2.2    0.59   0.05   0.09   0.03   0.0
XSEC_MUMU      4.00    3.60    0.75   0.140  0.006  0.012  0.004  0.0
XSEC_TAUTAU    4.0     4.3     1.1    0.300  0.006  0.010  0.004  0.0
AFB_MUMU       0.598   0.31    0.195  0.050  0.0    0.004  0.0    0.0
AFB_TAUTAU     0.598   0.18    0.27   0.020  0.0    0.004  0.0    0.0
\end{verbatim}
 
\subsubsection*{L3 results at 183 GeV}
 
\begin{verbatim}
* Exact centre-of-mass energy: 182.7  GeV
*
XSEC_QQ       24.3    24.7     0.8    0.38   0.05   0.08   0.03   0.0
XSEC_MUMU      3.47    3.09    0.35   0.059  0.006  0.012  0.004  0.0
XSEC_TAUTAU    3.47    3.62    0.40   0.059  0.006  0.010  0.004  0.0
AFB_MUMU       0.582   0.62    0.08   0.020  0.0    0.004  0.0    0.0
AFB_TAUTAU     0.582   0.53    0.105  0.020  0.0    0.004  0.0    0.0
\end{verbatim}
 
\subsubsection*{L3 results at 189 GeV}
 
\begin{verbatim}
* Exact centre-of-mass energy: 188.7  GeV
*
XSEC_QQ       22.2    23.1     0.4    0.28   0.05   0.07   0.03   0.0
XSEC_MUMU      3.22    2.92    0.16   0.059  0.006  0.012  0.004  0.0
XSEC_TAUTAU    3.22    3.18    0.21   0.069  0.006  0.010  0.004  0.0
AFB_MUMU       0.573   0.58    0.04   0.020  0.0    0.004  0.0    0.0
AFB_TAUTAU     0.573   0.44    0.06   0.020  0.0    0.004  0.0    0.0
\end{verbatim}
 
\subsubsection*{L3 results at 192 GeV}
 
\begin{verbatim}
* Exact centre-of-mass energy: 191.6 GeV
*
XSEC_QQ       21.334  22.38    1.020  0.180  0.032  0.045  0.019  0.010
XSEC_MUMU      3.112   2.54    0.390  0.087  0.009  0.018  0.006  0.004
XSEC_TAUTAU    3.112   2.93    0.480  0.059  0.005  0.009  0.003  0.003
AFB_MUMU       0.571   0.69    0.120  0.069  0.000  0.014  0.000  0.004
AFB_TAUTAU     0.571   0.52    0.120  0.049  0.000  0.010  0.000  0.003
\end{verbatim}
 
\subsubsection*{L3 results at 196 GeV}
 
\begin{verbatim}
* Exact centre-of-mass energy: 195.5 GeV
*
XSEC_QQ       20.212  20.14    0.580  0.152  0.027  0.038  0.016  0.008
XSEC_MUMU      2.972   3.05    0.250  0.097  0.010  0.020  0.007  0.005
XSEC_TAUTAU    2.972   3.22    0.300  0.069  0.006  0.010  0.004  0.004
AFB_MUMU       0.566   0.53    0.070  0.039  0.000  0.008  0.000  0.002
AFB_TAUTAU     0.566   0.44    0.090  0.049  0.000  0.010  0.000  0.003
\end{verbatim}
 
\subsubsection*{L3 results at 200 GeV}
 
\begin{verbatim}
* Exact centre-of-mass energy: 199.6 GeV
*
XSEC_QQ       19.133  19.09    0.570  0.152  0.027  0.038  0.016  0.008
XSEC_MUMU      2.837   2.85    0.240  0.087  0.009  0.018  0.006  0.004
XSEC_TAUTAU    2.836   2.97    0.300  0.069  0.006  0.010  0.004  0.004
AFB_MUMU       0.561   0.44    0.080  0.039  0.000  0.008  0.000  0.002
AFB_TAUTAU     0.561   0.46    0.100  0.049  0.000  0.010  0.000  0.003
\end{verbatim}
 
\subsubsection*{L3 results at 202 GeV}
 
\begin{verbatim}
* Exact centre-of-mass energy: 201.8 GeV
*
XSEC_QQ       18.593  19.33    0.890  0.152  0.027  0.038  0.016  0.008
XSEC_MUMU      2.768   2.97    0.350  0.097  0.010  0.020  0.007  0.005
XSEC_TAUTAU    2.767   2.81    0.420  0.007  0.001  0.001  0.000  0.000
AFB_MUMU       0.559   0.59    0.090  0.020  0.000  0.004  0.000  0.001
AFB_TAUTAU     0.559   0.47    0.130  0.078  0.000  0.016  0.000  0.004
\end{verbatim}
 
\subsubsection*{L3 results at 205 GeV}
 
\begin{verbatim}
* Exact centre-of-mass energy: 204.9 GeV
*
XSEC_QQ       17.872  18.46    0.590  0.133  0.024  0.033  0.014  0.007
XSEC_MUMU      2.675   2.37    0.220  0.068  0.007  0.014  0.005  0.004
XSEC_TAUTAU    2.675   2.93    0.320  0.069  0.006  0.010  0.004  0.004
AFB_MUMU       0.556   0.48    0.090  0.029  0.000  0.006  0.000  0.002
AFB_TAUTAU     0.556   0.56    0.090  0.049  0.000  0.010  0.000  0.003
\end{verbatim}
 
\subsubsection*{L3 results at 207 GeV}
 
\begin{verbatim}
* Exact centre-of-mass energy: 206.5 GeV
*
XSEC_QQ       17.518  17.87    0.440  0.123  0.022  0.031  0.013  0.007
XSEC_MUMU      2.629   2.24    0.170  0.058  0.006  0.012  0.004  0.003
XSEC_TAUTAU    2.629   2.34    0.210  0.079  0.007  0.011  0.005  0.004
AFB_MUMU       0.554   0.54    0.060  0.020  0.000  0.004  0.000  0.001
AFB_TAUTAU     0.554   0.61    0.070  0.088  0.000  0.018  0.000  0.004
\end{verbatim}
 
\subsection*{OPAL}
 
\subsubsection*{OPAL results at 130 GeV}
 
\begin{verbatim}
* Exact centre-of-mass energy: 130.12 GeV
*
XSEC_QQ       83.078  79.30    3.8    1.25   0.52   0.47   0.20   0.54   
XSEC_MUMU      8.453   7.63    1.14   0.16   0.05   0.03   0.02   0.26
XSEC_TAUTAU    8.450   6.83    1.40   0.18   0.05   0.03   0.02   0.16
AFB_MUMU       0.705   0.40    0.15   0.02   0.0    0.004  0.0    0.0
AFB_TAUTAU     0.705   0.80    0.22   0.01   0.0    0.004  0.0    0.0
\end{verbatim}
 
\subsubsection*{OPAL results at 136 GeV}
 
\begin{verbatim}
* Exact centre-of-mass energy: 136.08 GeV
*
XSEC_QQ       66.875  66.30    3.3    1.04   0.43   0.40   0.17   0.47
XSEC_MUMU      7.298  10.37    1.31   0.16   0.07   0.03   0.03   0.23
XSEC_TAUTAU    7.295   7.32    1.39   0.19   0.05   0.02   0.02   0.15
AFB_MUMU       0.685   0.71    0.12   0.01   0.0    0.002  0.0    0.0
AFB_TAUTAU     0.684   0.86    0.20   0.01   0.0    0.003  0.0    0.0
\end{verbatim}
 
\subsubsection*{OPAL results at 161 GeV}
 
\begin{verbatim}
* Exact centre-of-mass energy: 161.34 GeV
*
XSEC_QQ       33.606  35.20    2.00   0.73   0.16   0.22   0.09   0.07
XSEC_MUMU      4.419   4.49    0.67   0.07   0.02   0.02   0.01   0.11
XSEC_TAUTAU    4.418   6.22    1.01   0.17   0.03   0.02   0.02   0.05
AFB_MUMU       0.609   0.45    0.14   0.01   0.0    0.005  0.0    0.0
AFB_TAUTAU     0.609   0.56    0.14   0.01   0.0    0.005  0.0    0.0
\end{verbatim}
 
\subsubsection*{OPAL results at 172 GeV}
 
\begin{verbatim}
* Exact centre-of-mass energy: 172.12 GeV
*
XSEC_QQ       27.566  26.80    1.80   0.57   0.13   0.16   0.07   0.05
XSEC_MUMU      3.790   3.56    0.59   0.08   0.02   0.02   0.01   0.11
XSEC_TAUTAU    3.789   3.85    0.78   0.11   0.02   0.02   0.01   0.06
AFB_MUMU       0.590   0.55    0.15   0.01   0.0    0.005  0.0    0.0
AFB_TAUTAU     0.590   0.56    0.19   0.01   0.0    0.005  0.0    0.0
\end{verbatim}
 
\subsubsection*{OPAL results at 183 GeV}
 
\begin{verbatim}
* Exact centre-of-mass energy: 182.69 GeV
*
XSEC_QQ       24.237  23.50    0.72   0.35   0.08   0.15   0.06   0.06
XSEC_MUMU      3.445   3.463   0.264  0.045  0.012  0.013  0.009  0.105
XSEC_TAUTAU    3.445   3.315   0.301  0.103  0.012  0.011  0.008  0.028
AFB_MUMU       0.576   0.543   0.071  0.011  0.0    0.004  0.0    0.0
AFB_TAUTAU     0.576   0.683   0.088  0.002  0.0    0.004  0.0    0.0
\end{verbatim}
 
\subsubsection*{OPAL results at 189 GeV}
 
\begin{verbatim}
* Exact centre-of-mass energy: 188.635 GeV
*
XSEC_QQ       22.188  21.99    0.37   0.09   0.04   0.09   0.03   0.03
XSEC_MUMU      3.206   3.142   0.145  0.033  0.005  0.005  0.004  0.007
XSEC_TAUTAU    3.206   3.445   0.211  0.085  0.006  0.002  0.004  0.020
AFB_MUMU       0.569   0.548   0.039  0.004  0.0    0.002  0.0    0.002
AFB_TAUTAU     0.569   0.591   0.054  0.008  0.0    0.001  0.0    0.010
\end{verbatim}
 
\subsubsection*{OPAL results at 192 GeV}
 
\begin{verbatim}
* Exact centre-of-mass energy: 191.590 GeV
*
XSEC_QQ       21.276  22.23    0.94   0.08   0.07   0.08   0.03   0.02
XSEC_MUMU      3.097   2.857   0.344  0.030  0.008  0.004  0.003  0.005
XSEC_TAUTAU    3.097   3.167   0.503  0.078  0.009  0.001  0.004  0.015
AFB_MUMU       0.566   0.341   0.102  0.004  0.0    0.002  0.0    0.002
AFB_TAUTAU     0.566   0.813   0.138  0.005  0.0    0.001  0.0    0.012
\end{verbatim}
 
\subsubsection*{OPAL results at 196 GeV}
 
\begin{verbatim}
* Exact centre-of-mass energy: 195.526 GeV
*
XSEC_QQ       20.154  19.78    0.55   0.08   0.04   0.07   0.02   0.02
XSEC_MUMU      2.961   2.932   0.215  0.031  0.007  0.004  0.004  0.004
XSEC_TAUTAU    2.961   2.893   0.298  0.072  0.007  0.001  0.003  0.011
AFB_MUMU       0.562   0.683   0.055  0.004  0.0    0.002  0.0    0.002
AFB_TAUTAU     0.562   0.373   0.103  0.013  0.0    0.001  0.0    0.005
\end{verbatim}
 
\subsubsection*{OPAL results at 200 GeV}
 
\begin{verbatim}
* Exact centre-of-mass energy: 199.522 GeV
*
XSEC_QQ       19.112  18.89    0.54   0.08   0.05   0.06   0.02   0.01
XSEC_MUMU      2.833   2.772   0.207  0.029  0.007  0.004  0.003  0.003
XSEC_TAUTAU    2.833   3.136   0.304  0.077  0.007  0.001  0.004  0.010
AFB_MUMU       0.558   0.637   0.059  0.004  0.0    0.002  0.0    0.001
AFB_TAUTAU     0.558   0.700   0.077  0.006  0.0    0.001  0.0    0.006
\end{verbatim}
 
\subsubsection*{OPAL results at 202 GeV}
 
\begin{verbatim}
* Exact centre-of-mass energy: 201.636 GeV
*
XSEC_QQ       18.596  18.54    0.77   0.08   0.05   0.05   0.02   0.01
XSEC_MUMU      2.768   2.363   0.280  0.025  0.006  0.004  0.003  0.003
XSEC_TAUTAU    2.768   2.954   0.430  0.072  0.008  0.001  0.004  0.009
AFB_MUMU       0.556   0.489   0.100  0.004  0.0    0.002  0.0    0.001
AFB_TAUTAU     0.556   0.440   0.130  0.010  0.0    0.001  0.0    0.004
\end{verbatim}
 
\subsubsection*{OPAL results at 205 GeV}
 
\begin{verbatim}
* Exact centre-of-mass energy: 204.881 GeV
*
XSEC_QQ       17.847  18.18    0.52   0.08   0.04   0.05   0.02   0.01
XSEC_MUMU      2.674   2.885   0.210  0.030  0.007  0.004  0.003  0.004
XSEC_TAUTAU    2.674   2.721   0.283  0.067  0.006  0.001  0.003  0.011
AFB_MUMU       0.553   0.512   0.063  0.004  0.0    0.002  0.0    0.002
AFB_TAUTAU     0.553   0.575   0.092  0.009  0.0    0.001  0.0    0.006
\end{verbatim}
 
\subsubsection*{OPAL results at 207 GeV}
 
\begin{verbatim}
* Exact centre-of-mass energy: 206.561 GeV
*
XSEC_QQ       17.479  16.81    0.39   0.08   0.04   0.04   0.02   0.02
XSEC_MUMU      2.627   2.766   0.158  0.029  0.006  0.004  0.003  0.005
XSEC_TAUTAU    2.627   2.782   0.219  0.068  0.006  0.001  0.003  0.013
AFB_MUMU       0.552   0.508   0.050  0.004  0.0    0.002  0.0    0.002
AFB_TAUTAU     0.552   0.472   0.075  0.010  0.0    0.001  0.0    0.005
\end{verbatim}

\section{Differential Cross-Section for Muon- and Tau-Pair Final States}
\label{app:2F:mt}

The following lists show for each centre-of-mass energy point (rounded
in GeV) the LEP-combined differential lepton-pair cross-sections (DC)
for $\mumu$ (MM) and $\tautau$ (TT) final states in 10
$\cos\theta$-bins ($1-10$) of constant width 0.2, comparing the LEP
average value and its total error with the SM prediction.  Also shown
is the overall $\chidf$ and the bin-by-bin $\chidf$ contribution.  The
overall matrix of correlation coefficients and inverse covariance
matrix are available at the LEPEWWG web site: {\tt
http://www.cern.ch/LEPEWWG}.

\begin{verbatim}
 Total chi2/NDF =  352.156/320
 183_DC_MM_1     average =  0.197 +-  0.183   SM=  0.547   chi2/NDF =   0.688/1
 183_DC_MM_2     average =  0.589 +-  0.163   SM=  0.534   chi2/NDF =   0.717/1
 183_DC_MM_3     average =  0.807 +-  0.174   SM=  0.627   chi2/NDF =   2.204/1
 183_DC_MM_4     average =  1.033 +-  0.197   SM=  0.823   chi2/NDF =   0.211/1
 183_DC_MM_5     average =  1.178 +-  0.236   SM=  1.121   chi2/NDF =   0.014/1
 183_DC_MM_6     average =  1.778 +-  0.276   SM=  1.521   chi2/NDF =   0.007/1
 183_DC_MM_7     average =  2.143 +-  0.315   SM=  2.020   chi2/NDF =   0.777/1
 183_DC_MM_8     average =  2.690 +-  0.367   SM=  2.619   chi2/NDF =   4.165/1
 183_DC_MM_9     average =  2.916 +-  0.420   SM=  3.314   chi2/NDF =   1.199/1
 183_DC_MM_10    average =  4.368 +-  0.529   SM=  4.096   chi2/NDF =   0.254/1
 183_DC_TT_1     average =  0.302 +-  0.351   SM=  0.548   chi2/NDF =   1.439/1
 183_DC_TT_2     average =  0.206 +-  0.240   SM=  0.535   chi2/NDF =   1.677/1
 183_DC_TT_3     average =  0.198 +-  0.230   SM=  0.627   chi2/NDF =   1.127/1
 183_DC_TT_4     average =  0.542 +-  0.254   SM=  0.823   chi2/NDF =   0.176/1
 183_DC_TT_5     average =  1.364 +-  0.302   SM=  1.121   chi2/NDF =   0.206/1
 183_DC_TT_6     average =  1.519 +-  0.350   SM=  1.521   chi2/NDF =   0.045/1
 183_DC_TT_7     average =  1.583 +-  0.389   SM=  2.020   chi2/NDF =   0.403/1
 183_DC_TT_8     average =  2.296 +-  0.450   SM=  2.619   chi2/NDF =   0.095/1
 183_DC_TT_9     average =  3.954 +-  0.574   SM=  3.313   chi2/NDF =   0.321/1
 183_DC_TT_10    average =  4.156 +-  0.919   SM=  4.095   chi2/NDF =   0.263/1
 189_DC_MM_1     average =  0.614 +-  0.080   SM=  0.532   chi2/NDF =   4.079/3
 189_DC_MM_2     average =  0.420 +-  0.065   SM=  0.514   chi2/NDF =   1.836/3
 189_DC_MM_3     average =  0.530 +-  0.069   SM=  0.595   chi2/NDF =   0.702/3
 189_DC_MM_4     average =  0.651 +-  0.077   SM=  0.772   chi2/NDF =   2.544/3
 189_DC_MM_5     average =  1.064 +-  0.089   SM=  1.044   chi2/NDF =  10.239/3
 189_DC_MM_6     average =  1.313 +-  0.111   SM=  1.411   chi2/NDF =   1.906/3
 189_DC_MM_7     average =  2.038 +-  0.123   SM=  1.872   chi2/NDF =   1.168/3
 189_DC_MM_8     average =  2.158 +-  0.139   SM=  2.426   chi2/NDF =   0.374/3
 189_DC_MM_9     average =  2.954 +-  0.158   SM=  3.072   chi2/NDF =   2.558/3
 189_DC_MM_10    average =  3.795 +-  0.216   SM=  3.799   chi2/NDF =   0.853/3
 189_DC_TT_1     average =  0.389 +-  0.123   SM=  0.532   chi2/NDF =   7.662/3
 189_DC_TT_2     average =  0.379 +-  0.093   SM=  0.515   chi2/NDF =   5.211/3
 189_DC_TT_3     average =  0.485 +-  0.089   SM=  0.595   chi2/NDF =  10.195/3
 189_DC_TT_4     average =  0.809 +-  0.100   SM=  0.772   chi2/NDF =   0.944/3
 189_DC_TT_5     average =  0.848 +-  0.118   SM=  1.044   chi2/NDF =   0.139/3
 189_DC_TT_6     average =  1.323 +-  0.139   SM=  1.411   chi2/NDF =   7.994/3
 189_DC_TT_7     average =  1.989 +-  0.154   SM=  1.872   chi2/NDF =   2.494/3
 189_DC_TT_8     average =  2.445 +-  0.179   SM=  2.426   chi2/NDF =   0.841/3
 189_DC_TT_9     average =  2.467 +-  0.225   SM=  3.071   chi2/NDF =   2.313/3
 189_DC_TT_10    average =  4.111 +-  0.357   SM=  3.798   chi2/NDF =   7.763/3
 192_DC_MM_1     average =  0.481 +-  0.198   SM=  0.524   chi2/NDF =   6.372/2
 192_DC_MM_2     average =  0.384 +-  0.173   SM=  0.504   chi2/NDF =   1.804/2
 192_DC_MM_3     average =  0.788 +-  0.186   SM=  0.579   chi2/NDF =   2.816/2
 192_DC_MM_4     average =  0.581 +-  0.212   SM=  0.748   chi2/NDF =   0.388/2
 192_DC_MM_5     average =  1.324 +-  0.248   SM=  1.008   chi2/NDF =   2.698/2
 192_DC_MM_6     average =  1.187 +-  0.292   SM=  1.360   chi2/NDF =   3.178/2
 192_DC_MM_7     average =  1.932 +-  0.334   SM=  1.803   chi2/NDF =   6.530/2
 192_DC_MM_8     average =  2.080 +-  0.379   SM=  2.337   chi2/NDF =   0.245/2
 192_DC_MM_9     average =  3.003 +-  0.429   SM=  2.960   chi2/NDF =   2.441/2
 192_DC_MM_10    average =  3.083 +-  0.552   SM=  3.662   chi2/NDF =   2.378/2
 192_DC_TT_1     average =  0.014 +-  0.325   SM=  0.524   chi2/NDF =   1.103/2
 192_DC_TT_2     average =  0.355 +-  0.247   SM=  0.505   chi2/NDF =   2.256/2
 192_DC_TT_3     average =  0.479 +-  0.245   SM=  0.580   chi2/NDF =   1.130/2
 192_DC_TT_4     average =  0.762 +-  0.278   SM=  0.748   chi2/NDF =   2.704/2
 192_DC_TT_5     average =  0.816 +-  0.331   SM=  1.008   chi2/NDF =   0.540/2
 192_DC_TT_6     average =  1.609 +-  0.385   SM=  1.360   chi2/NDF =   0.055/2
 192_DC_TT_7     average =  1.810 +-  0.433   SM=  1.803   chi2/NDF =   0.026/2
 192_DC_TT_8     average =  2.059 +-  0.491   SM=  2.337   chi2/NDF =   0.688/2
 192_DC_TT_9     average =  2.643 +-  0.599   SM=  2.959   chi2/NDF =   1.439/2
 192_DC_TT_10    average =  2.575 +-  0.935   SM=  3.661   chi2/NDF =   6.306/2
 196_DC_MM_1     average =  0.535 +-  0.119   SM=  0.512   chi2/NDF =   3.633/2
 196_DC_MM_2     average =  0.485 +-  0.103   SM=  0.491   chi2/NDF =   1.848/2
 196_DC_MM_3     average =  0.668 +-  0.111   SM=  0.560   chi2/NDF =   0.766/2
 196_DC_MM_4     average =  0.484 +-  0.126   SM=  0.718   chi2/NDF =   1.473/2
 196_DC_MM_5     average =  0.802 +-  0.147   SM=  0.964   chi2/NDF =   1.659/2
 196_DC_MM_6     average =  1.507 +-  0.172   SM=  1.298   chi2/NDF =   2.480/2
 196_DC_MM_7     average =  1.657 +-  0.197   SM=  1.720   chi2/NDF =   1.467/2
 196_DC_MM_8     average =  2.303 +-  0.223   SM=  2.229   chi2/NDF =   0.450/2
 196_DC_MM_9     average =  2.949 +-  0.253   SM=  2.824   chi2/NDF =   0.068/2
 196_DC_MM_10    average =  3.272 +-  0.325   SM=  3.495   chi2/NDF =   1.622/2
 196_DC_TT_1     average =  0.810 +-  0.211   SM=  0.513   chi2/NDF =   2.172/2
 196_DC_TT_2     average =  0.738 +-  0.147   SM=  0.491   chi2/NDF =   2.311/2
 196_DC_TT_3     average =  0.524 +-  0.141   SM=  0.560   chi2/NDF =   9.697/2
 196_DC_TT_4     average =  0.688 +-  0.162   SM=  0.718   chi2/NDF =   0.718/2
 196_DC_TT_5     average =  0.976 +-  0.195   SM=  0.964   chi2/NDF =   1.445/2
 196_DC_TT_6     average =  0.977 +-  0.225   SM=  1.298   chi2/NDF =   0.257/2
 196_DC_TT_7     average =  1.648 +-  0.252   SM=  1.719   chi2/NDF =   3.406/2
 196_DC_TT_8     average =  1.965 +-  0.289   SM=  2.228   chi2/NDF =   0.535/2
 196_DC_TT_9     average =  2.269 +-  0.357   SM=  2.823   chi2/NDF =   1.278/2
 196_DC_TT_10    average =  3.346 +-  0.557   SM=  3.494   chi2/NDF =   0.714/2
 200_DC_MM_1     average =  0.558 +-  0.113   SM=  0.501   chi2/NDF =   1.899/2
 200_DC_MM_2     average =  0.376 +-  0.098   SM=  0.478   chi2/NDF =   3.670/2
 200_DC_MM_3     average =  0.799 +-  0.105   SM=  0.541   chi2/NDF =   2.306/2
 200_DC_MM_4     average =  0.817 +-  0.118   SM=  0.689   chi2/NDF =   2.762/2
 200_DC_MM_5     average =  1.105 +-  0.139   SM=  0.922   chi2/NDF =   1.269/2
 200_DC_MM_6     average =  1.462 +-  0.162   SM=  1.239   chi2/NDF =   0.517/2
 200_DC_MM_7     average =  1.849 +-  0.185   SM=  1.640   chi2/NDF =   0.217/2
 200_DC_MM_8     average =  2.122 +-  0.211   SM=  2.126   chi2/NDF =   5.430/2
 200_DC_MM_9     average =  2.947 +-  0.239   SM=  2.694   chi2/NDF =   0.365/2
 200_DC_MM_10    average =  3.474 +-  0.306   SM=  3.336   chi2/NDF =   0.435/2
 200_DC_TT_1     average =  0.489 +-  0.201   SM=  0.501   chi2/NDF =   0.340/2
 200_DC_TT_2     average =  0.619 +-  0.141   SM=  0.478   chi2/NDF =   7.623/2
 200_DC_TT_3     average =  0.528 +-  0.137   SM=  0.541   chi2/NDF =   0.040/2
 200_DC_TT_4     average =  0.628 +-  0.155   SM=  0.689   chi2/NDF =   0.631/2
 200_DC_TT_5     average =  1.067 +-  0.186   SM=  0.922   chi2/NDF =   2.966/2
 200_DC_TT_6     average =  1.130 +-  0.214   SM=  1.239   chi2/NDF =   1.361/2
 200_DC_TT_7     average =  1.871 +-  0.240   SM=  1.640   chi2/NDF =   0.687/2
 200_DC_TT_8     average =  2.043 +-  0.274   SM=  2.125   chi2/NDF =   0.684/2
 200_DC_TT_9     average =  2.777 +-  0.339   SM=  2.694   chi2/NDF =   1.916/2
 200_DC_TT_10    average =  3.437 +-  0.523   SM=  3.336   chi2/NDF =   0.841/2
 202_DC_MM_1     average =  1.137 +-  0.162   SM=  0.495   chi2/NDF =   3.111/2
 202_DC_MM_2     average =  0.295 +-  0.139   SM=  0.471   chi2/NDF =   2.215/2
 202_DC_MM_3     average =  0.506 +-  0.149   SM=  0.531   chi2/NDF =   3.903/2
 202_DC_MM_4     average =  0.455 +-  0.169   SM=  0.674   chi2/NDF =   0.372/2
 202_DC_MM_5     average =  0.860 +-  0.197   SM=  0.900   chi2/NDF =   1.540/2
 202_DC_MM_6     average =  1.010 +-  0.230   SM=  1.208   chi2/NDF =   0.967/2
 202_DC_MM_7     average =  1.749 +-  0.264   SM=  1.599   chi2/NDF =   6.636/2
 202_DC_MM_8     average =  1.844 +-  0.299   SM=  2.072   chi2/NDF =   2.847/2
 202_DC_MM_9     average =  2.268 +-  0.339   SM=  2.627   chi2/NDF =   0.898/2
 202_DC_MM_10    average =  3.396 +-  0.435   SM=  3.254   chi2/NDF =   0.873/2
 202_DC_TT_1     average =  0.968 +-  0.287   SM=  0.495   chi2/NDF =  10.336/2
 202_DC_TT_2     average =  0.322 +-  0.189   SM=  0.471   chi2/NDF =   2.713/2
 202_DC_TT_3     average =  0.420 +-  0.194   SM=  0.531   chi2/NDF =   0.236/2
 202_DC_TT_4     average =  0.731 +-  0.220   SM=  0.674   chi2/NDF =   1.905/2
 202_DC_TT_5     average =  0.922 +-  0.263   SM=  0.900   chi2/NDF =   2.804/2
 202_DC_TT_6     average =  0.789 +-  0.300   SM=  1.208   chi2/NDF =   0.094/2
 202_DC_TT_7     average =  1.953 +-  0.341   SM=  1.599   chi2/NDF =   2.468/2
 202_DC_TT_8     average =  1.838 +-  0.386   SM=  2.072   chi2/NDF =   4.162/2
 202_DC_TT_9     average =  3.129 +-  0.479   SM=  2.626   chi2/NDF =   9.918/2
 202_DC_TT_10    average =  3.186 +-  0.747   SM=  3.254   chi2/NDF =   1.368/2
 205_DC_MM_1     average =  0.621 +-  0.113   SM=  0.485   chi2/NDF =   2.027/2
 205_DC_MM_2     average =  0.385 +-  0.098   SM=  0.461   chi2/NDF =   0.169/2
 205_DC_MM_3     average =  0.382 +-  0.104   SM=  0.517   chi2/NDF =   4.554/2
 205_DC_MM_4     average =  0.443 +-  0.118   SM=  0.654   chi2/NDF =   0.774/2
 205_DC_MM_5     average =  0.891 +-  0.137   SM=  0.870   chi2/NDF =   1.913/2
 205_DC_MM_6     average =  1.205 +-  0.160   SM=  1.166   chi2/NDF =   1.383/2
 205_DC_MM_7     average =  1.614 +-  0.183   SM=  1.542   chi2/NDF =   5.186/2
 205_DC_MM_8     average =  1.663 +-  0.209   SM=  1.998   chi2/NDF =   0.393/2
 205_DC_MM_9     average =  2.097 +-  0.237   SM=  2.534   chi2/NDF =   0.449/2
 205_DC_MM_10    average =  3.318 +-  0.306   SM=  3.140   chi2/NDF =   6.351/2
 205_DC_TT_1     average =  0.363 +-  0.203   SM=  0.486   chi2/NDF =   6.520/2
 205_DC_TT_2     average =  0.562 +-  0.137   SM=  0.461   chi2/NDF =   0.697/2
 205_DC_TT_3     average =  0.603 +-  0.135   SM=  0.517   chi2/NDF =   4.695/2
 205_DC_TT_4     average =  0.443 +-  0.154   SM=  0.654   chi2/NDF =   0.276/2
 205_DC_TT_5     average =  0.397 +-  0.179   SM=  0.870   chi2/NDF =   0.237/2
 205_DC_TT_6     average =  1.242 +-  0.209   SM=  1.166   chi2/NDF =   0.132/2
 205_DC_TT_7     average =  1.522 +-  0.237   SM=  1.542   chi2/NDF =   0.804/2
 205_DC_TT_8     average =  1.846 +-  0.268   SM=  1.998   chi2/NDF =   1.367/2
 205_DC_TT_9     average =  2.045 +-  0.330   SM=  2.533   chi2/NDF =   3.717/2
 205_DC_TT_10    average =  4.671 +-  0.520   SM=  3.140   chi2/NDF =   1.658/2
 207_DC_MM_1     average =  0.518 +-  0.087   SM=  0.481   chi2/NDF =   0.388/2
 207_DC_MM_2     average =  0.496 +-  0.075   SM=  0.456   chi2/NDF =   0.051/2
 207_DC_MM_3     average =  0.473 +-  0.079   SM=  0.510   chi2/NDF =   5.634/2
 207_DC_MM_4     average =  0.781 +-  0.089   SM=  0.643   chi2/NDF =   5.052/2
 207_DC_MM_5     average =  0.795 +-  0.104   SM=  0.855   chi2/NDF =   2.185/2
 207_DC_MM_6     average =  0.995 +-  0.121   SM=  1.145   chi2/NDF =   0.627/2
 207_DC_MM_7     average =  1.630 +-  0.139   SM=  1.515   chi2/NDF =   0.808/2
 207_DC_MM_8     average =  2.247 +-  0.159   SM=  1.963   chi2/NDF =   4.025/2
 207_DC_MM_9     average =  2.491 +-  0.179   SM=  2.489   chi2/NDF =   4.407/2
 207_DC_MM_10    average =  2.995 +-  0.231   SM=  3.086   chi2/NDF =   1.136/2
 207_DC_TT_1     average =  0.272 +-  0.145   SM=  0.481   chi2/NDF =   0.134/2
 207_DC_TT_2     average =  0.412 +-  0.106   SM=  0.456   chi2/NDF =   6.521/2
 207_DC_TT_3     average =  0.534 +-  0.104   SM=  0.510   chi2/NDF =   0.745/2
 207_DC_TT_4     average =  0.563 +-  0.118   SM=  0.644   chi2/NDF =   0.133/2
 207_DC_TT_5     average =  0.683 +-  0.140   SM=  0.855   chi2/NDF =   5.976/2
 207_DC_TT_6     average =  1.443 +-  0.161   SM=  1.145   chi2/NDF =   1.658/2
 207_DC_TT_7     average =  1.351 +-  0.181   SM=  1.514   chi2/NDF =   1.519/2
 207_DC_TT_8     average =  1.761 +-  0.207   SM=  1.962   chi2/NDF =   6.867/2
 207_DC_TT_9     average =  1.655 +-  0.255   SM=  2.489   chi2/NDF =   0.561/2
 207_DC_TT_10    average =  3.597 +-  0.399   SM=  3.085   chi2/NDF =   3.709/2
\end{verbatim}

\section{Differential Cross-Section for Electron-Positron Final States}
\label{app:2F:ee}

The following lists show for each centre-of-mass energy point (rounded
in GeV) the LEP-combined differential cross-sections (DC) for $\ee$
(EE) final states in 15 $\cos\theta$-bins ($1-15$ with bin boundaries
as defined in Tables~\ref{tab:ff:dsdceeres1} and
~\ref{tab:ff:dsdceeres2}), comparing the LEP average value and its
total error with the SM prediction.  Also shown is the overall
$\chidf$ and the bin-by-bin $\chidf$ contribution.  The overall matrix
of correlation coefficients and inverse covariance matrix are
available at the LEPEWWG web site: {\tt http://www.cern.ch/LEPEWWG}.

\begin{verbatim}
 Total chi2/NDF =  199.402/189
 189_DC_EE_1     average =  1.401 +-  0.161   SM=  1.590   chi2/NDF =   1.576/1
 189_DC_EE_2     average =  2.030 +-  0.160   SM=  1.816   chi2/NDF =   6.274/2
 189_DC_EE_3     average =  2.162 +-  0.170   SM=  2.162   chi2/NDF =   1.237/2
 189_DC_EE_4     average =  2.298 +-  0.186   SM=  2.681   chi2/NDF =   0.654/2
 189_DC_EE_5     average =  4.321 +-  0.230   SM=  3.906   chi2/NDF =   4.262/2
 189_DC_EE_6     average =  4.898 +-  0.348   SM=  5.372   chi2/NDF =   2.403/2
 189_DC_EE_7     average =  6.090 +-  0.404   SM=  6.892   chi2/NDF =   6.751/2
 189_DC_EE_8     average =  8.838 +-  0.476   SM=  9.610   chi2/NDF =   2.341/2
 189_DC_EE_9     average = 12.781 +-  0.576   SM= 13.345   chi2/NDF =   3.970/2
 189_DC_EE_10    average = 19.586 +-  0.707   SM= 19.445   chi2/NDF =   0.115/2
 189_DC_EE_11    average = 30.598 +-  0.895   SM= 30.476   chi2/NDF =   2.386/2
 189_DC_EE_12    average = 50.488 +-  1.135   SM= 51.012   chi2/NDF =   2.339/2
 189_DC_EE_13    average = 95.178 +-  1.520   SM= 95.563   chi2/NDF =   0.211/2
 189_DC_EE_14    average =211.427 +-  2.900   SM=212.390   chi2/NDF =   2.620/1
 189_DC_EE_15    average =679.146 +-  5.773   SM=689.989   chi2/NDF =   1.921/1
 192_DC_EE_1     average =  1.300 +-  0.364   SM=  1.539   chi2/NDF =   0.051/1
 192_DC_EE_2     average =  2.099 +-  0.419   SM=  1.754   chi2/NDF =   0.462/2
 192_DC_EE_3     average =  1.871 +-  0.385   SM=  2.091   chi2/NDF =   1.602/2
 192_DC_EE_4     average =  1.808 +-  0.422   SM=  2.604   chi2/NDF =   1.619/2
 192_DC_EE_5     average =  3.800 +-  0.519   SM=  3.778   chi2/NDF =   3.179/2
 192_DC_EE_6     average =  5.015 +-  0.891   SM=  5.205   chi2/NDF =   1.897/2
 192_DC_EE_7     average =  5.695 +-  0.976   SM=  6.692   chi2/NDF =   9.314/2
 192_DC_EE_8     average =  9.239 +-  1.175   SM=  9.242   chi2/NDF =   0.003/2
 192_DC_EE_9     average = 12.941 +-  1.414   SM= 12.800   chi2/NDF =   0.749/2
 192_DC_EE_10    average = 20.761 +-  1.807   SM= 18.776   chi2/NDF =   0.371/2
 192_DC_EE_11    average = 26.466 +-  2.074   SM= 29.471   chi2/NDF =   4.398/2
 192_DC_EE_12    average = 49.382 +-  2.671   SM= 49.338   chi2/NDF =   1.721/2
 192_DC_EE_13    average = 89.676 +-  3.615   SM= 92.079   chi2/NDF =   2.159/2
 192_DC_EE_14    average =204.579 +-  6.760   SM=206.087   chi2/NDF =   0.054/1
 192_DC_EE_15    average =655.724 +- 12.588   SM=669.173   chi2/NDF =   0.482/1
 196_DC_EE_1     average =  1.470 +-  0.261   SM=  1.483   chi2/NDF =   1.887/1
 196_DC_EE_2     average =  1.527 +-  0.221   SM=  1.695   chi2/NDF =   0.421/2
 196_DC_EE_3     average =  2.058 +-  0.250   SM=  2.000   chi2/NDF =   0.865/2
 196_DC_EE_4     average =  2.788 +-  0.284   SM=  2.498   chi2/NDF =   0.014/2
 196_DC_EE_5     average =  3.646 +-  0.318   SM=  3.610   chi2/NDF =   0.212/2
 196_DC_EE_6     average =  5.887 +-  0.521   SM=  4.999   chi2/NDF =   1.809/2
 196_DC_EE_7     average =  6.233 +-  0.591   SM=  6.406   chi2/NDF =   1.078/2
 196_DC_EE_8     average =  9.016 +-  0.694   SM=  8.832   chi2/NDF =   2.379/2
 196_DC_EE_9     average = 13.444 +-  0.856   SM= 12.326   chi2/NDF =   0.593/2
 196_DC_EE_10    average = 18.568 +-  0.977   SM= 18.039   chi2/NDF =  11.452/2
 196_DC_EE_11    average = 27.056 +-  1.223   SM= 28.300   chi2/NDF =   0.962/2
 196_DC_EE_12    average = 49.391 +-  1.619   SM= 47.362   chi2/NDF =   0.784/2
 196_DC_EE_13    average = 88.163 +-  2.154   SM= 88.473   chi2/NDF =   0.982/2
 196_DC_EE_14    average =197.369 +-  4.121   SM=198.250   chi2/NDF =   0.438/1
 196_DC_EE_15    average =637.846 +-  8.003   SM=642.688   chi2/NDF =   0.118/1
 200_DC_EE_1     average =  1.483 +-  0.245   SM=  1.420   chi2/NDF =   0.002/1
 200_DC_EE_2     average =  1.638 +-  0.214   SM=  1.623   chi2/NDF =   0.802/2
 200_DC_EE_3     average =  2.068 +-  0.227   SM=  1.885   chi2/NDF =   3.449/2
 200_DC_EE_4     average =  2.362 +-  0.250   SM=  2.409   chi2/NDF =   0.753/2
 200_DC_EE_5     average =  4.251 +-  0.313   SM=  3.435   chi2/NDF =   1.068/2
 200_DC_EE_6     average =  5.244 +-  0.506   SM=  4.770   chi2/NDF =   1.098/2
 200_DC_EE_7     average =  5.888 +-  0.571   SM=  6.157   chi2/NDF =   0.142/2
 200_DC_EE_8     average =  8.244 +-  0.667   SM=  8.471   chi2/NDF =   3.666/2
 200_DC_EE_9     average =  9.506 +-  0.736   SM= 11.773   chi2/NDF =   8.162/2
 200_DC_EE_10    average = 16.376 +-  0.920   SM= 17.262   chi2/NDF =   3.021/2
 200_DC_EE_11    average = 27.000 +-  1.214   SM= 27.117   chi2/NDF =   2.513/2
 200_DC_EE_12    average = 44.614 +-  1.537   SM= 45.607   chi2/NDF =   5.241/2
 200_DC_EE_13    average = 86.454 +-  2.060   SM= 85.143   chi2/NDF =   0.582/2
 200_DC_EE_14    average =190.962 +-  3.941   SM=190.786   chi2/NDF =   0.760/1
 200_DC_EE_15    average =604.986 +-  7.608   SM=617.718   chi2/NDF =   0.059/1
 202_DC_EE_1     average =  1.568 +-  0.368   SM=  1.401   chi2/NDF =   2.070/1
 202_DC_EE_2     average =  1.344 +-  0.276   SM=  1.579   chi2/NDF =   0.070/2
 202_DC_EE_3     average =  2.107 +-  0.345   SM=  1.836   chi2/NDF =   1.503/2
 202_DC_EE_4     average =  3.240 +-  0.406   SM=  2.361   chi2/NDF =   1.130/2
 202_DC_EE_5     average =  2.911 +-  0.394   SM=  3.356   chi2/NDF =   3.574/2
 202_DC_EE_6     average =  4.603 +-  0.628   SM=  4.669   chi2/NDF =   0.358/2
 202_DC_EE_7     average =  6.463 +-  0.861   SM=  6.017   chi2/NDF =   1.590/2
 202_DC_EE_8     average =  7.457 +-  0.957   SM=  8.320   chi2/NDF =   3.276/2
 202_DC_EE_9     average = 11.032 +-  1.113   SM= 11.554   chi2/NDF =   0.602/2
 202_DC_EE_10    average = 16.428 +-  1.338   SM= 16.891   chi2/NDF =   1.489/2
 202_DC_EE_11    average = 27.153 +-  1.643   SM= 26.583   chi2/NDF =   4.350/2
 202_DC_EE_12    average = 46.490 +-  2.214   SM= 44.786   chi2/NDF =   0.246/2
 202_DC_EE_13    average = 87.253 +-  2.887   SM= 83.473   chi2/NDF =   1.047/2
 202_DC_EE_14    average =189.026 +-  5.516   SM=186.904   chi2/NDF =   0.626/1
 202_DC_EE_15    average =599.860 +- 10.339   SM=605.070   chi2/NDF =   0.476/1
 205_DC_EE_1     average =  1.102 +-  0.205   SM=  1.355   chi2/NDF =   3.910/1
 205_DC_EE_2     average =  1.470 +-  0.195   SM=  1.539   chi2/NDF =   4.105/2
 205_DC_EE_3     average =  2.050 +-  0.231   SM=  1.786   chi2/NDF =   0.679/2
 205_DC_EE_4     average =  2.564 +-  0.255   SM=  2.280   chi2/NDF =   0.611/2
 205_DC_EE_5     average =  3.410 +-  0.300   SM=  3.253   chi2/NDF =   1.269/2
 205_DC_EE_6     average =  5.308 +-  0.472   SM=  4.479   chi2/NDF =   1.159/2
 205_DC_EE_7     average =  5.836 +-  0.571   SM=  5.820   chi2/NDF =   1.925/2
 205_DC_EE_8     average =  7.996 +-  0.635   SM=  8.077   chi2/NDF =   0.869/2
 205_DC_EE_9     average = 10.607 +-  0.764   SM= 11.200   chi2/NDF =   0.581/2
 205_DC_EE_10    average = 14.729 +-  0.874   SM= 16.322   chi2/NDF =   1.139/2
 205_DC_EE_11    average = 26.189 +-  1.157   SM= 25.722   chi2/NDF =   0.829/2
 205_DC_EE_12    average = 43.124 +-  1.497   SM= 43.217   chi2/NDF =   0.942/2
 205_DC_EE_13    average = 79.255 +-  1.976   SM= 80.939   chi2/NDF =   0.758/2
 205_DC_EE_14    average =179.842 +-  3.838   SM=180.878   chi2/NDF =   3.902/1
 205_DC_EE_15    average =587.999 +-  7.527   SM=586.205   chi2/NDF =   2.437/1
 207_DC_EE_1     average =  1.440 +-  0.196   SM=  1.339   chi2/NDF =   0.019/1
 207_DC_EE_2     average =  1.426 +-  0.163   SM=  1.517   chi2/NDF =   1.800/2
 207_DC_EE_3     average =  1.889 +-  0.177   SM=  1.745   chi2/NDF =   0.809/2
 207_DC_EE_4     average =  2.156 +-  0.198   SM=  2.240   chi2/NDF =   4.511/2
 207_DC_EE_5     average =  3.215 +-  0.233   SM=  3.194   chi2/NDF =   2.133/2
 207_DC_EE_6     average =  4.434 +-  0.357   SM=  4.380   chi2/NDF =   4.019/2
 207_DC_EE_7     average =  6.393 +-  0.463   SM=  5.729   chi2/NDF =   1.649/2
 207_DC_EE_8     average =  6.951 +-  0.481   SM=  7.972   chi2/NDF =   1.727/2
 207_DC_EE_9     average = 11.221 +-  0.615   SM= 11.019   chi2/NDF =   1.981/2
 207_DC_EE_10    average = 15.933 +-  0.739   SM= 16.053   chi2/NDF =   1.275/2
 207_DC_EE_11    average = 25.676 +-  0.923   SM= 25.254   chi2/NDF =   5.712/2
 207_DC_EE_12    average = 42.075 +-  1.188   SM= 42.456   chi2/NDF =   0.527/2
 207_DC_EE_13    average = 77.611 +-  1.569   SM= 79.639   chi2/NDF =   0.550/2
 207_DC_EE_14    average =173.825 +-  3.002   SM=178.042   chi2/NDF =   0.026/1
 207_DC_EE_15    average =573.637 +-  6.024   SM=576.688   chi2/NDF =   3.200/1
\end{verbatim}

\chapter{Determination of the LEP Centre-of-Mass Energy Using
 Radiative-Return Events}
\label{mw:zret:appendix}

The LEP collaborations performed measurements of radiative Z boson
production, $\ee\to\mathrm{Z}+\gamma\to\ff+\gamma$, at centre-of-mass
energies well above the Z peak,
$\sqrt{s}=161-209~\GeV$~\cite{bib:mw:a-mw, bib:mw:d-zreturn,
  bib:mw:l-zreturn, bib:mw:o-zreturn}.  Events with pairs of
electrons, muons, taus, and hadronic jets were selected. The presence
of hard ISR photons, mostly emitted at small polar angles with respect
to the beam directions and recoiling against the di-fermion system,
led to typical event topologies with acollinear fermions measured in
the detector. Due to the photon emission, the mass of the two-fermion
system, $\sqrt{s'}$, is reduced to values less than $\sqrt{s}$. The
spectrum of $\sqrt{s'}$ exhibits a resonance peak around the Z boson
mass and allows a determination of $\MZ$. The determination of
$\sqrt{s'}$ furthermore involves the knowledge of the $\ee$
centre-of-mass mass energy, because in the kinematic reconstruction of
the $\mathrm{Z}+\gamma\to\ff+\gamma$ events, energy-momentum
conservation is imposed. Thus, a measurement of the Z boson mass in
radiative-return events, $\MZ^\ff$, is equivalent to determining the
average $\sqrt{s}$ of each analysed data
set. Figure~\ref{zret:fig:opal-l3} shows two examples of the
two-fermion mass spectra measured by the LEP experiments.

\begin{figure}[tbhp]
\centerline{
 \epsfig{file=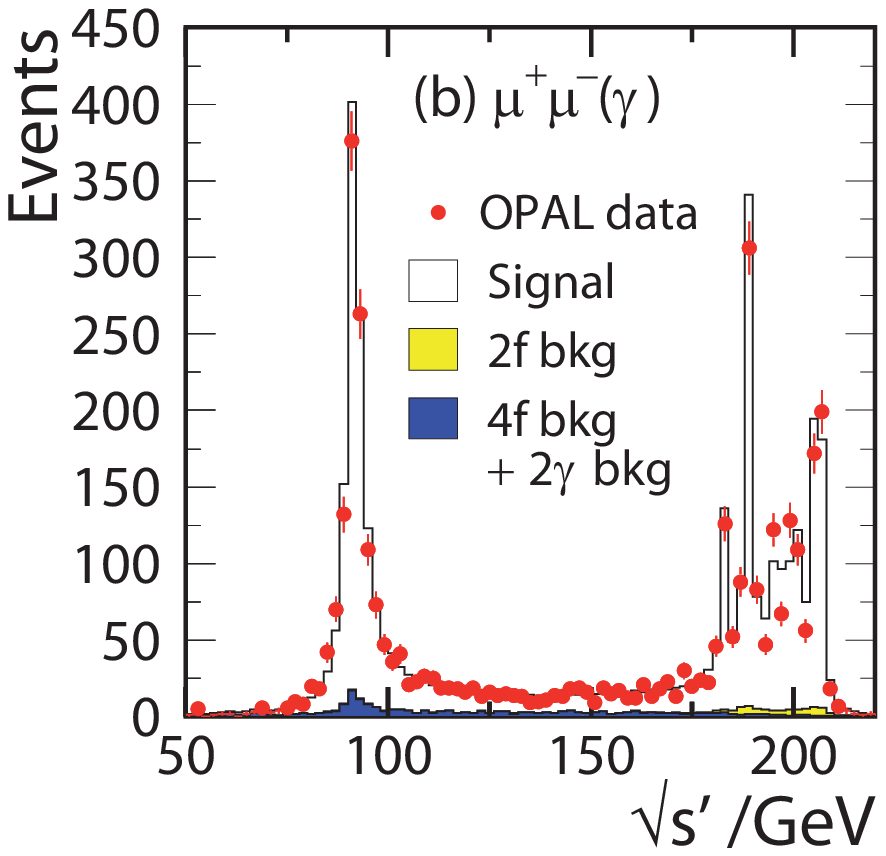,width=0.5\textwidth}
 \epsfig{file=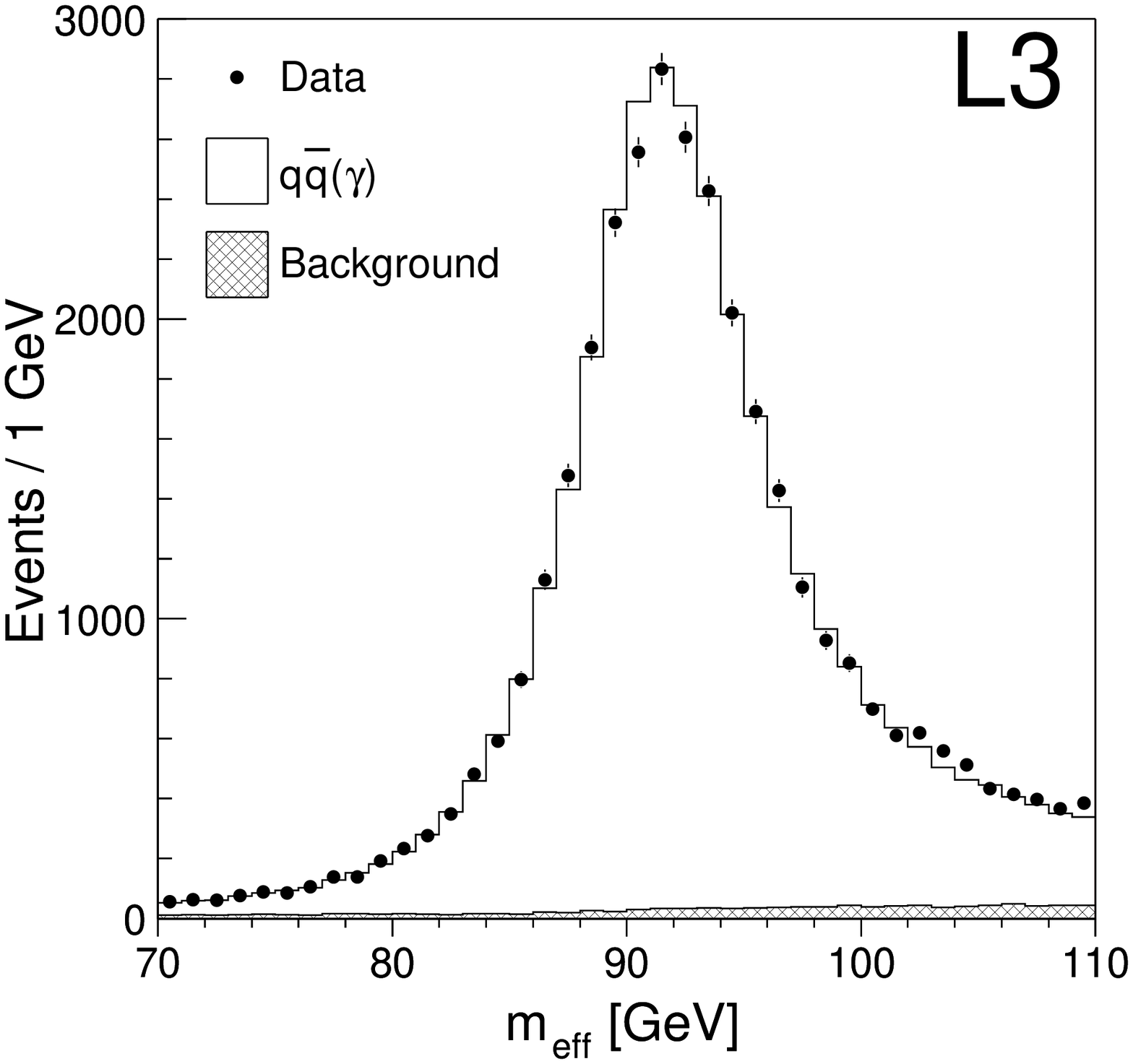,width=0.5\textwidth}
}
\caption[Reconstructed two-fermion mass spectrum.]  {Examples of
   reconstructed two-fermion mass spectra in the
   $Z+\gamma\to\mu^+\mu^-+\gamma$ channel (left) and in the hadronic
   channel (right), as measured by OPAL and by L3,
   respectively~\cite{bib:mw:o-zreturn}.  The data collected at
   different centre-of-mass energies is combined and compared to
   Monte-Carlo predictions using the nominal Z bosons
   mass~\cite{bib-Z-pole}. The Z resonance peak is clearly visible.}
\label{zret:fig:opal-l3}
\end{figure}

The ISR photons are either detected as isolated energy depositions in
the calorimeters compatible with an electromagnetic shower or as
missing momentum pointing along the beam directions. Typically, the
energy of the calorimeter shower is required to be larger than
$30-60~\GeV$. For hadronic final states, a kinematic fit is applied to
the event imposing energy and momentum conservation. In case the ISR
photons are not observed in the detector, the sum of the photon
momenta is assumed to either point along only one beam direction or
along both beam axes.  In the kinematic fit, usually both
possibilities are tested and the one obtaining the best fit results is
eventually chosen. In this way, the mass of the two-fermion system is
reconstructed with optimised precision. In leptonic final states,
$\sqrt{s'}$ is determined using the well-measured polar angles of the
leptons, according to the following equation:

\begin{eqnarray}
\label{equ:sprime}
\sqrt{s'}&=&\sqrt{1-\frac{2E_{\mathrm{ISR}}}{\sqrt{s}}}\;\mbox{, with}\\
E_{\mathrm{ISR}}&=&\sqrt{s}\frac{|\sin(\theta_1+\theta_2)|}{\sin\theta_1+\sin\theta_2+|\sin(\theta_1+\theta_2)|}\;.
\end{eqnarray}
The leptonic polar angles $\theta_1$ and $\theta_2$ are determined
either relative to the beam axis if no photon is measured in the
detector, or relative to the direction of the measured ISR photon.

After correcting for remaining background, the mass of the Z boson is
extracted either by applying a Monte-Carlo event reweighting based on
the corresponding matrix element of the signal process or by fitting
an analytical function describing the signal spectrum to the data. The
measured Z mass, $\MZ^\ff$, is then compared to the Z mass determined
in precision measurements at Z-pole energies~\cite{bib-Z-pole},
$\MZ=91.1875~\pm~0.0021~\GeV$. The comparison is converted into a
difference in terms of the centre-of-mass energy, $\Delta\sqrt{s}$,
between $\sqrt{s}$ derived from radiative return events and the
nominal centre-of-mass energy, $\sqrt{s}_\mathrm{LEP}$, determined by
the precise LEP energy calibration~\cite{bib:mw:lep-energy-1,
*bib:mw:lep-energy-2}:

\begin{equation}
  \Delta\sqrt{s} = 
  \sqrt{s}-\sqrt{s}_\mathrm{LEP}=\sqrt{s}\frac{\MZ^\ff-\MZ}{\MZ} \,.
\end{equation}
This observable is eventually used to combine the results of the four
LEP experiments.

Various sources of systematic uncertainties are studied and possible
correlations between them are taken into account in the combination
procedure. The dominant uncertainty is due to the modelling of the
fragmentation process in hadronic Z decays. An uncertainty of
$22~\MeV$ on $\Delta\sqrt{s}$ is derived from a comparison of
different fragmentation models implemented in the
\Pythia~\cite{JETSET}, \Herwig~\cite{HERWIG}, and
\Ariadne~\cite{ARIADNE} software packages.  The Monte-Carlo
predictions of the $\ee\to\ff+\gamma$ process are calculated using the
KK~v4.02~\cite{bib:ff:KK} Monte-Carlo generator. Theoretical
uncertainties in the description of ISR and FSR and missing higher
order corrections are estimated by reweighting events applying
different orders of $\alpha$ in the prediction and comparing it to the
${\cal O}(\alpha^2)$ calculations in the Coherent Exclusive
Exponentiation scheme. Furthermore, the effect of neglecting the
interference between ISR and FSR was studied. The total systematic
uncertainties due to modelling of ISR and FSR amounts to $7~\MeV$. The
uncertainty due to the prediction of the four-fermion background using
measured cross-sections as input is estimated to be $6~\MeV$. The
operational parameters of the LEP collider may also influence the
reconstruction of the two-fermion mass. In particular, the effects of
beam energy spread and a possible asymmetry in beam energy were
studied and found to influence $\Delta\sqrt{s}$ by less than
$3~\MeV$. The uncertainty on the nominal Z boson mass contributes with
less than $1~\MeV$. All these sources of systematic uncertainties are
assumed to be fully correlated between experiments.

Each experiment determined the influence of detector alignment, bias
in angular measurements, uncertainty of energy and momentum scale and
resolution in great detail. Control samples were selected in data to
measure the various detector and resolution effects. These
uncertainties are treated as uncorrelated between
experiments. Uncertainties from limited Monte-Carlo statistics also
contribute. If sources of systematic uncertainties affect data from
different data taking periods or different channels similarly also the
corresponding correlations are taken into account.  A summary of the
different sources of systematic uncertainties is listed in
Table~\ref{zret:tab:systematics}. The uncertainties due to Monte-Carlo
statistics and detector bias and resolution are uncorrelated between
experiments. Details of the systematic effects due to detector bias
and resolution combine several individual sources of uncertainty and
are discussed in the publications by the
experiments~\cite{bib:mw:a-mw, bib:mw:d-zreturn, bib:mw:l-zreturn,
  bib:mw:o-zreturn}. The total statistical and systematic
uncertainties on the LEP average for $\Delta\sqrt{s}$ are $40~\MeV$
and $36~\MeV$, respectively.

\begin{table}[t*]
\begin{center}
\begin{tabular}{| l || c |}
\hline
Source &  Uncertainty on \\
       & $\Delta\sqrt{s}$ $[\MeV]$ \\
\hline
\hline
Fragmentation            & 22 \\
ISR/FSR Modelling        &  7 \\
Four Fermion Background  &  6 \\
Z Mass                   &  1 \\
LEP Parameters           &  3 \\
\hline
Total Correlated         & 23 \\
\hline
\hline
Monte-Carlo Statistics   &  7 \\
Detector Bias and Resolution & 28 \\
\hline
Total Uncorrelated       & 29 \\
\hline
\hline
Total Systematics        & 37 \\
Total Statistical        & 40 \\ 
\hline
Total                    & 54 \\
\hline
\end{tabular}
\caption[Systematic uncertainties on LEP centre-of-mass energy shift]{
 \label{zret:tab:systematics}
Systematic and statistical uncertainties on the measurement of the LEP
centre-of-mass energy shift, $\Delta\sqrt{s}$. 
}
\end{center}
\end{table}

When combining all available LEP
data~\cite{bib:mw:a-mw, bib:mw:d-zreturn, bib:mw:l-zreturn,
  bib:mw:o-zreturn} with Z decays to hadrons, and to electron, muon,
and tau pairs, the difference is found to be:

\begin{equation}
  \Delta\sqrt{s}=-54\pm 54~\MeV\,,
\end{equation}
in good agreement with no shift with respect to the more precise
standard LEP energy calibration. There is also no significant shift
observed when analysing the leptonic and hadronic decay channels
separately. Furthermore, there is no significant dependence on the LEP
beam energy, respectively data taking periods, as illustrated in
Figure~\ref{zret:fig:results} and Table~\ref{zret:tab:results}.

\begin{table}[t*]
\begin{center}
\begin{tabular}{| c || r @{$\;\pm\;$} c @{$\;\pm\;$} l |}
\hline
Data set & \multicolumn{3}{c |}{$\Delta\sqrt{s}$ $[\MeV]$} \\
\hline
\hline
$\ee\to\mbox{hadrons}+\gamma$ &  -88    &      40     &     56   \\
$\ee\to\ell^+\ell^-+\gamma$   &  -10     &      80     &     26   \\
\hline
$\sqrt{s}=183~\GeV$       &   70    &      98     &     50   \\
$\sqrt{s}=189~\GeV$        &  -86    &      60     &     46   \\
$\sqrt{s}=192-202~\GeV$        &  -66    &      62     &     44   \\
$\sqrt{s}=205-209~\GeV$        &  -140    &      70     &     52   \\
\hline
All LEP data  &  -54    &      40     &     36   \\
\hline
\end{tabular}
\caption[LEP combined results on LEP centre-of-mass energy shift]{
\label{zret:tab:results}
Combined results of \Aleph, \Delphi, \Ltre, and \Opal\ on the
determination of the LEP centre-of-mass energy shift,
$\Delta\sqrt{s}$, with respect to the nominal value of $\sqrt{s}$. The
results are shown for the leptonic and hadronic final states, as well
as for the different data taking periods, together with the LEP
combined value. Statistical and systematic uncertainties are given
separately. }
\end{center}
\end{table}
  
\begin{figure}[p]
\begin{center}
\mbox{\epsfig{file=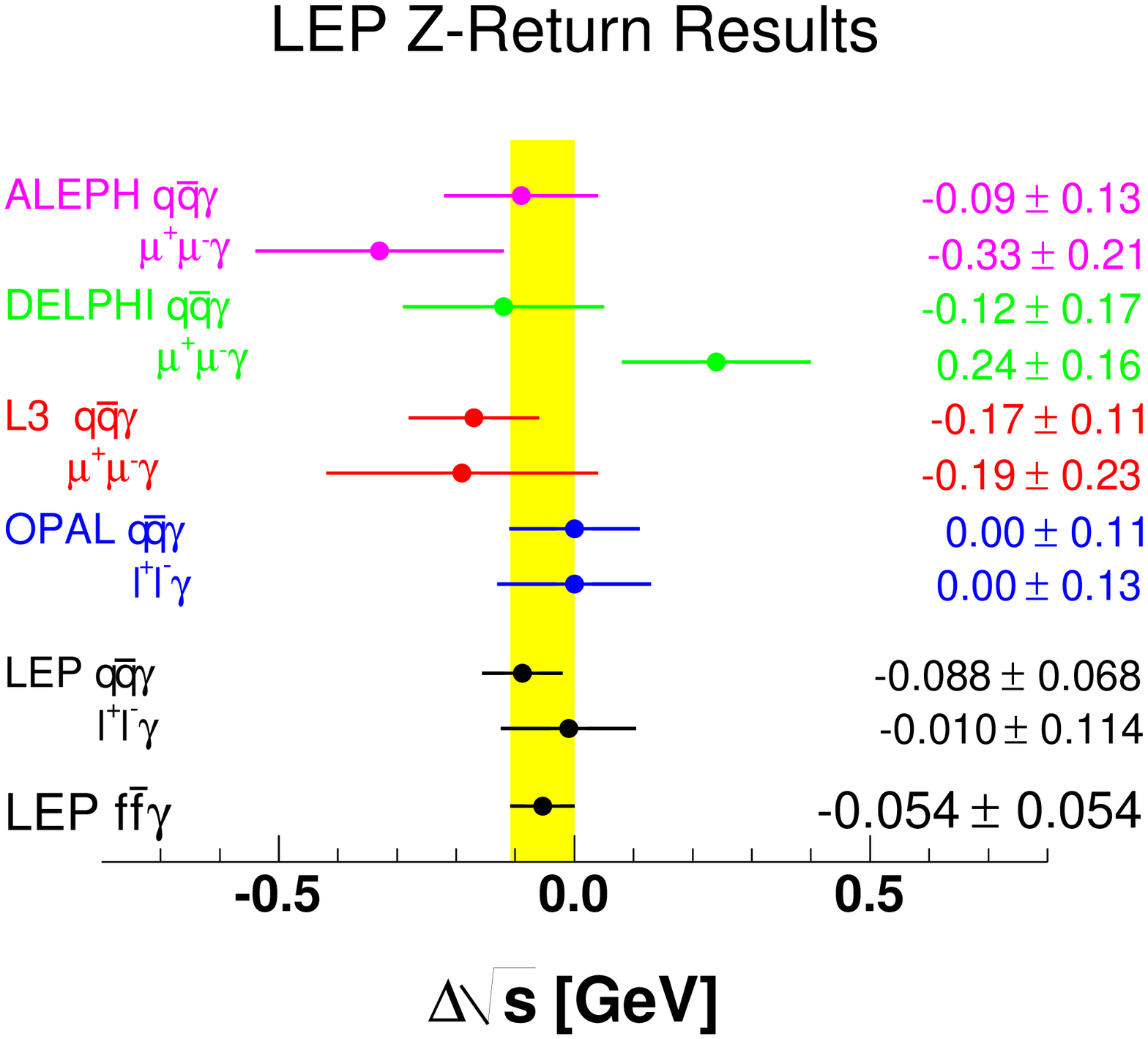,width=0.6\textwidth}}
\mbox{\epsfig{file=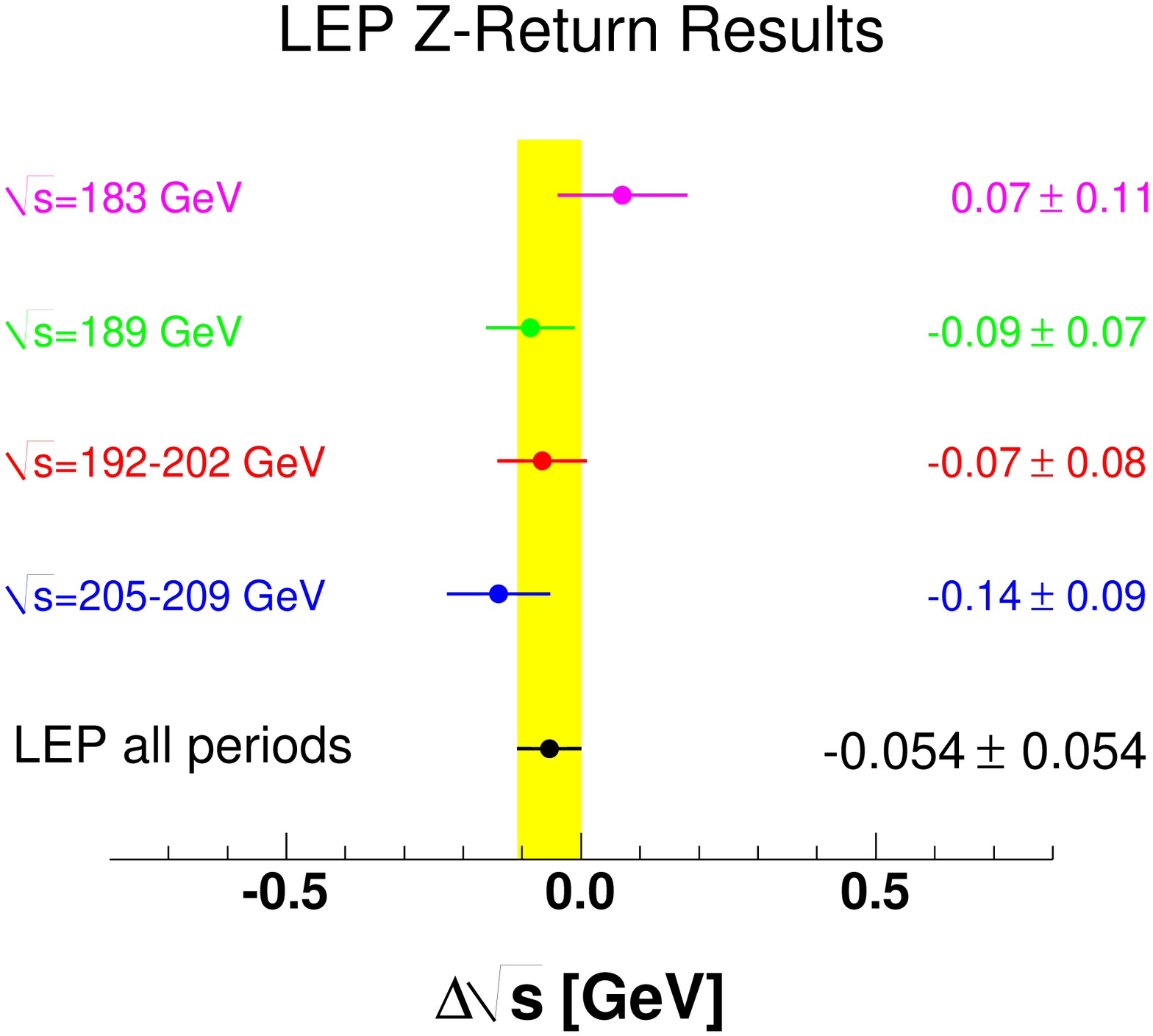,width=0.6\textwidth}}
\vskip -0.5cm
\caption{\label{zret:fig:results} Difference between the energy
 determined in Z-return events and the nominal LEP centre-of-mass
 energy, $\Delta\sqrt{s}$, for the different experiments and final
 states (top), and for the data taking periods with energies well
 above the W-pair threshold (bottom). The individual measurements as
 well as the LEP combined values take systematic uncertainties and
 their correlations into account.}
\end{center}
\end{figure}

\chapter{Tests of the Colour-Reconnection Combination Procedure}
\label{fsi:cr:appendix}

Here we report on the tests of the LEP combination procedure used to
combine the measurements of the LEP experiments on colour-reconnection
(CR). It is shown that the LEP combination procedure is able to
reproduce the combination of each experiment separately.

For each measurement, the dependence of the measured observable on the
model parameter $k_I$ is determined based on Monte-Carlo simulations.
For the particle-flow based measurements, the results are reported in
Table~\ref{fsi:cr:tab:pflowsk1}. The corresponding parameter values
for the phenomenological function shown in
Equation~\ref{eq:fsi:cr:pheno} are listed in
Table~\ref{fsi:cr:tab:rparams}.  The parametrisations of
$\delta_{\Delta\MW,i}(k_I)$ which are used to introduce systematic
uncertainties for the likelihood functions of the CR measurements from
$\Delta\MW$ by \Aleph, \Delphi, and \Opal\ are given by:

\begin{eqnarray}
\delta_{\Delta\MW,A}(k_I)&=&\left\{\begin{array}{l l}
0.416408 + (0.624184)^2 \cdot k_I      &, k_I\in[0.0,1.3)\\
1.227847 + (0.251441)^2 \cdot(k_I-1.3) &, k_I\in[1.3,2.5)\\
1.529576 + (0.750313)^2 \cdot(k_I-2.5) &, k_I\in[2.5,4.5)\\
3.030202 + (0.279341)^2 \cdot(k_I-4.5) &, k_I\in[4.5,6.0)\\
3.449214 + (0.600389)^2 \cdot(k_I-6.0) &, k_I\in[6.0,\infty)
\end{array} \right.\\
\delta_{\Delta\MW,D}(k_I)&=&0.233054 + (0.486925)^2\cdot k_I\\
\delta_{\Delta\MW,O}(k_I)&=&0.666308 + (0.483630)^2\cdot k_I\,.
\end{eqnarray}
A graphical comparison of the original input and the parametrised
$\Delta\chi^2$ distributions is displayed in
Figure~\ref{fsi:cr:fig:chi2ki} and shows good agreement.

\begin{table}[p]
  \begin{center}
    \begin{tabular}{| c || c | c |}
      \hline
      & \multicolumn{2}{c |}{\Rn($k_I$) for SK1 Model} \\
      $k_I$ & \Ltre & \Opal \\
      \hline
      \hline
0.10    & $0.8613 \pm 0.0037$ & $1.2816 \pm 0.0028$ \\
0.15    & $0.8598 \pm 0.0037$ & $1.2792 \pm 0.0028$ \\
0.20    & $0.8585 \pm 0.0037$ & $1.2759 \pm 0.0028$ \\
0.25    & $0.8561 \pm 0.0037$ & $1.2738 \pm 0.0028$ \\
0.35    & $0.8551 \pm 0.0037$ & $1.2683 \pm 0.0028$ \\
0.45    & $0.8509 \pm 0.0036$ & $1.2643 \pm 0.0028$ \\
0.60    & $0.8482 \pm 0.0036$ & $1.2575 \pm 0.0028$ \\
0.80    & $0.8414 \pm 0.0037$ & $1.2495 \pm 0.0028$ \\
0.90    & --                  & $1.2464 \pm 0.0028$ \\
1.00    & $0.8381 \pm 0.0036$ & $1.2420 \pm 0.0028$ \\
1.10    & --                  & $1.2389 \pm 0.0028$ \\
1.20    & --                  & $1.2355 \pm 0.0028$ \\
1.30    & --                  & $1.2326 \pm 0.0028$ \\
1.50    & $0.8318 \pm 0.0036$ & $1.2265 \pm 0.0028$ \\
1.75    & --                  & $1.2201 \pm 0.0028$ \\ 
2.00    & --                  & $1.2133 \pm 0.0028$ \\ 
2.50    & --                  & $1.2029 \pm 0.0028$ \\
3.00    & $0.8135 \pm 0.0036$ & $1.1942 \pm 0.0028$ \\
5.00    & $0.7989 \pm 0.0035$ & $1.1705 \pm 0.0028$ \\
10.00   & --                  & $1.1413 \pm 0.0028$ \\
30.00   & --                  & $1.1101 \pm 0.0028$ \\
60.00   & --                  & $1.0997 \pm 0.0028$ \\
100.00  & --                  & $1.0918 \pm 0.0028$ \\
10000.00& $0.7482 \pm 0.0033$ & $1.0780 \pm 0.0028$ \\
      \hline
    \end{tabular}
    \caption[SK1 Particle-Flow Predictions of L3 and OPAL]{
      \label{fsi:cr:tab:pflowsk1}
      Monte-Carlo predictions for the particle-flow parameter
      $R_N(k_I)$ provided for the SK1 model by \Ltre\ and \Opal.  }
  \end{center}
\end{table}

\begin{table}[p]
  \begin{center}
    \begin{tabular}{| c || c | c | c | c | c |}
      \hline
      Experiment & $a_1$ & $a_2$ & $a_3$ & $a_4$ & $b$ \\
      \hline
      \hline
      \Ltre& -12.1076& 2.03107&-0.23384 &-10.1780&1.18954\\
      \Opal& -0.26969& 0.20543&-0.06698 & 0.03388&10.8576\\
      \hline
    \end{tabular}
  \end{center}
  \caption[SK1 Particle-Flow Predictions of L3 and OPAL]{
    \label{fsi:cr:tab:rparams}
    Parameter sets used for the functional description of the
    particle-flow input provided for the SK1 model by \Ltre\ and \Opal.  }
\end{table}

The \Aleph\ input is available as a set of $\Delta\chi^2(k_I)$ values
including systematic uncertainties, which can be evaluated
directly. The result is shown in Figure~\ref{fsi:cr:fig:lep-input} and
the numerical analysis yields:

\begin{equation}
k_I=0.33^{+1.82}_{-0.33} \,.
\end{equation}
By construction, this agrees well with the original \Aleph\
result~\cite{bib:cr:ALEPH_MW}.  However, the \Aleph\ input does not
include BEC systematic uncertainties. To incorporate also this effect,
the correlated part of the systematic uncertainties is increased by
11\%. This value is derived from a dedicated study, not included in
the \Aleph\ publication.  The final result using only \Aleph\ data and
including BEC uncertainties is

\begin{equation}
k_I=0.34^{+1.86}_{-0.34}\,.
\end{equation}
The 68\% upper limit is about 3\% higher compared to the original
\Aleph\ input.

\Delphi\ provides $\Delta\chi^2(k_I)$ inputs from their $\Delta M_W$
and particle-flow analyses. In the \Delphi\ publication, both curves
are simply added neglecting correlations,
yielding~\cite{bib:cr:DELPHI_CR}:

\begin{equation}
k_I=2.2^{+2.5}_{-1.3}\,.
\end{equation}
As a cross-check, the same combination strategy is applied, i.e.,
assuming no correlations. When using a total $\chi^2(k_I)$ of:

\begin{equation}
\chi^2(k_I,c) = 
 \Delta\chi^2_{\Delta\MW,D,\mathrm{full}}(k_I)
+\Delta\chi^2_{\mathrm{p-flow},D,\mathrm{full}}(k_I)\,,
\end{equation}
the following result is obtained:

\begin{equation}
k_I=2.17^{+2.55}_{-1.33}\,,
\end{equation}
which is consistent with the combination performed by \Delphi.  The
LEP combination procedure allows a more refined treatment of
correlations.  Using a correlation coefficient of 0.50 between the
measurements, motivated by the full covariance matrix for the
particle-flow combination, the fit obtains:

\begin{equation}
k_I=2.12^{+2.61}_{-1.33} \,.
\end{equation}
This corresponds to a 5\% increase of the positive uncertainty and a
small shift of the minimum. The result is shown in
Figure~\ref{fsi:cr:fig:lep-input}.

The \Ltre\ input is provided in terms of $r(k_I)$ derived from
Monte-Carlo simulations and the measured $r_\mathrm{data}$ together
with measurement uncertainties. The fit results in:

\begin{equation}
k_I=0.76^{+1.89}_{-1.22}\,,
\end{equation}
and the corresponding $\Delta\chi^2$ curve is shown in
Figure~\ref{fsi:cr:fig:lep-input}.  The result obtained is different
from the \Ltre\ paper on CR~\cite{bib:cr:L3_CR}, where a value of
$k_I=0.08^{+1.02}_{-0.08}$ is derived. This is due to the difference
between the non-CR reference Monte-Carlo simulation used by \Ltre\ and
the common LEP Monte-Carlo sample used in this combination. These
Monte-Carlo samples were generated with different fragmentation and
hadronisation parameters; the former was tuned to describe \Ltre\ data
best. The difference in the $k_I$ result is fully compatible with the
systematic uncertainty assigned to hadronisation and fragmentation
effects.

\Opal\ measures $k_I$ in W-mass shift and particle-flow analyses. As
for \Delphi, a correlation coefficient of 0.50 is assumed between the
correlated uncertainties in both inputs. Both $\Delta\chi^2$ curves
and their combination are shown in Figure~\ref{fsi:cr:fig:lep-input}.
The combined fit yields

\begin{equation}
k_I=1.24^{+1.13}_{-0.77}\,.
\end{equation}
The 68\% C.L. upper limit of 2.37 is in good agreement with the \Opal\
result $k_I<2.3$ at 68\% C.L.~\cite{bib:cr:OPAL_MW}. \Opal\ presents
the results in terms of $p_\mathrm{reco}=0.43^{+0.15}_{-0.20}$ which
translates into $k_I\approx 1.3^{+1.1}_{-0.8}$, using the conversion
from $p_\mathrm{reco}$ to $k_I$ based on \Opal's Monte-Carlo
simulation.  Using only information from $\Delta M_W$, the fit obtains
$k_I=1.75^{+1.99}_{-1.28}$, also agreeing well with the \Opal\
publication: $k_I=1.7^{+2.0}_{-1.2}$.

In summary, the LEP combination procedure reproduces well the results
obtained by each collaboration, with the observed differences
explained by known systematic effects.

\begin{figure}[tbhp]
  \centerline{
    \epsfig{file=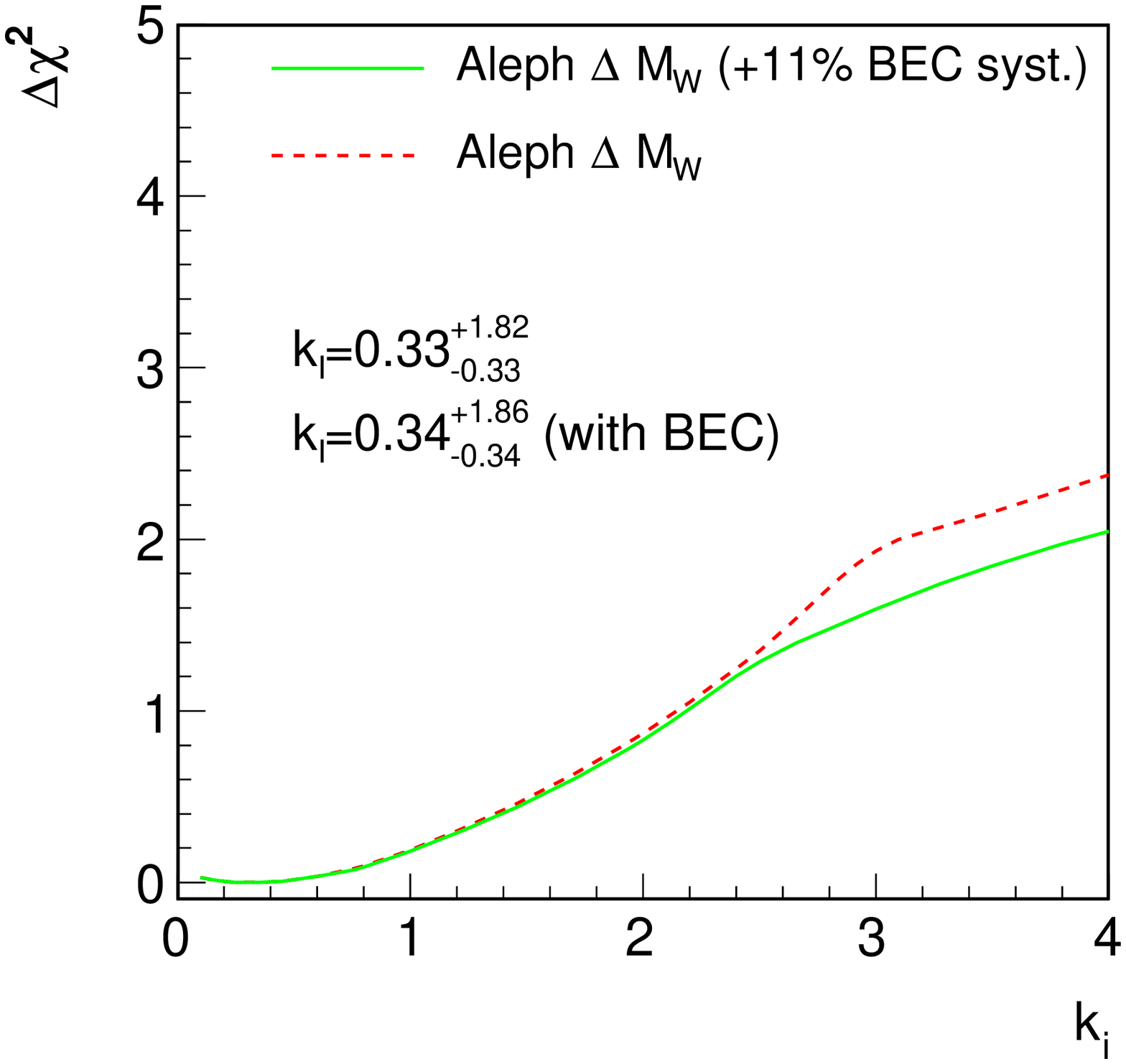,width=0.5\textwidth}
    \epsfig{file=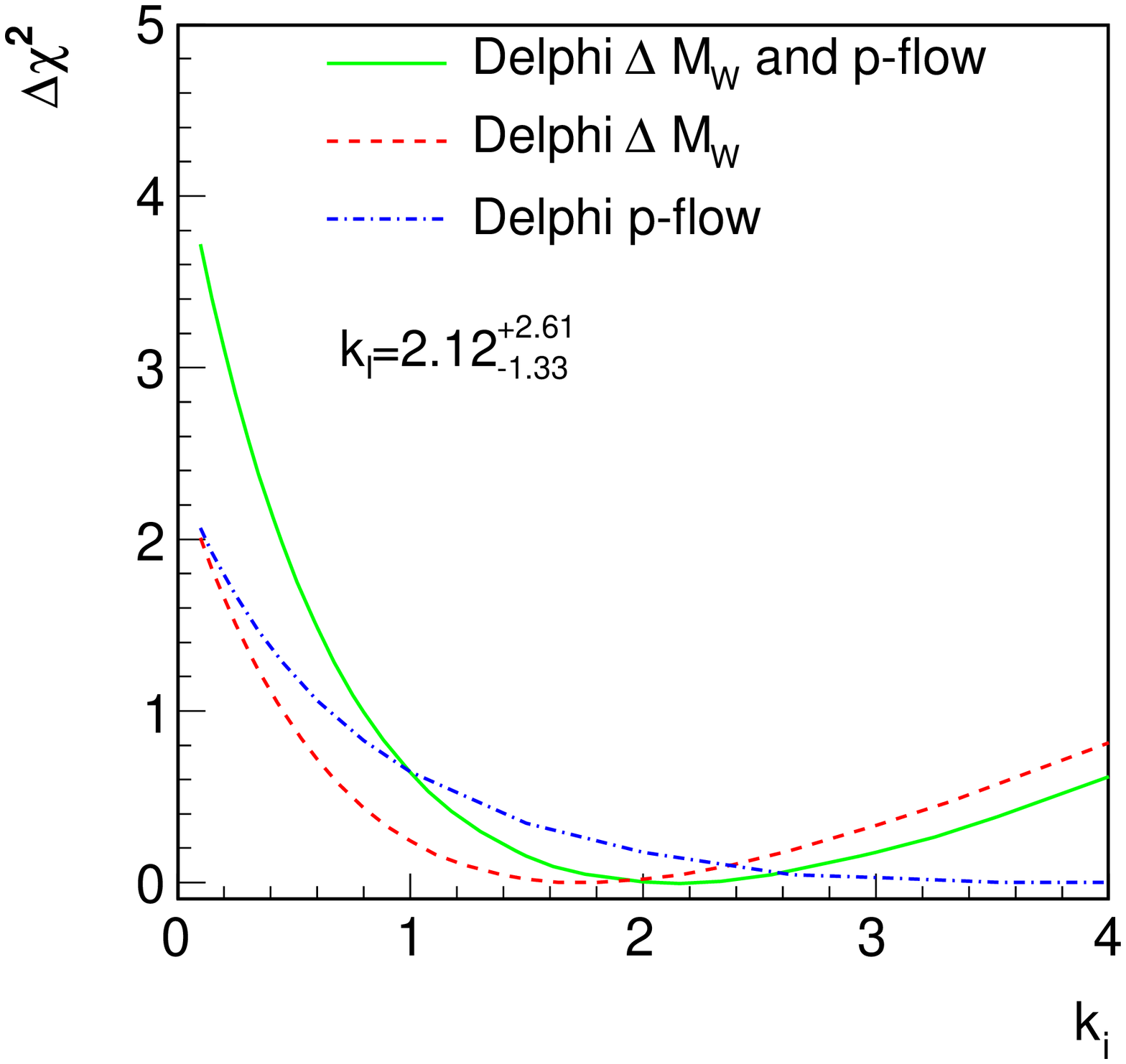,width=0.5\textwidth}
  }
  \centerline{
    \epsfig{file=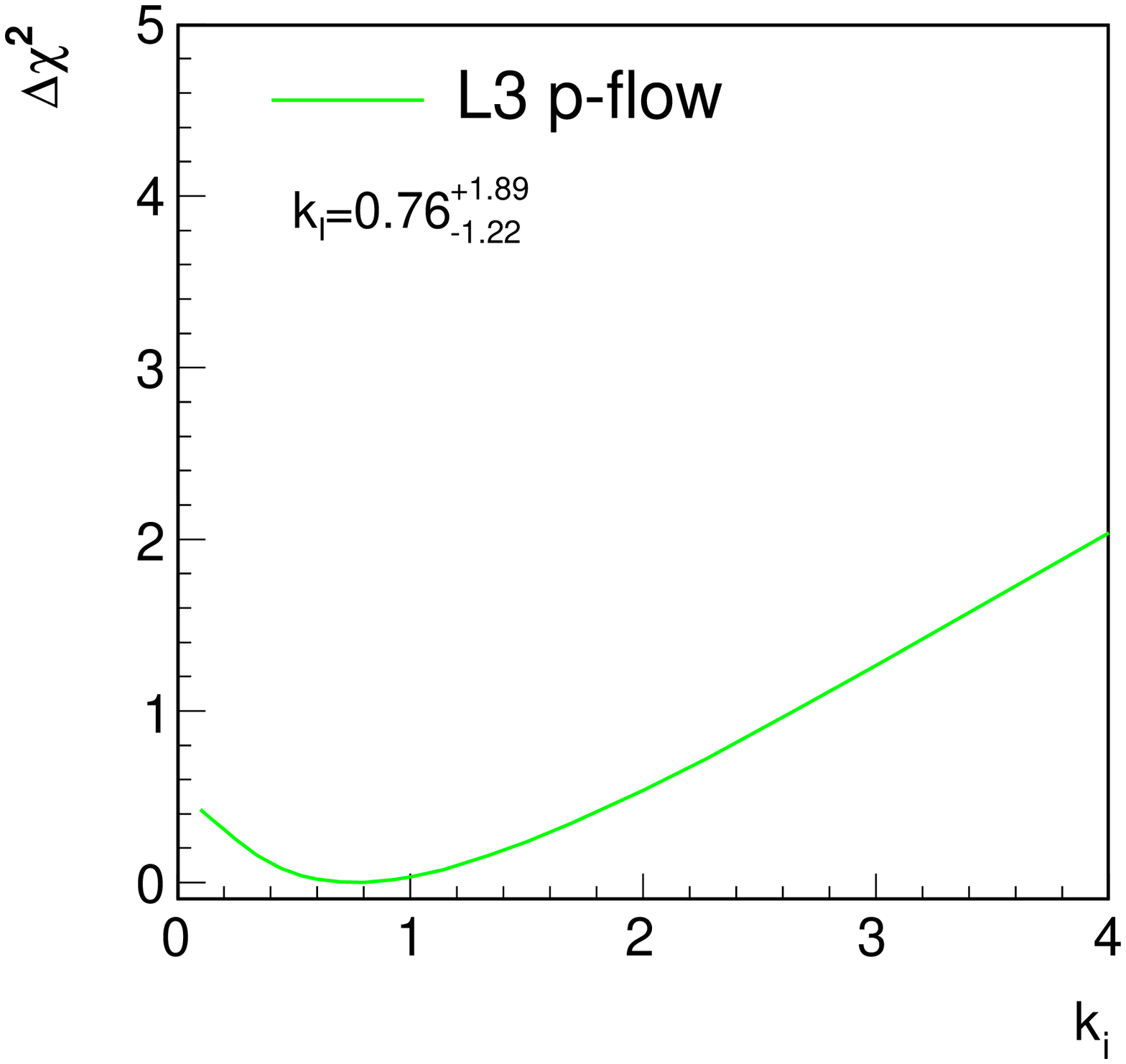,width=0.5\textwidth}
    \epsfig{file=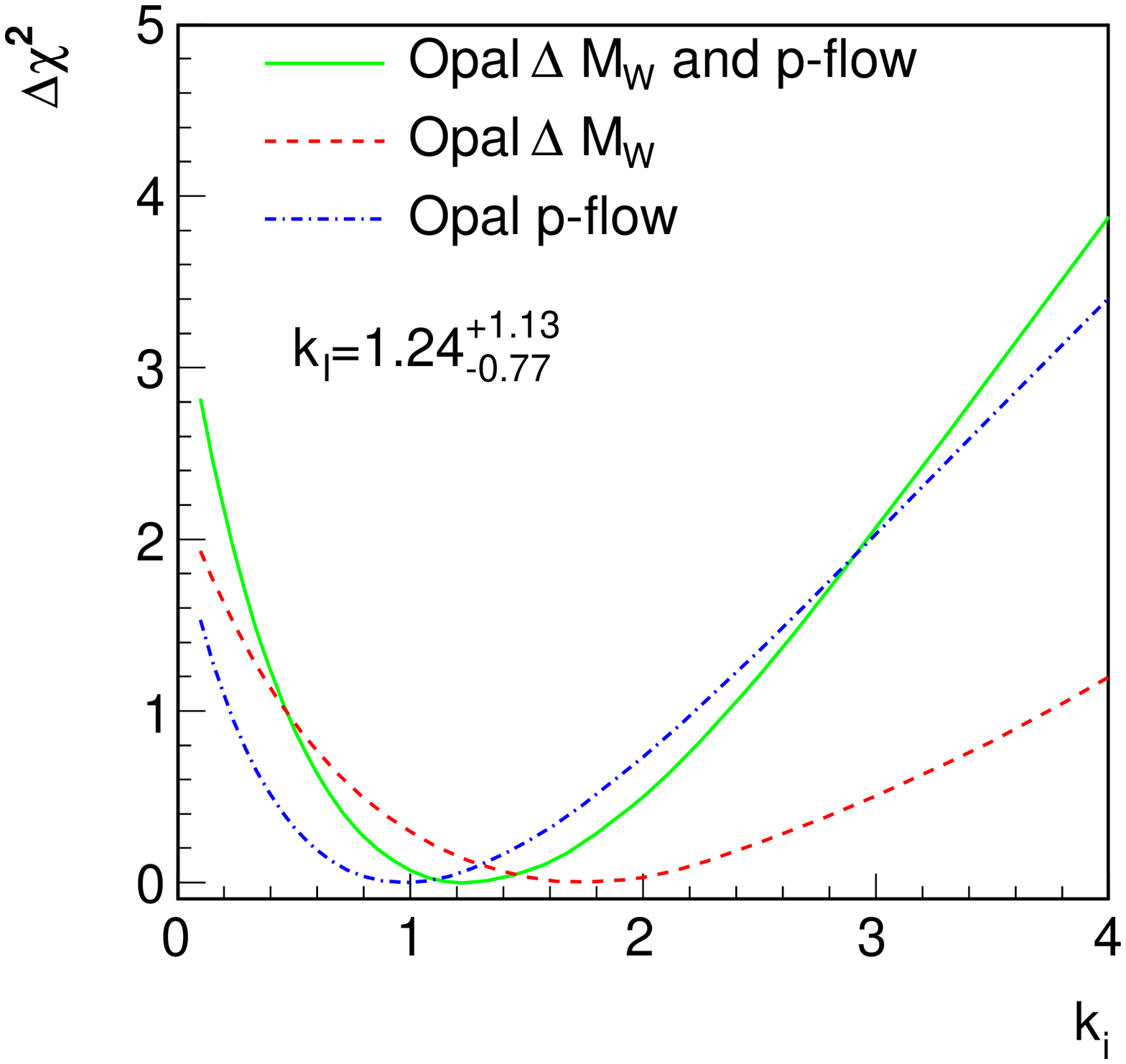,width=0.5\textwidth}
  }
 \caption[LEP input to the CR measurement.]  { LEP input to the CR
   measurement in terms of $\Delta\chi^2$ curves. The input data
   provided by the \Aleph\ experiment are shown as a dashed line and
   are compared to the data used in the LEP combination, where
   additional BEC systematic uncertainties are taken into account. The
   \Delphi\ and \Opal\ results from the analysis of the W-mass shift,
   $\Delta\MW$, and from the measurement of particle-flow are shown as
   dashed and dotted lines, respectively. The solid line represents
   the combined results taking correlations into account. The \Ltre\
   experiment provided input from the particle-flow measurement, also
   shown as a solid line.}
 \label{fsi:cr:fig:lep-input}
\end{figure}

\chapter{Detailed Inputs and Results of LEP Four-Fermion Averages}
\label{4f_sec:appendix}

Tables~\ref{4f_tab:WWmeasADLO}--\ref{4f_tab:rzeemeas} give the details of
the inputs and of the results for the calculation of LEP averages of
the measured four-fermion cross-sections and the corresponding ratios
of measured cross-sections to the theoretical predictions. For both
inputs and results, whenever relevant, the breakdown of the errors
into their various components is given in the tables.

For each measurement, the collaborations have privately provided
unpublished information which is necessary for the combination of the
LEP results, such as the expected statistical error or the split of
the total systematic uncertainty into correlated and uncorrelated
components.  Where necessary, minor re-arrangements with respect to
published results across error categories have been applied.

\begin{table}[hbtp]
\begin{center}
\begin{tabular}{|c|ccccc|c|c|c|}
\cline{1-8}
\roots & & & {\scriptsize (LCEC)} & {\scriptsize (LUEU)} & 
{\scriptsize (LUEC)} & & &
\multicolumn{1}{|r}{$\quad$} \\
(GeV) & $\sww$ & 
$\Delta\sww^\mathrm{stat}$ &
$\Delta\sww^\mathrm{syst}$ &
$\Delta\sww^\mathrm{syst}$ &
$\Delta\sww^\mathrm{syst}$ &
$\Delta\sww^\mathrm{syst}$ &
$\Delta\sww$ & 
\multicolumn{1}{|r}{$\quad$} \\
\cline{1-8}
\multicolumn{8}{|c|}
{\Aleph~\cite{4f_bib:aleww}} &
\multicolumn{1}{|r}{$\quad$} \\
\cline{1-8}
182.7 & 15.86 & $\pm$0.61 & $\pm$0.08 & $\pm$0.08 & $\pm$0.09 & $\pm$0.14& $\pm$0.63 & \multicolumn{1}{|r}{$\quad$} \\
188.6 & 15.78 & $\pm$0.34 & $\pm$0.07 & $\pm$0.05 & $\pm$0.09 & $\pm$0.12& $\pm$0.36 & \multicolumn{1}{|r}{$\quad$} \\
191.6 & 17.10 & $\pm$0.90 & $\pm$0.07 & $\pm$0.07 & $\pm$0.09 & $\pm$0.14& $\pm$0.90 & \multicolumn{1}{|r}{$\quad$} \\
195.5 & 16.60 & $\pm$0.52 & $\pm$0.07 & $\pm$0.06 & $\pm$0.09 & $\pm$0.12& $\pm$0.54 & \multicolumn{1}{|r}{$\quad$} \\
199.5 & 16.93 & $\pm$0.50 & $\pm$0.07 & $\pm$0.06 & $\pm$0.09 & $\pm$0.12& $\pm$0.52 & \multicolumn{1}{|r}{$\quad$} \\
201.6 & 16.63 & $\pm$0.70 & $\pm$0.07 & $\pm$0.07 & $\pm$0.09 & $\pm$0.13& $\pm$0.71 & \multicolumn{1}{|r}{$\quad$} \\
204.9 & 16.84 & $\pm$0.53 & $\pm$0.07 & $\pm$0.06 & $\pm$0.09 & $\pm$0.13& $\pm$0.54 & \multicolumn{1}{|r}{$\quad$} \\
206.6 & 17.42 & $\pm$0.41 & $\pm$0.07 & $\pm$0.06 & $\pm$0.09 & $\pm$0.13& $\pm$0.43 & \multicolumn{1}{|r}{$\quad$} \\
\cline{1-8}
\multicolumn{8}{|c|}
{\Delphi~\cite{4f_bib:delww}} &
\multicolumn{1}{|r}{$\quad$} \\
\cline{1-8}
182.7 & 16.07 & $\pm$0.68 & $\pm$0.09 & $\pm$0.09 & $\pm$0.08 & $\pm$0.15& $\pm$0.70 & \multicolumn{1}{|r}{$\quad$} \\
188.6 & 16.09 & $\pm$0.39 & $\pm$0.08 & $\pm$0.09 & $\pm$0.09 & $\pm$0.15& $\pm$0.42 & \multicolumn{1}{|r}{$\quad$} \\
191.6 & 16.64 & $\pm$0.99 & $\pm$0.09 & $\pm$0.10 & $\pm$0.09 & $\pm$0.16& $\pm$1.00 & \multicolumn{1}{|r}{$\quad$} \\
195.5 & 17.04 & $\pm$0.58 & $\pm$0.09 & $\pm$0.10 & $\pm$0.09 & $\pm$0.16& $\pm$0.60 & \multicolumn{1}{|r}{$\quad$} \\
199.5 & 17.39 & $\pm$0.55 & $\pm$0.09 & $\pm$0.10 & $\pm$0.09 & $\pm$0.16& $\pm$0.57 & \multicolumn{1}{|r}{$\quad$} \\
201.6 & 17.37 & $\pm$0.80 & $\pm$0.10 & $\pm$0.10 & $\pm$0.09 & $\pm$0.17& $\pm$0.82 & \multicolumn{1}{|r}{$\quad$} \\
204.9 & 17.56 & $\pm$0.57 & $\pm$0.10 & $\pm$0.10 & $\pm$0.09 & $\pm$0.17& $\pm$0.59 & \multicolumn{1}{|r}{$\quad$} \\
206.6 & 16.35 & $\pm$0.44 & $\pm$0.10 & $\pm$0.10 & $\pm$0.09 & $\pm$0.17& $\pm$0.47 & \multicolumn{1}{|r}{$\quad$} \\
\cline{1-8}
\multicolumn{8}{|c|}
{\Ltre~\cite{4f_bib:ltrww}} &
\multicolumn{1}{|r}{$\quad$} \\
\cline{1-8}
182.7 & 16.53 & $\pm$0.67 & $\pm$0.19 & $\pm$0.13 & $\pm$0.12 & $\pm$0.26& $\pm$0.72 & \multicolumn{1}{|r}{$\quad$} \\
188.6 & 16.17 & $\pm$0.37 & $\pm$0.11 & $\pm$0.06 & $\pm$0.11 & $\pm$0.17& $\pm$0.41 & \multicolumn{1}{|r}{$\quad$} \\
191.6 & 16.11 & $\pm$0.90 & $\pm$0.11 & $\pm$0.07 & $\pm$0.11 & $\pm$0.17& $\pm$0.92 & \multicolumn{1}{|r}{$\quad$} \\
195.5 & 16.22 & $\pm$0.54 & $\pm$0.11 & $\pm$0.06 & $\pm$0.10 & $\pm$0.16& $\pm$0.57 & \multicolumn{1}{|r}{$\quad$} \\
199.5 & 16.49 & $\pm$0.56 & $\pm$0.11 & $\pm$0.07 & $\pm$0.11 & $\pm$0.17& $\pm$0.58 & \multicolumn{1}{|r}{$\quad$} \\
201.6 & 16.01 & $\pm$0.82 & $\pm$0.11 & $\pm$0.06 & $\pm$0.12 & $\pm$0.17& $\pm$0.84 & \multicolumn{1}{|r}{$\quad$} \\
204.9 & 17.00 & $\pm$0.58 & $\pm$0.12 & $\pm$0.06 & $\pm$0.11 & $\pm$0.17& $\pm$0.60 & \multicolumn{1}{|r}{$\quad$} \\
206.6 & 17.33 & $\pm$0.44 & $\pm$0.12 & $\pm$0.04 & $\pm$0.11 & $\pm$0.17& $\pm$0.47 & \multicolumn{1}{|r}{$\quad$} \\
\cline{1-8}
\multicolumn{8}{|c|}
{\Opal~\cite{4f_bib:opaww}} &
\multicolumn{1}{|r}{$\quad$} \\
\cline{1-8}
182.7 & 15.45 & $\pm$0.61 & $\pm$0.10 & $\pm$0.04 & $\pm$0.05 & $\pm$0.12& $\pm$0.62 & \multicolumn{1}{|r}{$\quad$} \\
188.6 & 16.24 & $\pm$0.35 & $\pm$0.10 & $\pm$0.04 & $\pm$0.03 & $\pm$0.11& $\pm$0.37 & \multicolumn{1}{|r}{$\quad$} \\
191.6 & 15.93 & $\pm$0.86 & $\pm$0.10 & $\pm$0.04 & $\pm$0.03 & $\pm$0.11& $\pm$0.86 & \multicolumn{1}{|r}{$\quad$} \\
195.5 & 18.27 & $\pm$0.57 & $\pm$0.11 & $\pm$0.05 & $\pm$0.04 & $\pm$0.12& $\pm$0.58 & \multicolumn{1}{|r}{$\quad$} \\
199.5 & 16.29 & $\pm$0.54 & $\pm$0.11 & $\pm$0.04 & $\pm$0.03 & $\pm$0.12& $\pm$0.55 & \multicolumn{1}{|r}{$\quad$} \\
201.6 & 18.01 & $\pm$0.81 & $\pm$0.11 & $\pm$0.05 & $\pm$0.04 & $\pm$0.13& $\pm$0.82 & \multicolumn{1}{|r}{$\quad$} \\
204.9 & 16.05 & $\pm$0.52 & $\pm$0.11 & $\pm$0.04 & $\pm$0.04 & $\pm$0.12& $\pm$0.53 & \multicolumn{1}{|r}{$\quad$} \\
206.6 & 17.64 & $\pm$0.42 & $\pm$0.11 & $\pm$0.05 & $\pm$0.04 & $\pm$0.13& $\pm$0.44 & \multicolumn{1}{|r}{$\quad$} \\
\cline{1-8}
\end{tabular}
\caption[W-pair production cross-section]{W-pair production
cross-section (in pb) for different \CoM\ energies from the four LEP
experiments.  The first column contains the \CoM\ energy and the
second the measurements.  Observed statistical uncertainties are used
in the fit and are listed in the third column; when asymmetric errors
are quoted by the collaborations, the positive error is listed in the
table and used in the fit.  The fourth, fifth and sixth columns
contain the components of the systematic errors, as subdivided by the
collaborations into LEP-correlated energy-correlated (LCEC),
LEP-uncorrelated energy-uncorrelated (LUEU), LEP-uncorrelated
energy-correlated (LUEC).  The total systematic error is given in the
seventh column, the total error in the eighth.  }
\label{4f_tab:WWmeasADLO} 
\end{center}
\end{table}

\begin{table}[hbtp]
\begin{center}
\begin{tabular}{|c|ccccc|c|c|c|}
\hline
\multicolumn{9}{|c|}{LEP Averages } \\
\hline
\roots & & & {\scriptsize (LCEC)} & {\scriptsize (LUEU)} & 
{\scriptsize (LUEC)} & & &
\multicolumn{1}{|r|}{$\quad$} \\
(GeV) & $\sww$ & 
$\Delta\sww^\mathrm{stat}$ &
$\Delta\sww^\mathrm{syst}$ &
$\Delta\sww^\mathrm{syst}$ &
$\Delta\sww^\mathrm{syst}$ &
$\Delta\sww^\mathrm{syst}$ &
$\Delta\sww$ & $\chidf$ \\
\hline
\hline
182.7 & 15.92 & $\pm$0.33 & $\pm$0.10 & $\pm$0.04 & $\pm$0.04 & $\pm$0.11& $\pm$0.34 & 
 \multirow{8}{20.3mm}{$
   \hspace*{-0.3mm}
   \left\}
     \begin{array}[h]{rr}
       &\multirow{8}{8mm}{\hspace*{-4.2mm}26.6/24}\\
       &\\ &\\ &\\ &\\ &\\ &\\ &\\  
     \end{array}
   \right.
   $}\\
188.6 & 16.05 & $\pm$0.18 & $\pm$0.08 & $\pm$0.03 & $\pm$0.04 & $\pm$0.10& $\pm$0.21 & \\
191.6 & 16.42 & $\pm$0.46 & $\pm$0.08 & $\pm$0.04 & $\pm$0.04 & $\pm$0.10& $\pm$0.47 & \\
195.5 & 16.99 & $\pm$0.28 & $\pm$0.08 & $\pm$0.03 & $\pm$0.04 & $\pm$0.10& $\pm$0.29 & \\
199.5 & 16.77 & $\pm$0.27 & $\pm$0.08 & $\pm$0.03 & $\pm$0.04 & $\pm$0.10& $\pm$0.29 & \\
201.6 & 16.98 & $\pm$0.39 & $\pm$0.08 & $\pm$0.04 & $\pm$0.04 & $\pm$0.10& $\pm$0.40 & \\
204.9 & 16.81 & $\pm$0.27 & $\pm$0.08 & $\pm$0.03 & $\pm$0.04 & $\pm$0.10& $\pm$0.29 & \\
206.6 & 17.20 & $\pm$0.21 & $\pm$0.09 & $\pm$0.03 & $\pm$0.04 & $\pm$0.11& $\pm$0.24 & \\
\hline
\end{tabular}
\caption[W-pair production cross-section]{LEP combined W-pair production
cross-section (in pb) for different \CoM\ energies.  The first column
contains the \CoM\ energy and the second the measurements.  Observed
statistical uncertainties are used in the fit and are listed in the
third column; when asymmetric errors are quoted by the collaborations,
the positive error is listed in the table and used in the fit.  The
fourth, fifth and sixth columns contain the components of the
systematic errors, as subdivided by the collaborations into
LEP-correlated energy-correlated (LCEC), LEP-uncorrelated
energy-uncorrelated (LUEU), LEP-uncorrelated energy-correlated (LUEC).
The total systematic error is given in the seventh column, the total
error in the eighth.  The $\chidf$ of the fit is
also given in the ninth column.}
\label{4f_tab:WWmeasLEP} 
\end{center}
\end{table}

\begin{table}[hbtp]
\begin{center}
\hspace*{-0.3cm}
\renewcommand{\arraystretch}{1.2}
\begin{tabular}{|c|cccccccc|} 
\hline
\roots (GeV) 
      & 182.7 & 188.6 & 191.6 & 195.5 & 199.5 & 201.6 & 204.9 & 206.6 \\
\hline
182.7 & 1.000 & 0.145 & 0.065 & 0.104 & 0.105 & 0.076 & 0.104 & 0.130 \\
188.6 & 0.145 & 1.000 & 0.093 & 0.148 & 0.149 & 0.108 & 0.148 & 0.186 \\
191.6 & 0.065 & 0.093 & 1.000 & 0.066 & 0.067 & 0.048 & 0.066 & 0.083 \\
195.5 & 0.104 & 0.148 & 0.066 & 1.000 & 0.107 & 0.077 & 0.106 & 0.133 \\
199.5 & 0.105 & 0.149 & 0.067 & 0.107 & 1.000 & 0.078 & 0.106 & 0.134 \\
201.6 & 0.076 & 0.108 & 0.048 & 0.077 & 0.078 & 1.000 & 0.077 & 0.097 \\
204.9 & 0.104 & 0.148 & 0.066 & 0.106 & 0.106 & 0.077 & 1.000 & 0.132 \\
206.6 & 0.130 & 0.186 & 0.083 & 0.133 & 0.134 & 0.097 & 0.132 & 1.000 \\
\hline
\end{tabular}
\renewcommand{\arraystretch}{1.}
\caption[Correlations]{Correlation matrix for the LEP combined W-pair
cross-sections listed in Table~\protect\ref{4f_tab:WWmeasLEP}.
Correlations are all positive and range from 5\% to 19\%.}
\label{4f_tab:WWcorr} 
\end{center}
\end{table}

\begin{table}[hbtp]
\begin{center}
\hspace*{-0.3cm}
\renewcommand{\arraystretch}{1.2}
\begin{tabular}{|c|c|c|} 
\hline
\roots & \multicolumn{2}{|c|}{WW cross-section (pb)}                              \\
\cline{2-3} 
(GeV) & $\sww^{\footnotesize\YFSWW}$    
      & $\sww^{\footnotesize\RacoonWW}$ \\
\hline
182.7 & $15.361\pm0.005$ & $15.368\pm0.008$ \\
188.6 & $16.266\pm0.005$ & $16.249\pm0.011$ \\
191.6 & $16.568\pm0.006$ & $16.519\pm0.009$ \\
195.5 & $16.841\pm0.006$ & $16.801\pm0.009$ \\
199.5 & $17.017\pm0.007$ & $16.979\pm0.009$ \\
201.6 & $17.076\pm0.006$ & $17.032\pm0.009$ \\
204.9 & $17.128\pm0.006$ & $17.079\pm0.009$ \\
206.6 & $17.145\pm0.006$ & $17.087\pm0.009$ \\
\hline
\end{tabular}
\renewcommand{\arraystretch}{1.}
\caption[W-pair production cross-section]{ W-pair cross-section
predictions (in pb) for different \CoM\ energies, according to
\YFSWW~\protect\cite{\YFSWWref} and
\RacoonWW~\protect\cite{\RACOONWWref}, for $\Mw=80.35$~GeV.  The
errors listed in the table are only the statistical errors from the
numerical integration of the cross-section.}
\label{4f_tab:WWtheo} 
\end{center}
\end{table}

\begin{table}[hbtp]
\begin{center}
\begin{small}
\begin{tabular}{|c|cccccc|c|c|}
\hline
\roots & & & {\scriptsize (LCEU)} & {\scriptsize (LCEC)} & 
{\scriptsize (LUEU)} & {\scriptsize (LUEC)} & & \\
(GeV) & $\rww$ & 
$\Delta\rww^\mathrm{stat}$ &
$\Delta\rww^\mathrm{syst}$ &
$\Delta\rww^\mathrm{syst}$ &
$\Delta\rww^\mathrm{syst}$ &
$\Delta\rww^\mathrm{syst}$ &
$\Delta\rww$ &
$\chidf$ \\
\hline
\hline

\multicolumn{9}{|c|}{\YFSWW~\cite{\YFSWWref}}\\
\hline
182.7 & 1.037 & $\pm$0.021 & $\pm$0.000 & $\pm$0.006 & $\pm$0.003& $\pm$0.003 & $\pm$0.022&
\multirow{8}{20.3mm}{$
  \hspace*{-0.3mm}
  \left\}
    \begin{array}[h]{rr}
      &\multirow{8}{6mm}{\hspace*{-4.2mm}26.6/24}\\
      &\\ &\\ &\\ &\\ &\\ &\\ &\\  
    \end{array}
  \right.
  $}\\
188.6 & 0.987 & $\pm$0.011 & $\pm$0.000 & $\pm$0.005 & $\pm$0.002& $\pm$0.003 & $\pm$0.013&\\
191.6 & 0.991 & $\pm$0.028 & $\pm$0.000 & $\pm$0.005 & $\pm$0.002& $\pm$0.002 & $\pm$0.028&\\
195.5 & 1.009 & $\pm$0.016 & $\pm$0.000 & $\pm$0.005 & $\pm$0.002& $\pm$0.002 & $\pm$0.018&\\
199.5 & 0.985 & $\pm$0.016 & $\pm$0.000 & $\pm$0.005 & $\pm$0.002& $\pm$0.003 & $\pm$0.017&\\
201.6 & 0.994 & $\pm$0.023 & $\pm$0.000 & $\pm$0.005 & $\pm$0.002& $\pm$0.003 & $\pm$0.023&\\
204.9 & 0.982 & $\pm$0.016 & $\pm$0.000 & $\pm$0.005 & $\pm$0.002& $\pm$0.002 & $\pm$0.017&\\
206.6 & 1.003 & $\pm$0.013 & $\pm$0.000 & $\pm$0.005 & $\pm$0.002& $\pm$0.002 & $\pm$0.014&\\
\hline
Average & 
0.995 & $\pm$0.006 & $\pm$0.000 & $\pm$0.005 & $\pm$0.001& $\pm$0.003 & $\pm$0.008&
\hspace*{1.5mm}32.2/31\hspace*{-0.5mm}\\
\hline
\hline
\multicolumn{9}{|c|}{\RacoonWW~\cite{\RACOONWWref}}\\
\hline
182.7 & 1.036 & $\pm$0.021 & $\pm$0.001 & $\pm$0.007 & $\pm$0.003& $\pm$0.003 & $\pm$0.022&
\multirow{8}{20.3mm}{$
  \hspace*{-0.3mm}
  \left\}
    \begin{array}[h]{rr}
      &\multirow{8}{6mm}{\hspace*{-4.2mm}26.6/24}\\
      &\\ &\\ &\\ &\\ &\\ &\\ &\\  
    \end{array}
  \right.
  $}\\
188.6 & 0.988 & $\pm$0.011 & $\pm$0.001 & $\pm$0.005 & $\pm$0.002& $\pm$0.003 & $\pm$0.013&\\
191.6 & 0.994 & $\pm$0.028 & $\pm$0.001 & $\pm$0.005 & $\pm$0.002& $\pm$0.002 & $\pm$0.028&\\
195.5 & 1.011 & $\pm$0.017 & $\pm$0.001 & $\pm$0.005 & $\pm$0.002& $\pm$0.003 & $\pm$0.018&\\
199.5 & 0.987 & $\pm$0.016 & $\pm$0.001 & $\pm$0.005 & $\pm$0.002& $\pm$0.003 & $\pm$0.017&\\
201.6 & 0.997 & $\pm$0.023 & $\pm$0.001 & $\pm$0.005 & $\pm$0.002& $\pm$0.003 & $\pm$0.024&\\
204.9 & 0.984 & $\pm$0.016 & $\pm$0.001 & $\pm$0.005 & $\pm$0.002& $\pm$0.002 & $\pm$0.017&\\
206.6 & 1.007 & $\pm$0.013 & $\pm$0.001 & $\pm$0.005 & $\pm$0.002& $\pm$0.002 & $\pm$0.014&\\
\hline
Average & 
0.997 & $\pm$0.006 & $\pm$0.000 & $\pm$0.005 & $\pm$0.001& $\pm$0.003 & $\pm$0.008&
\hspace*{1.5mm}32.0/31\hspace*{-0.5mm}\\
\hline
\end{tabular}
\end{small}
\caption[Ratio of W-pair production cross-sections]{ Ratios of LEP
combined W-pair cross-section measurements to the expectations of the
considered theoretical models, for different \CoM\ energies and for
all energies combined.  The first column contains the \CoM\ energy,
the second the combined ratios, the third the statistical errors.  The
fourth, fifth, sixth and seventh columns contain the sources of
systematic errors that are considered as LEP-correlated
energy-uncorrelated (LCEU), LEP-correlated energy-correlated (LCEC),
LEP-uncorrelated energy-uncorrelated (LUEU), LEP-uncorrelated
energy-correlated (LUEC).  The total error is given in the eighth
column.  The only LCEU systematic sources considered are the
statistical errors on the cross-section theoretical predictions, while
the LCEC, LUEU and LUEC sources are those coming from the
corresponding errors on the cross-section measurements.  For the LEP
averages, the $\chidf$ of the fit is also given in the ninth column.}
\label{4f_tab:rWWmeas} 
\end{center}
\end{table}

 \renewcommand{\arraystretch}{1.2}
 \begin{table}[p]
 \begin{center}
 \begin{small}
 \hspace*{-0.0cm}
 \begin{tabular}{|l|cccc|c|c|c|}
 \cline{1-8}
 Decay & & & {\scriptsize (unc)} & {\scriptsize (cor)} & & & 
 3$\times$3 correlation \\
 channel & $\wwbr$ & 
 $\Delta\wwbr^\mathrm{stat}$ &
 $\Delta\wwbr^\mathrm{syst}$ &
 $\Delta\wwbr^\mathrm{syst}$ &
 $\Delta\wwbr^\mathrm{syst}$ &
 $\Delta\wwbr$ & 
 for $\Delta\wwbr$\\
 \cline{1-8}
 \multicolumn{8}{|c|}{\Aleph~\cite{4f_bib:aleww}}\\
 \hline
 \BWtoenu & 
 10.78 & $\pm$0.27 & $\pm$0.09 & $\pm$0.04 & $\pm$0.10 & $\pm$0.29 &
 \multirow{3}{47mm}{\mbox{$\Biggl(\negthickspace\negthickspace$
                      \begin{tabular}{ccc}
                       \phm1.000 &    -0.009 &    -0.332 \\
                          -0.009 & \phm1.000 &    -0.268 \\
                          -0.332 &    -0.268 & \phm1.000 \\
                      \end{tabular}
                      $\negthickspace\negthickspace\Biggr)$} } \\
 \BWtomnu & 
 10.87 & $\pm$0.25 & $\pm$0.07 & $\pm$0.04 & $\pm$0.08 & $\pm$0.26 & \\
 \BWtotnu & 
 11.25 & $\pm$0.32 & $\pm$0.19 & $\pm$0.05 & $\pm$0.20 & $\pm$0.38 & \\
 \hline
 \multicolumn{8}{c}{}\\
 
 \cline{1-8}
 \multicolumn{8}{|c|}{\Delphi~\cite{4f_bib:delww}}\\
 \hline
 \BWtoenu & 
 10.55 & $\pm$0.31 & $\pm$0.13 & $\pm$0.05 & $\pm$0.14 & $\pm$0.34 &
 \multirow{3}{47mm}{\mbox{$\Biggl(\negthickspace\negthickspace$
                      \begin{tabular}{ccc}
                       \phm1.000 &     0.030 &    -0.340 \\
                           0.030 & \phm1.000 &    -0.170 \\
                          -0.340 &    -0.170 & \phm1.000 \\
                      \end{tabular}
                      $\negthickspace\negthickspace\Biggr)$} } \\
 \BWtomnu & 
 10.65 & $\pm$0.26 & $\pm$0.06 & $\pm$0.05 & $\pm$0.08 & $\pm$0.27 & \\
 \BWtotnu & 
 11.46 & $\pm$0.39 & $\pm$0.17 & $\pm$0.09 & $\pm$0.19 & $\pm$0.43 & \\
 \hline
 \multicolumn{8}{c}{}\\
 
 \cline{1-8}
 \multicolumn{8}{|c|}{\Ltre~\cite{4f_bib:ltrww}}\\
 \hline
 \BWtoenu & 
 10.78 & $\pm$0.29 & $\pm$0.10 & $\pm$0.08 & $\pm$0.13 & $\pm$0.32 & 
 \multirow{3}{47mm}{\mbox{$\Biggl(\negthickspace\negthickspace$
                      \begin{tabular}{ccc}
                       \phm1.000 &    -0.016 &    -0.279 \\
                          -0.016 & \phm1.000 &    -0.295 \\
                          -0.279 &    -0.295 & \phm1.000 \\
                      \end{tabular}
                      $\negthickspace\negthickspace\Biggr)$} } \\
 \BWtomnu & 
 10.03 & $\pm$0.29 & $\pm$0.10 & $\pm$0.07 & $\pm$0.12 & $\pm$0.31 & \\
 \BWtotnu & 
 11.89 & $\pm$0.40 & $\pm$0.17 & $\pm$0.11 & $\pm$0.20 & $\pm$0.45 & \\
 \hline
 \multicolumn{8}{c}{}\\
 
 \cline{1-8}
 \multicolumn{8}{|c|}{\Opal~\cite{4f_bib:opaww}}\\
 \hline
 \BWtoenu & 
 10.71 & $\pm$0.25 & $\pm$0.09 & $\pm$0.06 & $\pm$0.11 & $\pm$0.27 & 
 \multirow{3}{47mm}{\mbox{$\Biggl(\negthickspace\negthickspace$
                      \begin{tabular}{ccc}
                       \phm1.000 &     0.135 &    -0.303 \\
                           0.135 & \phm1.000 &    -0.230 \\
                          -0.303 &    -0.230 & \phm1.000 \\
                      \end{tabular}
                      $\negthickspace\negthickspace\Biggr)$} } \\
 \BWtomnu & 
 10.78 & $\pm$0.24 & $\pm$0.07 & $\pm$0.07 & $\pm$0.10 & $\pm$0.26 & \\
 \BWtotnu & 
 11.14 & $\pm$0.31 & $\pm$0.16 & $\pm$0.06 & $\pm$0.17 & $\pm$0.35 & \\
 \hline
 \multicolumn{8}{c}{}\\
 
 \cline{1-8}
 \multicolumn{7}{|c}{LEP Average (without lepton universality assumption)}
 &\multicolumn{1}{c|}{}\\
 \hline
 \BWtoenu & 
 10.71 & $\pm$0.14 & $\pm$0.05 & $\pm$0.05 & $\pm$0.07 & $\pm$0.16 &
 \multirow{3}{47mm}{\mbox{$\Biggl(\negthickspace\negthickspace$
                      \begin{tabular}{ccc}
                        \phm1.000 & \phm0.136 &    -0.201 \\
                        \phm0.136 & \phm1.000 &    -0.122 \\
                           -0.201 &    -0.122 & \phm1.000 \\
                      \end{tabular}
                      $\negthickspace\negthickspace\Biggr)$} } \\
 \BWtomnu & 
 10.63 & $\pm$0.13 & $\pm$0.04 & $\pm$0.05 & $\pm$0.07 & $\pm$0.15 & \\
 \BWtotnu & 
 11.38 & $\pm$0.17 & $\pm$0.09 & $\pm$0.07 & $\pm$0.11 & $\pm$0.21 & \\
 \hline
 $\chidf$ & \multicolumn{1}{|c|}{6.3/9} & 
 \multicolumn{6}{c}{}\\
 \cline{1-2} 
 \multicolumn{8}{c}{}\\
 
 \cline{1-7} 
 \multicolumn{7}{|c|}{LEP Average (with lepton universality assumption)}
 &\multicolumn{1}{c}{}\\
 \cline{1-7} 
 \BWtolnu & 
 10.86 & $\pm$0.06 & $\pm$0.03 & $\pm$0.06 & $\pm$0.07 & $\pm$0.09 & 
 \multicolumn{1}{c}{}\\
 {\mbox{$\mathcal{B}(\mathrm{W}\rightarrow\mathrm{had.})$}}  & 
 67.41 & $\pm$0.18 & $\pm$0.10 & $\pm$0.17 & $\pm$0.20 & $\pm$0.27 & 
 \multicolumn{1}{c}{}\\
 \cline{1-7} 
 $\chidf$ & \multicolumn{1}{|c|}{15.4/11} &
 \multicolumn{6}{c}{}\\
 \cline{1-2} 
 \end{tabular}
 \vspace*{0.5cm}
 
 \end{small}
 \caption[W Branching fractions]{ W branching fraction measurements
 (in \%).  The first column contains the decay channel, the second the
 measurements, the third the statistical uncertainty.  The fourth and
 fifth column list the uncorrelated and correlated components of the
 systematic errors, as provided by the collaborations.  The total
 systematic error is given in the sixth column and the total error in
 the seventh.  Correlation matrices for the three leptonic branching
 fractions are given in the last column.}
 \label{4f_tab:Wbrmeas} 
 \end{center}
 \end{table}
 \renewcommand{\arraystretch}{1.}

\begin{table}[hbtp]
\begin{center}
\begin{small}
\begin{tabular}{|c|}
\hline
ALEPH~\cite{4f_bib:aleww} \\
\hline 
\end{tabular}
\\
\begin{tabular}{|c|c|c|}
\hline
$\sqrt{s}$ interval (GeV) & Luminosity (pb$^{-1}$) & Luminosity weighted $\sqrt{s}$ (GeV) \\
180-184 & 56.81 & 182.65 \\
\hline
\end{tabular}
\begin{tabular}{|c|c|c|c|c|c|c|c|c|c|c|}
\hline
cos$\theta_{\mathrm{W}-}$ bin $i$ & 1 & 2 & 3 & 4 & 5 & 6 & 7 & 8 & 9 & 10 \\
$\sigma_i$  (pb)                 & 0.216 & 0.498 & 0.696 & 1.568 & 1.293 & 1.954 & 2.486 & 2.228 & 4.536 & 6.088 \\
$\delta\sigma_i$(stat)  (pb)     & 0.053 & 0.137 & 0.185 & 0.517 & 0.319 & 0.481 & 0.552 & 0.363 & 0.785 & 0.874 \\
$\delta\sigma_i$(stat,exp.) (pb) & 0.263 & 0.276 & 0.309 & 0.341 & 0.376 & 0.415 & 0.459 & 0.523 & 0.597 & 0.714 \\
$\delta\sigma_i$(syst,unc)  (pb) & 0.012 & 0.018 & 0.017 & 0.025 & 0.023 & 0.021 & 0.036 & 0.047 & 0.047 & 0.066 \\
$\delta\sigma_i$(syst,cor)  (pb) & 0.004 & 0.003 & 0.003 & 0.003 & 0.003 & 0.004 & 0.004 & 0.003 & 0.004 & 0.006 \\
\hline
\end{tabular}

\begin{tabular}{|c|c|c|}
\hline
$\sqrt{s}$ interval (GeV) & Luminosity (pb$^{-1}$) & Luminosity weighted $\sqrt{s}$ (GeV) \\
184-194 & 203.14 & 189.05 \\
\hline
\end{tabular}
\begin{tabular}{|c|c|c|c|c|c|c|c|c|c|c|}
\hline
cos$\theta_{\mathrm{W}-}$ bin $i$ & 1 & 2 & 3 & 4 & 5 & 6 & 7 & 8 & 9 & 10 \\
$\sigma_i$  (pb)                 & 0.665 & 0.743 & 0.919 & 0.990 & 1.156 & 2.133 & 2.795 & 3.070 & 3.851 & 5.772 \\
$\delta\sigma_i$(stat)  (pb)     & 0.148 & 0.140 & 0.158 & 0.142 & 0.144 & 0.287 & 0.337 & 0.297 & 0.300 & 0.366 \\
$\delta\sigma_i$(stat,exp.) (pb) & 0.132 & 0.147 & 0.157 & 0.175 & 0.196 & 0.223 & 0.246 & 0.282 & 0.332 & 0.408 \\
$\delta\sigma_i$(syst,unc)  (pb) & 0.010 & 0.016 & 0.015 & 0.024 & 0.021 & 0.020 & 0.035 & 0.047 & 0.049 & 0.075 \\
$\delta\sigma_i$(syst,cor)  (pb) & 0.003 & 0.003 & 0.003 & 0.002 & 0.002 & 0.003 & 0.003 & 0.003 & 0.005 & 0.005 \\
\hline
\end{tabular}

\begin{tabular}{|c|c|c|}
\hline
$\sqrt{s}$ interval (GeV) & Luminosity (pb$^{-1}$) & Luminosity weighted $\sqrt{s}$ (GeV) \\
194-204 & 208.03 & 198.42 \\
\hline
\end{tabular}
\begin{tabular}{|c|c|c|c|c|c|c|c|c|c|c|}
\hline
cos$\theta_{\mathrm{W}-}$ bin $i$ & 1 & 2 & 3 & 4 & 5 & 6 & 7 & 8 & 9 & 10 \\

$\sigma_i$ (pb)                   & 0.802 & 0.475 & 0.886 & 0.972 & 1.325 & 1.889 & 2.229 & 3.581 & 4.428 & 6.380 \\
$\delta\sigma_i$(stat) (pb)       & 0.225 & 0.082 & 0.162 & 0.147 & 0.186 & 0.248 & 0.245 & 0.363 & 0.343 & 0.368 \\
$\delta\sigma_i$(stat,exp.) (pb)  & 0.124 & 0.134 & 0.149 & 0.167 & 0.188 & 0.214 & 0.241 & 0.281 & 0.338 & 0.433 \\
$\delta\sigma_i$(syst,unc) (pb)   & 0.007 & 0.013 & 0.012 & 0.021 & 0.018 & 0.016 & 0.032 & 0.046 & 0.049 & 0.082 \\
$\delta\sigma_i$(syst,cor) (pb)   & 0.003 & 0.002 & 0.002 & 0.002 & 0.002 & 0.002 & 0.002 & 0.003 & 0.003 & 0.004 \\
\hline
\end{tabular}

\begin{tabular}{|c|c|c|}
\hline
$\sqrt{s}$ interval (GeV) & Luminosity (pb$^{-1}$) & Luminosity weighted $\sqrt{s}$ (GeV) \\
204-210 & 214.62 & 205.90 \\
\hline
\end{tabular}
\begin{tabular}{|c|c|c|c|c|c|c|c|c|c|c|}
\hline
cos$\theta_{\mathrm{W}-}$ bin $i$ & 1 & 2 & 3 & 4 & 5 & 6 & 7 & 8 & 9 & 10 \\

$\sigma_i$ (pb)                   & 0.334 & 0.637 & 0.800 & 1.229 & 1.229 & 1.789 & 2.810 & 2.740 & 4.192 & 8.005 \\
$\delta\sigma_i$(stat) (pb)       & 0.072 & 0.136 & 0.148 & 0.224 & 0.176 & 0.237 & 0.351 & 0.246 & 0.306 & 0.474 \\
$\delta\sigma_i$(stat,exp.) (pb)  & 0.114 & 0.126 & 0.143 & 0.155 & 0.180 & 0.206 & 0.234 & 0.273 & 0.338 & 0.443 \\
$\delta\sigma_i$(syst,unc) (pb)   & 0.008 & 0.013 & 0.013 & 0.020 & 0.018 & 0.017 & 0.033 & 0.046 & 0.052 & 0.089 \\
$\delta\sigma_i$(syst,cor) (pb)   & 0.003 & 0.003 & 0.003 & 0.002 & 0.002 & 0.003 & 0.003 & 0.003 & 0.004 & 0.005 \\
\hline
\end{tabular}
\end{small}
\caption[Differential cross-section]{W$^{-}$ differential angular
cross-section in the 10 angular bins for the four chosen energy
intervals for the \Aleph\ experiment. For each energy range, the
measured integrated luminosity and the luminosity-weighted
centre-of-mass energy is reported.  The results per angular bin in
each energy interval are then presented: $\sigma_{i}$ indicates the
average of
d[$\sigma_{\mathrm{WW}}$(BR$_{e\nu}$+BR$_{\mu\nu}$)]/dcos$\theta_{\mathrm{W}^-}$
in the $i$-th bin of cos$\theta_{\mathrm{W}^-}$ with width 0.2.  The
values in each bin of the measured and expected statistical error and
of the systematic errors, LEP uncorrelated and correlated, are
reported as well. All values are expressed in pb. }
\label{4f_tab:dsdcost_aleph} 
\end{center}
\end{table}

\begin{table}[hbtp]
\begin{center}
\begin{small}
\begin{tabular}{|c|}
\hline
DELPHI~\cite{4f_bib:delww} \\
\hline 
\end{tabular}
\\
\begin{tabular}{|c|c|c|}
\hline
$\sqrt{s}$ interval (GeV) & Luminosity (pb$^{-1}$) & Luminosity weighted $\sqrt{s}$ (GeV) \\
180-184 & 51.63 & 182.65 \\
\hline
\end{tabular}
\begin{tabular}{|c|c|c|c|c|c|c|c|c|c|c|}
\hline
cos$\theta_{\mathrm{W}-}$ bin $i$ & 1 & 2 & 3 & 4 & 5 & 6 & 7 & 8 & 9 & 10 \\
$\sigma_i$  (pb)                 & 0.715 & 0.795 & 1.175 & 1.365 & 1.350 & 1.745 & 1.995 & 2.150 & 4.750 & 6.040 \\
$\delta\sigma_i$(stat)  (pb)     & 0.320 & 0.315 & 0.380 & 0.400 & 0.400 & 0.450 & 0.485 & 0.510 & 0.775 & 0.895 \\
$\delta\sigma_i$(stat,exp.) (pb) & 0.320 & 0.315 & 0.350 & 0.370 & 0.405 & 0.450 & 0.505 & 0.580 & 0.695 & 0.850 \\ 
$\delta\sigma_i$(syst,unc)  (pb) & 0.020 & 0.025 & 0.035 & 0.035 & 0.040 & 0.085 & 0.050 & 0.065 & 0.095 & 0.075 \\
$\delta\sigma_i$(syst,cor)  (pb) & 0.045 & 0.025 & 0.020 & 0.015 & 0.015 & 0.025 & 0.015 & 0.015 & 0.030 & 0.035 \\
\hline
\end{tabular}

\begin{tabular}{|c|c|c|}
\hline
$\sqrt{s}$ interval (GeV) & Luminosity (pb$^{-1}$) & Luminosity weighted $\sqrt{s}$ (GeV) \\
184-194 & 178.32 & 189.03 \\
\hline
\end{tabular}
\begin{tabular}{|c|c|c|c|c|c|c|c|c|c|c|}
\hline
cos$\theta_{\mathrm{W}-}$ bin $i$ & 1 & 2 & 3 & 4 & 5 & 6 & 7 & 8 & 9 & 10 \\
$\sigma_i$  (pb)                 & 0.865 & 0.760 & 0.990 & 0.930 & 1.330 & 1.460 & 1.675 & 2.630 & 4.635 & 5.400 \\
$\delta\sigma_i$(stat)  (pb)     & 0.180 & 0.170 & 0.185 & 0.180 & 0.215 & 0.225 & 0.240 & 0.300 & 0.405 & 0.455 \\
$\delta\sigma_i$(stat,exp.) (pb) & 0.165 & 0.170 & 0.180 & 0.200 & 0.215 & 0.240 & 0.270 & 0.320 & 0.385 & 0.490 \\
$\delta\sigma_i$(syst,unc)  (pb) & 0.020 & 0.020 & 0.035 & 0.035 & 0.040 & 0.085 & 0.050 & 0.060 & 0.100 & 0.085 \\
$\delta\sigma_i$(syst,cor)  (pb) & 0.040 & 0.020 & 0.020 & 0.015 & 0.015 & 0.020 & 0.015 & 0.015 & 0.025 & 0.035 \\
\hline
\end{tabular}

\begin{tabular}{|c|c|c|}
\hline
$\sqrt{s}$ interval (GeV) & Luminosity (pb$^{-1}$) & Luminosity weighted $\sqrt{s}$ (GeV) \\
194-204 & 193.52 & 198.46 \\
\hline
\end{tabular}
\begin{tabular}{|c|c|c|c|c|c|c|c|c|c|c|}
\hline
cos$\theta_{\mathrm{W}-}$ bin $i$ & 1 & 2 & 3 & 4 & 5 & 6 & 7 & 8 & 9 & 10 \\
$\sigma_i$ (pb)                  & 0.600 & 0.675 & 1.510 & 1.150 & 1.055 & 1.635 & 2.115 & 3.175 & 4.470 & 7.140 \\
$\delta\sigma_i$(stat) (pb)      & 0.155 & 0.160 & 0.215 & 0.190 & 0.185 & 0.225 & 0.255 & 0.320 & 0.385 & 0.500 \\
$\delta\sigma_i$(stat,exp.) (pb) & 0.150 & 0.160 & 0.170 & 0.180 & 0.200 & 0.230 & 0.260 & 0.310 & 0.380 & 0.505 \\
$\delta\sigma_i$(syst,unc) (pb)  & 0.015 & 0.020 & 0.030 & 0.035 & 0.035 & 0.085 & 0.045 & 0.055 & 0.105 & 0.100 \\
$\delta\sigma_i$(syst,cor) (pb)  & 0.025 & 0.015 & 0.015 & 0.015 & 0.015 & 0.015 & 0.010 & 0.015 & 0.025 & 0.030 \\
\hline
\end{tabular}

\begin{tabular}{|c|c|c|}
\hline
$\sqrt{s}$ interval (GeV) & Luminosity (pb$^{-1}$) & Luminosity weighted $\sqrt{s}$ (GeV) \\
204-210 & 198.59 & 205.91 \\
\hline
\end{tabular}
\begin{tabular}{|c|c|c|c|c|c|c|c|c|c|c|}
\hline
cos$\theta_{\mathrm{W}-}$ bin $i$ & 1 & 2 & 3 & 4 & 5 & 6 & 7 & 8 & 9 & 10 \\
$\sigma_i$ (pb)                  & 0.275 & 0.590 & 0.575 & 0.930 & 1.000 & 1.190 & 2.120 & 2.655 & 4.585 & 7.290 \\
$\delta\sigma_i$(stat) (pb)      & 0.120 & 0.145 & 0.140 & 0.170 & 0.175 & 0.195 & 0.255 & 0.290 & 0.385 & 0.505 \\
$\delta\sigma_i$(stat,exp.) (pb) & 0.145 & 0.150 & 0.160 & 0.175 & 0.195 & 0.220 & 0.250 & 0.300 & 0.380 & 0.520 \\
$\delta\sigma_i$(syst,unc) (pb)  & 0.015 & 0.020 & 0.025 & 0.035 & 0.035 & 0.085 & 0.045 & 0.055 & 0.110 & 0.110 \\
$\delta\sigma_i$(syst,cor) (pb)  & 0.020 & 0.015 & 0.010 & 0.010 & 0.015 & 0.010 & 0.010 & 0.010 & 0.020 & 0.030 \\
\hline
\end{tabular}
\end{small}
\caption[Differential cross-section]{W$^{-}$ differential angular
cross-section in the 10 angular bins for the four chosen energy
intervals for the \Delphi\ experiment. For each energy range, the
measured integrated luminosity and the luminosity-weighted
centre-of-mass energy is reported.  The results per angular bin in
each energy interval are then presented: $\sigma_{i}$ indicates the
average of
d[$\sigma_{\mathrm{WW}}$(BR$_{e\nu}$+BR$_{\mu\nu}$)]/dcos$\theta_{\mathrm{W}^-}$
in the $i$-th bin of cos$\theta_{\mathrm{W}^-}$ with width 0.2.  The
values in each bin of the measured and expected statistical error and
of the systematic errors, LEP uncorrelated and correlated, are
reported as well. All values are expressed in pb. }
\label{4f_tab:dsdcost_delphi} 
\end{center}
\end{table}

\begin{table}[hbtp]
\begin{center}
\begin{small}
\begin{tabular}{|c|}
\hline
L3~\cite{4f_bib:ltrww} \\
\hline 
\end{tabular}
\\
\begin{tabular}{|c|c|c|}
\hline
$\sqrt{s}$ interval (GeV) & Luminosity (pb$^{-1}$) & Luminosity weighted $\sqrt{s}$ (GeV) \\
180-184 & 55.46 & 182.68 \\
\hline
\end{tabular}
\begin{tabular}{|c|c|c|c|c|c|c|c|c|c|c|}
\hline
cos$\theta_{\mathrm{W}-}$ bin $i$ & 1 & 2 & 3 & 4 & 5 & 6 & 7 & 8 & 9 & 10 \\
$\sigma_i$  (pb)                 & 0.691 & 0.646 & 0.508 & 0.919 & 1.477 & 2.587 & 3.541 & 3.167 & 3.879 & 4.467 \\
$\delta\sigma_i$(stat)  (pb)     & 0.270 & 0.265 & 0.243 & 0.322 & 0.407 & 0.539 & 0.640 & 0.619 & 0.708 & 0.801 \\
$\delta\sigma_i$(stat,exp.) (pb) & 0.269 & 0.290 & 0.329 & 0.364 & 0.404 & 0.453 & 0.508 & 0.591 & 0.704 & 0.877 \\
$\delta\sigma_i$(syst,unc)  (pb) & 0.016 & 0.009 & 0.007 & 0.011 & 0.018 & 0.031 & 0.043 & 0.039 & 0.048 & 0.058 \\
$\delta\sigma_i$(syst,cor)  (pb) & 0.002 & 0.002 & 0.002 & 0.003 & 0.005 & 0.009 & 0.012 & 0.011 & 0.013 & 0.015 \\
\hline
\end{tabular}

\begin{tabular}{|c|c|c|}
\hline
$\sqrt{s}$ interval (GeV) & Luminosity (pb$^{-1}$) & Luminosity weighted $\sqrt{s}$ (GeV) \\
184-194 & 206.49 & 189.16 \\
\hline
\end{tabular}
\begin{tabular}{|c|c|c|c|c|c|c|c|c|c|c|}
\hline
cos$\theta_{\mathrm{W}-}$ bin $i$ & 1 & 2 & 3 & 4 & 5 & 6 & 7 & 8 & 9 & 10 \\
$\sigma_i$  (pb)                 & 0.759 & 0.902 & 1.125 & 1.320 & 1.472 & 1.544 & 2.085 & 2.870 & 4.144 & 6.022 \\
$\delta\sigma_i$(stat)  (pb)     & 0.128 & 0.151 & 0.173 & 0.190 & 0.209 & 0.213 & 0.254 & 0.303 & 0.370 & 0.459 \\
$\delta\sigma_i$(stat,exp.) (pb) & 0.115 & 0.137 & 0.160 & 0.180 & 0.205 & 0.223 & 0.262 & 0.304 & 0.367 & 0.461 \\
$\delta\sigma_i$(syst,unc)  (pb) & 0.017 & 0.013 & 0.015 & 0.015 & 0.017 & 0.018 & 0.024 & 0.034 & 0.048 & 0.074 \\
$\delta\sigma_i$(syst,cor)  (pb) & 0.003 & 0.003 & 0.004 & 0.005 & 0.005 & 0.005 & 0.007 & 0.010 & 0.014 & 0.021 \\
\hline
\end{tabular}

\begin{tabular}{|c|c|c|}
\hline
$\sqrt{s}$ interval (GeV) & Luminosity (pb$^{-1}$) & Luminosity weighted $\sqrt{s}$ (GeV) \\
194-204 & 203.50 & 198.30 \\
\hline
\end{tabular}
\begin{tabular}{|c|c|c|c|c|c|c|c|c|c|c|}
\hline
cos$\theta_{\mathrm{W}-}$ bin $i$ & 1 & 2 & 3 & 4 & 5 & 6 & 7 & 8 & 9 & 10 \\
$\sigma_i$ (pb)                  & 0.652 & 0.709 & 0.880 & 0.859 & 1.140 & 1.295 & 2.114 & 2.334 & 3.395 & 5.773 \\
$\delta\sigma_i$(stat) (pb)      & 0.105 & 0.123 & 0.146 & 0.155 & 0.179 & 0.192 & 0.255 & 0.264 & 0.333 & 0.442 \\
$\delta\sigma_i$(stat,exp.) (pb) & 0.092 & 0.117 & 0.140 & 0.164 & 0.184 & 0.209 & 0.245 & 0.288 & 0.354 & 0.459 \\
$\delta\sigma_i$(syst,unc) (pb)  & 0.014 & 0.010 & 0.011 & 0.010 & 0.013 & 0.015 & 0.024 & 0.027 & 0.040 & 0.071 \\
$\delta\sigma_i$(syst,cor) (pb)  & 0.002 & 0.002 & 0.003 & 0.003 & 0.004 & 0.004 & 0.007 & 0.008 & 0.012 & 0.020 \\
\hline
\end{tabular}

\begin{tabular}{|c|c|c|}
\hline
$\sqrt{s}$ interval (GeV) & Luminosity (pb$^{-1}$) & Luminosity weighted $\sqrt{s}$ (GeV) \\
204-210 & 217.30 & 205.96 \\
\hline
\end{tabular}
\begin{tabular}{|c|c|c|c|c|c|c|c|c|c|c|}
\hline
cos$\theta_{\mathrm{W}-}$ bin $i$ & 1 & 2 & 3 & 4 & 5 & 6 & 7 & 8 & 9 & 10 \\
$\sigma_i$ (pb)                  & 0.678 & 0.578 & 0.768 & 1.052 & 1.620 & 1.734 & 1.873 & 2.903 & 4.638 & 7.886 \\
$\delta\sigma_i$(stat) (pb)      & 0.111 & 0.114 & 0.140 & 0.168 & 0.212 & 0.226 & 0.238 & 0.302 & 0.394 & 0.534 \\
$\delta\sigma_i$(stat,exp.) (pb) & 0.089 & 0.117 & 0.141 & 0.164 & 0.186 & 0.216 & 0.251 & 0.303 & 0.387 & 0.528 \\
$\delta\sigma_i$(syst,unc) (pb)  & 0.015 & 0.008 & 0.010 & 0.012 & 0.019 & 0.020 & 0.021 & 0.034 & 0.054 & 0.097 \\
$\delta\sigma_i$(syst,cor) (pb)  & 0.002 & 0.002 & 0.003 & 0.004 & 0.006 & 0.006 & 0.006 & 0.010 & 0.016 & 0.027 \\
\hline
\end{tabular}
\end{small}
\caption[Differential cross-section]{W$^{-}$ differential angular
cross-section in the 10 angular bins for the four chosen energy
intervals for the \Ltre\ experiment. For each energy range, the
measured integrated luminosity and the luminosity-weighted
centre-of-mass energy is reported.  The results per angular bin in
each energy interval are then presented: $\sigma_{i}$ indicates the
average of
d[$\sigma_{\mathrm{WW}}$(BR$_{e\nu}$+BR$_{\mu\nu}$)]/dcos$\theta_{\mathrm{W}^-}$
in the $i$-th bin of cos$\theta_{\mathrm{W}^-}$ with width 0.2.  The
values in each bin of the measured and expected statistical error and
of the systematic errors, LEP uncorrelated and correlated, are
reported as well. All values are expressed in pb. }
\label{4f_tab:dsdcost_l3} 
\end{center}
\end{table}

\begin{table}[hbtp]
\begin{center}
\begin{small}
\begin{tabular}{|c|}
\hline
OPAL~\cite{4f_bib:opaww} \\
\hline 
\end{tabular}
\\
\begin{tabular}{|c|c|c|}
\hline
$\sqrt{s}$ interval (GeV) & Luminosity (pb$^{-1}$) & Luminosity weighted $\sqrt{s}$ (GeV) \\
180-184 & 57.38 & 182.68 \\
\hline
\end{tabular}
\begin{tabular}{|c|c|c|c|c|c|c|c|c|c|c|}
\hline
cos$\theta_{\mathrm{W}-}$ bin $i$ & 1 & 2 & 3 & 4 & 5 & 6 & 7 & 8 & 9 & 10 \\
$\sigma_i$  (pb)                 & 0.462 & 0.910 & 1.101 & 1.247 & 1.910 & 2.291 & 2.393 & 2.871 & 3.851 & 4.746 \\
$\delta\sigma_i$(stat)  (pb)     & 0.228 & 0.298 & 0.313 & 0.333 & 0.408 & 0.451 & 0.461 & 0.507 & 0.602 & 0.689 \\
$\delta\sigma_i$(stat,exp.) (pb) & 0.276 & 0.286 & 0.296 & 0.328 & 0.353 & 0.396 & 0.444 & 0.502 & 0.599 & 0.735 \\
$\delta\sigma_i$(syst,unc)  (pb) & 0.008 & 0.013 & 0.013 & 0.020 & 0.018 & 0.017 & 0.033 & 0.046 & 0.052 & 0.089 \\
$\delta\sigma_i$(syst,cor)  (pb) & 0.003 & 0.003 & 0.003 & 0.002 & 0.002 & 0.003 & 0.003 & 0.003 & 0.004 & 0.005 \\
\hline
\end{tabular}

\begin{tabular}{|c|c|c|}
\hline
$\sqrt{s}$ interval (GeV) & Luminosity (pb$^{-1}$) & Luminosity weighted $\sqrt{s}$ (GeV) \\
184-194 & 212.37 & 189.04 \\
\hline
\end{tabular}
\begin{tabular}{|c|c|c|c|c|c|c|c|c|c|c|}
\hline
cos$\theta_{\mathrm{W}-}$ bin $i$ & 1 & 2 & 3 & 4 & 5 & 6 & 7 & 8 & 9 & 10 \\
$\sigma_i$  (pb)                 & 0.621 & 0.980 & 1.004 & 1.125 & 1.193 & 1.944 & 2.190 & 2.696 & 3.622 & 5.798 \\
$\delta\sigma_i$(stat)  (pb)     & 0.135 & 0.160 & 0.158 & 0.165 & 0.168 & 0.213 & 0.228 & 0.256 & 0.305 & 0.401 \\
$\delta\sigma_i$(stat,exp.) (pb) & 0.139 & 0.145 & 0.154 & 0.167 & 0.180 & 0.202 & 0.230 & 0.267 & 0.326 & 0.417 \\
$\delta\sigma_i$(syst,unc)  (pb) & 0.008 & 0.013 & 0.013 & 0.020 & 0.018 & 0.017 & 0.033 & 0.046 & 0.052 & 0.089 \\
$\delta\sigma_i$(syst,cor)  (pb) & 0.003 & 0.003 & 0.003 & 0.002 & 0.002 & 0.003 & 0.003 & 0.003 & 0.004 & 0.005 \\
\hline
\end{tabular}

\begin{tabular}{|c|c|c|}
\hline
$\sqrt{s}$ interval (GeV) & Luminosity (pb$^{-1}$) & Luminosity weighted $\sqrt{s}$ (GeV) \\
194-204 & 190.67 & 198.35 \\
\hline
\end{tabular}
\begin{tabular}{|c|c|c|c|c|c|c|c|c|c|c|}
\hline
cos$\theta_{\mathrm{W}-}$ bin $i$ & 1 & 2 & 3 & 4 & 5 & 6 & 7 & 8 & 9 & 10 \\
$\sigma_i$ (pb)                  & 0.651 & 0.678 & 0.834 & 1.397 & 1.543 & 1.994 & 1.844 & 2.422 & 4.168 & 7.044 \\
$\delta\sigma_i$(stat) (pb)      & 0.147 & 0.145 & 0.153 & 0.191 & 0.200 & 0.224 & 0.219 & 0.256 & 0.344 & 0.472 \\
$\delta\sigma_i$(stat,exp.) (pb) & 0.140 & 0.148 & 0.156 & 0.168 & 0.185 & 0.204 & 0.238 & 0.282 & 0.353 & 0.478 \\
$\delta\sigma_i$(syst,unc) (pb)  & 0.008 & 0.013 & 0.013 & 0.020 & 0.018 & 0.017 & 0.033 & 0.046 & 0.052 & 0.089 \\
$\delta\sigma_i$(syst,cor) (pb)  & 0.003 & 0.003 & 0.003 & 0.002 & 0.002 & 0.003 & 0.003 & 0.003 & 0.004 & 0.005 \\
\hline
\end{tabular}

\begin{tabular}{|c|c|c|}
\hline
$\sqrt{s}$ interval (GeV) & Luminosity (pb$^{-1}$) & Luminosity weighted $\sqrt{s}$ (GeV) \\
204-210 & 220.45 & 205.94 \\
\hline
\end{tabular}
\begin{tabular}{|c|c|c|c|c|c|c|c|c|c|c|}
\hline
cos$\theta_{\mathrm{W}-}$ bin $i$ & 1 & 2 & 3 & 4 & 5 & 6 & 7 & 8 & 9 & 10 \\
$\sigma_i$ (pb)                  & 0.496 & 0.606 & 0.453 & 0.989 & 1.116 & 1.919 & 2.303 & 2.874 & 4.573 & 7.129 \\
$\delta\sigma_i$(stat) (pb)      & 0.122 & 0.129 & 0.111 & 0.151 & 0.158 & 0.206 & 0.227 & 0.256 & 0.335 & 0.442 \\
$\delta\sigma_i$(stat,exp.) (pb) & 0.123 & 0.133 & 0.140 & 0.149 & 0.164 & 0.185 & 0.215 & 0.258 & 0.331 & 0.458 \\
$\delta\sigma_i$(syst,unc) (pb)  & 0.008 & 0.013 & 0.013 & 0.020 & 0.018 & 0.017 & 0.033 & 0.046 & 0.052 & 0.089 \\
$\delta\sigma_i$(syst,cor) (pb)  & 0.003 & 0.003 & 0.003 & 0.002 & 0.002 & 0.003 & 0.003 & 0.003 & 0.004 & 0.005 \\
\hline
\end{tabular}
\end{small}
\caption[Differential cross-section]{W$^{-}$ differential angular
cross-section in the 10 angular bins for the four chosen energy
intervals for the \Opal\ experiment. For each energy range, the
measured integrated luminosity and the luminosity-weighted
centre-of-mass energy is reported.  The results per angular bin in
each energy interval are then presented: $\sigma_{i}$ indicates the
average of
d[$\sigma_{\mathrm{WW}}$(BR$_{e\nu}$+BR$_{\mu\nu}$)]/dcos$\theta_{\mathrm{W}^-}$
in the $i$-th bin of cos$\theta_{\mathrm{W}^-}$ with width 0.2.  The
values in each bin of the measured and expected statistical error and
of the systematic errors, LEP uncorrelated and correlated, are
reported as well. All values are expressed in pb. }
\label{4f_tab:dsdcost_opal} 
\end{center}
\end{table}

\begin{table}[p]
\begin{center}
\begin{tabular}{|c|ccccc|c|c|}
\hline
\roots & & & {\scriptsize (LCEC)} & {\scriptsize (LUEU)} & 
{\scriptsize (LUEC)} & & 
\multicolumn{1}{|r|}{$\quad$} \\
(GeV) & $\szz$ & 
$\Delta\szz^\mathrm{stat}$ &
$\Delta\szz^\mathrm{syst}$ &
$\Delta\szz^\mathrm{syst}$ &
$\Delta\szz^\mathrm{syst}$ &
$\Delta\szz$ & 
$\Delta\szz^\mathrm{stat\,(exp)}$ \\
\hline
\multicolumn{8}{|c|}
{\Aleph~\cite{4f_bib:alezz}} \\
\hline
182.7 & 0.11 & $^{+0.16}_{-0.11}$ & $\pm$0.01 & $\pm$0.03 & $\pm$0.03 & $^{+0.16}_{-0.12}$ & $\pm$0.14 \\
188.6 & 0.67 & $^{+0.13}_{-0.12}$ & $\pm$0.01 & $\pm$0.03 & $\pm$0.03 & $^{+0.14}_{-0.13}$ & $\pm$0.13 \\
191.6 & 0.62 & $^{+0.40}_{-0.32}$ & $\pm$0.01 & $\pm$0.06 & $\pm$0.01 & $^{+0.40}_{-0.33}$ & $\pm$0.36 \\
195.5 & 0.73 & $^{+0.24}_{-0.21}$ & $\pm$0.01 & $\pm$0.06 & $\pm$0.01 & $^{+0.25}_{-0.22}$ & $\pm$0.23 \\
199.5 & 0.91 & $^{+0.24}_{-0.21}$ & $\pm$0.01 & $\pm$0.08 & $\pm$0.01 & $^{+0.25}_{-0.22}$ & $\pm$0.23 \\
201.6 & 0.71 & $^{+0.31}_{-0.26}$ & $\pm$0.01 & $\pm$0.08 & $\pm$0.01 & $^{+0.32}_{-0.27}$ & $\pm$0.29 \\
204.9 & 1.20 & $^{+0.27}_{-0.25}$ & $\pm$0.01 & $\pm$0.07 & $\pm$0.02 & $^{+0.28}_{-0.26}$ & $\pm$0.26 \\
206.6 & 1.05 & $^{+0.21}_{-0.20}$ & $\pm$0.01 & $\pm$0.06 & $\pm$0.01 & $^{+0.22}_{-0.21}$ & $\pm$0.21 \\
\hline
\multicolumn{8}{|c|}
{\Delphi~\cite{4f_bib:delzz}} \\
\hline
182.7 & 0.35 & $^{+0.20}_{-0.15}$ & $\pm$0.01 & $\pm$0.00 & $\pm$0.02 & $^{+0.20}_{-0.15}$ & $\pm$0.16 \\
188.6 & 0.52 & $^{+0.12}_{-0.11}$ & $\pm$0.01 & $\pm$0.00 & $\pm$0.02 & $^{+0.12}_{-0.11}$ & $\pm$0.13 \\
191.6 & 0.63 & $^{+0.36}_{-0.30}$ & $\pm$0.01 & $\pm$0.01 & $\pm$0.02 & $^{+0.36}_{-0.30}$ & $\pm$0.35 \\
195.5 & 1.05 & $^{+0.25}_{-0.22}$ & $\pm$0.01 & $\pm$0.01 & $\pm$0.02 & $^{+0.25}_{-0.22}$ & $\pm$0.21 \\
199.5 & 0.75 & $^{+0.20}_{-0.18}$ & $\pm$0.01 & $\pm$0.01 & $\pm$0.01 & $^{+0.20}_{-0.18}$ & $\pm$0.21 \\
201.6 & 0.85 & $^{+0.33}_{-0.28}$ & $\pm$0.01 & $\pm$0.01 & $\pm$0.01 & $^{+0.33}_{-0.28}$ & $\pm$0.32 \\
204.9 & 1.03 & $^{+0.23}_{-0.20}$ & $\pm$0.02 & $\pm$0.01 & $\pm$0.01 & $^{+0.23}_{-0.20}$ & $\pm$0.23 \\
206.6 & 0.96 & $^{+0.16}_{-0.15}$ & $\pm$0.02 & $\pm$0.01 & $\pm$0.01 & $^{+0.16}_{-0.15}$ & $\pm$0.17 \\
\hline
\multicolumn{8}{|c|}
{\Ltre~\cite{4f_bib:ltrzz}} \\
\hline
182.7 & 0.31 & $\pm$0.16 & $\pm$0.05 & $\pm$0.00 & $\pm$0.01 & $\pm$0.17 & $\pm$0.16 \\
188.6 & 0.73 & $\pm$0.15 & $\pm$0.02 & $\pm$0.02 & $\pm$0.02 & $\pm$0.15 & $\pm$0.15 \\
191.6 & 0.29 & $\pm$0.22 & $\pm$0.01 & $\pm$0.01 & $\pm$0.02 & $\pm$0.22 & $\pm$0.34 \\
195.5 & 1.18 & $\pm$0.24 & $\pm$0.04 & $\pm$0.05 & $\pm$0.06 & $\pm$0.26 & $\pm$0.22 \\
199.5 & 1.25 & $\pm$0.25 & $\pm$0.04 & $\pm$0.05 & $\pm$0.07 & $\pm$0.27 & $\pm$0.24 \\
201.6 & 0.95 & $\pm$0.38 & $\pm$0.03 & $\pm$0.04 & $\pm$0.05 & $\pm$0.39 & $\pm$0.35 \\
204.9 & 0.77 & $^{+0.21}_{-0.19}$ & $\pm$0.01 & $\pm$0.01 & $\pm$0.04 & $^{+0.21}_{-0.19}$ & $\pm$0.22 \\
206.6 & 1.09 & $^{+0.17}_{-0.16}$ & $\pm$0.02 & $\pm$0.02 & $\pm$0.06 & $^{+0.18}_{-0.17}$ & $\pm$0.17 \\
\hline
\multicolumn{8}{|c|}
{\Opal~\cite{4f_bib:opazz}}  \\
\hline
182.7 & 0.12 & $^{+0.20}_{-0.18}$ & $\pm$0.00 & $\pm$0.03 & $\pm$0.00 & $^{+0.20}_{-0.18}$ & $\pm$0.19 \\
188.6 & 0.80 & $^{+0.14}_{-0.13}$ & $\pm$0.01 & $\pm$0.05 & $\pm$0.03 & $^{+0.15}_{-0.14}$ & $\pm$0.14 \\
191.6 & 1.29 & $^{+0.47}_{-0.40}$ & $\pm$0.02 & $\pm$0.09 & $\pm$0.05 & $^{+0.48}_{-0.41}$ & $\pm$0.36 \\
195.5 & 1.13 & $^{+0.26}_{-0.24}$ & $\pm$0.02 & $\pm$0.06 & $\pm$0.05 & $^{+0.27}_{-0.25}$ & $\pm$0.25 \\
199.5 & 1.05 & $^{+0.25}_{-0.22}$ & $\pm$0.02 & $\pm$0.05 & $\pm$0.04 & $^{+0.26}_{-0.23}$ & $\pm$0.25 \\
201.6 & 0.79 & $^{+0.35}_{-0.29}$ & $\pm$0.02 & $\pm$0.05 & $\pm$0.03 & $^{+0.36}_{-0.30}$ & $\pm$0.37 \\
204.9 & 1.07 & $^{+0.27}_{-0.24}$ & $\pm$0.02 & $\pm$0.06 & $\pm$0.04 & $^{+0.28}_{-0.25}$ & $\pm$0.26 \\
206.6 & 0.97 & $^{+0.19}_{-0.18}$ & $\pm$0.02 & $\pm$0.05 & $\pm$0.04 & $^{+0.20}_{-0.19}$ & $\pm$0.20 \\
\hline
\end{tabular}
\caption[Z-pair cross-section]{Z-pair production cross-section (in pb)
at different energies from the four LEP experiments.  The first column
contains the LEP \CoM\ energy, the second the measurements and the
third the statistical uncertainty.  The fourth, the fifth and the
sixth columns list the different components of the systematic errors
as defined in Table~\ref{4f_tab:rWWmeas}.  The total error is given in
the seventh column, and the eighth column lists the symmetrised
expected statistical error for each of the four experiments.}
\label{4f_tab:ZZmeasADLO} 
\end{center}
\end{table}

\begin{table}[p]
\begin{center}
\begin{tabular}{|c|ccccc|c|c|}
\hline
\multicolumn{8}{|c|} {LEP} \\
\hline
\roots & & & {\scriptsize (LCEC)} & {\scriptsize (LUEU)} & 
{\scriptsize (LUEC)} & & 
\multicolumn{1}{|r|}{$\quad$} \\
(GeV) & $\szz$ & 
$\Delta\szz^\mathrm{stat}$ &
$\Delta\szz^\mathrm{syst}$ &
$\Delta\szz^\mathrm{syst}$ &
$\Delta\szz^\mathrm{syst}$ &
$\Delta\szz$ & $\chidf$ \\
\hline
\hline
182.7 & 0.22 & $\pm$0.08 & $\pm$0.02 & $\pm$0.01 & $\pm$0.01 & $\pm$0.08 & 
 \multirow{8}{20.3mm}{$
   \hspace*{-0.3mm}
   \left\}
     \begin{array}[h]{rr}
       &\multirow{8}{8mm}{\hspace*{-4.2mm}14.5/24}\\
       &\\ &\\ &\\ &\\ &\\ &\\ &\\  
     \end{array}
   \right.
   $}\\
188.6 & 0.66 & $\pm$0.07 & $\pm$0.01 & $\pm$0.01 & $\pm$0.01 & $\pm$0.07 & \\
191.6 & 0.67 & $\pm$0.17 & $\pm$0.01 & $\pm$0.03 & $\pm$0.01 & $\pm$0.18 & \\
195.5 & 1.00 & $\pm$0.11 & $\pm$0.02 & $\pm$0.02 & $\pm$0.02 & $\pm$0.12 & \\
199.5 & 0.95 & $\pm$0.12 & $\pm$0.02 & $\pm$0.02 & $\pm$0.02 & $\pm$0.12 & \\
201.6 & 0.81 & $\pm$0.17 & $\pm$0.02 & $\pm$0.02 & $\pm$0.01 & $\pm$0.18 & \\
204.9 & 0.98 & $\pm$0.12 & $\pm$0.01 & $\pm$0.02 & $\pm$0.02 & $\pm$0.13 & \\
206.6 & 1.00 & $\pm$0.09 & $\pm$0.02 & $\pm$0.02 & $\pm$0.02 & $\pm$0.09 & \\
\hline
\end{tabular}
\caption[Z-pair cross-section]{LEP combined Z-pair production
cross-section (in pb) at different energies.  The first column
contains the LEP \CoM\ energy, the second the measurements and the
third the statistical uncertainty.  The fourth, the fifth and the
sixth columns list the different components of the systematic errors
as defined in Table~\ref{4f_tab:rWWmeas}.  The total error is given in
the seventh column, and the eighth column lists the $\chidf$ of the
fit.}
\label{4f_tab:ZZmeasLEP} 
\end{center}
\end{table}

\begin{table}[hbtp]
\begin{center}
\hspace*{-0.3cm}
\renewcommand{\arraystretch}{1.2}
\begin{tabular}{|c|c|c|} 
\hline
\roots & \multicolumn{2}{|c|}{ZZ cross-section (pb)}  \\
\cline{2-3} 
(GeV) & $\szz^{\footnotesize\YFSZZ}$    
      & $\szz^{\footnotesize\ZZTO}$ \\
\hline
\hline
182.7 & 0.254[1] & 0.25425[2] \\
188.6 & 0.655[2] & 0.64823[1] \\
191.6 & 0.782[2] & 0.77670[1] \\
195.5 & 0.897[3] & 0.89622[1] \\
199.5 & 0.981[2] & 0.97765[1] \\
201.6 & 1.015[1] & 1.00937[1] \\
204.9 & 1.050[1] & 1.04335[1] \\
206.6 & 1.066[1] & 1.05535[1] \\
\hline
\end{tabular}
\renewcommand{\arraystretch}{1.}
\caption[Z-pair cross-section]{Z-pair cross-section predictions (in
pb) interpolated at the data \CoM\ energies, according to the
\YFSZZ~\protect\cite{ref:YFSZZ} and \ZZTO~\protect\cite{4f_bib:zzto}
predictions.  The numbers in brackets are the errors on the last digit
and arise from the numerical integration of the cross-section only.}
\label{4f_tab:ZZtheo} 
\end{center}
\end{table}

\begin{table}[hbtp]
\begin{center}
\begin{small}
\begin{tabular}{|c|cccccc|c|c|}
\hline
\roots & & & {\scriptsize (LCEU)} & {\scriptsize (LCEC)} & 
{\scriptsize (LUEU)} & {\scriptsize (LUEC)} & & \\
(GeV) & $\rzz$ & 
$\Delta\rzz^\mathrm{stat}$ &
$\Delta\rzz^\mathrm{syst}$ &
$\Delta\rzz^\mathrm{syst}$ &
$\Delta\rzz^\mathrm{syst}$ &
$\Delta\rzz^\mathrm{syst}$ &
$\Delta\rzz$ &
$\chidf$ \\
\hline
\hline
\multicolumn{9}{|c|}{\YFSZZ~\cite{ref:YFSZZ}}\\
\hline
182.7 & 0.857 & $\pm$0.307 & $\pm$0.018 & $\pm$0.068 & $\pm$0.041 & $\pm$0.040 & $\pm$0.320 &
\multirow{8}{20.3mm}{$
  \hspace*{-0.3mm}
  \left\}
    \begin{array}[h]{rr}
      &\multirow{8}{6mm}{\hspace*{-4.2mm}14.5/24}\\
      &\\ &\\ &\\ &\\ &\\ &\\ &\\  
    \end{array}
  \right.
  $}\\
188.6 & 1.007 & $\pm$0.104 & $\pm$0.020 & $\pm$0.019 & $\pm$0.022& $\pm$0.018 & $\pm$0.111&\\
191.6 & 0.859 & $\pm$0.220 & $\pm$0.017 & $\pm$0.013 & $\pm$0.032& $\pm$0.016 & $\pm$0.224&\\
195.5 & 1.118 & $\pm$0.127 & $\pm$0.023 & $\pm$0.021 & $\pm$0.025& $\pm$0.019 & $\pm$0.134&\\
199.5 & 0.970 & $\pm$0.119 & $\pm$0.020 & $\pm$0.018 & $\pm$0.025& $\pm$0.016 & $\pm$0.126&\\
201.6 & 0.800 & $\pm$0.170 & $\pm$0.016 & $\pm$0.016 & $\pm$0.023& $\pm$0.012 & $\pm$0.174&\\
204.9 & 0.928 & $\pm$0.116 & $\pm$0.019 & $\pm$0.013 & $\pm$0.019& $\pm$0.014 & $\pm$0.121&\\
206.6 & 0.938 & $\pm$0.085 & $\pm$0.019 & $\pm$0.014 & $\pm$0.017& $\pm$0.016 & $\pm$0.091&\\
\hline
Average & 
0.960 & $\pm$0.045 & $\pm$0.008 & $\pm$0.017 & $\pm$0.009& $\pm$0.015 & $\pm$0.052&
\hspace*{1.5mm}17.4/31\hspace*{-0.5mm}\\
\hline
\hline
\multicolumn{9}{|c|}{\ZZTO~\cite{4f_bib:zzto}}\\
\hline
182.7 & 0.857 & $\pm$0.307 & $\pm$0.018 & $\pm$0.068 & $\pm$0.041 & $\pm$0.040 & $\pm$0.320 &
\multirow{8}{20.3mm}{$
  \hspace*{-0.3mm}
  \left\}
    \begin{array}[h]{rr}
      &\multirow{8}{6mm}{\hspace*{-4.2mm}14.5/24}\\
      &\\ &\\ &\\ &\\ &\\ &\\ &\\  
    \end{array}
  \right.
  $}\\
188.6 & 1.017 & $\pm$0.105 & $\pm$0.021 & $\pm$0.019 & $\pm$0.022& $\pm$0.019 & $\pm$0.113&\\
191.6 & 0.865 & $\pm$0.222 & $\pm$0.018 & $\pm$0.014 & $\pm$0.033& $\pm$0.016 & $\pm$0.226&\\
195.5 & 1.118 & $\pm$0.127 & $\pm$0.023 & $\pm$0.021 & $\pm$0.025& $\pm$0.019 & $\pm$0.134&\\
199.5 & 0.974 & $\pm$0.120 & $\pm$0.020 & $\pm$0.018 & $\pm$0.025& $\pm$0.016 & $\pm$0.126&\\
201.6 & 0.805 & $\pm$0.171 & $\pm$0.016 & $\pm$0.016 & $\pm$0.023& $\pm$0.012 & $\pm$0.174&\\
204.9 & 0.934 & $\pm$0.117 & $\pm$0.019 & $\pm$0.013 & $\pm$0.019& $\pm$0.013 & $\pm$0.122&\\
206.6 & 0.948 & $\pm$0.085 & $\pm$0.019 & $\pm$0.014 & $\pm$0.017& $\pm$0.016 & $\pm$0.092&\\
\hline
Average & 
0.966 & $\pm$0.046 & $\pm$0.008 & $\pm$0.017 & $\pm$0.009& $\pm$0.015 & $\pm$0.052&
\hspace*{1.5mm}17.4/31\hspace*{-0.5mm}\\
\hline
\end{tabular}
\end{small}
\caption[Ratio of Z-pair cross-sections]{Ratios of LEP combined Z-pair
cross-section measurements to the expectations, for different \CoM\
energies and for all energies combined.  The first column contains the
\CoM\ energy, the second the combined ratios, the third the
statistical errors.  The fourth to seventh columns contain the sources
of systematic errors as defined in Table~\ref{4f_tab:rWWmeas}.
The total error is given in the eighth column.  The only LCEU
systematic sources considered are the statistical errors on the
cross-section theoretical predictions, while the LCEC, LUEU and LUEC
sources are those coming from the corresponding errors on the
cross-section measurements.  For the LEP averages, the $\chidf$ of the
fit is also given in the ninth column.}
\label{4f_tab:rZZmeas} 
\end{center}
\end{table}

\begin{table}[p]
\vspace*{-0.7cm}
\begin{center}
\begin{small}
\begin{tabular}{|c|ccccc|c|c|}
\cline{1-8}
\roots & & & {\scriptsize (LCEC)} & {\scriptsize (LUEU)} & 
{\scriptsize (LUEC)} & & \\
(GeV) & $\swenh$ & 
$\Delta\swenh^\mathrm{stat}$ &
$\Delta\swenh^\mathrm{syst}$ &
$\Delta\swenh^\mathrm{syst}$ &
$\Delta\swenh^\mathrm{syst}$ &
$\Delta\swenh$ & 
$\Delta\swenh^\mathrm{stat\,(exp)}$ \\
\hline
\multicolumn{8}{|c|}
{\Aleph~\cite{4f_bib:alesw}} \\
\hline
182.7 & 0.44 & $^{+0.29}_{-0.24}$ & $\pm$0.01 & $\pm$0.01 & $\pm$0.01 & $^{+0.29}_{-0.24}$ & $\pm$0.26 \\
188.6 & 0.33 & $^{+0.16}_{-0.14}$ & $\pm$0.02 & $\pm$0.01 & $\pm$0.01 & $^{+0.16}_{-0.15}$ & $\pm$0.16 \\
191.6 & 0.52 & $^{+0.52}_{-0.40}$ & $\pm$0.02 & $\pm$0.01 & $\pm$0.01 & $^{+0.52}_{-0.40}$ & $\pm$0.45 \\
195.5 & 0.61 & $^{+0.28}_{-0.25}$ & $\pm$0.02 & $\pm$0.01 & $\pm$0.01 & $^{+0.28}_{-0.25}$ & $\pm$0.25 \\
199.5 & 1.06 & $^{+0.30}_{-0.27}$ & $\pm$0.02 & $\pm$0.01 & $\pm$0.01 & $^{+0.30}_{-0.27}$ & $\pm$0.24 \\
201.6 & 0.72 & $^{+0.39}_{-0.33}$ & $\pm$0.02 & $\pm$0.01 & $\pm$0.02 & $^{+0.39}_{-0.33}$ & $\pm$0.34 \\
204.9 & 0.34 & $^{+0.24}_{-0.21}$ & $\pm$0.02 & $\pm$0.01 & $\pm$0.02 & $^{+0.24}_{-0.21}$ & $\pm$0.25 \\
206.6 & 0.64 & $^{+0.21}_{-0.19}$ & $\pm$0.02 & $\pm$0.01 & $\pm$0.02 & $^{+0.21}_{-0.19}$ & $\pm$0.19 \\
\hline
\multicolumn{8}{|c|}
{\Delphi~\cite{4f_bib:delsw}} \\
\hline
182.7 & 0.11 & $^{+0.30}_{-0.11}$ & $\pm$0.02 & $\pm$0.03 & $\pm$0.08 & $^{+0.31}_{-0.14}$ & $\pm$0.30 \\
188.6 & 0.57 & $^{+0.19}_{-0.18}$ & $\pm$0.02 & $\pm$0.04 & $\pm$0.08 & $^{+0.21}_{-0.20}$ & $\pm$0.18 \\
191.6 & 0.30 & $^{+0.47}_{-0.30}$ & $\pm$0.02 & $\pm$0.03 & $\pm$0.08 & $^{+0.48}_{-0.31}$ & $\pm$0.43 \\
195.5 & 0.50 & $^{+0.29}_{-0.26}$ & $\pm$0.02 & $\pm$0.03 & $\pm$0.08 & $^{+0.30}_{-0.27}$ & $\pm$0.27 \\
199.5 & 0.57 & $^{+0.27}_{-0.25}$ & $\pm$0.02 & $\pm$0.02 & $\pm$0.08 & $^{+0.28}_{-0.26}$ & $\pm$0.25 \\
201.6 & 0.67 & $^{+0.39}_{-0.35}$ & $\pm$0.02 & $\pm$0.03 & $\pm$0.08 & $^{+0.40}_{-0.36}$ & $\pm$0.35 \\
204.9 & 0.99 & $^{+0.32}_{-0.30}$ & $\pm$0.02 & $\pm$0.05 & $\pm$0.08 & $^{+0.33}_{-0.31}$ & $\pm$0.28 \\
206.6 & 0.81 & $^{+0.22}_{-0.20}$ & $\pm$0.02 & $\pm$0.04 & $\pm$0.08 & $^{+0.23}_{-0.22}$ & $\pm$0.20 \\
\hline
\multicolumn{8}{|c|}
{\Ltre~\cite{4f_bib:ltrsw-1, 4f_bib:ltrsw-2, 4f_bib:ltrsw-3}} \\
\hline
182.7 & 0.58 & $^{+0.23}_{-0.20}$ & $\pm$0.03 & $\pm$0.03 & $\pm$0.00 & $^{+0.23}_{-0.20}$ & $\pm$0.21 \\
188.6 & 0.52 & $^{+0.14}_{-0.13}$ & $\pm$0.02 & $\pm$0.02 & $\pm$0.00 & $^{+0.14}_{-0.13}$ & $\pm$0.14 \\
191.6 & 0.84 & $^{+0.44}_{-0.37}$ & $\pm$0.03 & $\pm$0.03 & $\pm$0.00 & $^{+0.44}_{-0.37}$ & $\pm$0.41 \\
195.5 & 0.66 & $^{+0.24}_{-0.22}$ & $\pm$0.02 & $\pm$0.03 & $\pm$0.00 & $^{+0.25}_{-0.23}$ & $\pm$0.21 \\
199.5 & 0.37 & $^{+0.22}_{-0.20}$ & $\pm$0.01 & $\pm$0.02 & $\pm$0.00 & $^{+0.22}_{-0.20}$ & $\pm$0.22 \\
201.6 & 1.10 & $^{+0.40}_{-0.35}$ & $\pm$0.05 & $\pm$0.05 & $\pm$0.00 & $^{+0.40}_{-0.35}$ & $\pm$0.35 \\
204.9 & 0.42 & $^{+0.25}_{-0.21}$ & $\pm$0.02 & $\pm$0.03 & $\pm$0.00 & $^{+0.25}_{-0.21}$ & $\pm$0.25 \\
206.6 & 0.66 & $^{+0.19}_{-0.17}$ & $\pm$0.02 & $\pm$0.03 & $\pm$0.00 & $^{+0.20}_{-0.18}$ & $\pm$0.20 \\
\hline
\multicolumn{7}{|c|}
{LEP} & $\chidf$ \\
\hline
182.7 & 0.42 & $\pm$0.15 & $\pm$0.02 & $\pm$0.02 & $\pm$0.01 & $\pm$0.15 & 
 \multirow{8}{20.3mm}{$
   \hspace*{-0.3mm}
   \left\}
     \begin{array}[h]{rr}
       &\multirow{8}{8mm}{\hspace*{-4.2mm}13.2/16}\\
       &\\ &\\ &\\ &\\ &\\ &\\ &\\  
     \end{array}
   \right.
   $}\\
188.6 & 0.47 & $\pm$0.09 & $\pm$0.02 & $\pm$0.01 & $\pm$0.01 & $\pm$0.09 & \\
191.6 & 0.56 & $\pm$0.25 & $\pm$0.02 & $\pm$0.02 & $\pm$0.02 & $\pm$0.25 & \\
195.5 & 0.60 & $\pm$0.14 & $\pm$0.02 & $\pm$0.01 & $\pm$0.02 & $\pm$0.14 & \\
199.5 & 0.65 & $\pm$0.14 & $\pm$0.02 & $\pm$0.01 & $\pm$0.02 & $\pm$0.14 & \\
201.6 & 0.82 & $\pm$0.20 & $\pm$0.03 & $\pm$0.02 & $\pm$0.02 & $\pm$0.20 & \\
204.9 & 0.54 & $\pm$0.15 & $\pm$0.02 & $\pm$0.02 & $\pm$0.02 & $\pm$0.15 & \\
206.6 & 0.69 & $\pm$0.11 & $\pm$0.02 & $\pm$0.02 & $\pm$0.02 & $\pm$0.12 & \\
\hline
\end{tabular}
\end{small}
\caption[Single-W cross-section]{Single-W hadronic production
cross-section (in pb) at different energies.  The first column
contains the LEP \CoM\ energy, and the second the measurements.  The
third column reports the statistical error, and the fourth to the
sixth columns list the different systematic uncertainties.  The labels
LCEC, LUEU and LUEC are defined in Table~\ref{4f_tab:rWWmeas}. The
seventh column contains the total error and the eighth lists the
symmetrised expected statistical error for the three LEP measurements,
and, for the LEP combined value, the $\chidf$ of the fit.}
\label{4f_tab:WevHADmeas} 
\end{center}
\renewcommand{\arraystretch}{1.}
\end{table}

\begin{table}[p]
\vspace*{-0.7cm}
\begin{center}
\begin{small}
\begin{tabular}{|c|ccccc|c|c|}
\cline{1-8}
\roots & & & {\scriptsize (LCEC)} & {\scriptsize (LUEU)} & 
{\scriptsize (LUEC)} & & \\
(GeV) & $\swent$ & 
$\Delta\swent^\mathrm{stat}$ &
$\Delta\swent^\mathrm{syst}$ &
$\Delta\swent^\mathrm{syst}$ &
$\Delta\swent^\mathrm{syst}$ &
$\Delta\swent$ & 
$\Delta\swent^\mathrm{stat\,(exp)}$ \\
\hline
\multicolumn{8}{|c|}
{\Aleph~\cite{4f_bib:alesw}} \\
\hline
182.7 & 0.60 & $^{+0.32}_{-0.26}$ & $\pm$0.02 & $\pm$0.01 & $\pm$0.01 & $^{+0.32}_{-0.26}$ & $\pm$0.29 \\
188.6 & 0.55 & $^{+0.18}_{-0.16}$ & $\pm$0.02 & $\pm$0.01 & $\pm$0.01 & $^{+0.18}_{-0.16}$ & $\pm$0.18 \\
191.6 & 0.89 & $^{+0.58}_{-0.44}$ & $\pm$0.02 & $\pm$0.01 & $\pm$0.02 & $^{+0.58}_{-0.44}$ & $\pm$0.48 \\
195.5 & 0.87 & $^{+0.31}_{-0.27}$ & $\pm$0.03 & $\pm$0.01 & $\pm$0.02 & $^{+0.31}_{-0.27}$ & $\pm$0.28 \\
199.5 & 1.31 & $^{+0.32}_{-0.29}$ & $\pm$0.03 & $\pm$0.01 & $\pm$0.02 & $^{+0.32}_{-0.29}$ & $\pm$0.26 \\
201.6 & 0.80 & $^{+0.42}_{-0.35}$ & $\pm$0.03 & $\pm$0.01 & $\pm$0.02 & $^{+0.42}_{-0.35}$ & $\pm$0.38 \\
204.9 & 0.65 & $^{+0.27}_{-0.23}$ & $\pm$0.03 & $\pm$0.02 & $\pm$0.02 & $^{+0.27}_{-0.23}$ & $\pm$0.27 \\
206.6 & 0.81 & $^{+0.22}_{-0.20}$ & $\pm$0.03 & $\pm$0.02 & $\pm$0.02 & $^{+0.22}_{-0.20}$ & $\pm$0.22 \\
\hline
\multicolumn{8}{|c|}
{\Delphi~\cite{4f_bib:delsw}} \\
\hline
182.7 & 0.69 & $^{+0.41}_{-0.23}$ & $\pm$0.02 & $\pm$0.04 & $\pm$0.08 & $^{+0.42}_{-0.25}$ & $\pm$0.33 \\
188.6 & 0.75 & $^{+0.22}_{-0.20}$ & $\pm$0.02 & $\pm$0.04 & $\pm$0.08 & $^{+0.23}_{-0.22}$ & $\pm$0.20 \\
191.6 & 0.40 & $^{+0.54}_{-0.31}$ & $\pm$0.02 & $\pm$0.03 & $\pm$0.08 & $^{+0.55}_{-0.33}$ & $\pm$0.48 \\
195.5 & 0.68 & $^{+0.33}_{-0.28}$ & $\pm$0.02 & $\pm$0.03 & $\pm$0.08 & $^{+0.34}_{-0.38}$ & $\pm$0.30 \\
199.5 & 0.95 & $^{+0.33}_{-0.29}$ & $\pm$0.02 & $\pm$0.03 & $\pm$0.08 & $^{+0.34}_{-0.30}$ & $\pm$0.29 \\
201.6 & 1.24 & $^{+0.51}_{-0.42}$ & $\pm$0.02 & $\pm$0.04 & $\pm$0.08 & $^{+0.52}_{-0.43}$ & $\pm$0.41 \\
204.9 & 1.06 & $^{+0.36}_{-0.30}$ & $\pm$0.02 & $\pm$0.05 & $\pm$0.08 & $^{+0.37}_{-0.32}$ & $\pm$0.33 \\
206.6 & 1.14 & $^{+0.26}_{-0.23}$ & $\pm$0.02 & $\pm$0.04 & $\pm$0.08 & $^{+0.28}_{-0.25}$ & $\pm$0.23 \\
\hline
\multicolumn{8}{|c|}
{\Ltre~\cite{4f_bib:ltrsw-1, 4f_bib:ltrsw-2, 4f_bib:ltrsw-3}} \\
\hline
182.7 & 0.80 & $^{+0.28}_{-0.25}$ & $\pm$0.04 & $\pm$0.04 & $\pm$0.01 & $^{+0.28}_{-0.25}$ & $\pm$0.26 \\
188.6 & 0.69 & $^{+0.16}_{-0.14}$ & $\pm$0.03 & $\pm$0.03 & $\pm$0.01 & $^{+0.16}_{-0.15}$ & $\pm$0.15 \\
191.6 & 1.11 & $^{+0.48}_{-0.41}$ & $\pm$0.02 & $\pm$0.04 & $\pm$0.01 & $^{+0.48}_{-0.41}$ & $\pm$0.46 \\
195.5 & 0.97 & $^{+0.27}_{-0.25}$ & $\pm$0.02 & $\pm$0.02 & $\pm$0.01 & $^{+0.27}_{-0.25}$ & $\pm$0.25 \\
199.5 & 0.88 & $^{+0.26}_{-0.24}$ & $\pm$0.02 & $\pm$0.03 & $\pm$0.01 & $^{+0.26}_{-0.24}$ & $\pm$0.25 \\
201.6 & 1.50 & $^{+0.45}_{-0.40}$ & $\pm$0.03 & $\pm$0.04 & $\pm$0.02 & $^{+0.45}_{-0.40}$ & $\pm$0.38 \\
204.9 & 0.78 & $^{+0.29}_{-0.25}$ & $\pm$0.02 & $\pm$0.03 & $\pm$0.01 & $^{+0.29}_{-0.25}$ & $\pm$0.29 \\
206.6 & 1.08 & $^{+0.21}_{-0.20}$ & $\pm$0.02 & $\pm$0.03 & $\pm$0.01 & $^{+0.21}_{-0.20}$ & $\pm$0.23 \\
\hline
\multicolumn{7}{|c|}
{LEP} & $\chidf$ \\
\hline
182.7 & 0.70 & $\pm$0.17 & $\pm$0.03 & $\pm$0.02 & $\pm$0.02 & $\pm$0.17 &
 \multirow{8}{20.3mm}{$
   \hspace*{-0.3mm}
   \left\}
     \begin{array}[h]{rr}
       &\multirow{8}{8mm}{\hspace*{-4.2mm}8.1/16}\\
       &\\ &\\ &\\ &\\ &\\ &\\ &\\  
     \end{array}
   \right.
   $}\\
188.6 & 0.66 & $\pm$0.10 & $\pm$0.02 & $\pm$0.02 & $\pm$0.01 & $\pm$0.10 & \\
191.6 & 0.81 & $\pm$0.27 & $\pm$0.02 & $\pm$0.02 & $\pm$0.02 & $\pm$0.28 & \\
195.5 & 0.85 & $\pm$0.16 & $\pm$0.02 & $\pm$0.01 & $\pm$0.02 & $\pm$0.16 & \\
199.5 & 1.05 & $\pm$0.15 & $\pm$0.02 & $\pm$0.01 & $\pm$0.02 & $\pm$0.16 & \\
201.6 & 1.17 & $\pm$0.23 & $\pm$0.03 & $\pm$0.02 & $\pm$0.02 & $\pm$0.23 & \\
204.9 & 0.80 & $\pm$0.17 & $\pm$0.02 & $\pm$0.02 & $\pm$0.02 & $\pm$0.17 & \\
206.6 & 1.00 & $\pm$0.13 & $\pm$0.03 & $\pm$0.02 & $\pm$0.02 & $\pm$0.14 & \\
\hline
\end{tabular}
\end{small}
\caption[Single-W cross-section]{Single-W total production
cross-section (in pb) at different energies.  The first column
contains the LEP \CoM\ energy, and the second the measurements.  The
third column reports the statistical error, and the fourth to the
sixth columns list the different systematic uncertainties.  The labels
LCEC, LUEU and LUEC are defined in Table~\ref{4f_tab:rWWmeas}.  The
seventh column contains the total error and the eighth lists the
symmetrised expected statistical error for the three LEP measurements,
and, for the LEP combined values, the $\chidf$ of the fit.}
\label{4f_tab:WevTOTmeas} 
\end{center}
\renewcommand{\arraystretch}{1.}
\end{table}

\begin{table}[hbtp]
\begin{center}
\hspace*{-0.3cm}
\renewcommand{\arraystretch}{1.2}
\begin{tabular}{|c|c|c|c|c|c|} 
\hline
\roots & \multicolumn{3}{|c|}{We$\nu \rightarrow $qqe$\nu$ cross-section (pb)} 
& \multicolumn{2}{|c|}{We$\nu$ total cross-section (pb)} \\
\cline{2-6} 
(GeV) & $\swenh^{\footnotesize\Grace}$    
      & $\swenh^{\footnotesize\WPHACT}$ 
      & $\swenh^{\footnotesize\WTO}$ 
      & $\swent^{\footnotesize\Grace}$
      & $\swent^{\footnotesize\WPHACT}$  \\
\hline
\hline
182.7 & 0.4194[1] & 0.4070[2] & 0.40934[8] & 0.6254[1] & 0.6066[2] \\
188.6 & 0.4699[1] & 0.4560[2] & 0.45974[9] & 0.6999[1] & 0.6796[2] \\ 
191.6 & 0.4960[1] & 0.4810[2] & 0.4852[1] &  0.7381[2] & 0.7163[2] \\ 
195.5 & 0.5308[2] & 0.5152[2] & 0.5207[1] &  0.7896[2] & 0.7665[3] \\ 
199.5 & 0.5673[2] & 0.5509[3] & 0.5573[1] &  0.8431[2] & 0.8182[3] \\ 
201.6 & 0.5870[2] & 0.5704[4] & 0.5768[1] &  0.8718[2] & 0.8474[4] \\ 
204.9 & 0.6196[2] & 0.6021[4] & 0.6093[2] &  0.9185[3] & 0.8921[4] \\ 
206.6 & 0.6358[2] & 0.6179[4] & 0.6254[2] &  0.9423[3] & 0.9157[5] \\ 
\hline
\end{tabular}
\renewcommand{\arraystretch}{1.}
\caption[Single-W cross-section]{Single-W hadronic and total
cross-section predictions (in pb) interpolated at the data \CoM\
energies, according to the \Grace~\protect\cite{\GRACEref},
\WPHACT~\protect\cite{\WPHACTref} and \WTO~\protect\cite{\WTOref}
predictions.  The numbers in brackets are the errors on the last digit
and arise from the numerical integration of the cross-section only.}
\label{4f_tab:Wentheo} 
\end{center}
\end{table}

\begin{table}[hbtp]
\begin{center}
\begin{small}
\begin{tabular}{|c|cccccc|c|c|}
\hline
\roots & & & {\scriptsize (LCEU)} & {\scriptsize (LCEC)} & 
{\scriptsize (LUEU)} & {\scriptsize (LUEC)} & & \\
(GeV) & $\rwev$ & 
$\Delta\rwev^\mathrm{stat}$ &
$\Delta\rwev^\mathrm{syst}$ &
$\Delta\rwev^\mathrm{syst}$ &
$\Delta\rwev^\mathrm{syst}$ &
$\Delta\rwev^\mathrm{syst}$ &
$\Delta\rwev$ &
$\chidf$ \\
\hline
\hline
\multicolumn{9}{|c|}{\Grace~\cite{\GRACEref}}\\
\hline
182.7 & 1.122 & $\pm$0.266 & $\pm$0.001 & $\pm$0.041 & $\pm$0.029 & $\pm$0.026 & $\pm$0.272 &
\multirow{8}{20.3mm}{$
  \hspace*{-0.3mm}
  \left\}
    \begin{array}[h]{rr}
      &\multirow{8}{6mm}{\hspace*{-4.2mm}8.1/16}\\
      &\\ &\\ &\\ &\\ &\\ &\\ &\\  
    \end{array}
  \right.
  $}\\
188.6 & 0.936 & $\pm$0.142 & $\pm$0.001 & $\pm$0.033 & $\pm$0.022 & $\pm$0.024 & $\pm$0.149 &\\
191.6 & 1.094 & $\pm$0.370 & $\pm$0.001 & $\pm$0.030 & $\pm$0.026 & $\pm$0.028 & $\pm$0.373 &\\
195.5 & 1.081 & $\pm$0.199 & $\pm$0.001 & $\pm$0.028 & $\pm$0.017 & $\pm$0.023 & $\pm$0.203 &\\
199.5 & 1.242 & $\pm$0.183 & $\pm$0.001 & $\pm$0.028 & $\pm$0.017 & $\pm$0.022 & $\pm$0.187 &\\
201.6 & 1.340 & $\pm$0.258 & $\pm$0.001 & $\pm$0.031 & $\pm$0.021 & $\pm$0.023 & $\pm$0.261 &\\
204.9 & 0.873 & $\pm$0.185 & $\pm$0.001 & $\pm$0.025 & $\pm$0.020 & $\pm$0.020 & $\pm$0.189 &\\
206.6 & 1.058 & $\pm$0.138 & $\pm$0.001 & $\pm$0.026 & $\pm$0.019 & $\pm$0.021 & $\pm$0.143 &\\
\hline
Average & 
1.058 & $\pm$0.068 & $\pm$0.000 & $\pm$0.029 & $\pm$0.008& $\pm$0.022 & $\pm$0.078&
\hspace*{1.5mm}12.2/24\hspace*{-0.5mm}\\
\hline
\hline
\multicolumn{9}{|c|}{\WPHACT~\cite{\WPHACTref}}\\
\hline
182.7 & 1.157 & $\pm$0.274 & $\pm$0.001 & $\pm$0.043 & $\pm$0.030 & $\pm$0.027 & $\pm$0.281 &
\multirow{8}{20.3mm}{$
  \hspace*{-0.3mm}
  \left\}
    \begin{array}[h]{rr}
      &\multirow{8}{6mm}{\hspace*{-4.2mm}8.1/16}\\
      &\\ &\\ &\\ &\\ &\\ &\\ &\\  
    \end{array}
  \right.
  $}\\
188.6 & 0.965 & $\pm$0.146 & $\pm$0.001 & $\pm$0.034 & $\pm$0.023 & $\pm$0.024 & $\pm$0.154 &\\
191.6 & 1.128 & $\pm$0.382 & $\pm$0.001 & $\pm$0.031 & $\pm$0.027 & $\pm$0.029 & $\pm$0.385 &\\
195.5 & 1.115 & $\pm$0.206 & $\pm$0.001 & $\pm$0.029 & $\pm$0.017 & $\pm$0.023 & $\pm$0.210 &\\
199.5 & 1.280 & $\pm$0.188 & $\pm$0.001 & $\pm$0.029 & $\pm$0.018 & $\pm$0.022 & $\pm$0.193 &\\
201.6 & 1.380 & $\pm$0.265 & $\pm$0.001 & $\pm$0.032 & $\pm$0.022 & $\pm$0.024 & $\pm$0.269 &\\
204.9 & 0.899 & $\pm$0.191 & $\pm$0.001 & $\pm$0.026 & $\pm$0.020 & $\pm$0.020 & $\pm$0.195 &\\
206.6 & 1.089 & $\pm$0.142 & $\pm$0.001 & $\pm$0.027 & $\pm$0.020 & $\pm$0.022 & $\pm$0.148 &\\
\hline
Average & 
1.090 & $\pm$0.070 & $\pm$0.000 & $\pm$0.030 & $\pm$0.008& $\pm$0.023 & $\pm$0.080&
\hspace*{1.5mm}12.2/24\hspace*{-0.5mm}\\
\hline
\end{tabular}
\end{small}
\caption[Ratio of single-W cross-sections]{Ratios of LEP combined
total single-W cross-section measurements to the expectations, for
different \CoM\ energies and for all energies combined.  The first
column contains the \CoM\ energy, the second the combined ratios, the
third the statistical errors.  The fourth to seventh columns contain
the sources of systematic errors, as defined in
Table~\ref{4f_tab:rWWmeas}.
The total error is given in the eighth column.  The only LCEU
systematic sources considered are the statistical errors on the
cross-section theoretical predictions, while the LCEC, LUEU and LUEC
sources are those arising from the corresponding errors on the
cross-section measurements.}
\label{4f_tab:rwenmeas} 
\end{center}
\end{table}

\clearpage

\begin{table}[p]
\vspace*{-0.0cm}
\begin{center}
\begin{small}
\begin{tabular}{|c|ccccc|c|c|}
\cline{1-8}
\roots & & & {\scriptsize (LCEC)} & {\scriptsize (LUEU)} & 
{\scriptsize (LUEC)} & & \\
(GeV) & $\szee$ & 
$\Delta\szee^\mathrm{stat}$ &
$\Delta\szee^\mathrm{syst}$ &
$\Delta\szee^\mathrm{syst}$ &
$\Delta\szee^\mathrm{syst}$ &
$\Delta\szee$ & 
$\Delta\szee^\mathrm{stat\,(exp)}$ \\
\hline
\multicolumn{8}{|c|}
{\Aleph~\cite{4f_bib:alesw}} \\
\hline
182.7 & 0.27 & $^{+0.21}_{-0.16}$ & $\pm$0.01 & $\pm$0.02 & $\pm$0.01 & $^{+0.21}_{-0.16}$ & $\pm$0.20 \\
188.6 & 0.42 & $^{+0.14}_{-0.12}$ & $\pm$0.01 & $\pm$0.03 & $\pm$0.01 & $^{+0.14}_{-0.12}$ & $\pm$0.12 \\
191.6 & 0.61 & $^{+0.39}_{-0.29}$ & $\pm$0.01 & $\pm$0.03 & $\pm$0.01 & $^{+0.39}_{-0.29}$ & $\pm$0.29 \\
195.5 & 0.72 & $^{+0.24}_{-0.20}$ & $\pm$0.01 & $\pm$0.03 & $\pm$0.01 & $^{+0.24}_{-0.20}$ & $\pm$0.18 \\
199.5 & 0.60 & $^{+0.21}_{-0.18}$ & $\pm$0.01 & $\pm$0.03 & $\pm$0.01 & $^{+0.21}_{-0.18}$ & $\pm$0.17 \\
201.6 & 0.89 & $^{+0.35}_{-0.28}$ & $\pm$0.01 & $\pm$0.03 & $\pm$0.01 & $^{+0.35}_{-0.28}$ & $\pm$0.24 \\
204.9 & 0.42 & $^{+0.17}_{-0.14}$ & $\pm$0.01 & $\pm$0.03 & $\pm$0.01 & $^{+0.17}_{-0.15}$ & $\pm$0.17 \\
206.6 & 0.70 & $^{+0.17}_{-0.15}$ & $\pm$0.01 & $\pm$0.03 & $\pm$0.01 & $^{+0.17}_{-0.15}$ & $\pm$0.14 \\
\hline
\multicolumn{8}{|c|}
{\Delphi~\cite{4f_bib:delsw}} \\
\hline
182.7 & 0.56 & $^{+0.27}_{-0.22}$ & $\pm$0.01 & $\pm$0.06 & $\pm$0.02 & $^{+0.28}_{-0.23}$ & $\pm$0.24 \\
188.6 & 0.64 & $^{+0.15}_{-0.14}$ & $\pm$0.01 & $\pm$0.03 & $\pm$0.02 & $^{+0.16}_{-0.14}$ & $\pm$0.14 \\
191.6 & 0.63 & $^{+0.40}_{-0.30}$ & $\pm$0.01 & $\pm$0.03 & $\pm$0.03 & $^{+0.40}_{-0.30}$ & $\pm$0.32 \\
195.5 & 0.66 & $^{+0.22}_{-0.18}$ & $\pm$0.01 & $\pm$0.02 & $\pm$0.03 & $^{+0.22}_{-0.19}$ & $\pm$0.19 \\
199.5 & 0.57 & $^{+0.20}_{-0.17}$ & $\pm$0.01 & $\pm$0.02 & $\pm$0.02 & $^{+0.20}_{-0.17}$ & $\pm$0.18 \\
201.6 & 0.19 & $^{+0.21}_{-0.16}$ & $\pm$0.01 & $\pm$0.02 & $\pm$0.01 & $^{+0.21}_{-0.16}$ & $\pm$0.25 \\
204.9 & 0.37 & $^{+0.18}_{-0.15}$ & $\pm$0.01 & $\pm$0.02 & $\pm$0.02 & $^{+0.18}_{-0.15}$ & $\pm$0.19 \\
206.6 & 0.69 & $^{+0.16}_{-0.14}$ & $\pm$0.01 & $\pm$0.01 & $\pm$0.03 & $^{+0.16}_{-0.14}$ & $\pm$0.14 \\
\hline
\multicolumn{8}{|c|}
{\Ltre~\cite{4f_bib:ltrzee}} \\
\hline
182.7 & 0.51 & $^{+0.19}_{-0.16}$ & $\pm$0.02 & $\pm$0.01 & $\pm$0.03 & $^{+0.19}_{-0.16}$ & $\pm$0.16 \\
188.6 & 0.55 & $^{+0.10}_{-0.09}$ & $\pm$0.02 & $\pm$0.01 & $\pm$0.03 & $^{+0.11}_{-0.10}$ & $\pm$0.09 \\
191.6 & 0.60 & $^{+0.26}_{-0.21}$ & $\pm$0.01 & $\pm$0.01 & $\pm$0.03 & $^{+0.26}_{-0.21}$ & $\pm$0.21 \\
195.5 & 0.40 & $^{+0.13}_{-0.11}$ & $\pm$0.01 & $\pm$0.01 & $\pm$0.03 & $^{+0.13}_{-0.11}$ & $\pm$0.13 \\
199.5 & 0.33 & $^{+0.12}_{-0.10}$ & $\pm$0.01 & $\pm$0.01 & $\pm$0.03 & $^{+0.13}_{-0.11}$ & $\pm$0.14 \\
201.6 & 0.81 & $^{+0.27}_{-0.23}$ & $\pm$0.02 & $\pm$0.02 & $\pm$0.03 & $^{+0.27}_{-0.23}$ & $\pm$0.19 \\
204.9 & 0.56 & $^{+0.16}_{-0.14}$ & $\pm$0.01 & $\pm$0.01 & $\pm$0.03 & $^{+0.16}_{-0.14}$ & $\pm$0.14 \\
206.6 & 0.59 & $^{+0.12}_{-0.10}$ & $\pm$0.01 & $\pm$0.01 & $\pm$0.03 & $^{+0.12}_{-0.11}$ & $\pm$0.11 \\
\hline
\multicolumn{7}{|c|}
{LEP} & $\chidf$ \\
\hline
182.7 & 0.45 & $\pm$0.11 & $\pm$0.01 & $\pm$0.02 & $\pm$0.01 & $\pm$0.11 & 
 \multirow{8}{20.3mm}{$
   \hspace*{-0.3mm}
   \left\}
     \begin{array}[h]{rr}
       &\multirow{8}{8mm}{\hspace*{-4.2mm}13.0/16}\\
       &\\ &\\ &\\ &\\ &\\ &\\ &\\  
     \end{array}
   \right.
   $}\\
188.6 & 0.53 & $\pm$0.07 & $\pm$0.01 & $\pm$0.01 & $\pm$0.01 & $\pm$0.07 &  \\
191.6 & 0.61 & $\pm$0.15 & $\pm$0.01 & $\pm$0.02 & $\pm$0.01 & $\pm$0.15 &  \\
195.5 & 0.55 & $\pm$0.09 & $\pm$0.01 & $\pm$0.01 & $\pm$0.01 & $\pm$0.10 &  \\
199.5 & 0.47 & $\pm$0.09 & $\pm$0.01 & $\pm$0.02 & $\pm$0.01 & $\pm$0.10 &  \\
201.6 & 0.67 & $\pm$0.13 & $\pm$0.01 & $\pm$0.01 & $\pm$0.01 & $\pm$0.13 &  \\
204.9 & 0.47 & $\pm$0.10 & $\pm$0.01 & $\pm$0.01 & $\pm$0.01 & $\pm$0.10 &  \\
206.6 & 0.65 & $\pm$0.07 & $\pm$0.01 & $\pm$0.01 & $\pm$0.01 & $\pm$0.08 &  \\
\hline
\end{tabular}
\end{small}
\caption[Single-Z cross-sections]{Single-Z hadronic production
cross-section (in pb) at different energies.  The first column
contains the LEP \CoM\ energy, and the second the measurements.  The
third column reports the statistical error, and the fourth to the
sixth columns list the different systematic uncertainties.  The labels
LCEC, LUEU and LUEC are defined in Table~\ref{4f_tab:rWWmeas}.  The
seventh column contains the total error and the eighth lists the
symmetrised expected statistical error for each of the three LEP
experiments, and, for the LEP combined value, the $\chidf$ of the
fit.}
\label{4f_tab:Zeemeas} 
\end{center}
\renewcommand{\arraystretch}{1.}
\end{table}

\begin{table}[hbtp]
\begin{center}
\hspace*{-0.3cm}
\renewcommand{\arraystretch}{1.2}
\begin{tabular}{|c|c|c|} 
\hline
\roots & \multicolumn{2}{|c|}{Zee cross-section (pb)}  \\
\cline{2-3} 
(GeV) & $\szee^{\footnotesize\WPHACT}$    
      & $\szee^{\footnotesize\Grace}$ \\
\hline
\hline
182.7 & 0.51275[4] & 0.51573[4] \\
188.6 & 0.53686[4] & 0.54095[5] \\
191.6 & 0.54883[4] & 0.55314[5] \\
195.5 & 0.56399[5] & 0.56891[4] \\
199.5 & 0.57935[5] & 0.58439[4] \\
201.6 & 0.58708[4] & 0.59243[4] \\
204.9 & 0.59905[4] & 0.60487[4] \\
206.6 & 0.61752[4] & 0.60819[4] \\
\hline
\end{tabular}
\renewcommand{\arraystretch}{1.}
\caption[Zee hadronic cross-sections]{Zee hadronic cross-section
predictions (in pb) interpolated at the data \CoM\ energies, according
to the \WPHACT~\protect\cite{\WPHACTref} and
\Grace~\protect\cite{\GRACEref} predictions.  The numbers in brackets
are the errors on the last digit and arise from the numerical
integration of the cross-section only.}
\label{4f_tab:Zeetheo} 
\end{center}
\end{table}

\begin{table}[hbtp]
\begin{center}
\begin{small}
\begin{tabular}{|c|cccccc|c|c|}
\hline
\roots & & & {\scriptsize (LCEU)} & {\scriptsize (LCEC)} & 
{\scriptsize (LUEU)} & {\scriptsize (LUEC)} & & \\
(GeV) & $\rzee$ & 
$\Delta\rzee^\mathrm{stat}$ &
$\Delta\rzee^\mathrm{syst}$ &
$\Delta\rzee^\mathrm{syst}$ &
$\Delta\rzee^\mathrm{syst}$ &
$\Delta\rzee^\mathrm{syst}$ &
$\Delta\rzee$ &
$\chidf$ \\
\hline
\hline
\multicolumn{9}{|c|}{\Grace~\cite{\GRACEref}}\\
\hline
182.7 & 0.871 & $\pm$0.214 & $\pm$0.000 & $\pm$0.020 & $\pm$0.035 & $\pm$0.025 & $\pm$0.219 &
\multirow{8}{20.3mm}{$
  \hspace*{-0.3mm}
  \left\}
    \begin{array}[h]{rr}
      &\multirow{8}{6mm}{\hspace*{-4.2mm}13.0/16}\\
      &\\ &\\ &\\ &\\ &\\ &\\ &\\  
    \end{array}
  \right.
  $}\\
188.6 & 0.982 & $\pm$0.120 & $\pm$0.000 & $\pm$0.022 & $\pm$0.023 & $\pm$0.024 & $\pm$0.126 &\\
191.6 & 1.104 & $\pm$0.272 & $\pm$0.000 & $\pm$0.019 & $\pm$0.027 & $\pm$0.025 & $\pm$0.276 &\\
195.5 & 0.964 & $\pm$0.163 & $\pm$0.000 & $\pm$0.016 & $\pm$0.024 & $\pm$0.025 & $\pm$0.167 &\\
199.5 & 0.809 & $\pm$0.160 & $\pm$0.000 & $\pm$0.018 & $\pm$0.030 & $\pm$0.023 & $\pm$0.165 &\\
201.6 & 1.126 & $\pm$0.219 & $\pm$0.000 & $\pm$0.023 & $\pm$0.024 & $\pm$0.021 & $\pm$0.222 &\\
204.9 & 0.769 & $\pm$0.157 & $\pm$0.000 & $\pm$0.019 & $\pm$0.019 & $\pm$0.021 & $\pm$0.160 &\\
206.6 & 1.062 & $\pm$0.119 & $\pm$0.000 & $\pm$0.018 & $\pm$0.018 & $\pm$0.024 & $\pm$0.124 &\\
\hline
Average & 
0.955 & $\pm$0.057 & $\pm$0.000 & $\pm$0.019 & $\pm$0.009 & $\pm$0.023 & $\pm$0.065&
\hspace*{1.5mm}17.1/23\hspace*{-0.5mm}\\
\hline
\hline
\multicolumn{9}{|c|}{\WPHACT~\cite{\WPHACTref}}\\
\hline
182.7 & 0.876 & $\pm$0.215 & $\pm$0.000 & $\pm$0.020 & $\pm$0.035 & $\pm$0.025 & $\pm$0.220 &
\multirow{8}{20.3mm}{$
  \hspace*{-0.3mm}
  \left\}
    \begin{array}[h]{rr}
      &\multirow{8}{6mm}{\hspace*{-4.2mm}13.0/16}\\
      &\\ &\\ &\\ &\\ &\\ &\\ &\\  
    \end{array}
  \right.
  $}\\
188.6 & 0.990 & $\pm$0.120 & $\pm$0.000 & $\pm$0.022 & $\pm$0.023 & $\pm$0.025 & $\pm$0.127 &\\
191.6 & 1.112 & $\pm$0.274 & $\pm$0.000 & $\pm$0.020 & $\pm$0.027 & $\pm$0.026 & $\pm$0.277 &\\
195.5 & 0.972 & $\pm$0.164 & $\pm$0.000 & $\pm$0.016 & $\pm$0.025 & $\pm$0.025 & $\pm$0.168 &\\
199.5 & 0.816 & $\pm$0.161 & $\pm$0.000 & $\pm$0.019 & $\pm$0.030 & $\pm$0.023 & $\pm$0.167 &\\
201.6 & 1.135 & $\pm$0.221 & $\pm$0.000 & $\pm$0.023 & $\pm$0.024 & $\pm$0.021 & $\pm$0.224 &\\
204.9 & 0.776 & $\pm$0.158 & $\pm$0.000 & $\pm$0.019 & $\pm$0.019 & $\pm$0.021 & $\pm$0.162 &\\
206.6 & 1.067 & $\pm$0.120 & $\pm$0.000 & $\pm$0.018 & $\pm$0.018 & $\pm$0.024 & $\pm$0.125 &\\
\hline
Average & 
0.962 & $\pm$0.057 & $\pm$0.000 & $\pm$0.020 & $\pm$0.009 & $\pm$0.024 & $\pm$0.065&
\hspace*{1.5mm}17.0/23\hspace*{-0.5mm}\\
\hline
\end{tabular}
\end{small}
\caption[Ratio of single-Z hadronic cross-sections]{Ratios of LEP
combined single-Z hadronic cross-section measurements to the
expectations, for different \CoM\ energies and for all energies
combined.  The first column contains the \CoM\ energy, the second the
combined ratios, the third the statistical errors.  The fourth to
seventh columns contain the sources of systematic errors as defined in
Table~\ref{4f_tab:rWWmeas}.
The total error is given in the eighth column.  The only LCEU
systematic sources considered are the statistical errors on the
cross-section theoretical predictions, while the LCEC, LUEU and LUEC
sources are those arising from the corresponding errors on the
cross-section measurements.  For the LEP averages, the $\chidf$ of the
fit is also given in the ninth column.}
\label{4f_tab:rzeemeas} 
\end{center}
\end{table}

\chapter{Constraints on the Standard Model}
\label{chap:sm}

\section{Introduction}

The experimental measurements used here to place constraints on the
Standard Model (SM) consist of three groups: (i) the final Z-pole
results measured in electron-positron collisions by the ALEPH, DELPHI,
L3, OPAL and SLD experiments, as combined in
Reference~\cite{bib-Z-pole}; (ii) the mass and width of the W boson
measured at {\LEPII} and described earlier in this report; and (iii)
the measurements of the mass of the top quark and the mass and width
of the W boson at the Tevatron collider.

The measurements allow checks of the validity of the SM and, within
its framework, to infer valuable information about its fundamental
parameters.  The accuracy of the Z-boson and W-boson measurements
makes them sensitive to the mass of the top quark $\Mt$, and to the
mass of the Higgs boson $\MH$ through electroweak loop corrections.
While the leading $\Mt$ dependence is quadratic, the leading $\MH$
dependence is logarithmic.  Therefore, the inferred constraints on
$\Mt$ are much stronger than those on $\MH$.

In a first step, the predictions for the mass and width of the W boson
based on measurements performed at lower Z-pole centre-of-mass
energies (\LEPI, SLC, etc)~\cite{bib-Z-pole} are compared to the
direct measurements performed at {\LEPII} and the Tevatron.  The
comparison between prediction and direct measurement is also performed
for the mass of the top quark.  Finally, all measurements are used to
infer constraints on the Higgs boson of the minimal SM.

This analysis updates our previous analysis~\cite{bib-Z-pole}.
Similar analyses of this type are presented in
References~\cite{PDG2010, Gfitter-1, *Gfitter-2}, obtaining equivalent
results when accounting for the different sets of measurements
considered.

\section{Measurements}

The measured quantities considered here are summarised in
Table~\ref{tab-SMIN}.  The predictions of these observables are also
shown in this table, derived from the results of the SM fit to the
combined high-$Q^2$ measurements described in the last column of
Table~\ref{tab-BIGFIT}.  The measurements obtained at the Z pole by
the LEP and SLC experiments ALEPH, DELPHI, L3, OPAL and SLD and their
combinations, reported in parts a), b) and c) of Table~\ref{tab-SMIN},
are final and published~\cite{bib-Z-pole}.

The measurements of the W-boson mass published by
CDF~\cite{CDF-MW-PRL90, *CDF-MW-PRD90, *CDF-MW-PRL95, *CDF-MW-PRD95,
*CDF-MW-2000, *CDF2MWPRL, *CDF2MWPRD, CDF2MWPRL2012} and D0~\cite{
D0-MW:PRL1996, *D0-MW:PRD1998a, *D0-MW:PRL1998, *D0-MW:PRD1998b,
*D0-MW:PRD2000, *D0-MW:PRL2000, *D0-MW:PRD2002, *D0-MW:PRL2009,
D0-MW:PRL2012}, and on the W-boson width published by
CDF~\cite{CDF-GW, CDF2GW} and D0~\cite{D0-GW, D02GW} are combined by
the Tevatron Electroweak Working Group, based on a detailed treatment
of correlated systematic uncertainties, with the result: $\MW =
80.387\pm0.016~\GeV$~\cite{PP-MW:2012} and $\GW =
2.050\pm0.058~\GeV$~\cite{PP-GW:2010}.  Combining these Tevatron
results with the final {\LEPII} results presented in
Chapter~\ref{chap:mw} of this report, $\MW=80.376\pm0.033~\GeV$ and
$\GW=2.195\pm0.083~\GeV$, the resulting world averages are:

\begin{eqnarray}
\MW & = & 80.385 \pm 0.015~\GeV\\
\GW & = &  2.085 \pm 0.042~\GeV\,,
\end{eqnarray}
and are used in the following.

For the mass of the top quark, $\Mt$, the published results from
CDF~\cite{ TEVEWWG2012-top-1, *TEVEWWG2012-top-3, *TEVEWWG2012-top-5,
*TEVEWWG212-top-7, *TEVEWWG2012-top-9a, *TEVEWWG2012-top-9b,
*TEVEWWG2012-top-11, *TEVEWWG2012-top-12} and D0~\cite{
TEVEWWG2012-top-2, *TEVEWWG2012-top-4, *TEVEWWG2012-top-6,
*TEVEWWG2012-top-8, *TEVEWWG2012-top-10a, *TEVEWWG2012-top-10b} are
combined by the Tevatron Electroweak Working Group with the result:
$\Mt=173.2\pm0.9~\GeV$~\cite{TeVEWWGtop-1107, *TeVEWWGtop-1207}.

In addition to these high-$Q^2$ results, the following results
measured in low-$Q^2$ interactions and reported in
Table~\ref{tab-SMpred} are considered: (i) the measurements of atomic
parity violation in caesium\cite{QWCs:exp:1, *QWCs:exp:2,
*QWCs:exp:2-e}, with the numerical result\cite{QWCs:theo:2003:new}
based on a revised analysis of QED radiative corrections applied to
the raw measurement; (ii) the result of the E-158 collaboration on the
electroweak mixing angle\footnote{ E-158 quotes in the
$\overline{\mathrm{MS}}$ scheme, evolved to $Q^2=\MZ^2$.  We add
0.00029 to the quoted value in order to obtain the effective
electroweak mixing angle~\cite{PDG2010}.}  measured in M{\o}ller
scattering~\cite{E158RunI, *E158RunI+II+III}; and (iii) the final
result of the NuTeV collaboration on neutrino-nucleon neutral to
charged current cross-section ratios~\cite{bib-NuTeV-final,
*bib-NuTeV-final-e}.

Using neutrino-nucleon data with an average $Q^2\simeq20~\GeV^2$, the
NuTeV collaboration has extracted the left- and right-handed couplings
combinations $\gnlq^2=4\gln^2(\glu^2+\gld^2) =
[1/2-\swsqeff+(5/9)\swsqsqeff]\rhon\rho_{\mathrm{ud}}$ and
$\gnrq^2=4\gln^2(\gru^2+\grd^2) =
(5/9)\swsqsqeff\rhon\rho_{\mathrm{ud}}$, with the $\rho$ parameters
for example defined in \cite{bib-PCLI}.  The NuTeV results for the
effective couplings are: $\gnlq^2=0.30005\pm0.00137$ and
$\gnrq^2=0.03076\pm0.00110$, with a correlation of $-0.017$.  While
the result on $\gnrq$ agrees with the $\SM$ expectation, the result on
$\gnlq$, relatively measured nearly eight times more precisely than
$\gnrq$, shows a deficit with respect to the expectation at the level
of 2.9 standard deviations~\cite{bib-NuTeV-final,
*bib-NuTeV-final-e}. A recent study finds that EMC-like isovector
effects are able to explain this difference~\cite{CBT:2009,
*BCLT:2009}.

An important quantity in electroweak analyses is given by the running
electromagnetic fine-structure constant, $\alpha(\MZ^2)$.  The
uncertainty in $\alpha(\MZ^2)$ arises from the contribution of light
quarks to the photon vacuum polarisation, $\dalhad$:

\begin{equation}
\alpha(\MZ^2) = \frac{\alpha(0)}%
   {1 - \Delta\alpha_\ell(\MZ^2) -
   \dalhad - 
   \Delta\alpha_{\mathrm{top}}(\MZ^2)} \,,
\end{equation}
where $\alpha(0)=1/137.036$.  The top contribution, $-0.00007(1)$,
depends on the mass of the top quark. The leptonic contribution is
calculated to third order~\cite{bib-alphalept} to be $0.03150$, with
negligible uncertainty.  For the hadronic contribution $\dalhad$, we
use the new result $0.02750\pm0.00033$~\cite{bib-BP11} listed in the
first row of Table~\ref{tab-SMIN}, which takes into account recent
results on electron-positron annihilations into hadrons at low
centre-of-mass energies measured by the BES, CMD-2, KLOE and BABAR
collaborations.  The reduced uncertainty of $0.00033$ still causes an
error of 0.00012 on the $\SM$ prediction of $\swsqeffl$ and of 0.08 on
the fitted value of $\log(\MH)$, while the $\alfmz$ values presented
here are stable against a variation of $\alpha(\MZ^2)$ in the interval
quoted.  Several evaluations of $\dalhad$ exist which are more theory
driven~\cite{bib-Swartz, *bib-Zeppe, *bib-Alemany, *bib-Davier,
*bib-alphaKuhn, *bib-jeger99, *bib-Erler, *bib-ADMartin,
*bib-Troconiz-Yndurain, *bib-Hagiwara:2003,
*bib-Troconiz-Yndurain-2004, *bib-Teubner-2010, *bib-Teubner-2011,
bib-Davier-2010, *bib-Davier-2011}, resulting in a much reduced
uncertainty.  To show the effect of the $\alpha(\MZ^2)$ uncertainty on
the results, we also use the most recent of these evaluations,
$0.02757\pm0.00010$~\cite{bib-Davier-2010, *bib-Davier-2011}.

\begin{table}[p]
\begin{center}
\renewcommand{\arraystretch}{1.10}
\begin{tabular}{|ll||r|r|r|r|}
\hline
 && \mcc{Measurement with}  &\mcc{Systematic} & \mcc{Standard-} & \mcc{Pull} \\
 && \mcc{Total Error}       &\mcc{Error}      & \mcc{Model fit} &            \\
\hline
\hline
&&&&& \\[-3mm]
& $\dalhad$\cite{bib-BP11}
                & $0.02750 \pm 0.00033$ &         &0.02759& $-0.3$ \\
&&&&& \\[-3mm]
\hline
a) & \underline{\LEPI}   &&&& \\
   & line-shape and      &&&& \\
   & lepton asymmetries: &&&& \\
&$\MZ$ [\GeV{}] & $91.1875\pm0.0021\pz$
                & ${}^{(a)}$0.0017$\pz$ &91.1874$\pz$ & $ 0.0$ \\
&$\GZ$ [\GeV{}] & $2.4952 \pm0.0023\pz$
                & ${}^{(a)}$0.0012$\pz$ & 2.4959$\pz$ & $-0.3$ \\
&$\shad$ [nb]   & $41.540 \pm0.037\pzz$ 
                & ${}^{(b)}$0.028$\pzz$ &41.478$\pzz$ & $ 1.7$ \\
&$\Rl$          & $20.767 \pm0.025\pzz$ 
                & ${}^{(b)}$0.007$\pzz$ &20.742$\pzz$ & $ 1.0$ \\
&$\Afbzl$       & $0.0171 \pm0.0010\pz$ 
                & ${}^{(b)}$0.0003\pz & 0.0164\pz     & $ 0.7$ \\
&+ correlation matrix~\cite{bib-Z-pole} &&&& \\
&                                             &&&& \\[-3mm]
&$\tau$ polarisation:                         &&&& \\
&$\cAl~(\ptau)$ & $0.1465\pm 0.0033\pz$ 
                & 0.0016$\pz$ & 0.1481$\pz$ & $-0.5$ \\
                      &                       &&&& \\[-3mm]
&$\qq$ charge asymmetry:                      &&&& \\
&$\swsqeffl(\Qfbhad)$
                & $0.2324\pm0.0012\pz$ 
                & 0.0010$\pz$ & 0.231439     & $ 0.8$ \\
&                                             &&&& \\[-3mm]
\hline
b) & \underline{SLD} &&&& \\
&$\cAl$ (SLD)   & $0.1513\pm 0.0021\pz$ 
                & 0.0010$\pz$ & 0.1481$\pz$ & $ 1.6$ \\
&&&&& \\[-3mm]
\hline
c) & \underline{{\LEPI}/SLD Heavy Flavour} &&&& \\
&$\Rbz{}$        & $0.21629\pm0.00066$  
                 & 0.00050     & 0.21579     & $ 0.8$ \\
&$\Rcz{}$        & $0.1721\pm0.0030\pz$
                 & 0.0019$\pz$ & 0.1723$\pz$ & $-0.1$ \\
&$\Afbzb{}$      & $0.0992\pm0.0016\pz$
                 & 0.0007$\pz$ & 0.1038$\pz$ & $-2.9$ \\
&$\Afbzc{}$      & $0.0707\pm0.0035\pz$
                 & 0.0017$\pz$ & 0.0742$\pz$ & $-1.0$ \\
&$\cAb$          & $0.923\pm 0.020\pzz$
                 & 0.013$\pzz$ & 0.935$\pzz$ & $-0.6$ \\
&$\cAc$          & $0.670\pm 0.027\pzz$
                 & 0.015$\pzz$ & 0.668$\pzz$ & $ 0.1$ \\
&+ correlation matrix~\cite{bib-Z-pole} &&&& \\
&                                              &&&& \\[-3mm]
\hline
d) & \underline{{\LEPII} and Tevatron} &&&& \\
&$\MW$ [\GeV{}] ({\LEPII}, Tevatron)
& $80.385 \pm 0.015\pzz$ &      $\pzz$   & 80.377$\pzz$ & $ 0.5$ \\
&$\GW$ [\GeV{}] ({\LEPII}, Tevatron)
& $ 2.085 \pm 0.042\pzz$ &      $\pzz$   &  2.092$\pzz$ & $-0.2$ \\
&$\Mt$ [\GeV{}] (Tevatron~\cite{TeVEWWGtop-1107, *TeVEWWGtop-1207})
& $173.2\pm 0.9\pzz\pzz$ &      $\pzz$ &173.3$\pzz\pzz$ & $-0.1$ \\
\hline
\end{tabular}\end{center}
\caption[Measurements]{ Summary of high-$Q^2$ measurements included in the
  combined analysis of SM parameters. Section~a) summarises {\LEPI}
  averages, Section~b) SLD results ($\cAl$ includes $\ALR$ and the
  polarised lepton asymmetries), Section~c) the {\LEPI} and SLD heavy
  flavour results, and Section~d) electroweak measurements from
  {\LEPII} and the Tevatron.  The total errors in column 2 include the
  systematic errors listed in column 3; the determination of the
  systematic part of each error is approximate.  The $\SM$ results in
  column~4 and the pulls (difference between measurement and fit in
  units of the total measurement error) in column~5 are derived from
  the SM fit to all high-$Q^2$ data, see Table~\ref{tab-BIGFIT}
  column~4.\\ $^{(a)}$\small{The systematic errors on $\MZ$ and $\GZ$
  contain the errors arising from the uncertainties in the $\LEPI$
  beam energy only.}\\ $^{(b)}$\small{Only common systematic errors
  are indicated.}\\ }
\label{tab-SMIN}
\end{table}

An additional input parameter, not shown in Table~\ref{tab-SMIN}, is
the Fermi constant $G_F$, determined from the $\mu$ lifetime: $G_F =
1.16637(1) \cdot 10^{-5}~\GeV^{-2}$\cite{bib-Gmu-1, *bib-Gmu-2,
*bib-Gmu-3}.  New measurements of $G_F$ yield values which are in good
agreement~\cite{Chitwood:2007pa, *Barczyk:2007hp, *Webber:2010zf}.
The relative error of $G_F$ is comparable to that of $\MZ$; both
uncertainties have negligible effects on the fit results.

\section{Theoretical Uncertainties}

Detailed studies of the theoretical uncertainties in the SM
predictions due to missing higher-order electroweak corrections and
their interplay with QCD corrections had been carried out by the
working group on `Precision calculations for the $\Zzero$
resonance'\cite{bib-PCLI}, and later in References~\cite{BP:98}
and~\cite{PCP99}.  Theoretical uncertainties are evaluated by
comparing different but, within our present knowledge, equivalent
treatments of aspects such as resummation techniques, momentum
transfer scales for vertex corrections and factorisation schemes.  The
effects of these theoretical uncertainties are reduced by the
inclusion of higher-order corrections\cite{bib-twoloop-1,
*bib-twoloop-2, *bib-twoloop-3,
*bib-twoloop-4,bib-QCDEW-1,*bib-QCDEW-2} in the electroweak libraries
TOPAZ0~\cite{Montagna:1993py, *Montagna:1993ai, *Montagna:1996ja,
*Montagna:1998kp} and ZFITTER~\cite{\ZFITTERref}.

The use of the higher-order QCD corrections\cite{bib-QCDEW-1,
*bib-QCDEW-2} increases the value of $\alfmz$ by 0.001, as expected.
The effect of missing higher-order QCD corrections on $\alfmz$
dominates missing higher-order electroweak corrections and
uncertainties in the interplay of electroweak and QCD corrections. A
discussion of theoretical uncertainties in the determination of
$\alfas$ can be found in References~\cite{bib-PCLI}
and~\cite{bib-SMALFAS-1, *bib-SMALFAS-2, *bib-SMALFAS-3}, with a more
recent analysis in Reference~\cite{Stenzel:2005sg} where the
theoretical uncertainty is estimated to be about 0.001 for the
analyses presented in the following.

The complete (fermionic and bosonic) two-loop corrections for the
calculation of $\MW$~\cite{Twoloop-MW}, and the complete fermionic
two-loop corrections for the calculation of
$\swsqeffl$~\cite{Twoloop-sin2teff} have been calculated.  Including
three-loop top-quark contributions to the $\rho$ parameter in the
limit of large $\Mt$~\cite{Threeloop-rho}, efficient routines for
evaluating these corrections have been implemented since version 6.40
in the semi-analytical program ZFITTER.  The remaining theoretical
uncertainties are estimated to be $4~\MeV$ on $\MW$ and 0.000049 on
$\swsqeffl$.  The latter uncertainty dominates the theoretical
uncertainty in the SM fits and the extraction of constraints on the
mass of the Higgs boson presented below. For a consistent treatment,
the complete two-loop calculation for the partial Z decay widths
should be calculated.

The theoretical uncertainties discussed above are not included in the
results presented in Tables~\ref{tab-BIGFIT} and~\ref{tab-SMpred}.  At
present the impact of theoretical uncertainties on the determination
of $\SM$ parameters from the precise electroweak measurements is small
compared to the error due to the uncertainty in the value of
$\alpha(\MZ^2)$, which is included in the results.

\section{Standard-Model Analyses}

\subsubsection{Strong Coupling Constant}

Of the measurements listed in Table~\ref{tab-SMIN}, $\Rl$ is the one
most sensitive to QCD corrections.  For $\MZ=91.1875~\GeV$ and
imposing $\Mt=173.2\pm0.9~\GeV$~\cite{TeVEWWGtop-1107,
*TeVEWWGtop-1207} as a constraint, $\alfas=0.1223\pm0.0038$ is
obtained.  Alternatively,
$\slept\equiv\shad/\Rl=2.0003\pm0.0027~\nb$~\cite{bib-Z-pole}, which
has higher sensitivity to QCD corrections and less dependence on
$\MH$, yields: $\alfas=0.1179\pm0.0030$.  The central values obtained
increase by 0.0013 and 0.0010, respectively, when changing $\MH$ from
$100~\GeV$ to $300~\GeV$.  These results on $\alfas$, as well as those
reported in the next section, are in good agreement with both
independent measurements of $\alfas$ and the world average
$\alfmz=0.1184\pm 0.0007$~\cite{Bethke:2009}.

\subsubsection{Electroweak Analyses}

In the following, several different SM analyses as reported in
Table~\ref{tab-BIGFIT} are discussed.  The $\chi^2$ minimisation is
performed with the program MINUIT~\cite{MINUIT}, and the predictions
are calculated with ZFITTER~6.43 as a function of the five SM input
parameters $\dalhad$, $\alfmz$, $\MZ$, $\Mt$ and $\LOGMH$ which are
varied simultaneously in the fits; see~\cite{bib-Z-pole} for details
on the fit procedure.  The somewhat large $\chidf$ for all of
these fits is caused by the large dispersion in the values of the
leptonic effective electroweak mixing angle measured through the
various asymmetries at {\LEPI} and SLD~\cite{bib-Z-pole}.
Following~\cite{bib-Z-pole} for the analyses presented here, this
dispersion is interpreted as a fluctuation in one or more of the input
measurements, and thus we neither modify nor exclude any of them.  A
further significant increase in $\chidf$ is observed when the
low-$Q^2$ NuTeV results are included in the analysis.

To test the agreement between the Z-pole data~\cite{bib-Z-pole}
({\LEPI} and SLD) and the SM, a fit to these data is performed.  This
fit differs from the corresponding analysis reported in
Reference~\cite{bib-Z-pole} in that the new result for
$\dalhad$~\cite{bib-BP11}, reported in Table~\ref{tab-SMIN}, is used.
The result is shown in Table~\ref{tab-BIGFIT}, column~1. The indirect
constraints on $\MW$ and $\Mt$ are shown in Figure~\ref{fig:mtmW},
compared with the direct measurements.  Also shown are the SM
predictions for Higgs masses between 114 and 1000~\GeV.  The indirect
and direct results on $\MW$ and $\Mt$ are in good agreement. In both
cases, a low value of the Higgs-boson mass is preferred.

For the fit shown in column~2 of Table~\ref{tab-BIGFIT}, the direct
measurement of $\Mt$ from the Tevatron experiments is included, in
order to obtain the best indirect determination of $\MW$.  The result
is also shown in Figure~\ref{fig-mhmw}.  The indirect determination of
the W-boson mass, $80.363\pm0.020~\GeV$, is in good agreement with the
direct measurements at {\LEPII} and the Tevatron, $\MW=
80.385\pm0.015~\GeV$.  For the fit shown in column~3 of
Table~\ref{tab-BIGFIT} and Figure~\ref{fig-mhmt}, the direct $\MW$ and
$\GW$ measurements from {\LEPII} and the Tevatron are included instead
of the direct $\Mt$ measurement, in order to obtain the constraint
$\Mt= 178^{+11}_{-8}~\GeV$, in good agreement with the much more
precise direct measurement of $\Mt = 173.2\pm0.9~\GeV$.

The best constraints on $\MH$ are obtained when all high-$Q^2$
measurements are used in the fit.  The results of this fit are shown
in column~4 of Table~\ref{tab-BIGFIT}.  The predictions of this fit
for observables measured in high-$Q^2$ and low-$Q^2$ reactions are
listed in Tables~\ref{tab-SMIN} and~\ref{tab-SMpred}, respectively.
In Figure~\ref{fig-chiex} the observed value of $\Delta\chi^2 \equiv
\chi^2 - \chi^2_{\mathrm{min}}$ as a function of $\MH$ is plotted for
this fit including all high-$Q^2$ results.  The solid curve is the
result using ZFITTER, and corresponds to the last column of
Table~\ref{tab-BIGFIT}.  The shaded band represents the uncertainty
due to uncalculated higher-order corrections, as estimated by ZFITTER.
Also shown is the result (dashed curve) obtained when using $\dalhad$
of Reference~\cite{bib-Davier-2010, *bib-Davier-2011}.

\begin{table}[p]
\renewcommand{\arraystretch}{1.1}
  \begin{center}
\begin{tabular}{|c||c|c|c|c|c|}
\hline
&     - 1 -              &      - 2 -             &    - 3 -               &     - 4 -             \\
& all Z-pole             & all Z-pole data        & all Z-pole data        & all Z-pole data       \\
& data                   &    plus   $\Mt$        & plus $\MW$, $\GW$      & plus $\Mt,\MW,\GW$    \\
\hline
\hline
$\Mt$\hfill[\GeV] 
& $173^{+13 }_{-10}$     & $173.2^{+0.9}_{-0.9}$  & $178.1^{+10.9}_{-7.8}$  & $173.3^{+0.9}_{-0.9}$  \\
$\MH$\hfill[\GeV] 
& $118^{+203}_{-64}$     & $122^{+59}_{-41}$      & $148^{+237}_{-81}$     & $ 94^{+29}_{-24}$      \\
$\log_{10}(\MH/\GeV)$  
& $2.07^{+0.43}_{-0.34}$ & $2.09^{+0.17}_{-0.18}$ & $2.17^{+0.41}_{-0.35}$ & $1.97^{+0.12}_{-0.13}$ \\
$\alfmz$          
& $0.1190\pm 0.0027$     & $0.1191\pm0.0027$      & $0.1190\pm 0.0028$     & $0.1185\pm 0.0026$     \\
\hline
$\chidf$ ($P$)
& $16.0/10~(9.9\%)$      & $16.0/11~(14\%)$       & $16.5/12~(17\%)$       & $16.9/13~(21\%)$       \\
\hline
\hline
$\swsqeffl$
& $\pz0.23149$           & $\pz0.23149$           & $\pz0.23144$           & $\pz0.23139$ \\[-1mm]
& $\pm0.00016$           & $\pm0.00016$           & $\pm0.00014$           & $\pm0.00011$ \\
$\swsq$     
& $\pz0.22334$           & $\pz0.22332$           & $\pz0.22298$           & $\pz0.22305$ \\[-1mm]
& $\pm0.00062$           & $\pm0.00039$           & $\pm0.00026$           & $\pm0.00023$ \\
$\MW$\hfill[\GeV]
& $80.362\pm0.032$       & $80.363\pm0.020$       & $80.381\pm0.013$       & $80.377\pm0.012$   \\
\hline
\end{tabular}
\end{center}
\caption[Results]{ Results of the fits to: (1) all Z-pole data ({\LEPI} and
  SLD), (2) all Z-pole data plus the direct $\Mt$ determination, (3)
  all Z-pole data plus the direct $\MW$ and $\GW$ determinations, (4)
  all Z-pole data plus the direct $\Mt,\MW,\GW$ determinations (i.e.,
  all high-$Q^2$ results).  As the sensitivity to $\MH$ is
  logarithmic, both $\MH$ as well as $\log_{10}(\MH/\GeV)$ are quoted.
  The bottom part of the table lists derived results for $\swsqeffl$,
  $\swsq$ and $\MW$.  See text for a discussion of theoretical errors
  not included in the errors above.  }
\label{tab-BIGFIT}
\renewcommand{\arraystretch}{1.0}
\end{table}

\begin{table}[p]
\begin{center}
\renewcommand{\arraystretch}{1.10}
\begin{tabular}{|ll||r||r|r|l|}
\hline
 && {Measurement with}  & {Standard-Model} & {Pull}  \\
 && {Total Error}       & {High-$Q^2$ Fit} & {    }  \\
\hline
\hline
&APV~\cite{QWCs:theo:2003:new}
                &                        &                      &       \\
\hline
&$\QWCs$        & $-72.74\pm0.46\pzz\pz$ & $-72.909\pm0.025\pzz$& $0.4$ \\
\hline
\hline
&M{\o}ller~\cite{E158RunI, *E158RunI+II+III}
                &                        &                      &        \\
\hline
&$\swsqMSb$     & $0.2330\pm0.0015\pz$   & $0.23110\pm0.00011$  & $1.3$  \\
\hline
\hline
&$\nu$N~\cite{bib-NuTeV-final, *bib-NuTeV-final-e}
                &                        &                      &        \\
\hline
&$\gnlq^2$      & $0.30005\pm0.00137$    & $0.30397\pm0.00013$  & $2.9$  \\
&$\gnrq^2$      & $0.03076\pm0.00110$    & $0.03011\pm0.00002$  & $0.6$  \\
\hline
\end{tabular}\end{center}
\caption[Low-$Q^2$ measurements]{ Summary of measurements performed in
  low-$Q^2$ reactions: atomic parity violation, $e^-e^-$ M{\o}ller
  scattering and neutrino-nucleon scattering.  The SM results and the
  pulls (difference between measurement and fit in units of the total
  measurement error) are derived from the SM fit including all
  high-$Q^2$ data (Table~\ref{tab-BIGFIT}, column~4) with the Higgs
  mass treated as a free parameter.}
\label{tab-SMpred}
\end{table}

The 95\% one-sided confidence level upper limit on $\MH$ (taking the
band into account) is $152~\GeV$.  When the 95\% C.L. lower limit on
$\MH$ of 114.4~\GeV{} obtained from direct searches at
{\LEPII}~\cite{LEPSMHIGGS} is included, the upper limit increases from
$152~\GeV$ to $171~\GeV$.  

Given the direct measurements of the other four SM input parameters,
each observable is equivalent to a constraint on the mass of the SM
Higgs boson. These constraints are compared in
Figure~\ref{fig-higgs-obs}.  For very low Higgs-masses, the
constraints are qualitative only as the effects of real
Higgs-strahlung, neither included in the experimental analyses nor in
the SM calculations of expectations, may become
sizeable~\cite{Kawamoto:2004pi}.  Besides the measurement of the W
mass, the most sensitive measurements are the asymmetries, \ie,
$\swsqeffl$.  A reduced uncertainty for the value of $\alpha(\MZ^2)$
would therefore result in an improved constraint on $\log\MH$ and thus
$\MH$, as already shown in Figure~\ref{fig-chiex}.

Direct searches for the Higgs boson of the SM are currently performed
at the Tevatron and the LHC.
In summer 2012, the combined Higgs-boson analyses of the Tevatron
experiments CDF and D0 excluded the mass ranges of $100-103~\GeV$ and
$147-180~\GeV$ and reported evidence for a new particle with a
combined significance of about three standard
deviations~\cite{TeVNPHWG:2012-Summer, *Tevatron-Higgs:2012-Summer}.
At the same time, using both 2011 and some 2012 data, the LHC
collaborations ATLAS and CMS excluded the mass regions of
$110-122~\GeV$ and $128-600~\GeV$ and both reported independently the
observation of a new particle in Higgs-boson searches with a
significance of five or more standard
deviations~\cite{ATLAS-HIGGS-2012-Summer, *CMS-HIGGS-2012-Summer}.
The electroweak precision data are well compatible with the hypothesis
that the new particle, observed with a mass in the range of
$125-126~\GeV$, is the Higgs boson of the SM, as is also evident from
Figures~\ref{fig:mtmW} to~\ref{fig-higgs-obs}. If the new particle is
not the Higgs boson of the SM, the results of electroweak fits such as
those presented here may be unreliable because in that case the new
particle is not considered in the calculation of electroweak radiative
corrections.

\begin{figure}[htbp]
\begin{center}
\includegraphics[width=0.9\linewidth]{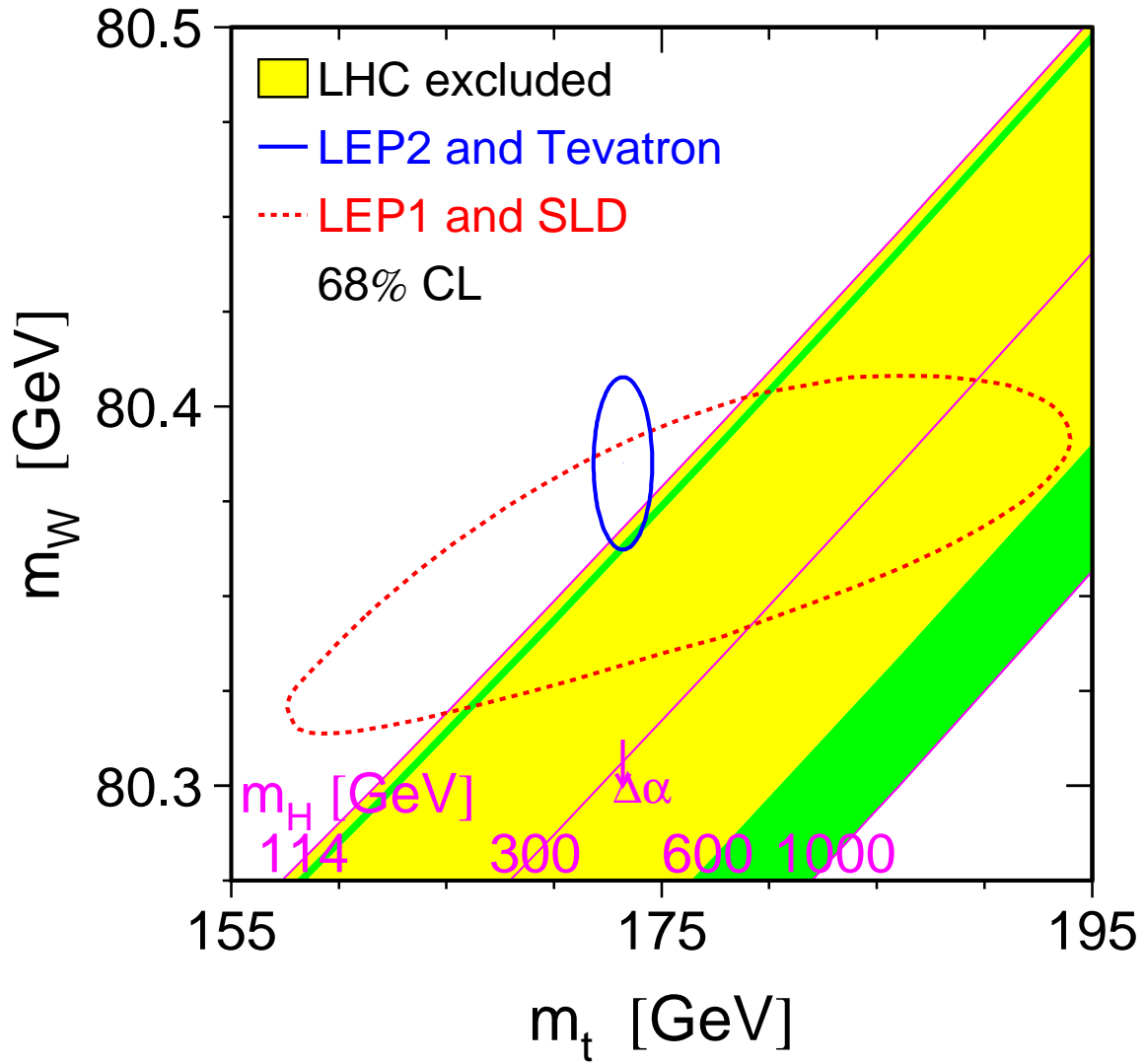}
\caption[Contour]{ The comparison of the indirect constraints on $\MW$
  and $\Mt$ based on $\LEPI$/SLD data (dashed contour) and the direct
  measurements from the $\LEPII$/Tevatron experiments (solid contour).
  In both cases the 68\% CL contours are plotted.  Also shown is the
  SM relationship for the masses as a function of the Higgs mass in
  the region favoured by theory ($<1000~\GeV$) and allowed by direct
  searches (dark green bands).  The arrow labelled $\Delta\alpha$
  shows the variation of this relation if $\alpha(\MZ^2)$ is changed
  by plus/minus one standard deviation. This variation gives an
  additional uncertainty to the SM band shown in the figure.}
\label{fig:mtmW}
\end{center}
\end{figure}
\begin{figure}[htbp]
\begin{center}
\includegraphics[width=0.9\linewidth]{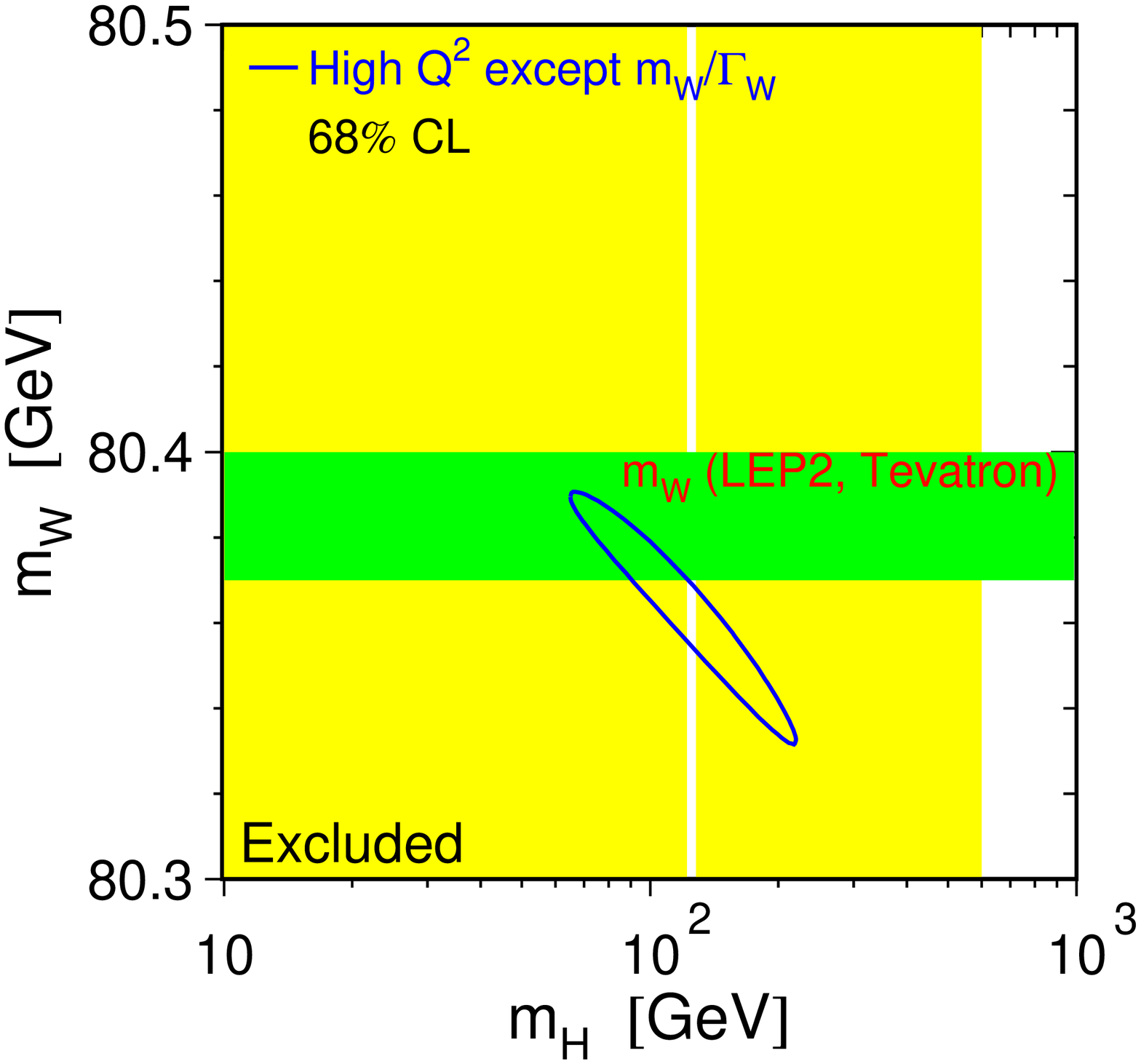}
\end{center}
\vspace*{-0.6cm}
\caption[Contour]{ The 68\% confidence level contour in $\MW$ and
  $\MH$ for the fit to all data except the direct measurement of
  $\MW$, indicated by the shaded horizontal band of $\pm1$ sigma
  width.  The vertical bands show the 95\% CL exclusion ranges on
  $\MH$ from the direct searches.  }
\label{fig-mhmw}
\end{figure}
\begin{figure}[htbp]
\begin{center}
\includegraphics[width=0.9\linewidth]{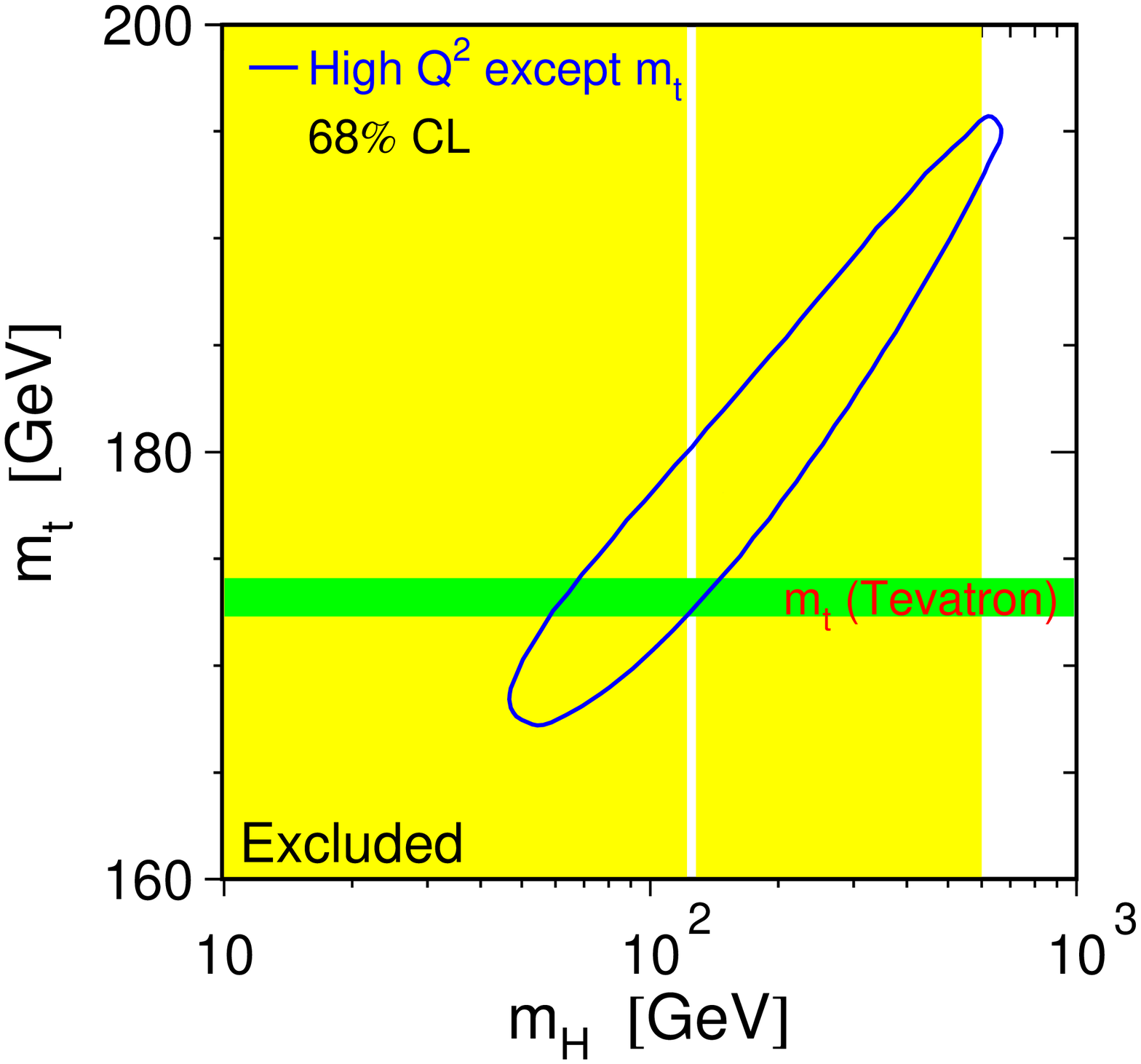}
\end{center}
\vspace*{-0.6cm}
\caption[Contour]{ The 68\% confidence level contour in $\Mt$ and
  $\MH$ for the fit to all data except the direct measurement of
  $\Mt$, indicated by the shaded horizontal band of $\pm1$ sigma
  width.  The vertical bands show the 95\% CL exclusion ranges on
  $\MH$ from the direct searches.  }
\label{fig-mhmt}
\end{figure}
\begin{figure}[htbp]
\begin{center}
\includegraphics[width=0.9\linewidth]{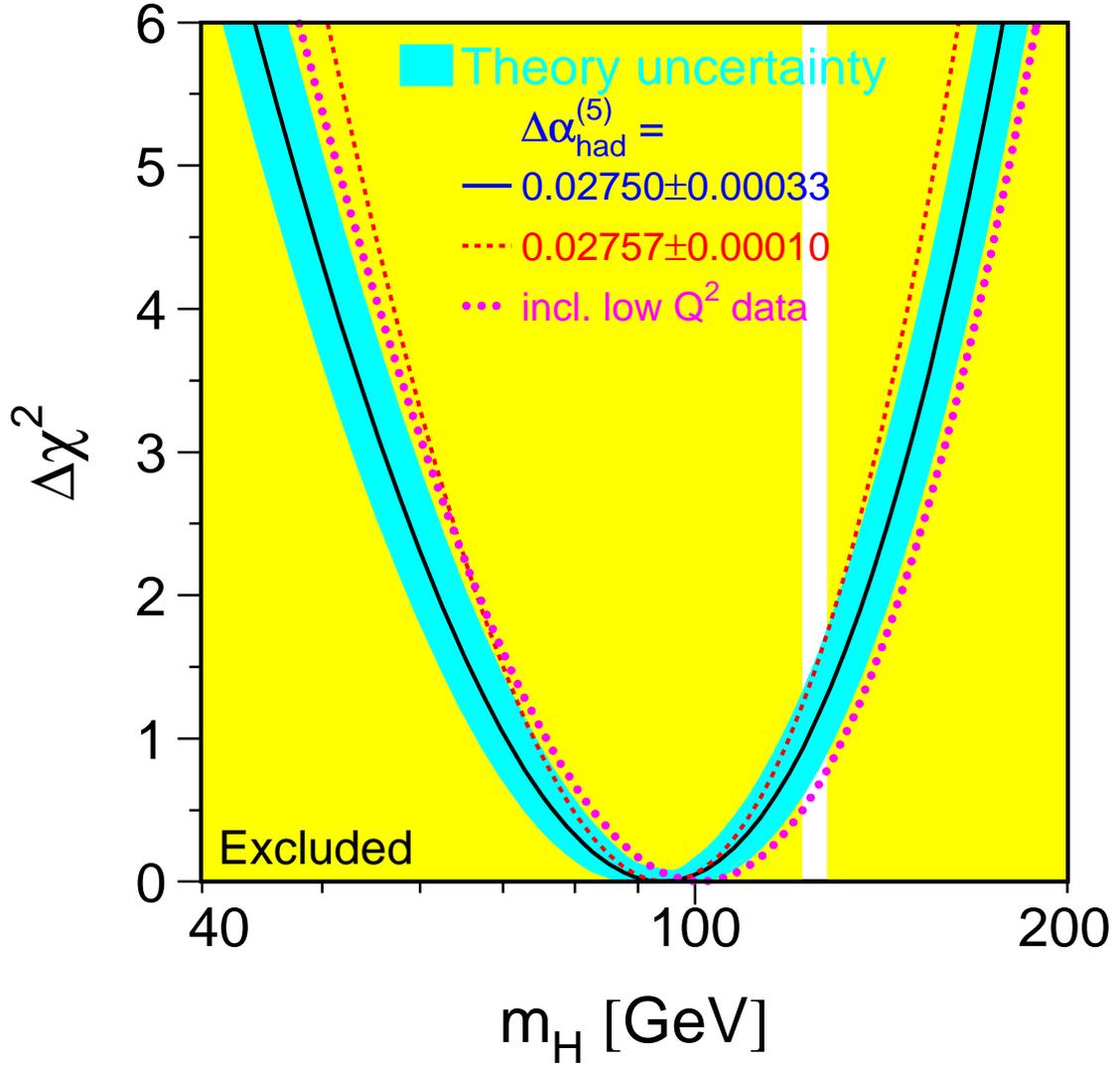}
\end{center}
\vspace*{-0.6cm}
\caption[Higgs blue band]{ $\Delta\chi^{2}=\chi^2-\chi^2_{min}$ {\it
  vs.} $\MH$ curve.  The line is the result of the fit using all
  high-$Q^2$ data (last column of Table~\protect\ref{tab-BIGFIT}); the
  band represents an estimate of the theoretical error due to missing
  higher order corrections.  The vertical bands show the 95\% CL
  exclusion ranges on $\MH$ from the direct searches.  The dashed curve
  is the result obtained using the evaluation of $\dalhad$ from
  Reference~\cite{bib-Davier-2010, *bib-Davier-2011}. The dotted curve
  corresponds to a fit including also the low-$Q^2$ data from
  Table~\ref{tab-SMpred}. }
\label{fig-chiex}
\end{figure}

\begin{figure}[p]
\vspace*{-1.0cm}
\begin{center}
\includegraphics[height=18cm]{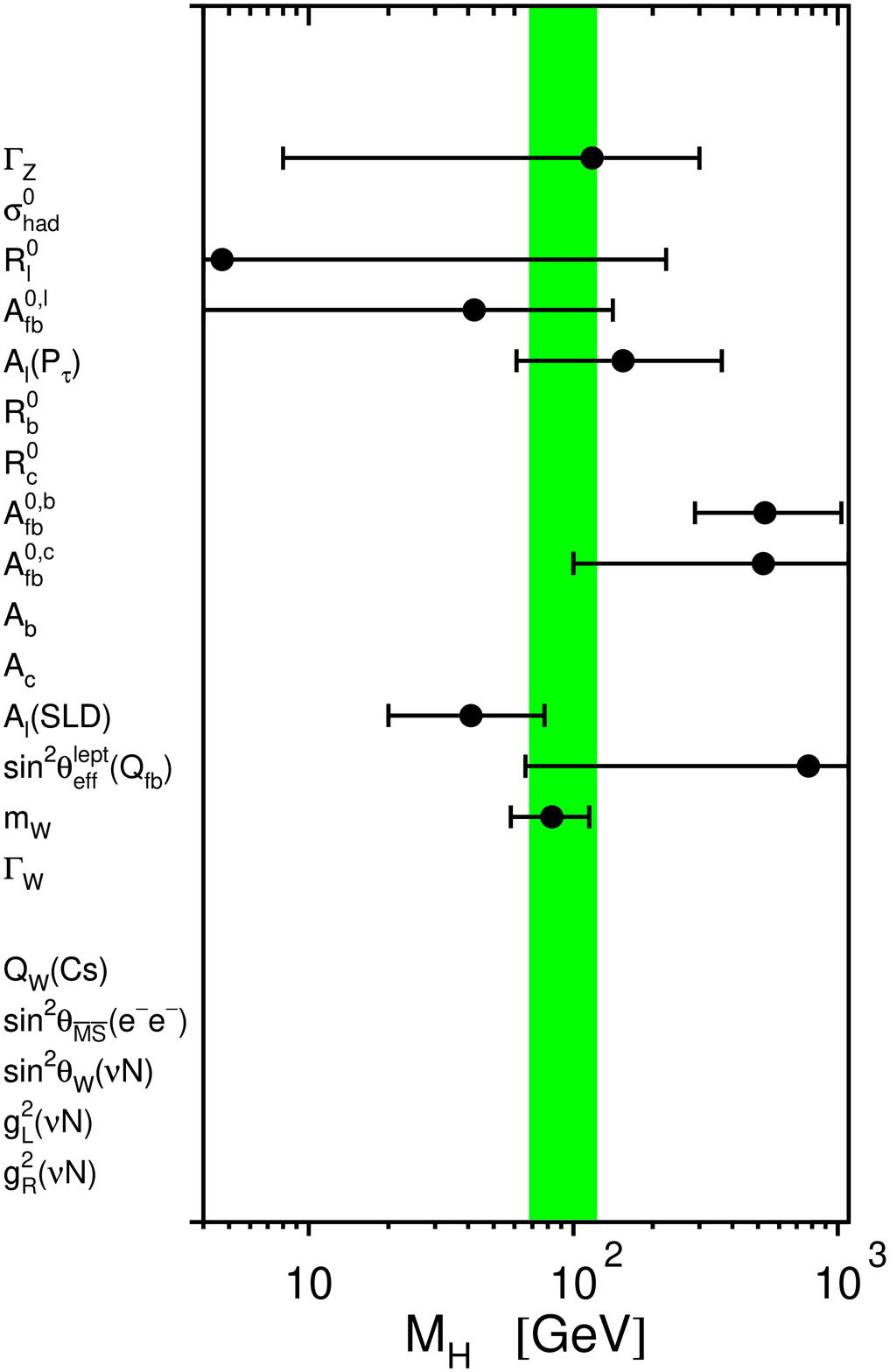}
\end{center}
\vspace*{-0.6cm}
\caption[Higgs Constraints]{ Constraints on the mass of the Higgs
boson from each observable. The Higgs-boson mass and its 68\% CL
uncertainty is obtained from a five-parameter SM fit to the
observable, constraining $\dalhad=0.02750\pm0.00033$,
$\alfmz=0.118\pm0.003$, $\MZ=91.1875\pm0.0021~\GeV$ and
$\Mt=173.2\pm0.9~\GeV$.  Because of these four common constraints the
resulting Higgs-boson mass values are highly correlated.  The shaded
band denotes the overall constraint on the mass of the Higgs boson
derived from all observables including the above four SM parameters as
reported in the last column of Table~\ref{tab-BIGFIT}.  Results are
only shown for observables whose measurement accuracy allows to
constrain the Higgs-boson mass on the scale of the figure.  }
\label{fig-higgs-obs}
\end{figure}

\bibliographystyle{Lep2Rep}

\bibliography{lep2rep}

\begin{mcbibliography}{100}
\bibitem{YFSWWack}
The \YFSWW\ cross-sections at 155--215 GeV have been kindly provided by the
  authors\relax
\relax
\bibitem{RACOONWWack}
The \RacoonWW\ cross-sections at 155--215 GeV have been kindly provided by the
  authors\relax
\relax
\bibitem{KORALWack}
The \KoralW\ cross-sections at 155--215 GeV have been kindly provided by the
  authors. They have actually been computed using \YFSWW~\cite{\YFSWWref},
  switching off non-leading $\oa$ radiative corrections and the screening of
  the Coulomb correction, to reproduce the calculation from \KoralW\relax
\relax
\bibitem{WPHACTack}
The \WPHACT\ cross-sections at 160--210 GeV have been kindly provided by
  A.~Ballestrero\relax
\relax
\bibitem{WTOack}
The \WTO\ cross-sections at 160--210 GeV have been kindly provided by the
  author\relax
\relax
\end{mcbibliography}

\end{document}